# THE INTERNATIONAL LINEAR COLLIDER

Technical Design Report | Volume 4: Detectors

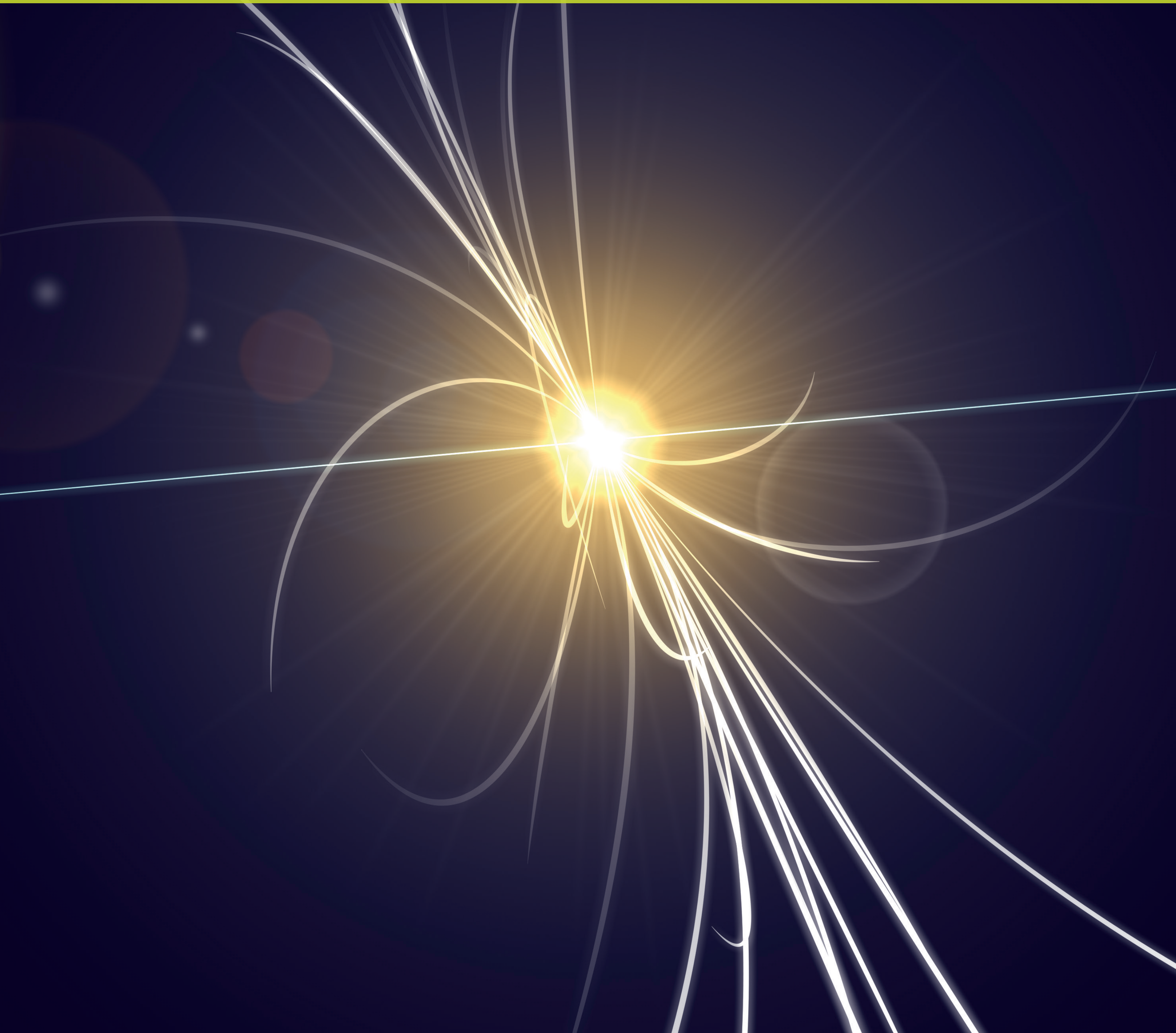

# The International Linear Collider

# Technical Design Report

2013

international linear collider

Volume 1: Executive Summary

Volume 2: Physics

Volume 3: Accelerator

Part I: R&D in the Technical Design Phase

Volume 3: Accelerator

Part II: Baseline Design

Volume 4: Detectors

# Volume 4

# Detectors

Editors

Ties Behnke, James E. Brau, Philip Burrows, Juan Fuster, Michael Peskin,
Marcel Stanitzki, Yasuhiro Sugimoto, Sakue Yamada, Hitoshi Yamamoto

# Acknowledgements

We acknowledge the support of BMWF, Austria; MinObr, Belarus; FNRS and FWO, Belgium; NSERC, Canada; NSFC, China; MPO CR and VSC CR, Czech Republic; Commission of the European Communities; HIP, Finland; IN2P3/CNRS, CEA-DSM/IRFU, France; BMBF, DFG, Helmholtz Association, MPG and AvH Foundation, Germany; DAE and DST, India; ISF, Israel; INFN, Italy; MEXT and JSPS, Japan; CRI(MST) and MOST/KOSEF, Korea; FOM and NWO, The Netherlands; NFR, Norway; MNSW, Poland; ANCS, Romania; MES of Russia and ROSATOM, Russian Federation; MON, Serbia and Montenegro; MSSR, Slovakia; MICINN-MINECO and CPAN, Spain; SRC, Sweden; STFC, United Kingdom; DOE and NSF, United States of America.

# Contents

# Introduction to the Detailed Baseline Design report

For the past five years, the global physics and detector community of linear collider physicists has continued to advance the physics studies and detector developments toward the International Linear Collider (ILC). This report presents the outcome of this recent large and successful phase in four individual parts. In Part I, an outline of the physics motivation is presented first, in Chapter 1, drawn from the detailed physics volume [1] accompanying this volume. With the recent discovery at the LHC of a boson with Higgs Boson-like properties this physics program initially emphasises studies of this particle. The physics summary makes a more general note of the open questions in particle physics and the potential role of the ILC in addressing these questions. Why is Nature asymmetric? What is the nature of dark matter? Why do baryons dominate anti-baryons in the universe? These are the questions that define the frontier of particle physics and the TeV scale holds the potential to provide illumination on each of them, with the Higgs Boson likely playing an essential role.

Following the physics summary, Chapter 1 contains the detector performance requirements and the main challenges for detectors in realising this level of performance. Each detector capability has high performance level goals set by physics needs. The vertex sensors provide excellent flavour tagging. The trackers target precise recoiling mass measurements in the Higgs-strahlung process. Calorimetry must separate di-jet decays of the W and Z. The precision expected of the ILC detectors is unprecedented and specific detector benchmark reactions have been defined to demonstrate this precision can be achieved. Two detectors are planned for the ILC and the motivation for this is explained. With this in mind, two specific detector concepts with complementary designs have been developed and studied. While significant progress has been reached in developing these detector designs and the technical validation through R&D, significant work is still needed to bring the technical designs to a similar state of maturity and construction readiness as the collider.

Following this overview of physics and detectors, Chapter 1 concludes with a description of the process that has guided the global effort through Letters of Intent (LOIs), to the validation of two detector concept groups, leading finally to completion of the Detailed Baseline Designs of SiD and ILD presented later in this report.

The two detector groups have worked together successfully on many common aspects of the ILC. Chapter 2 presents some details of the common efforts, including detector R&D, software tool development and generator sample production, machine detector interface, beam instrumentation, engineering tools, and detector costing.

ILC detector R&D goals have been addressed by many collaborations formed to address the diverse needs. Many of these collaborations contribute to both detector concepts. The vertex detector R&D aims to develop the fine pitched, low mass sensor demanded by the ILC physics goals. Several sensor technologies are under development, applying the monolithic active pixel sensor (MAPS) approach. Fine Pixel Charge Coupled Devices are also being developed. Mechanical design, a critical aspect of achieving the performance goals, is also being pursued. Silicon tracking and Time Projection

Chamber (TPC) R&D have been carried out to support the two complementary approaches for tracking of SiD and ILD. Many options for calorimetry have received R&D effort. Silicon-tungsten as well as scintillator strips with silicon photodiodes are the approaches being developed for the ECAL of the detectors. The Hadron Calorimeter options which have been developed include glass RPCs and Scintillator pads, as well as a number of other technologies that are described. The development of particle flow analysis (PFA) has been a central consideration and a guide for the calorimeter R&D. Forward calorimetry has been a specialised subject with dedicated R&D.

The two detector groups have developed independent software frameworks, but they have cooperated on many tools and projects. For example, the generator samples for physics studies have been produced jointly. A common event data model, LCIO, has been adopted. PandoraPFA and LCFIPlus have been applied by both detector concept groups.

Work on the machine detector interface (MDI) has been an area of close collaboration between the two detector concept groups, as well as with the GDE machine physicists. This includes work on push-pull, detector shielding, installation, and collider hall designs.

An effort has been made to develop and apply common engineering tools, including an engineering data management system. Likewise, common considerations have been made in estimating the costs of each of the detector designs.

Parts II and III present the details of each of the detector concept studies, SiD and ILD. Since submitting the Letters of Intent in 2009, both detectors designs have been updated. Their subsystem technologies have benefited from substantial R&D. Some engineering studies have been possible. The reconstruction software and simulation models have improved and been applied to the specified benchmark reactions.

SiD is a compact, cost-constrained detector made possible by silicon tracking in a 5 Tesla magnetic field produced by a superconducting solenoid. Silicon detectors enable time-stamping on single bunch crossings to provide robust performance. The ILD concept evolved from two similar concepts: GLD and LDC. The ILD design results in a large detector optimised for resolution and track separation, with flexibility for operation at energies up to the TeV range. The ILD tracker is a Time Projection Chamber (TPC) providing continuous tracking for excellent pattern recognition and dE/dx capability. ILD employs a large, 3.5 Tesla superconducting solenoid. Both detector designs employ low-mass high-resolution vertex detectors, highly granular calorimeters optimised for particle flow analysis and operate with a triggerless readout. The designs have been developed in concert with the design of a push-pull system and adequate experimental hall space, as well as a realistic installation scheme. Both are self shielding in order to allow occupancy in the collider hall by one detector group while the other is accumulating collider beam interactions. Results of the simulation studies by each detector concept group of the benchmark reactions are presented in Parts II (SiD) and III (ILD).

Finally, this report ends with a brief concluding statement and a comment on future directions in Part IV. The detectors presented here are mature concepts, backed by detailed R&D studies, with very limited engineering considerations so far. It is time for increased emphasis on engineering, further optimisation, while R&D studies continue to advance the detector technologies.

# Volume 4

## Detectors

# Part I

## Detectors at the ILC:

## Challenges, Coordination and R&D

# Chapter 1
# The physics and detector challenges of the ILC

This initial chapter introduces the background for the ILC physics and detector efforts of the past five years. First, the physics motivation is outlined, highlighting the precision measurements of the Higgs Boson candidate that was recently discovered at the LHC. Next, the detector challenges and performance requirements are described, including machine backgrounds, beam instrumentation, and the motivation for two detectors, as well as the benchmark processes defined to demonstrate detector performance. The chapter ends with a description of the process that guided the ILC physics and detector work to its current state of maturity.

## 1.1 Physics program of the International Linear Collider

In the Physics Volume of this report, we have described the goals of the experimental program of the ILC in full detail. In this section, we review those goals and the experimental program that they call for.

### 1.1.1 Physics goals of the ILC

Among the great mysteries of elementary particle physics, there are three that are likely to be solved by new information from experiments at the TeV energy scale. These concern the three areas in which the Standard Model of particle physics is incomplete as the theory of nature: First, though the Standard Model incorporates a simple phenomenological model of spontaneous symmetry breaking through its Higgs field, the Standard Model gives no understanding of this symmetry breaking. It does not provide a mechanism for the phenomenon or even predict the mass scale at which it occurs. Second, the Standard Model does not provide a particle to describe the "dark matter" that makes up 80% of the mass in the universe. Third, the Standard Model does not provide a mechanism to generate the baryon-antibaryon asymmetry of the universe.

The discovery by the ATLAS and CMS experiments of the "Higgs-like particle" near 125 GeV—and the exclusion of the possibility that the Higgs boson could be at higher mass—gives us a direct path by which experiments can clarify the origin of the symmetry breaking of the electroweak interactions. It has long been appreciated that an electron-positron collider operating in the centre-of-mass energy range of 250 GeV to 1 TeV would be an ideal instrument for the precision study of the Higgs boson. The discovery of the new particle now allows us to map out a specific program of experiments. This program accesses all of the Higgs boson production reactions shown in Figure I-1.1.

The Higgs boson program of the ILC begins at the energy of 250 GeV, near the peak of the cross section for $e^+e^- \to Zh$. The presence of a $Z$ boson at the energy appropriate to recoil tags the Higgs boson events. This allows direct measurement of the Higgs boson branching ratios. The ILC detectors can identify and separate the various predicted Higgs decays, including the two-jet hadronic decays to $b\bar{b}$, $c\bar{c}$, and $gg$. The $Z$ tag also allows the ILC experiments to measure the branching ratio to invisible modes, and also to unexpected models with exotic long-lived particles. Measurement of the peak in the $Z$ recoil energy also gives a precise determination of the Higgs boson mass.

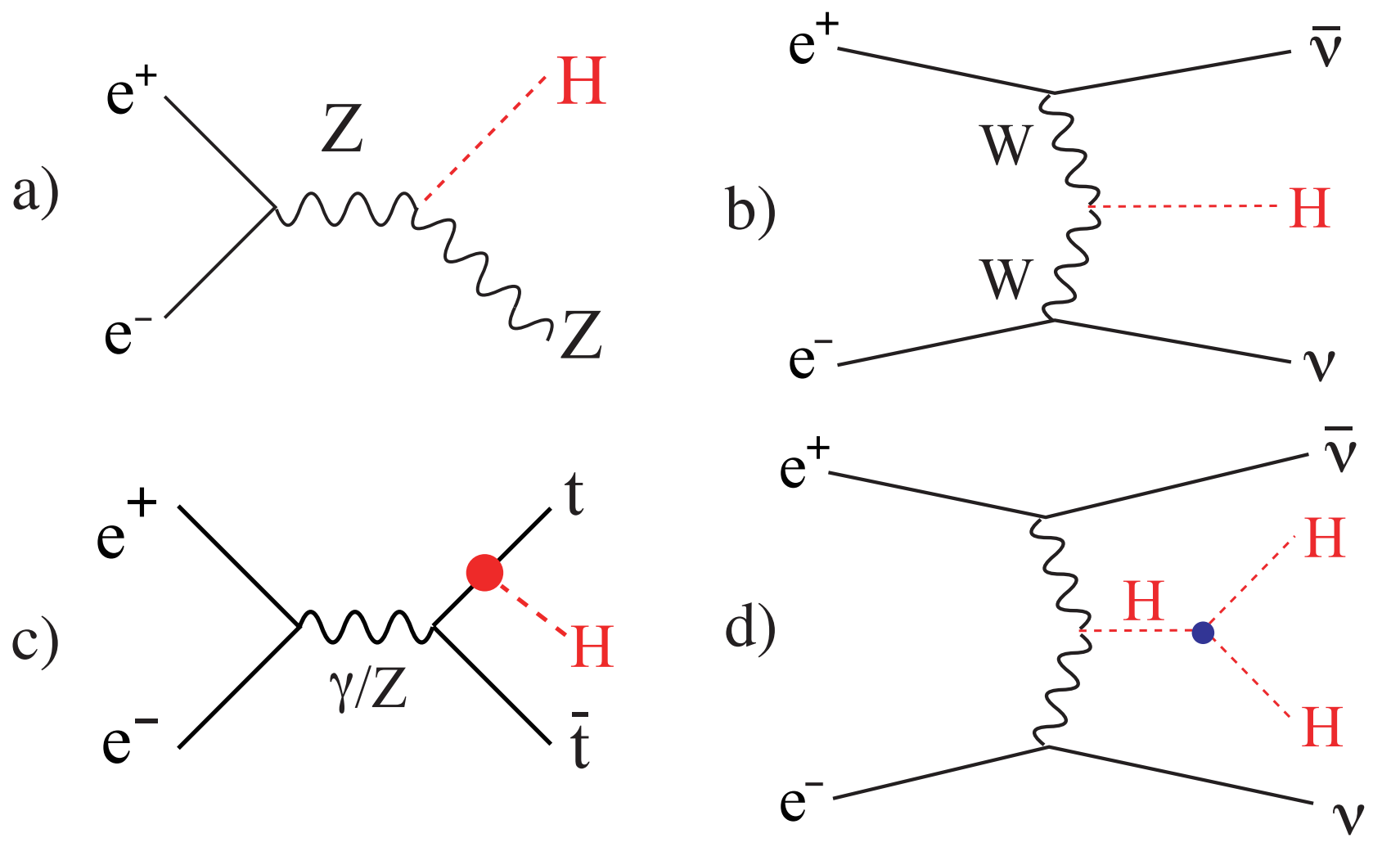

**Figure I-1.1** Representative Feynman diagrams for the major Higgs production processes at the ILC: (a) $e^+e^- \to Zh$; (b) $e^+e^- \to \nu\bar{\nu}h$; (c) $e^+e^- \to t\bar{t}h$; (d) $e^+e^- \to \nu\bar{\nu}hh$.

At higher energy, the $WW$ fusion process of Higgs production, $e^+e^- \to \nu\bar{\nu}h$, turns on. Measurement of this process at the full ILC energy of 500 GeV gives a model-independent precision measurement of the total Higgs boson width. Experiments at 500 GeV also allow first measurements of the Higgs boson coupling to $t\bar{t}$ and of the Higgs boson self-coupling. At a centre-of-mass energy of 1000 GeV, all of the Higgs boson production reactions are fully accessible and the Higgs boson branching ratios can be studied with even higher precision.

A complete review of the Higgs boson program of the ILC, with numerical estimates of the experimental capabilities and comparison to the expectations for the LHC, can be found the Physics Volume of the ILC Technical Design Report [1] (Chapter 2).

Models that repair the incompleteness of the Standard Model and give dynamical explanations for electroweak symmetry breaking necessarily contain additional particles beyond the Higgs boson. These might be the particles of an extended Higgs boson sector, or exotic partners of the quarks, leptons, and gauge bosons. For many of these particles, there are strong arguments that their masses lie in the ILC energy range. New particles beyond the Standard Model have not yet been discovered at the LHC, but there is still great opportunity to discover such particles when the LHC operates at 14 TeV. The discovery of new strongly interacting particles with TeV masses by the LHC could well point to additional new particles with only electroweak interactions that lie in the ILC energy range.

The discussion in the Physics Volume of the ILC Technical Design Report [1] reviews the current picture of new physics models, incorporating what we have learned from the LHC measurements at 7 and 8 TeV, and surveys the opportunities that these models offer for the ILC experiments. For any new particle in the ILC energy range, the ILC provides a rich program to clarify its properties. The ILC experiments will be able to measure the masses with high precision, determine the electroweak quantum numbers and measure any associated mixing angles, and measure the decay branching ratios in a model-independent way.

In models in which the Higgs boson is composite or a part of a complex new sector, the interactions that lead to the light Higgs boson must also leave their imprint on the Standard Model particles, especially on the top quark and the $W$ and $Z$ bosons that couple to it most strongly. The ILC experiments offer powerful capabilities to measure the electroweak couplings of the quarks, leptons, and gauge bosons. The estimates of the precision expected for probes of electroweak couplings and discussions of the importance of these measurements to the more general question of the origin of electroweak symmetry breaking are given in in Ref. [1] (Chapters 3,4 and 5). They will supersede the precision of the existing data and enable the study of new physics at energy scales beyond the centre-of-mass energy of the ILC.

**Table I-1.1**
Major physics processes to be studied at the ILC, together with the lowest centre-of-mass energy at which they can be studied. Shown are the reaction, the process to be studied, and a key indicating which polarisation scenario would bemost advantageous. For more information see the text.

| Energy | Reaction | Physics Goal | Pol. |
|---|---|---|---|
| 91 GeV | $e^+e^- \to Z$ | ultra-precision electroweak | *A* |
| 160 GeV | $e^+e^- \to WW$ | ultra-precision $W$ mass | *H* |
| 250 GeV | $e^+e^- \to Zh$ | precision Higgs couplings | *H* |
| | $e^+e^- \to t\bar{t}$ | top quark mass and couplings | *A* |
| 350–400 GeV | $e^+e^- \to WW$ | precision $W$ couplings | *H* |
| | $e^+e^- \to \nu\bar{\nu}h$ | precision Higgs couplings | *L* |
| | $e^+e^- \to f\bar{f}$ | precision search for $Z'$ | *A* |
| | $e^+e^- \to t\bar{t}h$ | Higgs coupling to top | *H* |
| 500 GeV | $e^+e^- \to Zhh$ | Higgs self-coupling | *H* |
| | $e^+e^- \to \tilde{\chi}\tilde{\chi}$ | search for supersymmetry | *B* |
| | $e^+e^- \to AH, H^+H^-$ | search for extended Higgs states | *B* |
| | $e^+e^- \to \nu\bar{\nu}hh$ | Higgs self-coupling | *L* |
| | $e^+e^- \to \nu\bar{\nu}VV$ | composite Higgs sector | *L* |
| 700–1000 GeV | $e^+e^- \to \nu\bar{\nu}t\bar{t}$ | composite Higgs and top | *L* |
| | $e^+e^- \to \tilde{t}\tilde{t}^*$ | search for supersymmetry | *B* |

Many models of dark matter give as its origin a new stable particle with its mass in the hundred-GeV range. For such models, it would be ideal to collect experimental measurements of the properties of the particle and use these to predict the cosmic density, for comparison to astrophysical observations. A number of examples of models of new physics in which the ILC measurements are sufficiently detailed to make this comparison possible are also shown in Ref. [1] (Chapters 7 and 8).

Models of baryogenesis based on new physics at the TeV scale require new parameters of CP violation in a Higgs boson sector that is necessarily extended beyond that of the Standard Model. Experimental tests of these models require detailed studies of these new Higgs particles. The capabilities of the ILC experiments of carrying out these measurements is described in Ref. [1] (Chapters 6 and 8).

The ILC thus offers a rich experimental program that addresses the most important open issues in elementary particle physics.

### 1.1.2 Energy and beam polarisation of the ILC

The discussion above of the ILC program on the Higgs boson emphasised the ability of the ILC to run at any energy within its range that might give the greatest physics potential. This is a unique advantage of a linear collider. The accelerator can run with only minor modifications at any energy below its design energy, with instantaneous luminosity roughly proportional to the energy. If higher energy is needed, it is only necessary to make the main linac longer. There are limits, of course, but the ILC is designed to run effectively over a very broad range in energy.

The flexibility of the ILC in energy is described in Table I-1.1, which summarizes the most important reactions that will be studied by the ILC experiments at a range of its possible energy settings. The Higgs boson reactions described in the previous section come into play in an orderly way as the energy of the collider is increased.

The last column of Table I-1.1 describes the importance of polarization in the ILC program. Another important advantage of a linear collider is that it preserves the polarization of an electron or positron beam during the acceleration process. Polarization has central importance in electroweak physics above the $Z$. At these high energies at which the ILC operates, it becomes obvious that the left- and right-handed spinning electrons are different elementary particles with distinct electroweak quantum numbers. In particular, the left- and right-handed electrons couple differently to the $SU(2)$ and $U(1)$ components of the Standard Model gauge group, so the different polarized reactions access different slices of the electroweak interaction. This increases the power of the ILC in several different respects.

In explanations to follow, $P(-)$ and $P(+)$ are the polarizations for the $e^-$ and $e^+$ beams, with, for example, $P(-) = -1$ corresponding to 100% left-handed polarization.

*H*- At the minimum, polarization can be used to enhance the event rate. In $e^+e^-$ annihilation, an electron annihilates a positron of the opposite helicity. The ILC offers beam polarization both for electrons and for positrons. Thus, it is possible to tune the electron and positron polarization to be opposite ($e^-_L e^+_R$ or $e^-_R e^+_L$), enhancing the probability of an annihilation. The increase in the effective luminosity is

$$\mathcal{L}/\mathcal{L}_0 = 1 - P(-)P(+) \; , \tag{I-1.1}$$

giving $\mathcal{L}/\mathcal{L}_0 = 1.24$ for ∓80% $e^-$, ±30% $e^+$ polarisation.

*A*- At the $Z$ resonance, in the precision measurement of the electroweak couplings of the top quark, and in precision measurement of $e^+e^- \to f\bar{f}$, the beam polarisation asymmetry is itself an observable containing crucial physics information. The effective polarisation for annihilation reactions is enhanced by the presence of positron polarisation,

$$P_{eff} = \frac{P(-) - P(+)}{1 - P(-)P(+)} \; . \tag{I-1.2}$$

giving $P_{eff} = 89\%$ for ∓80% $e^-$, ±30% $e^+$ polarisation.

*L*- Certain Standard Model processes, especially at high energy, occur dominantly from the $e^-_L e^+_R$ polarisation state. Polarising to this state enhances the rates for such processes by

$$\mathcal{L}/\mathcal{L}_0 = (1 - P(-))(1 + P(+)) \; , \tag{I-1.3}$$

or $\mathcal{L}/\mathcal{L}_0 = 2.34$ for −80% $e^-$, +30% $e^+$ polarisation.

*B*- Conversely, new physics searches at high energy benefit from suppression of the $e^-_L e^+_R$ state to suppress Standard Model backgrounds from $WW$ production and $WW$ fusion processes.

The flexibility of the ILC in the choice of energy and polarisation is exploited in the physics analyses described in the Physics Volume and in the benchmarking analyses presented in this volume. It is a very important advantage of the ILC design that the precise energy and polarisation settings can be chosen year by year in response to ILC discoveries and complementary information from the LHC program.

## 1.2 Experimentation at the ILC

The arguments in the previous sections show that the ILC is highly motivated on theoretical grounds. But there is another justification for the ILC from a different point of view. The ILC experiments will be carried out with the most precise detectors ever built for general studies of particle interactions at high energy. They will give us an unprecedented view of the dynamics of the Standard Model. These capabilities will drive the detailed study of the Higgs boson and of any other particles that appear in the ILC energy range.

The ILC detectors will improve on the detectors built for LEP and for LHC in the precision of their tracking and calorimetry.

The ILC beams provide an environment so benign that it is possible to design detectors with minimal material in the tracking volume. The angular coverage of the tracking will be enhanced compared to the LEP experiments. The calorimetry will make use of the strategy of particle flow to reduce the uncertainty in calorimetric di-jet mass measurements by a factor of two over what has been achieved at LEP and LHC. These improvements are driven by physics requirements, to obtain the Higgs boson mass at the highest precision, and to discriminate the $W$ and $Z$ bosons in the hadronic decays. They will also bring improvements to event reconstruction in many other aspects of QCD and electroweak measurements.

The ILC detectors will feature pixel vertex detectors that give unprecedented capability to tag displaced vertices from $b$, $c$, and $\tau$ decays. At a hadron collider, the large rates of QCD events make it difficult to exclude light quarks without sacrificing tagging efficiency.

Finally, the set of physical observables available in $e^+e^-$ annihilation at high energy is intrinsically richer as cross sections and beam polarisation asymmetries contain independent essential pieces of information on the electroweak couplings of the particles under study. In addition, particles with masses above the $Z$ mass have order-1 spin asymmetries in their weak decays. The full structure of these decays can be studied by the detailed event reconstructions available at the ILC.

All of these capabilities can be brought to bear, in particular, in the precision study of the Higgs boson. We have argued above the that study of this particle will be the next major exploration in elementary particle physics, the most direct route that we have now to answering the great questions of the TeV energy scale. The ILC experiments will reveal the Higgs boson in high-precision, low-background observations that encompass all of the major couplings of this particle. It is these experiments that will truly bring the Higgs boson to light.

## 1.3 Detector challenges and performance requirements

The ILC detectors face new challenges that require significant advances in collider detector performance. The physics goals described in the previous section drive this exceptional performance. The machine environment is benign by LHC standards, enabling designs and technologies that are unthinkable at the LHC. However, the ILC environment poses its own set of background issues that must be overcome. The payoff will be physics studies with unprecedented precision.

The ILC provides a broad spectrum of physics opportunities, which the detector must be prepared to address. These include Higgs Factory, Giga-Z, Top Yukawa couplings, di-boson production, SUSY, and other new physics often motivated by alternative models. Each of these creates its own particular set of requirements. In general, the detectors have been designed to cover the requirements for all such possibilities, over the full range of energy operations. The initial machine is planned for a capability of up to 500 GeV, with energy variability down to 200 GeV, and special running at the Z-pole as well. The upgrade of the energy would bring the operation up to 1 TeV.

The physics opportunities place a premium on high resolution jet energy reconstruction and di-jet mass performance. Consequently, calorimetry must advance beyond current state of the art, and Particle Flow Algorithm (PFA) calorimetry has been developed to meet this challenge. This technique of energy reconstruction makes use of the fact that many of the energy deposits in the calorimeter (on average about 65% of jet energies) are generated by charged tracks, which are very well measured by the tracker. Separation of such deposits from those generated in the calorimeter by neutral particles (photons and neutral hadrons) results in a much better energy measurement of jets. A calorimeter that can isolate and measure separately each individual particle contribution results in an optimal precision when the neutral energy measured in the calorimeter is combined with the charged energy measured in the tracker. The dominant limit comes from confusion within the calorimeter between the individual particle contributions. This motivates the high granularity of the electromagnetic and hadron calorimeters. New detector technologies and new reconstruction algorithms enable the needed precision of 3 to 4 percent for 100 GeV jets, set by the requirement to separate W and Z di-jet final states to be reached.

The requirements on charged track momentum resolution are driven by Higgs studies, particularly through the Higgs-strahlung process, where the recoiling Higgs is reconstructed from the associated Z boson decaying into a lepton pair. In order to realise this extremely high accuracy, the ILC R&D program has been developing high field magnets and high precision/low mass trackers. The requirement set by the recoiling Higgs reconstruction of $\Delta p/p^2$ of $5 \times 10^{-5}$ (GeV/c)$^{-1}$ has been achieved.

Flavour tagging, as well as quark charge tagging, will be available at an unprecedented level of performance as a result of the development of a new generation of vertex detectors. New sensor designs have been developed to address the modest, but challenging, ILC backgrounds. The soft beamstrahlung pairs create high occupancies that demand fast readouts, requiring extra power. These factors must be accommodated with very low mass detectors and supports. This is a significant challenge. However, the low duty cycle of the ILC permits power pulsing, which reduces the heat load and the need for cooling.

Muon systems are required for identification, as the inner tracker provides adequate tracking precision for muon momentum measurements. The iron flux return for the detector magnetic field supplies the material needed to identify muons, and also leads to a self-shielded detector.

Significant soft $e^+e^-$ pairs are produced at the interaction point from the beam collision induced beamstrahlung (see machine backgrounds below); the interaction region layout has been designed to guide these charged background particles out of the detector. Optimally, the direction of the magnetic field along the beamline must be directed parallel to the outgoing beam, which passes through the detector off-axis to the main solenoid field direction. This optimal configuration can be achieved through the superposition of the conventional solenoidal field from the detector with a dipole field, produced by adding some dedicated dipole windings to the detector solenoid. Such a so-called Detector Integrated Dipole (DID) becomes effective once the crossing-angle increases beyond a few mrad. For historical reasons, this configuration with the field aligned with the outgoing beam is called anti-DID.

The very forward calorimetry must contend with high backgrounds primarily from the soft $e^+e^-$ pairs that are guided through the detector. These high radiation loads with bunch crossings every few hundred nanoseconds complicate the very forward calorimeter designs. The high energy singly-produced electrons and positrons buried in the large number of pairs must be detected.

Table I-1.2 summarises several selected benchmark physics processes and fundamental measurements that make particular demands on one subsystem or another, and set the primary requirements for detector performance.

Time stamping to an individual bunch train is important for reducing the overlap of events. Two-photon events contribute a particularly strong source of such backgrounds, increasing with centre-of-mass energy. The ILC time structure with its fraction of a per cent duty cycle (1 millisecond bunch trains at 5 Hz) makes power pulsing a possible and desirable feature for many of the detector subsystems, significantly reducing heat load. Nevertheless, powering the readout electronics of each subsystem, such as the front-end readout chip of the silicon tracker, is a challenge. The readout chips require high current at low voltage with large conductor mass. In order to reduce this mass, power delivery based on serial power or capacitive DC-DC conversion is being studied. In addition, the pulsed power system must deliver quiescent currents.

**Table I-1.2.** Detector performance needed for key ILC physics measurements.

| Physics Process | Measured Quantity | Critical System | Physical Magnitude | Required Performance |
|---|---|---|---|---|
| $Zhh$<br>$Zh \to q\bar{q}b\bar{b}$<br>$Zh \to ZWW^*$<br>$\nu\bar{\nu}W^+W^-$ | Triple Higgs coupling<br>Higgs mass<br>$B(h \to WW^*)$<br>$\sigma(e^+e^- \to \nu\bar{\nu}W^+W^-)$ | Tracker and Calorimeter | Jet Energy Resolution<br>$\Delta E/E$ | 3% to 4% |
| $Zh \to \ell^+\ell^- X$<br>$\mu^+\mu^-(\gamma)$<br>$Zh + h\nu\bar{\nu} \to \mu^+\mu^- X$ | Higgs recoil mass<br>Luminosity weighted $E_{\mathrm{cm}}$<br>BR($h \to \mu^+\mu^-$) | $\mu$ detector<br>Tracker | Charged particle Momentum Resolution<br>$\Delta p_t/p_t^2$ | $5 \times 10^{-5}(GeV/c)^{-1}$ |
| $Zh, h \to b\bar{b}, c\bar{c}, b\bar{b}, gg$ | Higgs branching fractions<br>b-quark charge asymmetry | Vertex | Impact parameter | $5\mu m \oplus$<br>$10\mu m/p(\mathrm{GeV/c})\sin^{3/2}\theta$ |
| SUSY, eg. $\tilde{\mu}$ decay | $\tilde{\mu}$ mass | Tracker<br>Calorimeter<br>$\mu$ detector | Momentum Resolution<br>Hermeticity | |

## 1.3.1 Machine backgrounds

While benign by LHC standards, for optimal performance of the detectors backgrounds must be carefully studied. A variety of processes create beam induced backgrounds in the detectors [202]. The main sources are:

**Beamstrahlung**

The passage of the two tightly focused beams near each other results in

- disrupted beam;
- photons, radiated into a very narrow cone in the forward direction, where those striking components result in significant backgrounds;
- electron-positron pairs, radiated into the forward direction and steered by the collective field of the opposing beam and the central magnetic field of the detector solenoid.

**Synchrotron Radiation**

The non-Gaussian tail of each beam passing through, but off axis, the final focusing elements of the beam delivery system near the interaction point generates synchrotron radiation entering the detector.

**Muons**

The non-Gaussian tail of each beam interacts with collimators defining the aperture of the beam line, generating muons, which are transported through the tunnel to the detector.

**Neutrons**

Interactions producing neutrons may arise from:

- Beamstrahlung induced $e^+e^-$ pairs which strike beam line components;
- Disrupted beam particles which strike beam line components;
- Backscatter of neutrons from primary beams and beamstrahlung which strike beam dumps.

**Hadrons and muon**

Electron - Positron pairs created by $\gamma\gamma$ interactions.

Each source has its own character.

**Pair Background**

Large numbers of $e^+e^-$ pairs created at the interaction point primarily follow the outgoing beams, with the detector solenoid controlling their motion. Some are produced with large enough transverse momenta to enter detector components. Also, the pairs create secondary particles by interacting with detector or collider components. These secondary particles can enter the detector and are another important background source.

**Photon Background**

The beam-beam interaction at the IP also produces a large number of photons, mostly radiated in the forward direction. While carrying a large amount of energy, they mostly leave with the outgoing beam. However, like the pairs, some generate secondary particles when interacting with forward components, and represent another important source of background.

**Synchrotron Radiation Background**

Synchrotron radiation photons produced in wakefield-induced beam scattering in the upstream machine elements represent another potential source. An optimised collimation system can control this source.

**Beam Halo Muon Background**

Muons are produced upstream of the detector when the beam halo interacts with collimators, generating high energy electromagnetic showers. Many muons can be created, and are then relatively easily transported to the detector, generating spurious horizontal tracks.

**Summary of Backgrounds**

The background sources have been investigated in various studies. For example, the beam-beam interaction and pair generation, radiative Bhabhas, disrupted beams and beamstrahlung photons for the 500 GeV ILC were studied with GUINEAPIG [333]. Also, the $\gamma\gamma$ hadronic cross section was approximated in the Peskin-Barklow scheme [2]. Based on these studies densities of particles which will reach the different sun-detectors have been estimated. Table I-1.3 summarises these estimates.

**Table I-1.3** Background sources for the nominal 500 GeV beam parameters.

| Source | #particles per bunch | $< \mathrm{E} >$ (GeV) |
|---|---|---|
| Disrupted primary beam | $2 \times 10^{10}$ | 244 |
| Bremstrahlung photons | $2.5 \times 10^{10}$ | 244 |
| $e^+e^-$ pairs from beam-beam interactions | 75k | 2.5 |
| Radiative Bhabhas | 320k | 195 |
| $\gamma\gamma \to$ hadrons/muons | 0.5 events/1.3 events | – |

## 1.3.2 Beam Instrumentation

Precise knowledge of beam parameters is critical to the ILC physics program. Luminosity, beam energy, and polarisation are measured by instrumentation close to the main detectors, which are described in more detail in Chapter 2.

*Luminosity measurement:* Accurate knowledge of the luminosity is required, both the energy-integrated luminosity, as well as the luminosity as a function of energy, dL/dE. Low-angle Bhabha scattering detected by dedicated calorimeters can provide the necessary precision for the integrated luminosity. Acollinearity and energy measurements of Bhabha events in the polar angle region from 120-400 mrad can be used to extract dL/dE.

*Beam energy measurement:* Beam energy measurements with an accuracy of (100-200) parts per million are needed for the determination of particle masses, including $m_{top}$ and $m_{Higgs}$. Energy measurements both upstream and downstream of the collision point are foreseen by two different techniques to provide redundancy and reliability of the results.

*Polarisation measurements:.* Precise measurements of parity-violating asymmetries require polarisation measurements with a precision of 0.25% or better. High statistics Giga-Z running requires polarimetry at the 0.1% level. The primary polarisation measurement comes from dedicated Compton polarimeters detecting the backscattered electrons and positrons. The best accuracy is achieved by implementing polarimeters both upstream and downstream of the IR.

### 1.3.3 Two detectors

The scientific productivity of collider facilities, such as the Tevatron, LEP, HERA, and the LHC, has benefited from independent operation of multiple experiments. This leads to operation of detectors with complementary strengths, cross-checking and confirmation of results, reliability, insurance against mishaps, competition between collaborations, as well as increased number of involved scientific personnel, all contributing to enhanced scientific success. Such complementary efforts benefit from independent software systems and differing philosophies and methodologies. There are numerous historical examples where this complementarity was essential, such as the inability of UA2 to confirm the mistaken claim for top quark discovery by UA1. Therefore, the ILC has been designed to enable two experimental detectors to share one interaction region.

Through the process described elsewhere in this report, two detector designs have been developed with complementary features. These detectors are described in detail in the following section, where first SiD is described followed by ILD. Both experiments are designed to achieve the ILC precision measurements and searches for new physics.

SiD is a compact, cost-constrained detector made possible with a 5 Tesla magnetic field and silicon tracking. Silicon enables time-stamping on single bunch crossings to provide robust performance. The highly granular calorimeter is optimised for particle flow analysis.

The ILD design results in a large detector optimised for good energy and momentum resolution, with flexibility for operation at energies up to the TeV range. It employs a highly granular calorimeter, with minimal material between the interaction point and the calorimeter. The tracker is a Time Projection Chamber (TPC) providing continuous tracking for excellent pattern recognition and dE/dx capability.

Each detector can be alternately be moved to the beam line to operate with collisions by way of the push-pull system. The push-pull system is designed with the detectors and associated infrastructure arranged to enable quick movement of each detector into and out of the interaction region. The details of this have been studied, and resulted in specific layouts. The alignment of detectors must be re-established after each movement, and procedures are being developed for this.

Figure I-1.2 shows the display of two simulated events showing two different configurations as they will be seen by the two detectors.

## 1.4 Physics benchmarks studies

Benchmark reaction processes have been defined for the detector groups to assess the performance level of each detector. A set was first produced for the detector Letters of Intent (LOIs) [3], and later a few additional processes were added for the Detailed Baseline Design efforts. Benchmark studies can demonstrate the performance of the overall detector concept, as well as quantify the physics reach of the experiment.

The LOI benchmarks [3] were designed to measure and demonstrate the performance of the overall detector concept at energies up to 500 GeV. The benchmarks specified for the DBD were defined to assess the detector performance up to 1 TeV.

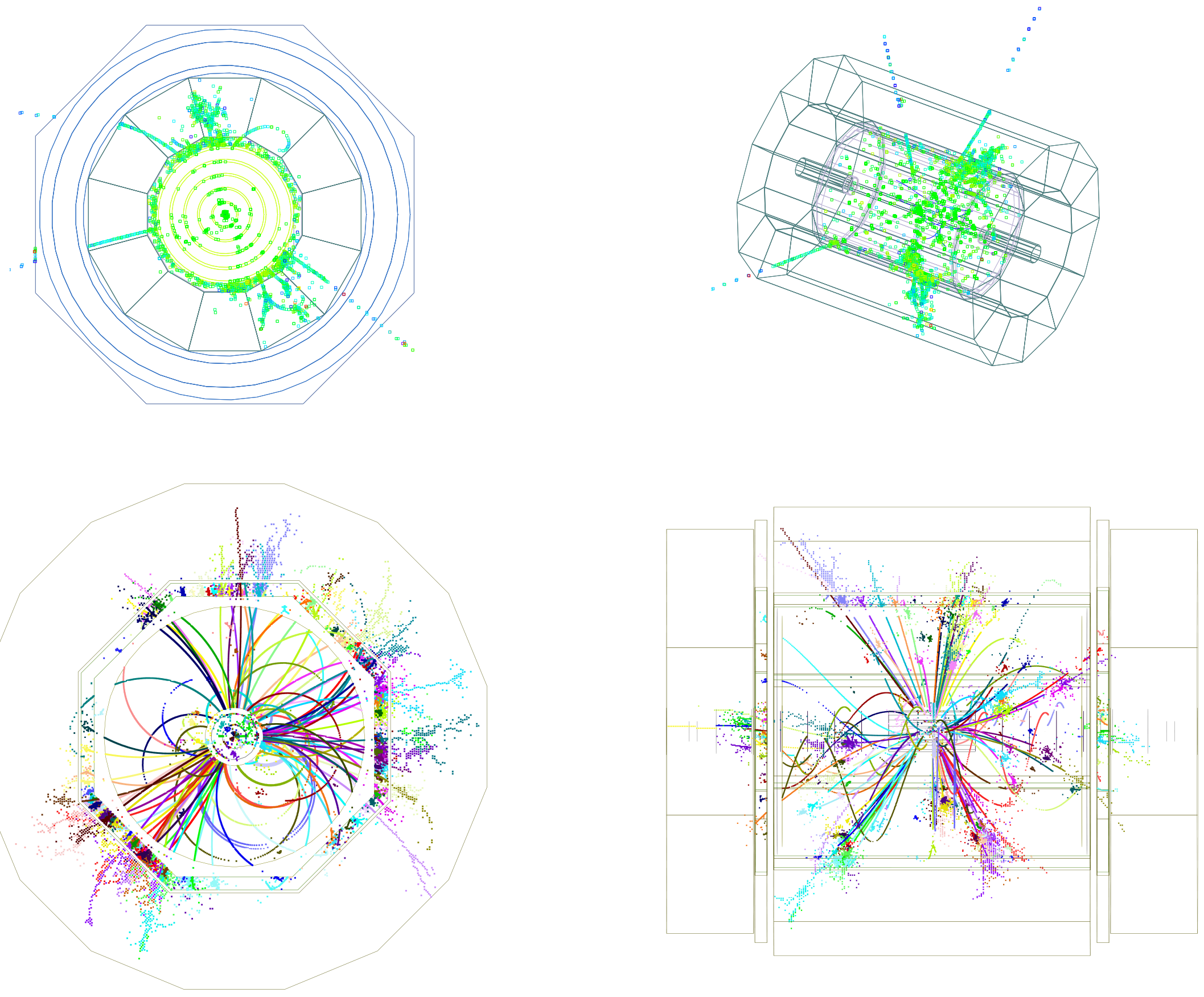

**Figure I-1.2.** Two event displays for: (up) $e^+e^- \to Zh; Z \to \mu^+\mu^-; h \to b\bar{b}$ at 250 GeV from SiD and (down) $e^+e^- \to t\bar{t}h; t\bar{t} \to 6q; h \to b\bar{b}$ from ILD at 1 TeV.

### 1.4.1 Definition of the first set of benchmark processes (250, 500 GeV) for the LOI

The first set of benchmark reactions was defined for the LOI [3]. For each reaction, the performance for both 250 fb$^{-1}$ for $\sqrt{s} = 250$ GeV and 500 fb$^{-1}$ for 500 GeV was requested.

1. $e^+e^- \to Zh$, $Z \to e^+e^-$ or $\mu^+\mu^-, h \to X$ ($M_h = 120$ GeV, $\sqrt{s} = 250$ GeV), measuring the Higgs mass and the cross section. These processes test: a.- momentum resolution, b.- material distribution in the detector, in particular in the tracker, and c.- photon ID;

2. $e^+e^- \to Zh$, $h \to c\bar{c}$ and $\mu^+\mu^-$, $Z \to \nu\bar{\nu}$ ($M_h = 120$ GeV, $\sqrt{s} = 250$ GeV), measuring the BR($h \to c\bar{c}$) and the BR($h \to \mu^+\mu^-$). These final states check: a.- heavy flavour tagging, secondary vertex reconstruction, b.- multi jet final state, c-tagging in jets, uds anti-tagging (particle ID), and c.- anti-tagging by studying the $h \to gg$ channel;

3. $e^+e^- \to Zh$, $h \to c\bar{c}$, $Z \to q\bar{q}$ ($M_h = 120$ GeV, $\sqrt{s} = 250$ GeV), measuring BR($h \to c\bar{c}$). In addition to the charm tagging, this final state tests the confusion resolution capability;

4. $e^+e^- \to \tau^+\tau^-$ ($\sqrt{s} = 500$ GeV), measuring efficiency and purity, as well as cross section, forward-backward asymmetry (A$_{FB}$) and P$_\tau$ ($\tau$ polarisation). These channels test: a.- tau reconstruction, aspects of particle flow, b.- $\pi^0$ reconstruction, and c.- tracking of very close-by tracks;

5. $e^+e^- \to t\bar{t}$, $t \to bW^+$, $W^+ \to q\overline{q'}$ ($M_{top} = 175$ GeV, $\sqrt{s} = 500$ GeV), measuring cross section, forward-backward asymmetry (A$_{FB}$), and m$_{top}$. This tests the following: a.- multi jet final states, dense jet environment, b.- particle flow, c.- b-tagging inside a jet, d.- lepton tagging in hadronic events (b-ID), and e.- tracking in a high multiplicity environment;

6. $e^+e^- \rightarrow \chi^+\chi^-$ / $\chi_0^2\ \chi_0^2$ ($\sqrt{s} = 500$ GeV), for SUSY parameter point 5 of Table 1 of [4], measuring cross section and masses.These channels test the following: a.- particle flow (WW, ZZ separation), and b.- multi-jet final states.

The above reactions represent a minimum number of physics processes that were studied for the LOI and in fact they are far from all physics studies envisioned but are representative tests of the detector capabilities. The next reactions are of very high importance for the physics reach of the ILC project. However they were considered less relevant to the optimisation of the detector parameters, or had overlap with other reactions included in the list above. These are:

1. $e^+e^- \rightarrow Zhh$, While this reaction is very challenging for the particle flow performance, it has a very small cross-section, and as such not well suited for a detector study or optimisation. However, it is a very important physics goal;

2. secondary vertex reconstruction and quark charge measurement. This reaction is very important for the optimisation of the vertex detector. It relies on very sophisticated vertexing tools to be fully implemented;

3. low mass difference between SUSY states (low ΔM SUSY). This tests the detector in the very forward direction, including the electron veto capability in the extreme forward region.

### 1.4.2 Definition of the second set of benchmark processes (1 TeV) for the DBD

A supplementary set of processes was defined for the DBD. Motivated by the important open questions regarding the scaling of the detector performance to the higher energy of 1 TeV, a few reactions were chosen based on their usefulness in studying this.

These processes were to be carried out with event samples of 1 ab$^{-1}$. The electron and positron polarisation was assumed to be consistent with the estimate from the GDE for 1 TeV, close to 80% and 20%, respectively The sample should be equally divided between the configurations (-/+) and (+/-) [5].

1. $e^+e^- \rightarrow \nu\bar{\nu}h$ at $\sqrt{s} = 1$ TeV, where $h$ is a Standard Model Higgs boson of mass 125 GeV, in the final states $h \rightarrow \mu^+\mu^-$, $b\bar{b}$, $c\bar{c}$, gg, WW*. The goal is to measure the cross section times branching ratio for these reactions;

2. $e^+e^- \rightarrow W^+W^-$ at $\sqrt{s} = 1$ TeV, considering both hadronic and leptonic (e, $\mu$) decays of the W. The goal is to use the value of the forward W pair production cross section to measure in situ the effective left-handed polarisation $(1 - P_{e^-})(1 + P_{e^+})/4$ for each of two polarisation configurations;

3. $e^+e^- \rightarrow t\bar{t}h$ at $\sqrt{s} = 1$ TeV, where $h$ is a Standard Model Higgs boson of mass 125 GeV, in the final state $h \rightarrow b\bar{b}$. The reaction involves final states with eight jets and final states with six jets, one lepton, and missing energy. The goal is to measure the top Yukawa coupling.

## 1.5 The Physics and Detector Study of the International Linear Collider

The physics and detector studies matured from a Letter-of-Intent (LOI) process started in 2007. As the plan to develop a technical design for the ILC unfolded, the ILC Steering Committee (ILCSC) [6] recognised the importance of defining detailed detector concepts so that they could be considered in the design of the ILC, addressing issues of the interaction region and demonstrating the feasibility and the capability of pursuing physics at ILC, and they initiated the LOI process. The framework and various milestones of the process are described briefly in this introduction. More details can be found in the Interim Report [7].

### 1.5.1 Call for LOIs

In October 2007, the ILCSC announced a call for Letters of Intent to produce reference designs for two detectors for the ILC [8]. When the GDE published the ILC Reference Design Report in summer 2007, there were four detector concepts in its detector volume. The call for LOIs was intended to lead the community to form two capable groups that would develop their concepts to a technically advanced stage and produce detailed baseline designs at the same time as the GDE's accelerator Technical Design Report. The submitted LOIs were reviewed by an advisory body called the International Detector Advisory Group (IDAG). In order to conduct the LOI procedure, the ILCSC appointed Sakue Yamada as Research Director, who was to set up a management structure and to recruit the IDAG members.

### 1.5.2 The management formation

With consultation and agreement of the ILCSC and the steering body of each region, the Research Director invited the three co-chairs of the World Wide Study (WWS) [9, 10] to join the management team as the regional contacts. This management structure ensured good communication with the detector community during the tenure of the Research Director, a period of continued R&D and preparation for realization of the ILC.

Jim Brau from North America, Francois Richard from Europe and Hitoshi Yamamoto from Asia joined the management by January 2008. Later, from February 2011, Juan Fuster took over the role of the European regional contact.

The IDAG was formed with the approval of ILCSC as listed in Table I-1.4.

**Table I-1.4** Members of the international detector advisory group, IDAG

| | | |
|---|---|---|
| **Exp. & Det.** | Michael Danilov | ITEP |
| | Michel Davier (Chair) | LAL/Paris Sud |
| | Paul Grannis | Stony Brook |
| | Dan Green | FNAL |
| | Dean Karlen | Victoria |
| | Sun-Kee Kim | SNU |
| | Tomio Kobayashi | Tokyo |
| | Weiguo Li | IHEP |
| | Richard Nickerson | Oxford |
| | Sandro Palestini | CERN |
| **Phenomenology** | Rohini Godbole | IIS |
| | Christophe Grojean | CEA-Irfu/CERN |
| | JoAnne Hewett | SLAC |
| **Accelerator** | Eckhard Elsen | DESY |
| | Tom Himel | SLAC |
| | Nobu Toge | KEK |

### 1.5.3 Organisation of detector activity

The organisation of the detector activity is illustrated in Figure I-1.3. It has a decision making body, various working bodies, links to the outer community for necessary cooperation and a central management. Much of the detector R&D and physics simulation was carried within the concept groups, which cooperate with various R&D collaborations. Each group designed a detector system following its concept and carried out the R&D work outlined in its LOI. In this process it was thought that cooperation among the separate concept groups would be important.

There were certain tasks like the push-pull studies that could be pursued only with close cooperation among the concept groups and with the accelerator's beam delivery system team. Also, sharing of commonly needed tasks was encouraged in order to optimise the outcomes. In order to facilitate such cooperation and communication among the concept groups, there were common task groups (CTG) consisting of members from all the concept groups and, where necessary, members

**Figure I-1.3**
ILC detector management organisation

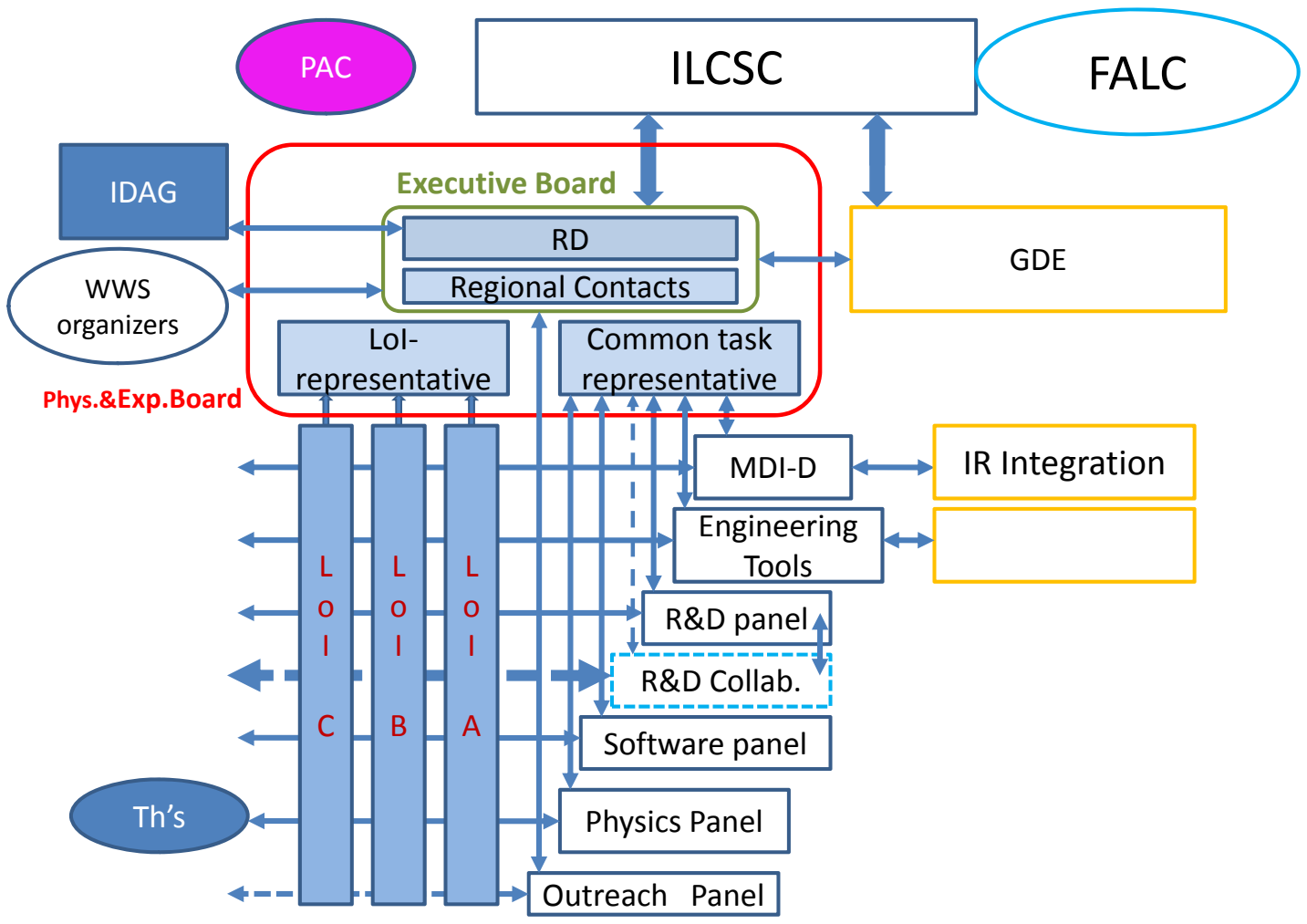

from the wider detector R&D and/or theory communities.

Three LOI groups were identified by a call for Expression of Interest (EOI) announced in February 2008. They were ILD (International Large Detector), SiD (Silicon Detector) and Fourth. The Physics and Experiment Board (PEB) is the decision-making body, which started out with representatives of the LOI groups and the management members.

The five common task groups were set up by end of May 2008; the machine-detector interface (MDI) group, engineering tools group, detector R&D group, software group and physics group. The MDI group established a link to the beam delivery system group of the GDE and had to start immediately. For the detector R&D common task group, several horizontal R&D collaborations, such as calorimetry or tracking, were asked to send representatives to maintain good communication. The physics common task group included additional theory members. The PEB became filled by Summer 2008 with the conveners of the common task groups.

## 1.5.4 The LOIs and their validation

The Letters of Intent were submitted from the ILD, SiD and Fourth groups by the due date, March 2009 [11, 12, 13]. They were examined by the IDAG in detail. Following multiple interviews with the concept groups IDAG made its recommendations for validation in August 2009. The conclusion that ILD and SiD should be validated was presented by the RD to the ILCSC and was approved. During the ALCPG workshop in Albuquerque, USA, in September 2009, the IDAG examination and validation were reported by the chair [14].

The validation was a clearly defined milestone and the validated groups were authorised to proceeed toward the final goal, the completion of the detailed base line designs of their detector concepts. The organisation of the detector activity was modified accordingly.

IDAG monitored further the activities of the validated groups and also the activity of the common task groups. ILCSC extended the mandate of the IDAG, keeping the same membership, through to the end of the DBD process.

### 1.5.5 Work for the detailed baseline designs

The validated groups prepared detailed plans with milestones for key items to reach the final goal. In order to guide this, the RD directorate and the IDAG agreed on nine crucial items to be included in the planning such as; completion of R&D for critical detector components for their feasibility proof, defining a detailed baseline design of the detector system, completing basic integration of the baseline design accounting for insensitive zones, setting up a realistic model of the detector for physics simulation, completion of studies of a push-pull scheme and integration into the interaction region, making physics simulations for the new set of benchmark reactions at 1 TeV, and improving the cost estimate.

The plan of each group was produced by October 2011 with a caveat that it was made under the assumption that necessary resources would become available in due course. It was agreed that all the efforts be made in order to satisfy the minimum requirement for each item. This was achieved as is described in this report. An indispensable element for the success of the project is the close cooperation between the two groups under the consensus that limited resources need to be shared efficiently wherever possible. Also, in some area of the studies cooperation with the CLIC detector group was helpful.

IDAG monitored the progress regularly twice a year, during the LC workshops and gave timely suggestions for improvement. At the meeting in April 2012, IDAG reviewed the outlined contents of DBD prior to drafting, and finally it reviewed the entire draft in October, 2012.

### 1.5.6 The common task groups

The common task groups aimed to solve issues which are common to the both concept groups. Each of the five common task groups had specific roles. The members were provided from the both concept groups and, where appropriate, also from the wider communities or detector R&D collaborations. Each CTG was represented in the Physics and Experiment Board. They made substantial achievements as described in the next chapter. Here a brief introduction is given for each common task group.

Machine Detector Interface group This group worked on all the matters related to the interplay of the detector with the accelerator. It had a close link to the accelerator colleagues working on the beam delivery system and the experimental hall design. One of its important objectives was a push-pull mechanism for the two detectors which allows quick and stable operation both for the detectors and the collider. The group contributed much in the discussions with the accelerator team on the design of the beam parameters, as well.

The members are: Karsten Buesser (convener), Phillip Burrows (deputy), Alain Hervé, Thomas Markiewicz, Marco Oriunno and Toshiaki Tauchi.

Engineering Tool Group This group worked to find a common engineering design tool, which could used by the two groups, and also with the accelerator team. Once a scheme was defined, the group tried to make it familiar to the both groups. This effort was very effective though given the limited resources available.

The members are: Catherine Clerc (convener), Alain Herve, Kurt Krempetz, Thomas Markiewicz, Marco Oriunno (deputy), and Hiroshi Yamaoka.

Detector R&D group The group facilitated various ways to promote detector R&D studies. It made a contact point to the various horizontal R&D collaborations, which are independent bodies but play crucial roles for developing the ILC detector components. In order to keep good communication with these collaborations, major R&D collaborations sent representatives to this group. One of the outcomes was a survey of the spin-off technologies that emerged from the ILC R&D activities.

The members are:Dhiman Chakraborty, Marcel Demarteau (convener), John Hauptman, Ronald

Lipton, Wolfgang Lohmann (deputy), Burkhard Schmidt, Aurore Savoy-Navarro, Felix Sefkow, Tohru Takeshita, Jan Timmermans, Andrew White, Marc Winter.

Software Group This group worked to prepare necessary simulation tools and to produce common event samples for the benchmark studies. Due to the limited human resources, sharing tasks for the new benchmark simulation was crucial. The group coordinated this in an effective way. It also had close contact with the CLIC simulation team.

The members are: Frank Gaede, Norman Graf (deputy), Tony Johnson, Akiya Miyamoto (convener).

Physics Group This group began its work to understand the physics issues associated with the case for the ILC and the priorities of its experimental program. The group was made of representatives of the concept groups, plus a number of interested theorists. The group also carried out various physics studied requested by the management. This group tracked the progress of the LHC experiments to investigate their implications to the ILC program. Most recently this group organised the writing of the Physics volume of this report.

The members are: Tim Barklow, Stewart Boogert, Seong Youl Choi, Klaus Desch, Keisuke Fujii(deputy), Youannning Gao, Heather Logan, Klaus Moenig, Andrei Nomerotski, Michael Peskin(convener), Aurore Savoy-Navarro, Georg Weiglein(deputy), Jae Yu.

### 1.5.7 Other working groups

In addition to the original five common task groups, more working groups were created subsequently to solve specific tasks in a relatively short period.

A typical case is the SB2009 working group, which was organised soon after the ALCPG workshop, 2009 in Albuquerque, to study the effects of the proposed SB2009 accelerator parameters on experiments. The members include representatives of the concept groups, related Common Task Groups (MDI, Software and Physics), and some theorists, and was convened by James Brau of the management. This group made a contact point with the accelerator team to communicate systematically and organise necessary works like simulations using the given beam parameters. It continued further to discuss with the accelerator group on the beam parameters for the simulation of benchmarks at 1 TeV. Details of its activity are described in the Interim Report[7].

There was a working group for the new benchmarks to be added for the DBD. It was led by the physics common task group and worked with the representatives of the two groups and the software common task group. It made a report on the list of priority- and work-sharing for each possible physics channel as introduced in the previous section.

We had a common costing working group to coordinate costing methodology between the two groups. The group also learned from the accelerator costing how to handle the inflation or the changes of currency exchange rates. More details of the group are presented in the next chapter.

The CLIC-ILC joint working group was initiated in early 2010 following the discussions of the ILCSC. It surveyed ongoing cooperation and looked for further synergies between the two linear collider detector activities. Before this working group was formed, there had already been much grass root cooperation since 2008. The cooperation has become more intensive since CLIC deployed the two ILC concepts for its detectors. There was an overlap of the members who prepare both CLIC Conceptual Design Report and the ILC DBD. More description can be found in the Interim Report [7].

## 1.6 The World Wide Study

The World Wide Study of Physics and Detectors for Future Linear $e^+e^-$ Colliders (WWS) was organised in 1998 to give voice to the community of physicists interested in the realisation of the linear collider. The WWS OC (organising committee) is a broadly representative formal committee selected by each region from the WWS members. Since its formation in 1998, the WWS, mostly represented by the WWS OC, initially served a number of roles:

- voice the views of the global linear collider physics community,
- formulate the physics justification for the linear collider and promote its case in the broader physics community,
- coordinate the work of the three regional linear collider studies,
- organise the program for the series of linear collider workshops (LCWS),
- serve as a physics and detectors subcommittee of the ILCSC.

The WWS has remained independent from any other organisation, with no official role in any project based organisation of the linear collider effort (GDE, Research Directorate, CLIC study group etc.). When the ILC Research Directorate was created in 2007 under the leadership of Sakue Yamada, many of the WWS ILC specific studies were integrated into the ILC Research Directorate and the role of the WWS narrowed somewhat. Nevertheless, many of the roles outlined above remained for the WWS.

The WWS represents a broad community of physicists interested in the physics and detectors of a linear collider. All linear collider options are addressed and promoted; the WWS provides a forum for the open comparison among possible directions. In particular, it provides the principal forum in which theorists and experimentalists discuss and elaborate the linear collider. It engages and motivates theorists to do studies critical to developing and explaining the scientific case for the linear collider. Additionally, the WWS gives a voice to the diverse set of universities and research institutes that must be mobilised in support of the linear collider. It also connects to the broader particle physics community that is not currently directly involved in the linear collider activities. Finally, it speaks with an independent voice, on issues of physics requirements, organisation, and other relevant issues, always with positive motivation.

The WWS has been a valuable resource to the overall effort, being drawn on when needed to provide additional services beyond those explicitly outlined above, such as providing the membership for the parameters committee that established the requirements for the ILC design and initiating the development of detector concepts which paved the way to the development of detector LoIs and detailed baseline designs presented in this report.

# Chapter 2
# Description of Common Tasks and Common Issues

In this chapter the common tasks and issues of the two detector concepts are presented. They are the outputs of common efforts between the two concept groups with a wider community as well as with the accelerator team. The joint approach was necessary and effective to share loads under limited resources for various common goals and also to contact the BDS and CFS teams of the accelerator group in a well coordinated manner. Those works were carried out mostly through the common task groups, which were organised for this purpose from the very beginning. For the detector R&D programs various cooperative relations were formed with many R&D collaborations. While each detector group collaborated with R&D collaborations depending on its detector component, there was also regular contact with major R&D collaborations through the detector R&D common task group in order to facilitate better communication. All these efforts were indispensable for the presented accomplishment of the two detector groups.

While each group will describe specific items relevant to the concept in its chapter, the most common items or very similar items are described in this section. Covered topics are common detector R&D, common simulation and software tools, machine-detector interface issues including the push-pull scheme, common engineering tools, beam instrumentation for the energy and polarisation measurements and detector costing methodology.

## 2.1 ILC Detector Research and Development

Because of the well-defined initial state, an electron-positron collider offers the possibility to carry out measurements with an unprecedented level of precision. To realise this exceedingly high level of measurement accuracy, stringent requirements are placed on the performance of the detectors. In some cases, this calls for innovative detector designs employing new detector technologies. Novel analysis techniques are also proposed. For example, the particle flow reconstruction introduced at LEP [15] has been developed to new levels of performance by addressing the challenges posed for the detectors. The operational conditions of the machine allow for a unique way to operate the detectors which, if demonstrated to be feasible, would provide significant advantages. This chapter first introduces some concepts common to both detectors. This is followed by a description of the research and development projects that are necessitated by the physics and as such are shared by the proposed experiments. Detector concept specific R&D is discussed in the respective chapters of the SiD and ILD sections in this report. It should be noted that the potential merits of this research and development work reaches far beyond the ILC for most efforts.

### 2.1.1 Overview of the overall detector strategies

Many of the interesting physics processes at the ILC will be characterised by multi-jet final states. The reconstruction of the invariant masses of two or more jets will provide a powerful tool for event reconstruction and event identification. The physics at the ILC requires a clean separation of the hadronic decays of the W- and Z-boson. An invariant mass resolution comparable to the gauge boson widths, i.e. $\sigma_m/m = 2.7\% \approx \Gamma_{\mathrm{W}}/\mathrm{m}_{\mathrm{W}} \approx \Gamma_{\mathrm{Z}}/\mathrm{m}_{\mathrm{Z}}$, leads to an effective separation of better than 3 $\sigma$ in the mass peaks of the hadronic decays of the vector bosons. To achieve this unprecedented mass resolution, the ILC detectors have adopted the particle flow approach by combining calorimetry and tracking. In contrast to a purely calorimetric measurement, particle flow calorimetry requires the reconstruction of the four-vectors of all visible particles in an event. The jet energy reconstruction then proceeds as follows. First, the energy deposits in the calorimeter are identified. The momenta of all charged particles are then measured in the tracking detectors. The energy deposits in the calorimeter are then associated with the charged particle tracks. To reconstruct the jet energy, the measurement of the momentum in the tracker is used rather than the measurement of the energy in the calorimeter since the tracker has superior resolution and the energy deposits in the calorimeter associated with the tracks are removed. Only the energy measurements for photons and neutral hadrons are obtained from the calorimeters. In this manner, the poor hadronic energy resolution affects only about 10% of the energy in the jet and a jet energy resolution of about $0.19/\sqrt{\mathrm{E(GeV)}}$ would be ideally obtained. In practice, this level of performance cannot be achieved as it is impossible to perfectly associate all energy deposits with the correct particles. For example, if part of a charged hadron shower is identified as a separate cluster, the energy is associated with a neutron and effectively double-counted as it is already accounted for by the track momentum. Similarly, when a photon and hadronic particle are close together, hits originating from one could be assigned to the other and not be accounted for. These effects are called "confusion" and represent the limiting factor in particle flow calorimetry. The concept of particle flow calorimetry relies crucially on the ability to correctly assign calorimeter energy deposits to the correct reconstructed particle. This in turn drives the calorimeters to be highly granular, both longitudinally as well as transversely, a trademark of the ILC calorimeters. For ILC detectors, jet energy reconstruction has evolved into a very complex pattern recognition problem.

Those requirements have spurred the development of new technologies for calorimetry. Figure I-2.1 shows the "technology tree" of all technologies being pursued by the CALICE LC calorimetry collaboration for both electromagnetic as well as hadronic calorimeters. As an absorber medium, both tungsten and iron are being studied with both analogue and digital readout. A plethora of active media are being studied ranging from scintillators with novel SiPM readout, silicon based monolithic active pixel sensors (MAPS) to gaseous detectors such as Resistive Plate Chambers (RPCs) and Micromegas. Granularity of the readout can range from 3x3 cm$^2$ pixel size to 50x50 $\mu$m$^2$ for the MAPS-based electromagnetic calorimeter.

ILC detectors must have complementary properties operating as a single unit therefore an excellent tracking performance goes hand-in-hand with the concept of particle flow calorimetry. The tracking technologies being considered are either silicon- or gaseous-based. For the former, silicon strip technology is being considered but also highly pixelated silicon sensors are an option. A Time Projection Chamber (TPC) is the detector being considered for gaseous based tracking in which ionisation generated by a traversing charged particle will drift towards the endplate where the signal is amplified and processed. The Micromegas, GEM and CMOS technology are being studied within the Linear Collider TPC collaboration.

An important physics goal at the ILC is the identification of hadronic jets originating from heavy quarks. This is best achieved by a topological reconstruction of the displaced vertex structure and the kinematics associated with their decays. The ability to reconstruct the sequence of primary, secondary and tertiary vertices depends on the impact parameter resolution and the track reconstruction capability. These in turn are related to the single point resolution of the sensors, the location of the first measurement, and the overall occupancy in the detector. To meet the ILC requirements this leads to fine pitch, low-mass pixel vertex detectors as close to the interaction point as possible. The material budget desired is about 0.1% $X_0$ per layer for the vertex detector and less than 1% $X_0$ per layer for a silicon tracker. For a TPC the material budget is accumulated in the end plates and a material budget of 30% $X_0$ per endplate is the goal. The ILC concept detectors have not decided on a baseline technology for the vertex detector and all technology options are considered common and are described in more detail in the next section.

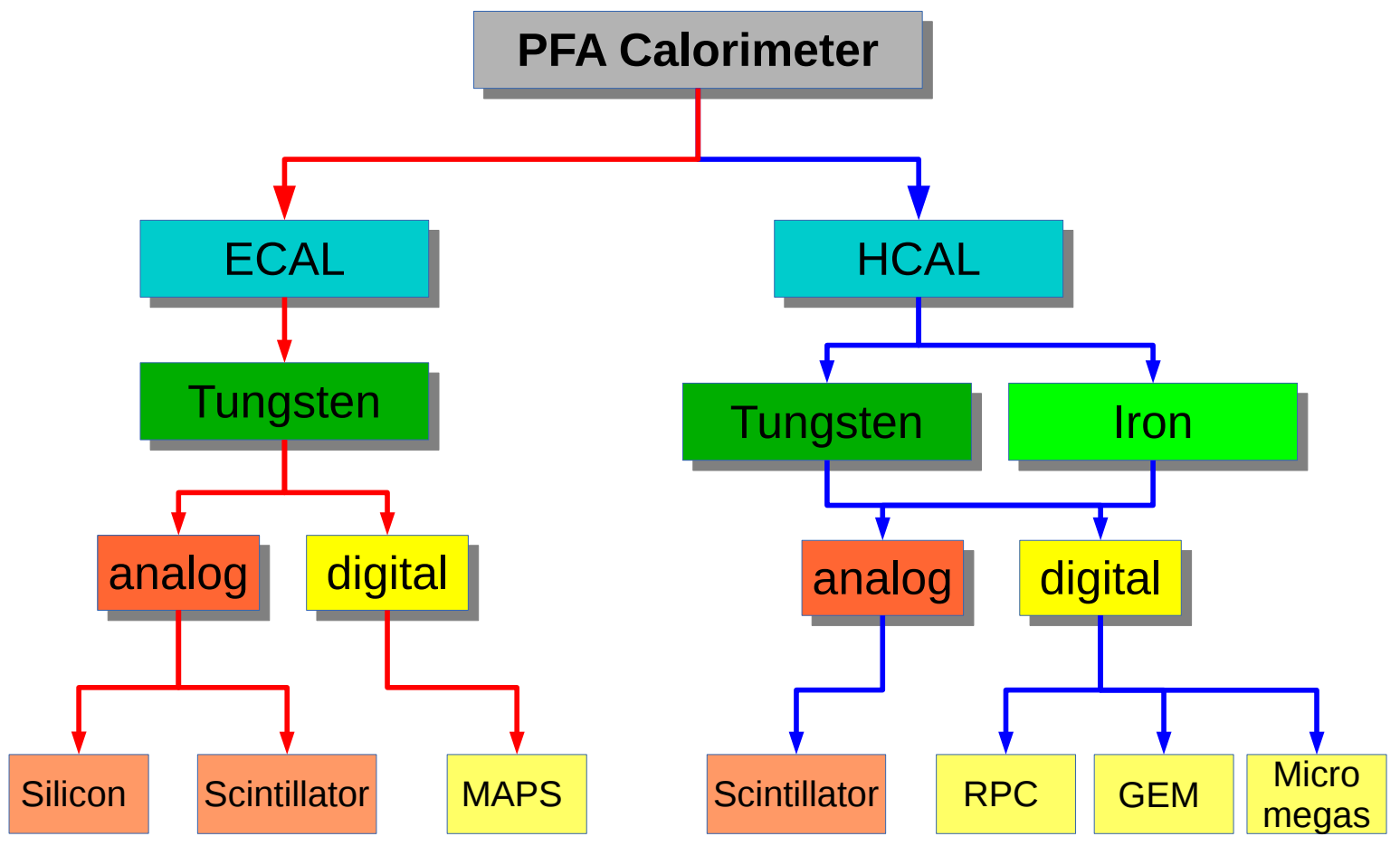

**Figure I-2.1**
Tree of electromagnetic calorimeter technologies -ECAL- (left) and hadron calorimeter technologies -HCAL- (right).

To achieve an ultra low-mass detector configuration, a unique feature of the ILC machine is exploited. The ILC time structure results in collisions of bunches at the interaction point every 308 ns. Bunch trains consisting of 2820 bunches in each beam pass through the interaction point five times per second. Consequently, the bunch trains are about one millisecond long, separated by about 199 milliseconds. The detector can thus be put in a quiescent state for 199 out of 200 milliseconds at the machine repetition rate of 5 Hz, since there are no interactions during this period. This is referred to as “power-pulsing”. Allowing for transient times, to turn the detector “on and off” and starve the system of power, power-pulsing could lead to an overall reduction in power consumption of nearly a factor of hundred. This feature is employed by both detector concepts. One of the most significant benefits of power-pulsing is that the vertex and tracking detector does not need active cooling. This significantly lowers the overall mass budget for these detectors, which is crucial for obtaining the required resolutions. It is expected that a heat load of about 20W for the barrel vertex detector can be removed using forced convection with dry air.

The feasibility of power-pulsing has already been demonstrated for several technologies. System tests at full magnetic field strengths and an evaluation of the mechanical stability under pulsed power still need to be performed. Test are being planned by both detector concepts of detectors with balanced load lines in high magnetic fields to measure the detector alignment stability.

### 2.1.2 Vertex Detector Technologies

Within the ILC many different pixel sensor technologies are being studied. Integration of the sensing node with front-end signal processing circuitry in a single unit is a key characteristic shared by many efforts. The aim is to go to very small pixel size for superior impact parameter resolution and minimisation of pattern recognition ambiguities, ultra-thin detectors to minimise the material budget and low power to eliminate the need for active cooling. The Monolithic Active Pixel Sensor (MAPS) technology implements a high density matrix of pixels with signal processing circuitry on the same substrate as the sensitive volume. The DEPFET technology implements a single active element within the active pixel. The CMOS Pixel Sensor (CPS) technology is able to integrate a full front-end signal processing chain with correlated double sampling in each pixel. In the 3D and Silicon-On-Insulator (SOI) technology the sensing and processing functionalities are separated in different layers. In the SOI technology, the sensing is provided by a high-resistive substrate connected through an insulating layer with the processing layer. In the 3D technology each function is fully separated in different wafers, processed independently, which are then bonded together. All of these technologies have the capability of delivering sensors less than 75 $\mu$m thick with better than 5 $\mu$m hit resolution and low power consumption.

Fine Pixel Charged Coupled Devices (FPCCDs) allow for particularly small pixels, $5\times5$ $\mu\text{m}^2$, which results in a sub-micron spatial resolution and excellent two-track separation capability. The fine segmentation also mitigates occupancy issues and thus allows for integration over numerous bunch-crossings. The CCDs have to be operated at cryogenic temperatures and the detector would need to be installed in a light foam cryostat. Prototype devices have been made and sub-micron position resolution has been achieved. The FPCCD instantaneous power consumption is moderate and a slow signal processing in between consecutive bunch trains can be envisaged.

The DEPFET concept integrates a p-MOS transistor in each pixel on the fully depleted, detector-grade bulk silicon [16]. Electrons, produced in the bulk through ionising radiation are collected in the internal gate and modulate the transistor current. Their low input capacitance ensures low noise operation and makes them suitable for a broad range of applications from collider detectors to X-ray astronomy. Even though the first formulation of the concept goes back to the 1980s, it took nearly two decades before progress in silicon processing capabilities at HEP institutes and industry allowed it to be possible to turn the idea into a complete detector concept. A DEPFET sensor generally is a $700\times250$ matrix of $50\times50$ $\mu\text{m}^2$ pixels. Auxiliary readout is integrated on the edge of the sensor, which allows thinning to 50 $\mu$m in the sensitive region, using an etching technique, retaining a frame which ensures the stiffness of the mechanical module. The sensors are read out in "rolling shutter"mode with a frame rate of 50 kHz. Many prototype devices were produced and tested to demonstrate the viability of the concept and to establish DEPFETs as a mature technology for vertex detectors [17].

A considerable effort is going into the development of MAPS detectors, notably the CMOS Pixel Sensors (CPS). Matrices with $1152\times576$ pixels, with 18.4 $\mu$m pitch with a column parallel read-out architecture with amplification and correlated double sampling inside each pixel have been demonstrated [18]. A spatial resolution approaching 3 $\mu$m has been achieved with binary charge encoding. The current architectures are being extended to address the critical ILC specifications. The Chronopixel sensor is a monolithic CMOS pixel sensor, which has the capability to record the time of each hit with sufficient precision to assign each hit to a specific bunch crossing of the collider. This reduces the backgrounds due to integration over many beam crossings to a virtually negligible level. Two prototype devices were built to date with a pixel size of $50\times50$ $\mu\text{m}^2$ in a 180 nm process. The design for the final device calls for a pixel size of $15\times15$ to $20\times20$ $\mu\text{m}^2$, which requires a 45 nm technology which is impractical for prototypes. Progress with the prototype devices is very good and indicate good signal to noise performance, reasonable power consumption, good circuit flexibility and adequate charge collection in the epitaxial layer in the presence of sophisticated front-end electronics.

In microelectronics, 3D technology refers to the stacking of multiple thin layers of circuitry with vertical interconnections between them. This area is developing rapidly as a way of increasing circuit density without the major retooling and investment needed for smaller feature size.The enabling technologies for 3D are wafer thinning, aligning and bonding, and the formation of Through-Silicon Vias (TSVs). Although the increased circuit density provided by multi-layer circuits is in itself an important application for High Energy Physics, it is the increased range of processing and interconnection options that the technology provides that has the most potential. Using these technologies arrays of chips can be bonded to sensors and electronics to form essentially monolithic arrays of sensors with no dead space between chips and with interconnections taken from the back rather than the edge of the IC. The Vertically Integrated Pixel (VIP) chip was conceived as a demonstration readout chip for the ILC vertex detector. The first version of the chip (VIP2a) was a three-tier device produced at MIT Lincoln Laboratory using a fully depleted SOI process [19]. A major success was that all of the interconnections between the circuit layers worked. A second iteration, the VIP2b was fabricated in the 3D process developed by Tezzaron/Global Foundries. This process uses a bulk 130 nm CMOS process with modifications to allow the top copper metal layer to be used for face-to-face wafer bonding, and to include vias that extend 6 $\mu$m into the bulk material. Initial testing of the 2D parts show excellent analog performance. Tests of the full functionality of the 3D chips are underway.

Active edge sensors are an outgrowth of work done to develop 3D sensors, which provide good charge collection combined with radiation hardness and yields sensors that are active over its full area. Combination of active edge technology with 3D integration can provide a technique for tiling complex shaped areas with fully active sensor arrays with no dead regions.

The development of vertex detector technologies within the framework of the ILC is a poster child for technology spin-off to other areas of the field and synergy with other scientific disciplines [20]. The DEPFET technology has been adopted by the BELLE-II collaboration at the KEK b-factory as the baseline design for their vertex detector; the MAPS sensors are developed for the STAR experiment and the next generation of sensors are being considered for the ALICE Inner Tracker System upgrade. The 3D technology is a candidate technology for the LHC tracker upgrades. Because of the extensive R&D carried out under the ILC umbrella, these projects had a mature technology available to be taken to the detector stage. Other applications of the DEPFET concept are X-ray imaging in space experiments and at the XFEL X-ray light source at DESY.

Most of the technologies discussed are making excellent progress toward the development of a high-performing pixel detector for the ILC. There are remaining challenges to be met in the areas of material budget, power and pixel size, but steady progress is being made. Moreover, these efforts benefit greatly from the fast-paced developments in the semi-conductor industry.

### 2.1.3 Tracking Detector

The two ILC concept detectors have a complementary approach to tracking. The ILD detector employs a hybrid tracking system consisting of a large-volume gaseous TPC tracking detector surrounded by silicon tracking layers. The SiD detector is based on silicon technology only. Since silicon is the only technology common to both detectors, some key common development aspects for silicon will be described here. The R&D on the TPC is fully described in the ILD section.

Tracking detectors have grown tremendously in size over the last generations of experiments. As observed with the evolution of vertex detector technologies, also here there is a trend toward integration. Both experiments propose hybrid-less silicon sensors. These sensors have integrated pitch adapters to route traces to a single readout ASIC mounted directly on the sensor. Power and clock signals are provided directly to the ASIC on the sensor. This research is closely related to the development of silicon tracking sensors themselves. Given the large areas involved, various strip layouts are being studied to improve the pattern recognition while simultaneously limiting the number of readout channels. Thinner sensors are being studied that meet the required signal to noise ratio for the expected radiation dose.

Since large areas are to be covered, special emphasis is placed on efficient mechanical designs that provide modularity for ease of construction, minimise the material budget and provide sufficient space points to allow efficient pattern recognition especially in the forward and backward regions. Modules are arranged in self-supporting, light and robust structures which not only serve as support structures but also provide for cooling, cabling, services and alignment. As already mentioned, the alignment and stability issues are particularly important for the ILC where the modules are power-pulsed.

Traditional detector designs with a short central solenoid are not well suited for precision measurements in the forward and backward regions and designs for collider experiments were always a compromise favouring precision in the central region. Maintaining good tracking performance over a large polar angle range is a challenge for a number of reasons: The momentum resolution is degraded by the much smaller lever arm perpendicular to the magnetic field. Likewise, the vertexing capabilities are degraded by the large distance between the first measurement and the interaction point. Furthermore, the pattern recognition in the forward region must cope with low momentum particles "looping" through the detector, and background levels at the ILC increase rapidly with decreasing radius. Innovative solutions are required, that would benefit the community at large, to improve the performance of forward tracking systems. Sensor research geared towards fully active low-mass sensors, integrated front-end electronics, greater granularity and light-weight support structures with optimised tiling schemes may prove to be a most promising approach to overcome some of the limitations inherent in traditional forward tracking systems.

### 2.1.4 Calorimetry

The ILD and SiD detector concepts are both based on the particle flow algorithm (PFA) approach. For the electromagnetic and hadronic calorimeters, the ECAL and HCAL respectively, this requires unprecedented granularity to resolve the topologies, and a compact design, in order to keep showers as confined as possible. Over the past years, those demands have spurred the development of new technologies for calorimetry, like the use of silicon diode arrays for large scale detectors, novel high-gain low-cost photo-sensors (SiPMs), 2D-segmented Resistive Plate Chambers (RPCs), or Micro-Pattern Gas amplification Detectors (MPGD). All of them rely crucially on highly integrated low power mixed circuit Application Specific Integrated Circuits (ASICs).

The principal role of the ECAL is to identify photons and measure their energy. For the particle flow jet reconstruction, but also for hadronic $\tau$ decays, the capability to separate photons from each other and from near-by hadrons is of primary importance. The large difference between electromagnetic radiation length and nuclear interaction length is thus one of the reasons for the choice of tungsten as absorber material, the other being its small Molière radius. Silicon pad diodes lead to the highest possible compactness (and effective Molière radius) and exhibit excellent stability of calibration. Scintillating strips with silicon photo-detector readout provide a similar effective segmentation and offer a less costly, but also somewhat less compact, option. Both technologies could be combined in order to reach a cost performance optimum.

The role of the HCAL is to separate the deposits of charged and neutral hadrons and to precisely measure the energy of the neutrals. Their contribution to the jet energy, around 10% on average, fluctuates widely from event to event, and the accuracy of the measurement is the dominant contribution to the particle flow resolution for jet energies up to about 100 GeV. For higher energies, the performance is dominated by confusion, and both topological pattern recognition and energy information are important for correct track cluster assignment. Stainless Steel has been chosen both for mechanical and calorimetric reasons. Due to its rigidity, a self-supporting structure without auxiliary supports (dead regions) can be realised. Moreover, in contrast to heavier materials, iron with its moderate ratio of hadronic interaction length ($\lambda_I = 17$ cm) to electromagnetic radiation length ($X_0 = 1.8$ cm) allows a fine longitudinal sampling in terms of $X_0$ with a reasonable number of layers in a given total hadronic absorption length, thus keeping the detector volume and readout channel count small. This fine sampling is beneficial both for the measurement of the sizeable electromagnetic energy part in hadronic showers as for the topological resolution of shower substructure, needed for particle separation and weighting. For the HCAL read-out, two options have been developed: one is based on scintillator tiles with silicon photo-sensors and analogue read-out electronics, and the other is based on gaseous devices with one or two-bit, so-called semi-digital readout but finer transverse segmentation. The relative merits of the more detailed energy or spatial information of either option for the particle flow reconstruction are the subject of the ongoing studies. The main gaseous technology pursued is glass resistive plate chambers (RPCs), but structures based on GEMs or Micromegas are being considered as alternatives. The latter both provide a better correlation of the charge signal with deposited energy, but are less advanced in the realisation of large area detectors.

A broad R&D effort has been carried out to test these technologies and validate the simulations and the PFA performance predictions. This involved test beam campaigns with large installations, and due to their high granularity, set world records in terms of their channel count, exceeding that of the largest LHC calorimeter systems. This was made possible by maximising the use of common infrastructure such as mechanical devices, ASIC architectures and DAQ systems, and working within a common software and analysis framework that facilitates combination and comparison of test beam data. Most of this effort was organised within the CALICE collaboration which currently involves 350 members from 57 participating institutes worldwide. In addition R&D towards a highly compact silicon tungsten ECAL was performed by the SiD collaboration, as well as first studies towards a 3D segmented total absorption calorimeter with dual readout.

The development of calorimeter prototypes was roughly organised in two steps, which in practice overlapped. Physics prototypes provide a proof-of-principle of the viability of a given technology in terms of construction, operation and performance. In addition they are used to collect the large data sets which are invaluable for testing shower simulation programs, and for the development of PFAs with real data.

In a second step, technological prototypes address issues of scaling, integration and cost optimisation. They are required for each technology, but many large-area and multi-layer issues can already be addressed with so-called demonstrators, that is, modules with the adequate functionality but not completely instrumented, or at real scale and hence limited in terms of full system tests. These are not yet pre-production prototypes, and many of the issues addressed are still generic for each particular concept.

In 2011 CALICE reached a major milestone and completed a seven year series of test beam campaigns with physics prototypes of all major ECAL and HCAL technologies. About 400 million physics events have been recorded at CERN and Fermilab, and have been stored on the grid for analysis. In the meantime demonstrators are under intensive tests for all options, and physics data taking of the first full-size technological prototype has started in 2012. This has been very timely, enabling a description of the detectors together with the ILC technical design report. Not all efforts, however, have progressed at the same speed, and several beam tests are still being carried out and the data has not been fully analysed yet. It is already very clear that the test beam campaigns have provided the community with an unprecedented data sample revealing hadronic showers in exquisite detail which will form the basis for an in-depth evaluation in the near future.

A highlight among the rich amount of test beam results is the application of a PFA to beam test data [21]. Two displaced showers measured in CALICE prototypes of an analogue hadron and an electromagnetic calorimeter were mapped into the ILD detector geometry and subsequently processed by the Pandora particle flow algorithm for event reconstruction. Figure I-2.2 shows the probability to recover the energy of a 10 GeV neutral hadron within three sigma of the detector resolution as a function of the distance to a 10 GeV and 30 GeV charged pion, compared with simulations using different physics lists in Geant4. The good agreement of data and simulations, in particular for the QGSP-BERT physics list, underlines the reliability of full detector simulations in predicting the particle flow performance of the detector system. QGSP-BERT simulations apply the Quark Gluon String (QGS) model for high energy interactions of protons, neutrons, pions, kaons and nuclei. The high energy interactions create exited nuclei, which are passed to the precompound (P) model, which models the nuclear de-excitation. For primary protons, neutrons, pions and kaons below about 10 GeV Bertini cascades are used. The Bertini model produces more secondary neutrons and protons than the low energy parametrised model and clearly gives good agreement with the experimental data.

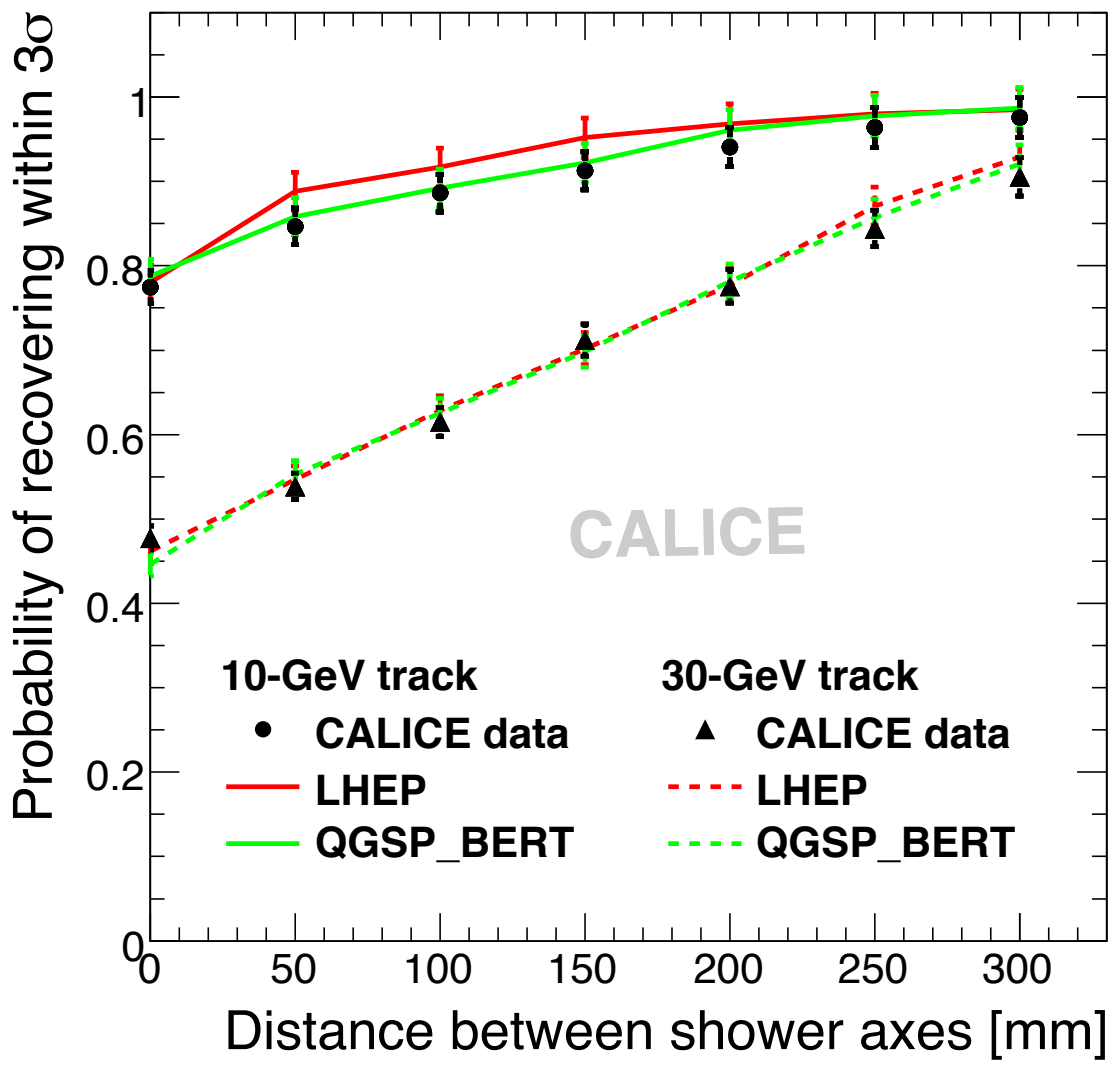

**Figure I-2.2**
Probability to recover the energy of a 10 GeV neutral hadron within three sigma of the detector resolution as a function of the distance from a 10 GeV (circles and continuous lines) and 30 GeV (triangles and dashed lines) charged hadron, respectively. Events are generated by mapping showers in the CALICE SiW ECAL, and AHCAL, into the ILD calorimeter system, and by reconstructing with PandoraPFA.

Similar studies will also be done using data taken with a digital HCAL (DHCAL) prototype tested in the same beam line and absorber structure with and without the ECAL in front. The prototype is instrumented with glass RPCs and has the front-end electronics embedded in the active layers to read its nearly 500,000 channels and provides the first possibility of an in-depth exploration of the digital approach to hadron calorimetry. Its superior imaging capabilities mark another highlight of ILC targeted calorimeter R&D and is illustrated in Figure I-2.3. Studies towards calibration and quantitative comparisons with simulations for the DHCAL are ongoing.

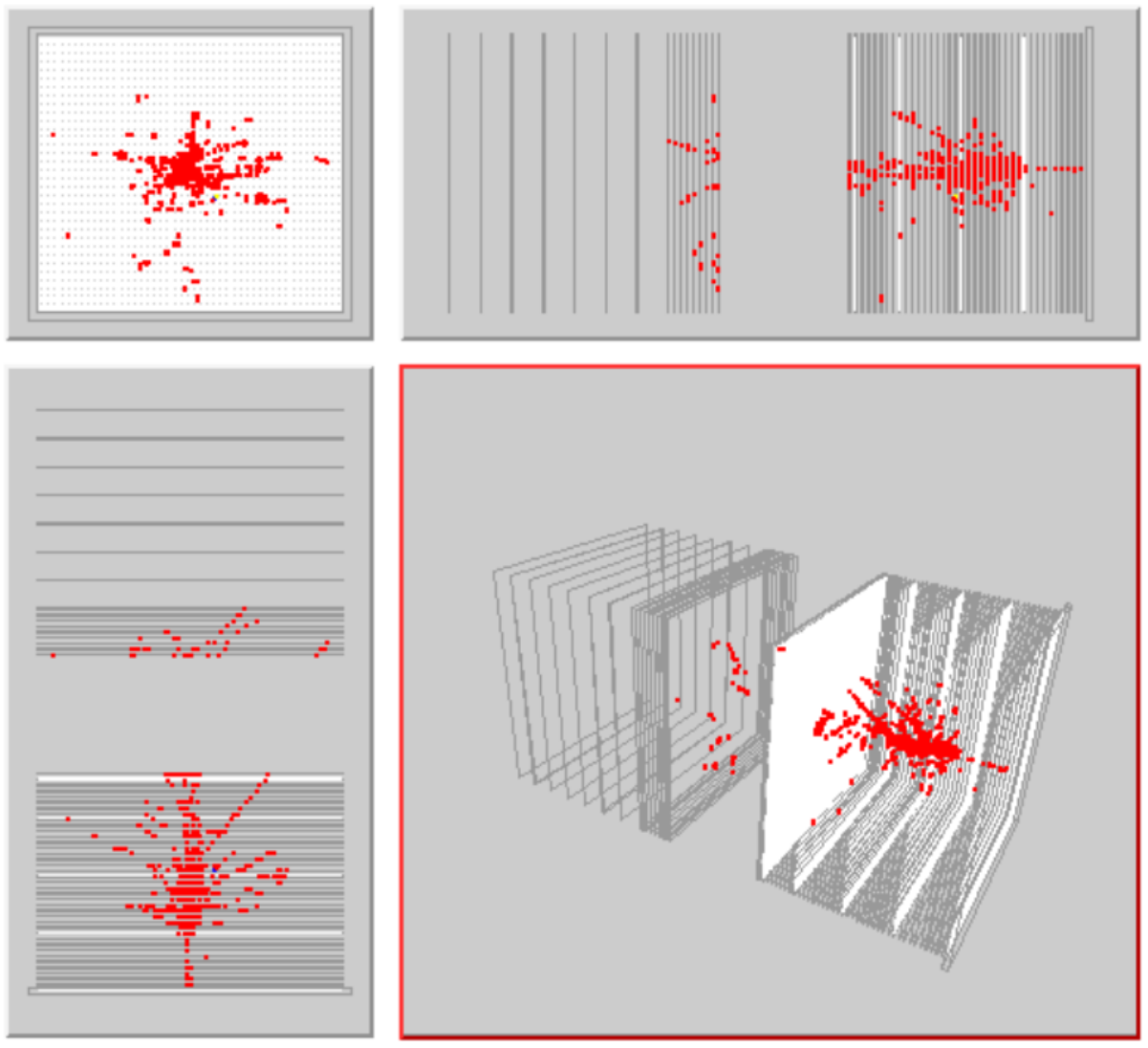

**Figure I-2.3**
Event display showing the interaction of a 10 GeV pion in the CALICE DHCAL with RPC read-out.

The optimisation of PFAs and the prediction of their performance relies on a realistic and detailed description of the propagation of hadronic showers in the calorimeters. It is one of the primary goals of the CALICE test beam campaigns to use the unprecedented granularity of the prototypes for detailed tests of the simulation models implemented in Geant4. Figure I-2.4 shows, as an example, the comparison of the average longitudinal shower profile for pions interacting in the SiW ECAL with a prediction using the QGSP-BERT physics list. The decomposition in terms of particles actually depositing energy is also shown. It shows that the detailed measurements provide specific information for the refinement of the models.

As a further example, the multiplicity of charged track segments reconstructed within hadronic showers is plotted as a function of the incoming particle energy and compared with model predictions. The agreement is not perfect, but still remarkable, given the level of detail probed, and far better than earlier versions of the simulation. This illustrates the progress towards the development of truly realistic Monte Carlo hadron shower simulations. The most recent simulations match the data within typically 5% , which qualifies them as a reliable tool for detector optimisation. This indicates that there has been significant progress with respect to the state of the art at the time when, for example, the LHC detectors were designed, and reflects the refinements based on the LHC calorimeter test beam series. The ILC-based calorimeter test beam data will provide the next step in providing more accurate simulations.

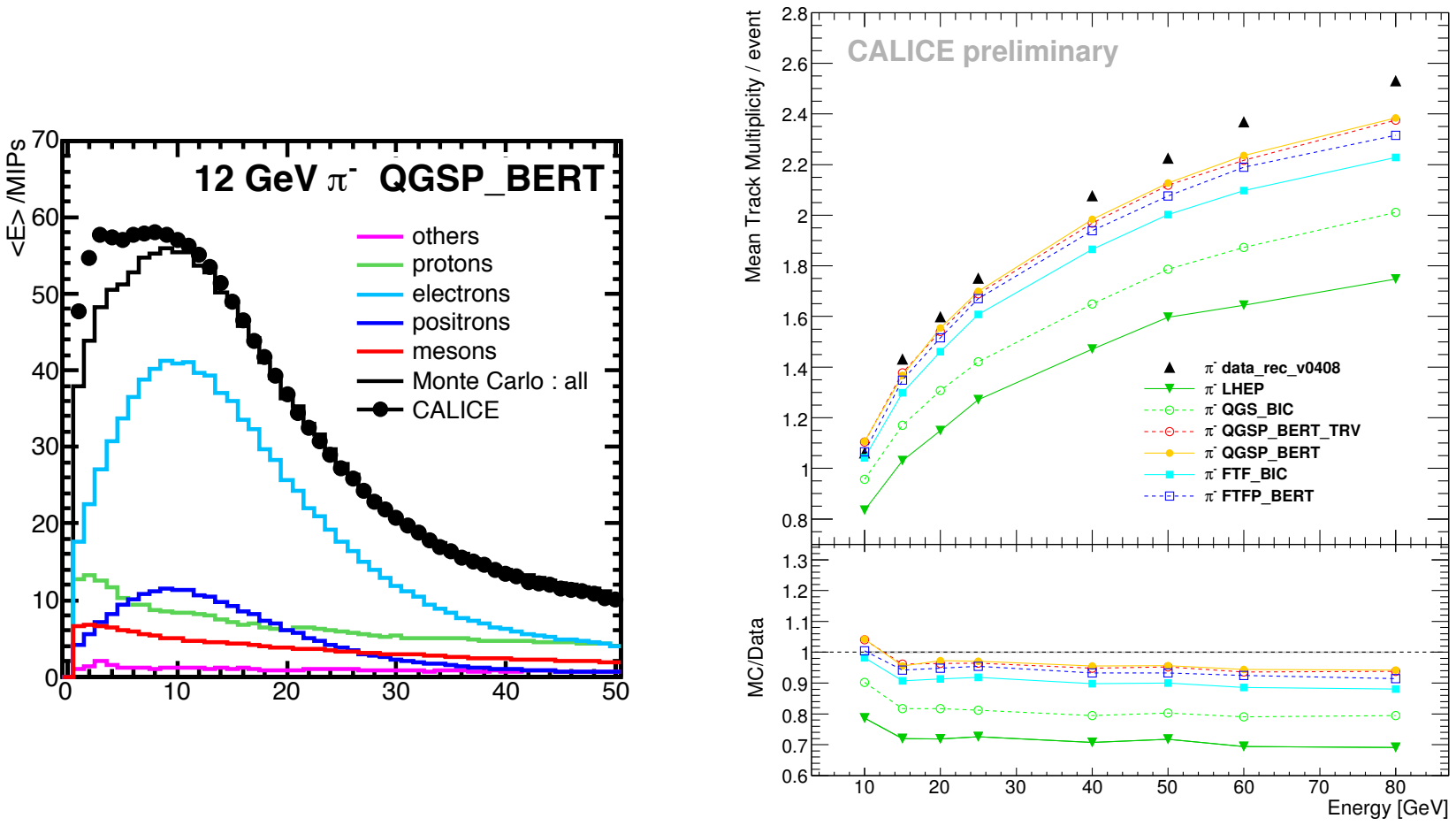

**Figure I-2.4**
Comparison of CALICE test beam data with simulations: longitudinal shower profile in the silicon tungsten ECAL, charged track multiplicity in the scintillator steel analogue HCAL.

A major achievement of the test beam results is that it demonstrated the viability of new technologies. Silicon pad diodes and scintillator tiles have shown a performance in terms of energy and topology resolution which quantitatively agree with simulations, and for scintillator strips and RPCs preliminary results indicate the same. The silicon pads benefit enormously from the intrinsic stability offered by this technology as demonstrated by the operation and analysis of a 10,000 channel calorimeter system. Scalable engineering solutions have recently been developed, with which highly compact structures can be realised for a collider detector, and first demonstrators have been exposed to beam. Within SiD, the KPiX chip, a 1024 channel mixed circuit ASIC which forms the heart of an ambitious ultra-compact silicon tungsten ECAL with direct hybrid-less chip-to-wafer bonding, has been developed and successfully tested.

The CALICE AHCAL was the first device that used the novel SiPM technology on a large scale, and its robustness and reliability has in the meantime encouraged other experiments, e.g. T2K, CMS and BELLE-II, to apply it in their detector upgrades. Correction procedures for intrinsic non-linearities and temperature sensitivities of the SiPMs were developed and successfully demonstrated.

Glass RPCs have for the first time been equipped with 2D segmented pad read-out and applied to calorimetry. The granularity of 1 cm$^2$ required the electronics to be embedded right from the first prototype. The 500,000 channel prototype has been operated stably at Fermilab and CERN and delivered first results. A second prototype with power-pulsed electronics has taken data in 2012. Power-cycling is one of the key strategies to minimise heat dissipation and cooling. The desire for more compact structures seemed to be contrary to these two goals, but it was demonstrated during the beam tests not to affect the performance.

Gas Electron Multipliers (GEMs) and Micromegas have demonstrated in tests with a few modules that they provide the necessary imaging capability, including low noise rates, and that they can be operated stably, as shown in tests deploying the small chamber configurations produced so far.

The data taking and analysis of the recorded data is far from complete and must continue, providing the basis for a full evaluation of the strengths and weaknesses of each technology for the application in a particle flow based detector.

In order to fully demonstrate integrated solutions, including a stable operation with on-detector zero-suppression, the second generation demonstrator units must be extended to systems large enough to record showers. In addition, there are still open issues at the system level, concerning power distribution, cooling, services and interfaces to be addressed with these second generation prototypes. The results so far, however, have clearly demonstrated that a particle flow calorimeter can be built, and deliver the predicted performance. Continued R&D is needed to complete the full program.

### 2.1.5 Forward Calorimetry

Special calorimeters are foreseen for the very forward regions of the ILC detectors. First, there is the so-called LumiCal to measure the luminosity with a precision of better than $10^{-3}$ using Bhabha scattering as reference process, $e^+e^- \to e^+e^-(\gamma)$. Then there is the BeamCal, positioned adjacent to the beampipe, to provide a bunch-by-bunch luminosity estimate and a determination of the beam parameters [22]. A third calorimeter, GamCal, about 100 m downstream of the detector, will assist in beam-tuning. A pair monitor positioned just in front of the BeamCal, which has a fast feedback system to the accelerator, will also be used for beam-tuning. These forward detectors, common to both experiments, have to withstand relatively high occupancies, requiring special front-end electronics and data transfer systems.

Monte Carlo simulations have been carried out to optimise the design of the forward region. In all calorimeters a robust electron and photon shower measurement is essential, making a small Molière radius preferable. Compact cylindrical sandwich calorimeters using tungsten absorber rings interspersed with finely radially segmented silicon or GaAs sensor planes are found to match the requirements. The LumiCal is used to measure precisely the polar angle of scattered electrons and positrons. It must be centred around the outgoing beam, with a precision requirements of 10 $\mu$m for the inner diameter of the acceptance radius and about 100 $\mu$m for the position with respect to the beam-line.

Due to the high occupancy created by the beamstrahlung and the two-photon processes, both calorimeters need a fast readout. Furthermore, the lower polar angle range of the BeamCal is exposed to a large flux of electrons, approaching one MGy per year. Hence, radiation hard sensors are needed.

A prototype of a silicon sensor for LumiCal is shown in Figure I-2.5 (left). They were manufactured by Hamamatsu using n-type silicon. The thickness is 350 $\mu$m and the strip pitch is 1.8 mm.

A possible sensor for the BeamCal is a high Ohmic GaAs sensor, shown in Figure I-2.5 (right) produced using the Liquid Encapsulated Czochralski method doped with a shallow donor and compensated with Cr as a deep acceptor. Sensors with several doping concentrations were exposed to a low energy high intensity electron beam up to a dose of 1 MGy. The leakage current per pad increased slightly with dose but was still at the level of 100 nA at room temperature. The charge collection efficiency is reduced at constant voltage by a factor of 10, but can be partially recovered by increasing the operation voltage. The challenge of BeamCal is to provide sensors that are radiation hard up to **Figure I-2.5** Prototypes sensors for LumiCal (left) and BeamCal (right).

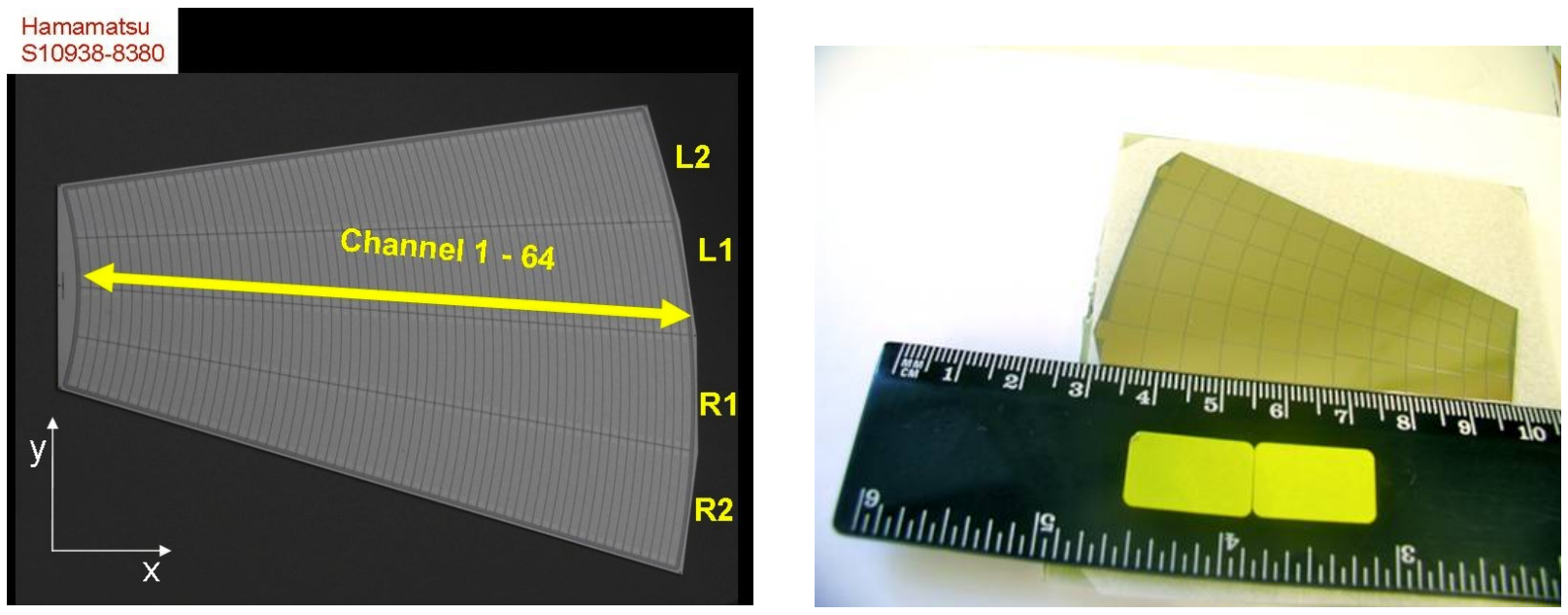

10 MGy of dose per year, a specification not unique to the ILC. Studies are being carried out in close collaboration with other experiments that also need very radiation hard sensors.

The pair monitor consists of one layer of silicon pixel sensors just in front of BeamCal to measure the distribution of the number of beamstrahlung pairs. GamCal is supposed to detect the photons from beamstrahlung for fast beam diagnostics.

Dedicated front-end and ADC ASICs have been designed in the 350 and 180 nm CMOS technology for the BeamCal and LumiCal. Dual gain charge sensitive amplifiers allow operation in calibration and standard data taking modes. The high amplification mode allows to measure the depositions of minimum ionising particles. Hence muons can be used from the beam halo or from annihilations for the calibration and sensor alignment studies. The low amplification mode will be used for the measurement of electromagnetic showers. Short shaping and conversion times allow readout or storage after each bunch crossing. The ASIC to be used for the BeamCal has in addition a fast analog adder to provide a fast feedback signal for beam tuning. In a test beam signal-to-noise ratios of better than 20 are measured for minimum ionising particles both for LumiCal and BeamCal sensors.

A critical aspect for the forward detectors is their very stringent alignment requirements. A novel laser position monitoring system has been designed, built and successfully tested to monitor the position of the two calorimeters with respect to the beam-pipe and the distance between them.

## 2.1.6 Beam Tests

A particularly important aspect of common detector development is the execution of beam tests. Beam tests have always played a critical role in the design and construction of new detectors. With the increase in sophistication of the experiments for high energy colliders, the importance of these tests has grown. Common beam instrumentation available for all detector development projects has proven to be extremely beneficial. Many benefits can be derived from the use of the same beam line instrumentation. Common instrumentation brings familiarity and provides for a larger user base that contributes towards further developing the peripheral instrumentation and software needed. Common instrumentation will also eliminate a source of uncertainty in the comparison of data between different technologies in the same beam line.

Within the ILC framework the EUDET project [23], aimed at providing infrastructures for research and development of novel detector technologies for the ILC, has been very successful. The main joint research activity was developing and improving test beam infrastructures in particular the commissioning of a high resolution pixel beam telescope.

A telescope consisting of six reference planes equipped with 18.4 $\mu$m pitch CMOS pixel sensors, called Mimosa26 developed by Strasbourg IPHC [18], was constructed. The Mimosa26 sensor is a fully digital sensor with binary readout. The sensors were thinned to 50 $\mu$m and a hit position resolution of better than 3 $\mu$m was achieved with a readout rate at the kHz level.

The excellent resolution, readout rate and data acquisition integration capabilities made the telescope a primary test beam tool for many groups. The commercial readout allowed easy cloning and by now four copies are running in addition to the original EUDET telescope. The user groups have extended far beyond the original ILC detector R&D groups and now include several LHC groups. Within the new European detector infrastructure project the test beam telescope will be further extended in terms of cooling infrastructure, read-out speed and precision. In order to provide a system optimised for the different requirements by the user community a combination of various pixel technologies is foreseen.

The EUDET program also provided the community with a large bore, high field superconducting solenoid for the studies of various micro-pattern gas detector technologies. This enabled the initiation of combined modular test beam campaigns. The word combined here refers to a beam test of various sub-detectors in one beam line at the same time. The term "modular" refers to the ability to exchange sub-detectors and replace them with one of different technology. Such test beam campaigns address the integrated nature of the ILC detectors, where the overall performance of the detector is determined by a critical, subtle interplay in performance of the subdetectors. Over the last years beam tests were carried out as collaborative efforts, not only within a horizontal R&D collaboration, but also between various ILC detector development groups. All ILC detector development projects have benefited tremendously from the creation of this common infrastructure. The advantages are self-evident and it behoves the community at large to further encourage and strengthen common infrastructure for combined test beams. Some crucial tests such as power-pulsing in a high magnetic field have not been carried out due to the limited availability of such infrastructure.

## 2.2 Common simulation and software tools

Software tools are the basis of detector benchmarkings. While the detector concept groups ILD and SID have developed their own independent software frameworks, they have also collaborated in developing a number of common software tools. Such tools include event generator programs and samples, event data models, file formats, and event reconstruction tools. These common tools and the summary of detector benchmark studies are presented in the following subsections.

### 2.2.1 Common generator samples

As the machine parameters and selected benchmark processes are common between the concepts, it was decided to perform benchmark studies based on the same event generator samples as a common effort between ILD and SiD. It was also decided that a system should be set up so that the effort could be shared between different sites. To this end, the event generation program, Whizard [24], was selected for the generation of events with up to 6-fermion final states and Physsim [25] for $t\bar{t}h$ and relevant background processes with more than 6-fermion and the common generator samples have been generated as follows.

#### 2.2.1.1 Overview of event generation

The generation framework based on Whizard was originally developed at SLAC for the LOI. Certain short comings of the LOI framework were remedied. By using the full CKM-matrix - rather than just it's diagonal elements - generating events with rare quark flavour combinations became possible. The treatment of $\tau$ polarisation, which for the LOI was only done for some particularly important cases ($e^+e^- \to \tau^+\tau^-$ and $e^+e^- \to \tilde{\tau}^+\tilde{\tau}^-$), was generalised to be applied to all $\tau$ modes. Finally, certain useful information (spin and colour-flow, energies of the initial particles) from the generator was output with the generated events, and hence made available for use in the physics analysis.

The requirement that the generated events should be usable both for ILD and SiD, and should be producible at different sites, implied that a set of well defined event-generator conditions must be documented. This would then enable the event-generation conditions to be propagated down the event processing stream, independently of the details of the implementation of the stream. The requirement was accomplished by demanding that each generation job should provide a set of meta-data describing the generated data.

GuineaPig [26] was used to simulate the beam-beam effects, based on a set of beam parameters defined by the GDE [27]. Such effects influence both the energy spectrum of the interacting initial particles, and the composition of the beams, notably how large the photon component is. The output from the GuineaPig simulation was used to create the needed spectra, which were then passed on to the event-generators.

The produced events were stored in STDHEP format [28], in files of a maximal size of 500 MB. At the end of a generation job, these files were uploaded to the grid, from where any user with a valid ILC grid-certificate could access them. In addition to the data-files themselves, the meta-data and the log-files are kept on the grid. Further details such as the various steering-files controlling any generation jobs were made available for inspection on the web [29]. In the case of Whizard, the integration grids - which are produced by evaluating the phase-space of the process, and subsequently used to efficiently generate un-weighted events - are also available for download on the web.

The source-code of the generators used, together with auxiliary programs needed (Pythia for fragmentation, Tauola for polarised tau-decays, STDHEP for the output, and Cernlib), with the beam-spectrum files for various machine configurations, and with installation procedures are maintained in an SVN repository, housed at CERN [30].

#### 2.2.1.2 Event generation by Whizard

The Whizard Monte Carlo was used for the generation of all $2 \to n$ processes, $n = 2, 3, 4, 5, 6$, where $n$ is the number final state fermions, and the two initial state particles are $e^+e^-$, $e^+\gamma$, $\gamma e^-$, or $\gamma\gamma$. It was also used for the generation of $e^+e^- \to f\bar{f}h$. Whizard provides a lowest order calculation of each $2 \to n$ sub-process, and simulates multiple photon radiation from the initial state electron and positron in the leading-logarithmic approximation. The luminosity weighted energy spectra of the initial state electron and positron including intrinsic machine energy spread and beamstrahlung effects can be supplied by the user through Fortran90 user interface subroutines. The spectra could be two dimensional to include a correlation between two beams. For sub-processes with $\gamma$'s in the initial state, Whizard provides a simulation of nearly real Weizsäcker-Williams photons, while the energy spectra and overall normalisation of beamstrahlung photons is supplied by the user.

The two dimensional luminosity weighted $e^+e^-$ energy spectra and beamstrahlung $\gamma\gamma$ energy spectra were calculated using the GuineaPig program [26]. GuineaPig output consisting of several million pairs of $e^+e^-$ or $\gamma\gamma$ energies were processed by a program that created Monte Carlo integration grid files. These grid files are read in by the Whizard user interface subroutines. They faithfully reproduce the underlying correlated two dimensional GuineaPig distributions, and can be used to generate large numbers of independent initial state $e^+e^-$ or $\gamma\gamma$ pairs. The $e^-\gamma$ and $\gamma e^+$ spectra were simulated using the one-dimensional $e^\pm$ and $\gamma$ distributions, and so correlations were not included for these initial states.

Final state QCD and QED showering of all final state quarks, and QED showers of final state muons and taus are simulated using the Pythia Monte Carlo program [31]. Pythia is also used for fragmentation and particle decay. Final state showering from electrons is normally switched off because Pythia does not use the correct showering $Q^2$ for most final state electron configurations.

The Higgs mass is set to 2 TeV unless the Higgs boson is a final state particle. When the Higgs is a final state particle its mass is set to 125 GeV, its branching fractions are set to the recommendations of the LHC Handbook on Higgs cross sections [32], and the decay is simulated using Pythia.

Gluon splitting is simulated using the the parton showering algorithm of Pythia, while amplitudes with a gluon propagator are switched off in the simulation of the $2 \to n$ sub-processes. This is the normal method for generating events in $e^+e^-$ collisions. Interference between QCD and electroweak amplitudes is not simulated in this arrangement, but is thought to be a 10% effect at most [33]. At some point in the future amplitudes with gluon propagators will be included in the $2 \to n$ sub-process simulation along with the matching algorithms that are needed to prevent double counting with QCD parton showering. However, this effort could not be completed in time for the DBD benchmarking studies.

#### 2.2.1.3 Event generation by Physsim

The study of the $t\bar{t}h$ benchmark process required generations of processes involving 8 or more fermion final states. Generations of such processes were not easy for Whizard, because very long CPU times were required to reach reasonable precisions of phase space integration due to the many channels involved. Therefore, the event generation of these processes were made by Physsim [25], which was used previous study on $t\bar{t}h$ coupling at 500 GeV [34].

Physsim calculates helicity amplitudes by Helas [35] and phase space integration and event generation are performed by the Bases/Spring package [36]. The processes generated by Physsim were, (1) $e^+e^- \to t\bar{t}h \to 6f + h$, (2) $e^+e^- \to t\bar{t}Z \to 6f + f\bar{f}$, and (3) $e^+e^- \to t\bar{t}g^* \to 6f + b\bar{b}$. Here $6f$ denotes 6 fermions decayed from $t\bar{t}$ system. The resonance effect in the $t\bar{t}$ system was not included. The decay of Higgs and the hadronisation of quarks were performed by Pythia with the same parameters as Whizard events, the $6f$ system and the remainder being hadronised independently. The colour flow effect between $6f$ and $b\bar{b}$ system in the process (3) was neglected. The algorithm to generate the initial state radiation, the hadronisation parameters and the version of Tauola used were same as those used for Whizard events in order to have the same event property as those generated by Whizard. The Feynman diagrams in the case of the $t\bar{t}h$ process is shown in I-2.6.

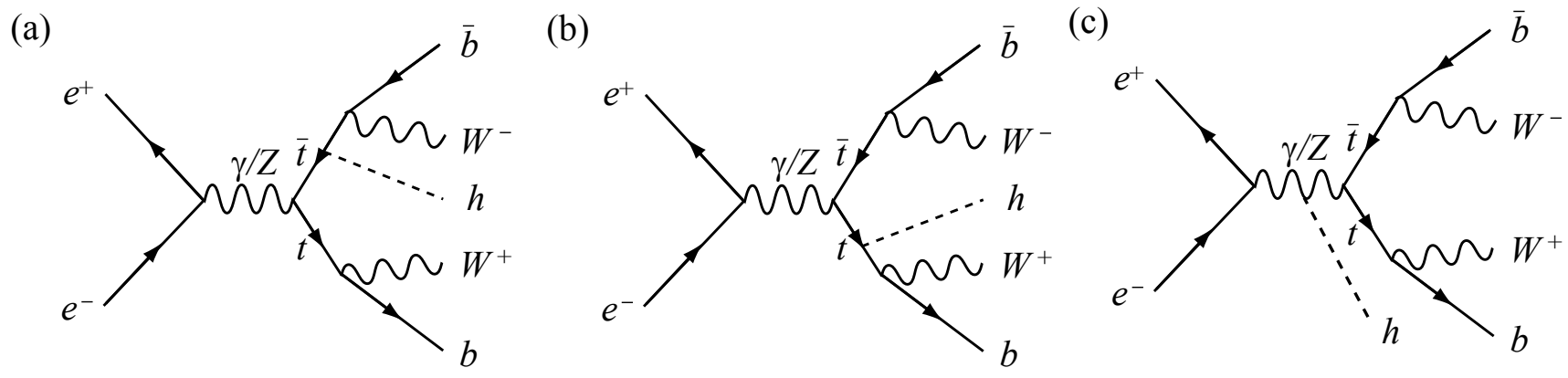

**Figure I-2.6** Feynman diagrams for the $e^+e^- \to t\bar{t}h$ process.

The events were generated separately depending on the initial beam helicity and the decay mode of $t\bar{t}$, either 6 quarks, 4 quarks plus lepton and neutrino, or 2 quarks ($b\bar{b}$), 2 leptons and 2 neutrinos. The $t\bar{t}h$ process was further divided by the decay mode of Higgs to $b\bar{b}$ or the rest.

#### 2.2.1.4 Hadronisation tuning

In the event-generation done for the LOI, hadronisation was done with the default settings of Pythia. However, it has been shown by the LEP collaborations that these setting do not describe the observations at LEP in all aspects.

**Table I-2.1** Predicted average numbers of various particle species in $e^+e^-$ collisions at 92 GeV, for default Pythia settings or OPAL settings compared to LEP data

| | Standard tune | OPAL tune | LEP combined data |
|---|---|---|---|
| All charged | 20.6246 | 20.5685 | 20.9400 $\pm$ 0.1900 |
| $\pi^0$ | 9.6814 | 9.8866 | 9.3800 $\pm$ 0.4500 |
| $\pi$ | 17.1178 | 17.5467 | 17.0500 $\pm$ 0.4300 |
| $K$ | 2.2879 | 2.1108 | 2.3600 $\pm$ 0.1100 |
| p | 1.2190 | 0.9110 | 0.9750 $\pm$ 0.0870 |
| n | 1.1661 | 0.8664 | - |
| $K^0_S$ | 1.1168 | 1.0150 | 1.0040 $\pm$ 0.0150 |
| $K^0_L$ | 1.1057 | 1.0164 | - |

In particular, the data indicates that the amounts of neutral, long-lived hadrons are over-estimated by Pythia. This has a direct influence on the expected jet-energy resolution, since more neutral hadrons in a jet implies higher risk for confusion-errors. Even in the absence of confusion, a higher neutral hadronic component deteriorates the jet- energy resolution, simply because neutral hadronic energy is the component measured with least precision.

The LEP collaborations were contacted, and asked to provide their best estimates of how to tune Pythia to the data. For technical reasons, the tunings provided by OPAL was chosen. Table I-2.1 shows a comparison between LEP data [37, 38], default Pythia and OPAL-tuned Pythia for particle multiplicities for a few selected particle species. Of particular importance are the numbers for protons and $K^0_S$, as these serve as proxies for the amount of neutrons and $K^0_L$, respectively, ie. for the amount of long-lived neutral hadrons expected. It is clearly seen that the default Pythia settings significantly over-estimate these particles - by about 15% - and hence the amount of neutral hadronic energy. Table I-2.2 shows the difference between the two settings at ILC energies. It can be observed that the difference has become even larger than at 92 GeV: Pythia with default settings predict about 25% more neutral hadrons than what the OPAL tune does

**Table I-2.2** Predicted average numbers of various particle species in $e^+e^- \to q\bar{q}$ $(q = uds)$ collisions at 500 GeV, for default Pythia settings or OPAL settings.

| | Standard tune | OPAL tune |
|---|---|---|
| All charged | 37.4267 | 37.4975 |
| $\pi^0$ | 17.2502 | 17.7834 |
| $\pi$ | 31.1060 | 32.3830 |
| $K$ | 3.7395 | 3.2706 |
| p | 2.5812 | 1.8439 |
| n | 2.5109 | 1.7778 |
| $K^0_S$ | 1.8006 | 1.6120 |
| $K^0_L$ | 1.8069 | 1.6119 |

### 2.2.1.5 Generator samples

A generated *process* is defined by an initial state and a final state. For the initial state this includes the polarisation of the incoming particles, as well as their nature. For the final state, it is defined by a combination of quarks and/or leptons, possibly in conjunction with one or more photons.

Processes were grouped into a physics oriented classification in order to reduce a number of processes. For example, The 4-fermion processes were classified as $ZZ$, $WW$, or mixed $WW$ and $ZZ$, according to the intermediate particles involved in the diagrams. The pure $ZZ$ class would typically contain processes with only one flavour present in the final state (e.g., $u\bar{u}u\bar{u}$), while the pure $WW$ class would contain more than 2 flavours (e.g., $u\bar{d}s\bar{c}$). The mixed class would be cases where both $ZZ$ and $WW$ diagrams could contribute, e.g.. $u\bar{d}\bar{u}d$. The processes with contributions from single-boson production were treated separately. In addition, there was a sub-division between full hadronic, semi-leptonic and fully leptonic final-states. This scheme reduced the several hundreds of unique channels for the 2- and 4-fermion samples to a few tens. Processes other than 2- and 4-fermion were also classified similarly.

This grouping of processes was implemented using two features in Whizard: particle aliases and process-grouping. Aliasing gives the possibility to assign aliases to groups of particles, which are then treated as a single entity. For instance, aliases were defined for up-type quarks ($u$ and $c$) and down-type quarks ($d$, $s$ and $b$). Process-grouping, on the other hand, groups individual processes together at the event-generation stage, so that a random mix of the selected processes are generated with the correct relative fractions.

The generated samples are summarised in Table I-2.3. In the table, $f$ is a quark, lepton, or an alias particle; $\gamma$'s in the initial state are due to beamstrahlung or initial state radiation. For each combinations of $e^{\pm}$ beam polarisation, samples of 1 ab$^{-1}$ were generated separately with fully polarised beams, except for a few exceptions; $e^+e^- \to e^+e^-$ process was generated as $e^+e^- \to e^+e^-\gamma$ with a kinematical cut on $e^+e^-$ invariant mass, opening angle and acoliniarity for $WW$ benchmark studies; The $t\bar{t}h$ and relevant 8-fermion samples include about 50k events which correspond to at least 8 ab$^{-1}$. The low $p_t$ $e^+e^-$ background events were generated by GuineaPig. The $\gamma\gamma \to$ mini-jets hadron events were generated by Pythia implemented in the Whizard framework using the same lumi-spectrum as other generators. Low $p_t$, high cross section, minimum bias $\gamma\gamma \to$ hadron events were generated based on the cross section model by M.Peskin [39] using either a phase space particle production model or the Pythia model for $\gamma\gamma \to$ hadrons, depending on whether the $\gamma\gamma$ centre of mass energy was less than or greater than 10 GeV.

**Table I-2.3**
Summary of the common generator samples.

| event-type | process |
|---|---|
| 1f | $e^{\pm}\gamma \to \gamma e$ |
| 2f | $e^+e^- \to f\bar{f}$ |
| 3f | $e^{\pm}\gamma \to (\mathrm{e\ or\ }\nu) + 2f$ |
| 4f | $e^+e^- \to 4f$ |
| 5f | $e^{\pm}\gamma \to (\mathrm{e\ or\ }\nu) + 4f$ |
| 6f | $e^+e^- \to 6f$ |
| aa_2f | $\gamma\gamma \to 2f$ |
| aa_4f | $\gamma\gamma \to 4f$ |
| aa_minijet | $\gamma\gamma \to$ hadron mini-jets |
| aa_lowpt | $\gamma\gamma \to$ low $\mathrm{p_t}$ hadrons |
| eepairs | beam induced low $\mathrm{p_t}$ $e^{\pm}$ pairs |
| higgs | $e^+e^- \to f\bar{f}h$ |
| tt$h$ | $e^+e^- \to t\bar{t}h, t\bar{t}Z,$ and $t\bar{t}g^*(g^* \to b\bar{b})$ |

These samples were generated without beam crossing angle, spread of interaction points, nor background overlay. These effects were taken into account at detector simulations or event reconstructions.

## 2.2.2 Common simulation and reconstruction tools

Besides the common generator tools and samples described above, both concepts have based their detailed simulation applications on the Geant4 [40] tool kit and share a common event data model and file format which is provided by LCIO [41]. At the reconstruction stage the pattern recognition and track fitting tasks are performed independently, whereas for the particle flow algorithm, the vertex finding and flavour tagging again common tools are used: PandoraPFA [42] and LCFIPlus [43] respectively. In the following we describe the common tools that have been developed in the context of the Linear Collider activities in more detail.

### 2.2.2.1 LCIO

The LCIO software package provides a common *Event Data Model* (EDM) and persistency solution for Linear Collider detector R&D. The development of LCIO started in 2003 and provides implementations in *C++*, *Java* and *Fortran* the languages used at the time. Using a common EDM and file format is a key requirement for easy sharing of software tools and algorithms across detector concepts and working groups.

In Figure I-2.7 the hierarchical EDM of LCIO is shown. It has been recently extended and improved as a preparation for the DBD. In particular the *Track* class has been extended to hold a number of *TrackStates* for the same set of *TrackerHits*, typically at the *Interaction Point*(IP), the first and last hit and at the entry point to the calorimeter. New classes for one dimensional measurements from Si-Strip detectors have been added in order to allow for an increased level of realism with respect to the LOI [44].

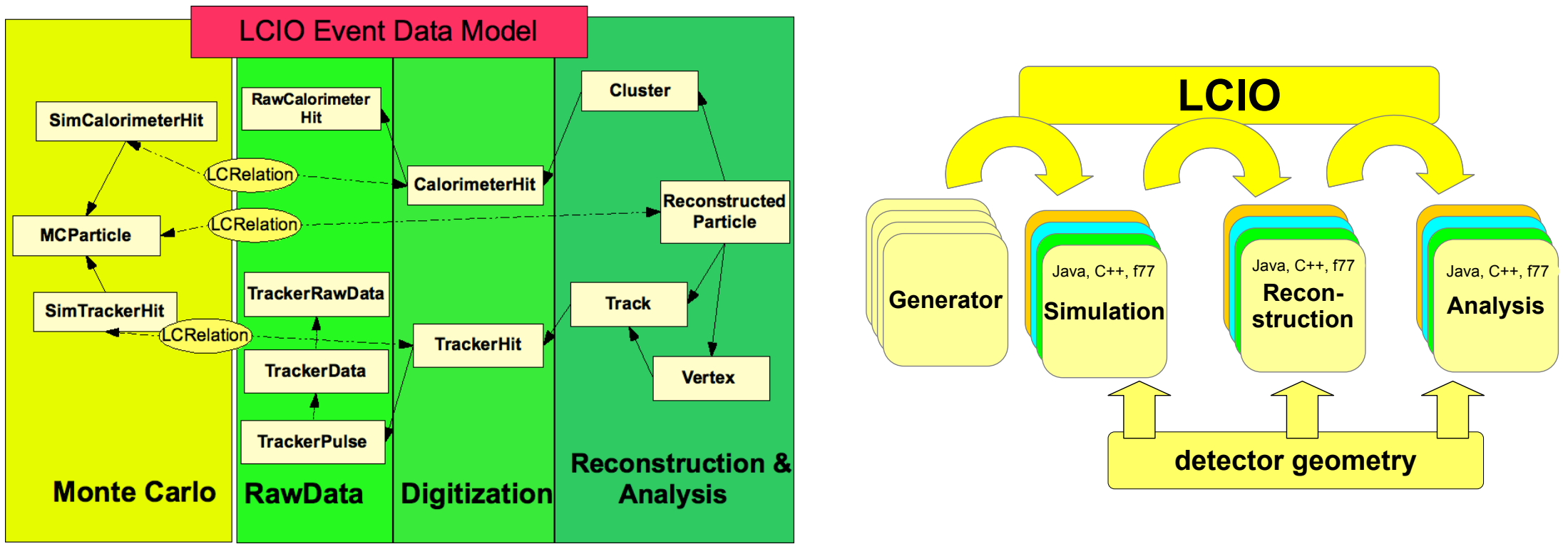

**Figure I-2.7.** Left: Schematic overview of the hierarchical event data model (EDM) of LCIO. Right: The common EDM and persistency scheme allow to exchange software tools even when they are written in different frameworks and languages.

## 2.2.2.2 PandoraPFA

PandoraPFA is an implementation of the *particle flow algorithm (PFA)*, which originally has been developed in the Marlin [45] framework for LC-like detectors. In a recent redesign of the framework it has been turned into a standalone library with interfaces to external software through well defined *Application Programming Interfaces* and essentially no external dependencies. Figure I-2.8 shows the layout of the new PandoraPFA tool kit. Both detector concepts have written client applications that interface the input data collections from LCIO, augmented with the corresponding geometry information of the detectors, to PandoraPFA and convert the output into the final collection of *ReconstructedParticles*, used for analysis. The actual algorithm will be described in the ILD and SiD specific sections.

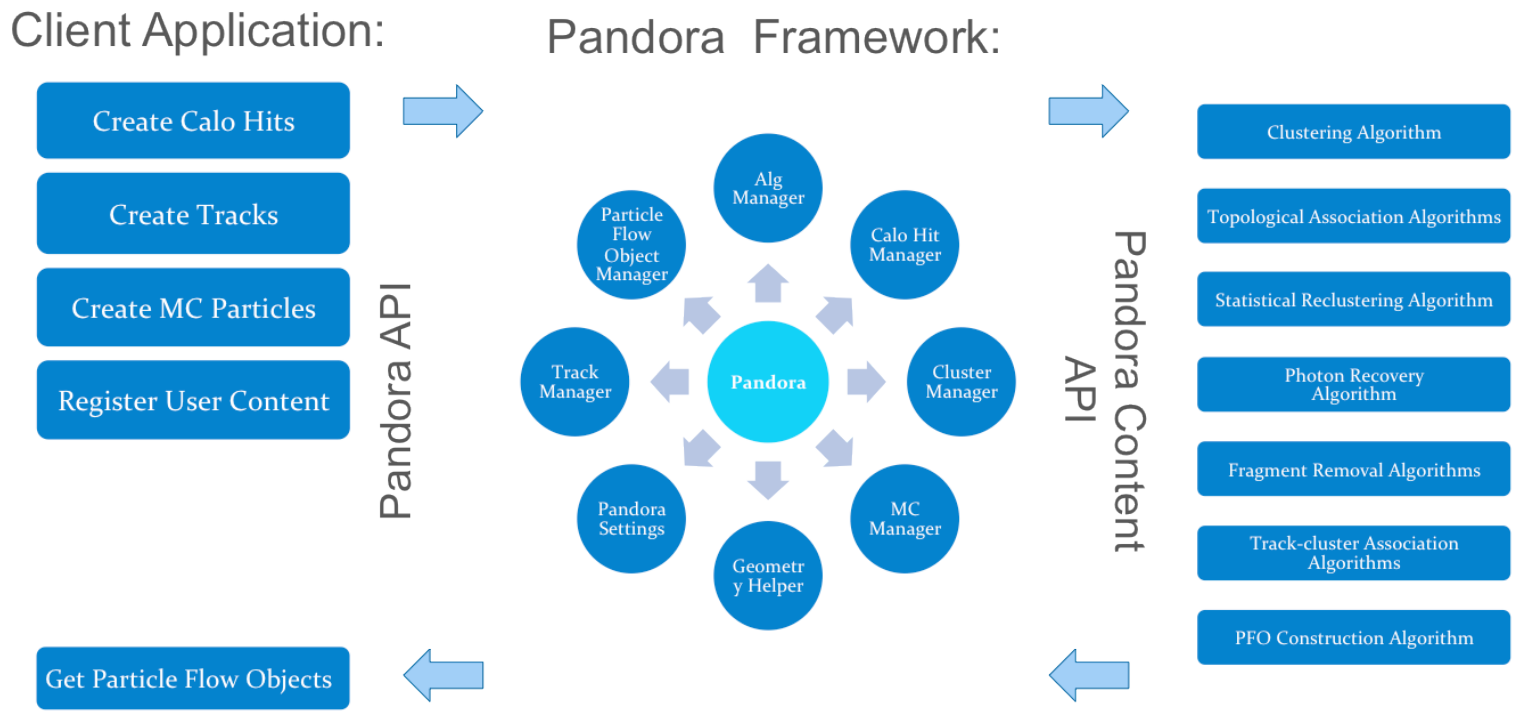

**Figure I-2.8** Schematic overview of the structure of the PandoraPFA algorithm showing the separation between the Client Application, the framework and the algorithms.

## 2.2.2.3 LCFIPlus

The LCFIVertex [46] software package had been developed for vertexing, flavour tagging and vertex charge reconstruction with an ILC vertex detector. It was originally developed for Z-pole physics and designed to find the vertex and tag the flavour of a jet, thus its algorithm is applied after reconstruction of a jet in an event. LCFIPlus [43] is a new Marlin package targeted for multi-jet events. In this package, vertices in an event are reconstructed before jet reconstruction so as to use found vertex information for jet reconstruction. The flavour tagging is made with the help of TMVA (*Toolkit for Multivariate Data Analysis with ROOT*). In addition to the standard variables, user specific variables for tagging can be easily introduced depending on the needs for the analysis. LCFIPlus is used by both detector concepts. The actual algorithms will be described in the ILD and SiD specific sections.

### 2.2.3 Summary of benchmark studies

The ILD and SiD detector concept groups made detector benchmark studies using the common generator samples described in the previous sub-section. While the details of software tools and analysis procedures are presented in the ILD and SiD chapters, a summary of them is presented in Table I-2.4. Note that the numbers quoted are indicative of the precisions achieved in full simulation studies from ILD and SiD.

The benchmark studies have been performed to investigate the physics capability of proposed detectors selecting a few numbers of channels. The implications of these estimates for the study of a 125 GeV/c$^2$ Higgs boson and ILC physics case are described in the physics volume.

**Table I-2.4.** Summary table of results from the benchmarking studies. In the luminosity column, the electron beam polarisation for eL (eR) is -80% (+80%) and the positron beam polarisation is +20%(-20%) at 1000 GeV and +30%(-30%) at other energies. For the earlier studies at 250 GeV, the Higgs boson mass was taken to be 120 GeV and the default Pythia 6.412 branching ratios were assumed. For the more recent studies at 1 TeV, the Higgs boson mass was taken to be 125 GeV and the branching ratios of ref.[32] were used.

| $\sqrt{s}$ (GeV) | L (ab$^{-1}$) eL/eR | Channel ($e^+e^- \to$) | Observable | Precision | Comment |
|---|---|---|---|---|---|
| 250 | 0.25/0 | $Zh, Z \to e^+e^-/\mu^+\mu^-$ | $\Delta\sigma/\sigma$ | 2.5 % | model indep. |
| | | | $\Delta m_h$ | 32 MeV | analysis |
| | | $Zh, h \to b\bar{b}$ | $\Delta(\sigma Br)/(\sigma Br)$ | 1.0% | |
| | | $Zh, h \to c\bar{c}$ | | 7.3% | |
| | | $Zh, h \to gg$ | | 8.9% | |
| | | $Zh, h \to \mu^+\mu^-$ | | 89% | |
| 500 | 0.25/0.25 | $t\bar{t} \to 6\text{-}jet$ | $\Delta\sigma/\sigma$ | 0.5% | |
| | | | $\Delta m_t$ | 40 MeV | |
| | | | $\Delta A_{FB}$ | 0.011 | |
| | | $\tau^+\tau^-$ | $\Delta A_{FB}$ | 0.21%$^\dagger$, 0.24%$^\ddagger$ | $^\dagger$eL, $^\ddagger$eR |
| | | | $\Delta\langle P_\tau\rangle$ | 1%(stat)⊕0.6%(stat) | |
| | 0.5/0 | $\tilde{\chi}_1^+\tilde{\chi}_1^-$ & $\tilde{\chi}_2^0\tilde{\chi}_2^0$ | $\Delta\sigma/\sigma(\tilde{\chi}_1^+\tilde{\chi}_1^-)$ | 0.6% | template fitting |
| | | $\to WW/ZZ + \not{E}$ | $\Delta\sigma/\sigma(\tilde{\chi}_2^0\tilde{\chi}_2^0)$ | 2.1% | |
| | | $\to 4\text{-}jet + \not{E}$ | $\Delta m(\tilde{\chi}_1^\pm)$ | 2.4 GeV | 2 parameter fit |
| | | | $\Delta m(\tilde{\chi}_2^0)$ | 0.9 GeV | |
| | | | $\Delta m(\tilde{\chi}_0^0)$ | 0.8 GeV | |
| | | $\tilde{\mu}_L^+\tilde{\mu}_L^- \to \mu^+\mu^- + \not{E}$ | $\Delta\sigma/\sigma$ | 2.5% | |
| | | | $\Delta m(\tilde{\mu}_L)$ | 0.5 GeV | |
| | | $\tilde{\tau}_1^+\tilde{\tau}_1^- \to \tau^+\tau^- + \not{E}$ | $\Delta m(\tilde{\tau}_1)$ | 0.1 GeV $\oplus 1.3\sigma_{LSP}$ | |
| | 2/0 | $Zhh$ | $\Delta\lambda/\lambda$ | 44% | |
| 1000 | 1/0 | $\nu\bar{\nu}h, h \to b\bar{b}$ | $\Delta(\sigma Br)/(\sigma Br)$ | 0.47 % | |
| | | $\nu\bar{\nu}h, h \to c\bar{c}$ | | 7.6 % | |
| | | $\nu\bar{\nu}h, h \to gg$ | | 3.1 % | |
| | | $\nu\bar{\nu}h, h \to WW^*$ | | 3.3 % | |
| | | $\nu\bar{\nu}h, h \to \mu^+\mu^-$ | | 32 % | |
| | 0.5/0.5 | $t\bar{t}h$ | $\Delta(\sigma Br)/(\sigma Br)$ | 8.7% | 8 jets ⊕ 6 jets |
| | | $W^+W^-$ | $\Delta\lvert P_{e^-}\rvert$ | 0.16% | full angle analysis |
| | | | $\Delta\lvert P_{e^+}\rvert$ | 0.23% | |
| | | | $\Delta P_{e^-}(L)_{\mathrm{eff}}$ | 0.11%$^\dagger$, 0.036%$^\ddagger$ | $^\dagger$eL, $^\ddagger$eR |
| | 2/0 | $\nu\bar{\nu}hh$ | $\Delta\lambda/\lambda$ | 18% | |

## 2.3 Machine Detector Interface

The Machine-Detector Interface (MDI) at the ILC covers all aspects that are of common concern to both detectors and to the machine. This usually covers topics like beam induced backgrounds, integration of the machine and detector elements in the Interaction Region (IR) as well as physics related beam instrumentation issues (e.g. polarisation and energy measurements). This section deals with those MDI topics that are of common concern to both detectors and that are not specific to the respective implementation of the IR: common assembly procedures, experimental area layouts, the push-pull system. Detector concept specific MDI topics are discussed in the respective chapters of the SiD and ILD sections in this report.

### 2.3.1 The push-pull concept

The ILC design foresees to have two detectors that share one interaction region in a push-pull operation scheme. In that scheme, one detector would take data, while the other one is waiting in the close-by maintenance position. At regular intervals, the data-taking detector is pushed laterally out of the interaction region, while the other detector is being pulled in. As the data taking intervals for each experiment should be short enough to avoid a potential discovery by one detector alone, the transition time for the exchange of the detectors needs to be short, i.e. in the order of one day, to keep the total integrated luminosity at the ILC high.

A time efficient implementation of the push-pull model of operation sets specific requirements and challenges for many detector and machine systems, in particular the IR magnets, the cryogenics, the alignment system, the beam line shielding, the detector design, and the overall integration. The minimal functional requirements and interface specifications for the push-pull IR have been successfully developed and published [47], to which all further IR design work on both the detectors and machine sides are constrained. The developments lead to a detailed design of the technical systems and the experimental area layout that follow detailed engineering specifications [48].

### 2.3.2 Detector motion system

The detector motion and support system has to be designed to ensure reliable push-pull operation allowing a hundred moves over the life of the experiment, while preserving internal alignment of the detectors internal components and ensuring accuracy of detector positioning. The motion system must be designed to preserve structural integrity of the collider hall floor and walls. Moreover, the motion and support system must be compatible with the tens of nanometre level vibration stability of the detector. In seismic regions the system must also be compatible with earthquake safety standards.

The detectors will be placed on platforms that preserve the detector alignment and will distribute the load evenly onto the floor (c.f. Figure I-2.9) The platform will carry also some of the detector services like electronic racks. Cables and supply lines will be routed to the platform in flexible cable chains that move in trenches underneath the platform itself.

An engineering study on a possible platform design has concluded that the flexure of the platform and the distortion of the cavern invert would sum to less than $\pm 2$ mm [49]. Two different types of transport systems are under study for the platform, air pads and high capacity rollers. In both cases, the platforms would be jacked onto the transport system to allow for the movement of a slightly undulated surface. In combination with a simple positive indexing mechanism, the platform with the detector can be positioned quickly within the required precision of 1 mm with respect to the beam axis. In parking or beam position, the platforms will be lowered on permanent supports. Trenches in the hall floor are needed for cable chains and for access to the platform undercarriage in case of maintenance.

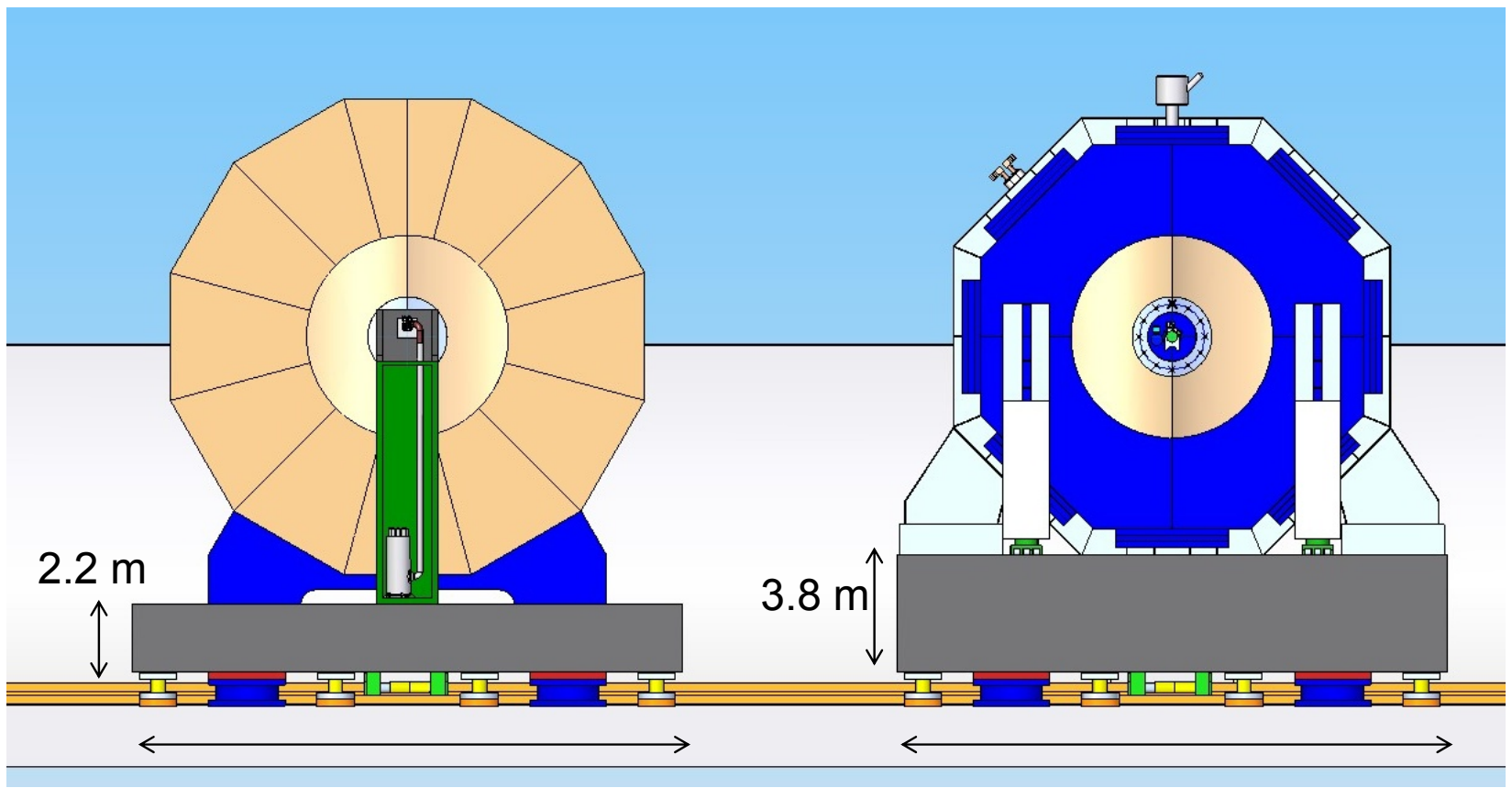

**Figure I-2.9**
Platform support concept for the push-pull system. Left - ILD; right - SiD

## 2.3.3 Shielding

### 2.3.3.1 Radiation

The ILC detectors are self-shielding with respect to ionising radiation that stems from maximum credible beam loss scenarios [50]. Additional shielding in the hall is necessary to fill the gap between the detector and the wall in the beam position. The design of this beam line shielding needs to accommodate both detectors, SiD and ILD, that are of significant size differences.

A common 'pac-man' design has been developed, where the movable shielding parts are attached to the wall of the detector hall - respectively to the tunnel stubs of the collider - and match to interface pieces that are borne by the experiments (c.f. Figure I-2.10).

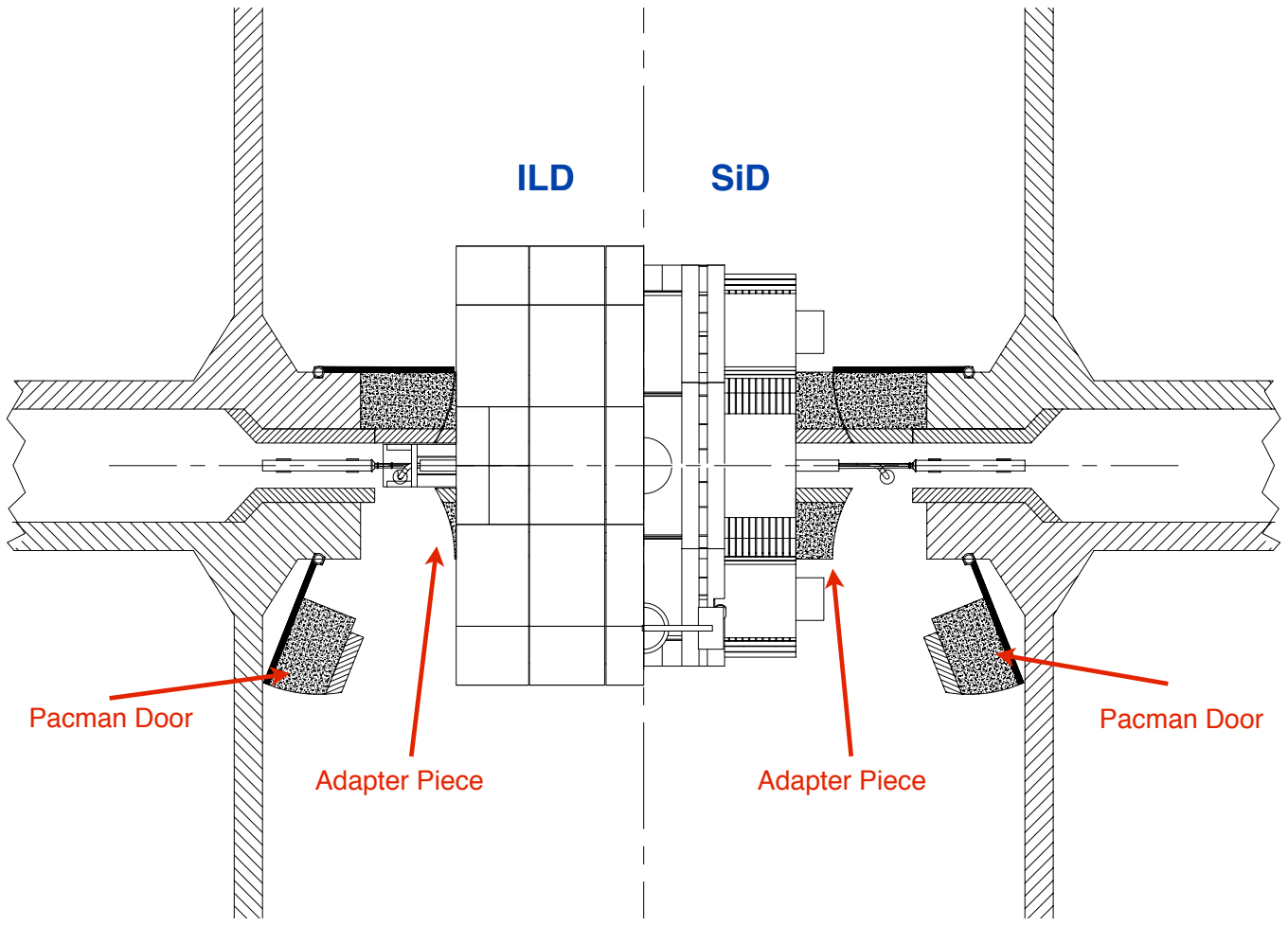

**Figure I-2.10**
Design of the beam line shielding compatible with two detectors of different sizes.

### 2.3.3.2 Magnetic fields

The magnetic stray fields outside the iron return yokes of the detectors need to be small enough to not disturb the other detector during operation or maintenance. A limit for the magnetic fields has been set to 5 mT at a lateral distance of 15 m from the beam line [141]. This allows the use of standard iron-based tools at the other detector. The design of the detector return yokes has been tested carefully for the fringe fields in simulations.

## 2.3.4 Detector installation schemes and timelines

The installation schemes of the detectors and the layout of the experimental areas on surface and underground depend on the geographical situation of the possible ILC sites. While the European and American sample sites assume a flat surface area, the Asian sample sites in Japan are located in the mountains where the requirements for the conventional facilities and buildings are different.

### 2.3.4.1 Flat surface ILC sites

In ILC sites with a flat surface, it is foreseen to have the underground experimental halls connected vertically with shafts to the surface area. In these conditions, the ILC detectors follow the assembly scheme that has been adopted by the CMS experiment at the LHC. The detectors will be pre-assembled, cabled and tested as much as possible in surface assembly buildings. The underground excavations and installations are thus done in parallel at the same time. Therefore the time schedule for the detector assembly, the civil construction, and the machine installation are mostly decoupled. Rather late in the construction period, about 1-2 years before the first beam is in the machine, the large detector parts will be lowered into the underground cavern through a large vertical shaft. The diameter of the shaft and the capacity of the temporary gantry crane for this procedure is defined by the largest detector part. This will be the central iron yoke ring of ILD with the mounted solenoid coil and installed barrel calorimeters. The big detector parts for both, ILD and SiD, can be loaded directly onto the respective platform. The final installation and commissioning of the detectors should then be performed in the maintenance areas of the underground cavern. Figure I-2.11 (top) shows a generic timeline for installation of the detectors in the flat surface sites.

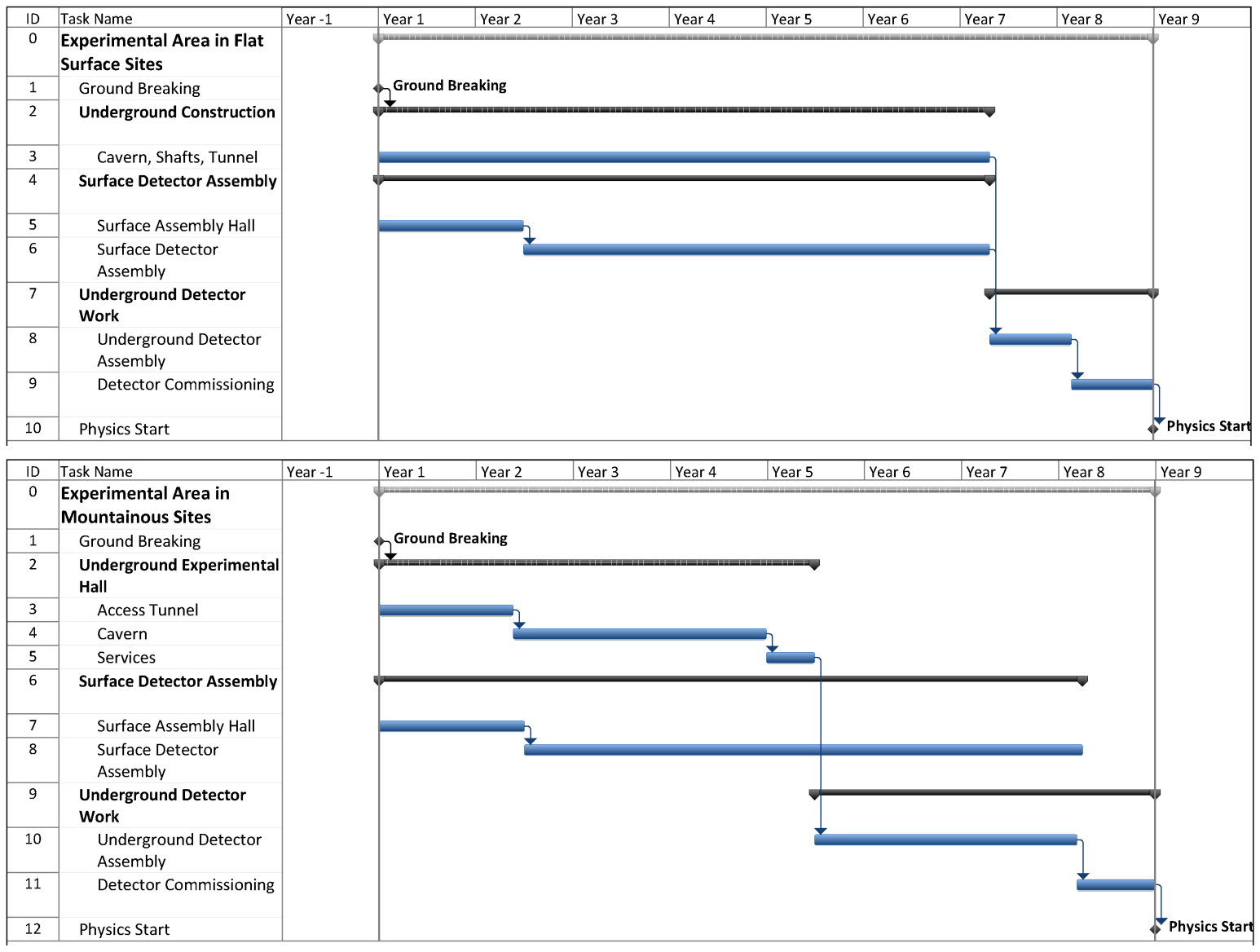

**Figure I-2.11** Generic detector assembly time lines for flat surface (top) and mountainous (bottom) ILC sites.

#### 2.3.4.2 Mountainous ILC sites

The ILC sites that are under study in Japan are in mountain regions. Therefore it is not possible to have vertical access shafts of ≈100 m length into the underground caverns. Instead, access will be provided by means of a horizontal access tunnel of ≈1 km length. The diameter of this tunnel will be given by the largest parts that need to be delivered into the experimental cavern in one piece. This would be the coil of the ILD detector solenoid that has a diameter of ≈9 m so the tunnel diameter would be in the order of 11 m. The transport system in the tunnel limits the mass of the parts to a maximum of ≈400 t.

Due to this boundary conditions, a modified detector installation scheme needs to be followed. In that case, still most parts of the detector would be pre-assembled and tested in the surface areas. However, more assembly work needs to be done underground. As for example the big yoke rings of ILD could not be transported through the tunnel, the assembly of the iron yoke needs to be done in the underground cavern. Also the installation of the solenoid and the calorimeters needs to be done in situ. Additional underground space and working time is needed in the mountain site cases of the ILC. Figure I-2.11 (bottom) shows a generic timeline for the installation of the detectors in the mountain sites. The timelines for the detector assemblies, the civil construction and the machine installation are interwoven.

### 2.3.5 Experimental area layout

The experimental area layouts for the different ILC sites need to fulfil the boundary conditions that are given by the installation schemes of the detectors, the needs for a safe and efficient running of the machine and both detectors in push-pull mode, and need to allow for efficient maintenance of the technical installations.

#### 2.3.5.1 Flat surface ILC sites

Figure I-2.12 shows the conceptual design of the underground experimental cavern for the flat surface ILC sites. The hall layout follows a Z-shape where the platforms transport the detectors perpendicular to the beam line. Each detector has a parking cavern where the detector could be opened for service and maintenance. One big 18 m diameter shaft enters the hall directly over the interaction point (IP). This shaft will be used for the initial assembly of both detectors. The large pre-assembled parts can be loaded directly onto the platforms. Two service shafts in the maintenance caverns will be used for services and for access in maintenance periods of one detector while the other one is taking data on the IP. Two smaller elevator shafts are foreseen for people and material transport as well as for safety egress.

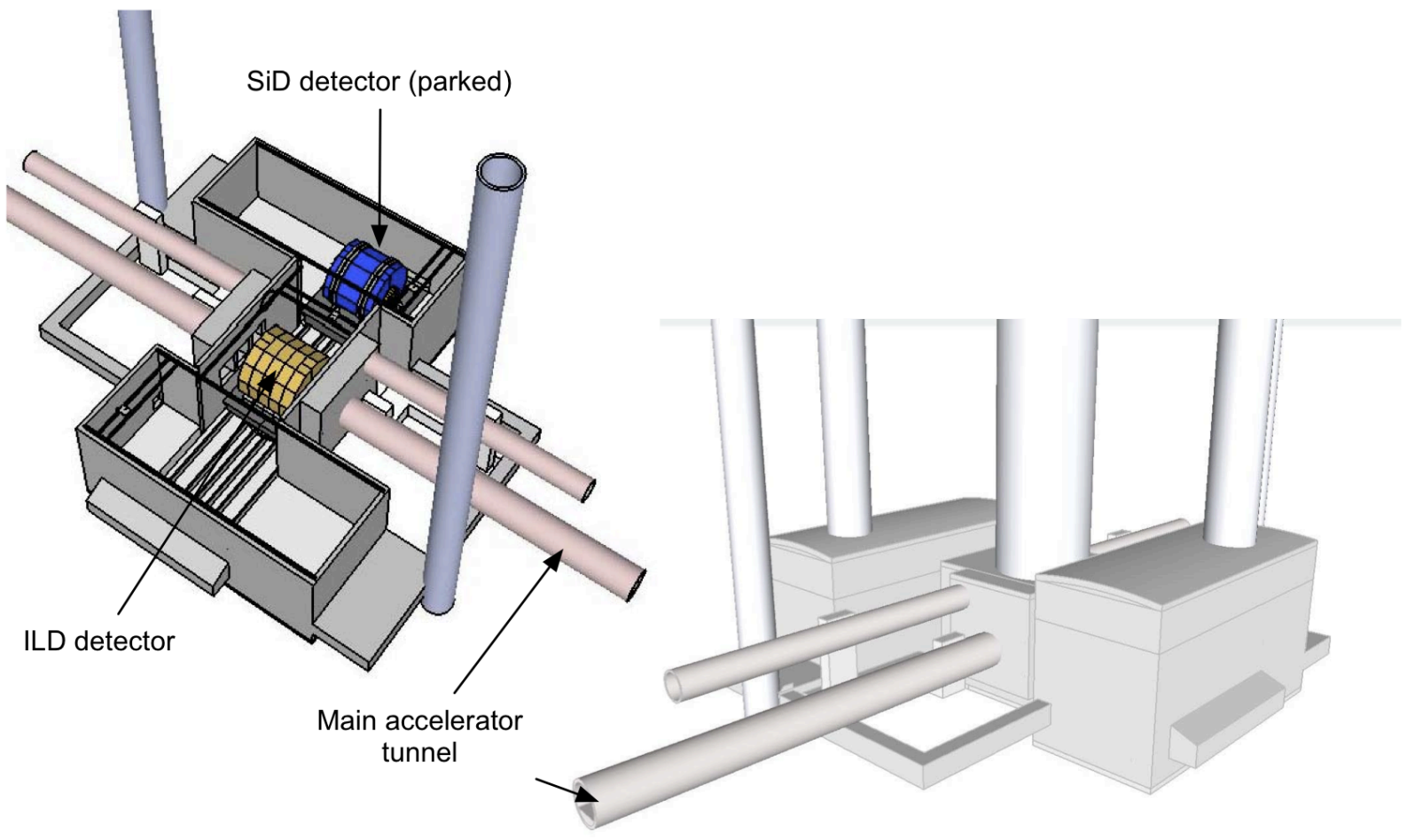

**Figure I-2.12** SiD and ILD in the experimental hall for the American (flat surface) ILC site.

#### 2.3.5.2 Mountainous ILC sites

The conceptual design of the experimental cavern for the mountainous ILC sites is shown in Figure I-2.13. The push-pull system will be very similar to the one in the flat surface case. Alcoves in the cavern enlarge the parking positions of the detectors to allow for the lateral opening and servicing of the detector parts. The access tunnel enters the hall twice, at the ILD and at the SiD side, to minimise the interference during the detector installation phase. The tunnel passes underneath the ILC beam line tunnel and extends towards the central region where the damping rings are located.

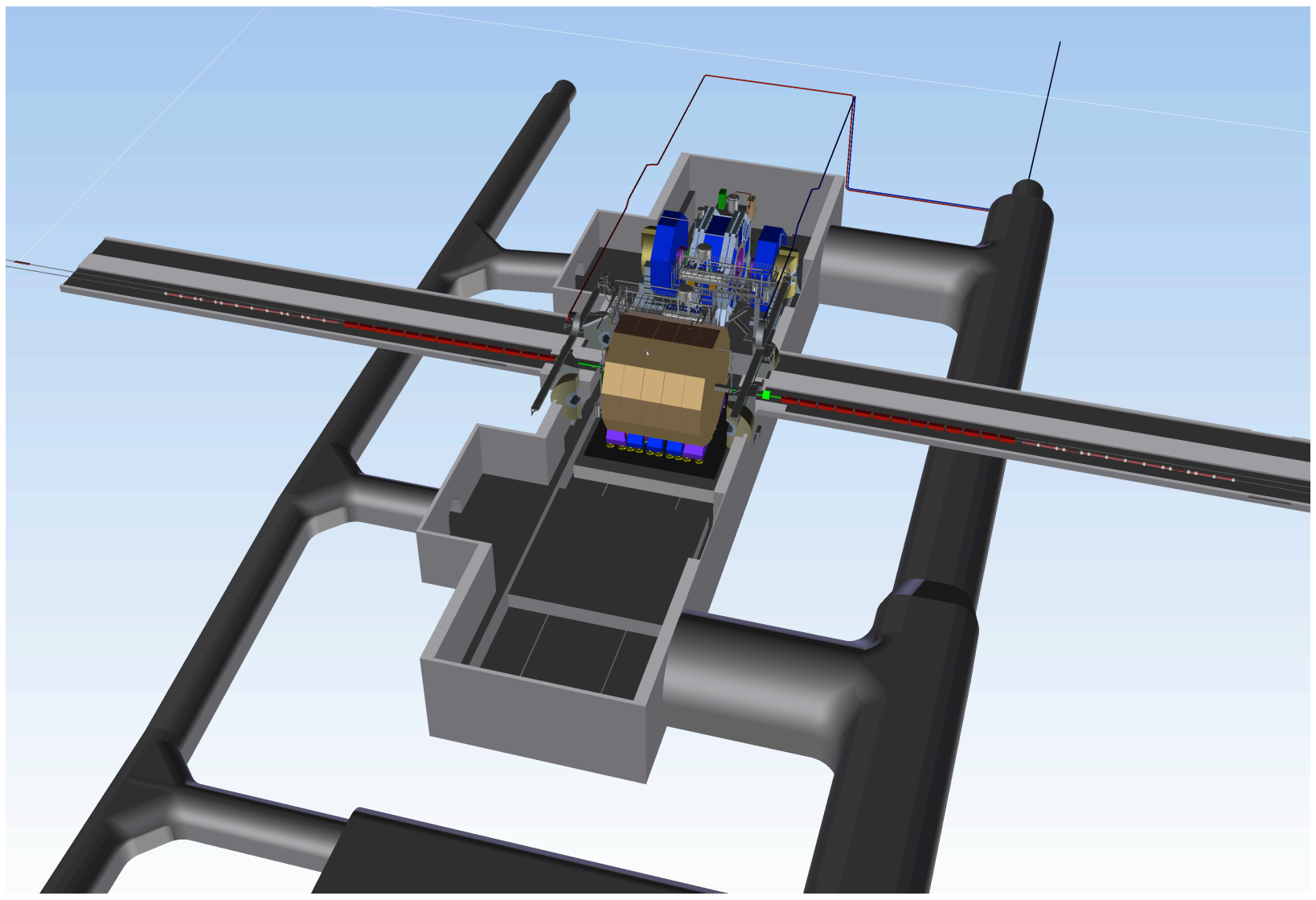

**Figure I-2.13** The experimental hall for mountainous ILC sites.

### 2.3.6 Detector services

A number of service and supply equipments needs to be established for the running and the maintenance of both detectors. The arrangement of the services depends on the technical requirements and can be sorted according to their proximity to the detector. Primary services should be located on the surface above the experimental hall (in the flat-surface sites) or in close-by service caverns (in the mountain site cases). They comprise of usually large and sometimes noisy facilities like water chillers, high voltage transformers, auxiliary power supplies, helium storage and compressors, and gas storage systems. Secondary services will be placed into the underground cavern in dedicated service areas. Examples are cooling water distributions, power supplies, gas mixture systems, power converters, and parts of the cryogenic system for the detectors. As the detectors will not be disconnected during the push-pull operations, all supplies that go directly to the detector will be run in flexible cable chains. The detectors will carry those services on-board that need to stay close, e.g. front-end electronics, patch panels, electronic containers.

Cryogenic helium for the superconducting solenoids and the QD0 magnets is foreseen to be supplied by a common system for both detectors. Two solutions are currently under study. In one, the liquid He is brought to the detectors via flexible cryogenic lines (c.f. Figure I-2.14). In that case, the cold boxes would be placed at service areas at the cavern walls. The second solution would place the cold boxes close to the detectors while gaseous He is supplied via flexible lines to the detector platforms. In each case, a re-cooler is placed on the platform of each detector for the 2K He supply of the QD0 magnets.

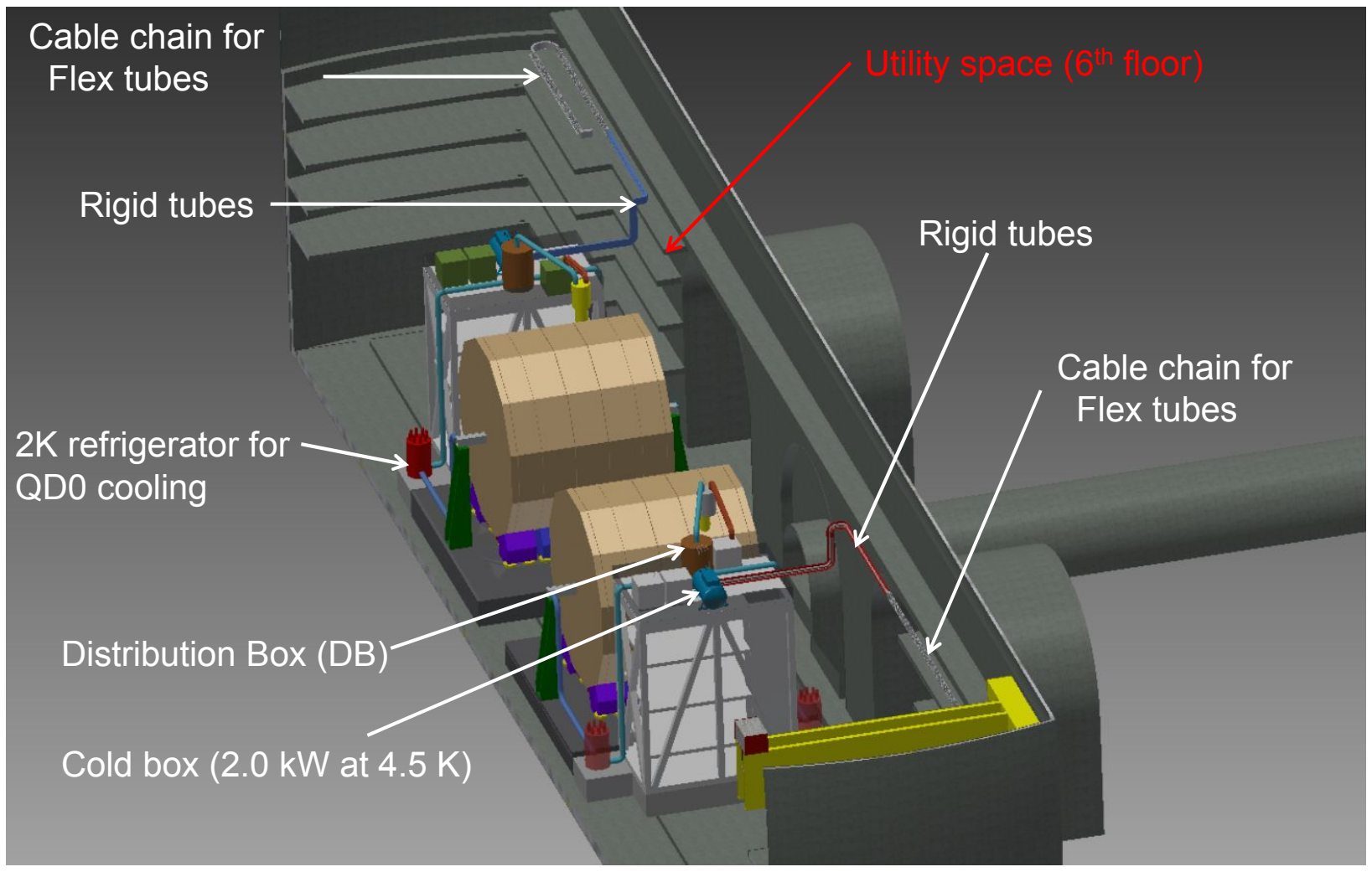

**Figure I-2.14** Common detector cryogenic system (study) with the cold boxes placed on service racks close to the detectors.

## 2.4 Beam Instrumentation

This section discusses the beam energy and polarisation measurement for the ILC. Since they are of crucial importance for the analysis of the $e^+e^-$ collision data, these devices are typically designed and operated by the detector collaborations. Located in the Beam Delivery System far away from the central detectors, they are shared between the two experiments and their cost is included in the accelerator costing. Additional information on these systems can be found in the ILC TDR and in [51].

### 2.4.1 Beam Energy Measurements

The ILC TDR design foresees redundant beam-based measurements of the incoming beam energy, capable of achieving a $10^{-4}$ accuracy, and of the energy spectrum of the disrupted beam after collisions. The measurements will be available in real time as a diagnostic tool to machine operators and will provide the basis for the determination of the luminosity-weighted centre-of-mass energy for physics analyses. Physics reference channels, such as a final state muon pair at the known $Z^0$ mass, are then foreseen to provide valuable cross checks of the collision scale, but only long after the data has been recorded.

#### 2.4.1.1 Upstream Energy Spectrometer

An energy spectrometer acts as a beam position monitor (BPM). It is located about $700$ m upstream of the interaction point, just after the energy collimation system. This spectrometer consists of four dipoles which introduce a fixed displacement of about 5 mm at the centre. Before, after and at the centre the beam line is instrumented with two or more cavity BPMs mounted on translation systems (so that the cavities can always be operated at their electromagnetic centre), shown in Figure I-2.15.

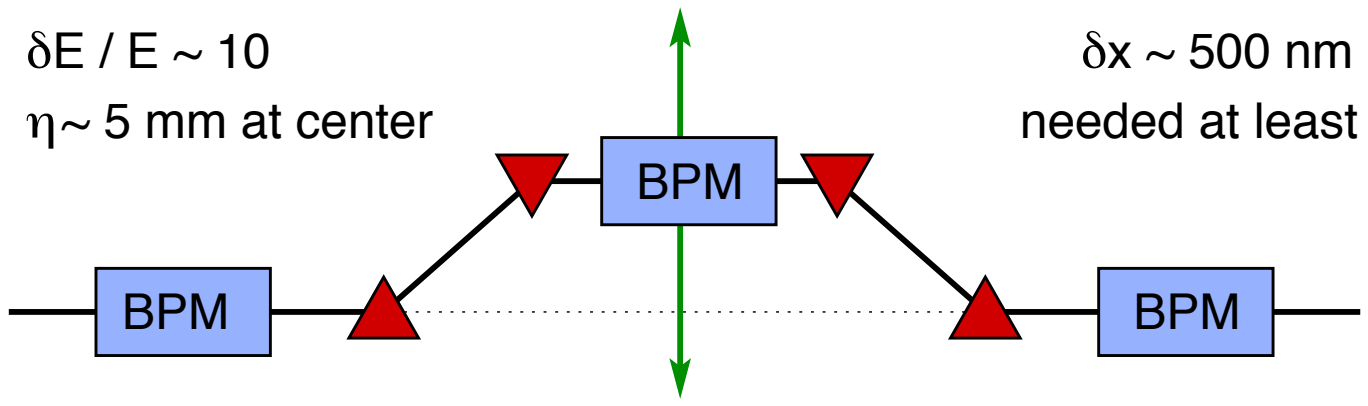

**Figure I-2.15** Schematic for the upstream energy spectrometer using beam position monitors.

With the four magnet chicane system, systematics effects produced by the magnets can be investigated, such as hysteresis and residual fields. The four magnet chicane also allows the spectrometer to be operated at different field strengths without disturbing the rest of the machine. It is important that the energy spectrometer be able to make precision energy measurements between $45.6$ GeV ($Z$-pole) and the highest ILC energy of 1 TeV. A precise measurement at $Z$-pole energies is of special importance since it defines the absolute energy scale. When operating the spectrometer with a fixed dispersion over the whole energy range, a BPM resolution of 0.5 $\mu$m is required.

A prototype test setup for such an energy spectrometer was commissioned in 2006 and 2007 in the T-474 experiment in the End Station A beamline at SLAC. The setup involved four dipole magnets and high-precision RF cavity BPMs in front, behind and in between the magnets. ESA test beams operated at 10 Hz with a bunch charge of $1.6 \cdot 10^{10}$ electrons, a bunch length of $500$ $\mu$m and an energy spread of $0.15\%$, i.e. with properties similar to ILC expectations. measurements normalised to the 5 mm dispersion (same dispersion as for the present ILC baseline energy spectrometer). The system turned out to be stable at the micron level over the course of one hour. When combining all the BPM stations to measure the precision of the orbit over the whole ESA-chicane beamline, a resolution of $0.8$ $\mu$m in $x$ and $1.2$ $\mu$m in $y$ was achieved [52]. This translates to a relative energy resolution of $5.5 \cdot 10^{-4}$ [53].

This result can be improved further by employing more precise BPMs. At high energies, the energy resolution is directly limited by the BPM resolution. Due to the fixed dispersion design, the running at lower energies, especially at the $Z$-pole, requires the chicane magnets to be operated at low fields, where the magnetic field measurement may not be accurate enough. A BPM resolution of $20$ nm would allow the chicane dipoles to be run at the same magnetic field for both the $Z$-pole and highest energy operation. This type of single shot accuracy has recently been demonstrated with the cavity BPM system at ATF2 [54].

### 2.4.1.2 Extraction Line Energy Spectrometer

The ILC Extraction-Line Spectrometer (XLS) design [55] is largely motivated by the experience of the Wire Imaging Synchrotron Radiation Detector (WISRD) at the SLC [56]. The WISRD measured the distance between two synchrotron stripes created by horizontal bend magnets which surrounded a precisely-measured dipole that provided a vertical bend proportional to the beam energy. The WISRD achieved a precision of $\Delta E_b/E_b \sim 2 \cdot 10^{-4}$ (200 ppm), where the limiting systematic errors were due to relative component alignment and magnetic field mapping.

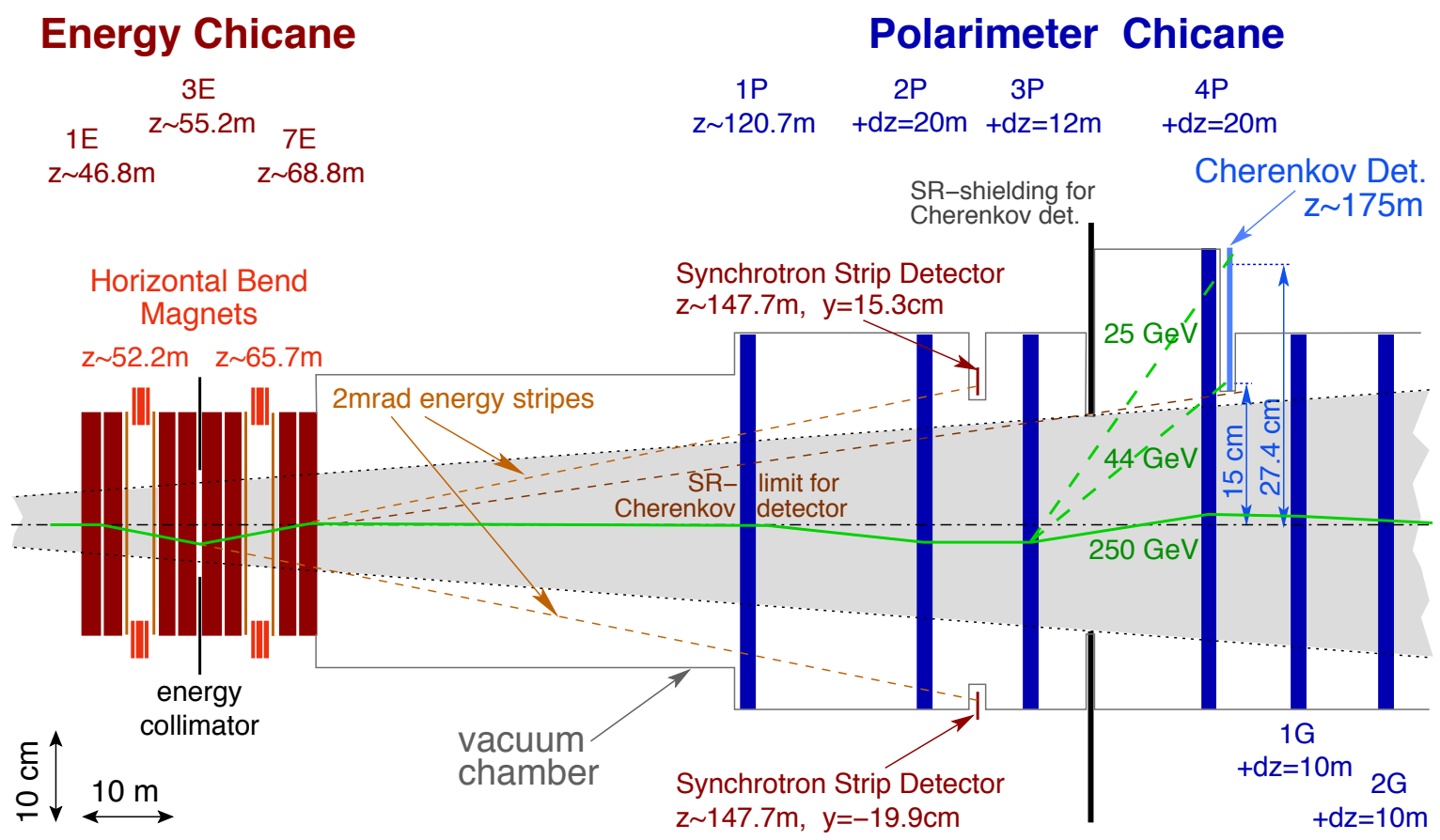

**Figure I-2.16** Schematic of the ILC extraction line diagnostics for the energy spectrometer and the Compton polarimeter.

The analysing dipole for the XLS is provided by a vertical chicane just after the capture quad section of the extraction line, about $55$ m downstream of the interaction point (see Figure I-2.16). The chicane provides a $\pm 2$ mrad vertical bend to the beam and in both legs of the chicane horizontal wiggler magnets are used to produce the synchrotron light needed to measure the beam trajectory. The optics in the extraction line is designed to produce a secondary focus about $150$ m downstream of the IP, which coincides with the centre of the polarimeter chicane and the Compton interaction point. The synchrotron light produced by the wigglers will also come to a vertical focus at this point, and position-sensitive detectors in this plane arrayed outside the beampipe will measure the vertical separation between the synchrotron stripes. With a total bend angle of $4$ mrad, and a flight distance of nearly $100$ m, the synchrotron stripes will have a vertical separation of $400$ mm, which must be measured to a precision of $40$ $\mu$m to achieve the target accuracy of $10^{-4}$. In addition to the transverse separation of the synchrotron stripes, the integrated bending field of the analysing dipole also needs to be measured and monitored to a comparable precision of $10^{-4}$. The distance from the analysing chicane to the detectors needs to only be known to a modest accuracy of 1 cm. For the XLS spectrometer, it has been proposed to use an array of radiation-hard $100$ $\mu$m quartz fibres. These fibres do not detect the synchrotron light directly, but rather detect Cherenkov radiation from secondary electrons produced when the hard photons interact with material near the detector. At ILC beam energies, the critical energy for the synchrotron radiation produced in the XLS wigglers is several tens of MeV, well above the pair-production threshold, and copious numbers of relativistic electrons can be produced with a thin radiator in front of the fibre array. The leading candidates for reading out these fibres are multi-anode PMs from Hamamatsu, similar in design to those used in scintillating fibre calorimeters. The advantage of this scheme over wires (as used in the SLC energy spectrometer) is to produce a reliable, passive, radiation-hard detector which does not suffer from cross talk or RF pickup, and still allows for easy gain adjustment and a large dynamic range. A more traditional wire-based detector could also be considered instead of (or in addition to) the quartz fibre detector. The energy spectrum of the beam after collision contains a long tail as a result of the beam-beam disruption in the collision process. This disrupted beam spectrum is not a direct measure of the collision energy spectrum, but it is produced by the same physical process, and direct observation of this disrupted tail will serve as a useful diagnostic for the collision process. The position-sensitive detector in the XLS is designed to measure this beam energy spectrum down to $50\%$ of the nominal beam energy. Near the peak, for a beam energy of $E_b = 250$ GeV, each $100$-micron fibre spans an energy interval of 125 MeV. Given a typical beam energy width of $0.2\%$, this means the natural width of the beam energy will be distributed across at least a handful of fibres, which will allow the centroid to be determined with a precision better than the fibre pitch, and some information about the beam energy width can be extracted as well.

## 2.4.2 Polarisation Measurements

The ILC TDR design foresees redundant beam-based measurements of the incoming beam polarisation and of the polarisation of the disrupted beam after collisions. The measurements will be available in real time as a diagnostic tool to machine operators and will provide the basis for the determination of the luminosity-weighted polarisation for physics analyses. Physics reference channels, such as $W$ pair production, are then foreseen to provide valuable cross checks of the luminosity-weighted polarisation scale, but only long after the data has been recorded. The systems have been designed to reach a final precision of $10^{-3}$ on the luminosity-weighted polarisation.

### 2.4.2.1 Upstream Polarimeter

The upstream Compton polarimeter is located at the beginning of the Beam Delivery System, upstream of the tuneup dump $1800$ m before the $e^+e^-$ IP. In this position it benefits from clean beam conditions and very low backgrounds. The upstream polarimeter configuration is shown in Figure I-2.17. It will provide fast and precise measurements of the polarisation before collisions. The beam direction at the Compton IP in both the vertical and horizontal must be the same as that at the IP within a tolerance of $\sim$ 50 $\mu$rad. The parameters for the upstream chicane and Cherenkov detector [57] were chosen such that the entire Cherenkov spectrum can be measured for all beam energies while still keeping the Cherenkov detector at a clearance of 2 cm with respect to the beam pipe.

The upstream polarimeter can be equipped with a laser similar to one used at the TTF/Flash source in operation at DESY. It can have the same pulse structure as the electron beam allowing measurements of every bunch. This permits fast recognition of polarisation variations within each bunch train as well as time-dependent effects that vary train-by-train. The statistical precision of the polarisation measurement is estimated to be $3\%$ for any two bunches with opposite helicity, leading to an average precision of $1\%$ for each bunch position in the train after the passage of only 20 trains (4 seconds). The average over two entire trains with opposite helicity will have a statistical error of $\Delta P/P = 0.1\%$.

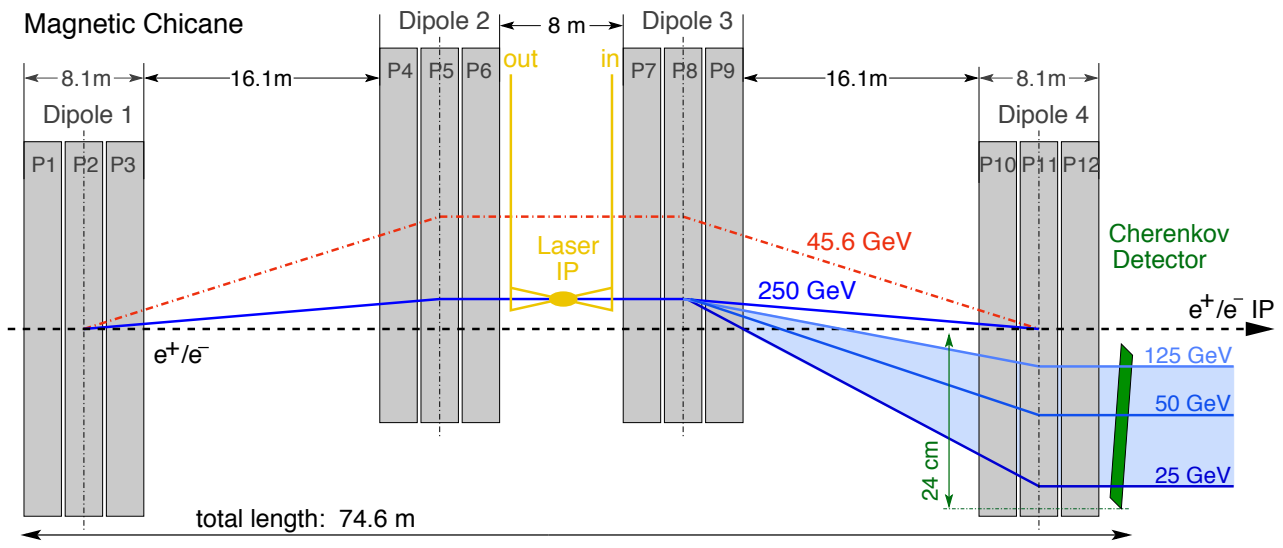

**Figure I-2.17** Schematic of the upstream polarimeter chicane.

## 2.4.2.2 Downstream Polarimeter

The downstream polarimeter, shown in Figure I-2.16, is located 150 m downstream of the IP in the extraction line and on axis with the IP and IR magnets. It can measure the beam polarisation both with and without collisions, thereby testing the calculated depolarisation due to collisions and the spin tracking. The downstream polarimeter chicane further accommodates a detector for the downstream energy spectrometer and provides magnetic elements for the GAMCAL system.

In order for the downstream Cherenkov detector to avoid the synchrotron radiation fan from the $e^+e^-$ IP (extending about $15$ cm from the beam pipe, see Figure I-2.16), the downstream dipole magnets are larger and have much higher fields. In addition, magnets 3P and 4P are operated at higher fields (compared to magnets 1P and 2P) in order to bend the scattered electrons further from the main beam axis. Therefore, two additional magnets (1G and 2G) are needed to bring the main beam back to its original trajectory.

The laser for the downstream polarimeter requires high pulse energies to overcome the substantially larger backgrounds in the extraction line. Three $5$ Hz laser systems will be used to generate Compton collisions for three out of $2800$ bunches in a train. Each laser is an all solid-state diode-pumped Nd:YAG, with a fundamental wavelength of $1064$ nm that will be frequency-doubled to $532$ nm. Each laser will sample one particular bunch in a train for a time interval of a few seconds to a minute, then select a new bunch for the next time interval, and so on in a pre-determined pattern. The Compton statistics are high with more than $1000$ Compton-scattered electrons per bunch in a detector channel at the Compton edge. With this design, a statistical uncertainty of less than $1\%$ per minute can be achieved for each of the measured bunches. This is dominated by fluctuations in Compton luminosity due to beam jitter and laser targeting jitter and to possible background fluctuations.

Background studies have been carried out for disrupted beam losses and for the influence of synchrotron radiation (SR). There are no significant beam losses for the nominal RDR ILC parameter set and beam losses still look acceptable for the nominal TDR beam parameters based on the low power option of the RDR. An SR collimator protects the Compton detector and no significant SR backgrounds are expected.

### 2.4.3 Luminosity weighted averages and correlations

There is a strong complementarity between beam-based instrumentation and physics reference reactions which are sensitive to collision energy and beam polarisation. While beam-based measurements generally provide higher statistics and more immediate feedback, physics reactions naturally provide a luminosity-weighted sampling of conditions over long time scales. In general, both pieces of information are necessary to achieve the physics goals of the ILC, and it is expected that a mixture of beam-based and physics reference reactions will be used to determine the collision energy spectrum and the beam polarisation values. One such reference reaction, namely the determination of the luminosity weighted long-term average of the beam polarisation for $W^+W^-$ production, has been chosen for benchmarking the ILD and SiD performance.

Even without the necessity of instrumentation to provide fast feedback for operations, beam-based instrumentation also provides crucial information to measure and constrain possible correlations between the collision energy, luminosity, and beam polarisation which are typically impossible to measure with the available statistics in physics reference reactions. Correlations between these parameters can arise due to long-term drifts in the machine, bunch-to-bunch variations along the bunch train, and even due to beam-beam interactions in the highly disrupted collision process.

One concrete example is the luminosity as a function of energy used at various steps in a threshold scan to measure the top quark mass. In addition to the luminosity-weighted average collision energy at each scan point, it is also necessary to know the shape of the luminosity spectrum at each point in detail, which includes any correlations between beam energy jitter and delivered luminosity. To achieve a relative accuracy of $\mathcal{O}(10^{-4})$ on the top quark mass, these correlations must be understood and controlled to a degree which requires detailed beam instrumentation to be able to measure these correlations directly. Similar arguments can be made for understanding beam polarisation, where direct correlations with energy and luminosity can arise due to the large spin precession of highly relativistic electrons in magnetic fields. Being able to correlate changes in the polarisation alignment due to final-focus orbit drifts with delivered luminosity or collision energy, for instance, may be an important systematic for high-precision measurements.

## 2.5 Common Engineering Tools

The design and integration of the ILD and SID detectors, together with the push-pull requirements asking for an unprecedented amount of infrastructures shared by the two experiments, call for common engineering tools enabling a consistent sharing of engineering documents like interfaces, radiation and magnetic field maps and specifications of CFS equipments.

ILC-EDMS (Engineering Data Management System) is a fully web based system which has the features answering these needs. It is promoted by the GDE and supported and operated at DESY ([58, 59, 60, 61] and links).

Among other key features, it will allow the international community to collaboratively design components using evolving CAD models and view the results using visualisation tools (see Figure I-2.18

All the data stored on EDMS have well defined life-cycle managed by the owner and shared with the relevant distribution lists. Documents can be kept as temporary, released or obsolete along the evolution of the project life.

The organisation of the ILC EDMS for detectors implies an efficient and logical description of the projects under a WBS (Work Breakdown Structure), defined under the responsibility of each concept group.

As a result, the ILC-EDMS include by now a level devoted to detectors, subdivided in one sub-level per detector, and specific workspace to manage the interaction between detectors and the civil engineering of the ILC facility. This node already contains some material for studies of the mechanism of push-pull, and the dimensioning of the hall and services. (see Figure I-2.19.)

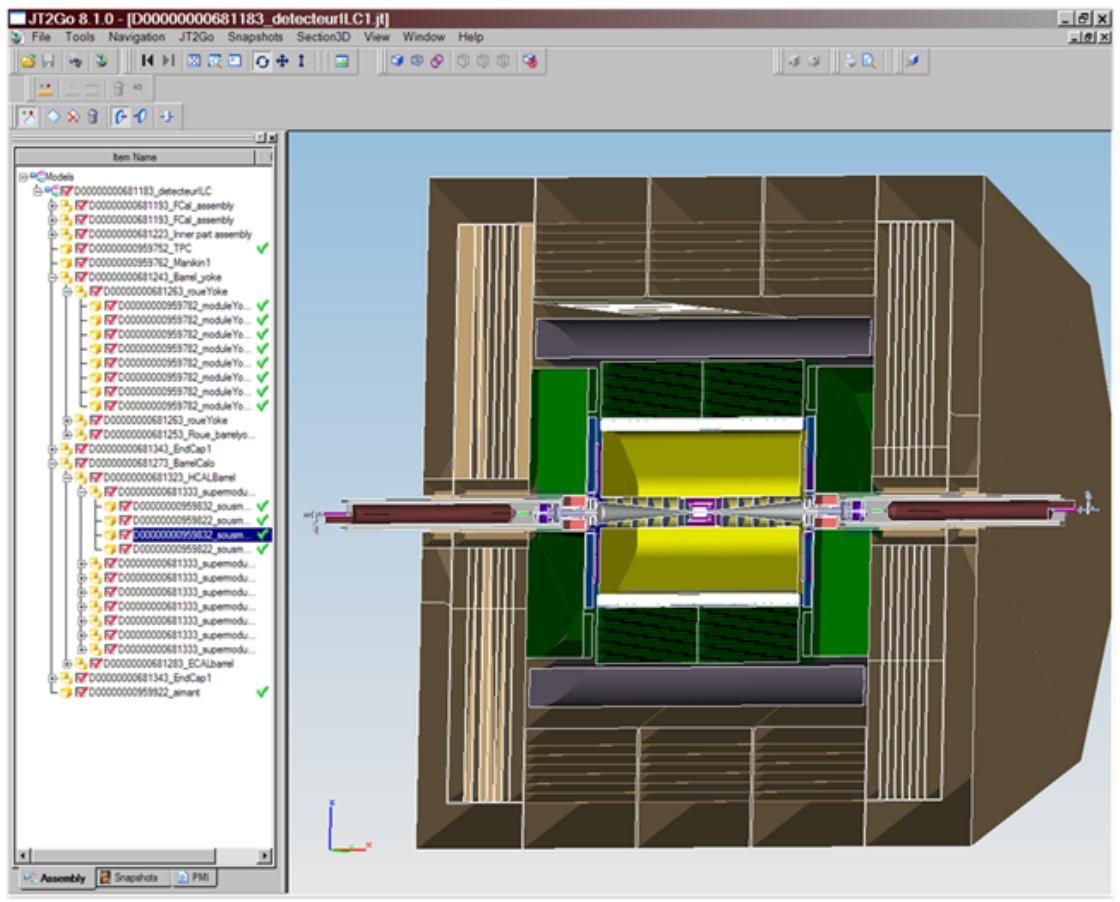

**Figure I-2.18**
Cross-sectional view of the ILD detector using the EDMS visualisation tool.

The detector top node in ILC-EDMS and the details for the two detectors SID and ILD is ready to operate. It will become an important and major tool for the future of the ILC detectors as a collaborative and management tool. It is already organised to efficiently follow the two detectors projects among each step of their life. The use of this EDMS provides the selection, definition and tracking of the mandatory documentation but the management of this documentation is a major issue and needs to be established [62].

## 2.6 Detector Costing Methodology

Costing is a key element of detector design and will become crucial for the detector approval and its construction. The performance of a detector, in general, is correlated with the cost, which normally will be bound by the resources. Thus, together with the technical aspects, realistic costing needs to be considered in the design work although the present designs are not meant for commitment for construction as announced by ILCSC in the call of LOIs.

The validated groups seriously worked on the cost estimation of their designed detectors. There are several difficulties, however, in the study which limit the precision and maturity of the cost estimation. First, it is still unknown when the detectors will be built. The prices of some raw materials vary with time and depending on the world economy. The variation may exceed the range of over-all inflation rate. Also technological advance or mass production could reduce the cost in a favourable direction. These make the long range extrapolation of the cost difficult. Second, there is another complexity for costing in that the detectors will be built by large international collaborations where the funding schemes and the costing methods are different among the participating institutions. Here also arises the question of currency exchange rates, which change often unstably, in estimating the total cost in one particular currency. The number of participating institutions and their counties will be very many, exceeding those of accelerator participation, and the unstable exchange rates requires careful handling in producing an total cost estimation, while in practice they will not cause a serious problem at the time of construction because most components will be provided by in-kind contribution. The HEP community has rich experience for building large collider detectors in big international collaborations handling these matters smoothly.

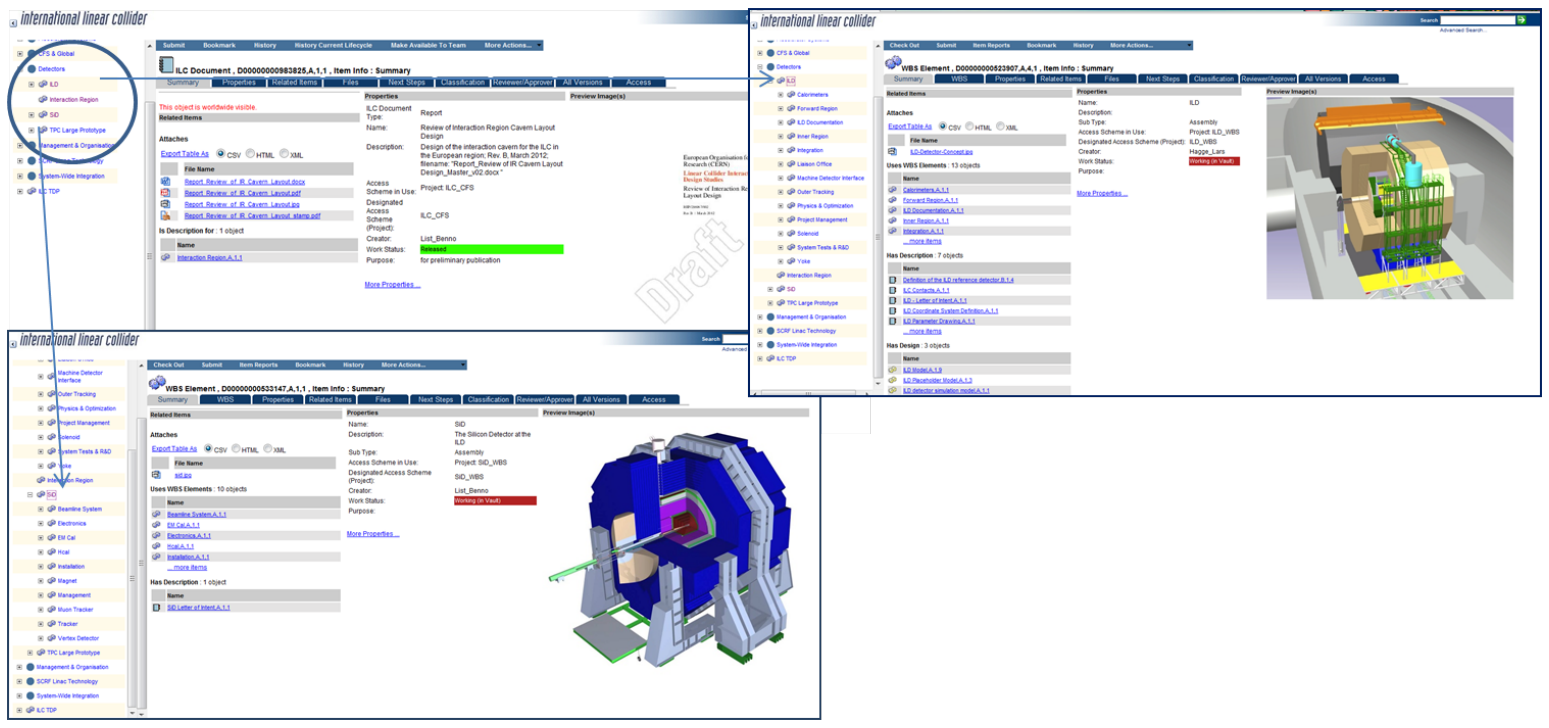

**Figure I-2.19**
The detector top node in the ILC-EDMS and the details for the two detectors SID and ILD

Both detector concepts made their best effort for costing under such circumstances. We also made an effort to make the costing method comparable between the two groups. However, there remain some differences between the groups on top of the above mentioned uncertainties. For the same reasons, the detector cost estimates should be compared with that of the accelerator with a caution. While we tried to present the detector costs in a similar way, they are less matured and the funding scheme will be different.

A common costing working group was formed for a close communication between the groups. The working group invited an adviser member from the accelerator team to deploy a consistent methodology to handle some of the issues, e.g. handling the exchange rate variations. One member from each detector group served in the CLIC detector costing, too. This was effective to coordinate indirectly between the costings of ILC detectors and CLIC detectors.

The group agreed about a few guidelines for the costing method and its presentation to be compatible as much as possible so that the delivered costs can be approximately compared even though the exact costing methods are not the same:

1. the detector cost will be provided in the unit of 2012 ILCU like the accelerator cost. Those component costs which were estimated in the past are converted to the 2012 cost by taking the appropriate inflation rate;
2. some costs of raw materials are fixed between the groups. While the items are not many, these cover a fair part of the total cost. They include tungsten for the calorimeter, two types of steel for the yoke, and the Si detector sensors. These raw material costs were used first for the CLIC detectors in the CLIC CDR;
3. the material cost and manpower cost are listed separately. The man power cost does not include in-house labour of participating Institutions;
4. where contingency needs to be indicated explicitly, it is listed separately. This depends on the country and SiD group followed this line;
5. the platform to be used for push-pull system is costed as a CFS component and is not included in the detector cost;
6. the electric power and the cooling water are assumed to be delivered in the experimental hall;
7. where currency exchange rates are needed, purchase power parity of different currencies of OECD is used as is done for the accelerator costing.

The groups have discussed costing of the large solenoids in some detail. While the both groups referred to the CMS magnet, the assumed construction models were different mainly depending on the past experiences. Nevertheless, each group understands how the other group estimates its magnet cost, and the given numbers look consistent.

Although the physics aims are similar, each group had its own design philosophy which lead to different designs of the detectors. For example there are different selection of the components, operation parameters and their sizes. In the costing section of each group, the subdivisions of the components is left free but the listed categorisation of the component costs are very similar allowing each item to be compared. Where differences are seen, the reason can be understood.

# Volume 4

# Detectors

## Part II

## SiD Detailed Baseline Design

# SiD Editors

Main Editors
P.N. Burrows, L. Linssen, M. Oreglia, M. Stanitzki, A. White

Vertex Detector
W. Cooper, R Lipton

Silicon Tracking
W. Cooper, M. Demarteau, T. Nelson

Calorimetry
R. Frey, A. White, L. Xia

Muon System
H. Band, G. Fisk

Superconducting Magnet System
W. Craddock, M. Oriunno

Engineering, Integration and the Machine Detector Interface
P. Burrows, T. Markiewicz

Forward Systems
T. Maruyama, B. Schumm

Electronics and DAQ
M. Breidenbach, G. Haller, M. Stanitzki

Simulation and Reconstruction
N. Graf, J. Strube

Benchmarking
D. Asner, T. Barklow, P. Roloff

Costs
M. Breidenbach

# SiD Introduction

The next generation of collider detectors, to study electroweak symmetry breaking and potential discoveries beyond the Standard Model, will demand a high level of precision in the measurement of physics processes. SiD was conceived as a fully integrated, unified design with the basic features of compactness, silicon-based tracking, fine-grained calorimetry and a high central magnetic field. Building on extensive experience with previous detectors, and exploiting major advances in sensors, materials, and electronics, this design has been developed for experiments at a future linear collider.

SiD is the result of many years of creative design by physicists and engineers, backed up by a substantial body of past and ongoing detector research and development. While each component has benefitted from continual development, the SiD design integrates these components into a complete system for excellent measurements of jet energies, based on the Particle Flow Algorithm (PFA) approach, as well as of charged leptons, photons and missing energy. The use of robust silicon vertexing and tracking makes SiD applicable to lepton colliders spanning a wide energy range, from a Higgs factory to multi-TeV machines. SiD has been designed in a cost-conscious manner, with the compact design that minimises the volumes of high-performing, high-value, components. The restriction on dimensions is offset by the relatively high central magnetic field from a superconducting solenoid.

This Detailed Baseline Design builds on the results presented in our earlier Letter of Intent [63]. We present an overview of the SiD Concept, its design philosophy, and the approach to the development of each component. We present detailed discussions of each of the SiD subsystems, an overview of the full GEANT4 description of SiD, the status of the tracking and calorimeter reconstruction algorithms, studies of subsystem performance based on these tools, results of physics benchmark analyses, an estimate of the cost of the detector, and an assessment of the research and development needed to provide the technical basis for an optimised SiD design. While detector and physics studies continue, we regard this document as a substantive starting point for the development of a full Technical Design Report.

# Chapter 1
# SiD Concept Overview

## 1.1 SiD Philosophy

SiD [63] is a general-purpose detector designed to perform precision measurements at a Linear Collider. It satisfies the challenging detector requirements that are described in the Common Section. SiD is based on the PFA paradigm, an algorithm by which the reconstruction of both charged and neutral particles is accomplished by an optimised combination of tracking and calorimetry. The net result is a significantly more precise jet energy measurement that results in a di-jet mass resolution good enough to distinguish between W and Z hadronic decays.

SiD (Figures II-1.1, II-1.2) is a compact detector based on a powerful silicon pixel vertex detector, silicon tracking, silicon-tungsten electromagnetic calorimetry (ECAL) and highly segmented hadronic calorimetry (HCAL). SiD also incorporates a high-field solenoid, iron flux return, and a muon identification system. The use of silicon sensors in the vertex, tracking and calorimetry enables a unique integrated tracking system ideally suited to particle flow.

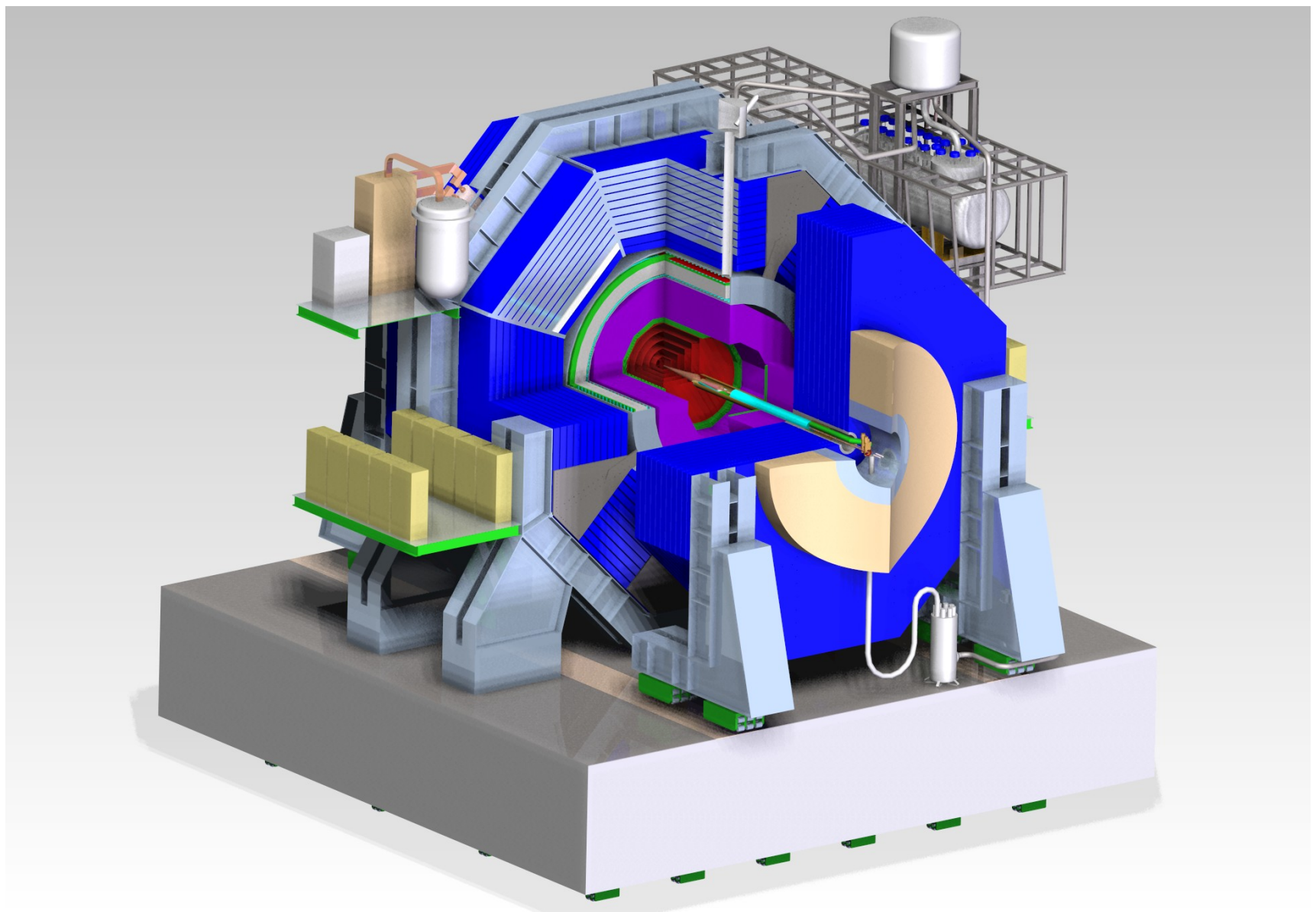

**Figure II-1.1**
SiD on its platform, showing tracking (red), ECAL (green), HCAL (violet) and flux return (blue).

The choice of silicon detectors for tracking and vertexing ensures that SiD is robust with respect to beam backgrounds or beam loss, provides superior charged particle momentum resolution, and eliminates out-of-time tracks and backgrounds. The main tracking detector and calorimeters are "live" only during each single bunch crossing, so beam-related backgrounds and low-$p_\mathrm{T}$ backgrounds from γγ processes will be reduced to the minimum possible levels. The SiD calorimetry is optimised for excellent jet energy measurement using the PFA technique. The complete tracking and calorimeter systems are contained within a superconducting solenoid, which has a 5 T field strength, enabling the overall compact design. The coil is located within a layered iron structure that returns the magnetic flux and is instrumented to allow the identification of muons.

All aspects of SiD are the result of intensive and leading-edge research aimed at achieving performance at unprecedented levels. At the same time, the design represents a balance between cost and physics performance. The key parameters of the SiD design are listed in Table II-1.1.

**Table II-1.1**
Key parameters of the baseline SiD design. (All dimension are given in cm).

| SiD Barrel | Technology | Inner radius | Outer radius | z extent |
|---|---|---|---|---|
| Vertex detector | Silicon pixels | 1.4 | 6.0 | $\pm$ 6.25 |
| Tracker | Silicon strips | 21.7 | 122.1 | $\pm$ 152.2 |
| ECAL | Silicon pixels-W | 126.5 | 140.9 | $\pm$ 176.5 |
| HCAL | RPC-steel | 141.7 | 249.3 | $\pm$ 301.8 |
| Solenoid | 5 Tesla SC | 259.1 | 339.2 | $\pm$ 298.3 |
| Flux return | Scintillator-steel | 340.2 | 604.2 | $\pm$ 303.3 |
| **SiD Endcap** | **Technology** | **Inner z** | **Outer z** | **Outer radius** |
| Vertex detector | Silicon pixels | 7.3 | 83.4 | 16.6 |
| Tracker | Silicon strips | 77.0 | 164.3 | 125.5 |
| ECAL | Silicon pixel-W | 165.7 | 180.0 | 125.0 |
| HCAL | RPC-steel | 180.5 | 302.8 | 140.2 |
| Flux return | Scintillator/steel | 303.3 | 567.3 | 604.2 |
| LumiCal | Silicon-W | 155.7 | 170.0 | 20.0 |
| BeamCal | Semiconductor-W | 277.5 | 300.7 | 13.5 |

## 1.2 Silicon-based Tracking

The tracking system is a key element of the ILC detector concepts. The particle flow algorithm requires excellent tracking with superb efficiency and two-particle separation. The requirements for precision measurements, in particular in the Higgs sector, place high demands on the momentum resolution at the level of $\delta(1/p_{\mathrm{T}}) \sim 2 - 5 \times 10^{-5}$/GeV/$c$.

Highly efficient charged particle tracking is achieved using the pixel detector and main tracker to recognise and measure prompt tracks, in conjunction with the ECAL, which can identify short track stubs in its first few layers to catch tracks arising from secondary decays of long-lived particles. With the choice of a 5 T solenoidal magnetic field, in part chosen to control the $\mathrm{e^+e^-}$ pair background, the design allows for a compact tracker design.

### 1.2.1 Vertex detector

To unravel the underlying physics mechanisms of new observed processes, the identification of heavy flavours will play a critical role. One of the main tools for heavy flavour identification is the vertex detector. The physics goals dictate an unprecedented spatial three-dimensional point resolution and a very low material budget to minimise multiple Coulomb scattering. The running conditions at the ILC impose the readout speed and radiation tolerance. These requirements are normally in tension. High granularity and fast readout compete with each other and tend to increase the power dissipation. Increased power dissipation in turn leads to an increased material budget. The challenges on the vertex detector are considerable and significant R&D is being carried out on both the development of the sensors and the mechanical support.

The SiD vertex detector uses a barrel and disk layout. The barrel section consists of five silicon pixel layers with a pixel size of $20 \times 20\ \mu\mathrm{m}^2$. The forward and backward regions each have four silicon pixel disks. In addition, there are three silicon pixel disks at a larger distance from the interaction point to provide uniform coverage for the transition region between the vertex detector and the outer tracker. This configuration provides for very good hermeticity with uniform coverage and guarantees excellent charged-track pattern recognition capability and impact parameter resolution over the full solid angle. This enhances the capability of the integrated tracking system and, in conjunction with the high magnetic field, makes for a very compact system, thereby minimising the size and costs of the calorimetry.

**Figure II-1.2**
SiD quadrant view.

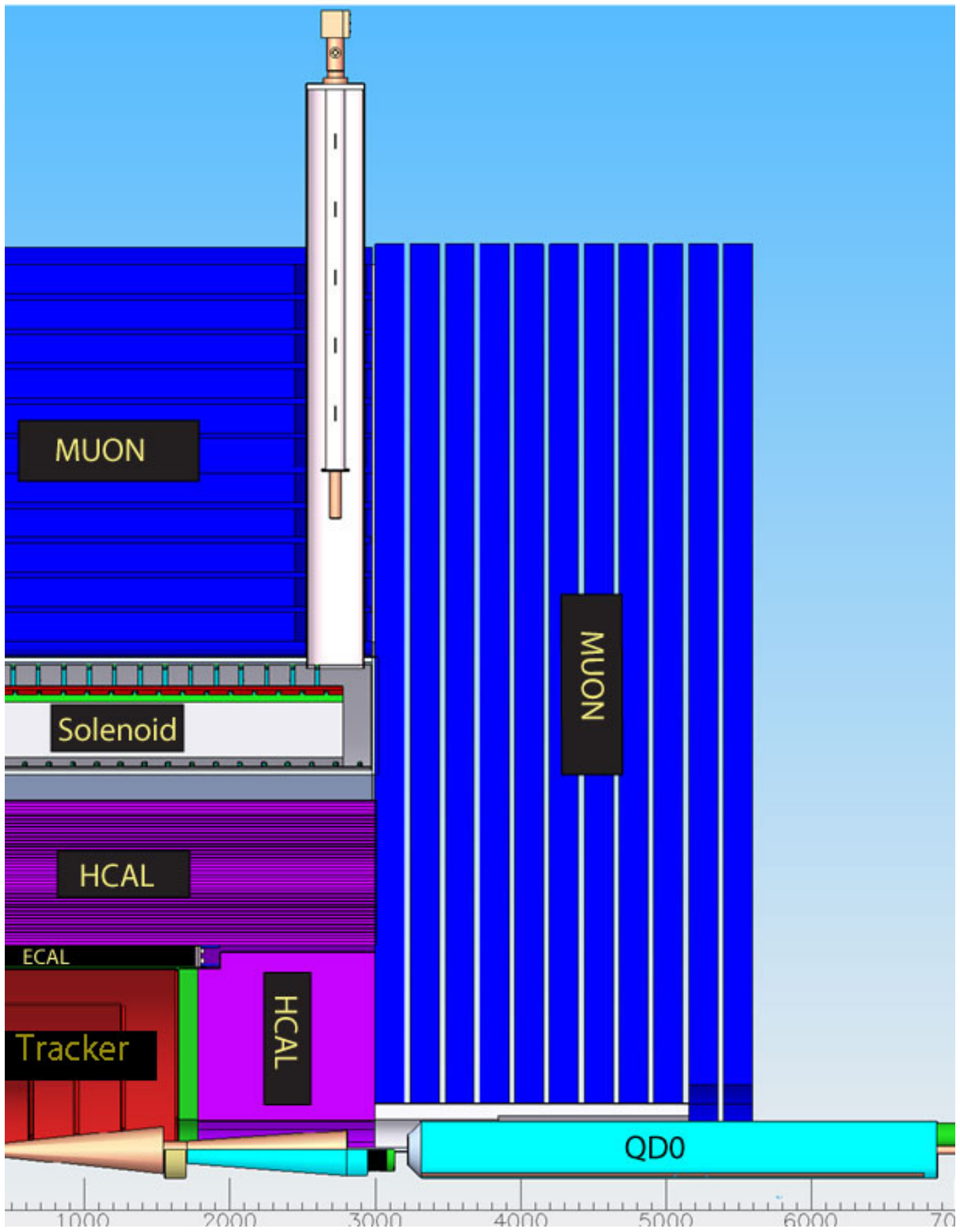

To provide for a very robust track-finding performance the baseline choice for the vertex detector is a sensor technology that provides time-stamping of each hit with sufficient precision to assign it to a particular bunch crossing. This significantly suppresses backgrounds.

Several technologies are being developed. One of them is a CMOS-based monolithic pixel sensor called Chronopixel. The main goal for the design is a pixel size of about $10 \times 10\ \mu\mathrm{m}^2$ with 99% charged-particle efficiency. Prototype devices have demonstrated that the concept works; what should be a fully functional chip is presently under test. More challenging is the 3D vertical integrated silicon technology, for which a full demonstration is also close.

Minimising the support material is critical to the development of a high-performance vertex detector. Different groups are studying an array of low-mass materials such as reticulated foams and silicon-carbide materials. An alternative approach that is being pursued very actively is the embedding of thinned, active sensors in ultra low-mass media. This line of R&D explores thinning active silicon devices to such a thickness that the silicon becomes flexible. The devices can then be embedded in, for example, Kapton structures, providing extreme versatility in designing and constructing a vertex detector.

Power delivery must be accomplished without exceeding the material budget and over heating the detector. The vertex detector design relies on power pulsing during bunch trains to minimise heating and uses forced air for cooling.

### 1.2.2 Main tracker

The main tracker technology of choice is silicon strip sensors arrayed in five nested cylinders in the central region and four disks following a conical surface with an angle of 5 degrees with respect to the normal to the beamline in each of the end regions. The geometry of the endcaps minimises the material budget to enhance forward tracking. The detectors are single-sided silicon sensors, approximately 10 × 10 $cm^2$ with a readout pitch of 50 $\mu$m. The endcaps utilise two sensors bonded back-to-back for small angle stereo measurements. With an outer cylinder radius of 1.25 m and a 5 T field, the charged track momentum resolution will be better than $\delta(1/p_{\mathrm{T}}) = 5 \times 10^{-5}/(\mathrm{GeV}/c)$ for high momentum tracks with coverage down to polar angles of 10 degrees.

The all-silicon tracking approach has been extensively tested using full Monte-Carlo simulations including full beam backgrounds. Besides having an excellent momentum resolution it provides robust pattern recognition even in the presence of backgrounds and has a real safety margin, if the machine backgrounds will be worse than expected.

## 1.3 Main calorimeters

The SiD baseline design incorporates the elements needed to successfully implement the PFA approach. This imposes a number of basic requirements on the calorimetry. The central calorimeter system must be contained within the solenoid in order to reliably associate tracks to energy deposits. The electromagnetic and hadronic sections must have imaging capabilities that allow both efficient track-following and correct assignment of energy clusters to tracks. These requirements imply that the calorimeters must be finely segmented both longitudinally and transversely. In order to ensure that no significant amount of energy can escape detection, the calorimetry must extend down to small angles with respect to the beampipe and must be sufficiently deep to prevent significant energy leakage. Since the average penetration depth of a hadronic shower grows with its energy, the calorimeter system must be designed for the highest-energy collisions envisaged.

In order to ease detector construction the calorimeter mechanical design consists of a series of modules of manageable size and weight. The boundaries between modules are kept as small as possible to prevent significant non-instrumented regions. The detectors are designed to have excellent long-term stability and reliability, since access during the data-taking period will be extremely limited, if not impossible.

The combined ECAL and HCAL systems consist of a central barrel part and two endcaps, nested inside the barrel. The entire barrel system is contained within the volume of the cylindrical superconducting solenoid.

The electromagnetic calorimeter has silicon active layers between tungsten absorber layers. The active layers use 5×5 $mm^2$ silicon pixels, which provide excellent spatial resolution. The structure has 30 layers in total, the first 20 layers having a thinner absorber than the last ten layers. This configuration is a compromise between cost, electromagnetic shower radius, sampling frequency, and shower containment. The total depth of the electromagnetic calorimeter is 26 radiation lengths ($\mathrm{X}_0$) and one nuclear interaction length.

The hadronic calorimeter has a depth of 4.5 nuclear interaction lengths, consisting of alternating steel plates and active layers. The baseline choice for the active layers is the glass resistive plate chamber, which has been extensively evaluated in testbeam campaigns at Fermilab and CERN. Two other technologies (GEM, and Micromegas) are currently being prototyped and evaluated as potential options for SiD.

## 1.4 Forward calorimeters

Two special calorimeters are foreseen in the very forward region: LumiCal for precise measurement, and BeamCal for fast estimation, of the luminosity. LumiCal and BeamCal are compact cylindrical electromagnetic calorimeters centred on the outgoing beam. They are based on 30 layers' depth of semiconductor-tungsten technology. BeamCal is placed just in front of the final focus quadrupole and LumiCal is aligned with the electromagnetic calorimeter endcap. LumiCal uses silicon sensor readout. It is a precision device with challenging requirements on the mechanics and position control. BeamCal is exposed to a large flux of low-energy electron-positron pairs originating from beamstrahlung. These depositions, useful for a bunch-by-bunch luminosity estimate and the determination of beam parameters, require radiation hard sensors. The detectors in the very forward region have to tackle relatively high occupancies, requiring dedicated front-end electronics.

The challenge for BeamCal is to find sensors that will tolerate about one MGy of dose per year. So far polycrystalline chemical vapour deposition (CVD) diamond sensors of area 1 $cm^2$ and larger sectors of GaAs pad sensors have been studied. Since large-area CVD diamond sensors are extremely expensive, they may be used for only the innermost part of BeamCal. At larger radii GaAs sensors appear to be a promising option. Sensor samples produced using the liquid encapsulated Czochralski method have been studied in a high-intensity electron beam.

For SiD, the main activities are the study of these radiation-hard sensors, development of the first version of the so-called Bean readout chip, and the simulation of BeamCal tagging for physics studies. SiD coordinates these activities with the FCAL R&D Collaboration.

## 1.5 Magnet Coil

The SiD superconducting solenoid is based on the CMS solenoid design philosophy and construction techniques, using a slightly modified CMS conductor as its baseline design. Superconducting strand count in the coextruded Rutherford cable was increased from 32 to 40 to accommodate the higher 5 T central field.

Many iron flux return configurations have been simulated in two dimensions so as to reduce the fringe field. An Opera 3D calculation with the Detector Integrated Dipole (DID) coil has been completed. Calculations of magnetic field with a 3D ANSYS program are in progress. These will have the capability to calculate forces and stress on the DID as well as run transient cases to check the viability of using the DID as a quench propagator for the solenoid. Field and force calculations with an iron endcap HCAL were studied. The field homogeneity improvement was found to be insufficient to pursue this option.

Conceptual DID construction and assembly methods have been studied. The solenoid electrical power system, including a water-cooled dump resistor and grounding, was established. Significant work has been expended on examining different conductor stabiliser options and conductor fabrication methods. This work is pursued as a cost- and time-saving effort for solenoid construction.

## 1.6 Muon System

The flux-return yoke is instrumented with position sensitive detectors to serve as both a muon filter and a tail catcher. The total area to be instrumented is very significant - several thousand square meters. Technologies that lend themselves to low-cost large-area detectors are therefore under investigation. Particles arriving at the muon system have seen large amounts of material in the calorimeters and encounter significant multiple scattering inside the iron. Spatial resolution of a few centimetres is therefore sufficient. Occupancies are low, so strip detectors are possible. The SiD baseline design uses scintillator technology, with RPCs as an alternative. The scintillator technology uses extruded scintillator readout with wavelength shifting fibre and SiPMs, and has been successfully demonstrated. Simulation studies have shown that nine or more layers of sensitive detectors yield adequate energy measurements and good muon-detection efficiency and purity.

## 1.7 The Machine-Detector Interface

The push-pull system for the two detectors was only conceptual at the time of LoI publication, but since then the engineering design has progressed significantly. A time-efficient implementation of the push-pull model of operation sets specific requirements and challenges for many detector and machine systems, in particular the interaction region (IR) magnets, the cryogenics, the alignment system, the beamline shielding, the detector design and the overall integration. The minimal functional requirements and interface specifications for the push-pull IR have been successfully developed and published [64, 65], to which all further IR design work on both the detectors and machine sides are constrained.

# Chapter 2
# SiD Vertex Detector

## 2.1 Introduction

The SiD vertex detector consists of a central barrel section with five silicon pixel layers and forward and backward disk regions, each with four silicon pixel disks. Three silicon pixel disks at large z provide uniform coverage for the transition region between the vertex detector and the silicon micro-strip based outer tracker. The barrel layers and disks are arranged to provide good hermeticity to $\cos(\theta) \approx 0.984$ and to guarantee good pattern recognition capability for charged tracking and excellent impact parameter resolution over the whole solid angle.

### 2.1.1 Vertex detector requirements

The physics goals of the ILC, particularly the need to separate bottom and charm vertices, drive the need for a very precise, light vertex detector. The time structure and low radiation background in the ILC provides an environment which allows us to consider very light, low power detector structures. The bunch structure, with a 1 ms long bunch train at 5 Hz, enables power pulsing of the electronics, providing a power saving of a factor of 50-100 for front-end analog power. Low power allows gas-based cooling, saving mass in cooling channels and associated structures. The vertex detector for SiD is designed to meet the following goals:

- Hit resolution better then 5 $\mu\mathrm{m}$ in the barrel
- Less than $0.3\%$ radiation length per layer
- Average power less than 130 $\mu\mathrm{W}/\mathrm{mm}^2$ in the barrel
- Single bunch time resolution.

These requirements then drive the design of the vertex system. The 5 $\mu\mathrm{m}$ resolution implies a pixel size of 17 $\mu\mathrm{m}$, larger if charge sharing is used to improve the resolution. Some CMOS MAPS devices, which collect charge by diffusion rather than drift, can utilise larger pixels because diffusion naturally spreads the charge.

The small radiation length per layer is driven by the need for precise three dimensional vertex resolution for heavy quark decays. This resolution has a direct effect on the efficiency for $b$ and $c$ hadron identification. For a device with less than $0.3\%$ radiation length per layer air cooling appears to be the only viable low-mass sensor cooling technique. Gas cooling places a limit on the average power based on the heat which can be removed by laminar flow of the cooling gas. We combine this with an effective duty factor of 50-100 to calculate the maximum average power in the barrel.

Timing resolution affects the number of overlapping events that occur when the detector is read out. Here there is a tradeoff between speed and front-end signal-to-noise and power. Fortunately, the low capacitance and high signal-to-noise ratio of a finely pixelated sensor allows for acceptable power dissipation for single-crossing ($\approx$ 300-700 ns) time resolution. Therefore our baseline design assumes single-crossing time-resolution.

## 2.2 Baseline Design

Given the significantly extended physics reach that can be achieved with superb vertex reconstruction – primary, secondary and tertiary – the vertex detector for SiD is proposed to be an all-silicon structure in a barrel-disk geometry. Side views of the vertex detector are shown in Figures II-2.1 and II-2.2.

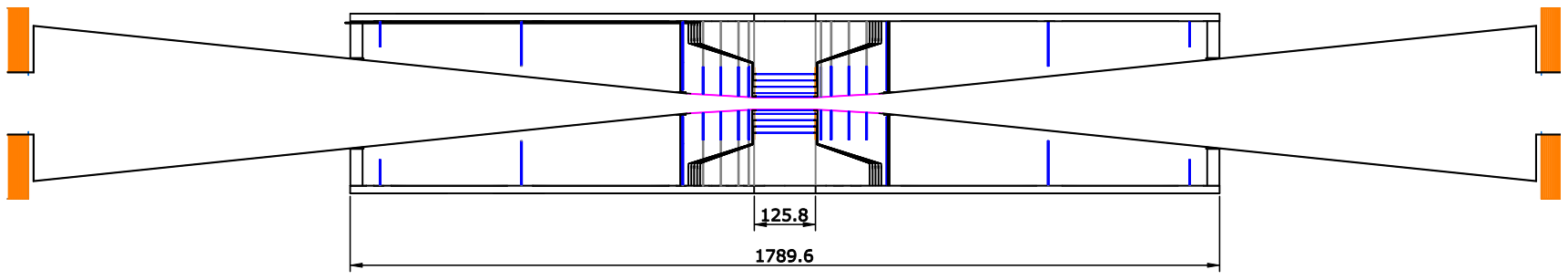

**Figure II-2.1**
Layout of the vertex and forward tracking region, including carbon-fibre support and forward cone. Dimensions are in mm

The geometry parameters of the vertex detectors are summarised in Table II-2.1. The five barrel sensor layers are arranged at radii ranging from 14 to 60 mm. The vertex detector also has four disk layers supported by carbon-fibre support disks at z positions ranging from about 72 to 172 mm. The innermost disk covers radii from 14 mm out to 71 mm. Forward tracking continues beyond the vertex detector proper with three additional small pixel disks, extending in $z$ from about 207 to 832 mm. The vertex barrel and inner endcaps have $\approx 20 \times 20$ $\mu$m pixels. The pixel size increases to $\approx 50 \times 50$ $\mu$m$^2$ for the forward tracker disks. The total area of the vertex barrels is $1.63 \times 10^5$ mm$^2$ and is $0.59 \times 10^5$ mm$^2$ for each set of 4 inner pixel disks and $1.96 \times 10^5$ mm$^2$ for each set of 3 forward pixel disks. The simulation described in the following chapters assumes $0.1\%$ radiation length per layer excluding cables and $20 \times 20$ $\mu$m pixels for the forward tracker disks.

**Table II-2.1**
The geometry parameters of the SiD vertex detector (Barrel, Disks and Forward Disks). Units are mm.

| Barrel | R | $z_{max}$ | |
|---|---|---|---|
| Layer 1 | 14 | 63 | |
| Layer 2 | 22 | 63 | |
| Layer 3 | 35 | 63 | |
| Layer 4 | 48 | 63 | |
| Layer 5 | 60 | 63 | |
| **Disk** | $R_{inner}$ | $R_{outer}$ | $z_{center}$ |
| Disk 1 | 14 | 71 | 72 |
| Disk 2 | 16 | 71 | 92 |
| Disk 3 | 18 | 71 | 123 |
| Disk 4 | 20 | 71 | 172 |
| **Forward Disk** | $R_{inner}$ | $R_{outer}$ | $z_{center}$ |
| Disk 1 | 28 | 166 | 207 |
| Disk 2 | 76 | 166 | 541 |
| Disk 3 | 117 | 166 | 832 |

### 2.2.1 Sensor Technology

There are a number of possible choices of sensor technology for the vertex detector, including 3D integrated sensors and readout chips [66], Silicon-on-insulator (SOI) [67], Monolithic Active Pixels (MAPS) [68, 69], hybrid pixels [70, 71], and DEPFETs [72].

All of these technologies have the capability of delivering sensors less than 75 $\mu$m thick with 5 $\mu$m hit resolution and low power consumption. They are also changing rapidly with advances in microelectronics. The vertex detector is physically small and SiD is designed to make insertion and removal of the vertex detector straightforward. These factors motivate postponing a decision on the details of sensor technology for the SiD vertex detector to a date as late as possible in the final design process. In this document we have chosen 3D technology to provide a definite reference for the detector design. Other choices would differ in details of the mechanical and electronic design of the vertex detector but would not affect the overall design philosophy. To achieve minimum mass in the barrel ladders we are exploring an all-silicon assembly as the baseline. Alternatives include foam-based ladders as explored in the PLUME collaboration [73], and carbon-fibre supports as prototyped at Fermilab.

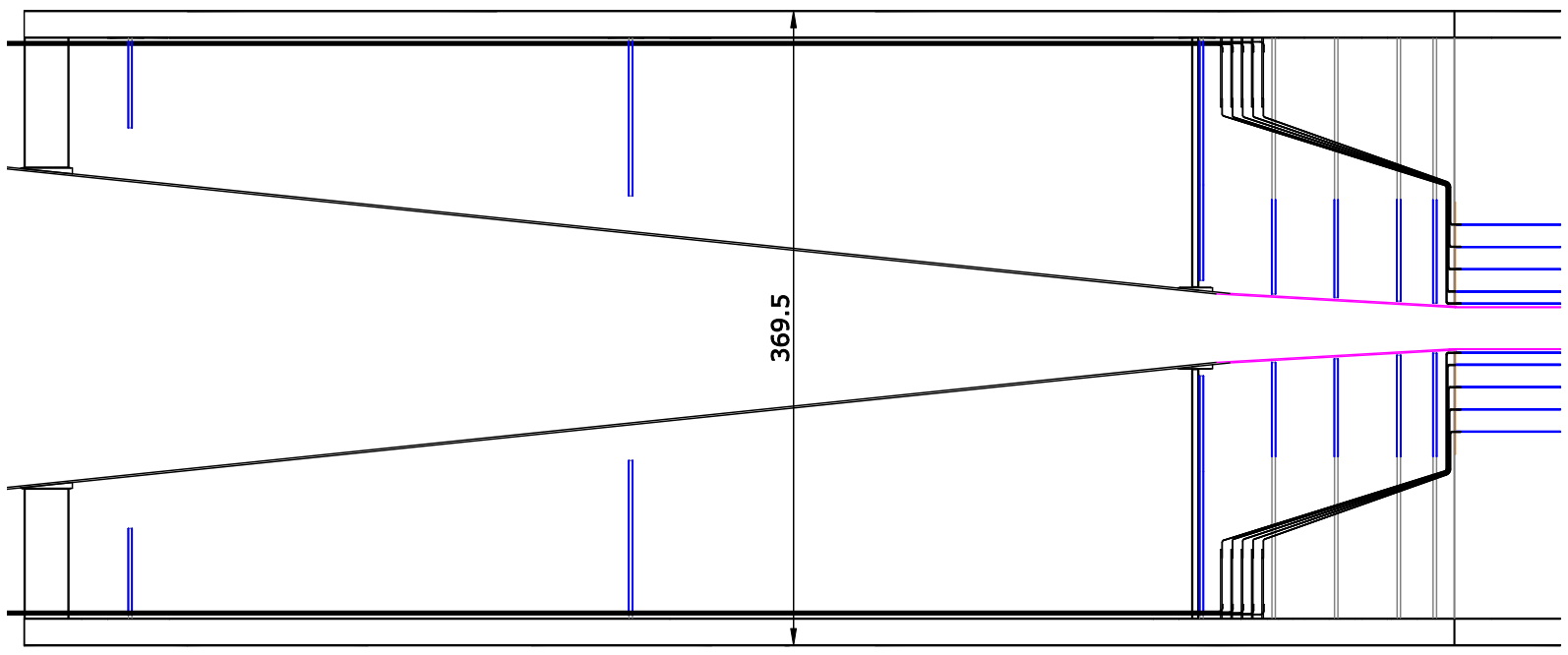

**Figure II-2.2** R-z view of the vertex detector and its support structure. Cable routes are also shown with DC-DC converters located on the support structure near the end of the first set of pixel disks. Dimensions are in mm.

## 2.3 3D-Based Module Design

In microelectronics, 3D technology refers to the stacking of multiple layers of circuitry with vertical interconnections between them. This area is developing rapidly as a way of increasing circuit density without the major re-tooling and investment needed for smaller feature sizes. The enabling technologies for 3D are wafer thinning, wafer bonding, and the formation of Through-Silicon Vias (TSVs). Although the increased circuit density provided by multi-layer circuits is in itself an important application for High Energy Physics, it is the increased range of processing and interconnection options provided by technology that offers the largest potential [74]. Using these technologies arrays of chips can be bonded to sensors and electronics to form essentially monolithic arrays of sensors with no dead space between chips and with interconnections taken from the back rather than the edge of the IC. Chip-to wafer technologies such as Direct Oxide Bonding (DBI) from Ziptronix [75] also promise lower cost and much finer pitch (4 $\mu m$ has been utilised for the Fermilab wafers) than conventional bump bonding. Heterogeneous layers of different technologies can be combined in a 3D stack to optimise overall sensor performance.

Combining 3D with active edge processing can result in fully active tiles which can be used to populate detector arrays in a variety of geometries with small dead regions. This is especially important for the forward disks where tiling will allow full coverage with minimal dead area.

### 2.3.1 Active Edge Devices

Active edge sensors are an outgrowth of work done to develop 3D silicon sensors, which provide good charge collection combined with radiation hardness. The technique utilises a deep reactive ion etch of silicon to create a nearly vertical trench with smooth edges. The high quality of the trench wall avoids charge generation normally associated with saw-cut edges [76]. The trenches are filled with doped polycrystalline silicon. Combination of active edge technology with 3D integration can provide a technique for tiling sensor arrays with low mass and high yield. Readout wafers are oxide bonded to sensor wafers with active edge processing. The resulting stack is thinned to expose the Through-Silicon-Vias and the handle wafer is removed by grinding and etching. This results in active tiles with coarse pitch bump-bond connections for readout. Using such tiles, large-area pixelated modules with complex shapes can be assembled with known good integrated sensor/readout dies and with large-pitch backside bump-bond interconnects. All fine-pitch bonds to the sensor are made using wafer-to-wafer oxide bonding. This is particularly useful for the pixel disks, where we want to populate an ≈18 cm radius disk with IC reticule ($\approx 2.5 \times 2.5$ cm$^2$) sized objects.

### 2.3.2 Barrel Sensor Interconnect to Readout

Any complex, pixelated device will require integration of sensors with readout chips whose size is limited by the reticle area of the CMOS process. There are several choices if we wish to fabricate a 12 cm long ladder. A "stitching" process modifies the reticules to allow reticule to reticule connections on the wafer, by double exposing an overlap region to form connections. The yield of the stitched array is the product of the individual yields. Active tiles can be bonded to a thin substrate which provides power and signal routing. There is a mass penalty associated with the backing structure. A third process, which we have chosen as the baseline, uses sensor wafers bonded to matched 3D wafers. The resulting stack is thinned and the readout and power connections are fabricated on the top aluminium layer of the readout IC layer. This results in minimum mass ladders with no additional material needed for support.

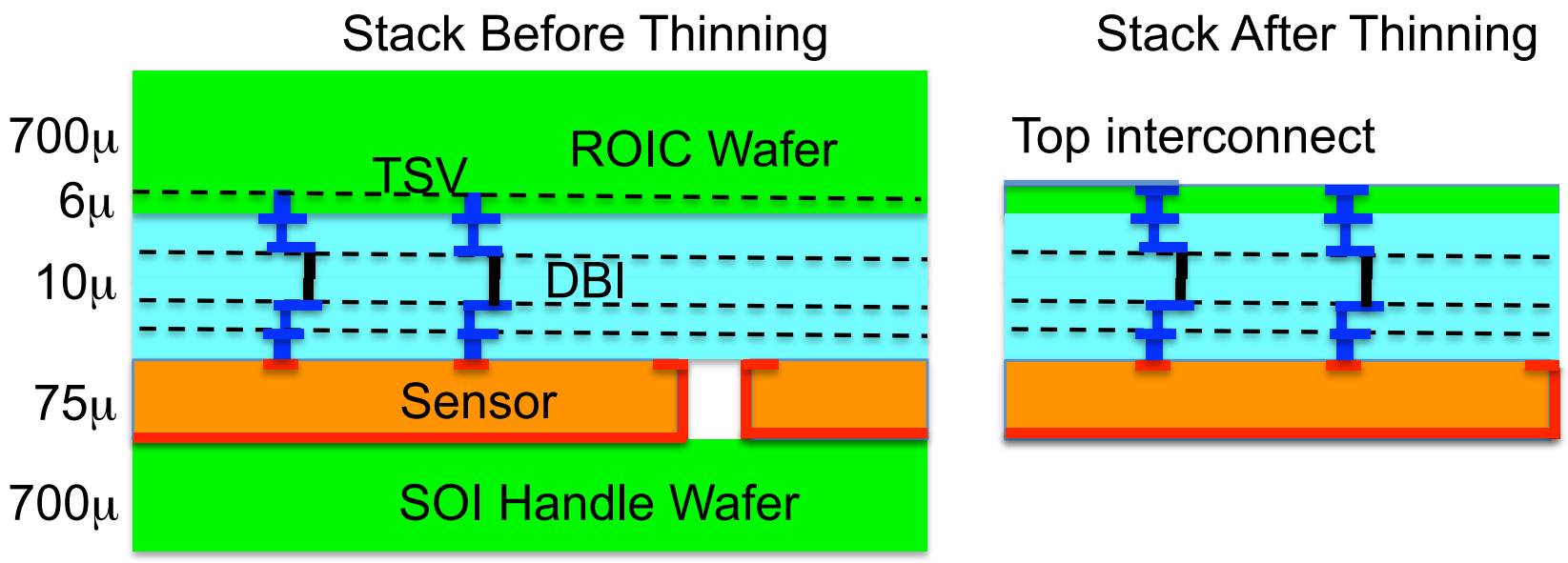

**Figure II-2.3**
Wafer stack structure before and after thinning and singulation.

Our design includes a number of features which have been or are being demonstrated, but full ladders have not yet been assembled. The design is based on active-edge SOI sensors bonded to readout chips with Through-Silicon-Vias [77]. Similar results can be obtained with SOI sensors utilising the handle wafer as a sensor, or MAPS-type designs.

A possible process would include:

1. Fabricate sensor wafer as an SOI stack with a 50 $\mu$m thick sensor bonded to a thick handle which will be removed after processing. Trenches are etched in the perimeter of sensors to provide the active edge.
2. Fabricate ReadOut Integrated Circuit (ROIC) Through-Silicon-Vias wafers with reticule pitch matched to the sensors.
3. Oxide bond ROICs to sensors with seed metal routing to match a smaller ROIC pixel pitch at the edges. This allows for the regions near the edges of reticules to be used for test structures and alignment.
4. Thin the stack to expose the Through-Silicon-Vias. Pattern the top layer to provide bussing to all power and readout connections. Form bump bond pads near the edges.
5. Etch the regions at the sensor periphery to singular the individual sensors.
6. Backgrind and etch the wafer to remove the handle.

Figure II-2.3 shows the wafer stack structure before and after thinning and singulation. This process is very similar to work currently being done at VTT and Ziptronix to demonstrate active-edge tile fabrication. The only significant difference for SiD would be the thickness of the sensor (50 vs. 200 $\mu$m) and bonding of multiple reticules to a single sensor. Alternatives, such as carbon-fibre or foam supports would simplify the process at some expense in mass.

### 2.3.3 Sensor tiling for disks

The ROIC/sensor bonding process for the four forward pixel disks is similar to that used for the barrel ladders except that single reticules are bonded and singulated. Each tile has a set of bump bonds distributed on the back side for power and readout interconnect. The tiles are bump bonded to a carbon-fibre backing plate co-cured with a Kapton circuit which provides routing to external connections. The four different inner radii of the disks would require four different reticule layouts. A optimised final design might utilise identical disks to minimise the varieties of layouts.

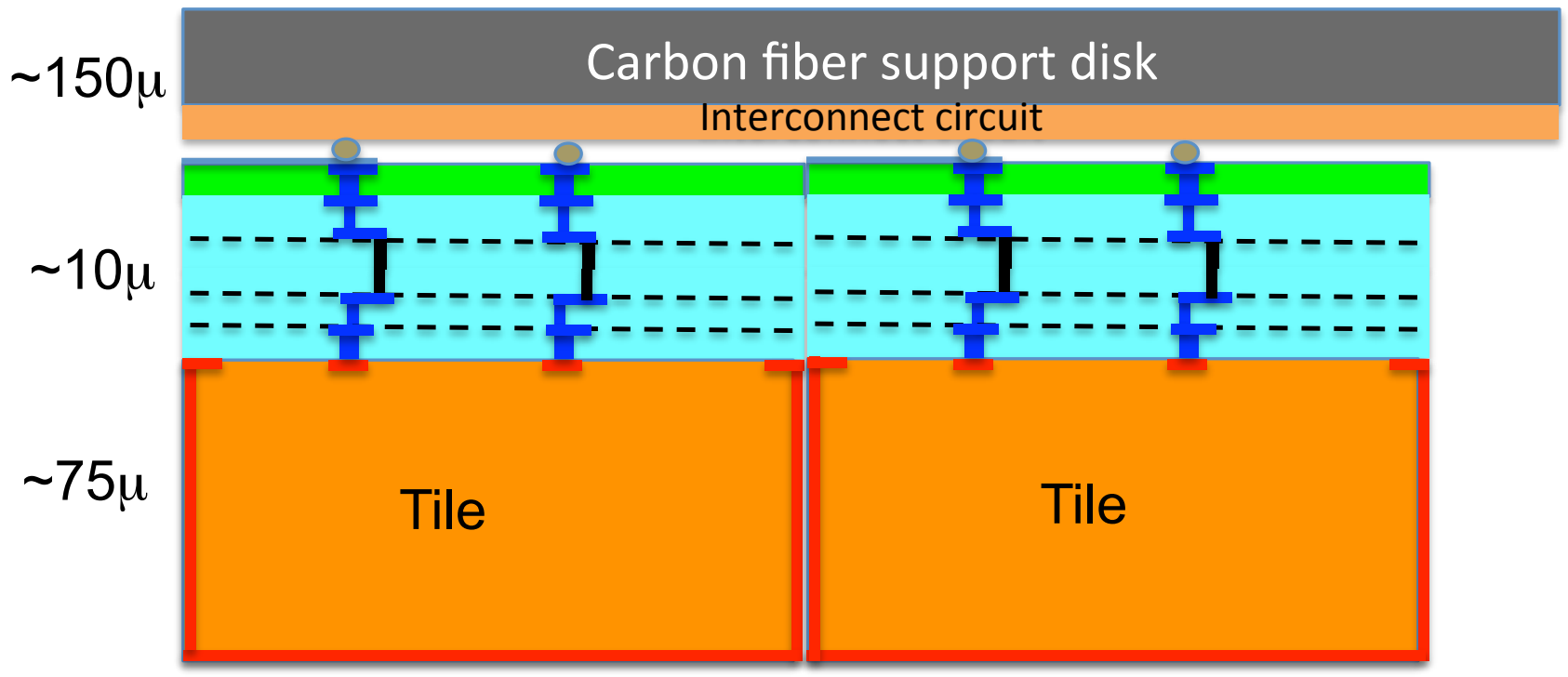

**Figure II-2.4**
Tiled structure used for the disk layers utilising a carbon-fibre backing disk.

A similar layout can be used for the forward pixel disks at large z. Figure II-2.5 shows a possible tiling which utilises only two reticule types. In this design the outer disk would use two rows of tiles, the middle would use four and the inner would use six. The active edge technique has the additional advantage that edges are formed by etching rather then saw cutting, so the trapezoid shapes can be fabricated easily.

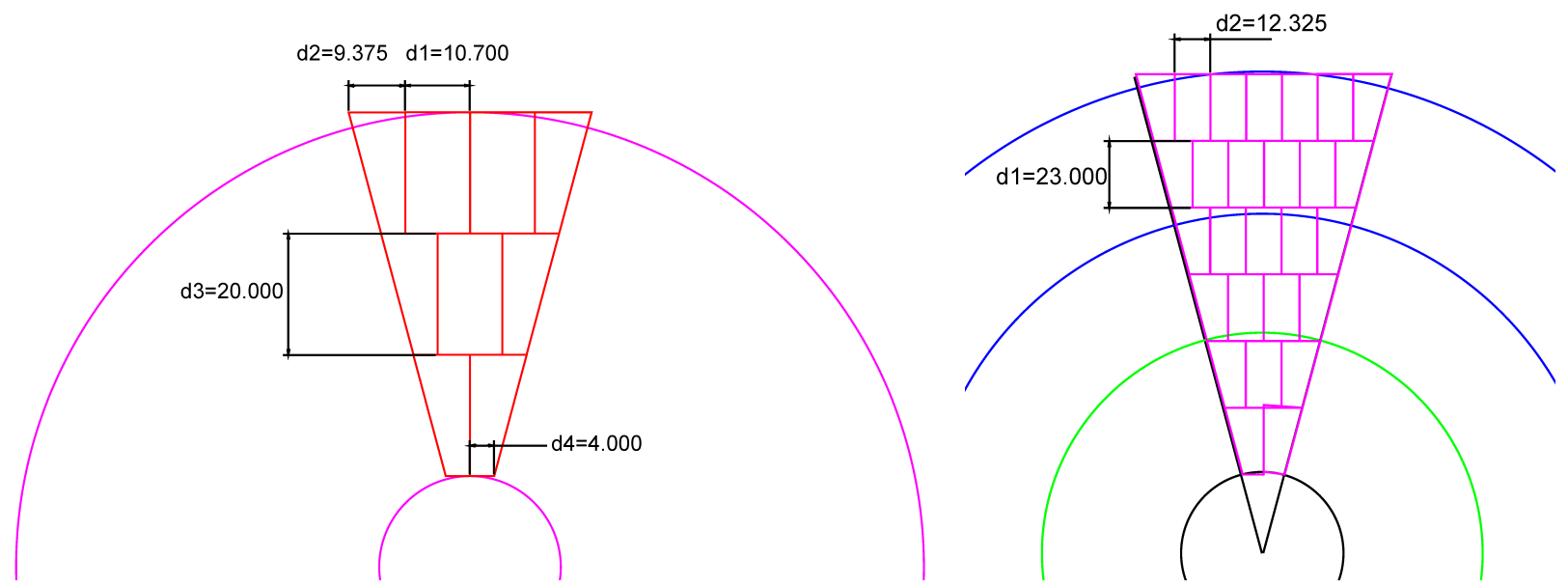

**Figure II-2.5**
Design of the reticule-based tiling for the innermost pixel disk (left) and for the disks at large z (right) with inner radii of 28 (black), 76 (green) and 117 (blue) mm and an outer radius of 166 mm.

## 2.4 Support structures and Integration

The vertex subsystem is supported by a double-walled carbon-fibre cylinder (Figure II-2.6) which serves multiple functions. In addition to supporting the vertex detector barrels and disks the cylinder stiffens the beampipe in the vertex region, serves as a cooling gas transport and manifold, and provides locations to mount cables and power converters.

To allow assembly about the beampipe and later servicing, the vertex detector is split at the level of the horizontal plane into top and bottom sub-assemblies. To accommodate the sensor geometry, the split line is offset between the right and left hemisphere. Once mated, the two sub-assemblies are supported from the beam pipe and stiffen the portion of the beampipe passing through them.

To prevent bending of the small-radius portion of the beampipe and ensure good stability of the vertex detector position, the outer vertex detector support cylinder is coupled to the beampipe at four longitudinal locations: $\pm$ 21.4 and $\pm$ 88.2 cm. The support cylinder is separated into top and bottom halves, as are all vertex detector structures. Inner and outer support cylinder walls are 0.26 mm thick. They are made from four plies of high-modulus carbon-fibre resin pre-preg. Wall separation is 15 mm.

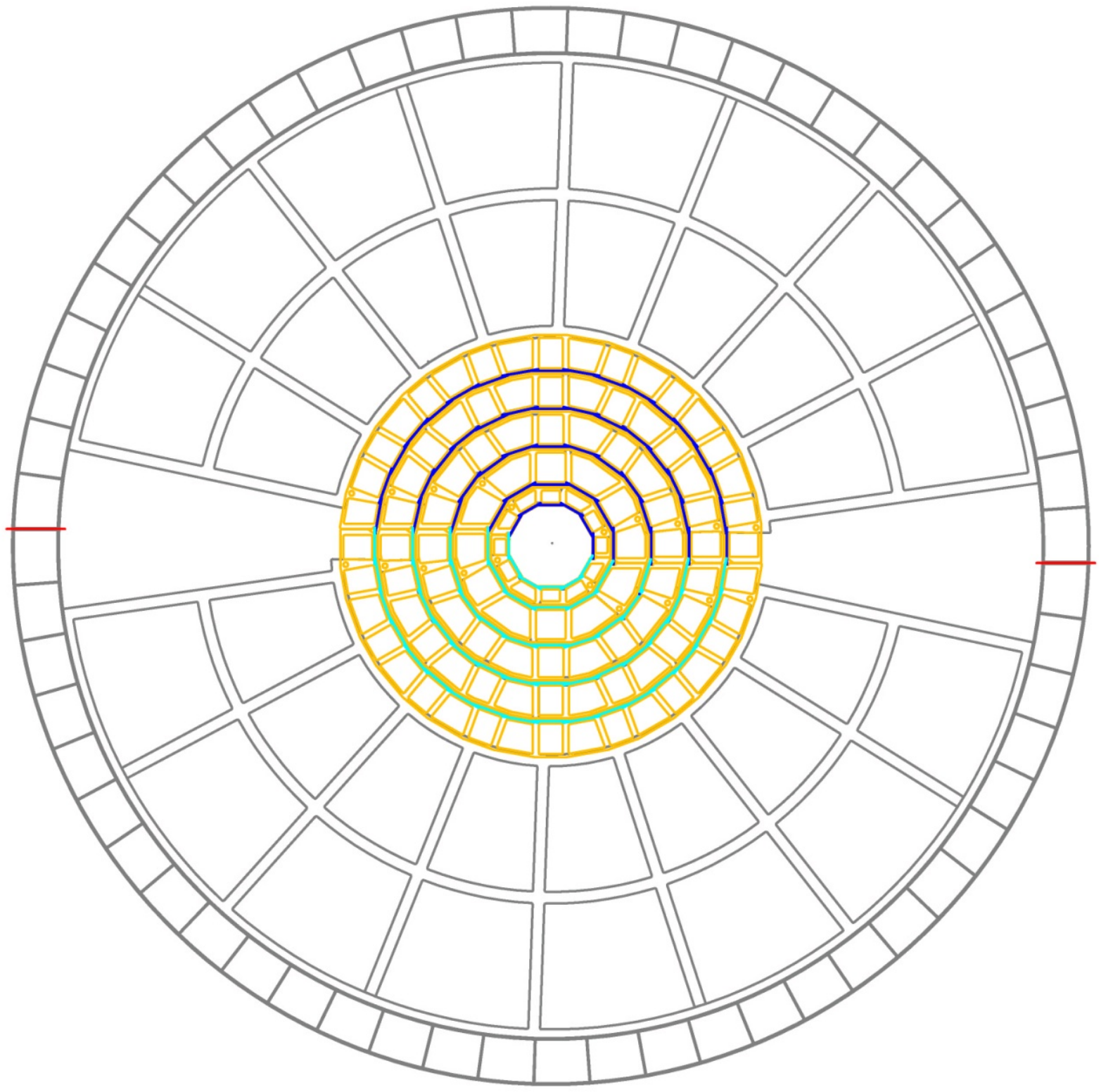

**Figure II-2.6**
End view of the vertex support cylinder showing ribs and cooling gas passages, internal stiffening web structures, and the barrel vertex detector. Top and bottom sections of the barrel are shown in blue and green.

During silicon tracker servicing the vertex detector and beampipe remain fixed while the outer silicon tracker rolls longitudinally. To allow for that motion and to permit the outer silicon tracker elements to be at the lowest possible radius, the outer radius of the vertex detector, including its support structures, has been limited to 18.5 cm. Additional space for any additional thermal insulation which might be needed, has been foreseen. To maximise the physics potential, the inner radius of vertex detector elements has been chosen to be as small as possible, while still being consistent with beam-related backgrounds and the beampipe profile. In the barrel region, the minimum radius to a sensor surface is 1.4 cm, governed by the beam backgrounds.

### 2.4.1 Power delivery

#### 2.4.1.1 Readout Considerations

The vertex detector readout is illustrated using the scheme with in-pixel storage of analog information and digital time stamps used both in the 3D-VIP or the Chronopixel chip [78]. In this scheme analog and digital information is stored within a pixel during the bunch train and read out between bunch trains. The pixel complexity is minimised by storing the address information on the periphery of the chip. Table II-2.2 summarises the power consumption of this readout scheme. Electrical connections of about one meter from the ladders to optical links installed on the support tube have been assumed. Assuming 32 bits are used per hit and 100 pF interconnect capacitance at 1.5 V, the local readout consumes 0.24 W of average power. If each of the 108 ladders is independently driven using a 200 MHz clock, the inner layers would dominate the readout time at 75 ms/ladder. The bit rate from the entire vertex detector is about 2 Gbit/s.

We base our estimates on the VIP chip, which utilises 6 $\mu\mathrm{A}$ per $24 \times 24\ \mu\mathrm{m}^2$ pixel in steady-state operation. An effective duty factor of 80 for power pulsing leads to an average power of $\approx$120 $\mu\mathrm{W}/\mathrm{mm}^2$. The average power in the vertex barrel is then 19.1 W. A similar calculation yields an average power of 1.37 W/disk.

**Table II-2.2**
Readout rates for the vertex detector.

| Layer | Ladders | hits/crossing | bits/train | Readout time (s) | Readout Power (W) | Analog Power (W) |
|---|---|---|---|---|---|---|
| 1 | 12 | 2000 | $1.8 \times 10^{8}$ | $7.5 \times 10^{-2}$ | 0.10 | 1.7 |
| 2 | 12 | 1200 | $1.1 \times 10^{8}$ | $4.5 \times 10^{-2}$ | 0.06 | 2.5 |
| 3 | 20 | 800 | $7.2 \times 10^{7}$ | $1.8 \times 10^{-2}$ | 0.04 | 3.7 |
| 4 | 28 | 450 | $4.1 \times 10^{7}$ | $7.2 \times 10^{-3}$ | 0.02 | 5.0 |
| 5 | 36 | 400 | $3.6 \times 10^{7}$ | $5.0 \times 10^{-3}$ | 0.02 | 6.2 |

#### 2.4.1.2 Pulsed Power and DC-DC conversion

The stringent power dissipation requirements for the vertex detector can be met by delivering power to the front-end electronics only when it is actually needed. The time structure of the ILC beam, with ≈1 ms bunch trains followed by ≈199 ms gaps, allows for power pulsing. Analog and digital circuits can be turned on and off selectively, taking into account capacitive rise and fall times of individual sub-circuits as well as the time to handle the actual data load. Bias levels, however, need to be maintained at all times.

Although the average power in the vertex detector is low, the instantaneous current during the bunch train can be quite high, especially in view of the low supply voltages in modern CMOS technologies. This results in either unacceptable voltage drops in the cables or the use of high cable masses to reduce resistance. To solve this problem, which is also a significant issue for the ATLAS and CMS detector upgrades, we plan to place DC-DC converters in the vicinity of the vertex detector. Moreover, adequate voltage regulation will be implemented in the power supply chain to avoid voltage spikes due to power pulsing. The design includes low-mass flex cables which are routed from the vertex barrel and disk modules to a location at the inner wall of the support cylinder

The powering of the SiD readout chips was studied with a one-step DC-DC buck converter providing the required voltage and current. Input to the converter is 12 V for an output of 1.2 V using an air core inductor. Test boards were constructed operating at 1 MHz with several different commercial DC-DC converter chips. Tests with ATLAS tracker silicon strip detectors [79, 80] indicated that the electrical noise is primarily electrostatic and can be shielded by a 20 $\mu\mathrm{m}$ Al foil. To further reduce the mass, higher frequency operation of the inductor and the buck converter is required. The portable platform industry (smartphones, tablets, etc) currently uses DC-DC converters operating up to 6 MHz, while 20 MHz converters are in the R&D stage.

### 2.4.2 Cooling

Cooling in the SiD vertex detector is based on forced convection with dry air. The flow for barrel cooling is assumed to be from one barrel end to the other. For a study of the heat dissipation and cooling, the average power dissipated in a sensor was taken to be131 $\mu\mathrm{W}/\mathrm{mm}^2$. That corresponds to a total power of 20 W for the five barrel layers. These numbers presume power pulsing. Power is assumed to be distributed uniformly over the sensor active surface and both sensor surfaces participate in heat removal. The supply air temperature was taken to be -15 $^0$C. For a given sensor, heat transferred inward through the carbon-fibre support was taken to be proportional to the surface contact between the sensor and carbon fibre. Thermal impedance through silicon, epoxy, and carbon-fibre laminate has been included, but turns out not to be very significant. The remaining heat was assumed to be transferred outward into the layer to layer gap. For flow and heat transfer calculations, the gap between barrel layers was taken to be 1 mm less than the nominal layer spacing and laminar flow was assumed.

In the gap between the innermost layer and the beam tube, flow is likely to be lower and temperature higher, once supply and return distribution patterns of air flow have been taken into account. Higher flow rate clearly improves the uniformity of sensor temperatures and reduces the difference between the sensor temperature and the cooling air. All flow rates which have been considered lead to temperature variations which should be acceptable for dimensional stability, which is crucial for high-precision vertexing. The time-dependent effects of power cycling remain to be investigated. Those depend on the thermal mass in the barrel and the details of the power cycling. The outer support cylinder of the vertex detector offers a natural thermal enclosure. Details of end openings in barrel membranes remain to be included. Those openings provide a mechanism for adjusting relative flow between barrel layers. A membrane between the outermost barrel layer and the vertex support cylinder will ensure that flow does not excessively bypass the barrel-to-barrel gaps. Similar calculations have been made to understand disk cooling. Those calculations are based on barrel results with a Reynold's number of 1800 (barrel flow = 20 g/s). Heat removal calculations for the first four disks at each end of the barrel assume the same power per unit area as the pixel sensors of the barrel. The result is a total power of 16.9 W for all eight disks and an air flow of 16.4 g/s.

We propose to deliver air via the outer support cylinder. To allow that, the two walls of the cylinder are separated by radially-oriented ribs running along the full cylinder length. The calculations assume ribs at 60 azimuths. Openings, each approximately 12.2 mm × 15 mm, at 18 z-locations in the inner cylinder wall distribute the flow to the various disk locations and to the barrel. At each azimuth, the cell through which flow passes was approximated by a rectangle of height 15 mm and width 18.2 mm. The wall thickness was assumed to be 0.26 mm for both cylinders and for all ribs.

The results gave a Reynold's number of 3105 in the portion of the cell which sees full flow, indicating turbulent flow. Since a portion of the flow exits the cell at each opening, the Reynold's number drops to 1725 at approximately z = 51.9 cm (a short distance inboard of the two outermost disks). While entrance effects may remain, the flow should gradually become laminar after that point. Supply and return connections to the outside world remain to be fully evaluated. With eight connections per end, each represented by a 20 mm × 40 mm rectangular passage, the Reynold's number is 12900 and flow is turbulent.

### 2.4.3 Cabling

We plan to utilise low-mass strip line cables based on aluminium conductor for signal communication and power distribution from the sensors to the DC-DC converter region. The inner ends of the cables will be wire bonded directly to the sensor ladders. In the DC-DC converter region signals will be converted to optical fibres. Power will be brought into the vertex region at $\approx 12$ V by aluminium cables which make the transition to copper outside the tracker volume. Varying Lorentz forces due to pulsing of the power are a particular concern. This is minimised by utilising balanced supply and return lines and twisted wires where appropriate. We will utilise a three-layer strip-line design with centre supply and outer return traces to minimise forces on the cables [81]. Tests of mechanical forces and vibration are planned utilising KPiX chips and a 7 T magnet available at Yale. The pulsed power frequency, 5 Hz, is significantly below the resonance frequency expected for major support structures.

## 2.5 R&D Status

### 2.5.1 Chronopixel

We have developed a design, in collaboration with SARNOFF Research Labs, for the Chronopixel devices that satisfy the ILC requirements [78]. The design of the ultimate device requires high resistivity silicon (5 k$\Omega$-cm) with a 15 $\mu$m thick epilayer and pixels of 10 × 10 to 15 × 15 $\mu$m which will require to use 45 nm technology. The 45 nm technology is currently too expensive for prototyping, so we foresee a series of prototypes that approach the ultimate design. The first prototype has been designed, fabricated and extensively tested. The second prototype has recently been fabricated and the testing of these devices is just getting started.

### 2.5.2 VIP 3D Chip

The Vertically Integrated Pixel (VIP) ASIC was conceived of as a demonstration readout chip for the ILC vertex detector [82]. The Lincoln Laboratory process has the advantage of very well-established wafer bonding and thinning, but the fully depleted SOI process is not well suited for analog applications and has larger feature size than advanced commercial processes. The final Fermilab designed ASIC (VIP2a) using this process was received and tested late in 2009. The analog front end of VIP2a, which was laid out using design rules modified at Fermilab based on failings of the earlier prototypes, worked well, as did all of the interconnections between circuit layers.

A second iteration, the VIP2b was fabricated in the 3D process developed by Tezzaron/Global Foundries. This process uses a bulk 0.13 $\mu m$ CMOS IC process with modifications to allow the top copper metal layer to be used for face-to-face wafer bonding, and to include vias that extend 6 $\mu m$ into the bulk material. After wafer bonding, one of a pair of wafers was back thinned to expose the deep vias, and metal pads deposited that are suitable for wire bonding or for further wafer bonding. We now have chips with successful 3D bonds between tiers. Initial testing of the 2D parts show excellent analog performance. Tests of the full functionality of the 3D chips are underway.

### 2.5.3 Thinned hybrid detector with high-density interconnect

An alternative option for a low-mass vertex detector combines thinned high-functionality readout ASICs with thin high-resistivity sensors, assembled using advanced low-mass interconnect technologies. For the ASIC the 65 nm deep sub-micron technology was successfully assessed, through the design and production of relevant pixel readout sub-circuits [83]. Subsequently a fully functional test chip has been designed in 65 nm technology. It comprises $64 \times 64$ pixels of $25 \times 25$ $\mu m$ size providing time-of-arrival and time-over-threshold functionality [84]. The ASIC foresees individual power pulsing of its analog and digital circuits. It has been submitted for production at the end of 2012. Development has been initiated towards low-mass fine-pitch flip-chip interconnect based on copper pillars. Module assembly is foreseen to make use of Through Silicon Vias (TSV) to carry interface signals to the backside of the pixel chip. This will offer 4-side buttable pixel chips, enabling the assembly of large-area pixel detectors with minimal dead space between individual pixel tiles. The TSV technology has meanwhile been applied successfully on Medipix3 chips manufactured using a 130 nm process [85].

### 2.5.4 Active Edge Tiles

Active tiles are central to the conceptual design of the forward disks. A program to demonstrate these devices is underway in collaboration with Fermilab, SLAC, and Cornell University. Sensors of $200$ $\mu m$ thickness are being fabricated on SOI wafers by VTT and planar dummy top wafers with tungsten contacts are being fabricated by Cornell. The two will be wafer-bonded by Ziptronix and this stack will be singulated and thinned by SLAC. We expect the VTT wafer to be complete by the end of 2012.

### 2.5.5 Critical R&D

By the conclusion of the current round of R&D, we expect to have demonstrated the basic sensor and IC technologies needed for SiD. The next logical step would be to develop a full sized ladder for the barrels and a wedge segment for the disks. We need to build prototype support structures, including the double walled outer cylinder and barrel and disk supports. We also would need to demonstrate the integration of ladders and wedges into barrels and disks, initially with one live and several dummy sensors. Finally, a full-sized prototype with heating elements would allow us to study air cooling and confirm flow and temperature calculations.

Studies of power delivery and cabling are critical. We would like to demonstrate a low-mass cabling system, including aluminium conductors, DC-DC conversion, and optical interconnects in the context of a full sized mechanical prototype. Again, many of the individual technologies have been demonstrated by the LHC experiments, the RHIC projects, or in ILC detector R&D, but a complete system has yet to be demonstrated.

## 2.6 Summary

The basic concepts in the SiD Vertex detector, low-mass mechanical designs, the split cylinder support structures, and the barrel/disk geometry are essentially unchanged from the SiD LOI. However, more detailed designs for cabling, power conversion, sensor technology, and mechanical supports and cooling are included in this report. Most of these components are, or will soon be, ready for the module prototype phase. At that point decisions would need to be made on tradeoffs such as the lower mass, but more challenging, all-silicon design vs a design which has carbon-fibre or foam supports and, ultimately, sensor technology.

# Chapter 3
# SiD Main Tracker

## 3.1 Introduction

The ILC physics goals impose performance requirements on the tracking system that exceed those met by any previous system. These are summarised in Table II-3.2. In particular, the need for excellent momentum resolution over a broad $p_{\mathrm{T}}$ spectrum creates significant design challenges. For high-$p_{\mathrm{T}}$ tracks superior position resolution and mechanical stability are required while for low- $p_{\mathrm{T}}$ tracks, an aggressive material budget is critical. Meanwhile, the need for high efficiency over a wide range of momenta and large solid angle motivates an integrated approach to tracking: the vertex detector, main tracker and calorimeter are designed to work in concert to achieve these goals robustly but with a narrow margin of extra layers that result in unnecessary material.

## 3.2 Baseline Design

The main tracker is a large all-silicon detector filling the space between the vertex detector and the electromagnetic calorimeter. It comprises five cylindrical barrel layers, with the four outer layers closed at the ends by conical, annular disks, as shown in Figure II-3.1.

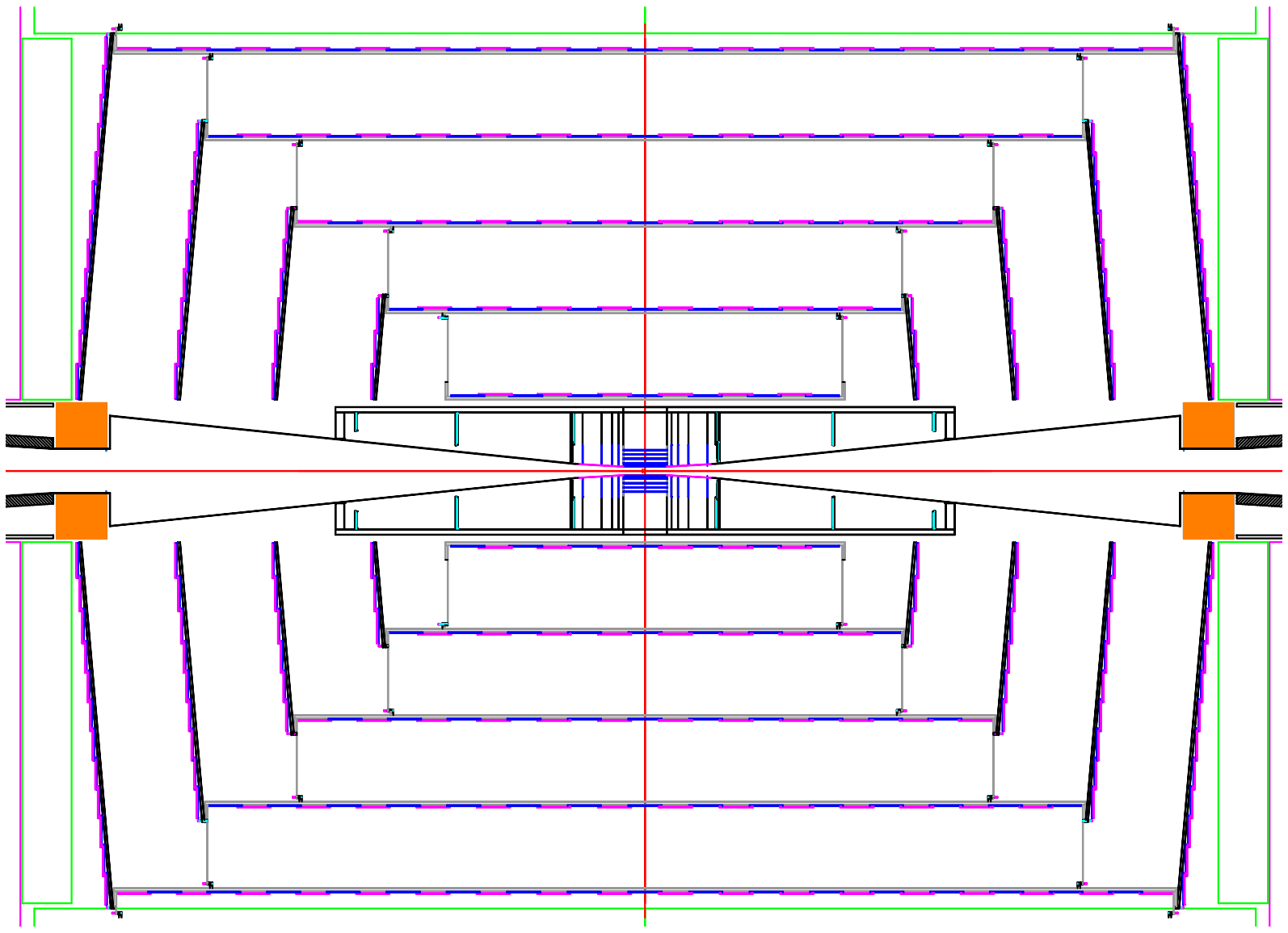

**Figure II-3.1**
$r-z$ view of the vertex detector and outer tracker.

In the baseline design the barrels are tiled with modules hosting a single silicon micro-strip sensor for axial-only measurement, while the disks are tiled with modules having a stereo pair of silicon micro-strip sensors. These cylinders are nested, connected by annular rings at the ends of each, to create a single unit supported from the ends of the electromagnetic calorimeter (ECAL). The design of the outer tracker is summarised in Table II-3.1 and more details of the design may be found in [63].

The coverage of the complete tracking system is shown in Figure II-3.2 as a function of the polar angle. At least six hits are measured for all tracks with a polar angle down to about 8°. For polar angles above 13° ten layers or more are traversed. The goals of the ILC physics program impose performance requirements on the tracking that exceed those met by any previous system and are summarised in Table II-3.2.

**Table II-3.1**
The layout of the main tracker.

| Barrel Region | R (cm) | Length of sensor coverage (cm) | Number of modules in $\phi$ | Number of modules in $z$ |
|---|---|---|---|---|
| Barrel 1 | 21.95 | 111.6 | 20 | 13 |
| Barrel 2 | 46.95 | 147.3 | 38 | 17 |
| Barrel 3 | 71.95 | 200.1 | 58 | 23 |
| Barrel 4 | 96.95 | 251.8 | 80 | 29 |
| Barrel 5 | 121.95 | 304.5 | 102 | 35 |
| **Disk Region** | **$z_{inner}$ (cm)** | **$R_{inner}$ (cm)** | **$R_{outer}$ (cm)** | **Number of modules per end** |
| Disk 1 | 78.89 | 20.89 | 49.80 | 96 |
| Disk 2 | 107.50 | 20.89 | 75.14 | 238 |
| Disk 3 | 135.55 | 20.89 | 100.31 | 438 |
| Disk 4 | 164.09 | 20.89 | 125.36 | 662 |

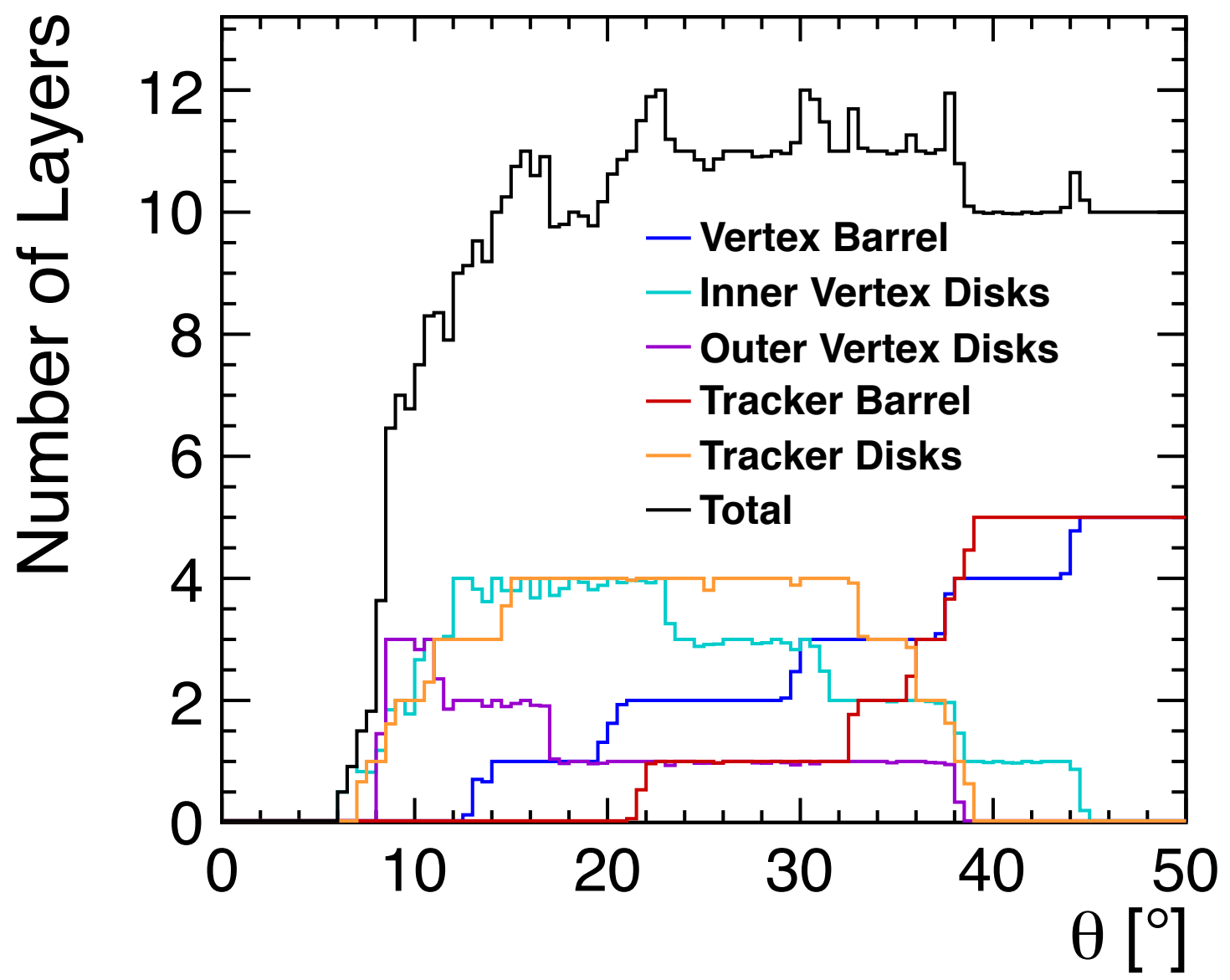

**Figure II-3.2**
The coverage of the SiD tracking system as a function of the polar angle $\theta$.

## 3.3 Baseline Design

The baseline design uses relatively conventional technologies to achieve the performance goals with low risk and minimal cost. The sensors are single-sided micro-strips. The barrel and disk supports, as well as the module supports, are composites of carbon fibre and low-density Rohacell 31 [86] foam. Low-mass hardware is fabricated in polyether ether ketone (PEEK).

There are, however, some key elements where novel solutions are required. The sensors, like those for the ECAL, employ a double-metal layer to route signals to bump-bonding arrays for readout by the KPiX ASIC [87, 88, 89]. As with the ECAL, traces on the second metal layer of the sensor connect power and signal lines on the KPiX chip to a readout cable that is also bump bonded to the face of the sensor. This arrangement eliminates the material and assembly complexity of hybrid circuit boards to host the readout electronics. The low power dissipation of KPiX makes gas cooling feasible, reducing further the required material. However, since KPiX achieves low power consumption through power pulsing with a duty cycle of approximately 1%, the instantaneous currents required to power the tracker are still large and requires a significant mass of conductor.

**Table II-3.2**
Performance goals for the main tracker.

| Parameter | Design Goal |
|---|---|
| coverage | hermetic above $\theta \sim 10^\circ$ |
| momentum resolution $\delta(1/p_\mathrm{T})$ | $\sim 2-5\times10^{-5}$/GeV/$c$ |
| material budget | $\sim 0.10-0.15\mathrm{X}_0$ in central region |
| | $\sim 0.20-0.25\mathrm{X}_0$ in endcap region |
| hit efficiency | $> 99\%$ |
| background tolerance | Full efficiency at 10× expected occupancy |

Concentrator boards located on the support rings at the ends of each barrel host DC-DC converters to transform high-voltage, low-current input power into low-voltage, high-current power for the modules, thus minimising the conductor and cross-sectional area required to deliver power into the boards from outside the tracking volume.

### 3.3.1 Barrels

A set of five cylindrical layers provides tracking coverage in the central portion of the detector. Each cylinder is formed from a sandwich of carbon fibre and Rohacell cured as a single unit, similar to those used in the DØ CFT and the ATLAS SCT [90]. The inherent rigidity of the cylinders allows for holes to be cut where allowed by module mounting locations to further reduce the average material experienced by passing particles without significantly compromising rigidity. However, it should be noted that such measures to reduce the material in the barrel and disk supports are not included in the current simulation or material estimates shown here. The outer surface of each cylinder is populated with PEEK mounting clips for the modules that allow the insertion and extraction of individual modules without the use of tools, facilitating module replacement without complete disassembly of the tracker. The normal to each module is tilted with respect to the radial direction to allow for overlap between modules that are adjacent in $\phi$ and partially compensate for the Lorentz direction. Adjacent modules in $z$ alternate between inner and outer mounting positions to provide longitudinal overlaps. Excluding overlaps, the material presented by a single barrel layer is approximately 0.9% $\mathrm{X}_0$ for tracks at normal incidence.

The modules themselves comprise a single sensor, read out via two bump-bonded KPiX ASICs and a short polyimide cable supported by a composite support frame. A picture of a prototype sensor and cable is shown in Figure II-3.4. The sensors are single-sided, poly-biased, AC-coupled, micro-strip sensors fabricated on 300 $\mu$m thick, p+ on n bulk, high resistivity silicon. The nominal sensor (readout) pitch is 25 (50) $\mu$m, with the intermediate strips capacitively coupled to the readout strips to improve single hit resolution. The KPiX chips bonded to the surface of the sensor, described more fully in Chapter 4, store time-stamped hits from the tracker exactly as for the ECAL sensors, for readout between bunch trains. Traces on the second metal layer of the sensor connect power and signal lines on the KPiX chip to a short readout cable, or pigtail, that is also bump-bonded to the face of the sensor. These copper-on-polyimide cables have tabs that provide bias voltage to the edges of the sensors and have micro-connectors that mate to the extension cables running along the surface of the cylinder to the concentrator boards located at each end. Great care has been taken to model the conductor required for each cable run and the stacks of cables required for each layer to arrive at realistic material estimates.

The back side of the sensor is glued to the face of a module support frame that comprises a pair of carbon composite sheets sandwiched around a thin sheet of Rohacell 31 [86]. This frame is approximately 50% void to reduce material and is passivated to isolate the carbon fibre from the high voltage on the rear of the sensor. A set of three spheres around the periphery of each frame provide a three-point kinematic mount to the mounting clips on the outer surface of the barrel cylinder. A small handle on each module provides a strain relief for the pigtail as it leaves the module and safe handling during assembly and installation.

### 3.3.2 Disks

The outer four barrel cylinders are partially closed at each end by slightly conical, annular disks that extend the coverage to the forward regions. These disks are fabricated using a carbon fibre and Rohacell 31 sandwich similar to that of the barrel cylinders. As with the barrel cylinders, the outer surfaces of the endcap disks are covered by a set of PEEK mounting clips that hold the disk modules. Adjacent modules in $\phi$ alternate between inner and outer mounting positions to provide overlap. The modules, mounted normal to the $z$ axis, step along the five degree slope of the cone to provide radial overlap. Excluding overlaps, the material budget for a single disk layer is approximately 1.3% $X_0$ for tracks at normal incidence.

The endcap modules are similar to those for the barrels, but have sensors on both sides of the module frames to provide a stereo measurement. The sensors on each side are identical trapezoids with strips parallel to one edge, and are technologically identical to those used in the barrel. A smaller sensor is used for small-radius portions of the disk, while a larger sensor is used in the larger-radius regions. As in the barrel, short pigtail cables bonded directly to the sensors connect to extension cables that transmit power from, and data, to concentrator boards mounted at the outer radius of each disk. The layout of the outer disk is shown in Figure II-3.3.

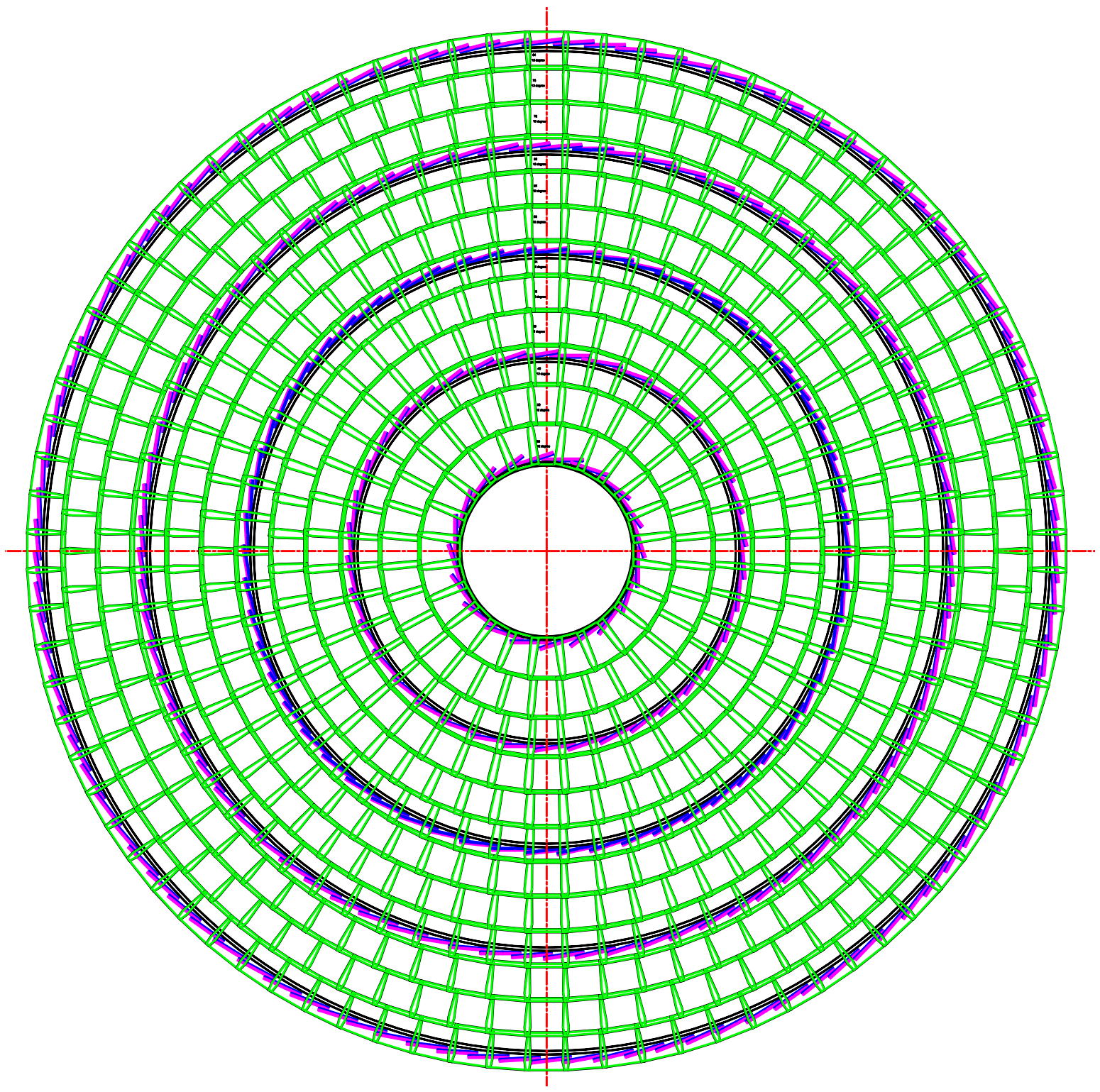

**Figure II-3.3**
R$\varphi$ projection view of the main tracker barrels (red) and disks (green).

### 3.3.3 Barrel and disk Integration

The barrel cylinders are nested, one inside the other, with spoked annular rings at the ends of each cylinder supporting it from the inside surface of the next cylinder outward. The outermost cylinder is mounted to the inside surface of the ECAL barrel. The disks that close the ends of the barrels mount to the inside circumference of these same rings, extending beyond the barrel radii to provide overlap between the barrels and disks. On the outer faces of these support rings are the concentrator boards that connect to individual modules. Each board hosts charge storage and DC-DC conversion to provide pulsed power to at least ten sensors, as well as distribution of clock and control signals, and electrical to optical conversion of output signals to concentrate the data. With high-voltage low-current power and optical transmission of data, the cable cross-section needed to service the concentrator boards for the entire detector is minimised, thereby improving the hermeticity at the barrel-disk transitions. In order to spread out the material of the concentrator boards and support rings, the barrel-disk transitions of the different layers are non-projective. The impact of the concentrator boards on the material budget can be seen in Section 10.3..

## 3.4 Critical R&D

The main tracker embraces conventional technologies where possible to minimise the risks and costs of the system and minimise the R&D necessary to bring it into production. However, there are several key areas where exploring new technologies, targeted at addressing specific performance limitations, is critical to meeting the performance goals. These technologies focus on minimising the tracker material necessary for good resolution for low-momentum particles while maintaining the mechanical stability. The key R&D projects involve the sensor readout, data transmission, power, cooling and the mechanical stability of the system.

Key to minimising the material in the tracker is the KPiX readout. Development of KPiX and the critical elements of the readout chain are of great importance, including bump bonding KPiX to sensors, development of sensors and cables, and development of the complete DAQ chain. Because the tracker is technologically identical to the ECAL in all of these respects, KPiX R&D for the tracker is undertaken together with that for the ECAL and is largely described in Chapter 4. However, with a very different set of requirements, the implementation for the tracker still differs in some respects that motivate tracker-specific R&D. This R&D has focused first on producing prototypes of a barrel module; that is the simplest module needed and solutions developed apply directly to the key issues for the disk modules.

With the requirement of full efficiency for minimum ionising particles and excellent single-hit precision, a signal-to-noise ratio in excess of 20 is required. This, in turn, sets the requirement for the noise performance of KPiX and necessitates sensors with very low readout capacitances and resistances. Prototype sensors (Figure II-3.4) were fabricated by the Hamamatsu Photonics Corporation and meet the noise performance requirements. Successive generations of KPiX chips have undergone improvements in noise performance and are now able to meet the goal, although testing of a fully assembled module will be required to verify the as-built noise performance. Assembly of a full module has been awaiting recently developed interconnect techniques for first ECAL prototypes, as described in Chapter 4.

The cable for the tracker differs somewhat from that required for the ECAL. It must have the lowest possible mass and the best possible noise performance, while servicing two KPiX chips simultaneously. A prototype cable that meets all of the requirements for the tracker has been produced (Figure II-3.4).

To speed development, this prototype cable was planned to be glued and then wire bonded to the sensor, rather than bump-bonded as called for in the design. However, a processing deficiency in the prototype sensors makes them susceptible to damage during wire bonding of the readout cable. The design and fabrication of a cable for bump bonding attachment, as successfully demonstrated in the ECAL, is under way. It will enable assembly and test of full prototype modules with the KPiX chips and sensors already in hand.

The other keys to the tracker design are low-mass support and cooling. While the module support frames are quite conventional, the techniques being considered for mounting these frames to the support cylinders and disks are somewhat novel and it is important to verify the details of these designs with prototypes before considering large-scale production. Testing with basic prototypes is being pursued so as to allow the design to evolve quickly. With a working prototype in hand, and using standard design guidelines, it should be possible to ensure success with a high degree of confidence. However, the final step of ensuring that these parts can be mass-produced will be expensive and must await further resources.

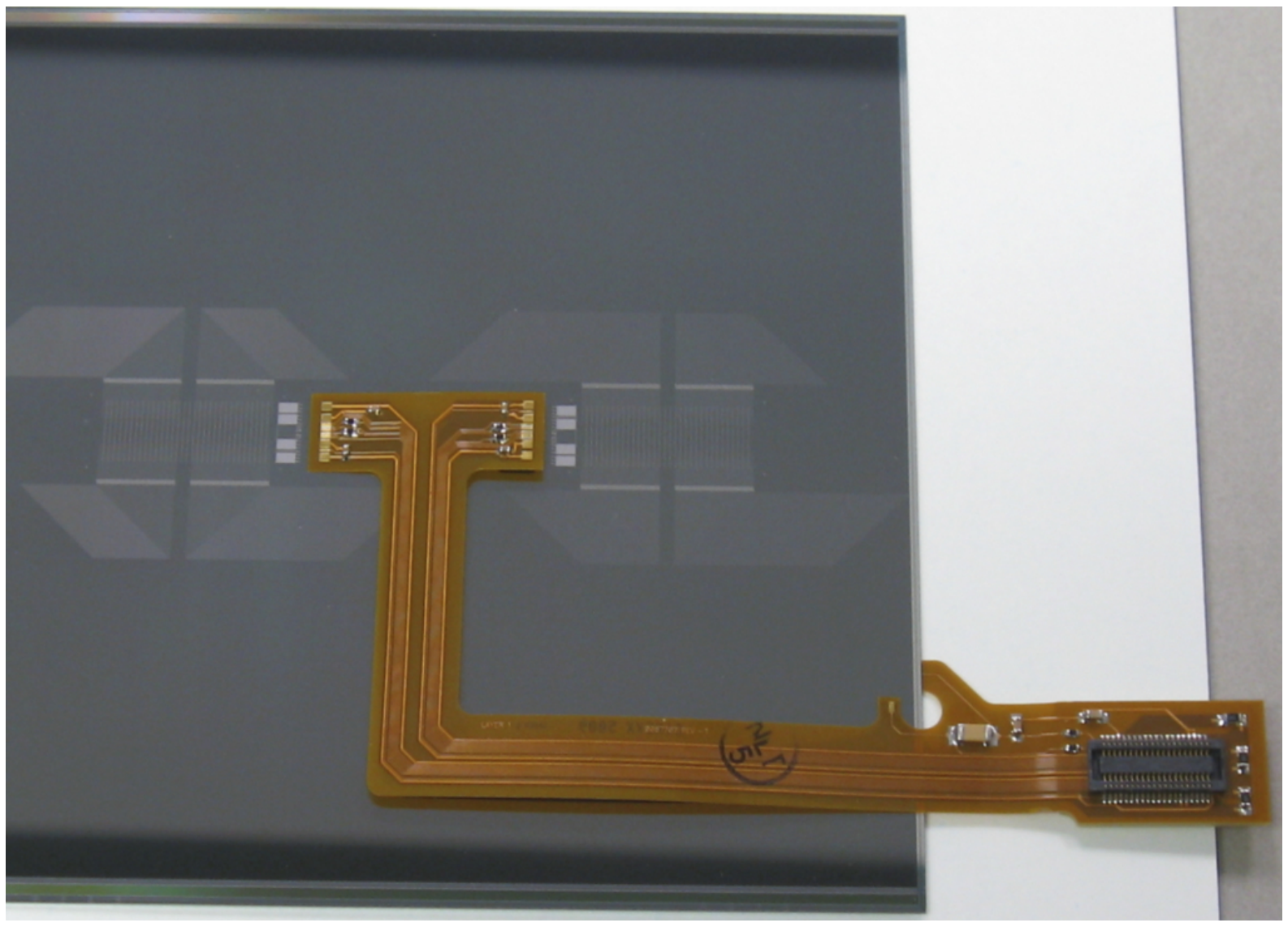

**Figure II-3.4** Prototypes of the barrel sensor and its pigtail cable shown together as they would be assembled. The bump bonding arrays for the KPiX chips and the double-metal fan outs can be seen on either side of the cable. The tab at the edge provides bias to the sensor.

Use of gas cooling depends principally on meeting the power consumption goals with KPiX, which has already been achieved [63]. The requirements for other cooling loads, such as those from the concentrator boards, can already be met with commercially available components. With gas velocities of approximately 1 cm/s, the impact on mechanical stability is negligible compared with other low-mass, gas-cooled silicon detectors being assembled for other experiments [91]. However, the requirement for hermetic coverage severely restricts gas flow in some parts of the detector, and further study is required to engineer the cooling system.

The main issue for the mechanical stability is Lorentz forces on the various elements of the tracker due to power-pulsing in the 5-Tesla magnetic field. In the barrel, conductors are largely parallel to the field, but the opposite is true in the disks. Development of cables with closely paired supply and return lines is a priority, and incorporation of this requirement into the next pigtail prototype is planned. Tests of modules inside a small-bore MRI magnet are being considered that would allow for collection of critical data on these effects. The rigidity of support structures should place any resonances well above the 5 Hz excitation frequency, but the design of the detector must take into account any harmonics.

Charge storage and high-voltage, low-current supply to the concentrator boards greatly reduce Lorentz effects on the supply of power from the outside, but present their own R&D challenges. Storing enough energy on the concentrator boards to provide power for the duration of a complete bunchtrain has become much more feasible due to industrial advances in high energy density capacitors. In fact, it appears likely that charge storage using a capacitor on each module may soon be feasible, which would all but eliminate Lorentz forces. Meanwhile, R&D into DC-DC conversion for the supply of future detectors has become an active field in recent years, with some work focused specifically on the needs of the ILC experiments [92, 93].

Efforts are ongoing that would significantly improve the performance of the outer tracker beyond that of the baseline design. One such effort considers the use of resistive charge sharing to determine the position of hits along strips to the precision of a few mm [63]. While instrumentation of both ends of each strip doubles the readout and the material budgets, cost, powering, and cooling constraints do not obviously exclude this option. Another topic of active investigation is whether the tracker could be built using monolithic active pixel sensors (MAPS). While this would clearly result in improved tracking performance, none of the technologies being investigated for the vertex detector can be convincingly scaled in power and cost to provide a solution for the outer tracker in the near future.

## 3.5 Performance

The tracking performance of the SIDLOI3 geometry has been studied using full event simulation and realistic event reconstruction of single-muon as well as di-jet events. In the reconstruction of each di-jet event a realistic number of hits from incoherent pairs and hadronic beam backgrounds are overlaid [94], corresponding to one bunch crossing at 1 TeV. This assumes that the time resolution of the tracking detectors is sufficient to separate hits from different bunch crossings..

The digitisation of the simulated tracker hits in the silicon detectors is performed using the SiSim package [95]. Diffusion of the deposited charge in the silicon is taken into account. A nearest-neighbour algorithm is used to identify the clusters which are input to the track finding. The seed tracker algorithm is used for the track finding and track fitting. This algorithm uses a strategy-based approach, where several sets of combinations of three layers define the possible seed layers for the track finding.

For the studies presented here "inside-out" tracking strategies are used. The two innermost vertex detector layers are excluded from seeding to mitigate the impact of the large number of hits from beam backgrounds on the track reconstruction time. Similarly, choosing as small a $\chi^2$ cut-off as possible in rejecting track candidates without compromising the track finding efficiency is essential in the presence of high-occupancy events.

In general a minimum of seven hits are required to define a track. In the barrel region this requirement is reduced to six hits to increase the track-finding efficiency for central low-momentum tracks. A secondary tracking algorithm using calorimeter stubs as seeds can be used to find those tracks with fewer hits from in-flight decays [96]. This algorithm is not used in the performance studies presented here.

### 3.5.1 Tracking Efficiency

The track-finding efficiency is defined as the fraction of the successfully reconstructed findable particles. The true match of the reconstructed track is determined by the majority of contributed hits. The findable particles are defined as those charged particles originating from within $\pm 5$ cm of the IP and travelling at least 5 cm. Any additional cuts are noted in the corresponding figures. In case of multiple interactions, only particles from the signal event are considered for the Track-finding efficiency. Due to the small total number of hits, falsely assigned hits have a significant impact on the reconstructed track parameters. Thus, an additional quality cut is introduced and only tracks with a maximum of one falsely assigned hit are counted as successfully reconstructed.

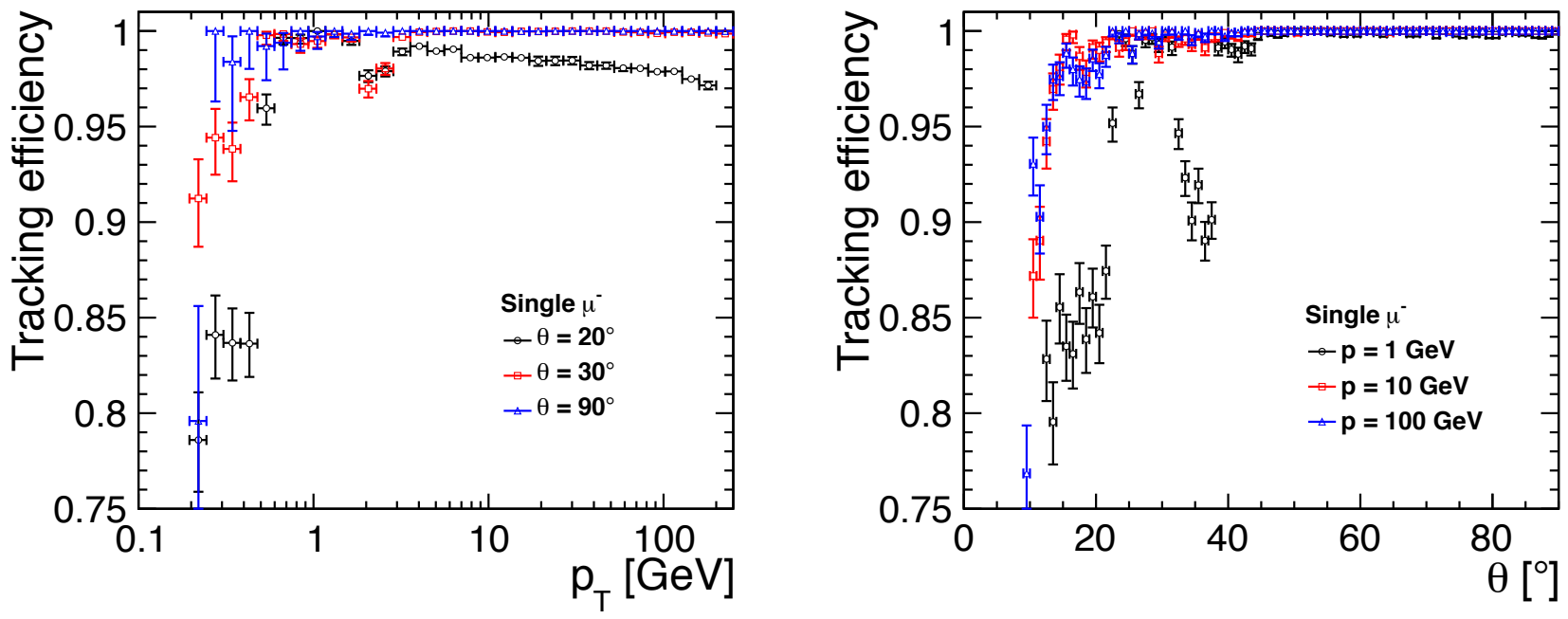

**Figure II-3.5** Tracking efficiency for single muon events in SIDLOI3 as function of the transverse momentum (left) and the polar angle (right).

For single muons the dependence of the track-finding efficiency on transverse momentum, $p_{\mathrm{T}}$, and polar angle, $\theta$, is shown in Figure II-3.5. The efficiency is nearly 100% for all tracks with a polar angle larger than $15°$ and a transverse momentum larger than 1 GeV. The efficiency for tracks below $15°$ deteriorates towards the detector acceptance of about $10°$, where it drops sharply. Requiring a minimum of seven hits reduces the efficiency for 1 GeV tracks in the region between $35°$ and $40°$ by about 10%. More central tracks are found efficiently for momenta as low as 0.2 GeV. Low-momentum tracks at polar angles below $25°$ suffer from an inefficiency of about 15% due to the higher material budget.

The track-finding efficiency has also been studied in di-jet decays of a $\mathrm{Z}'$ boson with mass of 1 TeV. The average tracking efficiency in these events, including beam induced backgrounds, is approximately 98%. The efficiency is almost constant for most polar angles and transverse momenta (Figure II-3.6). Similar to the performance for single-muon events there is a slightly reduced track finding efficiency for low-momentum tracks at a polar angle of around $40°$ and for very forward tracks of all momenta. In addition there is a drop in the efficiency for high-momentum forward tracks. These are typically in the centre of the jet and thus suffer most from confusion due to ghost hits in the stereo strip detectors.

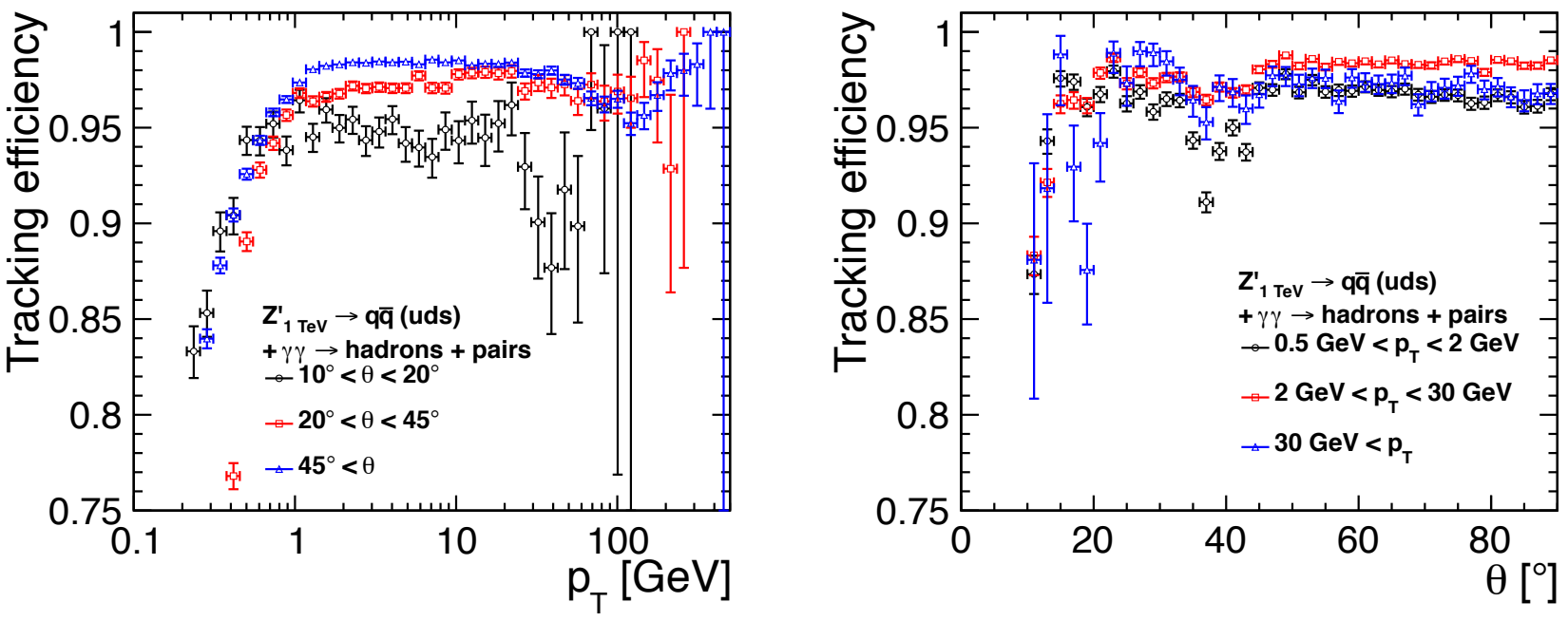

**Figure II-3.6**
Tracking efficiency in di-jet decays of a $\mathrm{Z}'$ particle with a mass of 1 TeV in SIDLOI3 as a function of the transverse momentum (left) and the polar angle (right) of the corresponding particle. Incoherent pairs and $\gamma\gamma \rightarrow \mathrm{hadrons}$ events corresponding to 1 bunch crossing are included.

In general the track-finding efficiency is limited by the total number of hits created by the corresponding particle and the local hit density (Figure II-3.7). The efficiency for particles within the acceptance that reach the calorimeters and thus necessarily pass through at least 10 layers is about 99%. The efficiency in dense jets is limited by the silicon micro-strip sizes. Particles which have any other hit closer than 100 $\mu\mathrm{m}$, which corresponds to twice the pitch of the readout in the strip detectors, have a reduced efficiency. For more isolated particles the efficiency is higher than 98%.

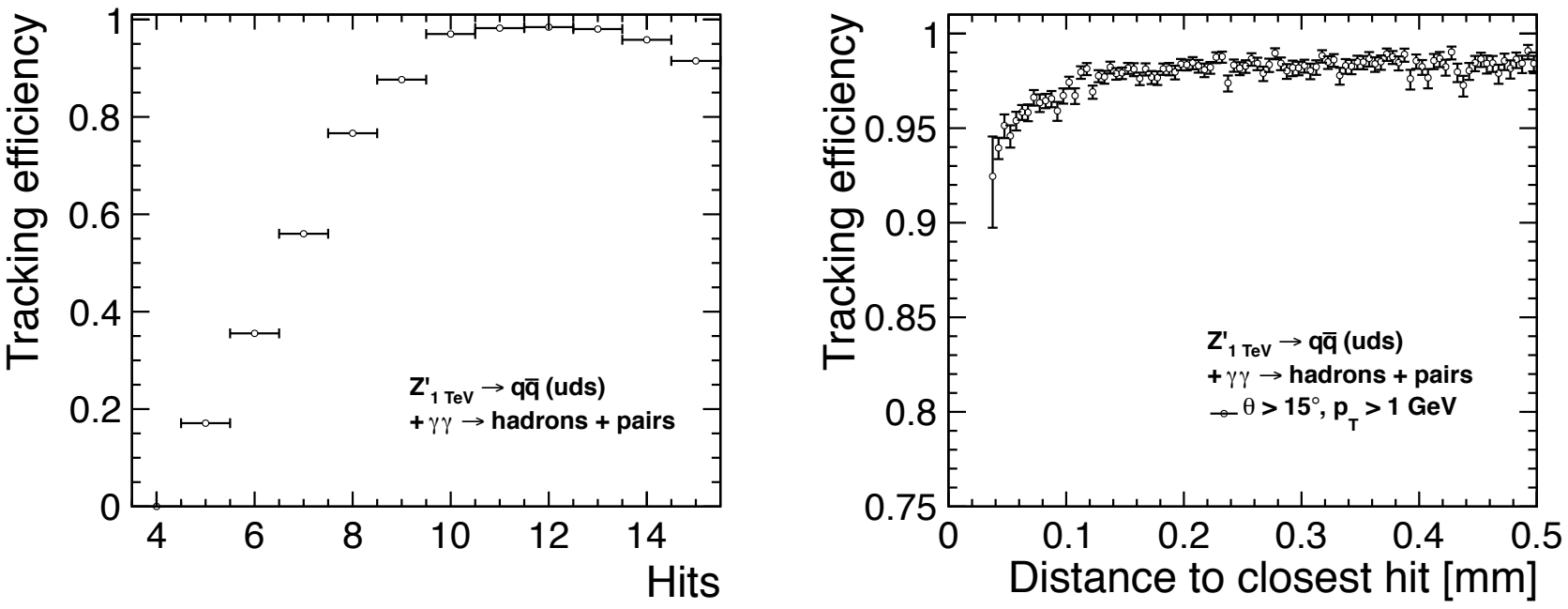

**Figure II-3.7**
Tracking efficiency in di-jet decays of a $\mathrm{Z}'$ particle with a mass of 1 TeV in SIDLOI3 as a function of the number of hits produced by the charged particle (left) and the distance to the closest hit from a different particle (right). Incoherent pairs and $\gamma\gamma \rightarrow \mathrm{hadrons}$ events corresponding to 1 bunch crossing are included.

### 3.5.2 Fake Rates

As mentioned above, the low number of tracking layers requires a very tight track purity definition. All reconstructed tracks with more than one falsely assigned hit are counted as fakes. This fake rate is shown in Figure II-3.8 for tracks in di-jet events including beam-induced backgrounds. Unlike the definition of tracking efficiency, these rates include tracks reconstructed from background particles. The fake rate is between 1% and 3%. High-momentum tracks are more likely to have more than one false hit assigned, since they are necessarily in the centre of a jet and suffer from higher local hit densities. The fake rate for tracks below $40°$ is lower by one order of magnitude than tracks in the central region. All tracker hits in the forward region have 3D information, which is not the case for the barrel strip detectors.

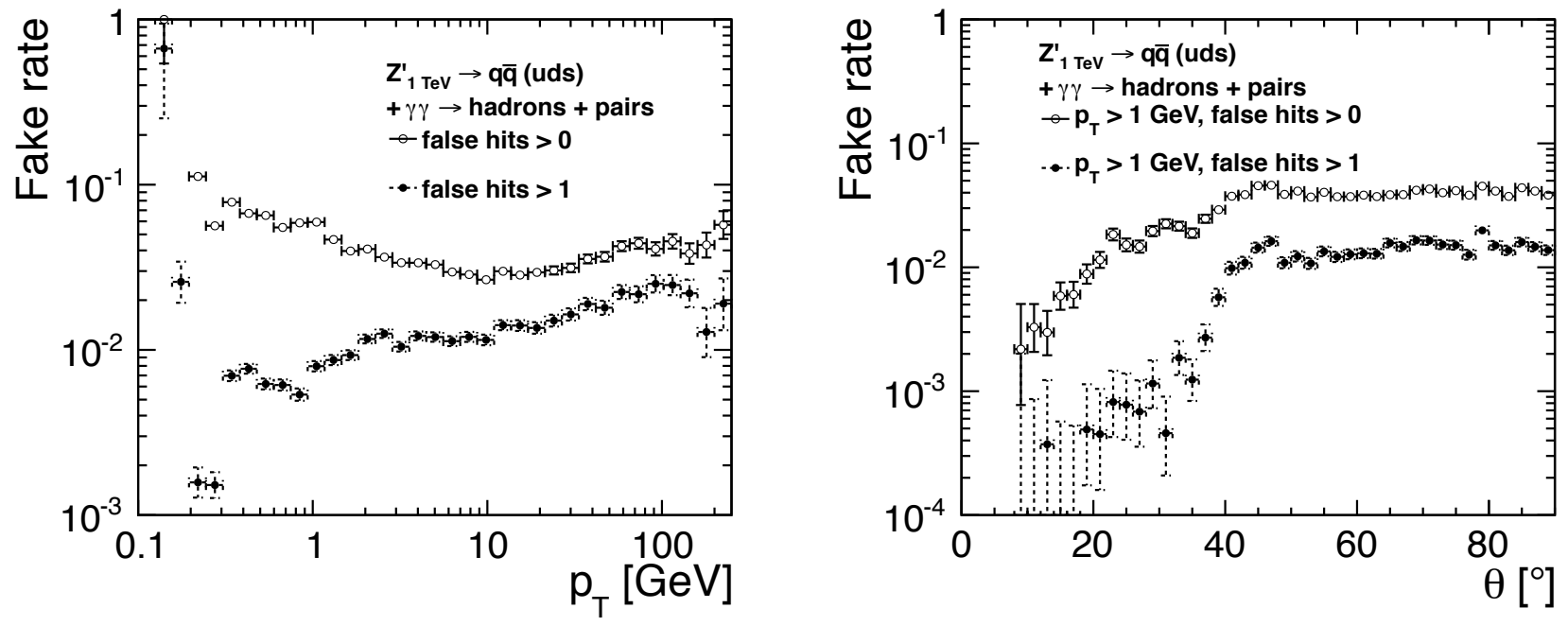

**Figure II-3.8** Fraction of reconstructed tracks with spurious hits in di-jet decays of a particle with a mass of 1 TeV in SIDLOI3 as a function of the transverse momentum (left) and the polar angle (right) of the reconstructed track. Incoherent pairs and $\gamma\gamma \to$ hadrons events corresponding to 1 bunch crossing are included.

### 3.5.3 Tracking Resolution

The normalised transverse momentum resolution achieved in the SIDLOI3 geometry is shown in Figure II-3.9 for single-muon events. The data points show the width of a Gaussian fit to the $\delta(p_\mathrm{T})/p_\mathrm{T}^2$ distribution of the corresponding reconstructed tracks. The dashed line represents a fit to the canonical parametrisation of the transverse momentum resolution:

$$\sigma(p_\mathrm{T})/p_\mathrm{T}^2 = a \oplus \frac{b}{p \sin\theta}. \tag{II-3.1}$$

Despite the ambitious material budget, the multiple scattering term given by $b$ dominates the momentum resolution for tracks up to 100-200 GeV. Whereas the momentum resolution for very forward tracks is limited by the short lever arm in the transverse projection, a momentum resolution of $\sigma(p_\mathrm{T})/p_\mathrm{T}^2 < 10^{-4}$ GeV$^{-1}$ is achieved for high-momentum tracks at polar angles larger than $30°$. Central tracks exceed a resolution of $\sigma(p_\mathrm{T})/p_\mathrm{T}^2 < 2 \times 10^{-5}$ GeV$^{-1}$.

The resolution on the transverse impact parameter, $d_0$, as well as the longitudinal impact parameter, $z_0$, is shown in Figure II-3.10. The $d_0$ resolution is better than a few $\mu$m for central tracks with a momentum exceeding a few GeV. For 1 GeV muons the $d_0$ resolution drops to about 10 $\mu$m for central tracks. In the forward region the resolution degrades by up to one order of magnitude at the acceptance limit of $\theta \approx 10°$. The $z_0$ resolution has a stronger dependence on the polar angle and, while similar to the $d_0$ resolution in the central region, it is about one order of magnitude worse for very forward tracks. In addition, the $z_0$ resolution for central tracks is limited by the lever arm of the straight line fit. In the current algorithm the strip hits in the barrel region are excluded from the straight line fit which results in a very short lever arm for central tracks. More details about the tracking performance can be found in [97].

Overall the silicon tracker shows excellent performance. Tracking efficiencies in excess of 99% are demonstrated over most of the momentum and acceptance range. An asymptotic momentum resolution of 1.46 $\cdot 10^{-5}$ and transverse impact parameter resolution better than 2 $\mu$m has been obtained. Even though the SiD tracker is very "thin" the studies show that the material budget still imposes limitations and a further reduction in mass would be beneficial. It is expected that some of the performance features can be mitigated through a further optimisation of the overall detector design. The physics results presented later are based on the tracking performance described here.

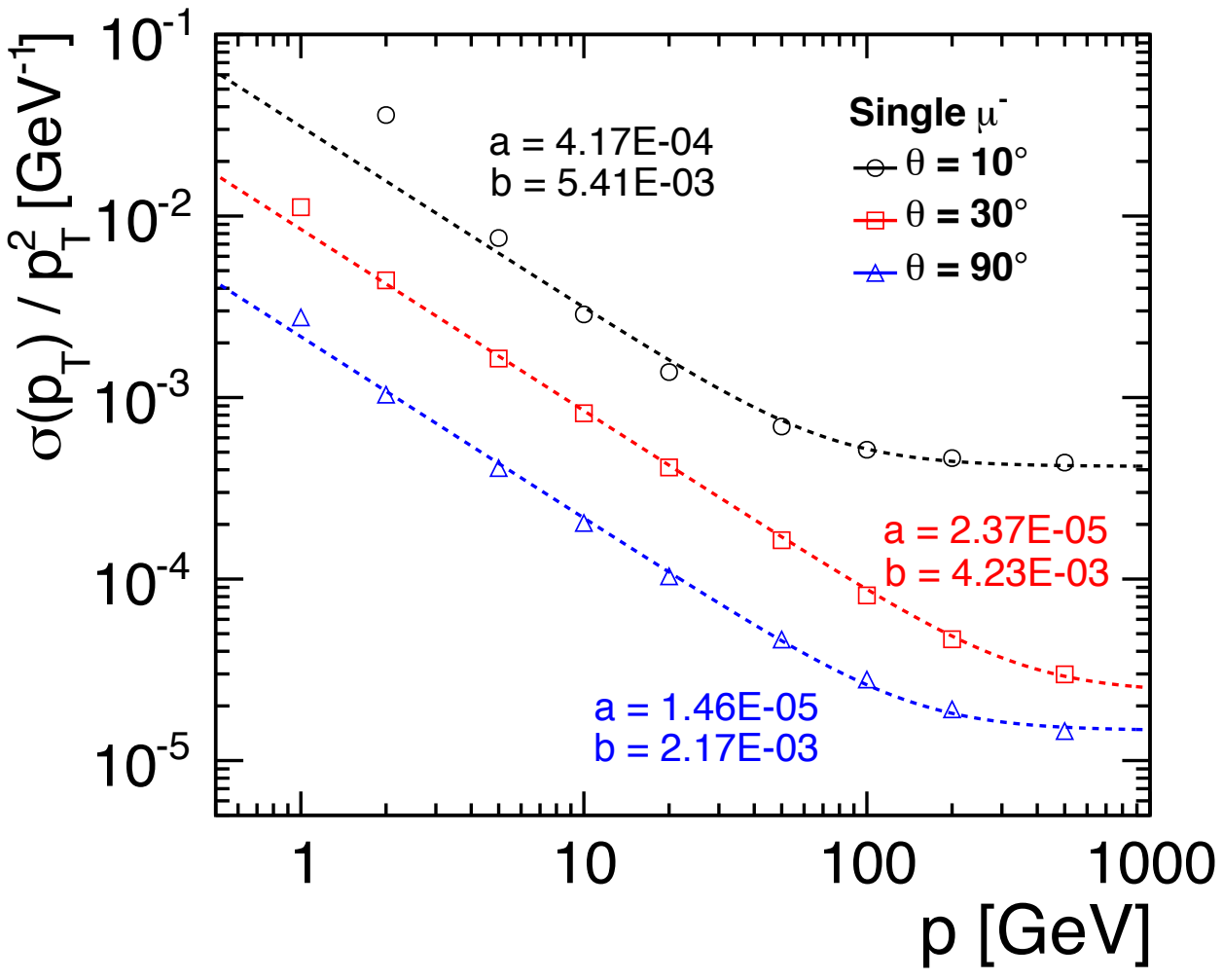

**Figure II-3.9** Normalised transverse momentum resolution for single-muon events in SIDLOI3 as function of momentum. The dashed lines indicate a fit to the parametrisation given in Equation II-3.1.

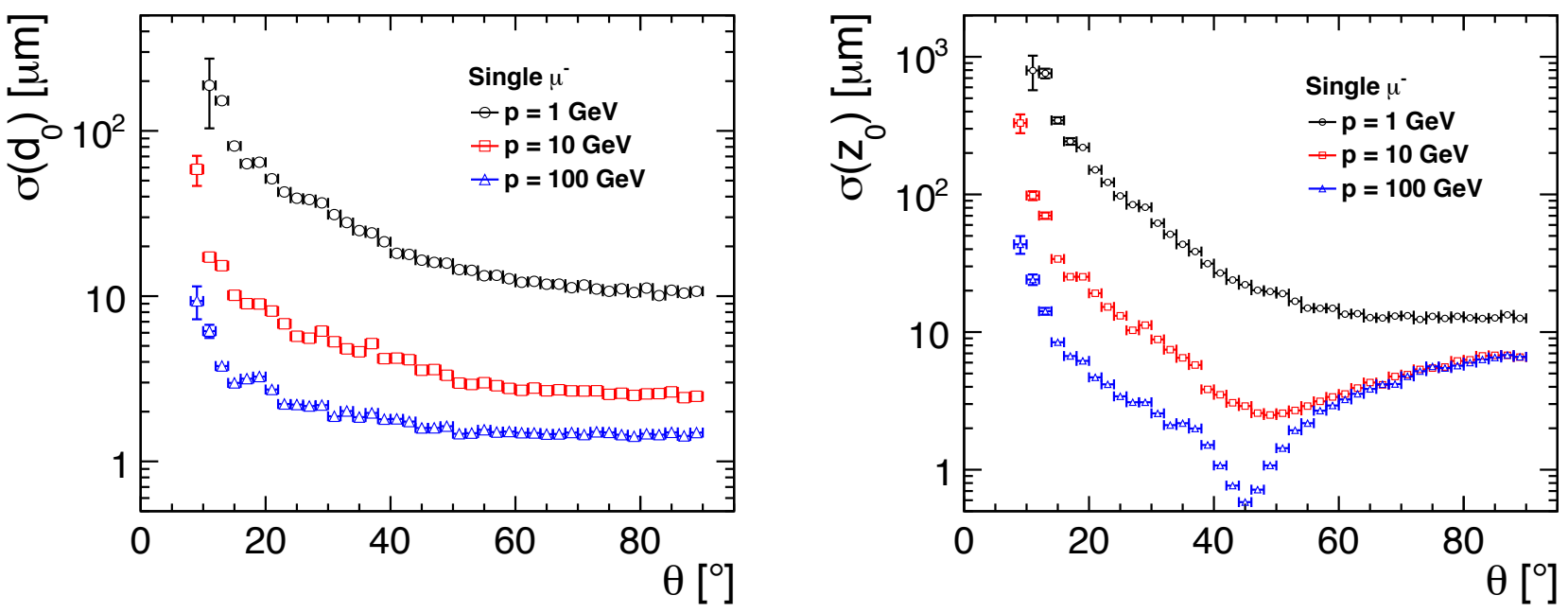

**Figure II-3.10** Impact parameter resolution $\sigma(d_0)$ (left) and $\sigma(z_0)$ (right) for single muon events in SIDLOI3 as function of the polar angle $\theta$.

## 3.6 Alignment

The alignment strategy for the SiD main tracker, vertex detector, and beam-pipe assemblies is based on:

1. a small number of robust, rigid elements;
2. precise positioning of smaller components during fabrication and assembly;
3. real-time monitoring of alignment changes, including during push-pull moves; and
4. track-based alignment for final precision.

It is expected to achieve $\approx$20 $\mu$m relative precision among outer tracker sensor modules in different layers after fabrication and assembly in the full detector. The final precision of a few $\mu$m is attained for individual sensor modules from track-based alignment, with real-time Frequency Scanned Interferometry (FSI) and laser-track monitoring providing both a bridge from the coarse to the fine alignment and a set of global corrections for time dependent structure motion and deformation [98].

The support structures for the vertex detector and main tracker have been designed to minimise distortions and maintain alignment. For the outer tracker, the double-walled support cylinders, concave disk support, and nested assembly with annular rings and kinematic mounts are intended to lead to a robust structure which can be treated as a single unit. Kinematic support from the central calorimetry is intended to minimise distortions of that structure under geometry changes of the calorimeters. For the vertex detector, double-walled support half cylinders are intended to preserve good internal alignment. Since the support structures deflect under beampipe loads, substantial R&D including measurements of prototypes will be necessary to confirm that this design will perform well.

Tracker alignment is expected to begin during fabrication and assembly. Sensor alignment within each main tracker module will be measured with respect to fiducials and mounting features of the module. Modules will be solidly anchored with stable relative position to stiff support cylinders and support disks, which are based upon carbon fibre laminate material. This material provides good thermal stability and should result in rigidity that is about 50 times higher than that of the CMS tracker. Predicted deflections of the support structures under gravity are below 10 $\mu\mathrm{m}$. Modules will be installed in groups with internal alignment of a group controlled to about 10 $\mu\mathrm{m}$. Reference features on each barrel and disk will allow the positions of each group of modules to be known with respect to the reference features to about 10 $\mu\mathrm{m}$. Hence the position and orientation of a given sensor should be known to approximately $10 \times \sqrt{3} = 17$ $\mu\mathrm{m}$. A large coordinate measuring machine (CMM) or equivalent laser-based equipment will be needed to achieve this accuracy.

Frequency scanned laser interferometry during assembly offers the potential for still better knowledge of alignment than the values above [99]. Moreover, tracker sensor modules slightly overlap within layers (and hence are tilted), which provides valuable linking of sensors within layers for track-based alignment.

We plan to use ball and cone mounts to mate barrels and disks with one another. This type of mount provides a reproducibility of 3 $\mu\mathrm{m}$. Again, a large CMM or laser-based equipment will be used to measure reference features on each object with a precision of about 10 $\mu\mathrm{m}$. This implies the precision to which individual sensors are known of $\approx$20 $\mu\mathrm{m}$, although individual groups of sensors will be known relative to one another with slightly better precision. Kinematic mounts will be used to support the outermost tracker barrel from the interior of ECAL. Support via kinematic mounts from some other portion of the detector has also been considered. All other outer tracker elements are then supported either directly or indirectly from the outermost barrel. If the kinematic mounts are designed correctly, push-pull operations may affect the absolute position of the main tracker, but should not affect its internal alignment.

The vertex detector is supported independently of the outer tracker. Outer support half-cylinders locate all vertex detector elements relative to one another. Relative alignment of elements within either top or bottom support cylinder is likely to be better than half-cylinder to half-cylinder alignment, which suggests that the two half-cylinders and detector elements they support may need to be treated as independent objects. The tracker would then be treated as three pieces: the outer tracker (including all barrel layers and disks), the upper half of the vertex detector, and the lower half of the vertex detector.

Alignment of the three pieces relative to one another will be monitored via FSI. A combination of frequency scanned interferometry and "laser-track" monitoring of relative sensor positions will monitor the internal alignment of the main tracker. After assembly, during data taking, and during push-pull operations, the FSI system will be run nearly continuously, providing "real time" measurement of global tracker distortions and of vibration amplitudes and frequencies (up to the Nyquist frequency of the FSI DAQ sampling). This type of monitoring may not be feasible for internal alignment of the two vertex detector halves due to constraints on the material budget.

A deformation monitoring system based on optical fibre sensing techniques is also under consideration [100]. Strain Optical Fibre Sensors (OFS) would be embedded in the carbon fibre supporting structures or/and sensor modules. The OFS would provide real-time strain information during the production, assembly, operation and push-pull operation of the instrumented tracker structures. From a detector integration point of view, using this kind of distributed monitoring requires only the embedding of 120 $\mu m$ diameter optical fibres in the carbon fibre composite; this means that it can be also considered as a suitable technology for the vertex detector.

### 3.6.1 Alignment Methods

The FSI system incorporates multiple interferometers fed by optical fibres from the same laser sources, where the laser frequency is scanned and fringes counted, to obtain a set of absolute lengths. This alignment method was pioneered by the Oxford group on the ATLAS experiment [98, 101, 102]. By defining $\mathcal{O}(100)$ "lines of sight" in the tracker system for absolute distance measurements, we will over-constrain the locations of fiducial points in space, allowing global distortions of the carbon-fibre support structure layers (translation, rotation, twist, bending, stretching, etc.) to be determined to the required precision. The real-time FSI measurements should allow for relevant time-dependent corrections to be applied when carrying out the final step of track-based alignment of individual silicon modules [99, 103].

With a test apparatus, the state of the art in precision DC distance measurements over distance scales of a meter under laboratory-controlled conditions has been reached and extended. Precisions better than 100 nm have been attained using a single tunable laser when environmental conditions are carefully controlled. Precisions under uncontrolled conditions (e.g., air currents, temperature fluctuations) were, however, an order of magnitude worse with the single laser measurements.

Hence for the tracker a dual-laser FSI system is foreseen that employs optical choppers to alternate the beams introduced to the interferometer by the optical fibres. By using lasers that scan over the same wavelength range but in opposite directions during the same short time interval, major systematic uncertainties can be eliminated. Bench tests have achieved a precision of 200 nm under highly unfavourable conditions using the dual-laser scanning technique [104].

A separate real-time alignment method with different systematic uncertainties will be provided by a "laser-track" system in which selected sensor modules are penetrated by laser beams to mimic infinite-momentum tracks. This method exploits the fact that silicon sensors have a weak absorption of infrared (IR) light. Consecutive layers of silicon sensors are traversed by IR laser beams.

The same sophisticated alignment algorithms as employed for track alignment with real particles can then be applied with arbitrarily high statistics to achieve relative alignment between modules to better than a few microns. This method employs the tracking sensors themselves, with only a minor modification to make them highly transparent to infrared light. A window with a diameter of few millimetres in the aluminium metallisation on the back of the sensor allows the IR beam to pass through. Since IR light produces a measurable signal in the silicon bulk, there is no need for any extra readout electronics. This alignment method has been implemented by both the AMS and CMS experiments [105, 106, 107].

The sensing element of the OFS monitor is a Fibre Bragg Grating (FBG) sensor operated as an optical strain gauge [100]. FBG sensors have many enhanced features with respect to traditional electrical strain gauges: no need for power or readout cabling, long term stability, immunity to electromagnetic fields and high voltage, as well as extreme temperature and radiation resistance. Concerning its application in tracker systems, one of the most important properties is its light weight since the actual FBG is "written" in a short section, only a few mm in length, of an optical fibre with a 125 $\mu m$ diameter. Multiplexing capabilities with many distributed FBG sensors on the same optical fibre are available; this technology also allows for long-range sensing, placing the readout unit well outside of the detector. The FBG sensor would be embedded in the carbon fibre structures supporting the modules and the module mechanics itself. The system is expected to reach local deformation sensitivities better than 1 $\mu m$ strain. The OFS monitor will provide very fast information on full structure deformations during the push-pull operations.

The final alignment of individual sensor modules will be track-based, using accumulated statistics from many detected tracks and constrained fitting to determine local position and orientation corrections for each module. The time to accumulate sufficient statistics for alignment of each individual module is expected, however, to be long enough to require continuous monitoring of global structure motions and deformations via the FSI and laser-track systems and to warrant robust, stable mechanical structures, as discussed above. Although six parameters are needed to describe a rigid module's position and orientation, the most critical parameter for microstrip planes is the offset of the module from nominal along the direction normal to the microstrips and in the module plane, since this is the coordinate measured most precisely by the strips. Expected translations in the orthogonal directions should have a negligible effect on tracking. Rotations of module planes about an axis parallel to the strips can lead to small biases in coordinate reconstruction, while rotation about an axis in the module plane and perpendicular to the strips should have negligible effects. To determine systematic offsets in the measured coordinate to a precision that is an order of magnitude smaller than the hit resolution requires $\mathcal{O}(100)$ tracks per module, assuming systematic variations in hit reconstruction for different strips in the same module are negligible. The sensor modules receiving the least number of tracks, i.e. $\cos(\theta) = 0$, outer barrel layer, are expected to be penetrated by $\mathcal{O}(10^4)$ tracks per month, making track-based alignment feasible for each separate data-taking epoch between push-pull moves. The fact that a large number of tracks produced will be back-to-back in the $x - y$ plane with approximately equal $p_{\mathrm{T}}$ values should enable more powerful constrained-fit determination of module offsets.

### 3.6.2 Push-Pull Considerations

Six rigid-body degrees of freedom are anticipated for main tracker alignment after a move of the detector: two transverse positions per end, an azimuth, and a $z$-position. Measurement data will be collected to monitor additional degrees of freedom corresponding to shape distortions which are expected to be quite small (twist, bending and stretching). The data will also be used to monitor long- and short-term instabilities of the rigid-body degrees of freedom. Twelve degrees of freedom are anticipated for the vertex detector alignment after a move: two transverse positions per barrel end, two transverse positions per support cylinder end, one azimuth per support cylinder end, and one $z$-position per support cylinder end. An additional four degrees of freedom (two transverse positions of the beampipe near each LumiCal) will be considered in estimates of support structure distortions.

During detector moves; alignment of the beampipe, the ends of the main tracker, and LumiCAL and BeamCAL will be monitored nearly continuously relative to the central calorimeter via frequency scanned interferometry. At the end of motion; alignment of the beampipe, LumiCAL and BeamCAL, and final quadrupoles will be adjusted relative to the main tracker and central calorimeter. The vertex detector is mounted from the beampipe and follows its motion. No adjustments to the position of the outer tracker are anticipated. Tune-up of beam position will be performed at low intensity while monitoring vertex detector and outer tracker backgrounds. The time required depends upon accelerator procedures.

During each move the FSI system will be operational and taking data continuously. At least six types of measurements are anticipated. The transverse and longitudinal positions of the ends of each outer tracker barrel layer at approximately eight azimuths will be measured. Also the transverse positions of each barrel layer for at least eight azimuths and additional $z$-locations along the layer will be determined as will be the overall length of each barrel layer for at least eight azimuths. The transverse and longitudinal positions of each disk near its outer periphery for at least eight azimuths will be evaluated as well as the beampipe transverse positions just inboard of each LumiCal location. Furthermore, the transverse and longitudinal positions of each vertex detector support cylinder at each end at approximately four azimuths will be assessed. Alarms will be set for any motion measured outside of what is expected. Consequently, electrical power will need to be maintained continuously for the laser system, and the optical bench will need to move with the detector. In addition, we envision measuring the strain of the structure during the move through the OFS method. Again, alarms would be set for measured values outside the expected range. Laser-track monitoring is also planned for a subset of the sensor modules. The OFS deformation monitoring system can be also operated continuously.

In summary, with the methodology described above, we expect to achieve a precision of 3 $\mu m$ or better on main tracker transverse coordinate offsets (barrels and disks) for an assumed hit precision of 7 $\mu m$ before and after a detector move. For the vertex detector, which is more demanding given an expected single hit resolution for two coordinates of better than 3 $\mu m$, the goal is a relative alignment precision of 1 $\mu m$ for coordinates transverse to the track.

# Chapter 4
# SiD Calorimetry

## 4.1 Introduction

The SiD baseline design uses a Particle Flow Algorithm (PFA) approach to Calorimetry. PFAs have been successfully applied to existing detectors, such as CDF, ZEUS, and CMS and have resulted in significant improvements of the jet energy resolution compared to methods based on calorimetric measurement alone. None of these detectors were originally designed with the application of PFAs in mind. The SiD design on the other hand considers a PFA approach necessary to reach the goal of obtaining a measurement uncertainty on the jet energy resolution of the order of 3% or better for jet energies above 100 GeV. SiD is therefore optimised for the PFA approach and the major challenge imposed on the calorimeter is the association of energy deposits with either charged or neutral particles impinging on the calorimeter. This results in several requirements on the its design:

- To minimise the lateral shower size of electromagnetic clusters the Molière radius of the ECAL must be minimised. This promotes efficient separation of electrons and charged hadron tracks.
- Both ECAL and HCAL must have imaging capabilities which allow assignment of energy cluster deposits to charged or neutral particles. This implies that the readout of both calorimeters needs to be finely segmented transversely and longitudinally.
- The calorimeters need to be inside the solenoid to be able to do track to cluster association; otherwise, energy deposited in the coil is lost and associating energy deposits in the calorimeter with incident tracks becomes problematic.
- The gap between the tracker and the ECAL should be minimised.
- The calorimeter needs to be extendable to small angles to ensure hermeticity, and be deep enough to contain hadronic showers.

Following is a description of the baseline designs and options for the ECAL and the HCAL. Also included are brief descriptions of alternative calorimeter technologies being considered by SiD.

## 4.2 Electromagnetic Calorimeter

### 4.2.1 Introduction

The major challenge imposed on the calorimeter by the application of PFAs is the association of individual particles with their energy depositions in the calorimeters. For the ECAL, this implies that electromagnetic showers be confined to small volumes in order to avoid overlaps. Effective shower pattern recognition is possible if the segmentation of readout elements is small compared to the showers. This level of transverse segmentation then also facilitates the separability of the electromagnetic showers from charged particle tracks due to un-interacted charged hadrons (and muons). The longitudinal segmentation is chosen not only to achieve the required electromagnetic energy resolution, but also to provide discrimination between electromagnetic showers and those hadrons which interact (typically deeper) in the $\approx 1$ interaction length of the ECAL. Finally, there should be a sufficient number of longitudinal readout layers to provide charged particle tracking in the ECAL. This is important not only for the PFA algorithms, but also to aid the tracking detectors, especially for tracks which do not originate from the IP.

The ECAL described in this section according to the qualitative description above is expected to have capabilites including:

- Measurement of beam-energy electrons and positrons (and photons) from (radiative) Bhabha scattering. This is sensitive to contact terms and the angular distribution provides important information on electroweak couplings, for example in interference terms between $Z$, $\gamma$, and a new $\mathrm{Z}'$. Precise Bhabha acollinearity distributions provides a key piece of the measurement of the luminosity spectrum [108], which is crucial for correct measurement of sharp threshold features in the annihilation cross section.
- electrons from $Q \to Q' e\nu$ (where Q = heavy quark).
- adequate electromagnetic energy resolution; the anticipated $\sim 0.17/\sqrt{E} \oplus 1\%$ is sufficient for this component in the PFA.

The imaging ECAL can also provide these more challenging measurements abilities compared to previous ECALs. These have already been demonstrated in simulation and in the PFA-based reconstruction:

- PFA reconstruction of photons in jets with high (95%) efficiency
- PFA tracking of charged particles in jets
- ECAL-assisted tracking (especially for decays of long-lived particles)
- $\pi^0$ reconstruction in $\tau$ decays. This is a crucial for identification of $\tau$ final states which are important for measuring $\tau$ polarisation, $P_\tau$.

Some other possibilities have not yet been fully demonstrated in simulation, but are under study and will be further studied:

- $\pi^0$ reconstruction in jets - this allows improvement of the EM component of jet energy [109]
- photon vertexing - the impact parameter resolution for photons of $\sim 1$ cm would be important for identifying decays where photons are the only visible decay products, such as predicted from some gauge-mediated SUSY-breaking models

In the following, we provide a description of the baseline ECAL. We then discuss the R&D program, including recent progress.

### 4.2.2 Global ECAL Design

A sampling ECAL provides adequate energy resolution for the ILC physics, as discussed above. Because of its small radiation length and Molière radius, as well as its mechanical suitability, we have chosen tungsten absorber/radiator. Due to practical considerations for ease of production of large plates and machining, the tungsten will be a (non-magnetic) alloy. This currently chosen alloy includes 93% W with radiation length 3.9 mm and Molière radius 9.7 mm. An additional benefit of tungsten is that it has a relatively large interaction length, which helps to ameliorate confusion between electromagnetic and hadron showers in the ECAL.

The longitudinal structure we have chosen has 30 total layers. The first 20 layers each have 2.50 mm tungsten thickness and $1.25$ mm readout gap. The last 10 layers each have 5.00 mm tungsten plus the same 1.25 mm readout gap. This configuration is a compromise between cost, shower radius, sampling frequency, and shower containment. The cost is roughly proportional to the silicon area, hence the total number of layers. We chose finer sampling for the first half of the total depth, where it has the most influence on electromagnetic resolution for showers of typical energy. However, as discussed below, an increase in sampling with fixed readout gaps has a detrimental effect on the shower radius. The total depth is 26 $X_0$, providing reasonable containment for high energy showers. Simulations in EGS4 and GEANT4 have shown the energy resolution $\Delta E/E$ for electrons or photons to be well described by $0.17/\sqrt{E} \oplus 1\%$.

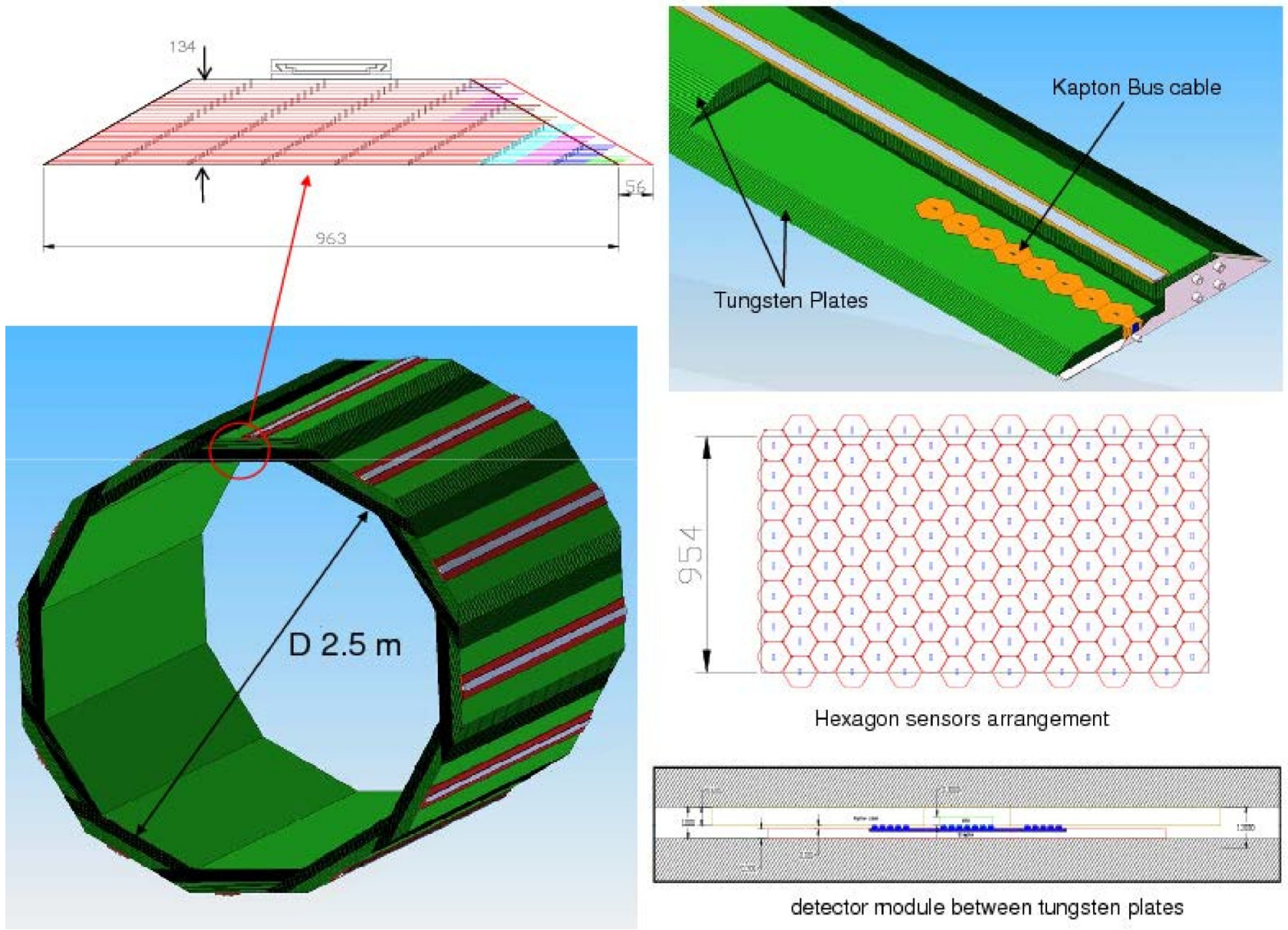

**Figure II-4.1**
Overall mechanical layout of the ECAL.

Silicon detectors are readily segmented. In the baseline design we have chosen (see description below), there is little penalty for segmenting the silicon sensors much more finely than typical shower radii. (The MAPS option takes this to the extreme.) As discussed above, the scale for this is set by the shower size, which we wish to be as small as feasible. A useful figure of merit for this is the Molière radius, which is 9 mm for pure tungsten. Since showers will spread in the material between tungsten layers, it is crucial to keep the readout gaps as small as possible. We can scale the shower radii by a simple factor to provide a figure of merit. In our case, this factor is $(2.50 + 1.25)/2.50 = 1.50$ for the crucial first 20 layers. We can then define the effective Molière radius, $\mathcal{R}$, as the Molière radius of the radiator multiplied by this factor. In our case, this is about $14$ mm. A crucial driving force in our design has been to provide as small a $\mathcal{R}$ as feasible, along with a transverse segmentation of the readout which is well below $\mathcal{R}$.

Table II-4.1 summarises the basic ECAL parameters. Figure II-4.1 shows the overall mechanical structure of the ECAL barrel, including detectors layout (for the baseline option) and readout gap.

**Table II-4.1**
Nominal parameters of the silicon-tungsten ECAL for SiD.

| | |
|---|---|
| inner radius of ECAL barrel | 1.27 m |
| maximum z of barrel | 1.76 m |
| longitudinal profile | 20 layers × 0.64 $X_0$ |
| | 10 layers × 1.30 $X_0$ |
| EM energy resolution | $0.17/\sqrt{E} \oplus 1\%$ |
| readout gap | $1.25$ mm (or less) |
| effective Molière radius ($\mathcal{R}$) | 14 mm |

Referring to Figure II-4.1, the construction of a barrel "wedge" module is carried out as follows. Because tungsten plates are only available with a maximum size of $1 \times 1$ m$^2$, the wedge assembly is done by interconnecting the plates with a screw-and-insert network, which transfers the load from the bottom of the stack to the rail. The design is self-supporting and it does not require additional material to provide the required stiffness. The assembly procedure for a single wedge is sequential with the sensors permanently captured in the gap between tungsten plates, which are specified to have high planarity, achieved at the vendor site by grinding. This specification has been verified on a batch of $15 \times 15$ cm$^2$ plates procured for the beam test module (see Section 4.2.3.2), which have planarity tolerances of $\pm 10$ $\mu$m, and have been confirmed by interviewing several tungsten vendors/producers. Because of the trapezoidal cross-section of the wedge, the assembly sequence is bottom up, with the wider plate at the base. The first layer of tungsten will be laid down on a jig tool to set the basic tolerances of the stack. Spacing inserts are placed at the locations of the cutouts at the sensor edges (see Figure II-4.2), followed by the sensors with flex cables.

The control of the gap tolerances relies on the flatness of the tungsten plate and on the spacers, which are individually quality-checked by metrology. The positioning tolerances of the sensor modules in the plane rely on the QC of spacers too, but also on the flex-cable, which will have mounting pads which mate with the inserts. The assembly of the sensors on the flex-cable will be done on a precision jig, which will guarantee the repeatability of the assigned tolerances. The second layer of tungsten will be overlaid on the sensors, once mechanical and electrical connection are tested. This process is repeated 30 times along the stack, which is the number of the active layers of a single wedge module. The last plate on the top has rails, which will allow the insertion and the support from the HCAL. Prior to insertion, each individual wedge will be equipped with a cold plate for thermal management, running along $z$ on one side of the wedge. The boxes on the two opposite sides at $\pm z$ contain the data concentrator electronics, which completes the assembly

### 4.2.3 Baseline Technology

In the baseline design, the ECAL readout layers are tiled by large, commercially feasible silicon sensors (presently from 15 cm wafers). The sensors are segmented into hexagonal pixels which are individually read out over the full range of charge depositions. The complete electronics for the pixels is contained in a single chip (the KPiX ASIC) which is bump bonded to the wafer. We take advantage of the low beam-crossing duty cycle ($10^{-3}$) to reduce the heat load using power pulsing, thus allowing passive thermal management within the ECAL modules. The realisation of this technology has been the subject of an intensive, ongoing R&D program.

The main parameters associated with the baseline technology choice are given in Table II-4.2. Some details of the design and R&D results are given below. Further details can be found in the references [110, 111].

**Table II-4.2**
Parameters of baseline silicon-tungsten ECAL and the MAPS option.

| | Baseline | MAPS option |
|---|---|---|
| pixel size | 13 mm$^2$ | $50\times$ 50 $\mu$m |
| readout gap | 1.25 mm (incl. 0.32 mm thick Si sensors) | similar |
| effective Molière radius | 14 mm | 14 mm |
| pixels per silicon sensor | 1024 | $1 \cdot 10^6$ |
| channels per KPiX chip | 1024 | - |
| dynamic range | $\sim 0.1$ to 2500 MIPs | 1 MIP |
| heat load | 20 mW per sensor | 20 mW per sensor |

Figure II-4.2 shows a sensor with 1024 pixels. Not shown in the drawing are the signal traces, part of the second layer metallisation of the sensors, which connect the pixels to a bump-bonding pad at the centre of the sensor for input to the KPiX readout chip. The pixels are DC-coupled to the KPiX, thus only two metallisation layers are required for the sensors. The pixels near the bump-bonding array at the centre are split to reduce capacitance from the large number of signal traces near the sensor centre. The electronic noise due to the resistance and capacitance of the traces has been minimised within the allowed trace parameters. The cutouts at the corners of the sensor are to accommodate mechanical stand-offs which support the gaps between the tungsten layers.

The lower-right image of Figure II-4.1 depicts a cross-sectional view of the readout gap in the vicinity of the centre of the sensor. The silicon sensor is about 320 $\mu$m thick. The KPiX is bump-bonded to the silicon sensor at an array of bump pads which are part of the second metallisation layer from sensor fabrication. This is a 32 $\times$ 32 array of bump bond pads. Polyimide (Kapton) flex cables connect near the centre of the sensors. The cables bring power and control signals into the KPiX chip and bring out the single digital output line for the 1024 channels.

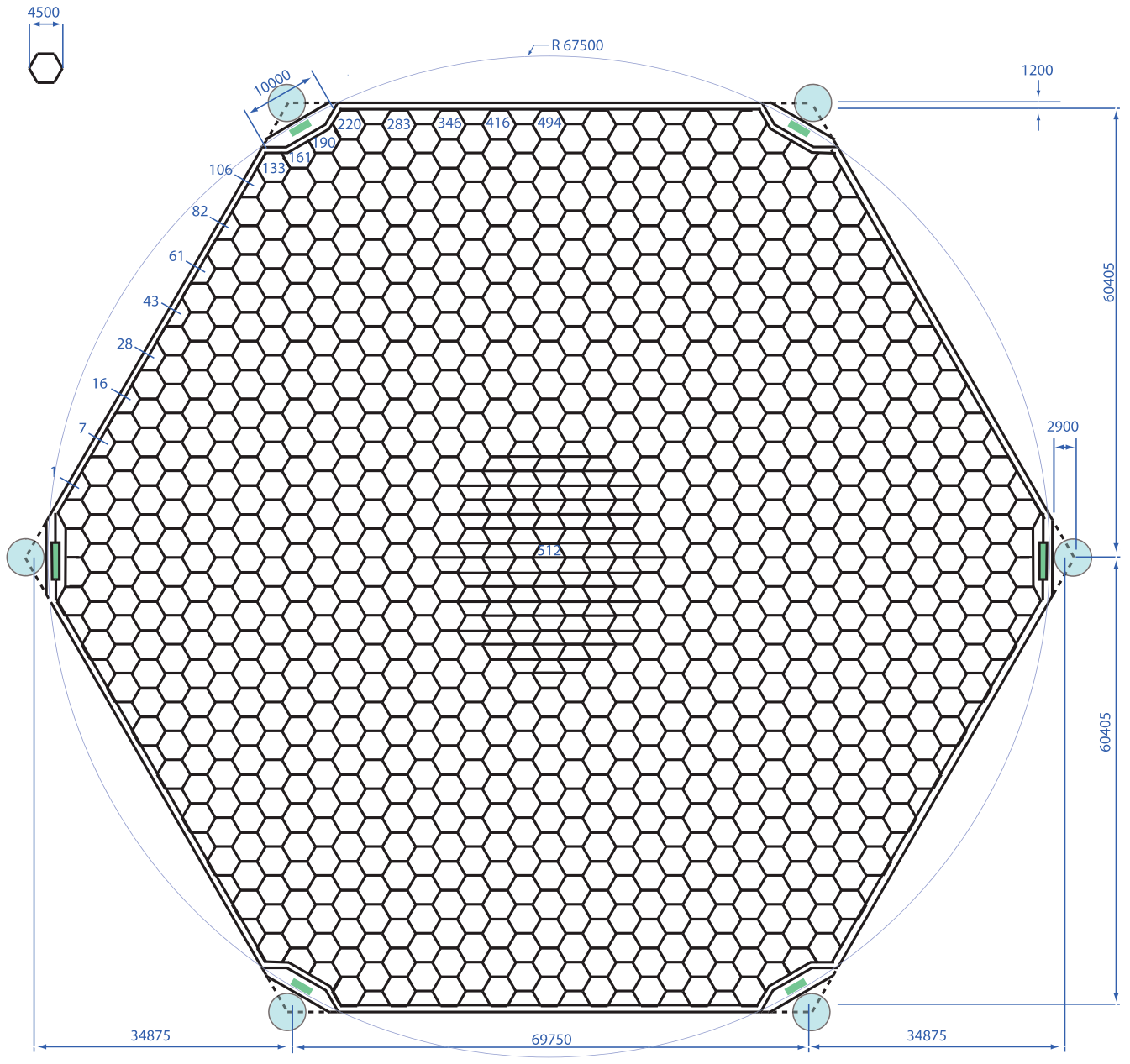

**Figure II-4.2** Drawing of a silicon sensor for the ECAL. The sensors are segmented into 1024 13 mm$^2$ pixels.

Thermal management is a crucial feature of this design. Our requirement is to hold the average power dissipation per wafer to less than 40 mW. This will allow the heat to be extracted purely passively, providing a highly compact and simple design, less subject to destructive failures. The ILC bunch structure allows for power pulsing. A factor 80-100 reduction in power consumption is obtained by switching off the most power hungry elements of the KPiX chip, e.g. the analog front end, for most of the interval between the bunch trains. The design of the KPiX chip yields an average power below 20 mW when power pulsing is applied. While we do not foresee the need for cosmic ray data, the power pulsing eliminates this possibility.

After several interactions with the R&D, in early 2012 a full 1024-channel KPiX was successfully bump-bonded to a sensor by IZM Company. Following this, a Kapton cable was successfully bump-bonded to the sensor assembly at UC Davis. The cable bonding uses a lower temperature solder than that used for the KPiX bonding. Figure II-4.3 shows the fully bonded assembly.

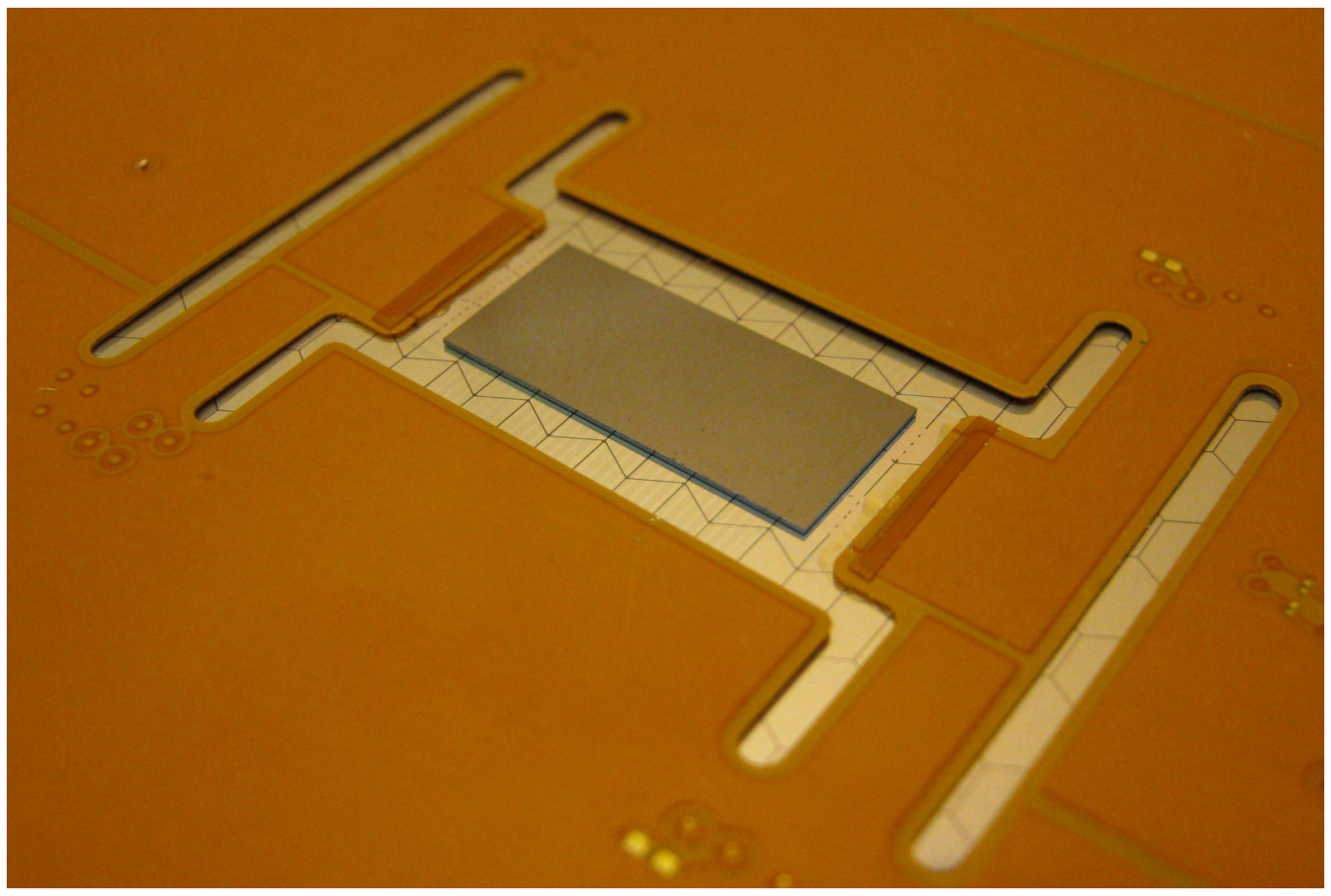

**Figure II-4.3** Photograph of the central region of a sensor. The KPiX chip is bump-bonded to the sensor and is visible through the central cutout of the Kapton cable. The slots in the Kapton allow for differential thermal expansion.

#### 4.2.3.1 Bonded sensor results

Initial bench tests of the bonded sensor of Figure II-4.3 have been carried out and the results are quite promising. A cosmic ray telescope was used to trigger KPiX and the charge of the pixel having the maximum charge was entered in the distributions shown in Figure II-4.4. The red distribution resulted when the ECAL sensor was placed within the telescope acceptance, while the blue distribution resulted when the sensor was outside the telescope acceptance. A Landau distribution (black) is fit to the red signal. The peak of the signal at about 4 fC is consistent with our expectation for MIPs passing through the fully-depleted 320 $\mu\mathrm{m}$ thick sensors.

With the highly integrated design we have chosen, a potential worry is crosstalk between channels. Figure II-4.4 indicates no evidence for crosstalk in any other channel when a large 500 fC signal is injected. The noise distribution is nicely fit by a Gaussian with RMS $0.2$ fC. This is to be compared with the 4 fC MIP signal. This noise level exceeds our requirements for the ECAL.

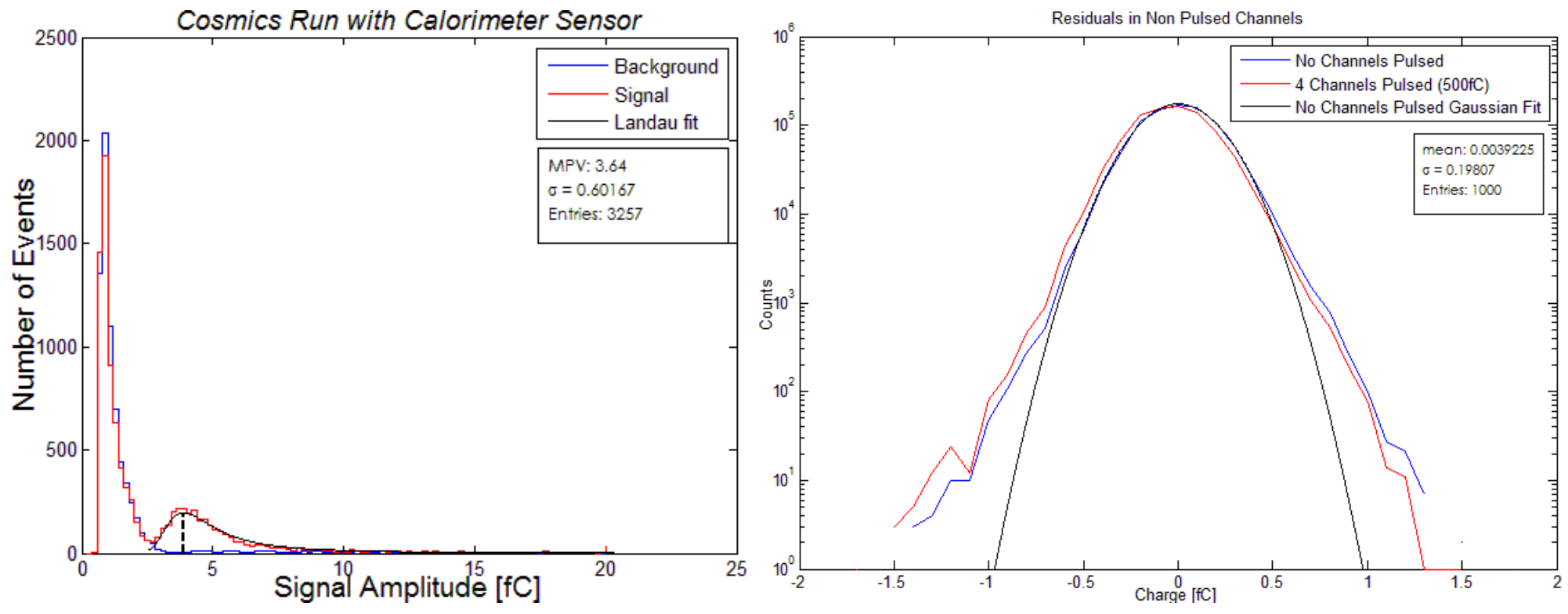

**Figure II-4.4.** Distribution of charge depositions in bonded sensor for cosmic ray triggered events (left). The MIP peak is clearly visible above the noise peak. Crosstalk test of bonded sensor (right). The charge distributions for all non-pulsed pixels are compared for a large pulse injection (red) and no pulse injection (blue). Also shown is a Gaussian fit with RMS $0.2$ fC.

#### 4.2.3.2 Prototype Module and Test Beam

Given the positive initial results of the first bonded sensors, we are moving forward with our plans to build a full-depth test module. This is shown in Figure II-4.5. The test stack is to have a width of one sensor, easily sufficient to contain electromagnetic showers. The longitudinal structure closely matches that of the SiD ECAL. The main difference is that we will have $1.5$ mm readout gaps for the test stack, rather than the nominal 1.25 mm gaps of the SiD design, in order to allow clearance for sensor assemblies to be slid in or out of the stack.

Since the operation of KPiX has been optimised for the bunch timing structure of the ILC, the optimal test facility would be a linear collider having a similar timing structure. Fortunately, SLAC is presently restoring a test beam capability at End Station A. We expect to have the sensors for the test module prepared and first data from this facility in 2013.

### 4.2.4 MAPS option

The Monolithic Active Pixel Sensor option [112] uses 50 $\times$50 $\mu\mathrm{m}^2$ silicon pixels as readout material. The main difference here is the usage of digital electromagnetic calorimetry where the ECAL is operated as a shower particle counter. The simulated performance [113] is illustrated in Figure II-4.6 where the potential advantages are clearly visible. These sensors could be manufactured in a commercial mixed-mode CMOS process using standard 300 mm wafers. This is an industrial and widely available process, so pricing for these wafers should be very competitive. We have also incorporated the usage of deep p-well implants and high resistivity epitaxial layers in this design, which was used to be rather non-standard for CMOS processes previously. This allows to include full CMOS functionality in a MAPS pixel, which has been possible before.

**Figure II-4.5** Schematic of test module to be tested in a beam. The module has a width of one sensor and a depth of 30 layers. The Kapton cables attached to each sensor feed concentrator boards, which in turn are connected to a mother board.

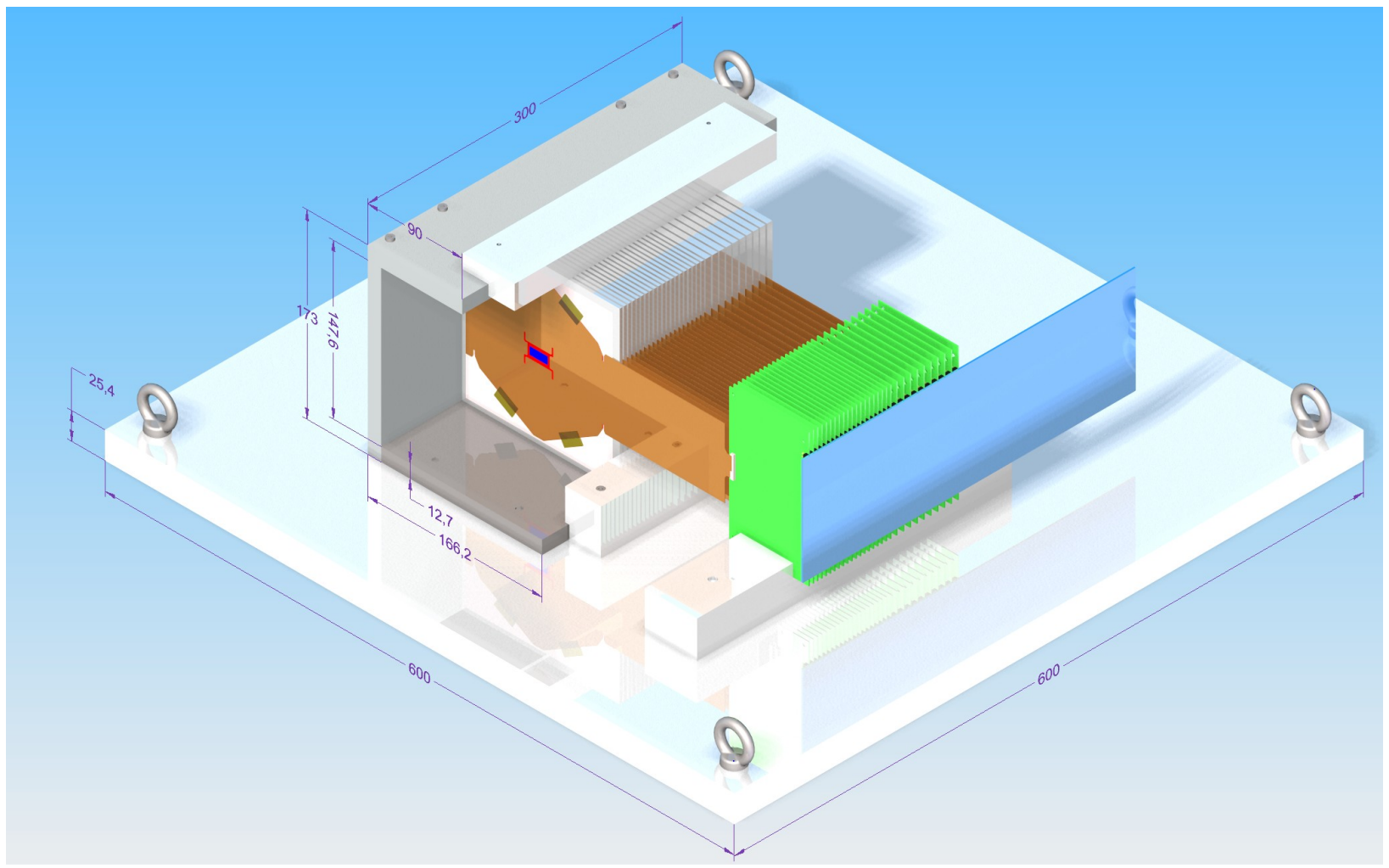

Three first-generation sensors for digital electromagnetic calorimetry, TPAC 1.0, 1.1 and 1.2 [113, 114] have been manufactured and tested. They consists of 168×168 pixels with the required size of 50×50 $\mu m$. The TPAC 1.2 will be described in more detail. It uses a the pre-Shaper architecture and consists of a charge preamplifier, a CR-RC shaper which generates a shaped signal pulse proportional to the amount of charge collected and a two-stage comparator which triggers the hit-flag. The sensor supports single-bunch time stamping with up to 13 bits. Each pixel has a 6 bit trim to compensate for pedestal variations and each pixel can be masked off individually. A bank of forty-two pixels shares nineteen memory buffers to store the hits during the bunch train. The sensor also supports power-pulsing already and is able to power off its front-end in the quiet time between bunch trains.

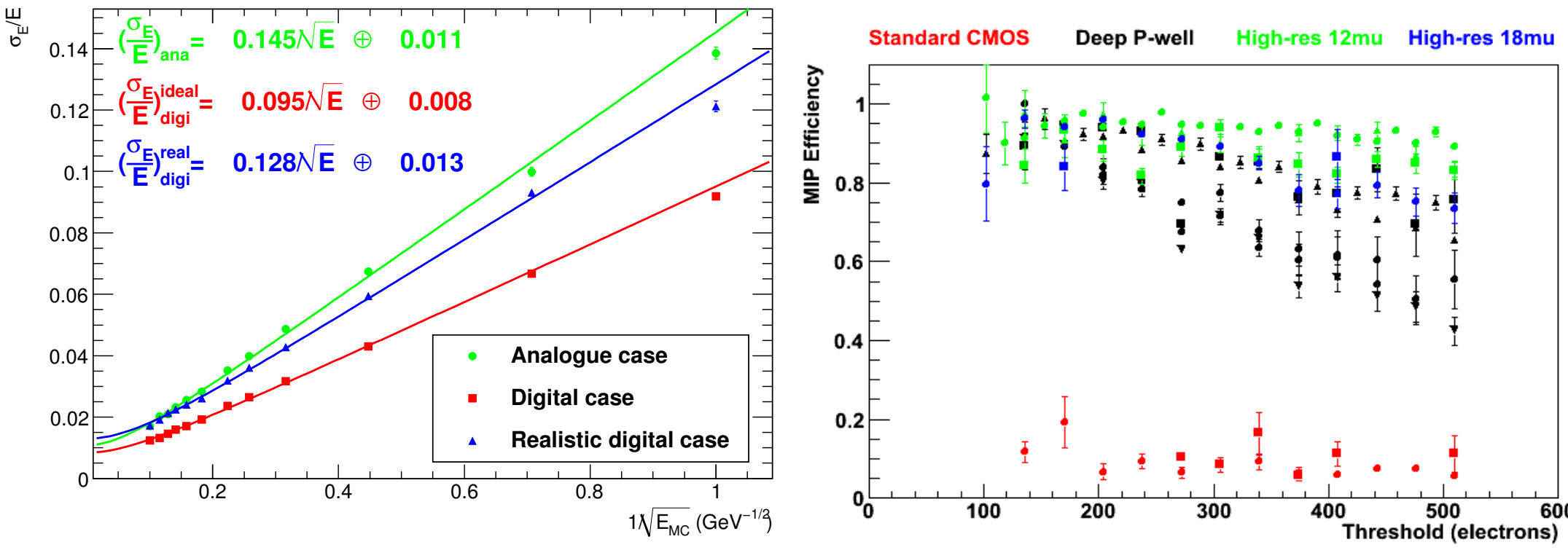

**Figure II-4.6.** Left: The energy resolution as a function of the incident energy for single electrons for both analog and digital readout using a GEANT4 simulation. The realistic digital cases includes effects of saturation and charge sharing, leading to a degradation of 35% [113]. Right: The MIP detection efficiency as a function of the comparator threshold for different process variants of TPAC 1.2 [114, 115]; no deep p-well implant (standard CMOS), deep p-well implant and high-resistivity epitaxial layer (all 12 $\mu m$) and 18 $\mu m$ high-resistivity layer

All sensors have been tested using sources and lasers [112, 113, 114, 115]. The TPAC 1.2 sensor was tested in test beams at CERN and DESY using a stack of six TPAC sensors. The minimum ionising particle efficiency was found by using the outer four TPAC planes to perform the track finding and then deriving the MIP detection efficiency of the two inner planes [114, 115]. This was done for a range of threshold values (see Figure II-4.6). The version without a deep p-well (using plain CMOS) shows a very low MIP efficiency. Including the deep p-well then increases the detection efficiency to around 80-85%. The addition of the high-resistivity epitaxial layer then makes TPAC close to a 100% efficient for minimum ionising particles.

The MAPS option is designed to fit in the same mechanical structure as the baseline option and we foresee a sensor size of 5 × 5 cm$^2$ (baseline) for a final system. As the active sensor area is less than 20 $\mu$m thick, it does allow back-thinning of the wafers down to 100 $\mu$m or less. The main parameters for the MAPS option are summarised in Table II-4.2.

### 4.2.5 Calibration and alignment

Silicon detectors are inherently insensitive to gain variations with time and should not have significant inter-pixel gain differences. Pixel to pixel gain differences in the electronic readout are calibrated by dedicated calibration circuitry within the KPiX chip. Perhaps the main calibration issue will be sensor to sensor gain differences. These are not expected to be large, but we are investigating different options for this calibration.

Alignment within ECAL modules and between modules should not be difficult to control with careful fabrication. Alignment to the inner detectors can be sufficiently established using charged particle tracks.

## 4.3 Hadronic Calorimeter

### 4.3.1 HCAL requirements

Within the PFA paradigm the role of the hadron calorimeter is to allow identification of the energy deposits from charged particles, and to measure the energy associated with neutral hadronic particles, such as neutrons and $K_L^0 s$. In this approach the challenge is to unambiguously identify energy deposits in the calorimeter as belonging to charged particles (and therefore to be ignored) or to neutral particles (and therefore to be measured). As a consequence, the optimal application of PFAs requires calorimeters with the finest possible segmentation of the readout. Further requirements imposed by the application of PFAs on the hadron calorimeter include:

- Operation in a (strong) magnetic field;
- Limitations on the thickness of the active element (to keep the coil radius as small as possible);
- Manageable accidental noise rate (to keep the confusion term manageable).

In general, the active elements need to satisfy standard performance criteria, such as reliability, stability, a certain rate capability and be affordable.

### 4.3.2 Description of the DHCAL concept

The PFA-based HCAL is a sandwich of absorber plates and instrumented gaps with active detector elements. The active detector element has very finely segmented readout pads, with 1×1 cm$^2$ size, for the entire HCAL volume. Each readout pad is read out individually, so the readout channel density is approximately 4×10$^5$/m$^3$. For the entire SiD HCAL, with $10^2$m$^3$ total volume, the total number of channels will be 4×10$^7$ which is one of the biggest challenges for the HCAL system. On the other hand, simulation suggests that, for a calorimeter with cell sizes as small as 1×1 cm$^2$, a simple hit counting is already a good energy measurement for hadrons. As a result, the readout of each channel can be greatly simplified and just record 'hit' or 'no hit' according to a single threshold (equivalent to a '1-bit' ADC). A hadron calorimeter with such simplified readout is called a Digital Hadron Calorimeter (DHCAL). In a DHCAL, each readout channel is used to register a 'hit', instead of measure energy deposition, as in traditional HCAL. In this context, gas detectors (such as RPC, GEM and Micromegas) become excellent candidates for the active element of a DHCAL. The SiD baseline design uses a DHCAL with RPC as the active element.

## 4.3.3 Global HCAL mechanical design

The SiD HCAL is located inside the magnet and surrounds the electromagnetic calorimeter, the latter being fixed to it. The HCAL internal and external radii are respectively: $R_{int}$=1417 mm and $R_{ext}$=2493 mm. The overall length is 6036 mm, centred on the interaction point.

The HCAL is divided into twelve identical azimuthal modules, as illustrated in Figure II-4.7. Each module has a trapezoidal shape and covers the whole longitudinal length. The chambers are inserted in the calorimeter along the Z-direction from both ends and can eventually be removed without taking out the absorber structure from the magnet. Special care of the detector layout has been taken into account to avoid a crack at $\theta$=90°.

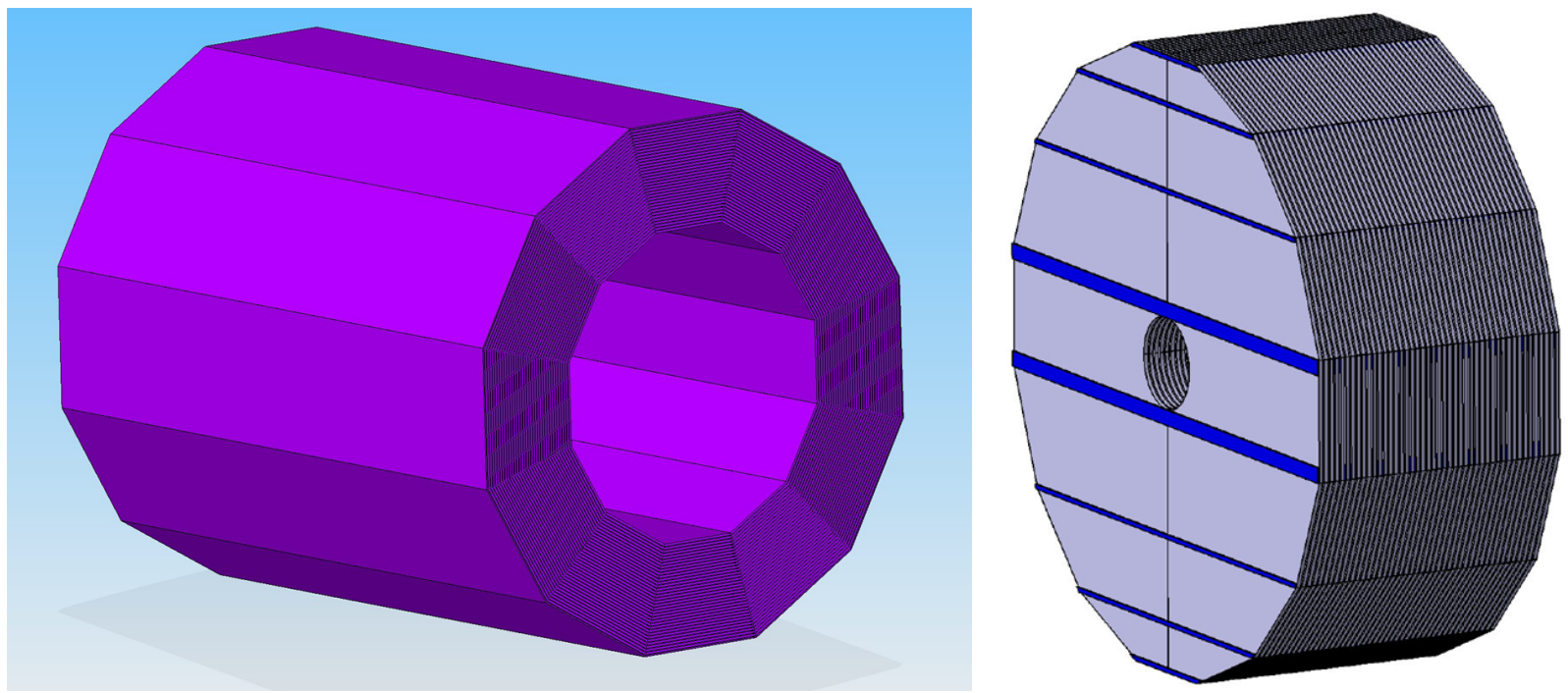

**Figure II-4.7** Cross-section of the HCAL barrel (left) and face and top views of the HCAL endcap (right).

The absorber plates are supported by several stringers fixed radially on both sides of the modules. Stringers of two consecutive modules are staggered in order to maximise the active detector area. Although the space between two consecutive modules is not instrumented, it is however filled by the absorber material. The barrel will be fixed on the magnet at 3 and 9 o'clock or 5 and 7 o'clock. Each endcap forms a plug that is inserted into an end of the barrel calorimeter. The layer structure of the end cap calorimeters is the same as for the barrel. Figure II-4.7 shows a view of one endcap.

## 4.3.4 Baseline technology

In the baseline design, the active element of the SiD DHCAL uses Resistive Plate Chambers (RPCs). RPC fulfils all the above mentioned requirements for a PFA DHCAL. For the standard two-glass plate design [116], a position resolution of a few hundred $\mu$m is typical and so a segmentation of the readout into pads of $1 \times 1$ cm$^2$ or smaller is technically meaningful. The design can be tuned to minimise the thickness. With two glass plates a layer thickness smaller than 8 mm appears achievable. If using the 1-glass design [116] an overall thickness of 1 mm less is conceivable. The noise rate for RPCs is in general extremely low, with values below 1 Hz/cm$^2$ for MIP detection efficiencies exceeding 90% [117].

RPCs with glass plates as resistive plates are reliable and operate stably. Long term tests showed no changes in performance [117]. The rate capability of RPCs is well understood [118] and is adequate for most of the solid angle of a colliding beam detector. In the forward region, where the rates are in general higher, RPCs using resistive plates with lower bulk resistivity or other high rate gas detectors might be required.

### 4.3.4.1 RPC chamber designs

Resistive Plate Chambers (RPCs) are gaseous detectors primarily in use for the large muon systems of colliding beam detectors. The detectors feature a gas volume defined by two resistive plates, typically Bakelite or glass.

The outer surface of the plates is coated with a layer of resistive paint to which a high voltage is applied. Depending on the high voltage setting of the chamber, charged particles crossing the gas gap initiate a streamer or an avalanche. These in turn induce signals on the readout strips or pads located on the outside of the plates.

Various chamber designs have been investigated [116] for the SiD DHCAL. Of these two are considered particularly promising: a two-glass and an one-glass plate design. Schematics of the two chamber designs are shown in Figure II-4.8. The thickness of the glass plates is 1.1 mm and the gas gap is maintained with fishing lines with a diameter of 1.2 mm. The overall thickness of the chambers, including layers of Mylar for high voltage protection but excluding readout board, is approximately 3.7 mm and 2.6 mm, respectively. The two-glass design is the current baseline, however, due to its attractive features, the one-glass design is being actively developed.

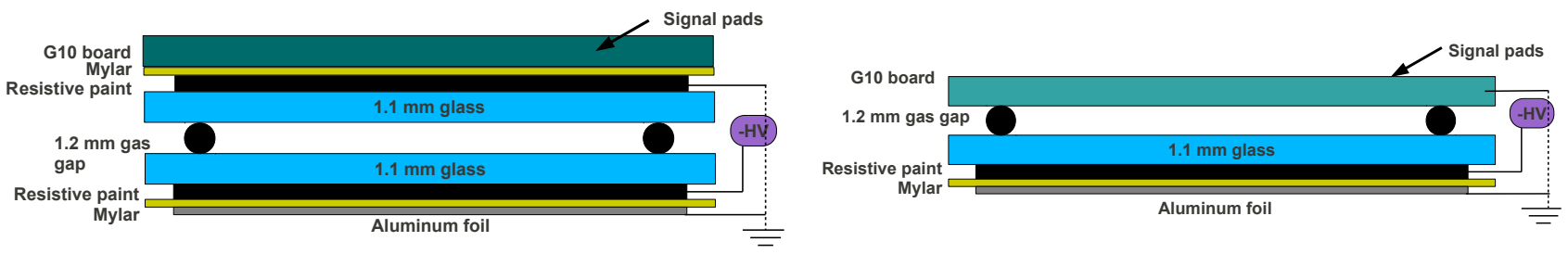

**Figure II-4.8** Schematic of the RPC design with two glass plates (left) and one glass plate (right). Not to scale.

### 4.3.4.2 Readout

The electronic readout system needs to be optimised for the readout of tens of millions of readout channels envisaged for the SiD hadron calorimeter operating at the future International Linear collider. Due to the high channel count, a front end ASIC and several layers of data concentration are considered necessary. In the R&D phase, a readout system was developed and constructed for the DHCAL prototype which handles nearly 500,000 readout channels. Even though the system was optimised for test beam operation and did not address all requirements for a realistic SiD DHCAL system, it achieved the very first embedded front-end readout for a calorimeter system and serves as a milestone towards the final engineering design.

A block diagram of the prototype DHCAL readout system is shown in Figure II-4.9. The electronics is divided into two parts: The "on-detector" electronics processes charge signals from the detector, collects data for transmission out, and acts as the interface for slow controls. The "back-end" electronics receives and processes the streams of data from the front-end electronics, and in turn passes it to the Data Acquisition (DAQ) system. It also has an interface to the timing and trigger systems.

A custom integrated circuit (ASIC) has been developed for the front-end. The ASIC chip, called DCAL performs, in addition to ancillary control functions, all of the front-end processing, including signal amplification, discrimination/comparison against threshold, recording the time of the hit, temporary storage of data, and data read out. It services 64 detector channels with a choice of two programmable gain ranges (~10 fC and ~100 fC sensitivity.)

The chips reside on front-end printed circuit boards that are embedded in the DHCAL active layer. There are 24 chips on a front-end board, servicing 1,536 channels. An FPGA based data concentrator (DCON) resides on the edge of the front-end board which collects data from the 24 DCAL chips and serves as the first level of data concentration. The DCON's send their data to the data collectors (DCOL's) through serial links. The DCOL's are located in VME crates and serves as the second level of data concentration. Each DCOL receives data from 12 DCON's and store the data into a large buffer for DAQ program to read via VME bus.

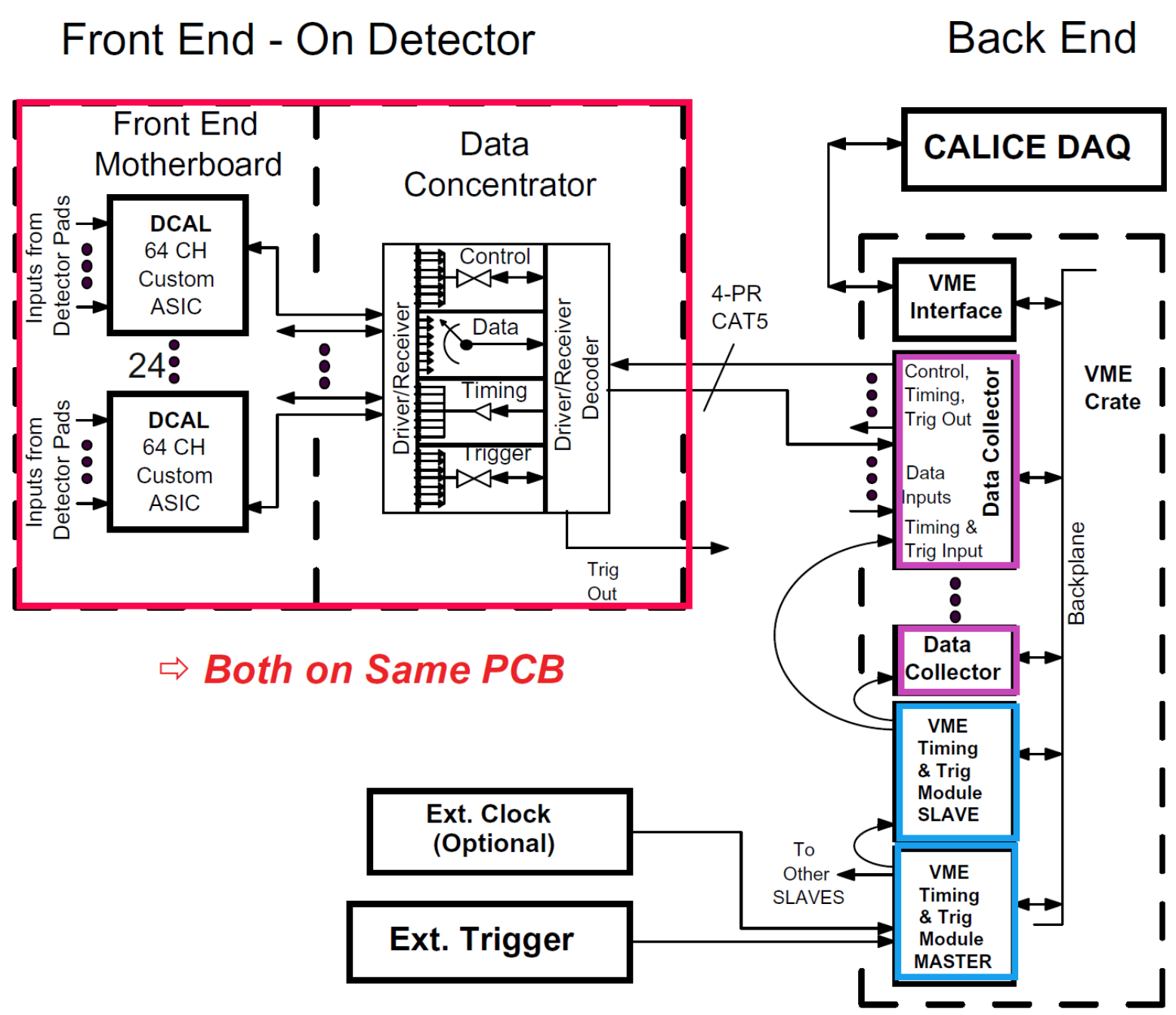

**Figure II-4.9** Block diagram of the readout system of the DHCAL prototype.

The system runs in two modes: triggered mode and triggerless (or self-triggered) mode. The first was designed for test beam runs with an external trigger. The second mode was designed for noise measurement, but was also found to be extremely useful in monitoring the RPC condition, collecting cosmic ray data for calibration and capturing all particles in a spill at a test beam.

The readout system for the DHCAL prototype is proven to be very reliable and has extremely low error rate. The front-end has very low noise. For all practical purposes, the noise coming from the front-end electronics can be safely ignored. As previously mentioned, the readout system for the DHCAL prototype did not address all requirements for a real DHCAL system in a colliding beam experiment. Further R&D is needed for the final readout design. There are two possibilities:

- Continue the development based on the success of the current DHCAL prototype readout system and focus on reducing power consumption, improving data transmission routing and optimising readout board thickness;
- Adopting KPiX readout for the DHCAL. The R&D needs to start from small scale chamber test and have several system level prototyping. This approach will be a longer development, however, it has the potential benefit of having a more uniform readout across the entire SiD detector;

### 4.3.4.3 Active layer design

In the baseline design, the barrel modules have a trapezoidal cross section. There are 40 layers in each module, each layer is 28 mm thick, consists of a 20 mm thick steel absorber and a 8 mm active layer for the RPC and its front-end readout. The innermost layer has a size of 740 mm $\times$ 6036 mm, and the outermost layer has a size of 1350 mm $\times$ 6036 mm.

Within each active layer, there are six RPCs that cover the entire area. Each RPC has a width 10 mm shorter than the full width of the layer. Along the beam direction, there are two different lengths for the RPCs: five at 966 mm and one at 1206 mm. With a 3 mm frame, the RPCs will have an active length of 960 mm or 1200 mm, which accommodates 96 or 120 readout pads of 1 $\times$ 1 cm$^2$ size. The six RPCs forms a row within a layer with the long RPC at the end. Neighbouring layers have the long RPC at alternating ends so that the dead areas between the RPCs are not lined up in all layers. The smallest RPC unit has a size of 730 mm × 966 mm, while the largest RPC has a size of 1340 mm × 1200 mm, which is still within comfortable range for RPC construction and handling.

The RPC sizes exceed the size of a reasonable PCB board, so each RPC will be read out by several boards that have 1 × 1 cm$^2$ pads on the RPC side and front-end components on the other side. The basic dimension of the readout board is 32×32 cm$^2$. The short RPC needs exactly three readout boards along its length, and the long RPC needs three standard boards and a 32×24 cm$^2$ board to fill the whole length. Several special boards, that are 32 cm or 24 cm long and have different widths, are needed to fill the entire width of the RPCs in different layers. Boards that are in the same row along beam direction are chained together using flex cables and are read out from both ends of the module.

The baseline design uses a two-glass RPC design, which has a thickness of 3.7 mm, including insulation material. The readout board has a total thickness of 3.8 mm, including the height of surface mount components. The total thickness of the active elements adds up to 7.5 mm, which leaves 0.5 mm tolerance to slide the RPC and the readout boards in and out of the 8 mm gap between the absorbers.

The RPCs run with negative high voltage. The high voltage side of the RPC faces the inner absorber, and the readout is on the ground side of the RPC and is close to the outer absorber. The RPCs leave 5 mm space along both sides of the gap which allow two 1/8" gas tubes and one thin high voltage cable to run into the gap along each side. They supply gas and high voltage to the two inner RPCs on the same half of the module, and the end RPC is directly accessible from the end of the module.

The endcap modules have a similar active layer design, except that all RPC's have direct access form the end of the modules which make gas and high voltage connections significantly easier.

#### 4.3.4.4 Results of prototype testing

The development of a hadron calorimeter based on the RPC technology progressed in several stages: a) Studies of various RPC designs, b) Construction and testing of a small scale calorimeter prototype, the Vertical Slice Test (VST), c) Construction of the DHCAL prototype, d) Testing of the DHCAL prototype in the Fermilab and CERN test beams. In the following we briefly report on the main results obtained in these stages.

<u>RPC tests</u> We choose to operate the RPCs in a saturated avalanche mode with an typical high voltage setting around 6.3 kV. The working gas has three components: Freon R134A (94.5%), isobutane (5.0%) and sulphur-hexafluoride (0.5%).

The size of the signal charges, the MIP detection efficiency and the pad multiplicity (as function of operating conditions) were measured with both cosmic rays and beam muons. As an example, Figure II-4.10 shows the pad multiplicity versus MIP detection efficiency [119]. Note the constant pad multiplicity at 1.1, independent of efficiency, for the one-glass design.

The RPCs in the Vertical Slice Test were also exposed to 120 GeV protons at various beam intensities to study their rate capability. The results show no loss of the MIP detection efficiency for rates below 100 Hz/cm$^2$. For higher rates, the efficiency drops exponentially in time (with a time constant depending on the beam intensity) until reaching a constant level.

The RPCs being tested have been operated continuously for over 18 months. Within the time period of these studies there was no evidence of long-term aging effects.

<u>DHCAL prototype and TCMT</u> The DHCAL prototype constitutes the first large scale hadron calorimeter with digital readout and embedded front-end electronics. It also utilised, for the first time, a pad-readout together with RPCs. The design of the DHCAL was based on preliminary work done with the Vertical Slice Test.

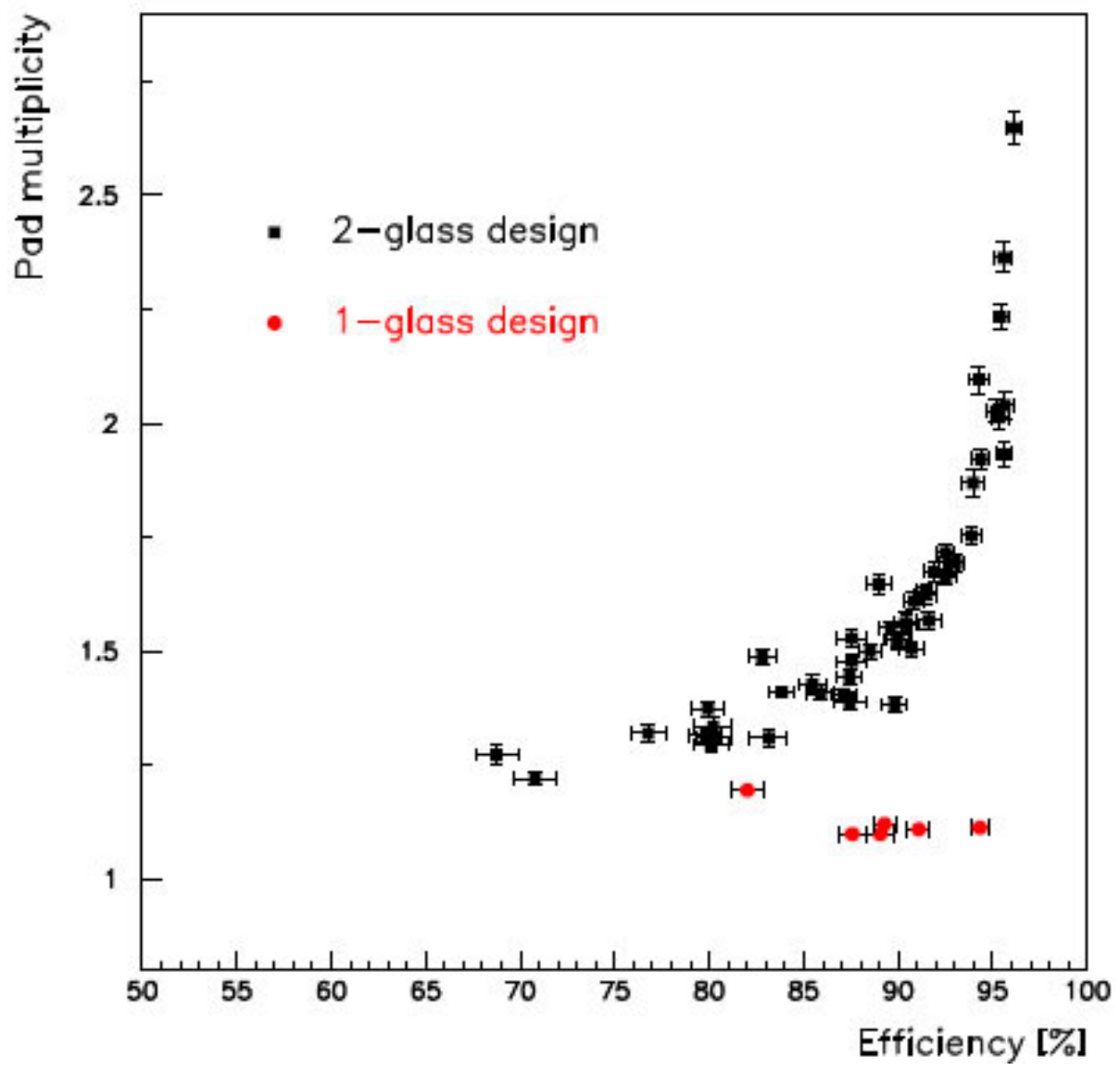

**Figure II-4.10**
Pad multiplicity versus MIP detection efficiency for 2-glass RPC and 1-glass RPC, measured with muon beam.

The DHCAL prototype consists of two parts: a 38-layer structure with 17.5 mm thick steel absorber plates and a 14-layer structure with eight 2.54 cm thick steel plates followed by six 10.0 cm thick steel plates. The former is commonly referred to as the DHCAL, or the Main Stack, and the latter is called the Tail Catcher and Muon Tracker (TCMT). These absorber structures were equipped with RPCs as active elements. Each layer measured approximately 1×1 $m^2$ and was inserted between neighbouring steel absorber plates. The Main Structure rested on a movable stage, which offered horizontal and vertical movements in addition to the possibility of rotating the entire stack. Figure II-4.11 shows photographs of the Main Structure and the TCMT.

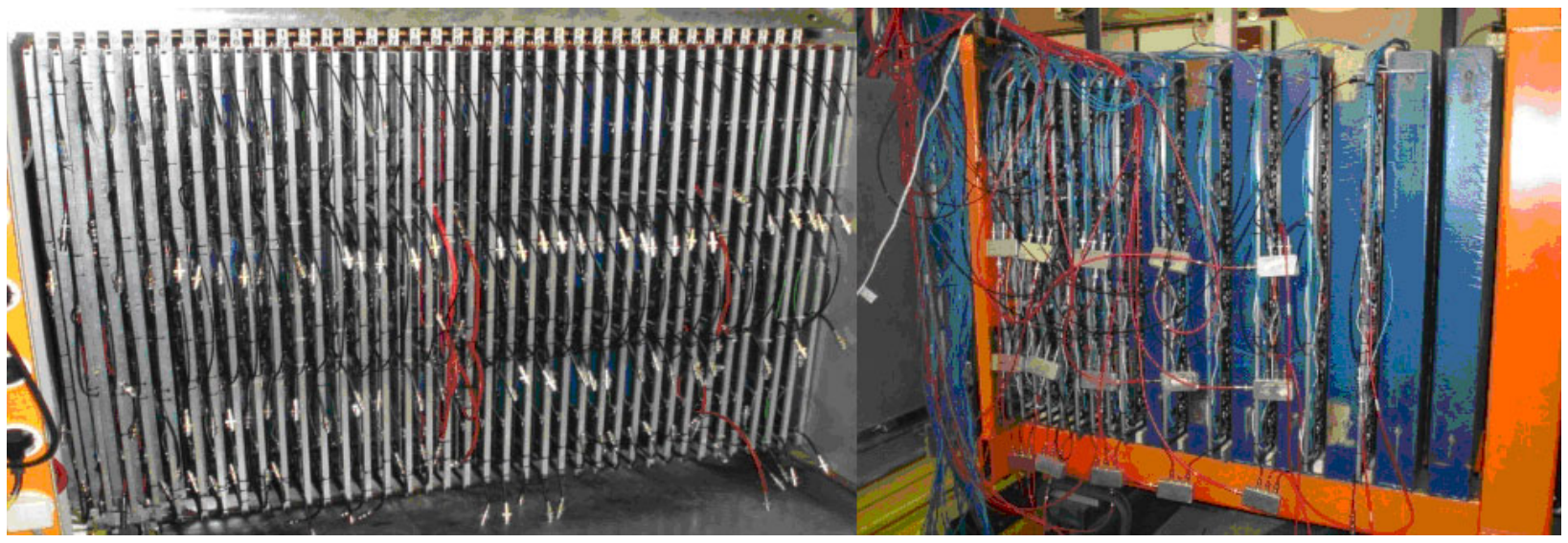

**Figure II-4.11**
Left: the DHCAL main stack (before cabling), right: the TCMT.

Each layer consists of three RPCs, each with an area of 32×96 $cm^2$ and stacked vertically on top of each other to create a 1×1 $m^2$ active area. Each RPC in turn is read out with two front-end boards, which covered the entire gas volume of the chambers. The three chambers and their boards are contained in a cassette structure providing the mechanical protection during transportation and installation.

The construction of the DHCAL prototype and TCMT started in fall 2008 and finished in early 2011.

DHCAL prototype test beam campaigns The DHCAL was tested extensively in the Fermilab test beam in various configurations: with the Scintillator Tail Catcher, with the Tail catcher equipped with RPCs, with or without the CALICE Silicon-Tungsten ECAL in front, and also with minimal absorber material between layers. In total, 9.4 million triggers were collected from muon beam and 14.4 million triggers are collected from secondary beam. Figure II-4.12 shows some events recorded by the DHCAL prototype and TCMT during the test beam campaigns.

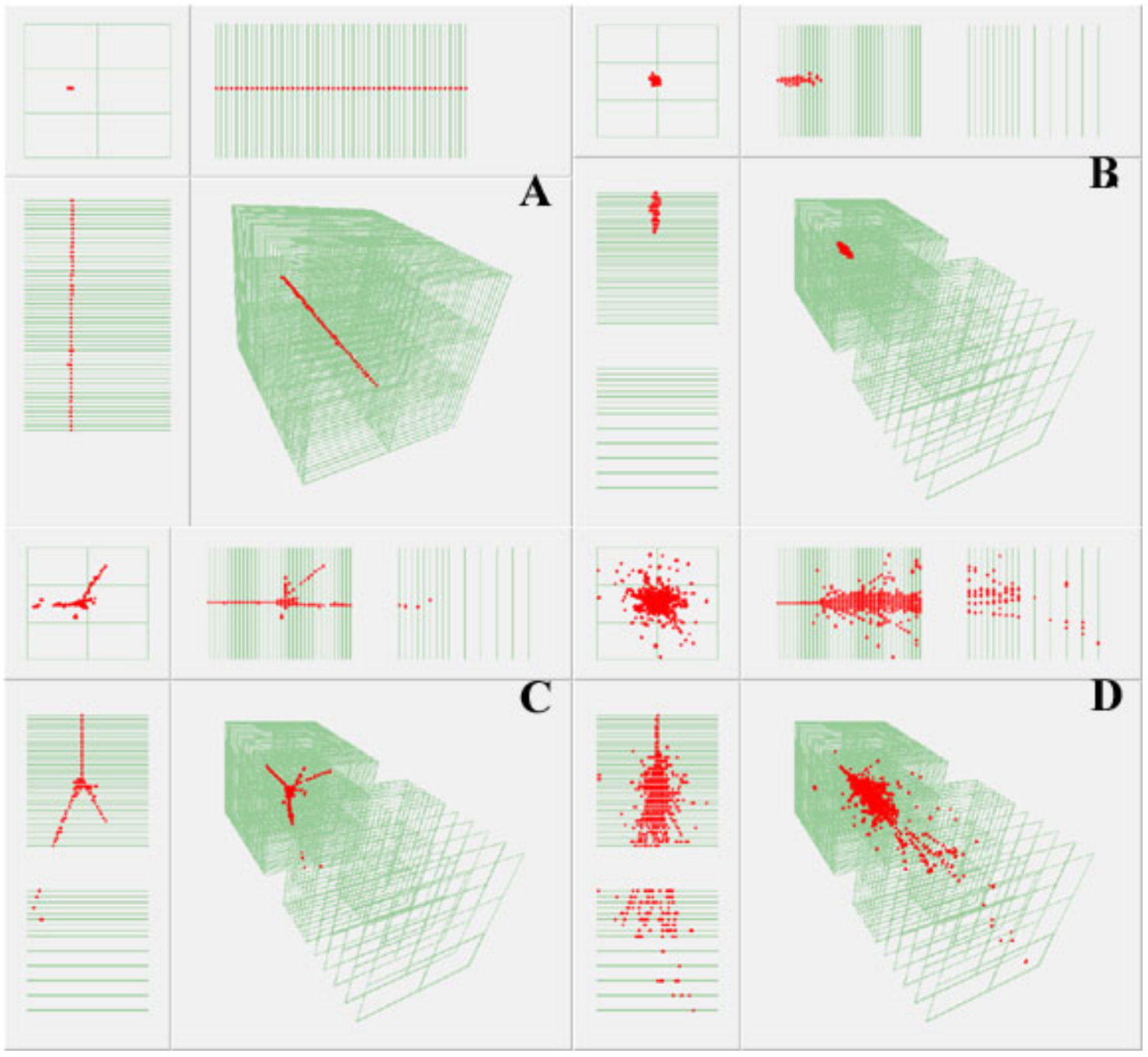

**Figure II-4.12** Events recorded by DHCAL prototype at the Fermilab test beam. A: a muon track; B: 8 GeV positron shower; C: 8 GeV pion shower; D: 120 GeV proton shower.

In Spring 2012 the DHCAL layers were transported to CERN and were inserted into a Tungsten absorber structure with 39 layers and a Steel tail catcher with 15 layers. Tests at both the Proton-Synchrotron and the Super-Proton-Synchrotron have been carried out. So far, 5 million muon events and 22 million secondary beam events have been collected.

### 4.3.5 DHCAL prototype performance

To first order the energy E of an incident particle is reconstructed as being proportional to the number N of pads hits. However, a non-vanishing noise rate and variations in the chamber efficiencies and average pad multiplicities need to be corrected for, such that the energy of an incident particle is reconstructed as

$$E = \alpha_{sample} \times (\sum_{i=0}^{n} N_i \cdot \frac{\varepsilon_0}{\varepsilon_i} \cdot \frac{\mu_0}{\mu_i} - N_{noise}) \qquad \text{(II-4.1)}$$

where the sum runs over all layers of the detector, $\alpha_{sample}$ is the sampling fraction which may depend on particle energy, $\varepsilon_0$ and $\mu_0$ are the average MIP detection efficiency and the average pad multiplicity of the detector, $\varepsilon_i$ and $\mu_i$ are the MIP detection efficiency and average pad multiplicity of layer i and $N_{noise}$ is the average contribution from noise. All these calibration parameters are carefully measured and monitored over time during the test beam campaigns for the DHCAL prototype. The DHCAL response for positrons and pions are measured at different beam momenta, and data analysis is still on-going. We show preliminary results for the DHCAL noise measurement, muon calibration, positron response and pion response.

#### 4.3.5.1 Noise measurement

The accidental noise rate was measured both with random triggers and with trigger-less acquisitions. Confirming our measurement with the Vertical Slice Test, the rate was found to be low, but to depend on the temperature of the stack and the ambient air pressure. For a given event, the accidental noise rate adds on average 0.01 to 0.1 hits in the entire DHCAL prototype, where one hit corresponds to about 60 MeV.

### 4.3.5.2 Muon calibration

At Fermilab, muons traversing the DHCAL were collected using the 32 GeV secondary beam, a 3 m long iron absorber and a trigger based on the coincidence of a pair of $1{\times}1$ m$^2$ Scintillator paddles located upstream and downstream of the detector.

Muon events were used to measure the local response of RPCs (efficiency and average pad multiplicity) in the DHCAL and TCMT. As an example, Figure II-4.13 (left) shows the MIP detection efficiency $\varepsilon_i$, the average pad multiplicity $\mu_i$ and the so-called calibration factors, $c_i{=}(\varepsilon_i\mu_i)/(\varepsilon_0\mu_0)$, as measured with two different techniques (tracks and track segments) versus layer number.

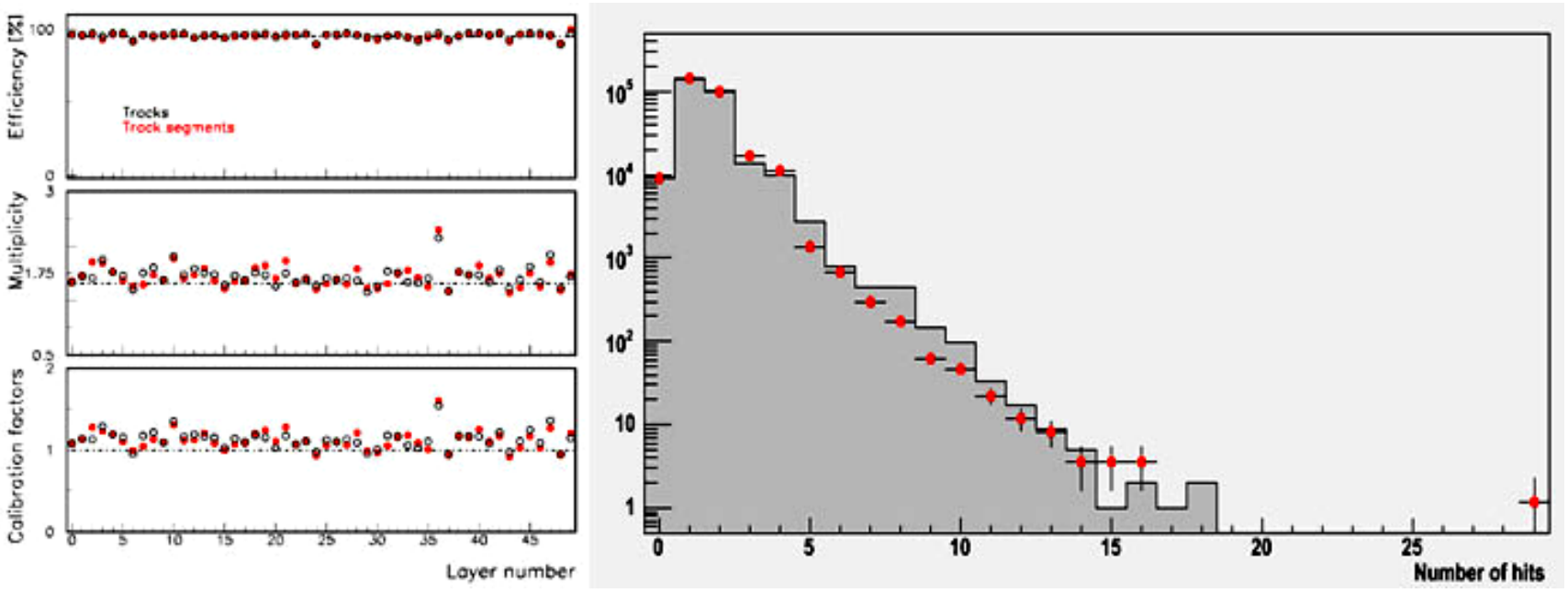

**Figure II-4.13**
Left: MIP detection efficiency, average pad multiplicity and calibration factors as function of layer number, as measured with both tracks and track segments; Right: response of a detector layer to muons averaged over the entire DHCAL with the histogram (data points) showing data (simulation).

The average response in clean regions of the stack, i.e. away from borders and fishing lines, was measured and is being used to tune the Monte Carlo simulation of the RPC response. Figure II-4.13 (right) shows a comparison of the measured and simulated RPC response.

### 4.3.5.3 Positron response

Secondary beam particles were collected at momentum points covering the range of 1 to 60 GeV at the Fermilab test beam. Data with the primary 120 GeV proton beam were also collected. The trigger (provided by the coincidence of two $19 \times 19$ cm$^2$ scintillator paddles positioned upstream of the DHCAL) accepted positrons and hadrons indiscriminately. The particles were later identified offline based on the information from the Cerenkov counters and shower shape.

The mean response of the DHCAL (before calibration) to identified positrons is shown in Figure II-4.14 (left). The response is fit with the nonlinear function N=a+bE$^m$. The fit describes the data well and is in accordance with the predictions in the VST results of positron showers [120]. In order to measure the electromagnetic energy resolution of the DHCAL the positron response need to be corrected for non-linearity. Figure II-4.14 (right) shows the electromagnetic energy resolution for both uncorrected and corrected values.

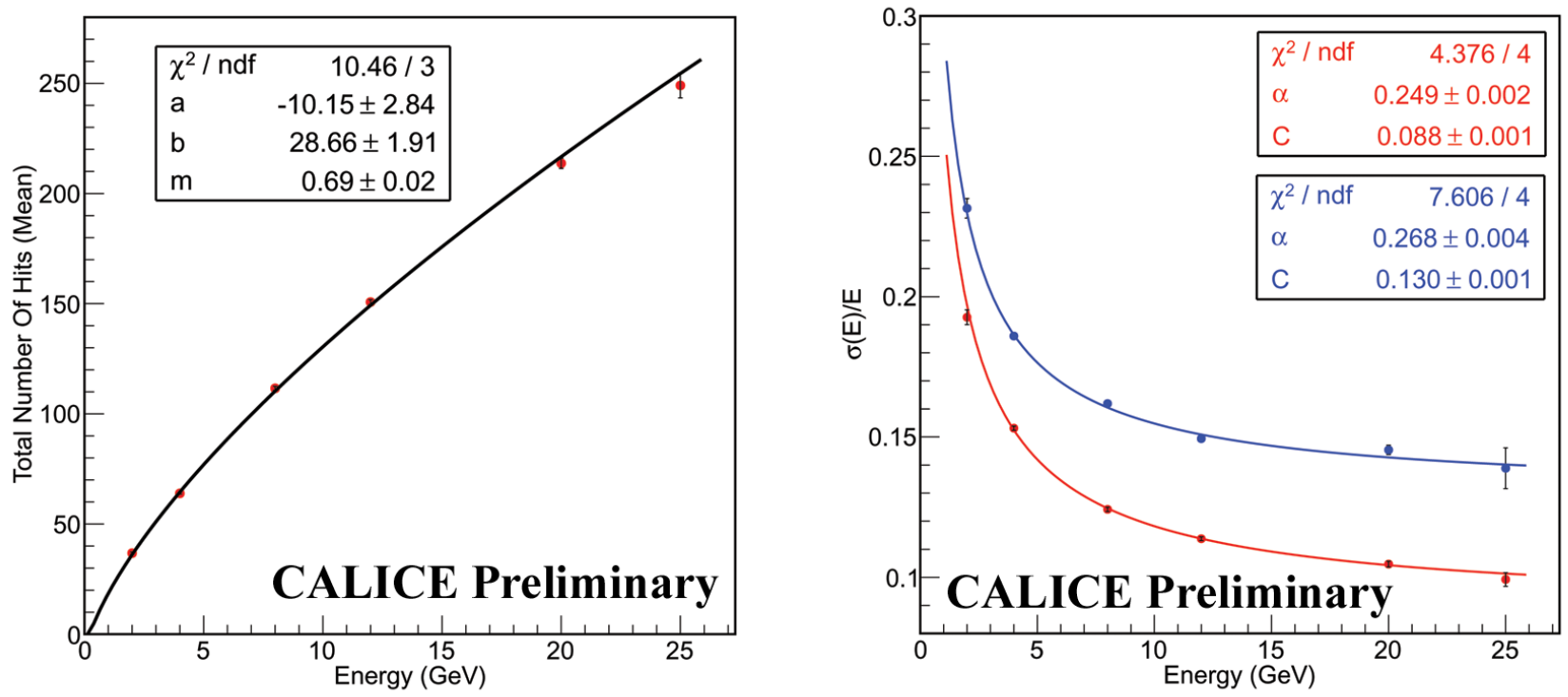

**Figure II-4.14**
Left: Mean response of DHCAL to positrons; Right: Non-linearity corrected (blue) and uncorrected (red) electromagnetic energy resolution for DHCAL.

#### 4.3.5.4 Pion response

The mean response of the DHCAL (before calibration) to pions is shown in Figure II-4.15 (left). The response is linear up to 25 GeV, and at 32 GeV, the response deviates from linear behaviour due to RPC response fluctuation and possible saturation effect. Therefore, 32 GeV data point is not included in the linear fit (N=aE where N is the total number of hits and E is the beam energy). Figure II-4.14 (right) shows the hadronic energy resolution of the DHCAL with the current particle identification algorithms. The fits represent the data well and for the longitudinally contained pions -that have no hits in the last two layers- a stochastic term of approximately 55% and a constant term of 7.5% is achieved. The measurements are within 1-2% of predictions based on the simulation of the large-size DHCAL prototype using the VST results [121].

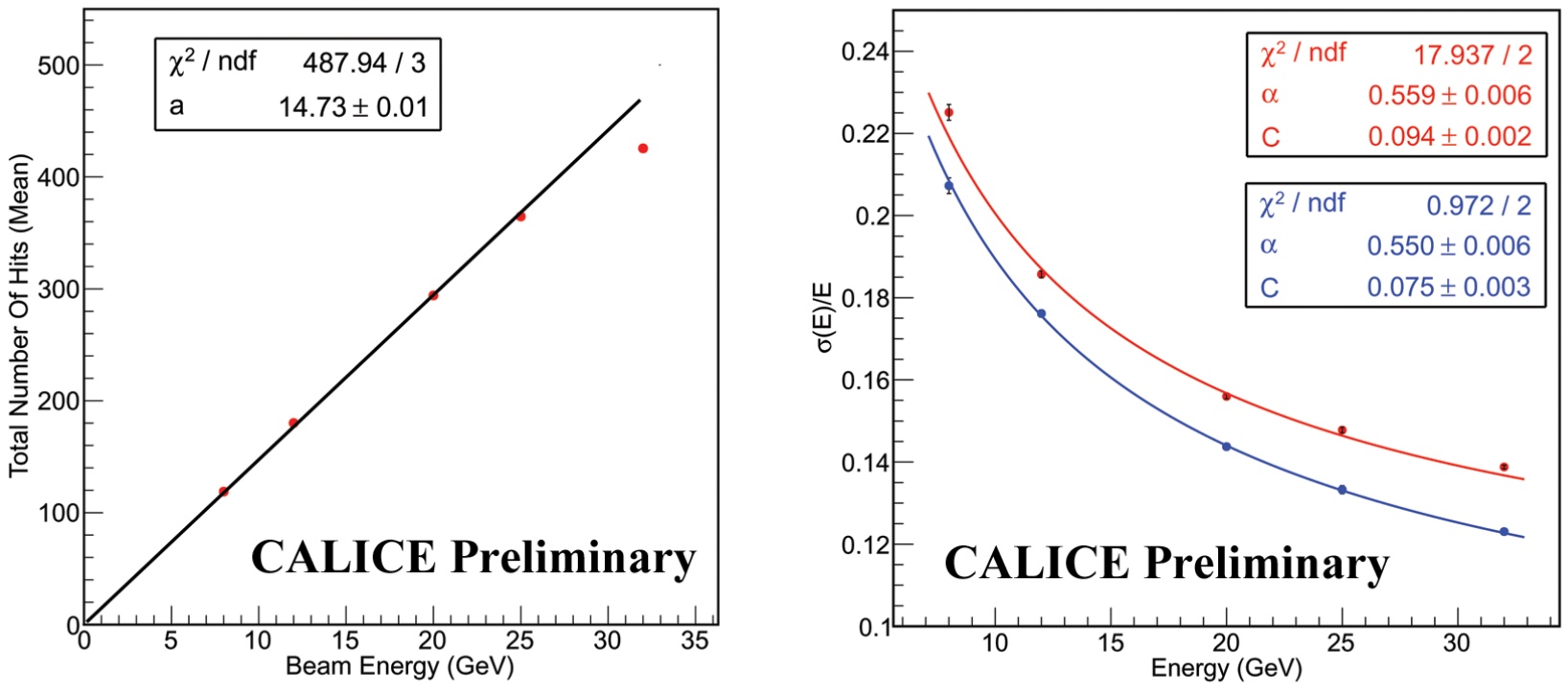

**Figure II-4.15**
Left: Mean response of DHCAL to pions; Right: hadronic energy resolution of DHCAL for all identified pions (red) and the longitudinally contained pions (blue).

With the data analysis still in a preliminary state, there are nevertheless a few conclusions to be drawn regarding an RPC-based digital hadron calorimeter:

- The RPC technology appears to satisfy the requirements of the active media of a highly segmented calorimeter.
- The dark rate in the DHCAL is very low and corresponds to a negligible amount of energy added to a single event.
- The response to positrons is as expected and consistent with predictions based on the VST [120]. As expected the response to positrons is non-linear, due to saturation effects introduced by the finite size of the readout pads.
- The response to pions is as expected and consistent with predictions based on the VST [121]. The response appears to be linear up to about 30 GeV.
- Using Tungsten absorbers instead of Steel plates leads to a reduced number of hits, by about 30%. To extend the range of the linear response beyond 20 GeV, a finer segmentation of the readout is required. It is conceivable that the application of software compensation techniques, which utilise the density of hits, is able to restore the linearity and improve the resolution at these higher particle energies.

### 4.3.6 R&D towards technical feasibility and optimisation

The DHCAL prototype was designed for proof of principle, R&D in several areas are still critical to demonstrate the technical feasibility and achieve design optimisation:

- The front-end of the DHCAL readout need to have significantly reduced power consumption, in order to avoid active cooling. Low power ASIC design techniques and power pulsing scheme are being considered to reduce the power dissipation by a factor of ~100.

- The digital part of the readout system need to be optimised for better data concentration and reduced number of data connections without sacrificing reliability. Several ideas, including token ring passing and wireless data link, are being considered.
- A novel 1-glass RPC design is being developed, which features distinct advantages, such as an average pad multiplicity close to unity, a thinner chamber, a higher rate capability and a generous insensitivity to the surface resistivity of the resistive paint. The feasibility of larger chambers based on this design needs to be established.
- The high rate RPC could be a nice solution for the forward region of the DHCAL. The group is currently collaborating with several other institutes in developing low resistivity glass and Bakelite material for high rate RPC.
- The group is pursuing the development of a realistic design of a DHCAL module. Several configurations are being considered.
- A high voltage distribution system is being developed which is capable of turning onoff, adjust voltage value, and monitoring the current of individual chambers from a single high voltage input source.
- In order to operate a large DHCAL system at a future colliding beam experiment, a gas recirculation system is needed for both cost and environmental considerations. Initial development has started.

## 4.3.7 Calibration of a Digital Hadron Calorimeter

The event record for the DHCAL will contain a list of hits and their location. The energy of a hadron $E_h$ will be reconstructed as described in formula II-4.1. In a running experiment, one needs to determine $\alpha_{sample}$, $\varepsilon_0$, $\mu_0$, and constantly monitor $\varepsilon_i$, $\mu_i$ and $N_{noise}$.

### 4.3.7.1 Sampling fraction and energy scale

The sampling fraction $\alpha_{sample}$ for charged hadrons can be measured by placing detector modules into a pion test beam of varying energies, which can also determine the energy dependence. The test beam data can also be used to validate a simulation procedure to reproduce the response of the modules to charged pions. The response of the modules to neutral hadrons will be simulated and the sampling term for neutral hadrons will be determined.

The overall energy scale of the jets reconstructed at the ILC will be cross checked using di-jet events and reconstructed $\mathrm{W}$ and $\mathrm{Z}$ boson masses. At $\sqrt{s} =$ 500 GeV with an integrated luminosity of $1\ \mathrm{fb}^{-1}$/day we expect to collect 2,800 (1,900) di-jet ($\mathrm{W}^+\mathrm{W}^-$) events/day. With enough statistics the dependence of the reconstructed jet energy on the electro-magnetic fraction of a jet or the fraction of neutral hadrons in a jet can be studied.

### 4.3.7.2 Monitoring of individual RPC's

Under fixed operating conditions (high voltage and threshold setting) the performance of RPCs depends on the ambient temperature, the atmospheric pressure and, for completeness, the gas flow. The last item only impacts the noise rate and the pad multiplicity. However, above a minimum gas flow these are seen to be constant and do not depend on variations of the flow. The performance of the RPCs does not depend on the ambient air humidity.

The dependence on the environmental conditions can be parameterised [117] as

$$\Delta\varepsilon = [-0.06 \cdot \Delta p(100Pa) + 0.3 \cdot \Delta T(^0C)]\%$$

$$\Delta\mu = [-0.25 \cdot \Delta p(100Pa) + 2.0 \cdot \Delta T(^0C)]\%$$

In the following we assume that the changes in performance are uniform within an entire chamber. Two methods will be employed to monitor the chamber's performance: one utilising track segments in events from ILC collisions and the other utilising cosmic rays.

- Track segment monitoring

  Imaging calorimeters offer the possibility to reconstruct individual track segments within hadronic showers [119] or in $e^+e^- \to \mu^+\mu^-$ events. Such track segments can be used to monitor the MIP detection efficiency $\varepsilon_i$ and the pad multiplicity $\mu_i$ of individual RPC's during the data taking period.

  It is estimated that a 3% measurement is achievable, either using track segment method or muon tracks, within approximately 5 days of running.

- Cosmic ray monitoring

  Cosmic rays are an ideal tool to monitor the performance of the chambers. With a crude estimate of the underground muon flux, horizontal chambers with an area of 2 $m^2$ obtain 1000 measurements per minute. The rate in vertical chambers will be reduced by say one order of magnitude. Nevertheless, the required precision of 3% can be obtained in less than one hour. However, if the front-end power is pulsed, this will lead to a reduction in duty cycle of up to a factor of 200. In this case, time estimate needs to be increased to approximately 1 week. Further studies are needed to understand the cooling needs of the DHCAL and to define the optimal duty factor, taking into account the need for monitoring the performance of the RPCs.

In long-term studies of prototype RPCs, the efficiency and pad multiplicity were seen to vary by $\pm 0.9\%$ and $\pm 5\%$, respectively. Applying corrections for the environmental conditions (i.e. ambient temperature and air pressure) based on the above mentioned equations reduces these variations to $\pm 0.8\%$ and $\pm 3\%$.

Based on detailed simulations of the response of RPCs the effect of uncertainties in the calibration on the measurement of single particle energies was estimated. The studies showed that, for instance, for 10 GeV $\pi^+$ the energy resolution degrades by approximately by $1\%$, if the entire module's response is smeared by a Gaussian distribution with a sigma of $3\%$. This is the worst case scenario, where the responses of all layers in a given module are $100\%$ correlated. If, on the other hand, all individual layers in a module fluctuate independently say by a Gaussian distribution with a sigma of $3\%$, the effect on the energy resolution is negligible.

#### 4.3.7.3 Measurement of the noise rate

The background rate can be measured utilising the self-triggered mode of the front-end readout. Measurements on the prototype chambers typically showed a background rate of 0.1 - 0.2 Hz/$cm^2$ at room temperature. As an example Figure II-4.16 shows the noise rate as function of high voltage setting for the same threshold as in the test beam.

Assuming a gate width of 200 ns and a total of 5 $\times 10^7$ readout channels, the expected noise rate at the ILC will be about 2 hits/event in the entire DHCAL. Assuming a calibration of 13.6 hits/GeV, as obtained in recent simulations of the DHCAL, the noise contribution corresponds in average to around 150 MeV/event and can be ignored for all practical purposes.

Beam related background rates, due to neutrons for instance, will be measured using bunch-crossing events and algorithms for separating energy deposits from $e^+e^-$ collisions and from beam backgrounds.

**Figure II-4.16** Noise rate as function of high voltage for six different chambers. The threshold was set at the default value of 110 counts. The default high voltage setting was 6.2 kV.

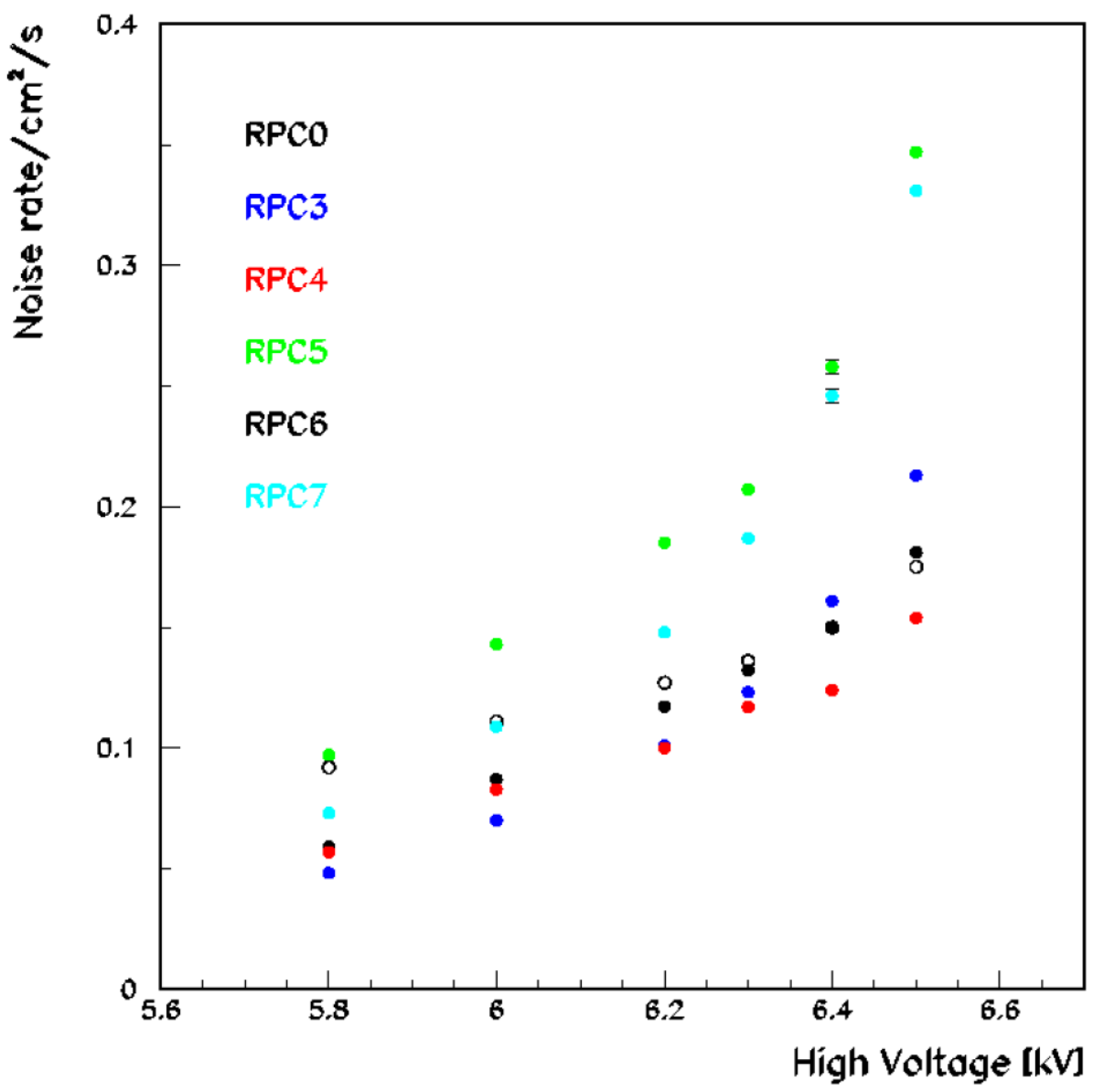

## 4.3.8 Alternative technologies

A number of viable alternative approaches to PFA-based hadronic calorimetry are also considered by SiD. Two, GEM and Micromegas, use advanced micro-pattern gas detector technologies benefitting from participation in the RD51 collaboration. A third approach uses small scintillator tiles readout by silicon photomultipliers. As for the baseline RPC approach, all these alternatives are the subject of development within the CALICE collaboration.

### 4.3.8.1 GEM

We have also been developing a digital hadronic calorimeter (DHCAL) using GEM as the sensitive gap detector technology. GEM detectors can provide flexible configurations which allow small anode pads for high granularity. It is robust and fast with only a few nano-second rise time, and has a short recovery time which allows a higher rate capability than other detectors. It operates at a relatively low voltage across the amplification layer, and can provide high gain using a simple gas ($ArCO_2$), which protects the detector from long term degradation issues, and is stable. The ionisation signal from charged tracks passing through the drift section of the active layer is amplified using a double GEM layer structure. The amplified charge is collected at the anode layer with pads at zero volts. The GEM design allows a high degree of flexibility with, for instance, possibilities for micro-strips for precision tracking layer(s) and variable pad sizes and shapes. Figure II-4.17 depicts how the double GEM approach can be incorporated into a DHCAL scheme.

**Figure II-4.17** Left: GEM DHCAL Concept. Centre: Drawing of the large-area GEM chamber. Right: Large-area GEM chamber under construction.

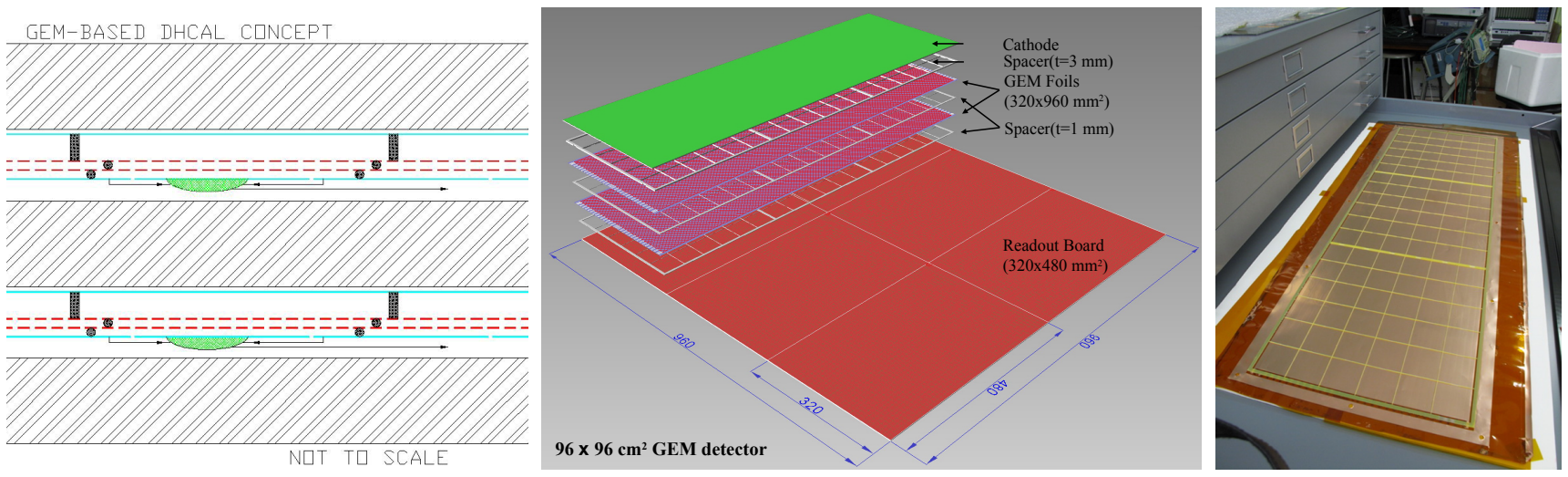

A number of double GEM chambers have been built and tested with cosmic rays, sources, and test beam. Stable operation has been achieved with 390 V across each GEM foil, leading to a gain of 11,000. The resulting typical MIP signal size is 40-50 fC, well above the noise level from the readout KPiX chip from SLAC.

The next step in developing the GEM approach to digital hadron calorimetry is the construction of a number of 1 m $\times$ 1 m layers for exposure as part of a 1 $m^3$ test beam stack. 1 m x 33 cm foils developed with CERN are being used to assemble double-GEM prototypes of the same size. Following this, a number of 1 m $\times$ 1 m layers will be assembled. Figure II-4.17 shows a schematic view of one of the 1 m $\times$ 33 cm layers under construction.

Single thick-GEMs [122] are also considered as an alternative to the double-GEM structure discussed above. A thick-GEM consists of a single circuit board about 0.5 mm thick and having holes of 200-400 $\mu m$ in diameter. An advantage of thick-GEMs is a possible reduction in overall DHCAL active layer thickness and easier handling and construction compared with regular thin foils.

#### 4.3.8.2 Micromegas

Introduction Digital calorimeters proposed for ILC or CLIC are expected to suffer from saturation due to the high particle multiplicity in the core of the showers. The resulting loss of linearity and resolution can in principle be mitigated if more than one threshold is used. A necessary condition for this approach to work is the proportionality between cell signals and the number of traversing particles. On average, this condition is met in Micro Pattern Gas Detectors like GEM and Micromegas because they are free of space charge effects.

Micromegas is a fast, position sensitive Micro Pattern Gas Detector operating in the proportional mode [123]. It functions in simple gas mixtures (*e.g.* $Ar/CO_2$) and at low fields and voltages ($<$ 500 V) and is thus extremely radiation hard. It is an alternative to RPCs that offers lower hit multiplicity and proportional signals well suited for a semi-digital readout. On the other hand, Micromegas suffers from discharges but efficient protections exist.

Micromegas chambers developed for the active part of a semi-digital HCAL (SDHCAL) consist of a commercially available 20 $\mu m$ thick woven mesh which separates the gas volume in a 3 mm drift gap and a 128 $\mu m$ amplification gap (so-called Bulk). Micromegas of 32 $\times$48 $cm^2$ acting as signal generating and processing units have been designed and fabricated. They were used to construct three chambers of 1 $m^2$ size which are described below.

Mechanical layout and assembly The 1 $m^2$ chamber features 9216 readout channels (1 $\times$1 $cm^2$ anode pads) and consists of six Printed Circuit Boards (PCB) of 32 $\times$48 $cm^2$ placed in the same gas chamber. Front-end chips and spark protection circuits are first soldered on the PCBs. Then a mesh is laminated on the opposite pad side of each PCB to obtain an Active Sensor Unit (ASU). Having 6 meshes instead of a single larger one decreases proportionally the energy that is released in the front-end electronics circuitry during a spark (Figure II-4.18).

Small spacers (1 mm wide, 3 mm high) are inserted between ASUs and support the cathode cover, defining precisely the drift gap. Plastic frames are closing the chamber sides, leaving openings for two gas pipes and flexible cables. The chamber is eventually equipped with readout boards and a patch panel for voltage distribution. The total thickness amounts to 9 mm which includes 2 mm for the steel cathode cover (part of the absorber), 3 mm of drift gap and 4 mm for PCB and ASICs. With this mechanical design, less than 2% of inactive area is achieved.

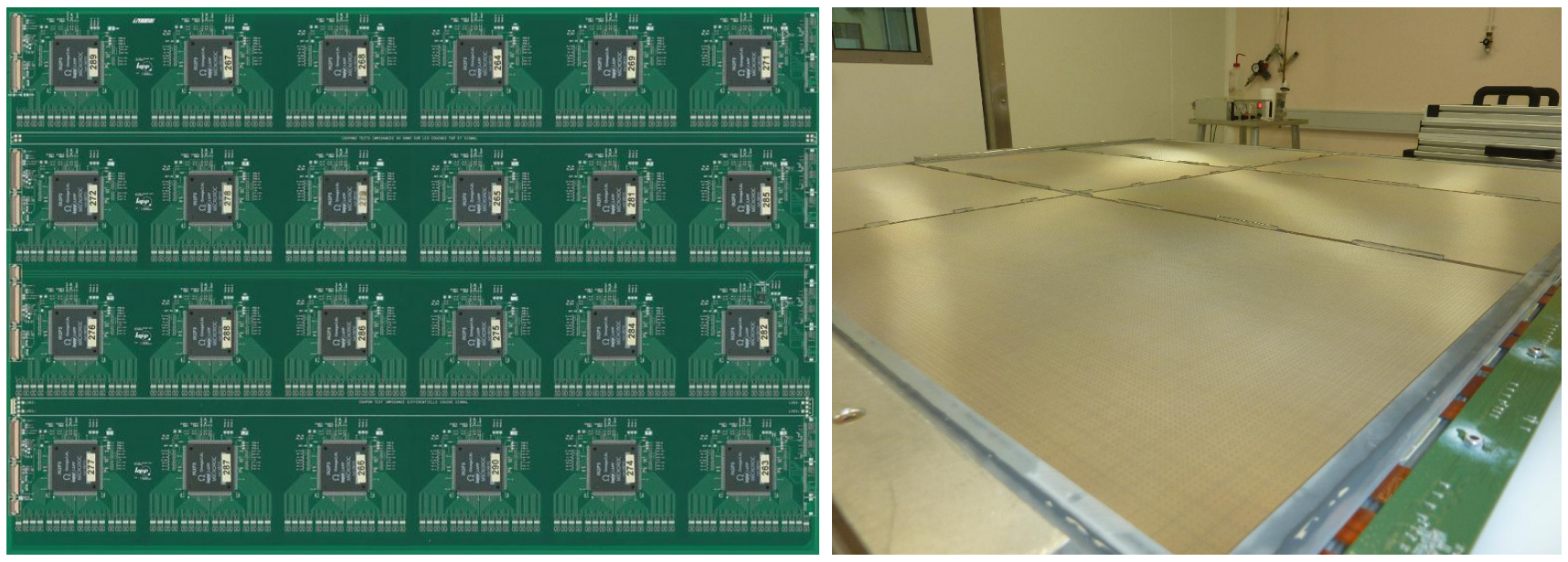

**Figure II-4.18** *One Active Sensor Unit (ASIC side) and the 1 $m^2$ prototype during assembly.*

Performance to MIPs  The response to minimum ionising particles (MIPs) was studied in a 150 GeV muon beam at CERN/SPS. The 1 $m^2$ chamber was flushed with a non-flammable mixture of $Ar/CF_4/iC_4H_{10}$ 95/3/2, the mesh voltages were varied between 300-420 V (gas gain $G$ of 100-8000). A profile of the beam recorded in internal trigger mode is shown in Figure II-4.18 (left) and indicates that the noise level can be kept low and uniform.

The strong dependence of the detection efficiency on the applied voltage is shown in Figure II-4.18 (centre). Thanks to the very low readout threshold (1-2 fC), a gas gain as low as $10^3$ (at 365 V) is sufficient to detect MIPs with an efficiency larger than 95%. Upon full exposure of two chambers, detailed efficiency maps over 8 $\times$8 $cm^2$ regions were produced revealing an efficiency of (96 $\pm$ 2)% (Figure II-4.18 (right)). Such a little variation indicates a good control of the chamber dimensions (gas gaps) as well as of the electronics parameters (gains, thresholds).

A benefit of Micromegas w.r.t. other gas detector technologies is the limited spatial extension of the avalanche signals. As a result of the little transverse diffusion experienced by the electrons in the gas (100-150 $\mu$m RMS), the hit multiplicity is below 1.15 up to 390 V ($G$ = 3000). At higher gains, neighbouring pads become sensitive to single electrons, increasing the multiplicity to 1.35 at 420 V ($G$ = 8000). There is however no reason to work in that regime as high MIP efficiency is reached at lower gains ($G$ = 1000).

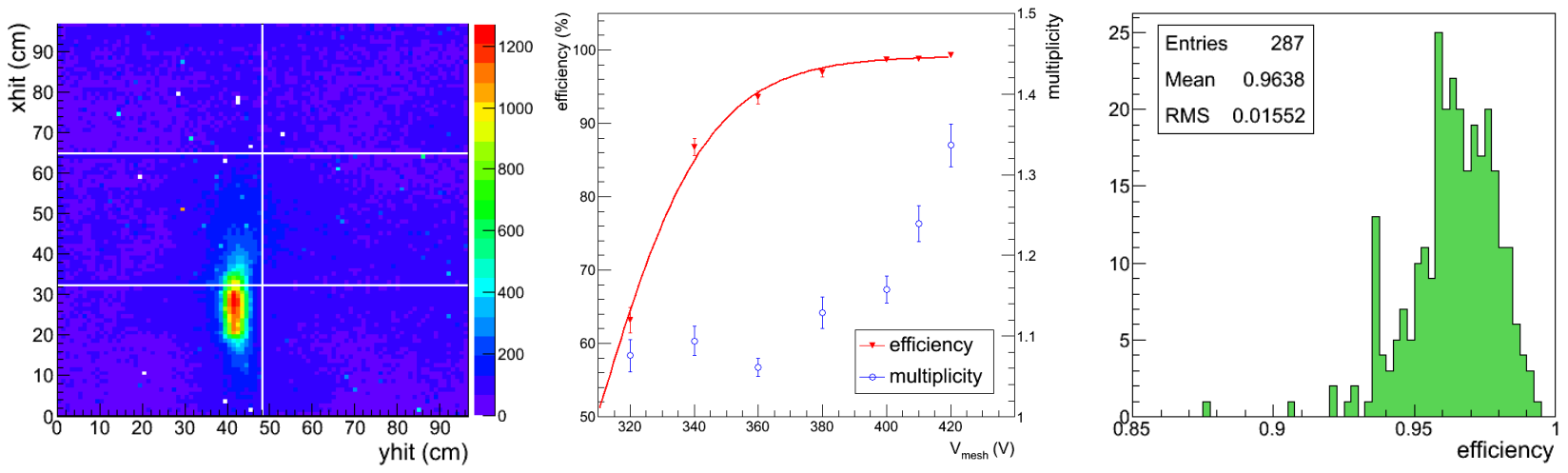

**Figure II-4.19**
*Muon beam profile using internal trigger mode (left) efficiency and pad multiplicity.*

Performance to pions  The response of the chamber to hadronic showers was studied with pions using first a 20 cm long block of iron (1 $\lambda_I$) upstream of the chamber. Later, two chambers were inserted inside the Fe/GRPC SDHCAL in the last two layers, behind 5 $\lambda_I$.

Directing a 150 GeV pion beam at the iron block, the distribution of the number of hits in the chamber was measured at mesh voltages of 325, 350 and 375 V ($G$ of about 350, 800 and 1700). The number of hit distributions, shown in Figure II-4.20 (left), exhibit a peak at $N_{hit}$ = 1 and a long tail from penetrating and showering pions respectively. The distributions at 350 and 375 V yield different efficiency to penetrating pions but remarkably, have a similar tail. Accordingly, a gas gain as low as 800 is sufficient to image most of the shower. Such a low working gas gain greatly improves the stability of the detector.

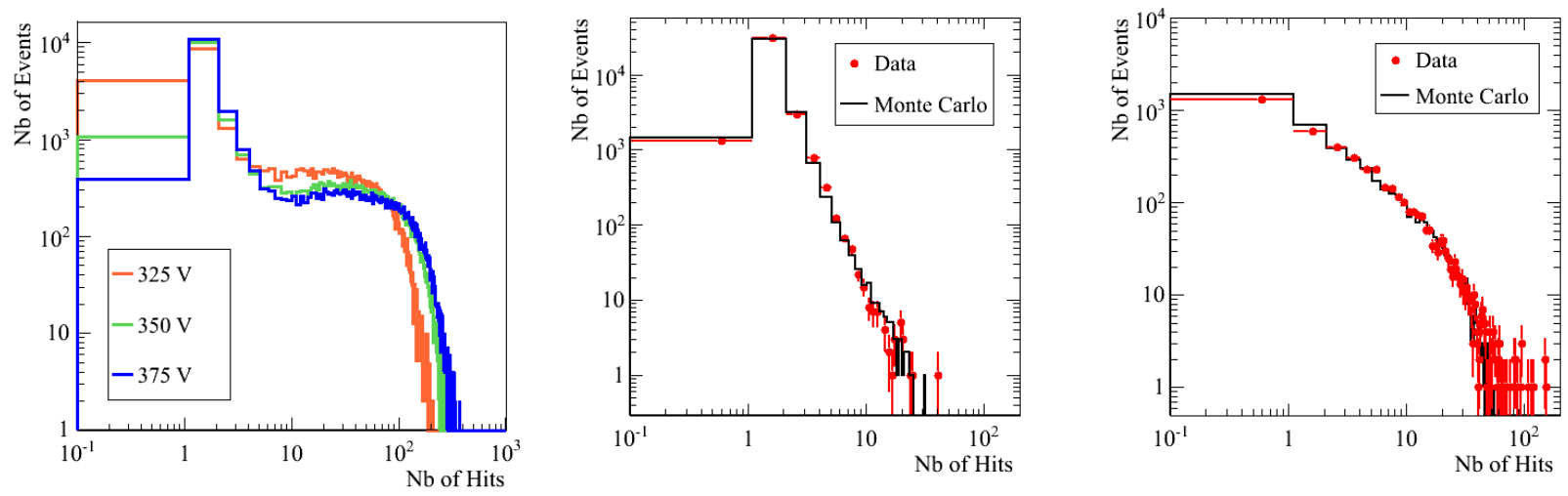

**Figure II-4.20**
Hit distribution from 150 GeV pions traversing a 20 cm thick iron block at various mesh voltages (left). Hit distributions from 100 GeV muons (centre) and pions (right) at layer 48 of the CALICE Fe/GRPC SDHCAL.

A good understanding of the detector is being achieved by comparing test beam data to Monte Carlo predictions. Preliminary results are presented in Figure II-4.20 (right) which shows the distribution after 5 $\lambda_I$ of Fe for 100 GeV pions. The readout threshold was tuned so as to reproduce the efficiency to muons (Figure II-4.20 (centre)). A satisfactory agreement is obtained for muons and pions meaning that the simulation is reliable. It should be stressed that no noise was introduced in the simulation, therefore, data are essentially free of noise.

#### 4.3.8.3 Scintillators

The CALICE Collaboration has been pursuing the design and prototyping of a fine granularity scintillator-based hadron calorimeter. This option capitalises on the marriage of proven detection techniques with novel photodetector devices. The main challenge for a scintillator-based calorimeter is the architecture and cost of converting light, from a large number of channels, to electrical signal. Studies demonstrate that small tiles (4-9 cm$^2$) interfaced to Silicon Photomultipliers (SiPMs)/Multi Pixel Photon Counter (MPPC) photodetectors [124], [125] offer an elegant solution. SiPM/MPPCs are multi-pixel photo-diodes operating in the limited Geiger mode. They have distinct advantage over conventional photomultipliers due to their small size, low operating voltages and insensitivity to magnetic fields. The *in situ* use of these photodetectors opens the doors to integration of the full readout chain to an extent that makes a high channel count scintillator calorimeter entirely plausible. Also, in large quantities the devices are expected to cost a dollar per channel making the construction of a full-scale detector instrumented with these photo-diodes financially feasible.

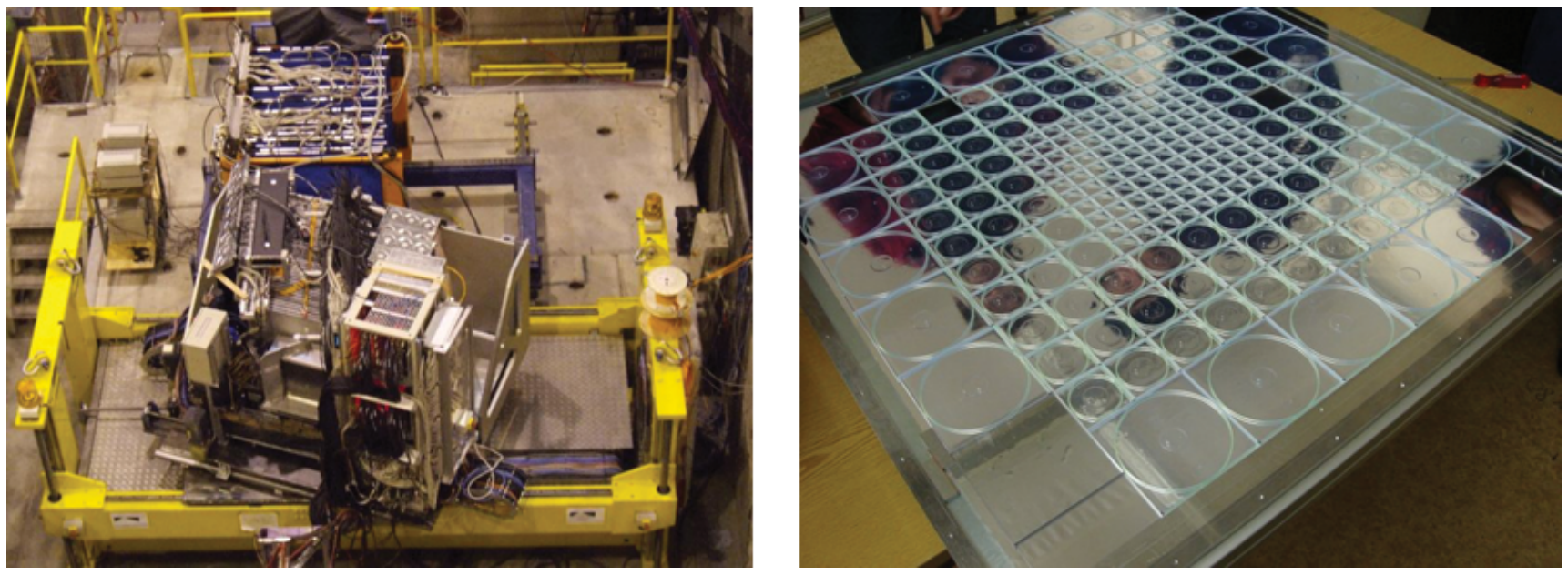

**Figure II-4.21**
CALICE test beam setup at CERN (left) and an active layer of the scintillator-SiPM prototype (right)

A $\sim$ 1 m$^3$ size scintillator-SiPM prototype [126] has been designed, constructed and exposed to a test beam during the 2006-2009 period at CERN and Fermilab (see Figure II-4.21). The active layers have subsequently been embedded in a tungsten stack which has collected data in the 2010-2011 period. Over numerous run periods the technology has proven to be versatile and robust, millions of electron, pion and proton events in the 2-180 GeV range were written to disk. Ongoing analysis of the data collected, has gone a long way in establishing the scintillator-SiPM option as a calorimeter technology (see Figure II-4.22), benchmarking hadron shower simulations [127] and testing the particle-flow paradigm using hadrons from real data [128].

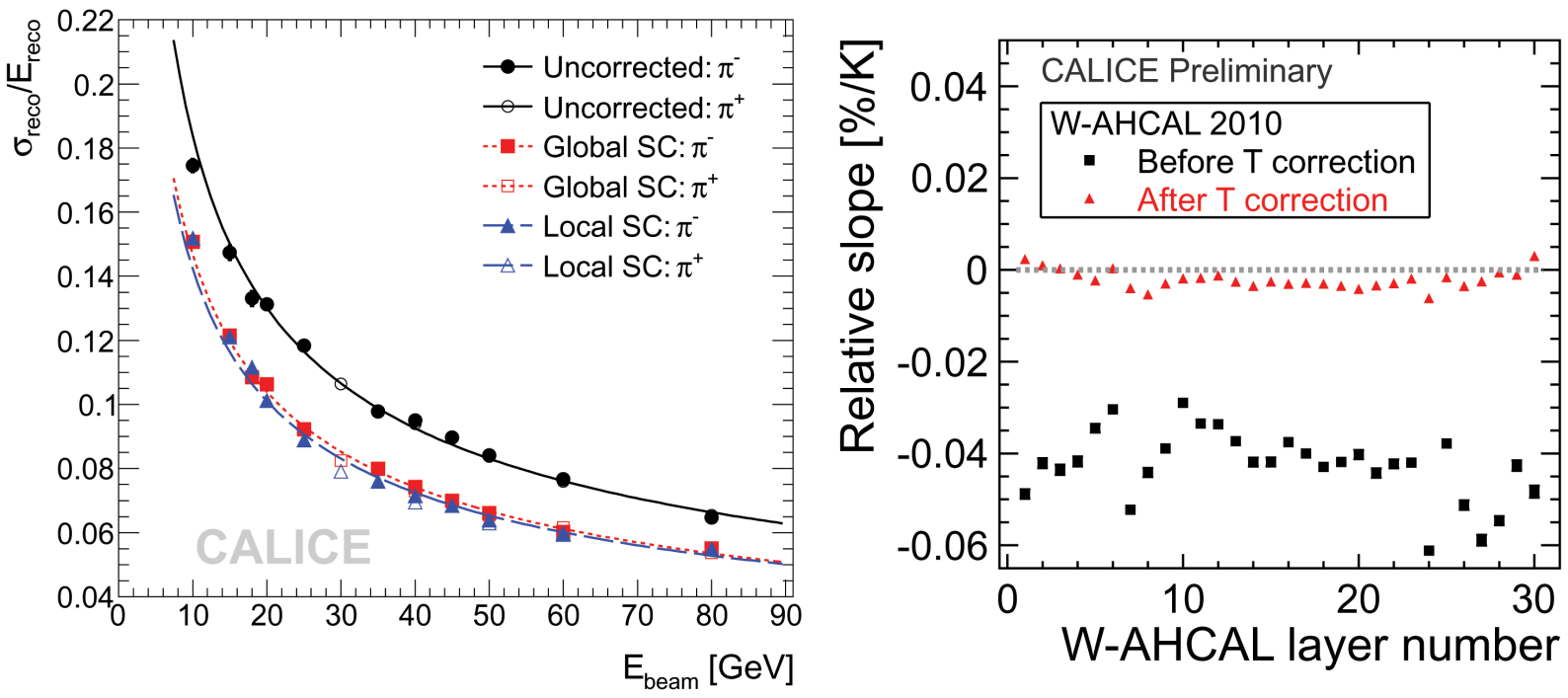

**Figure II-4.22**
Single pion resolution using simple energy sum and software compensation techniques (left) and slope of the SiPM response temperature dependence for AHCAL layers without and with temperature correction (right)

The focus of the current and future R&D effort is to demonstrate the scalability of this technology taking into account the stringent constraints on the power consumption and mechanical compactness through the development of an Integrated Readout Layer (IRL). In general for the IRL, it is proposed to have a printed circuit board (HCAL Base Unit or HBU) inside the detector which will support the scintillator tiles, connect to the silicon photodetectors and carry the necessary front-end electronics and signal/bias traces (see Figure II-4.23). This can however be achieved in a number of ways and a number of promising complementary approaches (e.g.fibre vs. direct or fibreless coupling of SiPMs to the tiles) have been developed in a coordinated fashion such that they can be characterised in a common electronics environment. This next generation front-end electronics carried aboard the HBUs is capable of self-triggering, precise time stamping, channel-by-channel bias control and a built-in LED calibration system. Commissioning of these readout slabs is at an advanced stage and is expected to expand into exposure in electron and hadron test beams in the near future.

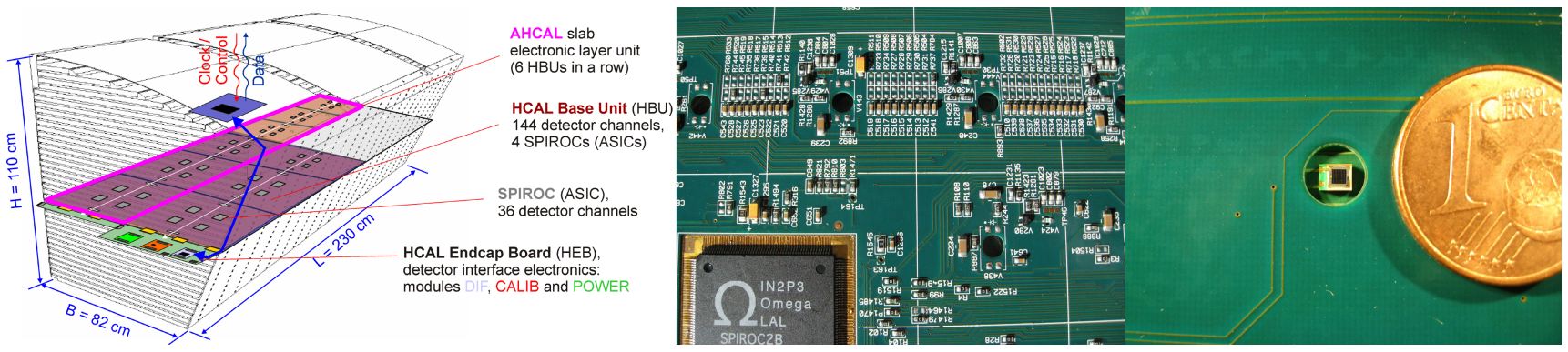

**Figure II-4.23** Conceptual design of a barrel wedge instrumented with IRL planes (left), a HBU prototype (centre) with a MPPC surface-mounted on it (right).

## 4.4 Summary

We have a silicon-tungsten electromagnetic calorimeter design that can satisfy the PFA requirements. We have developed a first-level mechanical design and have all the components and processes in place to construct and test a full-depth prototype. The technologies for PFA-based hadron calorimetry have seen significant development and testing since the submission of the SiD LOI. The construction of the cubic metre HCAL stack in the baseline RPC technology has provided much practical experience and confidence towards producing a full detector in this technology. The test beam data, providing unprecedented detail on hadronic showers, have shown that this is indeed a very promising technology, in the PFA context, for the SiD detector. We have also benefitted from the development of several alternative technologies.

# Chapter 5
# SiD Muon System

The SiD muon system is designed to identify muons from the interaction point with high efficiency and to reject almost all hadrons (primarily pions and kaons). The muon detectors will be installed in the gaps between steel layers of the solenoid flux return. The required position and rate capabilities of the detectors are modest and can be met by several different detector technologies. The baseline design uses double layers of extruded scintillator strips read out by silicon photomultipliers (SiPMs). Resistive plate chambers (RPCs) are also under consideration as an alternative design.

The SiD muon selection will combine information from the central tracker, calorimeter, and muon detectors to construct 3-dimensional tracks through the entire detector for each muon candidate. Candidates will be required to penetrate a number of interaction lengths consistent with the muon momentum. In addition, the observed number and position of hits along the fitted track length can be used to further discriminate against hadrons. The first layers of the muon system may also be useful as a tail-catcher for the hadronic calorimeter.

Muon systems characteristically cover large areas and are difficult to access or replace. Reliability and low cost are major requirements. Over 2.4 m of steel thickness will be required for the solenoid flux return, providing $> 13$ nuclear interaction lengths to filter hadrons emerging from the hadron calorimeter and solenoid. Since the central tracker will measure the muon candidate momentum with high precision, the muon system only needs sufficient position resolution to unambiguously match calorimeter tracks with muon tracks. Present studies indicate that a resolution of $\approx$ 2 cm is adequate. This can be achieved by two orthogonal layers of 4 cm wide extruded scintillators or RPC pickup strips.

**Figure II-5.1** Misidentification of pions as a function of the depth of the last hit muon layer.

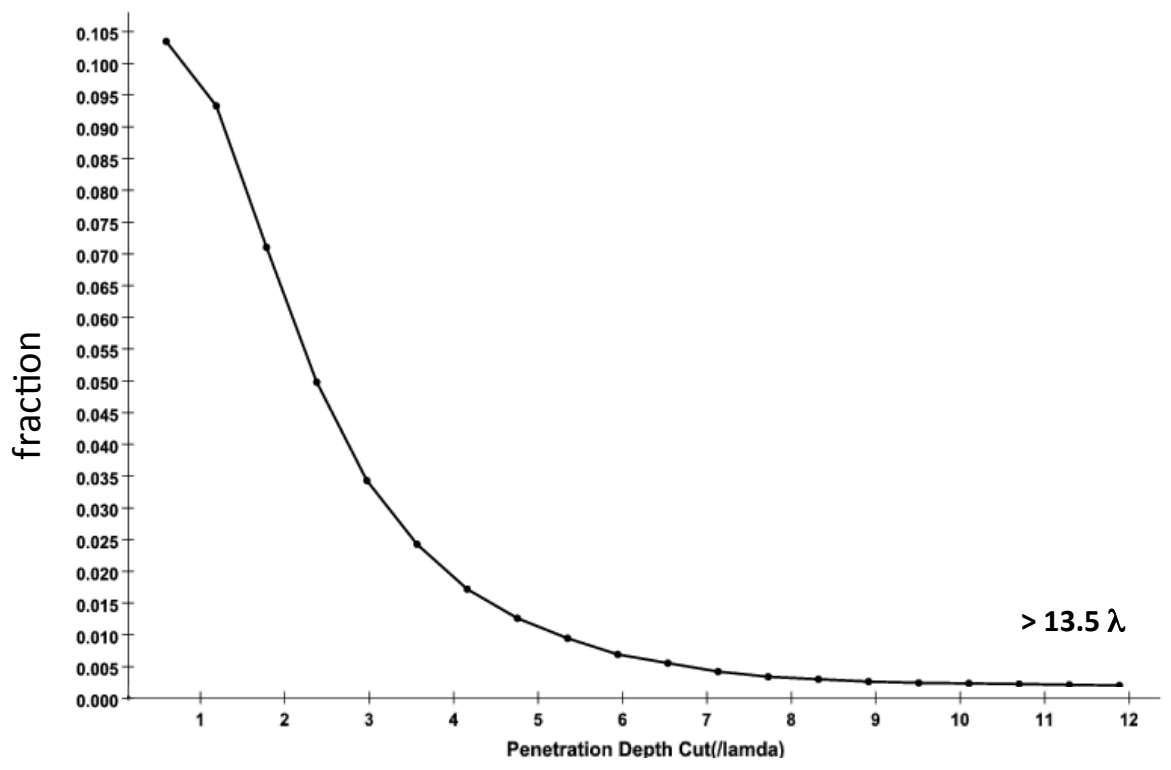

Full optimisation of the muon system design has not been completed. The total steel thickness is set by the solenoid flux return requirements. To check that the present design is thick enough we studied the misidentification rate of pions between 10 GeV and 50 GeV as a function of the depth in the flux return. As shown in Figure II-5.1, requiring that the track makes hits in some of the outer layers is sufficient to reduce the pion misidentification rate to 0.25%, consistent with the expected level of pion to muon decays. The present design, shown in Figure II-5.2 has ten layers in the barrel section and nine layers in each endcap. This provides a comfortable level of redundancy ($\geq 6$ layers) even in the region between the barrel and endcap. The optimum number of detector layers to cover the muon identification and tail catching functions was also studied for the CLIC case [129], where nine layers were found to be sufficient.

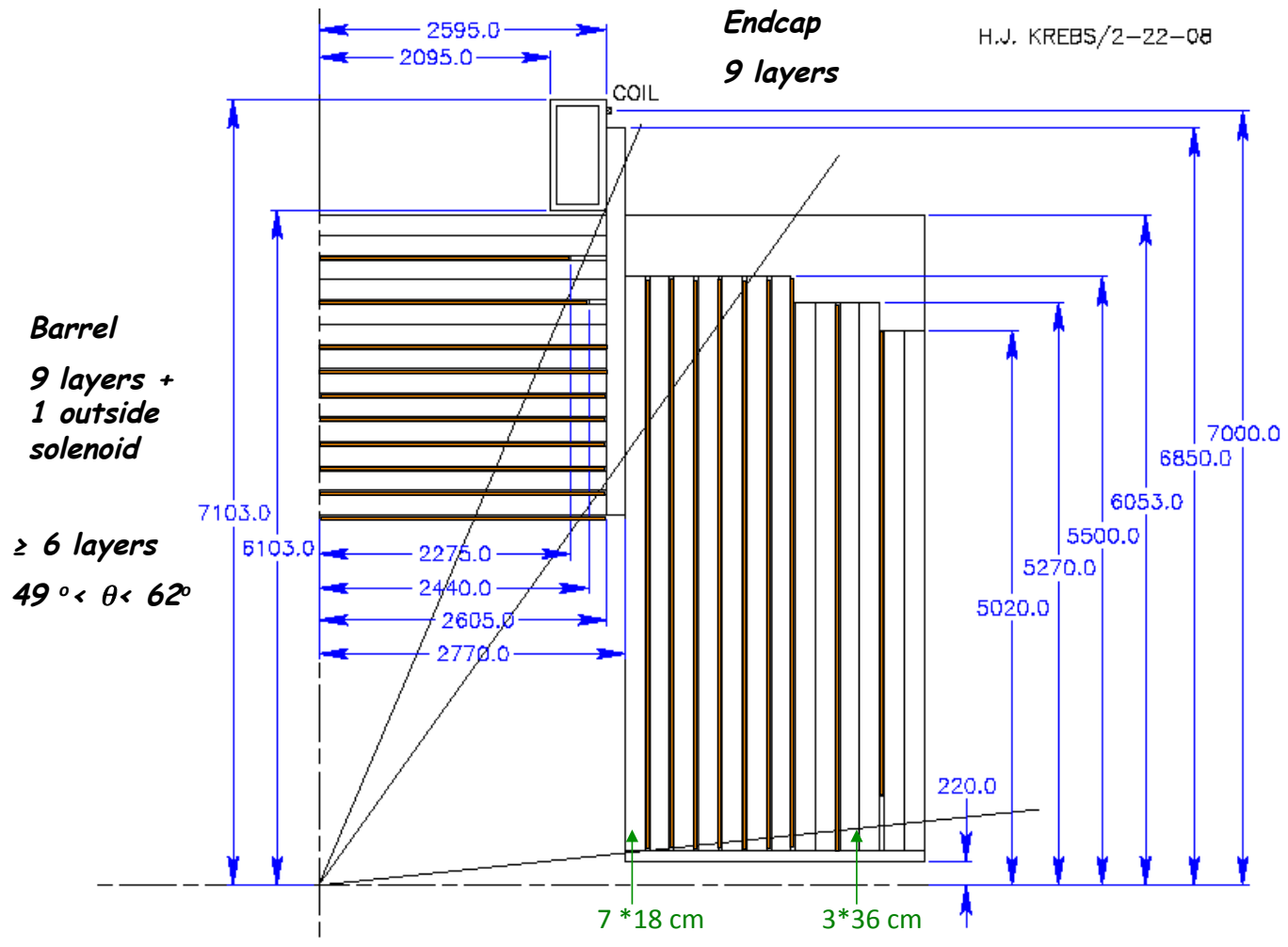

**Figure II-5.2** Quarter section view of the SiD steel flux return.

## 5.1 Backgrounds

Backgrounds in the muon system are expected to come primarily from beam losses upstream of the detector. The muon system is shielded from backgrounds generated at the collision point or along the internal beam lines by the calorimeters, which are greater than five absorption lengths thick. Therefore only penetrating backgrounds, such as high-energy muons or neutrons, affect the barrel muon detectors. Calculations [130] of the expected background from muons produced by collimators near the detector hall predict a rate of 0.8 muons/cm$^2$ per pulse train (1 ms) without muon spoilers, which is reduced to $3 \times 10^{-3}$/cm$^2$ per pulse train with the addition of muon spoilers. Physics backgrounds from two-photon processes producing hadrons or muon pairs significantly increase the expected signal rate in the endcap detectors near the beamline. At a radius of 22 cm the expected rate from hadrons and muons above 2 GeV is $\leq 0.04$/cm$^2$ per pulse train. The endcap detectors can also be hit by electromagnetic shower debris from local beam losses and may require additional shielding

## 5.2 Detector design

The muon system will start outside of the highly segmented electromagnetic and hadronic calorimeters and the 5 T solenoid cryostat at a radius of 3.3 m. In the design shown in Figure II-5.2 the barrel flux return is divided into seven layers of 18 cm steel and three layers of 36 cm steel in an octagonal barrel geometry. Endcaps of seven 18 cm thick steel octagons plus three 36 cm octagons will cap both ends of the barrel. The muon detectors will be inserted in the 4 cm gaps between the plates. In the barrel a detector layer is also inserted between the solenoid and the first steel plate. The size of the first barrel layer within each octant is approximately 2.9 m by 5.5 m, while the last layer is 4.7 m by 5.5 m. The total detector area needed in the barrel is $\approx$ 1600 m$^2$.

The endcap design is shown in Figure II-5.3 (left). Each octagonal layer is made from three steel plates bolted together. The spacers between layers are staggered as seen in Figure II-5.3 (right) to reduce projective cracks in the muon detection. The endcap detectors are subdivided by the spacers into rectangular or trapezoidal modules $\approx$ 1.8 m by 5.5 m. Each endcap has a total detector area of $\approx$ 1000 m$^2$.

**Figure II-5.3**
(left) Each layer of the endcap flux return is made of three vertical steel pieces bolted together. (right) Horizontal spacers separating the steel layers are offset by 0.7 m in alternate layers to avoid projective cracks.

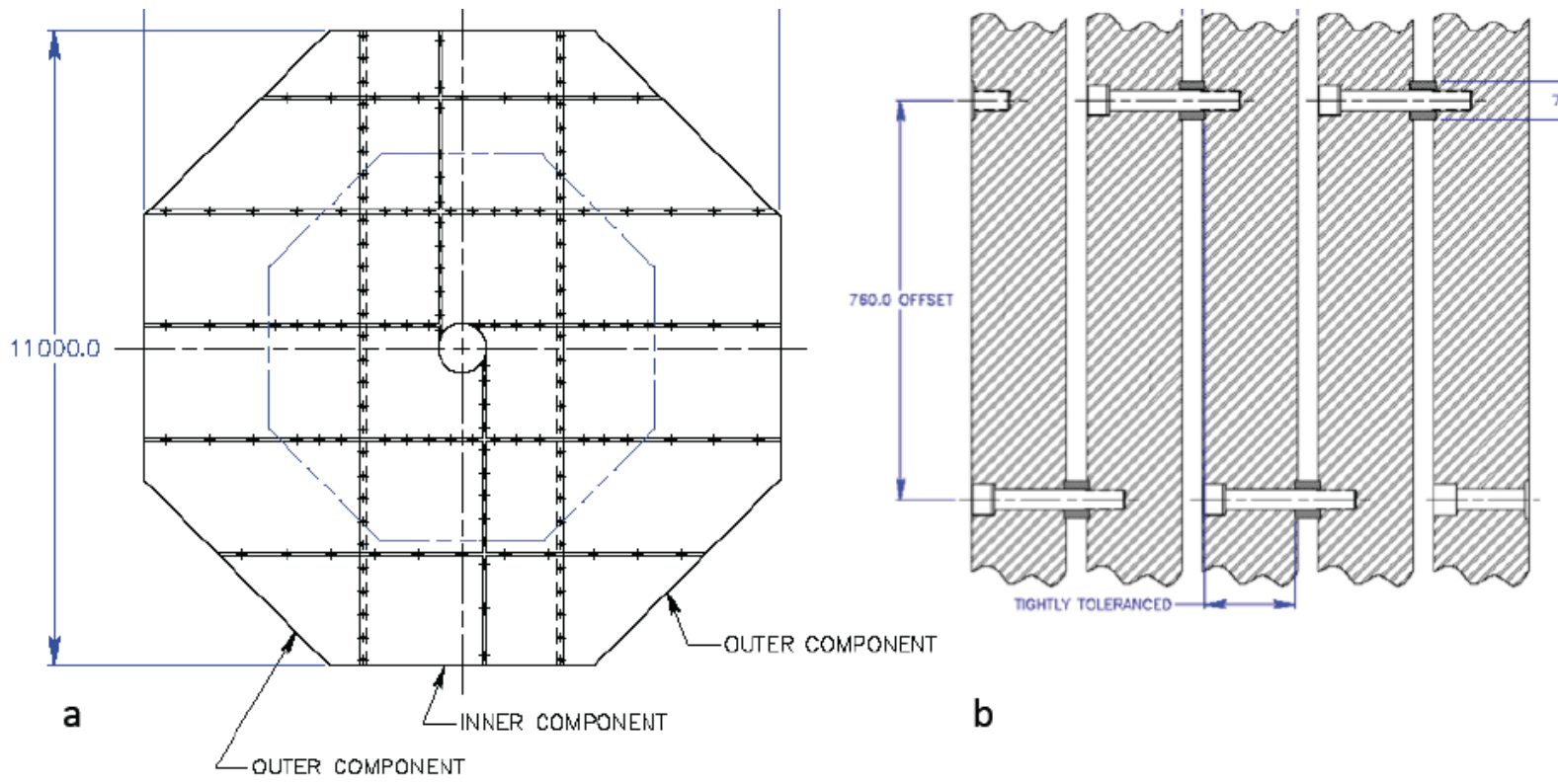

## 5.2.1 Scintillating strips

Extruded scintillating strips have been used in MINOS [131] and T2K [132] and are planned for $\mu 2e$ [133] and SuperB [134]. Wavelength shifting fibre is run down the centre of each strip. A $TiO_2$ reflective coating is co-extruded on the outside of the scintillator. The fibres extend out of the strips by $\sim 1$ cm and are readout by SiPMs

The baseline muon detector employs 1 cm thick by 4.1 cm wide scintillating strips, arranged in back-to-back twin-planes with perpendicular strips as shown in Figure II-5.4 (left). In the barrel strips in one plane are parallel to the beam direction (z-strips), while those in the adjacent plane are orthogonal ($r, \phi$ strips). These layers are glued together with aluminium sheets to form a rigid module. The aluminium sheets provide support while optically isolating the two layers. In the endcap, Figure II-5.4 (right), the gaps between the steel layers are broken up by rows of horizontal spacers. The vertical strips are short ($\approx 1.8$ m) while the horizontal strips are $\approx 5.5$ m long.

**Figure II-5.4**
(left) Each gap in the barrel flux return is filled by two orthogonal planes of scintillating strips glued to three aluminium sheets to form a rigid rectangular sandwich. (right) The endcap modules are $\approx 1.8$m high to slide between the spacers separating the endcap steel layers.

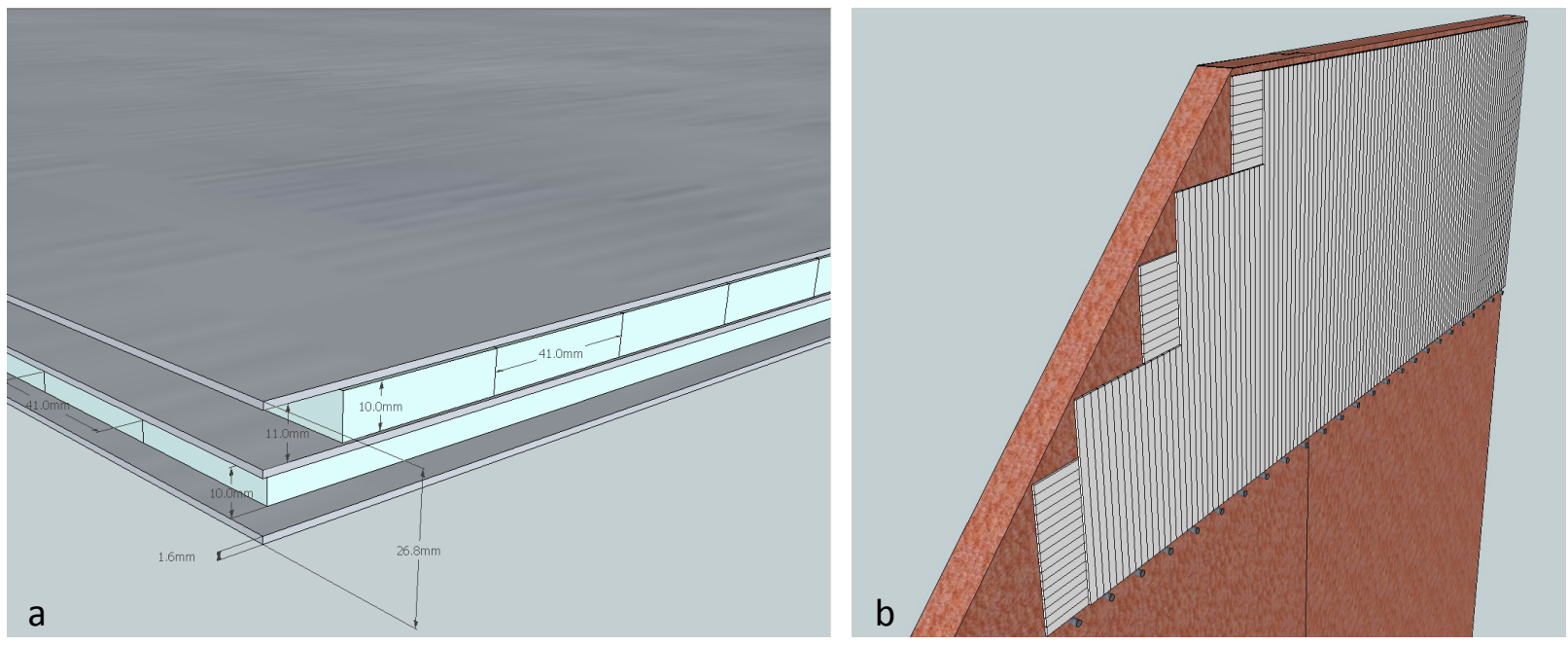

Particles emerging from $e^+e^-$ collisions at the interaction region create optical pulses via dE/dx when traversing the scintillator strips placed in the gaps of the barrel and endcap Fe return yokes. A fraction of the light is captured in a 1.2 mm diameter wave-length shifting (WLS) fibre located in a groove that runs along the length of the scintillator bar. The light travels through the WLS over 2 m to 5 m distance before reaching the input face of a Si photo-diode (SiPM) matrix, where it triggers an avalanche in one of the few-hundred micron-sized individual photo-diodes cells whose outputs are ganged together through individual output resistors to a common output. In our tests of candidate SiPMs for muon detection we have focused on devices with ≈700 cells with $40\ \mu\text{m} \times 40\ \mu\text{m}$ size fitting inside a 1.2 mm diameter circle [135]. As the cells in the Si matrix have good uniformity with similar areas and Si thicknesses, the summed avalanche signal output of the ganged cells is proportional to the number of cells hit. Therefore the devices can be calibrated adequately by using the individual photoelectron peaks in the summed signal of the SiPM. The calibration procedure makes use of peaks with one or two photoelectrons, as well as noise peaks. Signals from individual SiPMs are then sent on to receivers and collected for further digital processing.

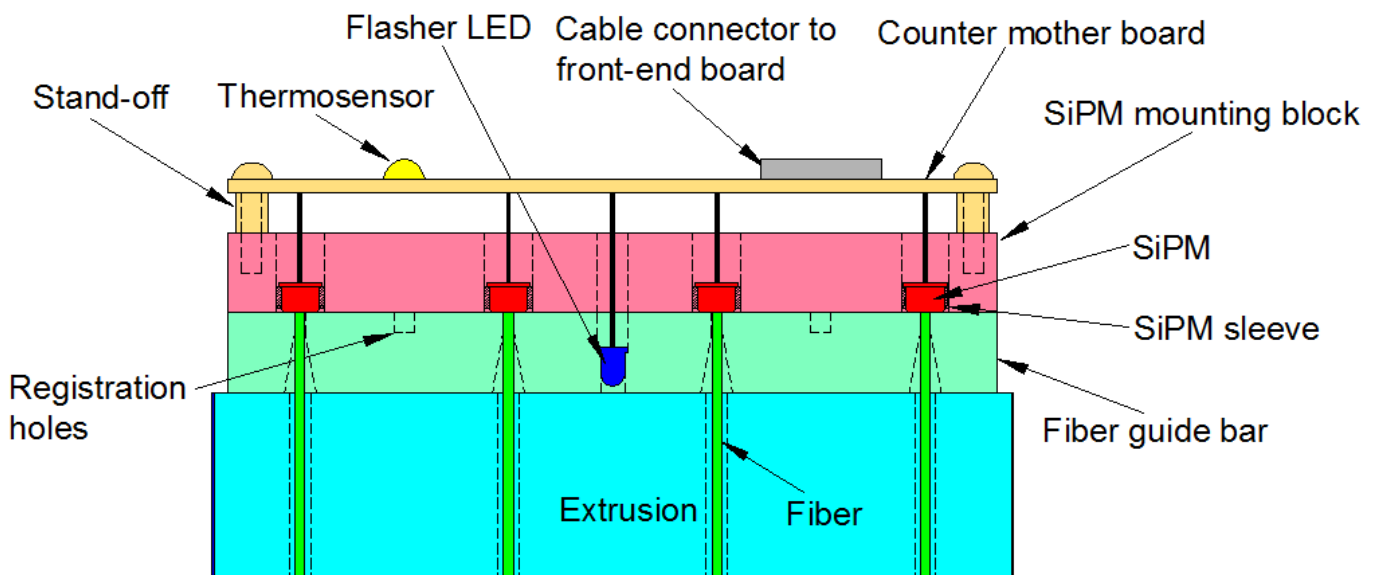

**Figure II-5.5**
SiPMs are positioned at the end of each fibre by a SiPM mounting block and fibre guide.

There are 7,451 axial barrel channels where both ends of the fibres are read for the barrel strips, making 14,902 readout channels. There are 10,810 (r, $\phi$) scintillator strips that add 10,810 single ended readout channels to make the barrel (B) channel count 25,712. The reason for not reading out both ends of the (r,$\phi$) strips is that there are more and consequently shorter WLS fibres (less occupancy) and less attenuation in these barrel channels. All the outer periphery ends of the Forward/Backward (F/B) channels are read (no double-ended readout). The F/B strip-scintillator planes add a 21,620 scintillator strips (21,620 channels of readout and electronics) to make a total of 25,712 central and 21,620 F/B in back-to-back quadrants for a total of 47,332 channels. The sum of WLS fibre and scintillator totals $\approx$ 164 km ($\approx$ 86 km for the ten barrel planes and $\approx$ 78 km for the nine endcap planes).

Recently the University of Virginia HEP group have developed small molded plastic parts that capture the detector end of the WLS fibre and accurately position it relative to the centre of the SiPM which has 600 Si pixels contained inside a circular area of 1.2 mm diameter. With this kind of connection of the polished signal fibre to the photodetector it should be possible to locate the readout devices on a separate long plastic or fibreglass strip that accepts the WLS ends for a plane or half plane of detectors as drawn in Figure II-5.5. A prototype strip/SiPM combination was tested in Fermilab Test Beam Experiment T995. Two 3.6 m long strips were connected by fibre to make an effective 7.2 m long strip. SiPMs were glued on both ends of the fibre. Beam was scanned along the length of the strip to study pulse height as a function of the distance from the SiPM. As seen in Figure II-5.6 the number of photoelectrons can be easily counted on either end of the strip even if the beam is placed near one end. The pedestal was quite small and stable. Requiring two or more photoelectrons eliminates nearly all of the noise signals.

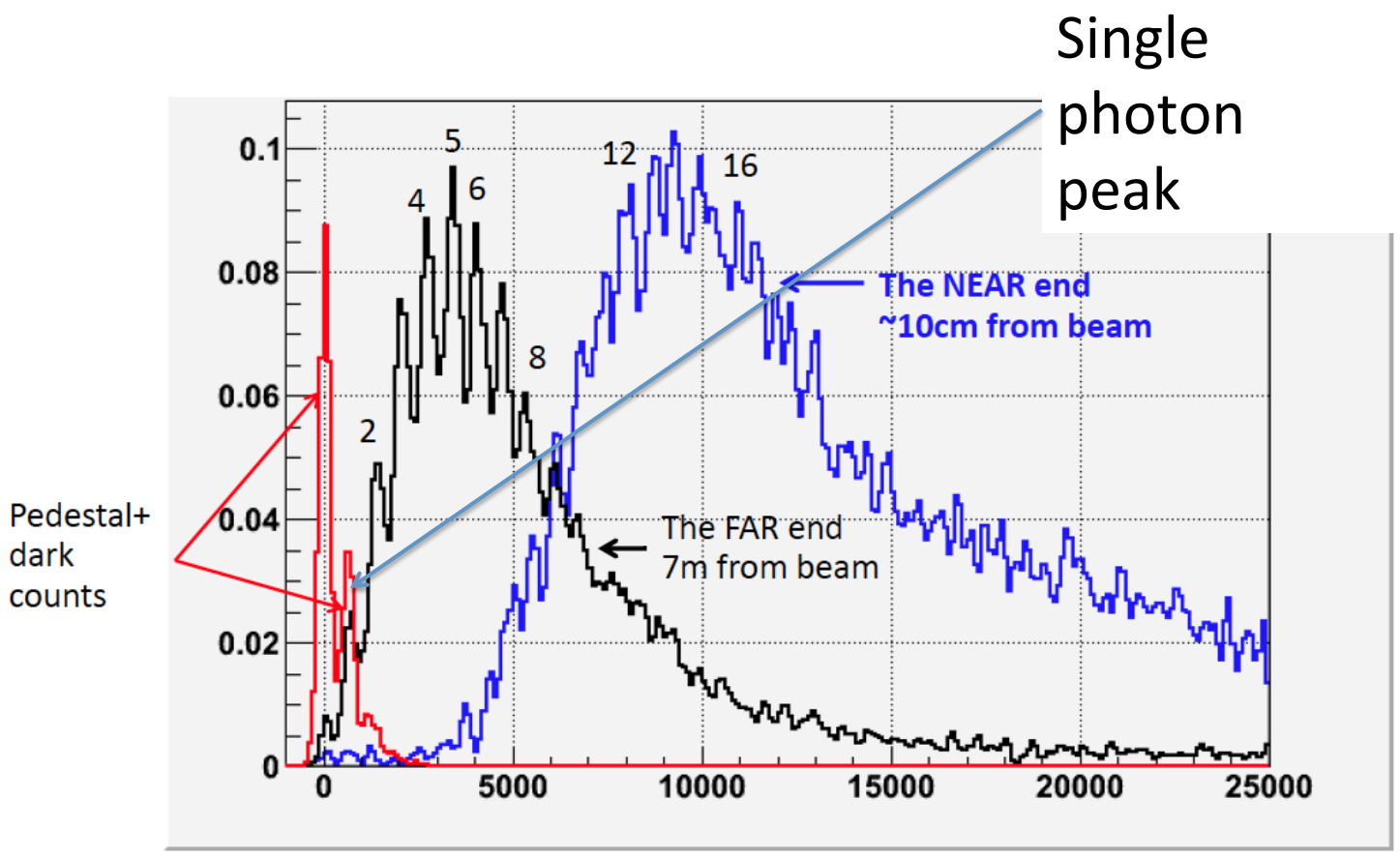

**Figure II-5.6**
Test beam data of two strips coupled by fibre to simulate a single long strip. Pulse height from the top strip (blue) and the bottom strip (black) are shown. The beam is 10 cm from the end of the top strip.

In Figure II-5.7 multiple beam positions in two different prototype strips were used to measure the attenuation of the light signal with fibre/strip length. The data can be modelled by the sum of two exponential fall-offs. Near the sensor, the attenuation length is $\sim$ 2.9 m. At 7 m from the sensor, the attenuation length is $\approx$ 6.5 m. Since the longest strips in SiD are less than six meters long the minimum expected pulse height is $> 5$ photoelectrons. With a threshold of two photoelectrons the scintillating strips with SiPM readout are very efficient.

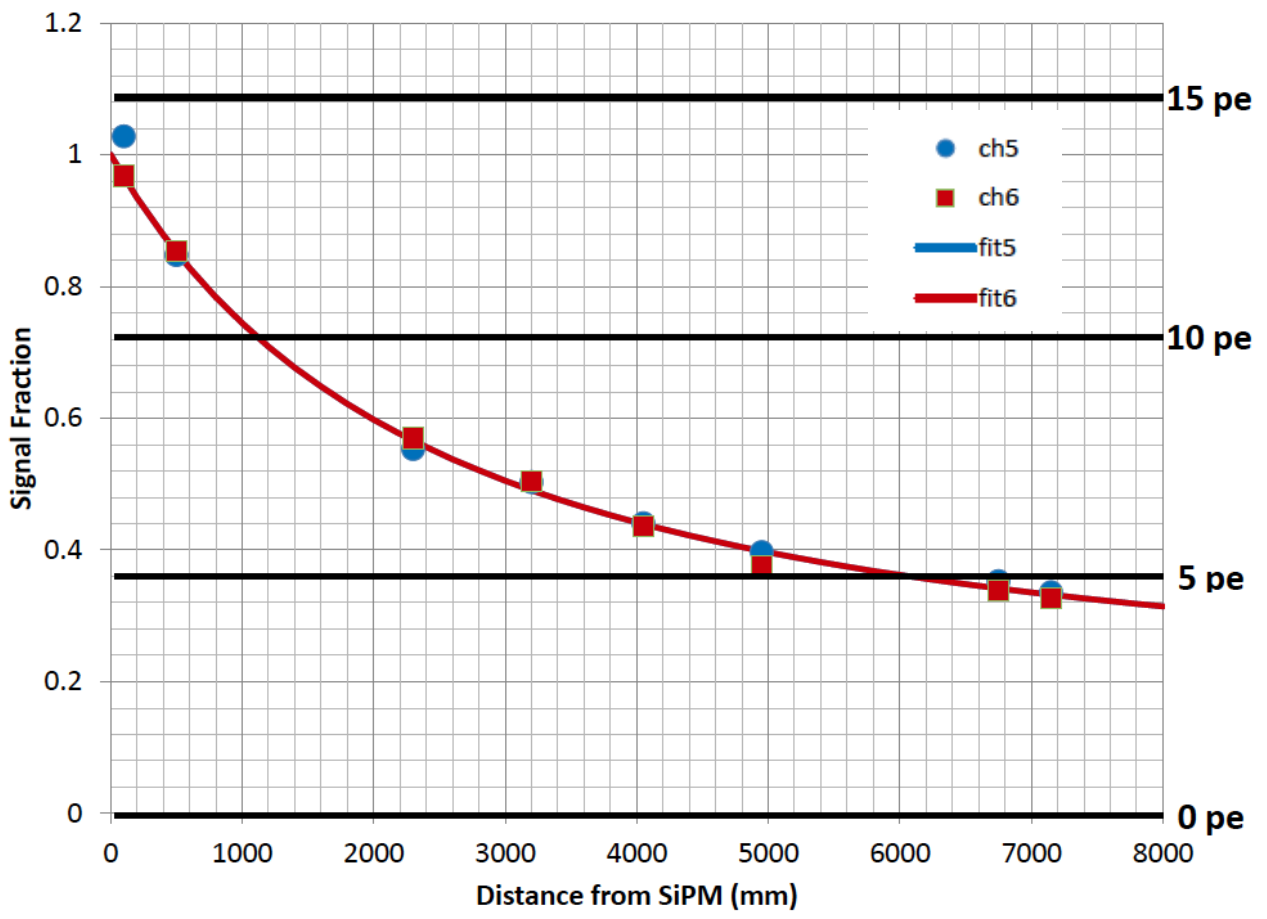

**Figure II-5.7**
The fraction of the total light collected by the SiPM as a function of the beam position along the strip for two different strips (ch 5 and 6).

## 5.2.2 Resistive Plate Chambers

Many large RPC systems have been built within the last ten years and understanding their performance will provide strong guidance for an SiD design. Several types of RPC construction have been used in high energy experiments. RPCs with Bakelite cathodes and anodes, initially reported in [136], found application in BaBar, CMS, ATLAS and a variety of cosmic ray and neutrino experiments. RPCs are inexpensive to build and can be easily constructed in a variety of shapes and sizes. There are, however, a few concerns about the use of RPCs in future experiments. RPCs use fluorocarbon gases which are regulated as greenhouse warming gases and require nontrivial gas delivery systems adding to operational costs. Further restrictions on the use of fluorocarbon gases are possible in the future. RPCs have also had reliability problems (BaBar was forced to replace its original RPCs and Belle had startup problems). However, significant progress has been made in understanding RPC aging mechanisms. The current ATLAS [137] and CMS [138] detectors, which run in avalanche mode, have shown good stability even at the high background rates expected at the LHC. The second generation BaBar RPCs [139] and the Belle RPCs preformed reliably at the low signal rates ($< 0.2$Hz/cm$^2$ ) expected for SiD detectors. RPCs are a viable detector alternative for the muon system, particularly if the RPC option is chosen for the hadron calorimeter. Bakelite was chosen for the RPCs because for the foreseen plate thickness of 2 mm, glass is significantly heavier than Bakelite and more brittle. Given the large-area chambers needed for the muon system, a Bakelite RPC system is most likely easier to construct and install, hence a conservative choice was made.

A RPC design for the muon detector planes would utilise two layers of single gap RPC HV chambers (1 $\times$ 2 m) with orthogonal readout strips on either side assembled into modules of the required size to fill each slot in the octagonal barrel or endcap. The chamber size would be varied so that joints between chambers do not align in the top and bottom layer. If the single gap RPC efficiency is 90%, then an average module efficiency of 93% can be achieved.

Close integration of the RPCs and front-end and digitisation electronics is necessary to minimise cabling and costs since the expected channel counts for the SiD detector are high (nearly one million for the muon system). One possible low cost solution that has been investigated is to adapt the KPiX chip, presently being developed for use in the SiD electromagnetic calorimeter, for use with RPCs. An RPC/KPiX interface board was designed and built to provide ribbon cable connections to a 64-channel KPiX chip (v7). The RPC strip signal is AC coupled to the KPiX input through a 5 nF blocking capacitor and a 2-stage diode protection network. Each strip is also tied to signal ground via a resistor external to the interface board.

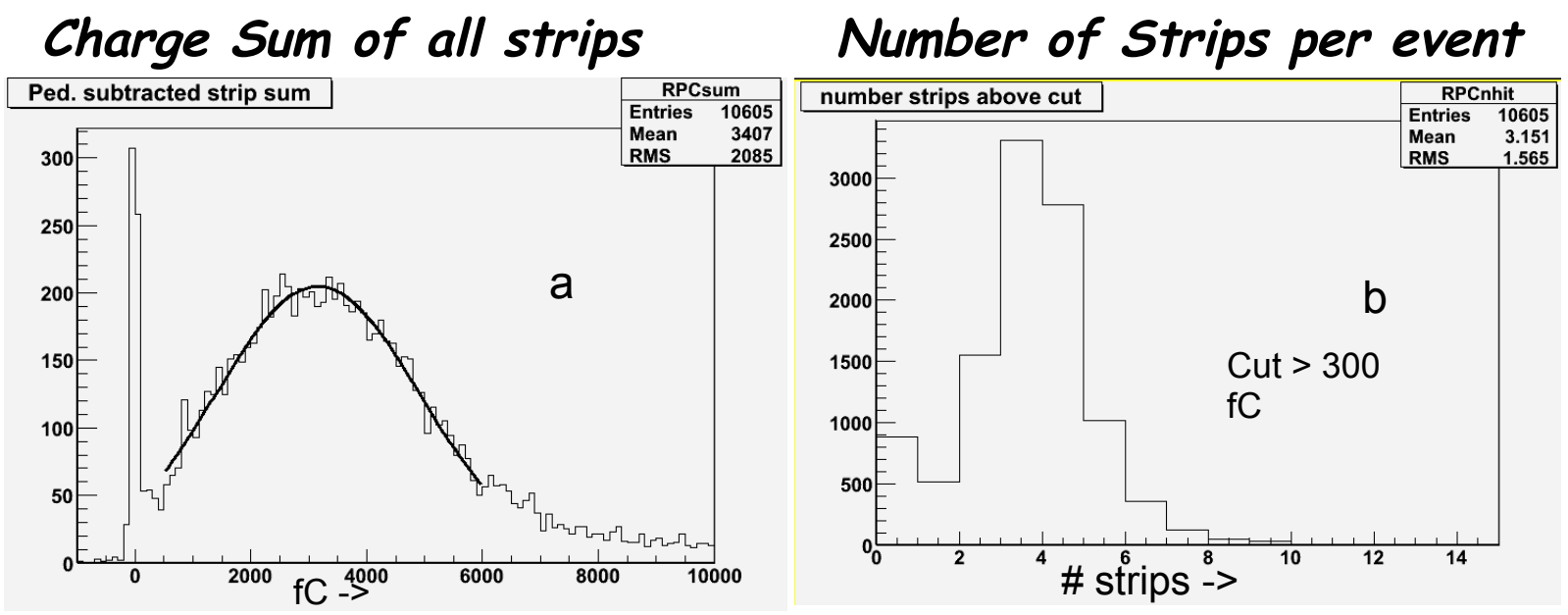

**Figure II-5.8** (left) Sum of the pulse heights in 13 RPC strips readout by a 64-channel KPiX chip (v7). The peak position of 3 pC and efficiency of $> 90\%$ are consistent with previous studies of avalanche mode RPCs. (right) The number of strips with a signal height above 300 fC per track.

A small (0.5 m $\times$ 0.5 m) test RPC with 13 strips was connected to the interface board by a 0.5 m cable. The chamber was operated at 9300 V in avalanche mode using a premix gas with composition of 75.5% Freon 134a, 19.4% Argon, 4.5% isobutane, and 0.5% SF6. The chamber efficiency had been previously measured to be $\approx 90\%$ using BaBar electronics. The signal sum of the 13 RPC strips on the HV ground side (positive signal) is shown in Figure II-5.8 (left). The sharp spike near zero is due to cosmic ray tracks that either missed the test RPC or to RPC inefficiency. The signal peak is centred at 3.8 pC with a width of 2.2 pC. The signal height is consistent with, but larger than, avalanche RPC signals measured by other groups ($\approx 1$ pC), which used avalanche gases with no Argon component. The avalanche gas contains 19.4% Argon and is expected to have a higher gas gain. The charge distribution in the RPC pickup strips was also studied. For each trigger, the strip with the maximum charge has typically less than half of the total charge in the event. The strip multiplicity was measured as a function of the discrimination threshold. With a threshold of 300 fC, about 92% of the cosmic triggers have one or more strips hit and the average strip multiplicity is 3.1, more than twice that observed with BaBar electronics. Although a good proof of principle, these tests show that further characterisation and optimisation of the interface board between the RPC and chip is needed

# Chapter 6
# SiD Superconducting Magnet System

## 6.1 Introduction

### 6.1.1 List of subsystems

The magnet subsystem consists of its own following subsystems:

1. A 6.8 m outside diameter × 6 m long 5 T superconducting solenoid with a separated iron plate flux return that is integral with the muon tracking system.
2. A superconducting 600 G Detector Integrated Dipole (DID) integrated with the solenoid.
3. A power supply, a DC contactor, a pressurised water cooled dump resistor, and a conventional mechanical dump switch that move with the detector.
4. A 1.5 kW helium liquefier and 5000 litre LHe storage dewar that supply 4.5 K LHe to both the solenoid and to a pair of 2 K cold boxes for each of the superconducting focusing magnets (QD0).
5. Interconnecting cold, warm, and vacuum plumbing lines including those to QD0, mounted directly on the detector.
6. Controls and instrumentation for the magnets and helium liquefier.

The shared resource ILC helium compressor system and the two superconducting QD0 focusing magnets with the internal design of their 2 K distribution boxes are not part of this subsection.

### 6.1.2 Design Philosophy

The superconducting solenoid is an expensive and technically challenging component. Its design is based on the successful 4 T CERN CMS superconducting solenoid, and thus a direct comparison is warranted in Table II-6.1 [140]. High purity aluminium superconductor stabilisation with indirect LHe cooling will be used. The CMS individual self supporting winding turn design philosophy is used for SiD, becoming even more important due to the higher 5 T field and the increased radial softness of six layers versus four layers. Figure II-6.1 shows a 3D cut-out with the principal elements of the SiD magnet.

The SiD solenoid has a stored energy per unit of cold mass of 12 kJ/kg, which is only slightly larger than the CMS value. The value of 12 kJ/kg is close to the upper bound at which such a large aluminium dominated magnet can be operated in a fail safe manner, in case the quench detection or energy extraction circuit were to fail. Upon such a failure, the average magnet temperature would reach 130 K. Engineering studies of the SiD solenoid indicate that the total volume of aluminium stabiliser/structure cannot be reduced by much with respect to the present baseline design.

**Table II-6.1**
SiD and CMS Superconducting Coil Comparison

| Quantity | SiD | CMS | Units |
|---|---|---|---|
| Central Field | 5.0 | 4.0 | T |
| Stored Energy | 1.59 | 2.69 | GJ |
| Stored Energy Per Unit Cold Mass | 12 | 11.6 | kJ/kg |
| Operating Current | 17.724 | 19.2 | kA |
| Inductance | 9.9 | 14.2 | H |
| Fast Discharge Voltage to Ground | 300 | 300 | V |
| Number of Layers | 6 | 4 | |
| Total Number of Turns | 1459 | 2168 | |
| Peak Field on Superconductor | 5.75 | 4.6 | T |
| Number of CMS superconductor strands | 40 | 32 | |
| % of Short Sample | 32 | 33 | |
| Temperature Stability Margin | 1.6 | 1.8 | K |
| Total Cold Mass of Solenoid | 130 | 220 | tonne |
| Number of Winding Modules | 2 | 5 | |
| $R_{min}$ Cryostat | 2.591 | 2.97 | m |
| $R_{min}$ Coil | 2.731 | 3.18 | m |
| $R_{max}$ Coil | 3.112 | 3.49 | m |
| $R_{max}$ Cryostat | 3.392 | 3.78 | m |
| $Z_{max}$ Cryostat | $\pm$ 3.033 | $\pm$ 6.5 | m |
| $Z_{max}$ Coil | $\pm$ 2.793 | $\pm$ 6.2 | m |
| Operating Temperature | 4.5 | 4.5 | K |
| Cooling Method | Forced flow | Thermosiphon | |

## 6.2 Magnetic Field and Forces

### 6.2.1 Requirements and Design

The SiD magnet system requires a 5 T central field, an alternating 600 G field along the axis from the DID, and a fringe field of less than 100 G at a metre distance from the outer iron surface [141]. An economic solution to the fringe field requirement has not yet been found. Two iron plates placed around the barrel and overlapping the doors with a combined thickness of 14 cm drops the 1 meter fringe field to 300 G. The 100 G at one meter is certainly achievable with the addition of sufficient iron and air gaps. Some components such as the expansion turbines inside the helium liquefier will most likely require additional local iron shielding.

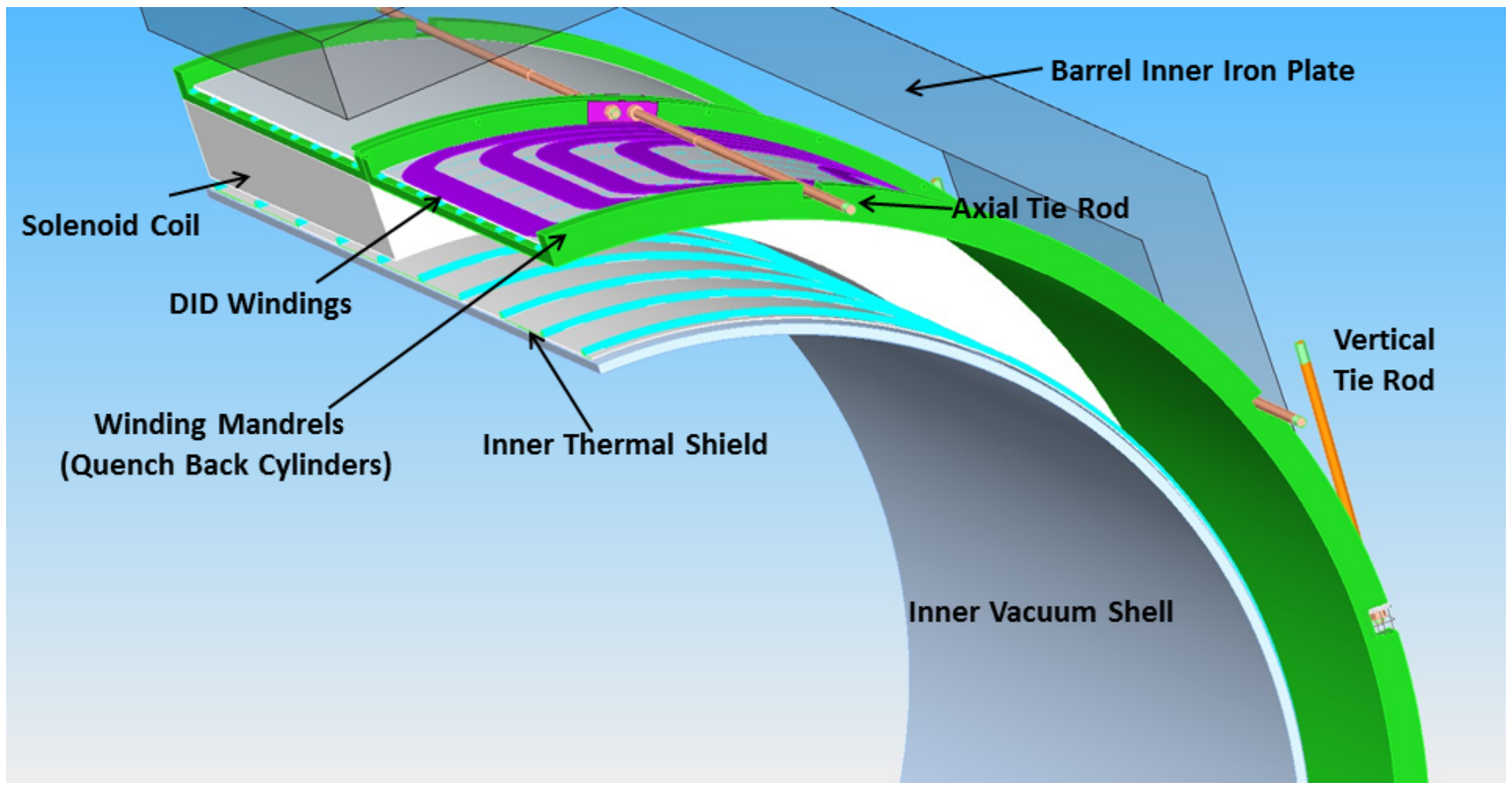

**Figure II-6.1**
Magnet section showing its principal elements.

Eleven 20 cm thick iron plates with 4 cm gaps form both the barrel and end wall portions of the flux return. There is also a 5 cm gap between the barrel and endcap that is partially filled with barrel iron connecting plates. These connecting plates are also part of the solenoid axial restraint system. The iron plates of the endcaps are held together with an iron cylinder on the inner diameter and top plates on the outer diameter.

### 6.2.2 Calculations

The results of two- and three-dimensional ANSYS magnetic field calculations of the magnet are shown in Figure II-6.2. The 3D ANSYS model also includes the DID, barrel/endcap iron gap details, and the cryogenic chimney and current lead penetration details. The DID coil position was calculated using OPERA 3D and custom codes. The ANSYS 3D model uses an edge element formulation and has seven million elements [142]. Advances in computation give a significant advantage to the SiD design as compared to prior CMS design work. The magnetic axial spring constant was found to be constant from 1 cm to 20 cm coil displacement. The axial magnetic force is maximum at full current; there is no iron saturation effect. An iron HCAL endcap was studied and rejected due to minimal improvement in field and field uniformity versus increased cost and complexity [143].

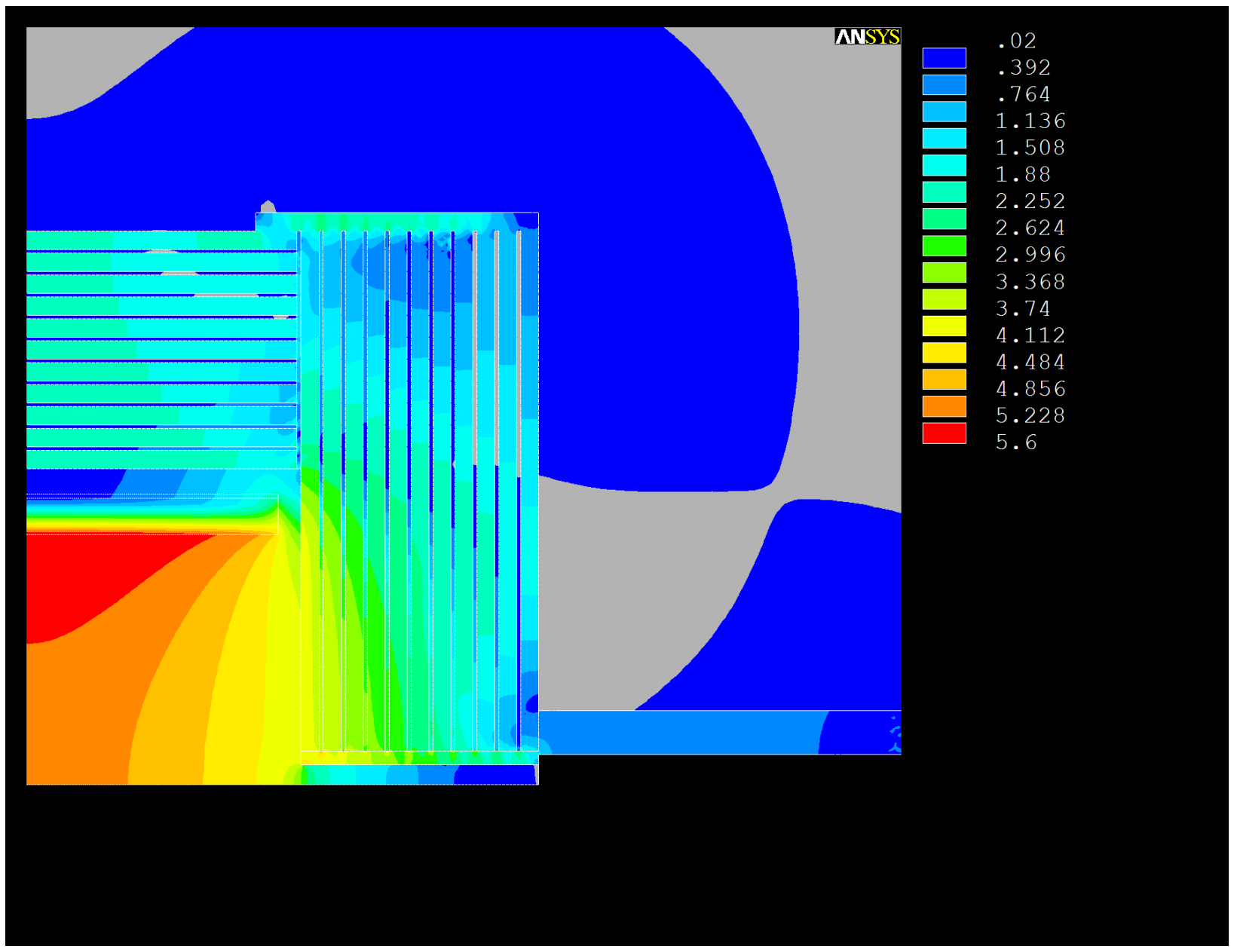

**Figure II-6.2** :2D Axisymmetric showing $B_{max}$. Only a small portion of the air is displayed. The gray/blue boundary is the 200 G line.

## 6.3 Mechanical Engineering

### 6.3.1 Solenoid Coil Production

The superconductor will be internally wound into two precision aluminium 5083-H321 mandrels using CMS winding procedures, including epoxy vacuum impregnation and mandrel joining techniques, and conductor splicing methods. The CMS coil winding experience will significantly reduce the SiD time and expense of winding line setup and commissioning. Coil winding and vacuum impregnation will take place at the vendors facility. The magnet will be shipped as two separate coils of 65 tonne each.

### 6.3.2 Integration of DID to solenoid

The Detector Integrated Dipole (DID) is mounted directly on top of the solenoid cooling tubes. The four separate 600 kA turn winding packages are sandwiched between a lower 3 mm Al sheet and an upper 5 mm Al sheet. Each package consists of five coils all electrically connected in series creating either a DID or anti-DID field. The coil packages are mounted directly on top of the solenoid LHe cooling loops by metal screws attached to the solenoid winding mandrel. Twenty two solenoid splices rest on top of the upper DID Al sheet and are supported by direct connection to the solenoid winding mandrel at the centre of the four DID winding packages. Conduction through the DID and direct physical connection through the DID centres establish cooling for the solenoid splices. All DID splices except for the two connections to the DID current leads are made in the space between the two aluminium sheets.

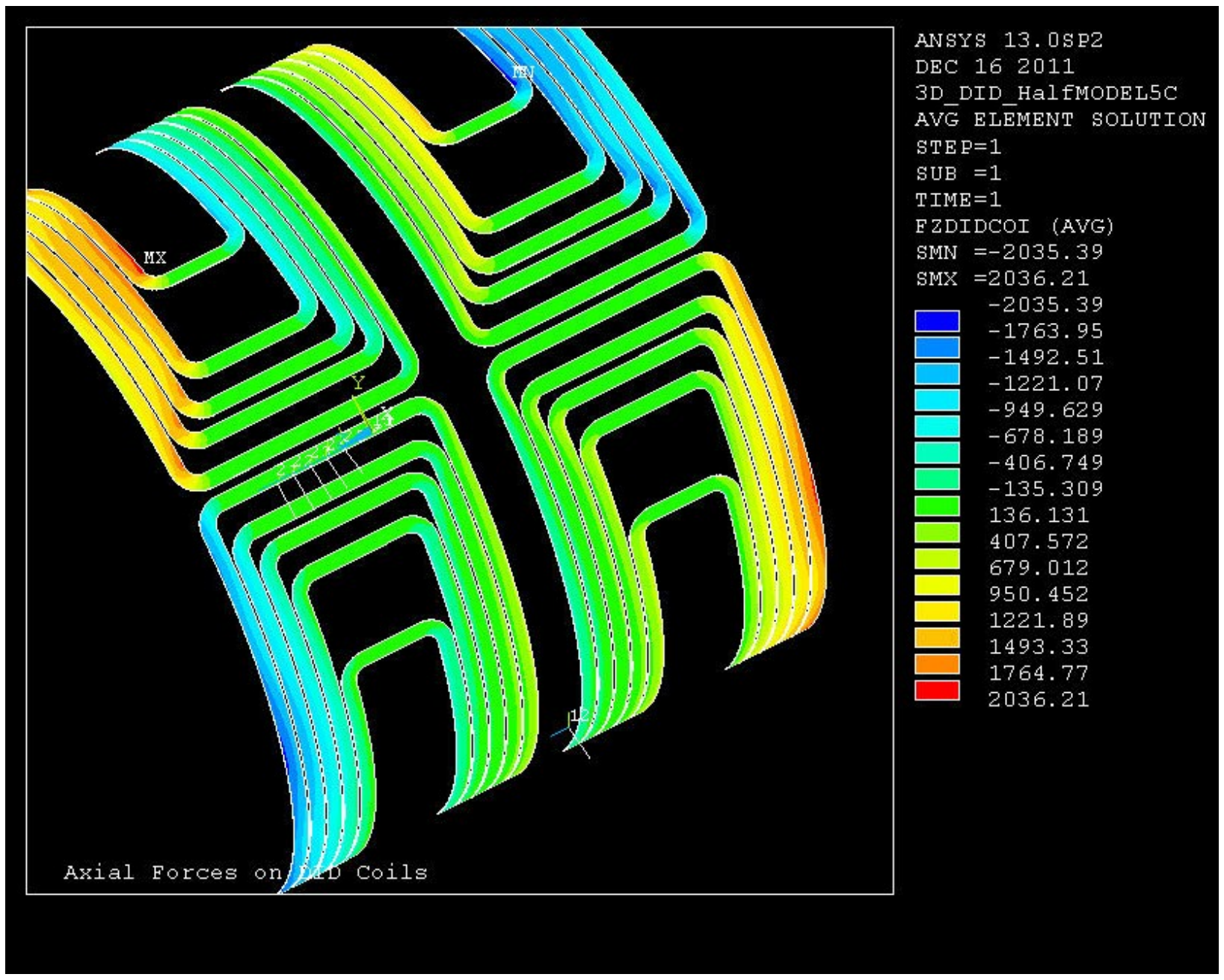

**Figure II-6.3**
DID coils showing axial forces; Only half of the coils are shown

## 6.3.3 Thermal shield, cooling tubes, current leads and standpipe

Both the inner and outer thermal shields are directly mounted on the inner and outer vacuum shells with Ti 15-3-3-3 studs and small diameter fibreglass epoxy tubes. The shields are fixed at the midplane and contract symmetrically from both ends. The shields are made from aluminium alloy with appropriate high resistance joints to reduce eddy current forces during solenoid fast discharge. Baseline design for shield and coil cooling loops is square aluminium tubing welded to the aluminium shells with transition to round stainless steel tubing. All stainless steel or bimetallic tubing that is generally more leak tight is an option to be studied. Some of these details are depicted in Figure II-6.1.

The cryostat, $\approx$ 60 K thermal shield, current leads, tie rods, and instrumentation will all be designed using standard cryogenic techniques. Current leads will be very similar to the CMS current leads. Two separate iron penetrations will be used, a 70 cm $\times$ 40 cm chimney for the current leads and 36 cm diameter chimney for the cryogenic plumbing. Vacuum pump down will take place through both chimneys.

## 6.3.4 Stress Analysis

ANSYS studies compared SiD and CMS solenoid stresses, deflections and forces. All stresses are evaluated after cool down and energisation. For this comparison the CMS conductor was used in the SiD analysis with results given in Table II-6.2. Note that all stress and deflections are very similar for the two coils.

**Table II-6.2**
Mechanical Comparison of the SiD and CMS Solenoid coils

| Quantity | SiD | CMS |
|---|---|---|
| Von Mises Stress in High Purity Aluminium (MPa) | 22.4 | 22 |
| Von Mises Stress in Structural Aluminium (MPa) | 165 | 145 |
| Von Mises Stress in Rutherford Cable (MPa) | 132 | 128 |
| Maximum Radial Displacement (mm) | 5.9 | 5 |
| Maximum Axial Displacement | 2.9 | 3.5 |
| Maximum Shear Stress on Insulation (MPa) | 22.6 | 21 |
| Radial Decentering Force (kN/cm) | 280 | 80 |
| Axial Decentering Force (kN/cm) | 1870 | 850 |

Cold mass tie rods will be segmented into three different systems based on direction (axial, vertical and radial) just as they were with CMS and BaBar. They will be manufactured from age hardened Inconel 718. Radial and vertical loads will be carried to the cryostat outer wall. Axial loads will be carried to the cryostat end plates. In all cases, the tie rod systems are substantially stiffer than the magnetic spring constant.

### 6.3.5 Vacuum Shell Design

The 304 stainless steel vacuum shells will be built according to the ASME (American Society of Mechanical Engineers) pressure vessel code design rules, but the cryostat will not be a coded vessel. Inner shell, outer shell and both end flanges are all 50 mm thick. In addition to the magnet weight and magnetic force loads, a detector weight of 450 tonnes, vacuum load and gravity self weight are imposed on the vacuum shell. The detector weight is carried by two linear rails on the inner shell. The total weight is transferred on two linear rails on the outer shell to the magnet iron. The solenoid axial decentering forces are transferred to the barrel/endcap spacer plates.

The linear and non-linear vacuum buckling ANSYS analysis has been completed and the primary stress results are summarised in Table II-6.3. Local peak stresses are much higher especially for the magnetic axial decentering case. These peak stress values and maximum primary stress values can easily be reduced to ASME Section VIII Div. 2 allowables by small additions of local reinforcing. The outer end plates will need radial rib reinforcing which is compatible with the detector cable routing. The ASME allowable stress is 138 MPa. A non-linear analysis gives a 0.62 MPa (6 atm.) vacuum buckling load [144].

**Table II-6.3** Cryostat vacuum shell maximal stress and deflection

| Load | Stress (MPa) | Deflection (mm) | Location of Max Stress |
|---|---|---|---|
| Axial Magnetic | 125 | 1.5 | Axial Support Pad |
| Detector Mass | 45 | 2.3 | Inner Vacuum Shell |
| Cold Mass + Radial Magnetic | 23 | 0.44 | Vertical Tie Rod Support Pad |
| Vacuum | 7.5 | 0.17 | Outer Shell |
| Gravity on Shell | Small | 0.11 | Both Shells |
| **All Loads Combined** | **190** | **3.5** | **Vacuum Shell End Plate** |

### 6.3.6 Assembly procedure

1. The coil mandrels are precision machined with welding of seamless end rings and cooling loops. The cooling loops are extensively leak tested.
2. The solenoid modules are wound with each layer in alternating direction.
3. The four DID coil modules are wound on a 3 mm thick Al sheet that is mounted onto a machined cylinder. The internal coil to coil splices for each of the four modules are completed. A 5 mm sheet is attached to the outer diameter of the DID coils.
4. The DID coils are vacuum impregnated. This is a higher temperature resin than the solenoid resin.
5. The DID coils are mechanically attached on top of the solenoid cooling loops with screws to the solenoid mandrel.
6. The Solenoid modules with attached DID coils are vacuum impregnated.
7. The two mating ends of the solenoid modules are precision machined.
8. The solenoid modules are stacked vertically and joined above ground at the detector site.
9. All 24 solenoid splices are completed above the DID. All DID module to module splices are completed
10. The axial tie rods are attached to the solenoid.

11. The inner and outer thermal shields are mounted to inner and outer vacuum shells.
12. The inner and outer vacuum shells are placed on the solenoid.
13. The vertical and radial tie rods are attached to the outer vacuum shell.
14. All internal plumbing and electrical connections are completed along with the mounting of the thermal shield end plates. Piping extends a short distance past the chimney opening. The solenoid lead ends and DID lead ends extend through the vacuum shell current lead opening and are wrapped in a loop.
15. Top and bottom vacuum end plates are welded.
16. All tie rods are tightened.
17. The completed magnet assembly is rotated horizontal on a shaft parallel to the ground using the overhead crane and two pulling cables.
18. The magnet is moved to the detector cavern and lowered vertically into the bottom half of the magnet iron.
19. The current leads and cryogenic chimney pipe assemblies are completed and welded.

## 6.4 Cryogenics

A helium refrigerator/liquefier of approximately 1.5 kW of 4.5 K refrigeration is located on the detector near the top. This choice means that the liquefier high pressure helium and compressor suction return lines must be flexible for push-pull operations. The QD0 2 K vacuum pumping lines must also be flexible. The liquefier supplies both forced flow 4.5 K saturated LHe and 40 to 80 K helium for the thermal shield and support rod thermal intercepts. The liquefier is a custom built commercial product whose detailed design and construction will be carried out by industry as part of the complete cryo plant procurement.

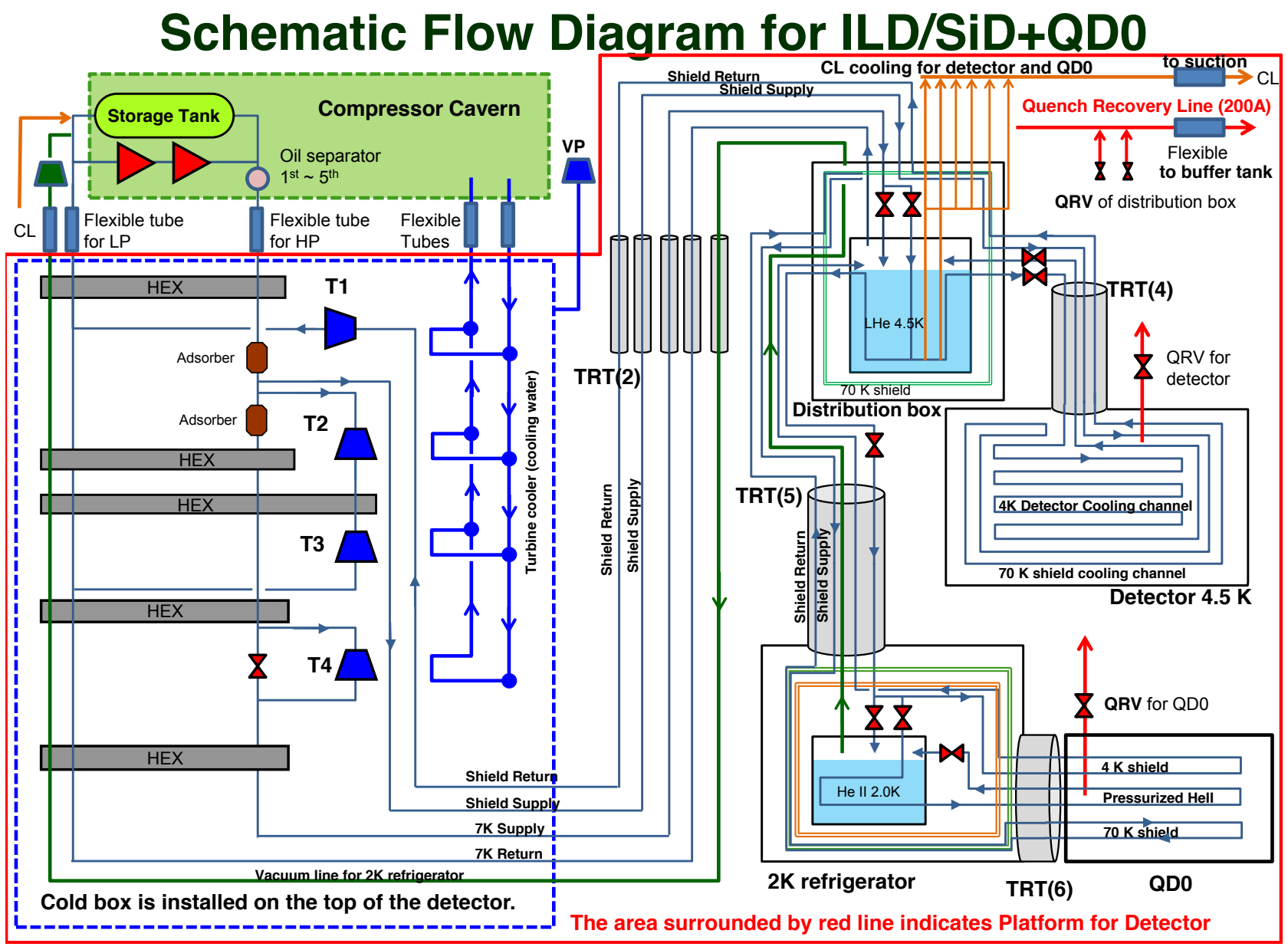

**Figure II-6.4** Cryogenic Flow Schematic

A 5000 litre LHe storage dewar is stationed next to the refrigerator liquefier. It serves as a pressure buffer for forced flow operation and as a LHe supply reservoir during liquefier down times. This technique was used successfully for a decade of running BaBar.

The detector valve box near the top of the detector is used to minimise flexible connections between detector and refrigerator. It also serves as the distribution point for supply of LHe to the two superconducting QD0 final focusing magnets 2 K cold boxes that are fixed on the detector. Figure II-6.4 is a flow diagram of the SiD cryogenic system.

## 6.5 Conductor

### 6.5.1 Solenoid baseline conductor (CMS)

A slightly modified CMS conductor is the SiD baseline design. The CMS conductor is fabricated by ebeam welding structural aluminium to the coextruded high purity Al/superconducting cable insert. A superconductor stability margin similar to CMS will be used requiring that the Rutherford cable be increased in size from 32 to 40 strands. In comparison to CMS, operating current as a % of critical current based on magnet peak field and temperature, improves from 33% to 32% for SiD.

However, the SiD conductor thermal margin will decrease from 1.8 K to 1.6 K. All magnet ANSYS finite element stress analysis to date has been with this conductor with overall dimensions of 21.6 mm $\times$ 64 mm.

### 6.5.2 Modified CMS Conductor Choices

Many other conductor designs are possible. One possibility is replacement of the high purity aluminium with an Al-0.1%Ni alloy that is stronger but still has good conductivity. This material was used for the ATLAS Central Solenoid conductor. Coextrusion tests of this alloy are currently being pursued by CERN [145]. Many other dilute aluminium alloys (e.g. scandium or binary elements) that form small intermetallic precipitates are possible but largely unexplored. Still other high purity reinforcement such as the standard $TiB_2$ grain refiner or carbon nanotubes are possibilities. Replacement of the structural aluminium with internal stainless steel rope would simplify conductor manufacture if a different method of coextrusion such as the ConKlad process could be industrialised for this size.

The ANSYS coupled transient electromagnetic and thermal diffusion model was used to evaluate conductor stability. With large size high purity aluminium stabilised superconducting conductors, current is slow to diffuse into the high purity aluminium during a temperature excursion reducing conductor stability. ANSYS results show that equivalent conductor stability is achieved when both types of CMS aluminium are replaced with a single aluminium that has one third the electrical and one third the thermal conductivity of the high purity aluminium.

### 6.5.3 DID conductor

The dipole coils are to be wound from a high purity aluminium and a CMS single superconducting strand co-extrusion. Two layers of 75 turns of 2.5 mm $\times$ 1.8 mm superconductor per winding are proposed. There will be 0.5 mm of fibreglass cloth between each turn and each layer. 800 A at 30% of $I_{\mathrm{critical}}$ is the operating point. The stored energy for an independently powered DID is in the range of 240 kJ. When coupled to the solenoidal field, the stored energy increases by $\approx$ 8 MJ. Because the stored energy is so small, the volume fraction of high purity aluminium to superconductor needed for safe energy extraction during a quench has been reduced from the CMS 12.4 to a ratio of 2.5. Forces on each of the four coils are rather large in sum but spread somewhat uniformly and are manageable (4100 kN radial and 7800 kN axial).

## 6.6 Electrical

### 6.6.1 Magnet Safety

The lower stored energy and inductance of the SiD magnet compared to the CMS case make it easier to protect in case of a quench. A conservative 300 V to centred tapped ground is chosen. Experimental tests and computer simulations show that the CMS quench propagation velocity around one complete turn is faster than the turn to turn quench propagation through the insulation. Because we have chosen the same conductor size and insulation thickness as CMS, with very similar electrical and thermal properties, peak temperatures will be less than the 80 K at CMS with dump resistor, but equal to the 130 K at CMS in case of a dump breaker failure. Both SiD and CMS safety rely on the winding mandrel serving as a quench back cylinder spreading the quench over the outer layer and absorbing some of the stored energy. Fast discharge of the DID as a solenoid quench propagator to reduce winding peak temperature and stress is a reasonable option. However, the detailed transient 3D ANSYS calculations remain to be done.

### 6.6.2 Power Supply, Dump Resistor and Dump Switch

The power supply, contactor, dump resistor and dump switch are attached on the side of the detector near the top. These three components are arranged to minimise the 18 kA bus lines. Power supply, DC contactor and mechanical dump switch are standard design components procured from outside vendors. The DC contactor allows for normal slow mode discharge and fast discharge. The power supply operates in only one quadrant, positive current and positive voltage. Therefore, more robust free wheeling diodes can be employed instead of the thyristors used at CMS and which permitted voltage control ramp down. The SiD magnet does not have a current reversal switch. The dump switch is a conventional commercially purchased double pole mechanical breaker with arc chutes. Both the positive and negative legs are mechanically ganged together ensuring that they open at the same time. The breaker and controls are housed in a steel box 2.6 m $\times$ 0.9 m $\times$ 1.5 m (high). The power supply is a standard water cooled power supply tailored to low inductance operation. Overall dimensions of this unit are 3.7 m $\times$ 1.0 m $\times$ 2.0 m (high).

A novel compact pressurised water cooled dump resistor will be used instead of a very large air cooled dump resistor such as the type used for CMS and other large superconducting magnets. An ASME coded vessel holding 3100 litre of water will rise to a conservative design value of 150 C at 0.48 MPa assuming the worse case of all 1.56 GJ deposited as sensible heat in the water of the resistor [146, 147]. Correct dimensioning of the stainless steel resistor element ensures that boiling heat transfer is only a third of the peak nucleate boiling flux at the metal/water interface. A 1.50 m diameter $\times$ 3.5 m tall cylindrical tank could be used. Internal connections will provide for both fast dump and normal slow dump modes. A centre tap grounding wire is attached to the electrical centre of the resistor.

# Chapter 7
# SiD Engineering, Integration and the Machine Detector Interface

## 7.1 Introduction

The ILC Reference Design Report (RDR) [148] was based on a site presumed to run ∼100 m below a topographically flat landscape. It specified the civil engineering parameters of a shared underground interaction region (IR) Hall accessed by two shafts symmetrically located around the beam line. More recently:

- A set of functional requirements for the design of the detectors and the IR was defined [141].
- The SiD and ILD detector concepts were validated.
- A platform similar to the CMS shaft plug was agreed to be the means of effecting the push-pull exchange of the detectors.
- A new cavern layout was designed featuring one shared 18 m diameter central shaft directly over the interaction point, serviced by a 4000 tonne gantry crane, separate assembly areas accessible to the sliding platform and separate garage areas for major detector component replacement, each serviced by an 8 m equipment shaft and a 5.6 m personnel elevator shaft. See Figure II-7.1.
- The possibility has arisen that the ILC would be built in a mountainous site where the IR would be accessed by a tunnel of limited diameter of length of order 1 km.

These features are described in more detail in the first part of this Volume (see 2.3) and in Volume 2 (Accelerator) of this TDR [149].

## 7.2 IR Hall Layout Requirements and SiD Assembly

The main subcomponents of SiD are its central barrel and its two endcaps. The majority of the SiD mass results from the flux return iron. The iron will be shipped to the ILC site from an industrial production facility in the form of sub-modules suitably sized (∼100 tonne) for road transportation. The solenoid coil will likewise be wound industrially and transported in sections, probably two, amenable to transport.

We expect the VXD, Tracker, ECAL, HCAL and muon system modules to be built at collaborating labs and universities and transported to the ILC site for final assembly. Table II-7.1 lists the masses and sizes of the SiD elements that determine the crane capacity and shaft size for installation.

**Table II-7.1** List of SiD element masses and sizes. For each barrel component the size given is the outer diameter × length ($z$), and for each endcap component it is length × outer diameter.

| Name | Mass ($10^3$ kg) | # Subcomponents | Mass ($10^3$ kg) | Size (m×m) |
|---|---|---|---|---|
| Barrel | 4160 | | | |
| ECAL | 60 | 12 | 5.0 | 2.8 × 3.5 |
| HCAL | 367 | 12 | 31.7 | 5 × 5.9 |
| Tracker | 3 | 1 | 3 | 2.5 × 3.3 |
| Coil | 180 | 2 | 90 | 6.8 × 5.9 |
| Magnet Yoke | 3360 | 8 | 420 | 12 × 5.9 |
| Yoke Arch Supports | 150 | 2 | 75 | 12 × 1 |
| Peripherals | 40 | | | |
| Each of Two Endcaps | 2450 | | | |
| ECAL | 10 | 1 | 10 | 0.15 × 2.5 |
| HCAL | 23 | 1 | 23 | 1.2 × 2.8 |
| Muon System | 30 | | | 2.6 × 12 |
| MDI Components | 10 | | | |
| Endcap Steel Plates | 2200 | 11 | 200 | 0.2 × 12 |
| Endcap Leg Supports | 140 | 2 | 70 | 2.6 × 6 |
| Infrastructure | 37 | | | |

## 7.2.1 Vertical Access (Europe and Americas sites)

Figure II-7.1 shows the layout of the IR Hall. This allows the 3 m thick SiD push-pull platform to be positioned directly under the gantry crane.

The service caverns allow for storage of the endcaps and unimpeded access to the barrel region for the initial installation or replacement of detector subcomponents. Access to the service caverns is through an 8 m diameter shaft serviced by a 40 tonne crane.

The vertical access assembly presumes that the SiD magnet, comprising the superconducting coil, iron barrel yoke and iron endcaps, will be pre-assembled and tested in an assembly hall above ground. Any detector subcomponents, notably the HCAL and ECAL, that are ready in time can be installed and tested above ground. Then SiD's three main subcomponents, the majority of the barrel and the two endcaps, will each be lowered as units onto the platform below.

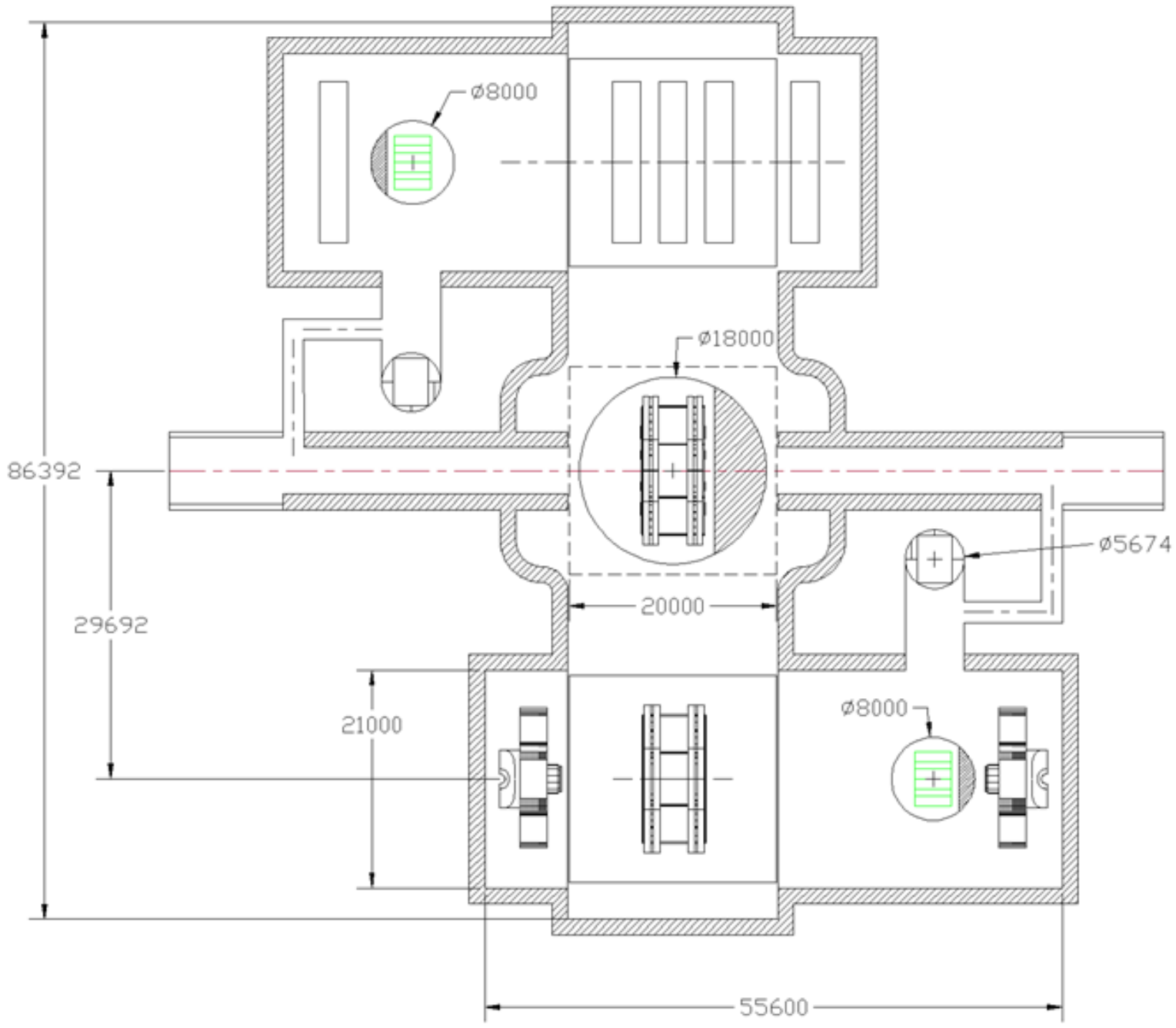

**Figure II-7.1** Layout of the IR Hall for vertical access, showing installation shafts and push-pull platforms.

The basic requirements for the assembly hall above ground are:

- A devoted crane with a minimum of 215 tonne main hook capacity, set by the largest subcomponent weight. The ILD and SiD cranes should roll on the same bridgework so that they can be used in tandem if the need arises.

- A steel reinforced concrete platform, upon which SiD will be assembled, which is structurally robust when supported on three sides as it slides over the 18 m diameter main access shaft to deliver the SiD barrel and endcaps to the gantry crane.
- A circa 4000 tonne capacity gantry that can lower the roughly 15 m $\times$ 5 m $\times$ 6 m 3500 tonne instrumented SiD Barrel and the two 11 m $\times$ 14 m $\times$ 6 m 2500 tonne endcaps onto the push-pull platform in the IR Hall.

It is foreseen that the surface assembly hall is aligned with its long axis parallel to the beam line. The construction platform will move in this direction as well. Its width is 20 m, approximately the width of the building, while the length will be large enough to comfortably house the barrel and the endcaps when open. The platform surface will be at floor grade and thus run in a track. The endcaps will move across the platform-floor junction on the rollers when required to mate with the barrel.

The SiD barrel, once lowered, will remain stationary on its platform. The endcaps, which must be routinely opened to service the detector, will move on a system of rollers guided by hardened rails. The current plan is to lower the endcaps first and to put them in their service caverns to await mating with the barrel. Once the barrel and endcaps have been lowered the main shaft and gantry crane are no longer needed.

The above-ground assembly sequence for a vertical access site can also be used for a horizontal access subterranean site. In the latter case the individual subcomponents are separately transported through an access tunnel of limited diameter to the IR Hall, where a 215 tonne bridge crane suffices for installation. In either case a plausible assembly sequence is:

- Assemble the two endcap leg supports on top of the platform.
- Transport each of the eleven 200 tonne endcap plates in three industrially manufactured segments to the crane and assemble into 11 m x 11 m octagonal plates. Mount each on the support legs and make plate to plate connections.
- Install muon chambers from the sides into each gap, and the endcap HCAL and ECAL to the innermost face.
- Assemble detector mounted PACMAN shielding on the endcaps.
- Once endcaps are completed move them to their alcoves.
- Assemble lower halves of barrel arch supports.
- Assemble industrially manufactured $\sim$100 tonne barrel steel stacked plate segments into sixteen $\sim$210 tonne half-wedges and use the crane to assemble the five lower barrel wedges, forming a cradle open at the top.
- Assemble the solenoid coil segments and DID coils into their cryostat and test at low current. Lift coil with fixture and thread into the cradle.
- Finish the remaining three barrel wedges, install muon system and finish with shear plates at each wedge-to-wedge junction.
- Thread solenoid with an assembly beam and mount the HCAL assembly spider onto it. Load each of the twelve 32 tonne HCAL wedges onto the spider and push into barrel on rollers.
- Repeat HCAL sequence with the much lighter ECAL.
- Thread in Tracker and VXD units when available.

The QD0 assembly (QD0, masks, FCAL) will need to be installed below ground. The platform will transport the endcaps to the alcove area, whence the assembly will be loaded from the rear.

### 7.2.2 Horizontal Access (Japan sites)

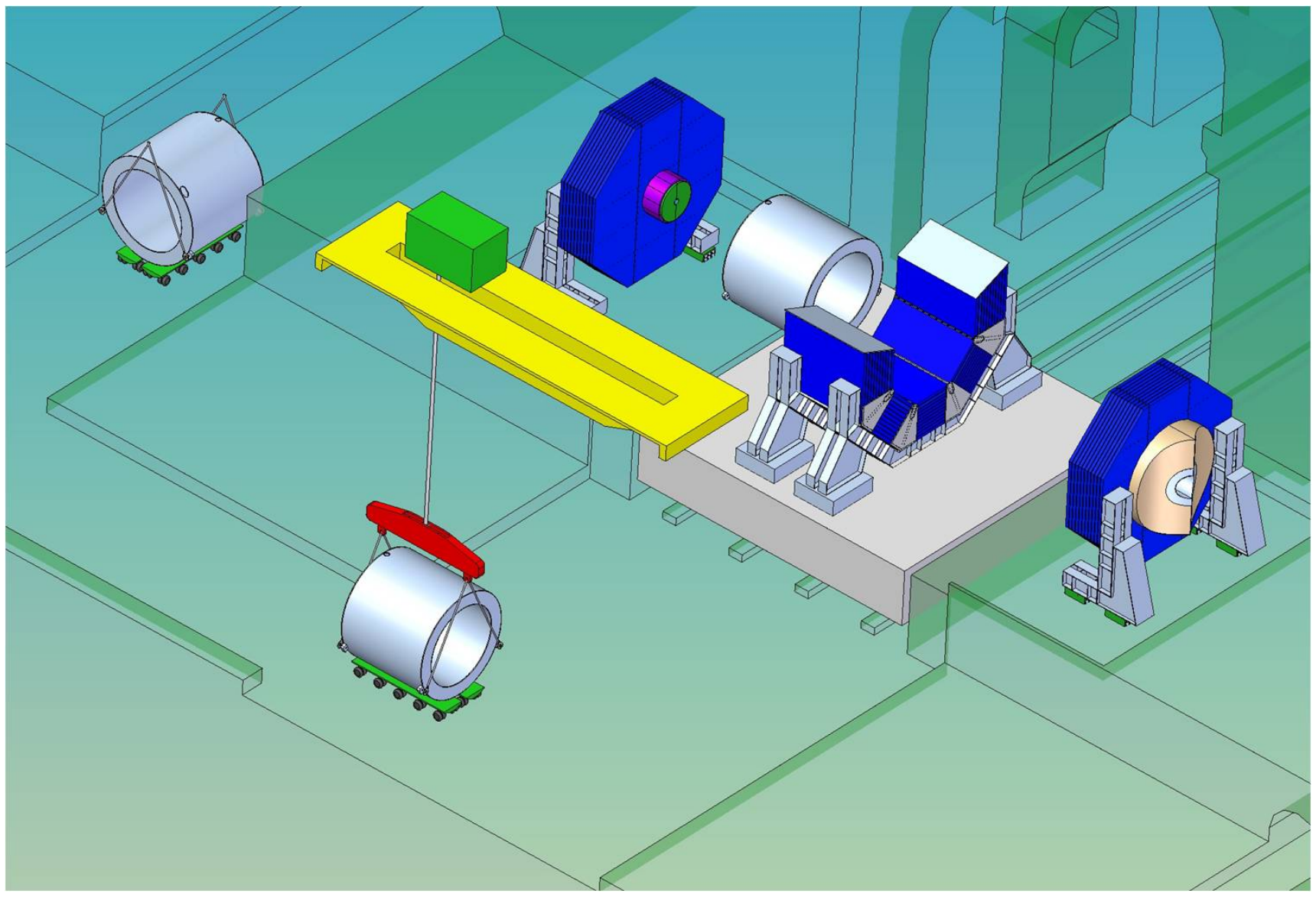

**Figure II-7.2** Transporting the largest detector element, the SiD solenoid, through the 11 m diameter access tunnel to the assembly area where the 215 tonne crane can lift it and place it within the SiD barrel.

The barrel and endcap installation procedures outlined above are directly applicable. One need only plan for the lengthier procedure of loading the heavy sub-elements onto the tunnel transport carts and their delivery to the IR Hall assembly area. The Japan site design specifies an 11 m diameter tunnel, which is sufficient to transport the largest element of SiD, its solenoid. Figure II-7.2 shows the SiD solenoid being transported around the final right-angle bend to the IR Hall where it is lifted by the 215 tonne crane and placed in the cradle formed by the lower elements of the SiD barrel yoke. Clearly, if the ILC schedule permits below-ground assembly of the detectors for the vertical access site, the diameter of the access shaft could be reduced from 18 m to 11 m.

### 7.2.3 Detector Access for Repairs

The upper part of Figure II-7.3 shows SiD with one of its endcaps opened by 2 m, sufficient to expose the FCAL region and the Tracker. This is the basic configuration for quick repair opportunities that may occur while SiD is on the beamline. In the lower part of Figure II-7.3 the endcap has been opened by 2.8 m, the maximum possible for SiD without having to disconnect the QD0 cryostat.

In this figure, the Tracker has been slid to one side to expose the VXD, a manoeuvre that would require the use of some portion of the tracker installation tooling. As such, it would probably be scheduled for a time when SiD is off the beamline. Repairs more major than replacement of a VXD module, such as replacement of the Tracker, barrel ECAL or barrel HCAL, will take place off the beamline.

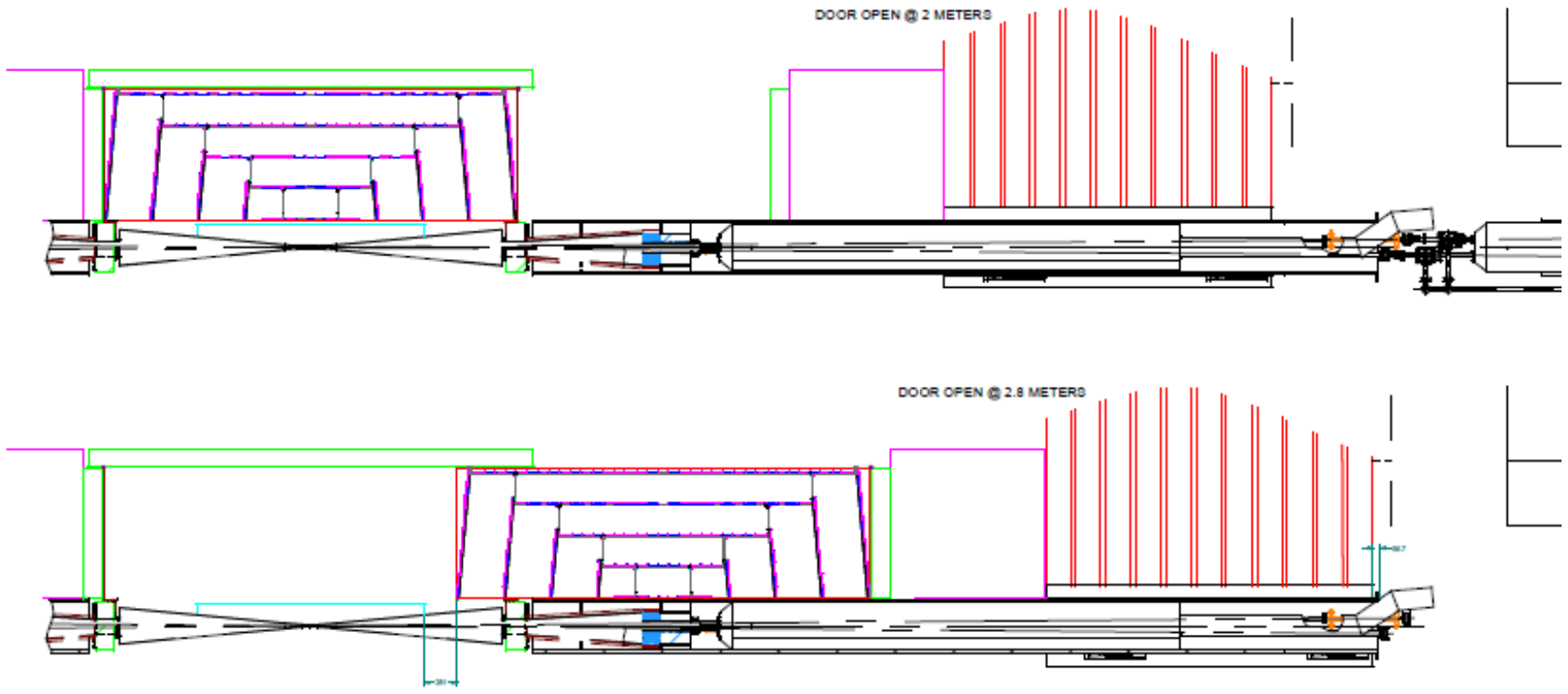

**Figure II-7.3** Upper: SiD with one of its endcaps opened by 2 m, sufficient to expose the FCAL region and the Silicon Tracker. Lower: the endcap has been opened by the maximum 2.8 m and the tracker has been slid to one side to expose the VXD for repair or replacement.

## 7.3 Detector Exchange Via a Sliding Platform

### 7.3.1 Introduction

Among the challenges to be addressed in the push-pull operation are the reproducibility of tight alignment to the beam at the $\pm 1$ mm level; the time requested to complete the swap cycle must be as low as reasonably achievable since it will reduce the integrated luminosity; umbilicals are needed to keep the detector connected to the DAQ and services such as cryogenics.

### 7.3.2 Platform

These requirements have been addressed by developing the concept of a reinforced concrete platform $20 \times 20$ m$^2$ in area and 3.8 m tall, with a total mass of $\sim$4500 tonne. To compensate for the different detector heights the SiD platform is thicker than that of ILD. Assuming the total SiD mass to be 9000 tonne, preliminary calculations have shown [150] that the maximum static deformation achievable is less than 1mm at the locations where the detector is supported by the platform. The construction will be very similar to the concrete slab designed for the CMS detector [151].

The movement system is designed for $\sim$14,000 tonne total mass of detector and platform. Two options are under consideration, air pads and Hillman rollers, each with hydraulic jacks above. For the air pads the expected friction is 1% and the total force required for the horizontal motion is 140 tonne. Assuming a maximum load capacity of 350 tonne for a single air pad, SiD will require the installation of 40 units under the platform. For the rollers the friction will be $\sim$3% and the force required for the horizontal motion will be $\sim$420 tonne, while only 14 units with 1000 tonne load capacity will be required. In both cases, the floor will need to be hardened with steel to prevent wear that would spoil the alignment performance.

A reliable linear guiding system built in the floor is also essential for air pads as well as for rollers. The force required in both cases for the horizontal motion can be comfortably provided by a set of hydraulic climbing jacks. Another set of hydraulic jacks will be placed at the beamline location of the platform to correct the final transverse alignment, if needed.

### 7.3.3 Vibration analysis and Luminosity Preservation

A structural dynamic model of the QD0 supported from SiD, including the platform, has been developed to calculate the free modes as well as the transfer function between the ground and the doublet. Using different ground vibration models available in the literature, that correspond to different accelerator sites, a maximum r.m.s. QD0 displacement of 20 nm has been calculated, more than a factor two below the maximum allowed. A campaign of experimental measurements of vibrations has been carried out to validate some key features of the model: the simulation of the reinforced concrete platform and correlation measurements between distant locations in the detector hall of CMS at CERN and SLD at SLAC. The reinforced concrete slab of CMS has been instrumented with geophones in various locations and the data have been used to benchmark a finite element model of the platform [152].

Good agreement between experimental data and simulation has been found with an internal damping ratio of 6.5%, somewhat higher than the values recommended for similar materials. The difference can be explained by the soil deformation and the presence of wheels, both of which were not included in the model. The measurements done at CMS and SLD have shown a good correlation at low frequencies between points at the two extreme sides of the cavern, i.e. the location of the final focus system [153].

### 7.3.4 Push Pull Detector Exchange Process and Time Estimate

The sequence of push-pull operations should allow a fast detector interchange to minimise loss of beamtime; realistically it should not take more than a few days. Defining as $t = 0$ the time when the beams have been dumped and the interlocks are released to allow the access of the technical personnel, the key steps are the opening of the PACMAN shielding, the breaking of the vacuum between the QD0 and the QF1, a reasonably fast horizontal movement from the IP to the garage position with an easy and reliable alignment system. The cryogenic system will stay on during the push-pull, with the umbilical able to accommodate the ~30 m movement requested. Figure II-7.4 summarises the sequence of steps and the minimum required time for the push-pull operation.

**Figure II-7.4** Summary chart of push-pull operational steps.

| Task | Duration (hour) | 8-hour shift | 8-hour shift | 8-hour shift | 8-hour shift |
|---|---|---|---|---|---|
| Secure ILC beams | 1 | | | | |
| Ramp magnets down | 3 | | | | |
| Open beamline shielding | 1 | | | | |
| Disconnect beamlines | 2 | | | | |
| Checkout detector transport system | 2 | | | | |
| Transport detector over 20 m | 2 | | | | |
| Transport other detector onto beamline | 2 | | | | |
| Connect beamline | 2 | | | | |
| Close beamline shileding | 1 | | | | |
| Check crude detector alignment and adjust | 2 | | | | |
| Ramp magnets up | 3 | | | | |
| Perform safety checks before beams | 1 | | | | |
| Start beam-based alignment | 10 | | | | |

## 7.4 Beampipe and Forward Region Design

### 7.4.1 Introduction to the Near Beamline Design

The SiD near-beamline design minimises the radial space required for the support and alignment of the final quadrupole lens QD0 to limit any loss of tracking and calorimeter acceptance. In the SiD design the silicon tracker slides over the QD0 support to expose the vertex detector for servicing (see Figure II-7.3).

### 7.4.2 Beampipe

The beampipe through the central portion of the vertex detector has been taken to be all-beryllium. Within the barrel region of the vertex detector, the beryllium beampipe forms a straight cylinder with inner radius of 1.2 cm and a wall thickness of 0.04 cm. At z = ±6.25 cm, a transition is made to a conical beampipe with a wall thickness of 0.07 cm. The half angle of the cone is 3.266°. Transitions from beryllium to stainless steel are made beyond the four inner vertex disks, at approximately z = ±20.5 cm. The initial stainless steel wall thickness is 0.107 cm; it increases to 0.15 cm at approximately z = ±120 cm. The half angle of the stainless steel cone is 5.329°. The inner profile of the beampipe is dictated by the need to avoid the envelope of $e^+e^-$ pairs from beamstrahlung.

### 7.4.3 LumiCal, BeamCal, Mask and QD0 Support and Alignment

The QD0 support tube (Figure II-7.5) is extended toward the IP to support the 220 kg LumiCal, the 507 kg 3 cm thick conical tungsten mask, the lightweight 13 cm thick 25 cm diameter borated polyethylene neutron absorber and the 136 kg BeamCal. The low-$z$ end of the support tube will be split along its centreline so that it can be opened to install the mask, absorber and BeamCal. The LumiCal will be bolted to the front end of the tube and be positioned so that it hangs 10 cm in front of the endcap ECAL when the detector is closed. While this choice complicates the vertex detector support system, it minimises any loss of acceptance between the LumiCal and the ECAL endcap. The loading of the support tube results in a deflection of 100 $\mu$m when the detector is closed, growing to 2.2 mm when the endcap is opened the nominal 2 m required to service the detector when on beamline, and 6 mm when the endcap is opened the maximum 2.8 mm allowed by the location of QF1 and the obstruction of the cryo-transfer line joining QD0 to its local 2 K refrigerator. A wedge mover system will need to act in conjunction with the endcap opening mechanism to keep the front end of the LumiCal fixed in space.

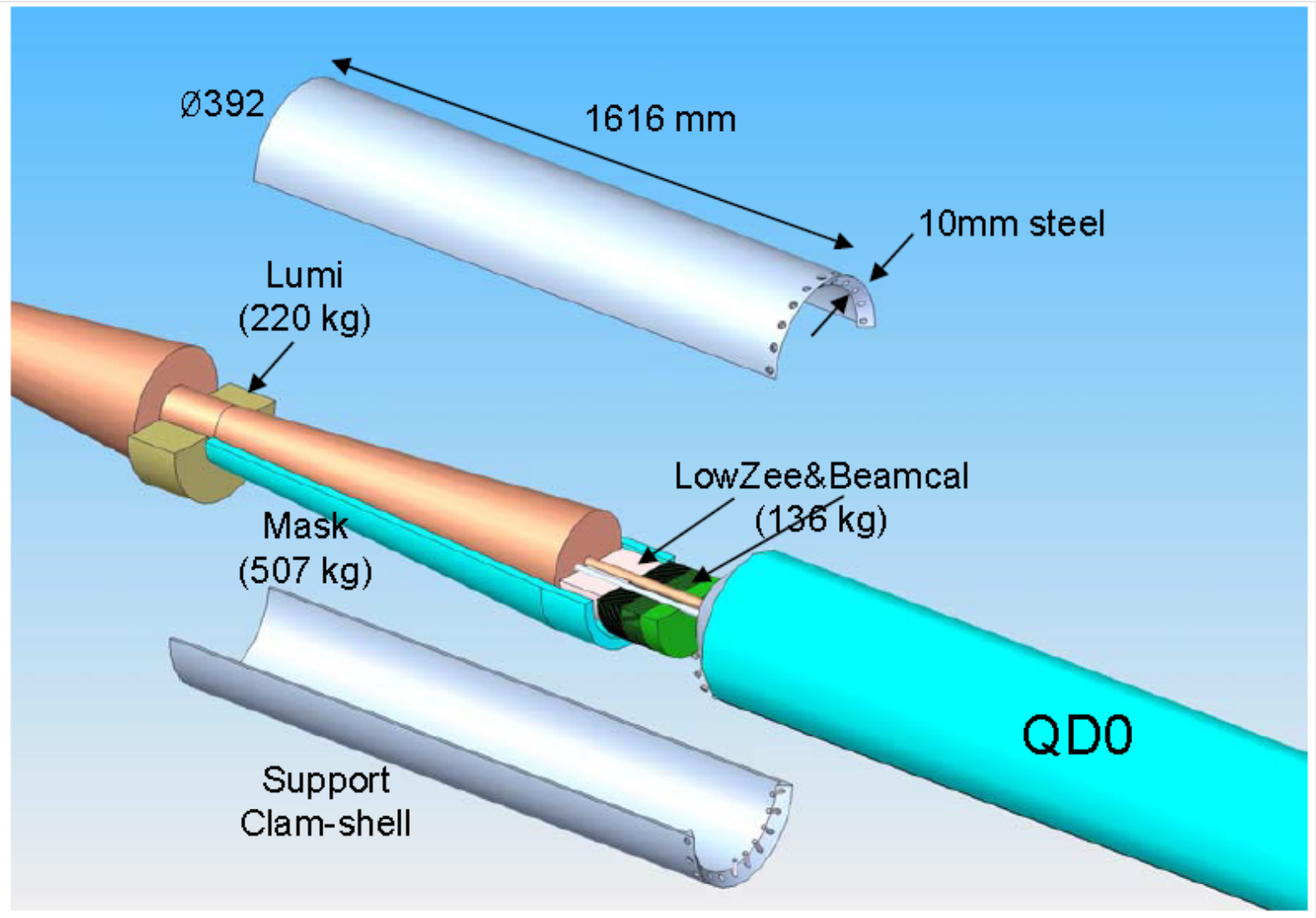

**Figure II-7.5** Detail of the LumiCal, mask and BeamCal which must be supported by the QD0 support tube and alignment system.

The beampipe through BeamCal terminates in a commercial flange. The conversion of the common beampipe to separate incoming and outgoing beampipe takes place in the 215 mm space between the back of BeamCal and the front face of the QD0 cryostat at 3.283 m from the IP.

### 7.4.4 Vacuum System and Performance

The vacuum requirements for the final focus and interaction region lengths of the beampipe have been specified [154] as 10 nTorr from 200-800 m from the IP, 1 nTorr in the region from 200 m up to the QD0 quadrupole and "much looser" than 1 nTorr between the QD0 cryostats. The region between QD0 and QF1 is evacuated to $< 10$ nTorr by the pumping action of the two cryostats and the 100 l/s ion pump on each beam line in front of QF1. Achieving 1 nTorr upbeam of QF1 will be a challenge. With a 20 mm diameter stainless beampipe and 50 l/s ion pumps every 2 m, the average pressure is 23 nTorr. Likely, either distributed pumping (antechamber, pumpscreens or NEG coatings) and/or larger diameter beampipes with bakeout facilities will be required to meet the 1 nTorr tolerance.

### 7.4.5 Feedback and BPMs

The intratrain feedback system is based on that described in the RDR [148]. A prototype system has been developed and tested with beam at ATF [155]. The parameters of the BPM and kicker required for ILC have been specified [156]. By combining a ground motion model with a set of transfer functions describing the vibrational effect of the magnet support system, in this case the SiD platform and detector, the reduction of luminosity loss can be studied [157, 158].

The left side of Figure II-7.6 shows the fractional loss of nominal luminosity as a function of the rms jitter of the opposing SD0/QD0 magnet systems when they are supported from SiD and the SF1/QF1 magnets are, like all the other magnets in the final focus, assumed to be rigidly attached to the ground. The feedback system limits the luminosity loss to 2% (4%) of the nominal value for rms motions up to 50 nm (200 nm), ~10 (~40) times the vertical spot size of beam at the IP.

The right side of Figure II-7.6 shows the contribution of mechanical jitter to the total jitter in the case where the ground motion model is that of the noisiest site studied (DESY near Hamburg). Even in this extreme case, the feedback system would limit luminosity loss to 2% with up to 17 nm of additional mechanical jitter, coming from, for example, vibrations induced by the liquefier or flow of liquid helium. A modelling program to ensure that the ground to magnet transfer function is correct is ongoing [159].

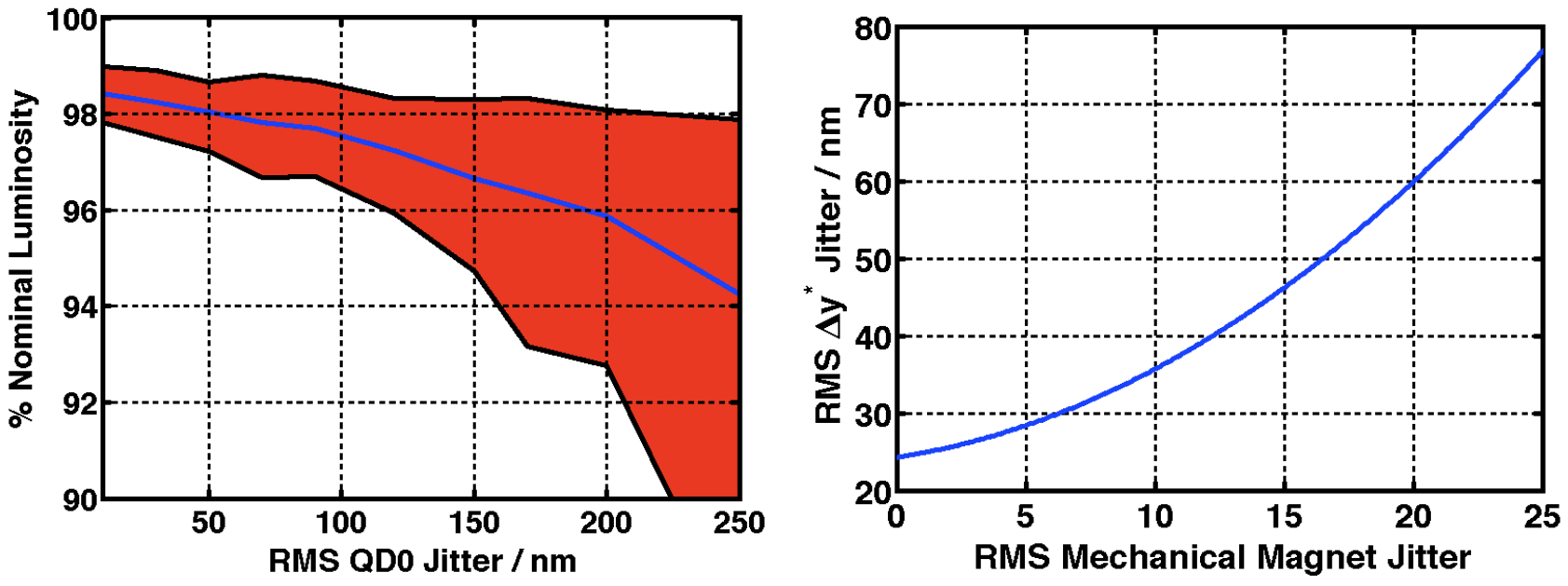

**Figure II-7.6** The fractional loss of nominal luminosity as a function of the rms $x$ and $y$ vibration of the SF1/QF1 and SD0/QD0 magnet systems (left). Contribution of mechanical jitter to overall vibration budget (right)

### 7.4.6 Frequency Scanning Interferometric (FSI) Alignment of QD0 and QF1

The FSI system incorporates multiple interferometers fed by optical fibres from the same laser sources, where the laser frequency is scanned and fringes counted, to obtain a set of absolute lengths [101, 103, 160].

To monitor the position of the QD0 cryostat to the required accuracy [141] of 50 $\mu\mathrm{m}$ in $x$, $y$, and 20 $\mu$rad in roll, pitch and yaw a network of "optical trusses" between beam launchers at known positions and reflectors placed on the QD0 cryostat is needed. Simulations [161], conservatively assuming 500 nm length accuracy, indicate that a network of four beam launchers placed on the front face of the QF1 cryostat, each of which sends a split beam to two of four similarly situated reflectors on the back end of the QD0 cryostat, and a similar network tying the inner edge of the innermost HCAL endcap to the front end of the QD0 cryostat, can achieve an precision of $\sim$1 $\mu\mathrm{m}$ in $x$ and $y$ and $\sim$1 $\mu$rad in all axis rotations. Schemes that can tie this network across the IP are important to develop.

# Chapter 8
# SiD Forward Systems

## 8.1 Forward Detector

The forward region is defined as polar angles $|\cos\theta| > 0.99$ ($\theta <$ 140 mrad), which is the angular region forward of the coverage of the SiD Endcap ECAL. The angular coverage is completed by two detectors, the Luminosity Calorimeter (LumiCal) and the Beam Calorimeter (BeamCal). As discussed in more detail below, the LumiCal is an annular calorimeter located approximately 1.6 m from the interaction point (IP), subtending angles between 40 mrad and 90 mrad. The BeamCal, the most forward of all the SiD subsystems, lies at a distance of approximately 2.8 m from the IP, subtending angles between 3 mrad and 40 mrad.

The instrumentation goals in this region are:

- Measurement of the integrated luminosity using small-angle Bhabha scattering (LumiCal) to an accuracy better than $10^{-3}$;
- Instantaneous luminosity measurement using beamstrahlung pairs (BeamCal);
- Extension of the calorimeter hermeticity into the small angles for physics searches (LumiCal and BeamCal);
- Provide a two-photon veto for new particle searches (BeamCal);
- Possible contribution to the determination of the luminosity spectrum by measuring the acolinearity angle of Bhabha scattering (LumiCal).

The detector challenges include good energy resolution, radiation hardness, interfacing with the final focus elements, high occupancy rate requiring special readout, and performing the physics measurements in the presence of the very high background in the forward direction (see Chapter 11.3.1).

### 8.1.1 Design criteria

#### 8.1.1.1 LumiCal Physics Requirements

The number of Bhabha events per bunch crossing for a detector with minimum and maximum polar angle coverage $\theta_{min}$ and $\theta_{max}$ (in mrad) is:

$$N = 0.5\mathrm{pb}\frac{\mathrm{L}}{\mathrm{R}}\int_{\theta_{\mathrm{min}}}^{\theta_{\mathrm{max}}}\frac{\mathrm{dcos}\theta}{\sin^4\theta/2} \sim 6\times10^{-6}\left(\frac{1}{\theta_{\mathrm{min}}^2}-\frac{1}{\theta_{\mathrm{max}}^2}\right)$$

for $\sqrt{s}$ =0.5 TeV, L=2$\times10^{34}\mathrm{cm}^{-2}\mathrm{s}^{-1}$, and bunch crossing rate R=$1.4\times10^4\mathrm{s}^{-1}$. Our goal is to measure the luminosity normalisation with an accuracy of several $10^{-4}$ for $\sqrt{s}$ =0.5 TeV. To do this one needs $\approx 10^8$ events collected over $\approx 10^7$ s, or about ten events per second. One can then calculate the absolute luminosity with $\approx 10\%$ statistical error every several minutes during the run. With a bunch crossing rate of $1.4\times10^4\mathrm{s}^{-1}$, we need $> 10^{-3}$ events per bunch crossing. To achieve this statistical accuracy, we start the fiducial region for the precision luminosity measurement well away from the beamstrahlung pair edge at $\theta$=20 mrad, with a fiducial region beginning at $\theta_{min}$=46 mrad, which gives $\approx 2 \times 10^{-3}$ events per bunch crossing.

#### 8.1.1.2 Luminosity precision and detector alignment

Since the Bhabha cross section is $\sigma \sim 1/\theta^3$, the luminosity precision can be expressed as

$$\frac{\Delta L}{L} = \frac{2\Delta\theta}{\theta_{min}},$$

where $\Delta\theta$ is a systematic error (bias) in polar angle measurement and $\theta_{min} = 46$ mrad is the minimum polar angle of the fiducial region. Because of the steep angular dependence, the precision of the minimum polar angle measurement determines the luminosity precision. To reach the luminosity precision goal of $10^{-3}$, the polar angle must be measured with a precision $\Delta\theta <$ 0.02 mrad and the radial positions of the sensors must be controlled within 30 $\mu$m relative to the IP.

#### 8.1.1.3 Monitoring the Instantaneous Luminosity with BeamCal

The colliding electron and positron bunches at the ILC generate large Lorentz forces, which cause radiation of gammas called beamstrahlung. Under the ILC Nominal beam parameters at $\sqrt{s} =$ 0.5 TeV, approximately 75k of the beamstrahlung photons convert into $\mathrm{e^+e^-}$ pairs. Since the number of pairs is directly proportional to the beam overlap, the instantaneous luminosity can be monitored to $\approx$10% per beam crossing by detecting pairs in the BeamCal.

#### 8.1.1.4 Dynamic range and MIP sensitivity

While minimum ionising particles (MIP) deposit 93 keV in a 320 $\mu$m-thick Si layer, a 250 GeV electron can deposit up to 160 MeV or 1700 MIP equivalents in a single cell near shower maximum. If we want a 100% MIP sensitivity, the S/N ratio for MIP should be greater than 10, and the dynamic range of the electronics needs to be at least 17,000. This dynamic range can be achieved by using a 10-bit ADC with two gain settings.

#### 8.1.1.5 Radiation hardness

The beamstrahlung pairs will hit the BeamCal, depositing 10 TeV of energy every bunch crossing. Sensor electronics could be damaged by the energy deposition, and sensor displacement damage could be caused by the resulting neutrons. The radiation dose varies significantly with radius, and a maximum dose of up to 100 MRad/year is expected near the beampipe. The main source of neutrons is from secondary photons in the energy range 5-30 MeV, which excite the giant nuclear dipole resonance, with subsequent de-excitation via the evaporation of neutrons. The neutron flux is approximately $5 \times 10^{13} \mathrm{n/cm^2}$ per year.

### 8.1.2 Baseline Design

The layout of the forward region is illustrated in Figure II-8.1. The LumiCal covers the polar angles from 40 mrad to 90 mrad, and the BeamCal from 3 mrad to 40 mrad.

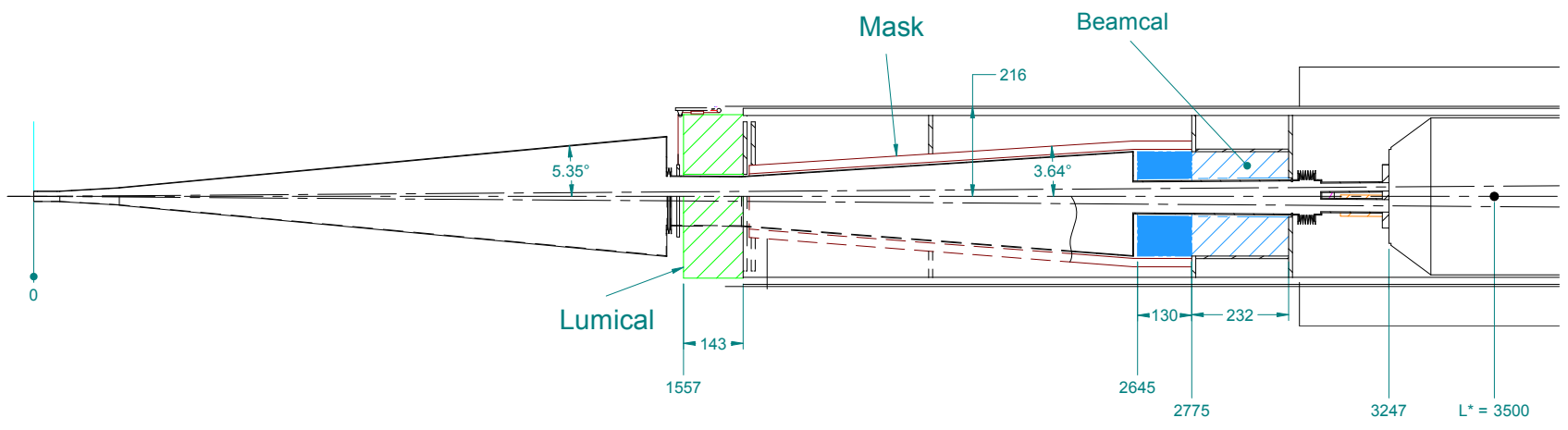

**Figure II-8.1** The SiD forward region.

The LumiCal consists of two cylindrical C-shaped modules surrounding the beampipe. The inner radius is 6 cm centred on the outgoing beam line with a horizontal offset of $\Delta x = 1.1$ cm (158 cm $\times$ 0.007). The inner radius is dictated by the requirement that no detector intercepts the intense beamstrahlung pairs, which are confined within 4 cm radius by the 5 Tesla solenoid field.

The longitudinal structure follows the ECAL design, consisting of 30 alternating layers of tungsten and silicon. The first 20 layers of tungsten each have a thickness equivalent to 2.5 mm (or 5/7 radiation length) of pure tungsten. The last 10 layers have twice this thickness, making a total depth of about 29 radiation length. The sensor is segmented with a $R - \phi$ geometry; a fine radial segmentation with 2.5 mm pitch is used to reach the luminosity precision goal of $10^{-3}$. The azimuthal division is 36 with each sensor covering 10 degrees. Table II-8.1 summarises the LumiCal parameters as well as those for the BeamCal, the description of which follows.

**Table II-8.1**
Forward Calorimeter Parameters

| Parameter | LumiCal | BeamCal |
|---|---|---|
| z Extent | 155.7 – 170.0 cm | 277.5 – 300.7 cm |
| Inner radius | 6.0 cm | 2.0 cm |
| Outer Radius | 20.0 cm | 13.5 cm |
| Instrumented | 42 – 110 mrad | 5 – 45 mrad |
| Fiducial | 46 – 86 mrad | — |
| Tungsten thickness | 2.5/5.0 mm (20/10 layers) | 2.5 mm |
| Sensor thickness | 320 $\mu$m | 320 $\mu$m |
| Radial division | 2.5 mm | 5.0 mm (2.5 mm $R > 7.5$ cm) |
| Azimuthal division | 36 segments | 5.0 mm |

The BeamCal consists of two cylindrical C-shaped modules split in half horizontally to accommodate the incoming and outgoing beam lines. The inner radius is 2 cm, centred on the outgoing beam line, and the outer radius is 13.5 cm. A second hole, of radius approximately 1 cm and displaced from the centre by approximately 5 cm, allows for the incoming beam line. The longitudinal structure consists of 50 alternating layers of tungsten and silicon. The tungsten thickness is 2.5 mm, making a total depth of 36 radiation lengths.

The inner region, at a radius of $R < 7.5$ cm, has a high signal rate from beamstrahlung pairs. The segmentation in this region is approximately $5\text{mm} \times 5\text{mm}$, which is roughly one half of the Molière radius. This segmentation is optimised so that tell-tale electrons or positrons from two-photon processes can be detected in the high beamstrahlung pair background. The outer region $R > 7.5$ cm is treated as a "far LumiCal" and has the same geometrical segmentation as the LumiCal.

Currently two electronic readout chips are being developed. The KPiX chip with 1024 channels is designed primarily for the ECAL. The chip has four hits per bunch train to be stored for each channel. The FCAL chip with 64 channels is designed to handle the 100% occupancy in the BeamCal. The chip has 2820 buffer space so that a complete bunch train can be stored.

Although the LumiCal occupancy is not 100%, the LumiCal region smaller than about 10 cm will have more than four hits per bunch train. Therefore, the LumiCal is foreseen to use the FCAL chip in the inner region and the KPiX chip in the outer region.

### 8.1.3 Forward Systems Development Work

In this section we present the recent developments on the forward systems that have been carried out in the framework of the SiD collaboration. These developments are a component of the overall R&D effort for linear collider forward systems.

**Table II-8.2**
BeamCal instrumentation ASIC specifications summary. Note that these prototype chip specifications are based on a now-outdated version of machine parameters; the next prototype will address the change.

| | |
|---|---|
| Input rate | 3.25 MHz during 0.87 ms, repeated every 200 ms |
| Channels per ASIC | 32 |
| Occupancy | 100% |
| Resolution | 10 bits for individual channels, 8 bits for fast feedback |
| Modes of operation | Standard data taking (SDT), Detector Calibration (DCal) |
| Input signals | 37 pC in SDT, 0.74 pC in DCal |
| Input capacitance | 40 pF (20-pF detectors and 20-pF wires) |
| Additional feature | Low-latency (1 μs) output |
| Additional feature | Internal pulser for electronics calibration |
| Radiation tolerance | 1 Mrad ($SiO_2$) total ionising dose |
| Power consumption | 2.19 mW per channel |
| Total ASIC count | 2,836 |

## 8.1.3.1 FCAL Electronics Development

The initial set of specifications for the BeamCal instrumentation ASIC is listed in Table II-8.2.

The Bean (BeamCal Instrumentation IC) prototype is a custom IC designed in a 180-nm CMOS process as a proof-of-concept to fulfil the BeamCal instrumentation specifications. The Bean block diagram, shown in Figure II-8.2, depicts the three channels of the prototype ASIC, as well as the adder that combines the outputs of all channels to provide a fast feedback signal. Each channel has a dual-gain charge amplifier, a precharger and calibration circuit, a filter, connecting buffers, and a dedicated analog-to-digital converter (ADC). The filter is only necessary in the calibration mode of operation (DCal mode), since in the standard data taking mode (SDT) the charge amplifier bandwidth is sufficient for filtering purposes. Future revisions of the Bean will be designed for a new set of machine specifications and will include additional channels, improved features, and a digital memory array.

Circuit description The charge amplifier was designed around a single-ended folded cascode amplifier with capacitive feedback. The feedback network has two selectable capacitors to implement the two gains for the SDT and DCal modes of operation. The feedback network also has a reset transistor that discharges the feedback capacitance in order to reset the charge amplifier between pulses, and a slow reset-release circuit that opens the reset transistor gradually in order to reduce the noise due to split doublets.

The charge amplifier and a dummy baseline generator are connected to the fully-differential ADC when in SDT mode, or to the fully-differential filter when in DCal mode, through level-shifting buffers. The filter is a switched-capacitor integrator that effectively reduces series noise by averaging eight samples of the charge amplifier output in the analog domain.

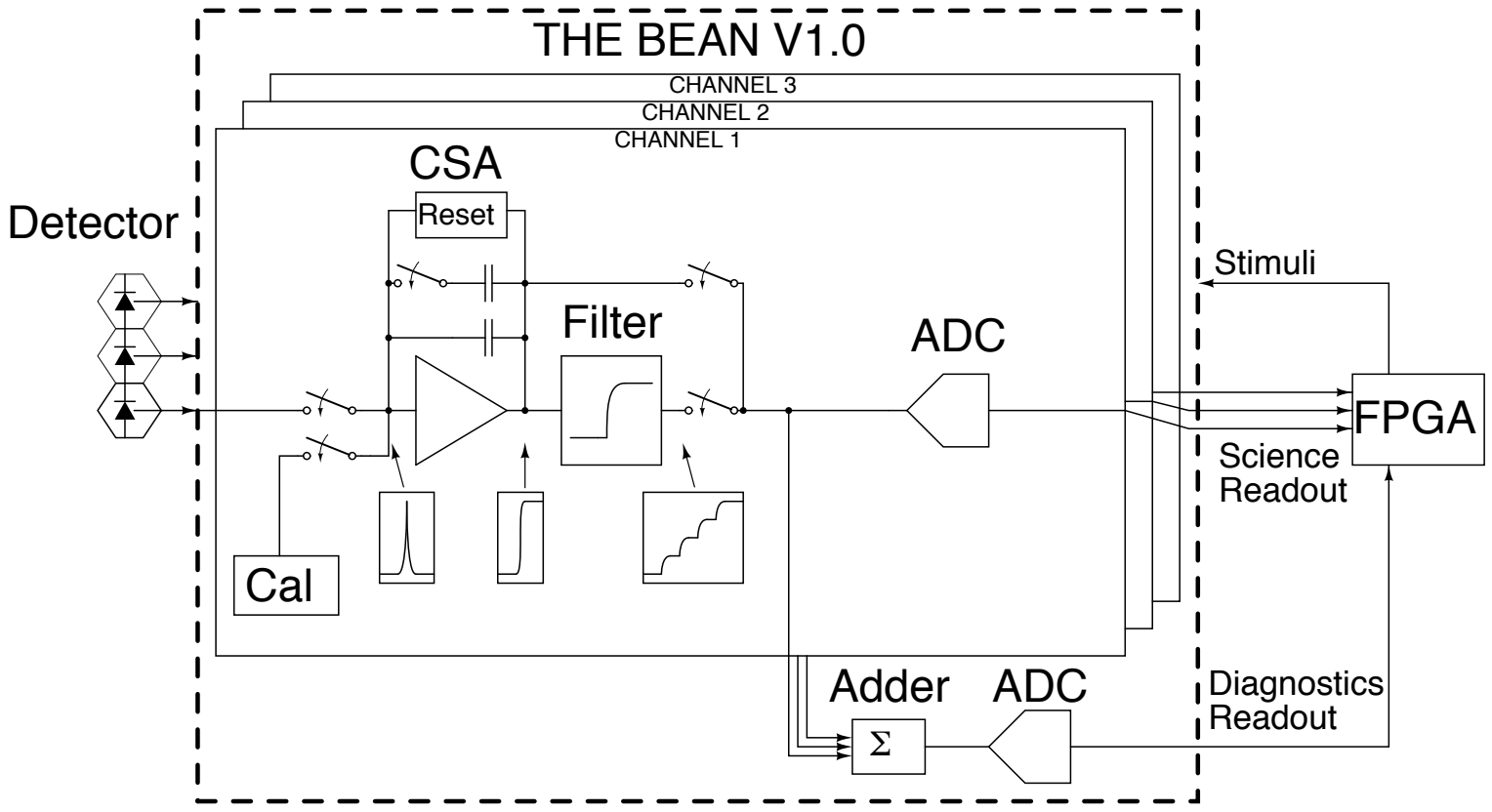

**Figure II-8.2**
The simplified block diagram of the Bean ASIC.

In order to provide the low-latency output[1] that combines the outputs of all channels in the chip, an analog adder is used. The adder operates in the sampled-data domain using switched capacitors, and can be easily scaled to include more channels. Both the signal and the adder output are digitised using a custom 10-bit successive approximation register (SAR) ADC. The converter samples the differential input voltage and, using an internal digital-to-analog converter (DAC), produces a voltage that tries to match the input voltage. Using a binary search algorithm for the internal DAC output voltage, on each conversion step the ADC produces the next significant bit of the digital output, starting from the most significant bit. The full conversion takes less than 250 ns to complete. The Bean die (Figure II-8.3) measures $2.4\,\mathrm{mm} \times 2.4\,\mathrm{mm}$. The channel pitch is 360 $\mu\mathrm{m}$ and includes generous power buses; four 1.8-V power supplies are required by the chip.

[1] The low-latency output will be used for beam tuning and diagnostics.

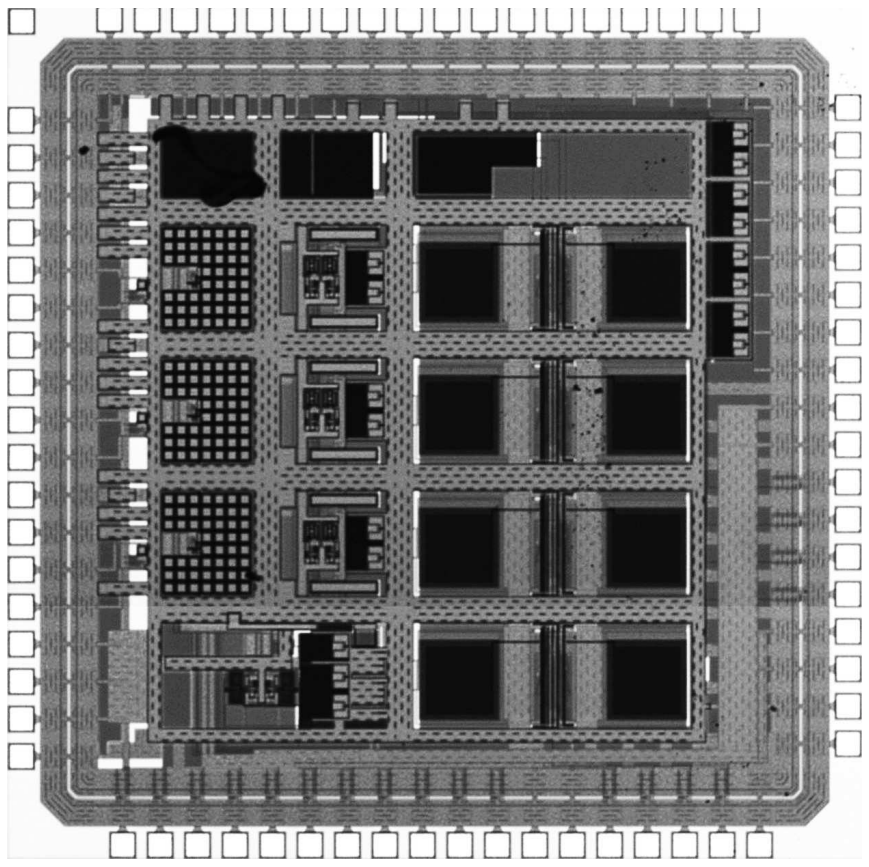

**Figure II-8.3**
The Bean microphotograph.

Test results The Bean ASIC was tested for linearity, crosstalk gain, adder operation and gain, bandwidth, weighting function, and noise. The chip linearity meets the specifications, with less than $1\%$ nonlinearity mainly due to the charge amplifier finite open-loop gain. The ADC nonlinearity contribution is negligible, except for a few missing codes due to the inductance in the reference lines. This problem will be fixed in future revisions of the chip. Figure II-8.4 shows the channel integral nonlinearity (INL) and differential nonlinearity (DNL) for the SDT mode.

The crosstalk, for either mode of operation, was measured to be less than $1.7\%$, and it is mostly due to indirect channel-to-channel coupling. The gain from each channel to the adder output was measured. Since there are three channels, the gain from each channel should be $0.33$. The measured gains range from $0.329$ to $0.345$, well within the expectations. The adder digital output is available in less than $1\,\mu\mathrm{s}$ from the pulse injection at the chip input. This low latency in providing the chip output is compatible with the fast feedback requirement specifications.

A 'chip bandwidth' test was performed to quantify the residual effect of an input pulse at the output measured in subsequent cycles. If the residual effect on subsequent cycles is null, then the chip can operate at the maximum speed without piling up data from different cycles. The bandwidth measurement was done by injecting a large input at a certain cycle, and measuring the output for that cycle and subsequent cycles. The test results show no evidence of memory effect in either mode of operation, which allows to operate the chip for $100\%$ occupancy.

From the chip weighting function and from the amplifier input-referred noise power spectral density and the detector leakage current, it is possible to compute the chip signal-to-noise ratio. The weighting functions were obtained through SPICE simulations, and then measured using the test setup described earlier. The measured weighting functions match the expectations, supporting the use of switched-capacitor filters.

The chip noise was measured in LSB units by using the histogram method. The capacitance at the chip input, mostly due to the test PCB, is higher than the expected input capacitance from the specifications, and consequently the noise measured is higher. In order to obtain fair measurements, noise was then estimated from the measured noise, scaling it down according to the ratio between actual and specified input capacitance. The noise estimation is 0.6 LSB in the SDT mode, and 1.41 LSB in the DCal mode. Most of the noise in the DCal mode is due to a design flaw in the filter amplifier, and will be fixed in future revisions of the chip.

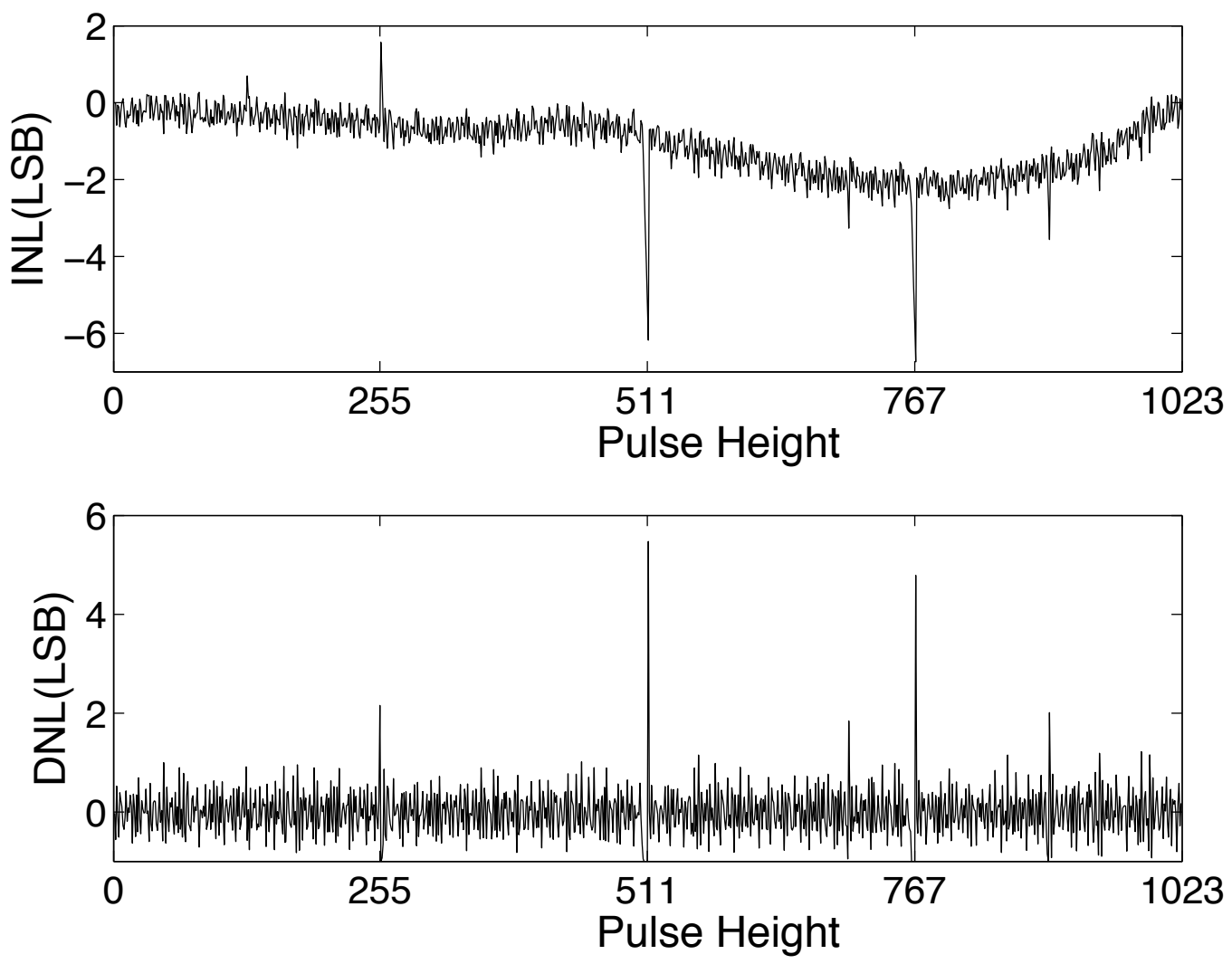

**Figure II-8.4**
The Bean integrated (INL) and differential (DNL) non-linearity in the standard data-taking (SDT) mode.

#### 8.1.3.2 Electromagnetic Radiation Damage Studies

The expected integrated dose of 100 MRad per year of electromagnetically-induced radiation poses a challenge to the design of the BeamCal. Standard n-bulk silicon diode sensors are not thought to be capable of withstanding such a dose without degrading to unacceptable levels of charge collection efficiency.

Prior studies [162] suggest substantially greater electromagnetic radiation tolerance for p-bulk sensors. However, particularly for p-type sensors for which damage from electromagnetic irradiation may be minimised, damage may be dominated by the hadronic component of the electromagnetic shower. Thus, a radiation-damage study of various silicon-sensor technologies is getting underway. This study will explore the charge-collection efficiency of both n- and p-type float-zone and magnetic Czochralski sensors exposed to electromagnetic showers radiation as well as that from a beam of pure electromagnetic particles, so that the two potential sources of radiation damage can be separated. GEANT4 simulations suggest a shower-maximum exposure rate of

$$1 \text{ MRad} \simeq \frac{0.8}{I_{beam}(\text{nA}) \cdot E_{beam}(\text{GeV})} \text{hours}$$

Even for a low-intensity beam, such as that of the SLAC ESTB testbeam facility, a four-hour run will expose a sample sensor to 100 MRad. An initial campaign of electromagnetic radiation damage studies is proposed for early 2013; if successful, the setup will be offered as a facility for the study of radiation hardness for other sensor technologies provided by the FCAL collaboration.

# Chapter 9
# SiD Electronics and DAQ

SiD has a coherent approach to its electronics architecture that is intended to satisfy the requirements of all subsystems. It is closely tied to the unique ILC timing structure with a long bunch train with 1 ms duration and then a period of 199 ms quiet time. The SiD electronics is designed to cope with up to 8192 bunches per train and a bunch spacing as small as 300 ns; this can easily satisfy the current ILC requirements of up to 2625 bunches per train and a bunch spacing of 344 ns [163]. A simplified block diagram of the SiD data-acquisition from the front-end electronics to the online-farm and storage system is shown in Figure II-9.1.

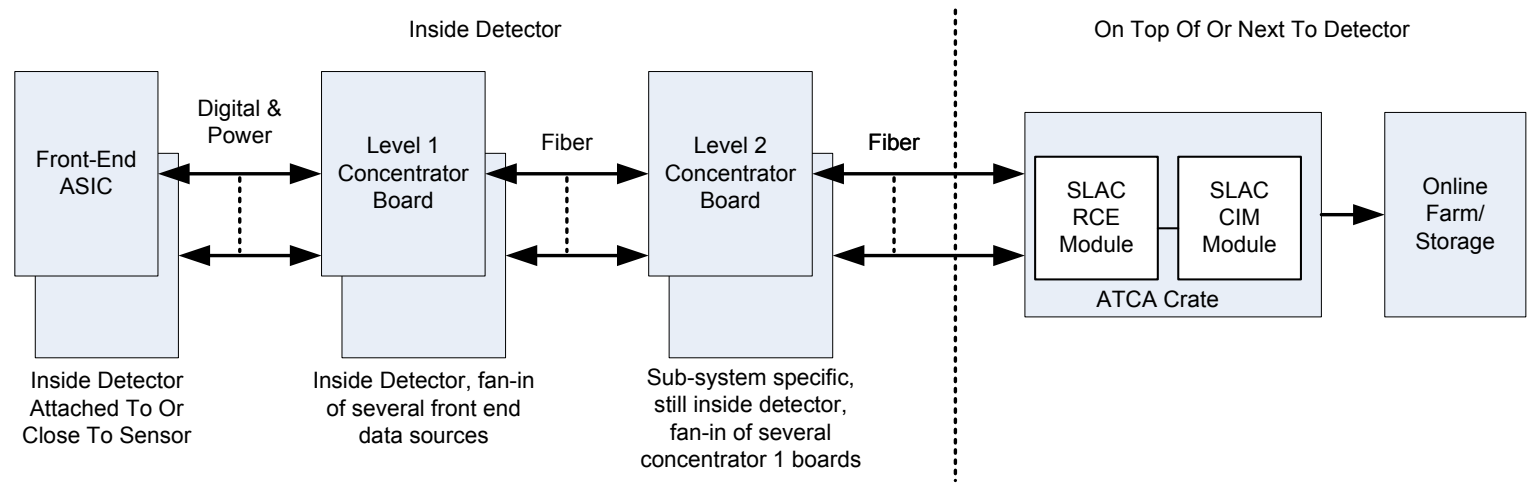

**Figure II-9.1** Simplified block-diagram of the SiD detector control and readout chain using the ATCA RCE and CIM modules (defined later in this chapter).

## 9.1 ASIC developments

All subsystems with the exception of the Vertex detector (for which the sensor technology has not been selected yet) and the BeamCal (which has approximately unit occupancy) are foreseen to be read out by variants of KPiX as the front-end Application-Specific Integrated Circuit (ASIC). For the BeamCal the Bean ASIC has been developed to address its specific requirements, see Section 8.1.3.1.

### 9.1.1 KPiX

KPiX [87, 88] is a multi-channel system-on-chip for self-triggered detection and processing of low-level charge signals. Figure II-9.2 shows a simplified block diagram of the KPiX, processing signals from 1024 input channels. The low level charge signal at the input is processed by the charge amplifier in two ranges with an automatic range switching controlled by the range threshold discriminator. The built-in calibration covers the full dynamic range of up to 10 pC. Leakage compensation is available for DC-coupled detectors and either internal or external trigger options can be selected.

Up to four sets of signals for each channel can be stored in one acquisition cycle corresponding to one ILC bunch train. The timestamp is stored using a 13-bit-deep counter, while the signal amplitude is first stored as a voltage on a capacitor before its subsequent digitisation using a Wilkinson-type ADC with 13-bit resolution. At the end of the acquisition and digitisation cycle nine words of digital information are available for each of the 1024 cells of the KPiX chip. The data are then read out serially from the KPiX before the next acquisition cycle starts. The power consumption for each individual channel is less than 20 μW. The latest version of KPiX has been manufactured using a 250 nm mixed-mode CMOS process and is currently being tested [88].

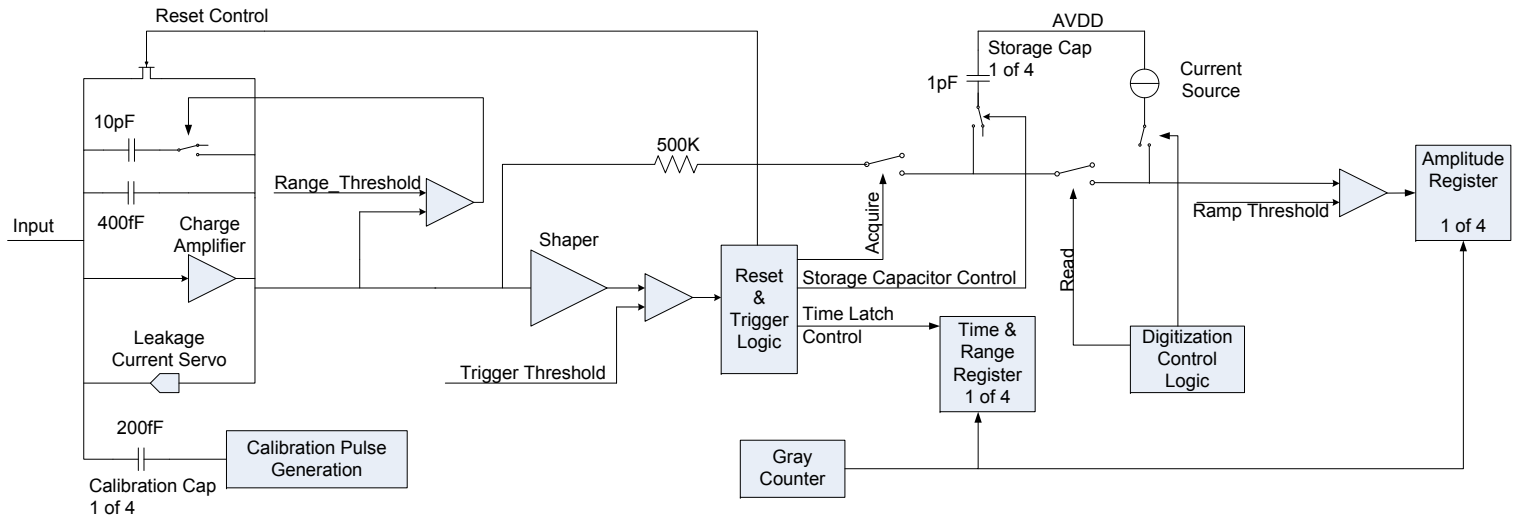

**Figure II-9.2** Simplified block diagram of one KPiX channel.

Table II-9.1 lists the currently foreseen number of KPiX ASICs for each subsystem. Main Tracker, ECAL, and HCAL use the 1024-channel version of the KPiX while the Muon subsystem uses a 64-channel KPiX version.

**Table II-9.1** Approximate count of KPiX ASICs for each subsystem.

| Sub-System | KPiX Count | Channels/KPiX |
|---|---|---|
| Main Tracker | 27464 | 1024 |
| ECAL | 102573 | 1024 |
| HCAL | 35071 | 1024 |
| Muons | 8839 | 64 |
| Total | 173947 | |

## 9.2 On-Detector Electronics

As illustrated in Figure II-9.1, several front-end ASICs (KPiX, Bean or Vertex ASICs) are connected to a Level-1 Concentrator (L1C) board using LVDS. The main functions of the L1C board are to fan out incoming signals and commands to the front-end modules and to bundle data from the front-end modules for transmission to the Level 2 Concentrator (L2C) boards. Additionally it can perform zero-suppression and sorting of the incoming event data. For example, for the ECAL Barrel 96 1024-channel KPiX chips would be served by eight front-end cables with twelve KPiX per L1C board, yielding a total of 821 L1C boards and 52 L2C boards (80000 KPiX, 96 per L1C board; 16 L1C boards for each L2C board). Figure II-9.3 shows a block diagram of the L1C board.

The Level 1 Concentrator boards are in turn connected via 3-Gbit/s fibres to the Level-2 Concentrator boards. Besides distributing signals to/from the L1C boards, the L2C boards merge and sort the data-streams of the incoming event data before transmission to the off-detector processor boards. Depending on the sub-system, the L2C boards are either located inside the detector or immediately outside. E.g. for the ECAL Barrel there are 52 L2C boards inside the detector volume.

## 9.3 Off-Detector Electronics

The L2C boards are connected via fibres to ATCA crates either on or next to the detector. ATCA (Advanced Telecommunications Computing Architecture) is the next generation communication equipment currently used by the telecommunication industry. It incorporates the latest trends in high-speed interconnect, processors and improved reliability, availability, and serviceability. Instead of parallel bus back-planes like VME, it uses high-speed serial communication and advanced switch technology within and between modules, redundant power, plus monitoring functions. For SiD the usage of 10 Gbit/s Ethernet is foreseen as the serial protocol.

Two custom ATCA boards, the Reconfigurable Cluster Element (RCE) Module and the Cluster Interconnect Module (CIM) were designed previously. Based on those two modules, a second generation RCE was built, as shown in Figure II-9.4 which combines the switch interconnect function of the CIM onto the RCE itself. A 96-port 10 Gbit/s Ethernet ASIC is placed on the RCE, providing communication between all the RCE modules in a crate and to destinations external to the crate with data rates up to 10 Gbit/s. The RCE modules connect via the backplane to the Rear Transition Modules (RTM) which interface via 48 3-Gbit/s fibre links to the sub-system L2C boards. The main ATCA board can hold up to five daughter-cards (shown in Figure II-9.4) each with a Virtex FPGA with two embedded PowerPC processors, four Gbytes of DDR3 memory, 8 Gbytes/sec cpu-data memory interface, four 10-Gbit/s Ethernet event data interfaces, and an open-source RTEMS (Real-Time Executive for Multiprocessor Systems) realtime operating system [164]. One ATCA crate can host up to 14 RCE boards, providing connections to $48 \times 14 = 672$ 3-Gbit/s fibre links into the detector for 2 Tbit/s IO.

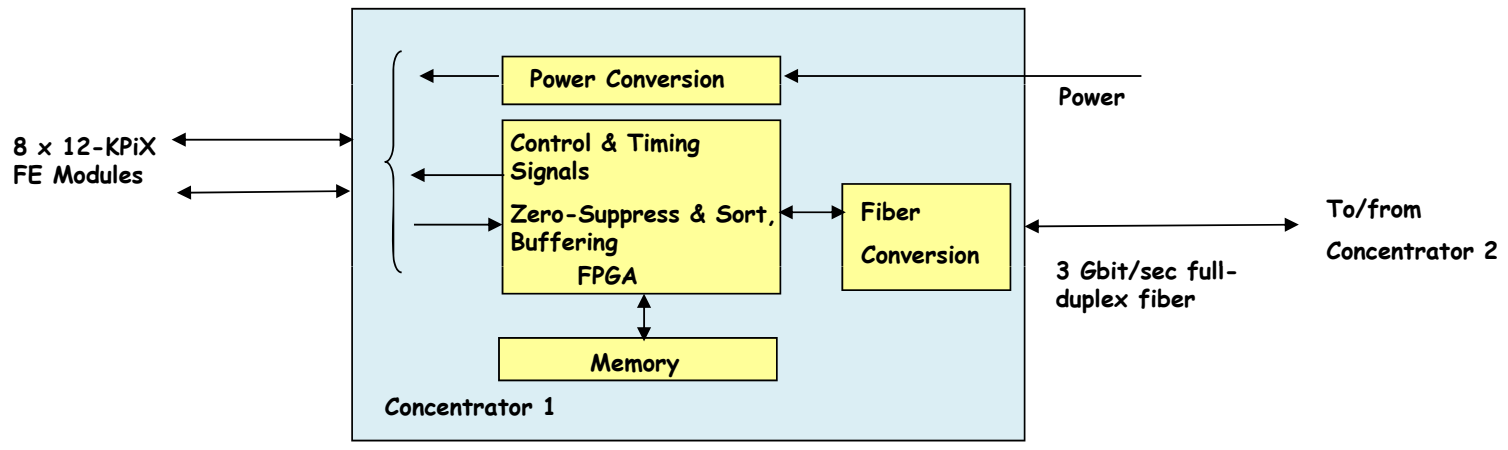

**Figure II-9.3** Block Diagram of the Level 1 Concentrator Board.

The maximum available data transfer rate is up to 520 Gbit/s, while the estimate for the complete SiD is approximately 320 Gbit/s including a factor of two safety margin. In principle, a partially loaded ATCA crate could serve the complete detector. However for partitioning reasons, the ability to run each of the subsystems completely independently during commissioning is highly desirable, and therefore a crate for each subsystem is planned.

The data are further sorted on an event-by-event basis in the ATCA system and then sent to the online processing system for potential further data reduction. Whether further data reduction is required is not determined yet, and the data may directly be forwarded to the offline system. Note that the event data are zero-suppressed in the sub-systems without the need for a global trigger system. All data produced in the front-ends above a programmable threshold is subsequently read out. For diagnostics and debugging, the DAQ includes the ability to assert calibration strobe and trigger signals, transmitted to the front-ends via the L2C and L1C boards using the fibres shown in Figure II-9.1.

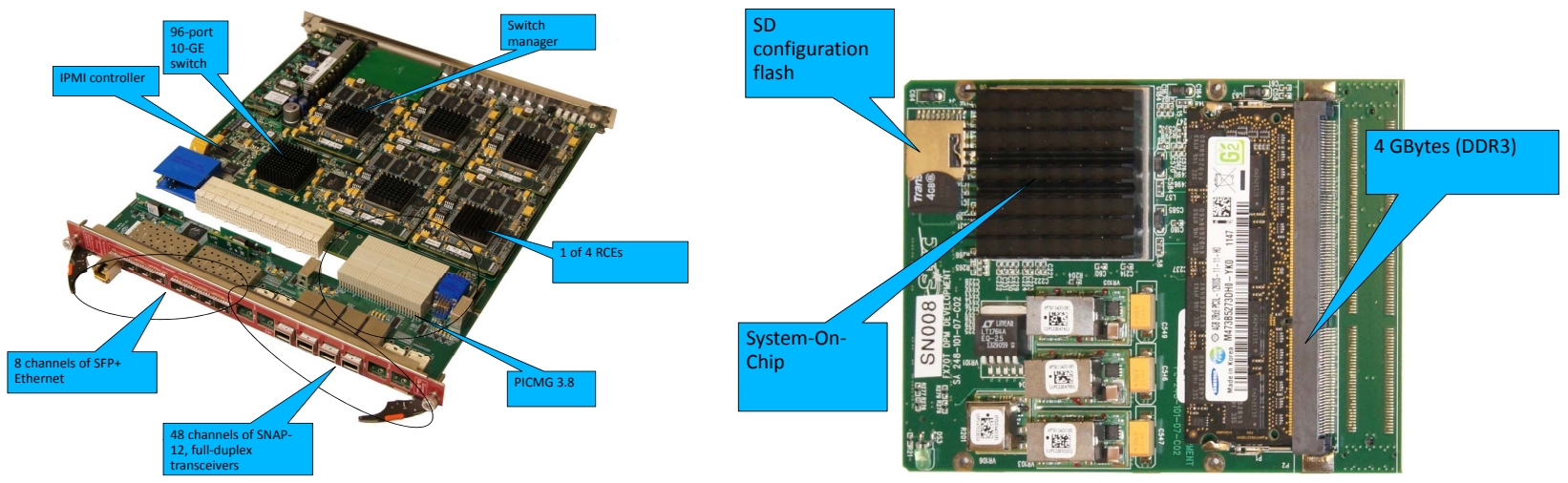

**Figure II-9.4** ATCA Reconfigurable Cluster Element (left) and RCE daughter module (right)

Power conversion circuits on the L2C and L1C boards supply the power to the front-ends, starting with 48 V or higher voltages from off-detector supplies and then using DC-DC converters. Alternatively, serial powering architectures are also under consideration. The power supplies will be located in several racks on, or next to, the detector. Environmental and health monitoring circuits are also included on the concentrator boards. In addition there may be additional monitoring boards in the detector, connected to RCE fibre interfaces. In addition there are crates of monitoring modules mounted in several racks on or next to the detector.

## 9.4 Overview of Electronics Channels and Expected Data Rates

Table II-9.2 provides a global overview of the electronic channel counts for SiD. For each subdetector the detector granularity, as currently used in the SIDLOI3 simulation model, and the approximate number of readout channels are listed. Preliminary studies have been carried out to estimate the average and maximum cell occupancies to be expected for a full bunch train at the ILC at 1 TeV [165]. The number of bits per hit provides a first estimate, and is based on a simplified approximation for individual hits with full on-detector zero-suppression and without on-detector clustering. It foresees some bits for addressing. The estimated data volume per bunch train is also listed. The beam parameters at 1 TeV presume 476 ns bunch separation and 2450 bunches per train. In the vertex and main tracker regions, it was found that the occupancy from $\gamma\gamma \rightarrow \mathrm{hadrons}$ events is typically more than one order of magnitude below the occupancy from incoherent pairs. The occupancies in the vertex detector and main tracker have therefore been calculated using incoherent pairs only. Average cluster sizes of 3.1 and 2.6 have been assumed for the pixel and strip detectors respectively.

Uncertainties in the simulation of the production rates and in the detector response have been estimated. As a result, safety factors of two for the background rates from $\gamma\gamma \rightarrow \mathrm{hadrons}$ and five for the ones from incoherent pairs have been included. The occupancy studies have used a uniform 5 T solenoid field along the z-direction. The use of a more realistic field map including anti-DID will alter the results for incoherent pairs in the most inner detector regions up to a few ten percent.

The simulations at 1 TeV show that expected occupancies in the first layers of Lumical are well above 400% over the full surface. Contrary to the 500 GeV case, it is therefore assumed that the Bean chip will be used throughout BeamCal and Lumical for 1 TeV running. The full 2820 Bean buffer depth is assumed to be read out for each cell and for each bunch train.

**Table II-9.2** Overview of read-out details for the various subdetectors. Occupancies and data volumes are for a full bunch train at 1 TeV and include beam-induced background as well as charge sharing between pixels/strips. Safety factors of five and two have been applied to the rates of incoherent pairs and $\gamma\gamma \rightarrow \mathrm{hadrons}$ respectively. BeamCal and Lumical are expected to be using the Bean chip with a buffer depth of 2820.

| | cell size ($\mathrm{mm}^2$) | number of channels ($10^6$) | av. to max. occ. (%) | approx. # bits per hit (bit) | data volume (Mbyte) |
|---|---|---|---|---|---|
| **VXD barrel** | 0.02×0.02 | 408 | 8 - 60 | 32 | 130 |
| **VXD disks inner** | 0.02×0.02 | 295 | 4 - 70 | 32 | 50 |
| **VXD disks outer** | 0.05×0.05 | 980 | 0.5 - 20 | 32 | 20 |
| **Main tracker barrel** | 0.05×100 | 16 | 33 - 300 | 32 | 20 |
| **Main tracker disks** | 0.05×100 | 11 | 4 - 500 | 32 | 2 |
| **ECAL barrel** | 3.5×3.5 | 72 | 2 - 45 | 40 | 7 |
| **ECAL endcap** | 3.5×3.5 | 22 | 33 - 2300 | 40 | 36 |
| **HCAL barrel** | 10×10 | 30 | 0.07 - 200 | 40 | 0.1 |
| **HCAL endcap** | 10×10 | 5 | 96 - 3600 | 40 | 24 |
| **LumiCal** | 2.5×var. | 0.061 | ≫100 | 16 | 340 |
| **BeamCal** | 2.5(5.0)×var. | 0.076 | ≫100 | 16 | 430 |

Including safety factors, the average hit occupancies in the muon barrel system amount to $7.5{\cdot}10^6/\mathrm{cm}^2$ per train, with a maximum of $0.08/\mathrm{cm}^2$ at the transition region to the endcap, due to particles passing through the HCAL barrel-endcap gap. To determine muon endcap occupancies the simulation model would need to be extended. Currently the model does not contain all material in the very forward region, such as QDO support tubes, which will shield backscattered particles. Overall the data from the muon system will have a very small impact on the data volume.The table shows that the occupancies can exceed 100% in several detector regions. The KPiX chip presently provides fast buffering of up to four hits per channel. The KPiX design can be adapted to contain a larger buffer depth if deemed required for high-occupancy regions.

# Chapter 10
# SiD Simulation and Reconstruction

## 10.1 Overview of the Simulation and Reconstruction Software

A large fraction of the software for the generation, simulation and reconstruction is shared between the ILD and SiD detector concepts (see 2.2). The generated events are simulated in the SiD detector by SLIC [166], a program encapsulating the functionality of the GEANT4 [167] tool kit, but providing the ability to define all aspects of the detector at runtime. The output information consists of primary charge deposition in the sensitive detectors providing the primary information regarding the energy deposition, hit position, time and Monte Carlo particle causing the energy deposition. At this level each of the physics events at 1 TeV is merged with a simulated event containing the equivalent of one bunch crossing of incoherent pair interactions. Additionally, hits and particles from $\gamma\gamma \to \mathrm{hadrons}$ events are merged with each physics event. The number of $\gamma\gamma \to \mathrm{hadrons}$ events follows a Poisson distribution with a mean of 4.1 (1.7) per bunch crossing at 1 TeV (500 GeV). Events produced for the 500 GeV study are not merged with incoherent pairs background.

The energy deposits in the active material of the detectors are then digitised into simulated hits using the `org.lcsim` reconstruction framework [168]. A more detailed description of the digitisation is given in Section 10.4. Pattern recognition and track fitting is the task of the SeedTracker algorithm, which has been used successfully in the benchmarking of SiD detector variants at a 500 GeV ILC [63] as well as at a 3 TeV CLIC [129]. The algorithms of the PANDORAPFA package [169] are responsible for the calorimetric reconstruction and the creation of particle flow objects (PFOs). In a first step, muons are identified, their hits removed from the calorimeters, and the remaining hits are clustered using a cone clustering algorithm. Charged particles are created through the positive match of a track with a cluster, where consistency of the measured energies is ensured through iterative re-clustering. The remaining clusters are assigned to neutral PFOs. A more detailed description of the particle identification is given in Section 10.6.1.

Vertices from secondary interactions are found by the LCFIPlus [170] flavour tagging package. The found vertices are then used in the jet clustering, which is described in more detail in Section 11.2.2.1.

## 10.2 The SiD DBD Production

The detector response simulation and event reconstruction was performed on the Worldwide LHC Computing Grid (WLCG) and the Open Science Grid (OSG). The ILCDIRAC [171, 172] tool was used for the full chain of bookkeeping, handling of meta data, automated job submission and monitoring. The jobs were submitted under the common ILC Virtual Organisation. Figure II-10.1 shows a distribution of the computing time used by country. Major contributors were CERN, various Grid sites in the UK (primarily RAL and Manchester), IN2P3 in France and Open Science Grid resources at FNAL and PNNL.

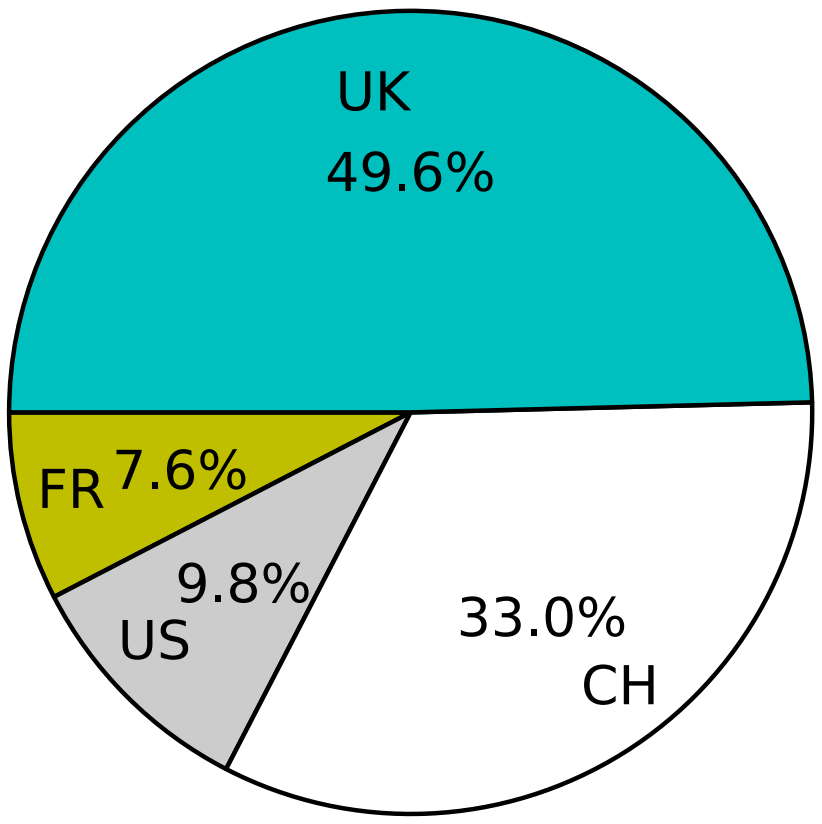

**Figure II-10.1**
CPU time used in the DIRAC mass production by country

## 10.3 Simulating the SiD Detector Geometry

### 10.3.1 The sidloi3 Model

The SIDLOI3 detector model reflects the design of SiD, as described in this document, with the Silicon-Tungsten ECAL and the RPC HCAL as the baseline. All of the tracker elements are modelled as planar silicon wafers with accompanying support structures. The geometry of the services (power and readout) is simplified, but reflects the gross amount and general distribution of the materials. The calorimeters are modelled as polygonal staves in the barrel region or planes in the endcaps, with interleaved readouts. For the complete details of the model as implemented in GEANT4 see [173].

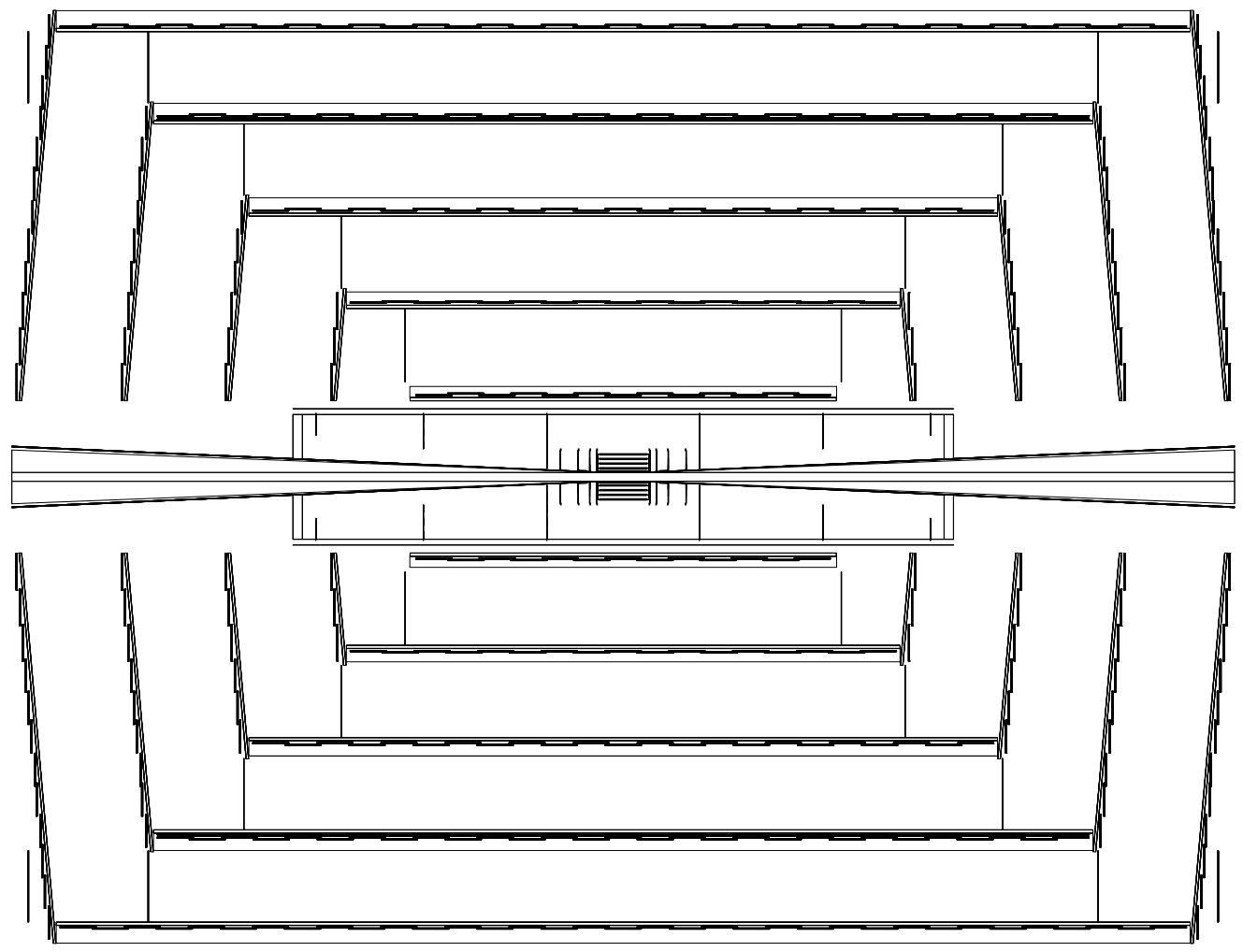

**Figure II-10.2**
R-z view of the tracking system as implemented in SIDLOI3 model. Some support and readout structures have been hidden to improve the visibility of the sensors.

A cross-section of the tracking detector is shown in Figure II-10.2. This is to be compared with Figure II-3.3 showing an engineering elevation view of the tracking system. An orthographic cutaway view of the central tracker as implemented in the SIDLOI3 model is shown in Figure II-10.3 (left). An orthographic cutaway view of the complete detector as implemented in the SIDLOI3 model is shown in Figure II-10.3 (right). The electromagnetic barrel calorimeter is modelled as a dodecagonal tube with overlapping staves. The hadron calorimeter barrel is composed of twelve symmetric staves. Finally, the octagonal layout of the magnetic flux return yoke, with its eleven layers of muon detection instrumentation is clearly visible.

Figure II-10.4 (left) shows the cumulative hadronic interaction lengths of the SiD detector elements as a function of the polar angle $\theta$, including the self-shielding provided by the thick mantle of

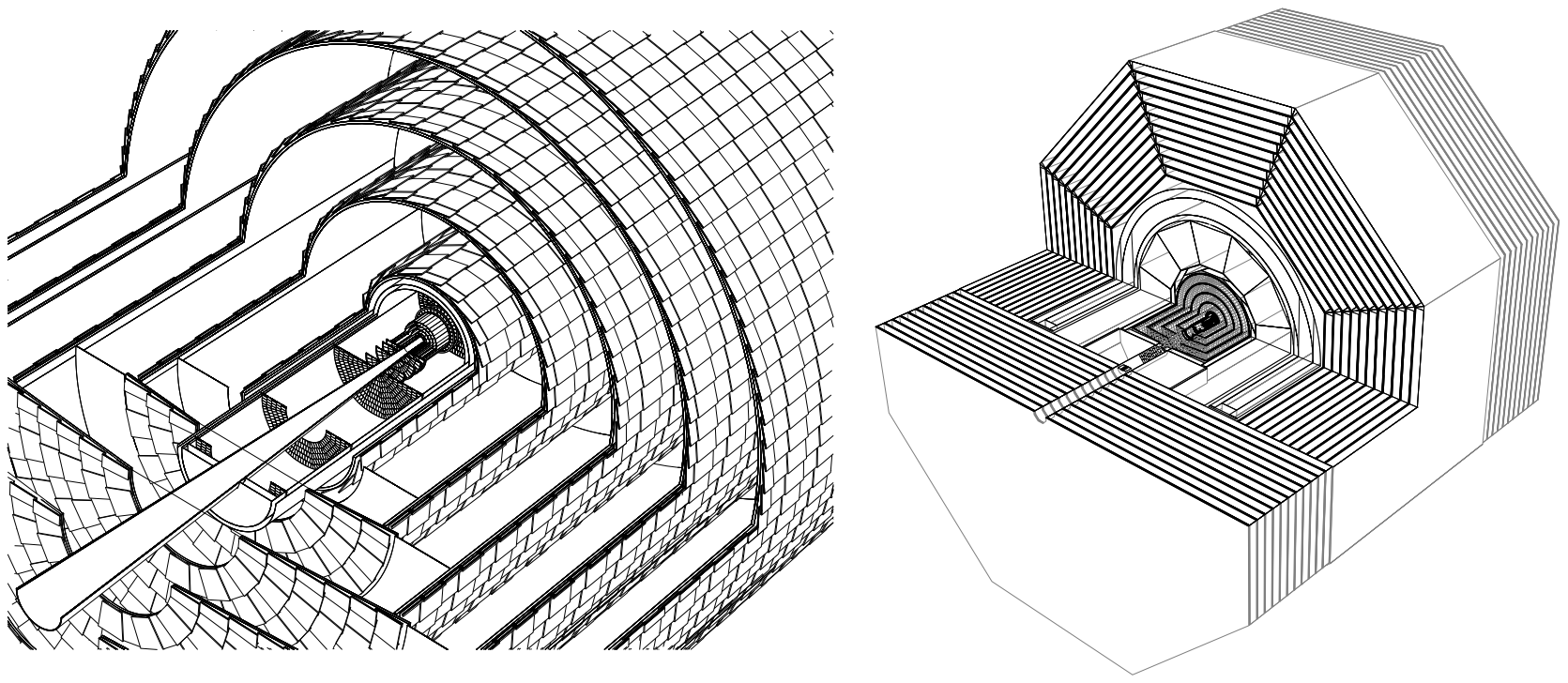

**Figure II-10.3**
Cutaway view of the tracking system as implemented in SIDLOI3 (left) and the complete detector (right). Some support and readout structures have been hidden to improve the visibility of the sensors (left) and the calorimeters.

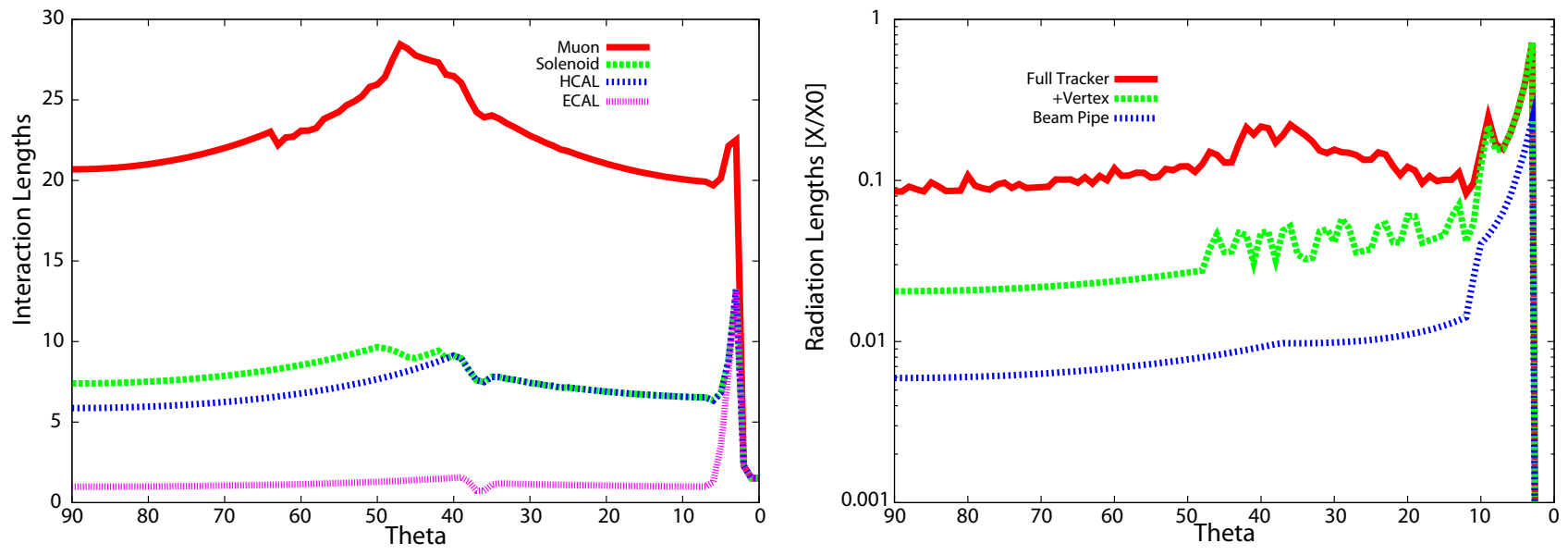

**Figure II-10.4**
The nuclear interaction lengths of SIDLOI3 (left) and the radiation lengths of SIDLOI3 tracking system (right) as a function of the polar angle $\theta$.

the flux-return steel, while Figure II-10.4 (right) displays cumulative material (expressed as a fraction of $X_0$) of the tracking region.

## 10.4 Simulation of the full Detector Response

The hits which are recorded and written out from the full Monte Carlo simulation contain the primary charge deposition. To simulate the response of a realistic, physical detector, this information needs to be converted to information which represents the electronic readout that would be collected from the detector. We refer to this process as hit digitisation.

### 10.4.1 Silicon Pixel and Strip Hit Digitisation

The silicon-based tracking detectors are precision devices with very high intrinsic spatial resolution. In order to realistically model their response, the effects of charge drift and diffusion in the silicon, as well as effects of pulse shaping and electronic noise need to be implemented in the simulation. The charge deposition for silicon strip detectors is simulated using an algorithm based on the CDF silicon sensor simulation. An extension of this model to pixels is used to model the response of the vertex detector elements.

### 10.4.2 Simulation of the Calorimeter Response

Calorimeters are designed to measure the energy of incident particles by inducing them to shower in the detector and to record the deposited energy. Because of the vast number of secondary particles produced when an incident particle showers, and because precise details of these secondary particles are unimportant to the energy measurement, we do not by default record primary charge depositions for each of them. Instead, we define volume elements in which we sum up the total amount of deposited energy, and record the earliest time of deposition from each separate incident particle. The energy in the ECAL is reconstructed by multiplying the energy deposited in the sensitive readout layers with sampling fractions, while in the digital HCAL, the energy is obtained from the number of cells with an energy deposition multiplied by a sampling fraction. These sampling fractions are determined from the response of SIDLOI3 to single muons, photons and $K^0_L$, respectively, at a variety of energies.

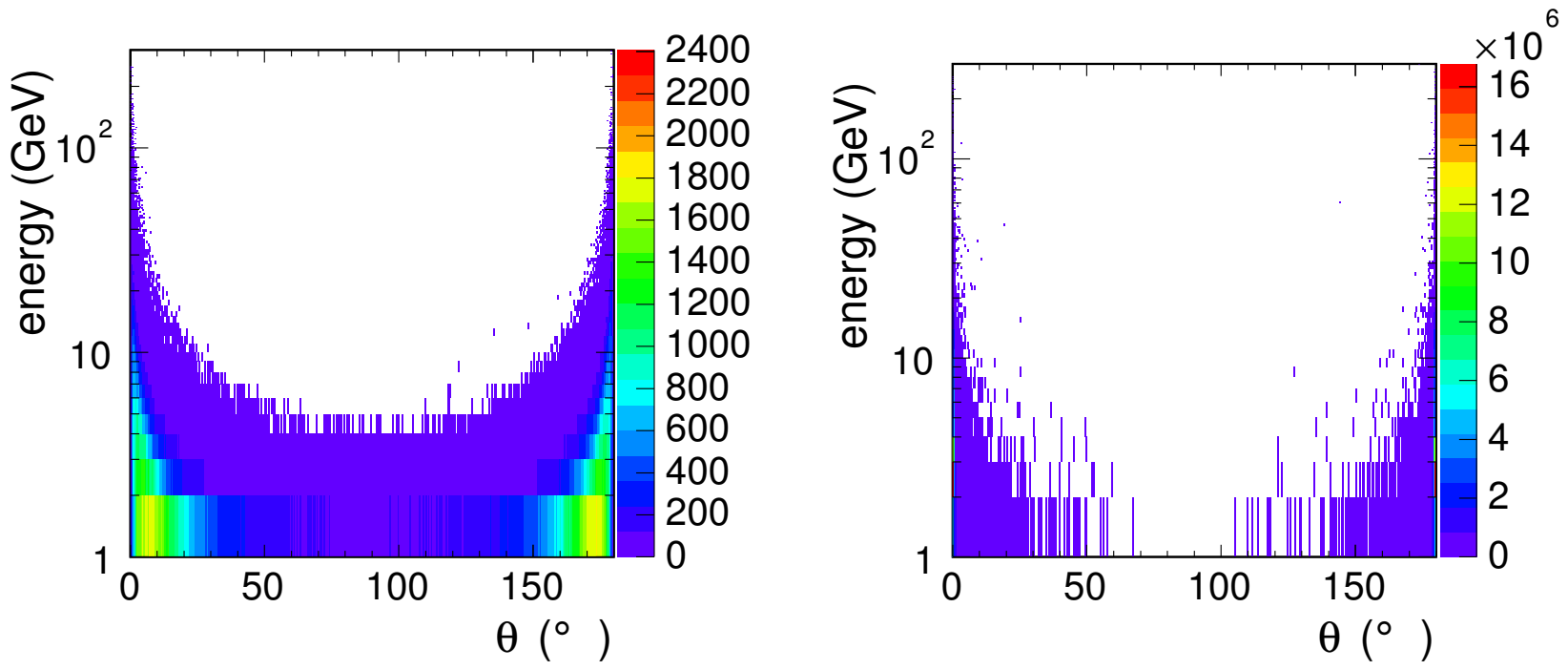

**Figure II-10.5** Kinematic properties of the machine-induced backgrounds from $\gamma\gamma \to \mathrm{hadrons}$ processes (left) and from incoherent pairs (right) at 1 TeV.

## 10.4.3 Clustering

The association of nearby strips, pixels or volume elements into a single hit is referred to as clustering. The signal sharing across readout elements can lead to improvements in the measurement precision and is therefore a crucial step in the reconstruction. The silicon strip and pixel signals are clustered using a nearest neighbour algorithm. Lookup tables are used to achieve approximately linear scaling of clustering time. Settable parameters are provided for noise, readout and clustering thresholds. Tracker hits are then created from these clusters. The position measurements (1D for strips, 2D for pixels) are derived from the energy-weighted centroids of the clusters, and the uncertainties are provided on a cluster-by-cluster basis. These hits are the input for the track finding. The algorithms used by PandoraPFA to cluster energy depositions in the calorimeters are described in more detail elsewhere [174].

## 10.5 Properties of Machine-Induced Background

Beamstrahlung processes in the ILC machine operated at 1 TeV result in a large rate of $\mathrm{e^+e^-}$ pairs at low transverse momenta that are predominantly produced in the forward region. These processes result in higher occupancies in the inner layers of the vertex detector and in the forward detectors. The main source of processes with higher transverse momentum are multiple $\gamma\gamma \to \mathrm{hadrons}$ events per bunch crossing.

The event samples for the compulsory DBD benchmarks are described in detail in Section 11. They were mixed with machine-induced background on an event-by-event basis. Samples at 1 TeV were mixed with an average of 4.1 $\gamma\gamma \to \mathrm{hadrons}$ events for each physics event and the 500 GeV processes were mixed with an average of 1.7 $\gamma\gamma \to \mathrm{hadrons}$ events. In addition, the 1 TeV samples include the detector response to an average of 450,000 low-momentum incoherent $\mathrm{e^+e^-}$ pairs. These processes follow a Gaussian distribution with a width of 225 $\mu\mathrm{m}$ in $z$ to account for the finite size of the ILC bunches. The physics process from the primary $\mathrm{e^+e^-}$ interaction was produced at $z = 0$.

The backgrounds were simulated in a separate step and merged with the primary physics process before the digitisation step outlined above. In order to keep the file size to a manageable level, particles that do not leave primary charge depositions in the sensitive detectors were dropped. The tool for the merging of the processes [94] has been developed for and tested in the CLIC CDR benchmarking analyses.

Figure II-10.5 shows two-dimensional distributions of the energy versus the polar angle of simulated particles produced in $\gamma\gamma \to \mathrm{hadrons}$ processes (left) and from incoherent $\mathrm{e^+e^-}$ pair production (right) at 1 TeV. While particles from incoherent pairs are predominantly in the forward region, particles from $\gamma\gamma \to \mathrm{hadrons}$ processes have significant energy also in the central region and reach the barrel calorimeters.

# 10.6 Detector Performance

## 10.6.1 Particle Identification: Photon, Electron and Muon ID

Particle identification (particle ID), and in particular lepton identification will be central to many physics studies at the ILC. Muons are identified and all of their hits removed before the calorimeter hits are clustered. The track-cluster agreement is optimised using various re-clustering strategies, which are guided by identifying the cluster as belonging to an electromagnetic or a hadronic interaction.

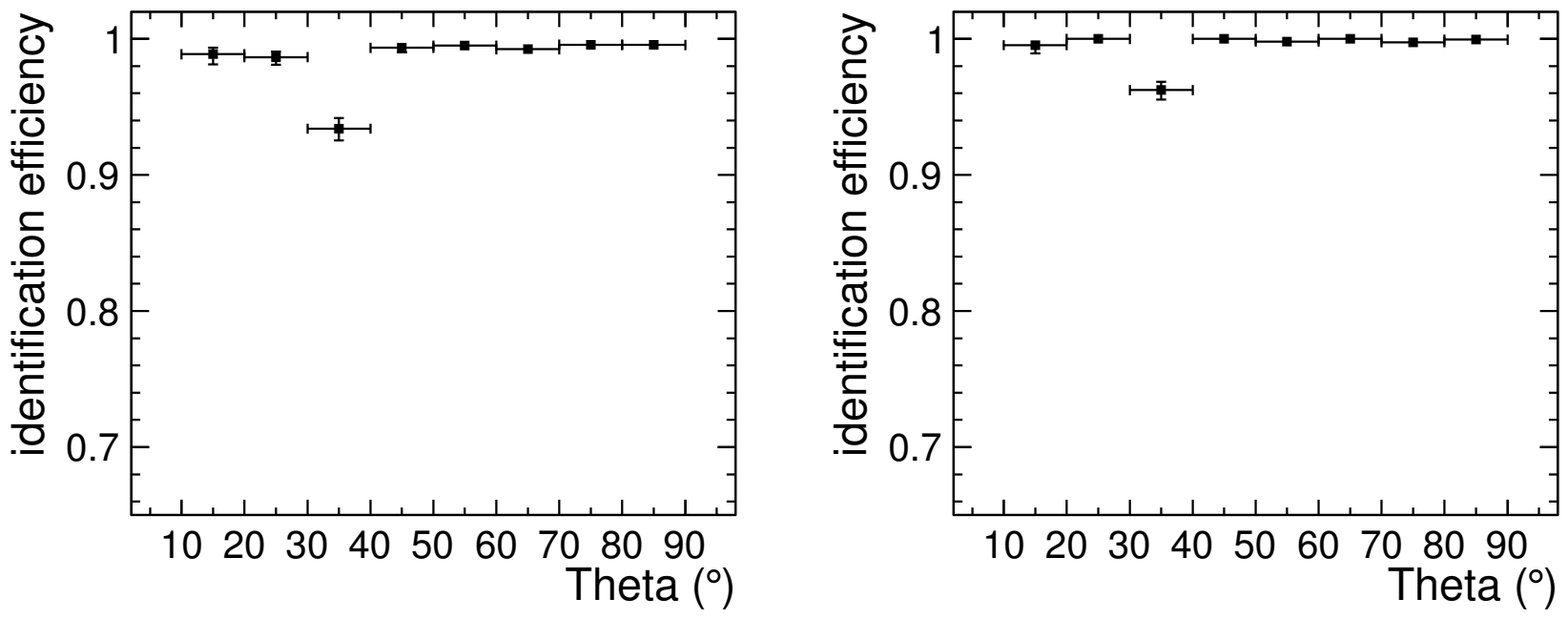

**Figure II-10.6** Particle identification efficiency for 10 GeV photons (left) and 100 GeV photons (right) as a function of the angle $\theta$.

The particle identification has been evaluated on samples of single photons, electrons, muons and pions of 10 GeV and 100 GeV, respectively (see Figures II-10.6, II-10.7 and II-10.8). In these plots, the reconstructed particle is required to have the same type as the generated particle, and for electrons and pions a loose energy cut of five times the resolution of the EM calorimeter above or below the energy of the generated particle is applied.

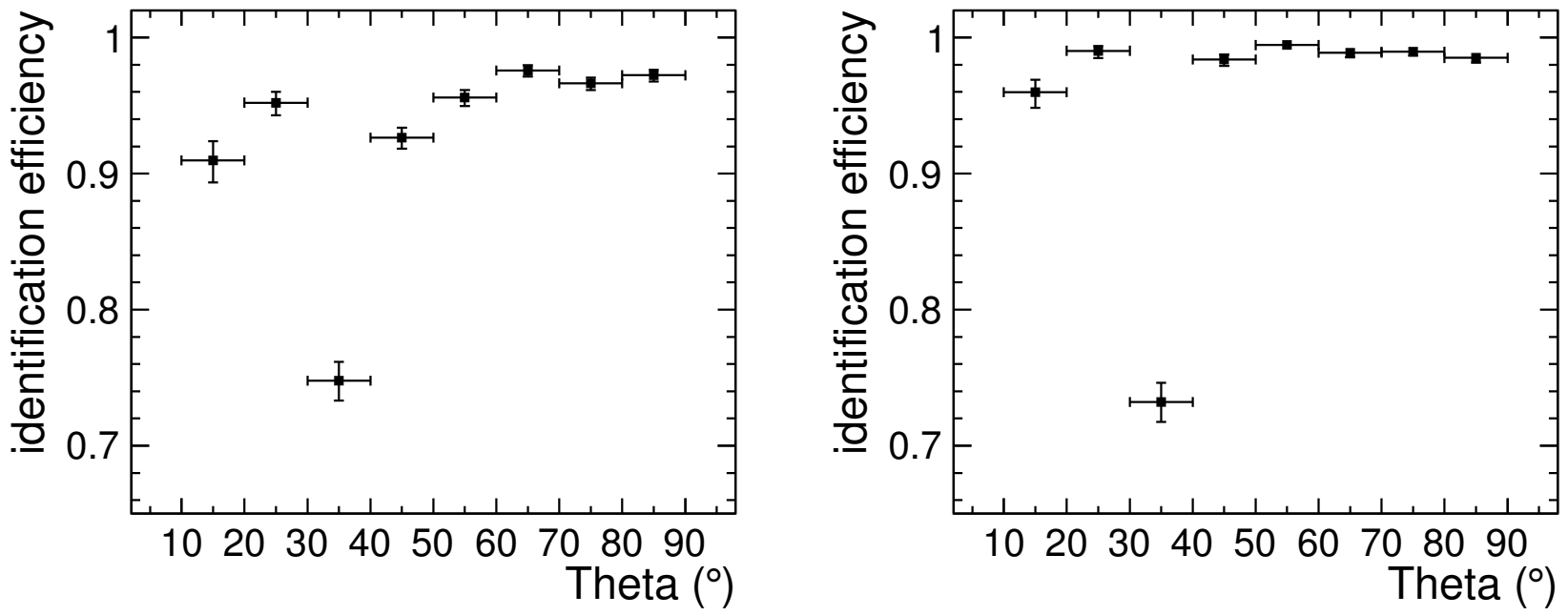

**Figure II-10.7** Particle identification efficiency for 10 GeV electrons (left) and 100 GeV electrons (right) as a function of the angle $\theta$.

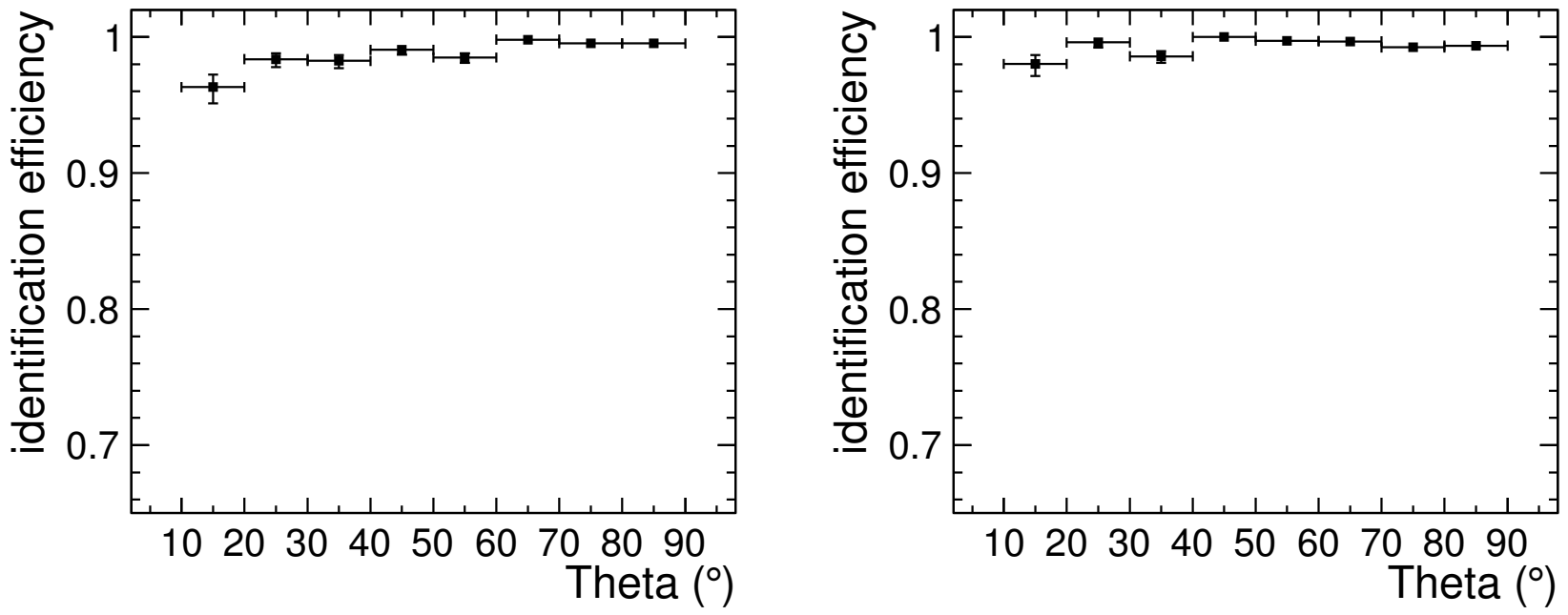

**Figure II-10.8** Particle identification efficiency for 10 GeV muons (left) and 100 GeV muons (right) as a function of the angle $\theta$.

The default for neutral particles is the neutron hypothesis, while charged particles are assigned the pion hypothesis by default. The muon identification efficiency is above 95% for values of the polar angle $10^\circ < \theta < 170^\circ$. The photon identification efficiency is above 99% over the same angular range, except for an inefficiency in the transition region between calorimeter barrel and endcap, which results in a dip in the bin $30° < \theta < 40°$. That same region results also in a dip of the electron efficiency, which is otherwise over 90% for 10 GeV electrons and above 98% for 100 GeV electrons.

The performance of the reconstruction as shown here has not been optimised for particle identification efficiency, but rather for jet energy resolution. We expect that a significant improvement can be achieved - particularly in the transition region between calorimeter barrel and endcap - with dedicated particle identification algorithms that are optimised for performance with the SiD digital HCAL.

### 10.6.2 Jet Flavour Tagging: Efficiency and Purity

**Figure II-10.9** Mis-identification efficiency of light quark jets (red points) and charm jets (green points) as beauty jets versus beauty identification efficiency in di-jets at $\sqrt{s} = 91$ GeV. The performance is shown without (left) and with background from $\gamma\gamma \to \mathrm{hadrons}$ events and incoherent pairs (right).

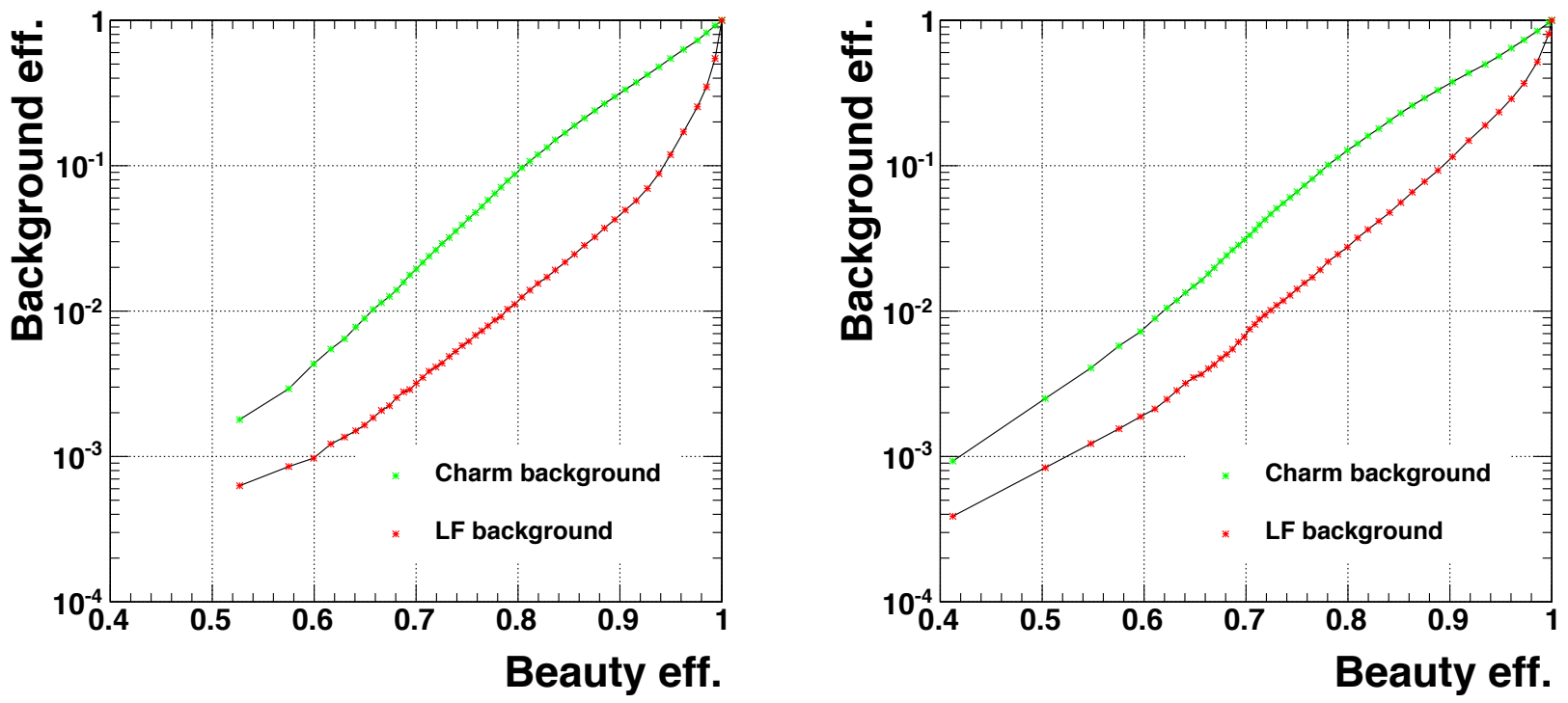

**(a)** without $\gamma\gamma \to \mathrm{hadrons}$ **(b)** with $\gamma\gamma \to \mathrm{hadrons}$

The ability to tag bottom and charm decays with high purity is a crucial aspect in the design of the vertex detector. Figure II-10.9 shows the b-tagging efficiency of a light quark sample (red curve) or a charm quark sample (green curve) versus the b-tagging efficiency of a bottom quark sample. The neural networks have been trained on a sample of di-jets at $\sqrt{s} = 91$ GeV and tested on a statistically independent sample.

#### 10.6.2.1 Vertex Resolution

One of the most important variables in jet flavour tagging networks is the decay length of the secondary vertices. The vertex resolution of the SiD vertex detector has been assessed in the context of the analysis of the top Yukawa coupling and in a sample of Z decays to light quarks at $\sqrt{s} = 91$ GeV. Figure II-10.10 (left) shows the position of the reconstructed primary vertex in events containing two isolated leptons and four b quarks. The physics interaction has been generated at the position (0, 0, 0), and the beam spot constraint has been turned off for the purpose of this study.

Figure II-10.10 (right) shows the resolution of the primary vertex position versus the number of tracks originating from the primary interaction.

**Figure II-10.10** Position of the reconstructed primary vertex (left) and resolution of the primary vertex position as a function of the number of tracks originating from that vertex (right).

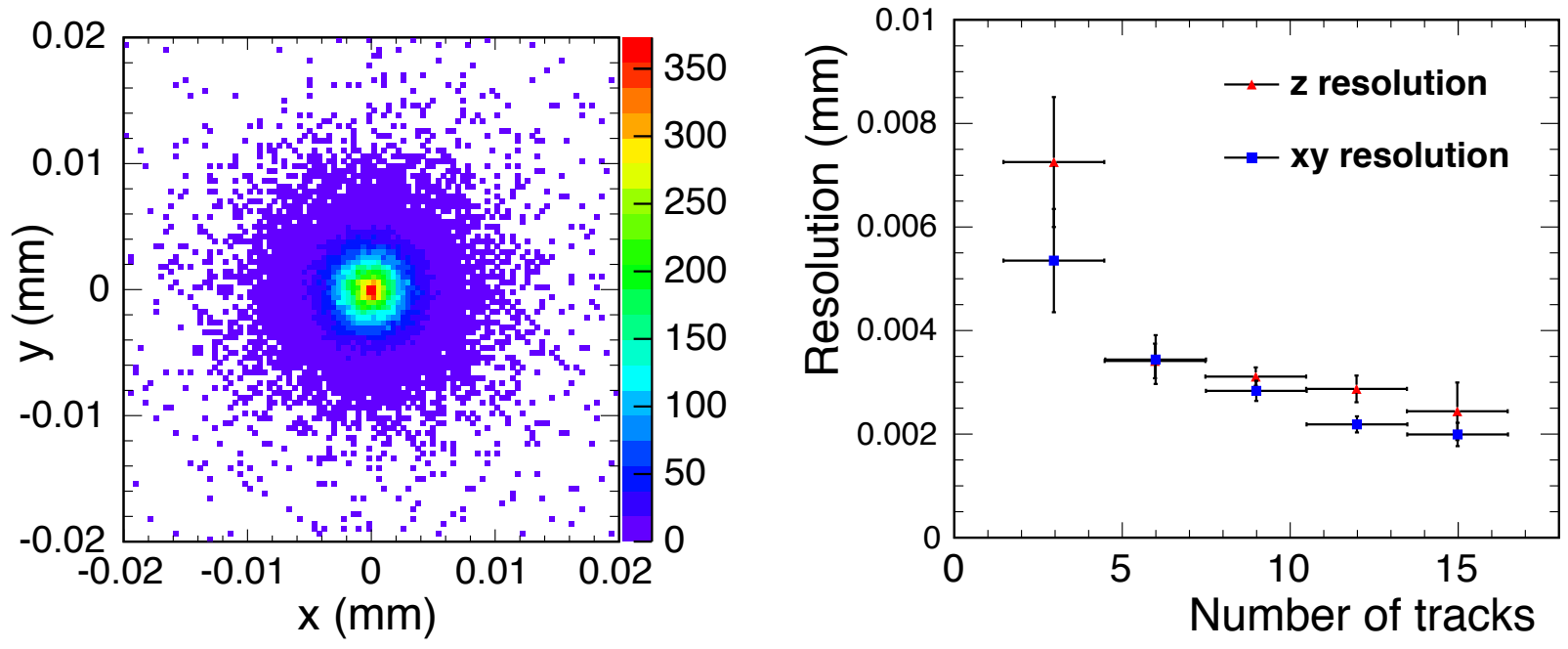

### 10.6.3 Energy and Mass Resolution

The design of the SiD detector concept has been optimised for jet energy resolution using the particle flow approach. This puts stringent requirements on the interplay of the various subdetectors and has led to the choice of calorimeters with a high degree of segmentation and transverse granularity. In addition, sophisticated reconstruction algorithms are necessary to obtain a jet energy resolution that allows to separate $\mathrm{W}$ and $\mathrm{Z}$ decays.

Figure II-10.11 (left) shows the PFA jet energy resolution without the effects from jet finding or background. To avoid a bias from possible tails, the rms90 value is computed to describe the energy or mass resolution of a particle flow algorithm. It is defined as the standard deviation of the distribution in the smallest range that contains 90% of the events. The events consist of $\mathrm{Z}'$ bosons of different masses decaying at rest to a pair of light quark jets. In these events the jet energy resolution is computed as the event energy resolution times $\sqrt{2}$, and the jet energy is half of $\sqrt{s}$ of the process.

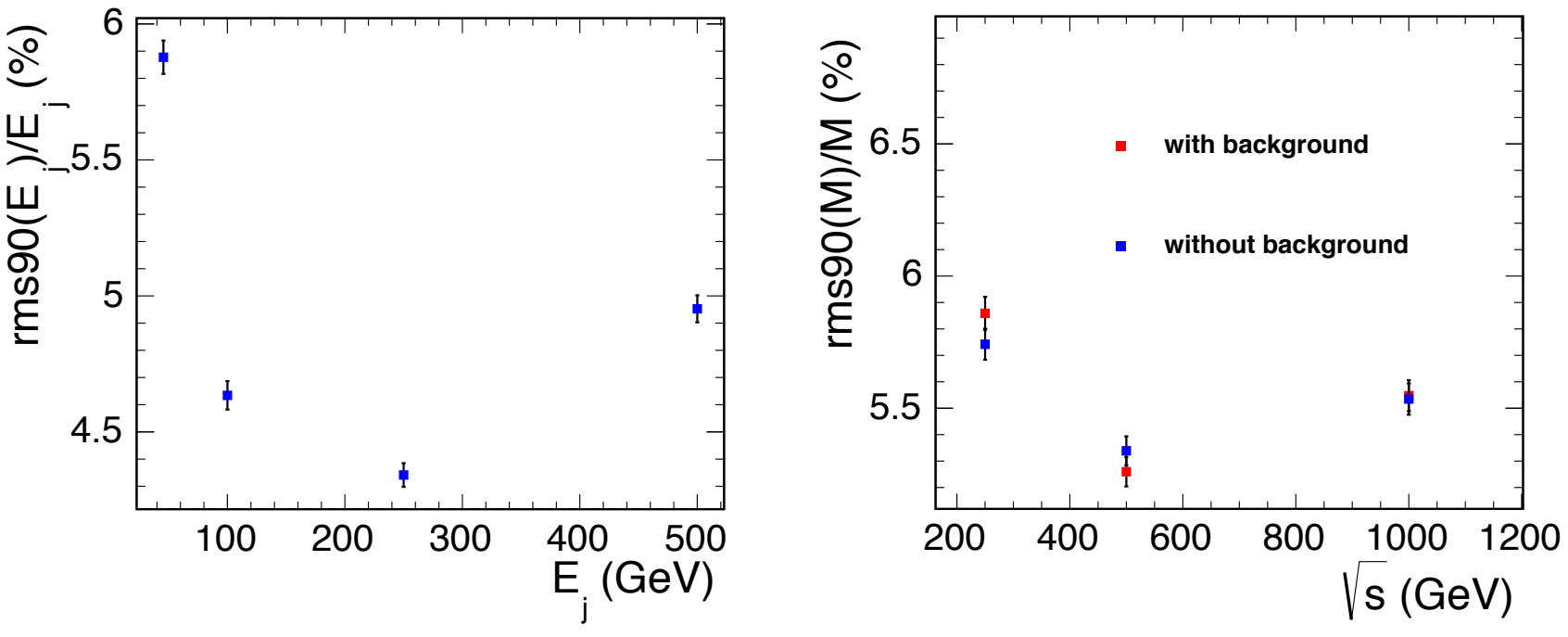

**Figure II-10.11** Left: Energy resolution of reconstructed $\mathrm{Z}'$ events of different masses decaying at rest to a pair of light quark jets. Right: Mass resolution of reconstructed $\mathrm{ZZ}$ events with and without the backgrounds at different values of $\sqrt{s}$. In these events, one $\mathrm{Z}$ boson decays invisibly and the other to a pair of jets.

Figure II-10.11 (right) shows the mass resolution of reconstructed $\mathrm{Z}$ bosons in $\mathrm{e}^+\mathrm{e}^- \rightarrow \mathrm{ZZ}$ events at different collision energies, where one $\mathrm{Z}$ decays to neutrinos, the other to two light quarks that give rise to two jets. The events have been clustered into two jets using the $k_t$ algorithm as implemented in the FastJet [175] package. The jets are combined to form a $\mathrm{Z}$ boson.

The value at each point in Figure II-10.11 (right) is computed as rms90(M)/M and given in per cent. The error bars indicate the error of the rms90 value of the distribution at each point. The addition of background from $\gamma\gamma \rightarrow \mathrm{hadrons}$ events and incoherent pairs results in only a minor change in resolution in these events. The difference can be explained by the additional background energy in the reconstructed jets balancing the small reconstruction bias towards lower energies in events without background.

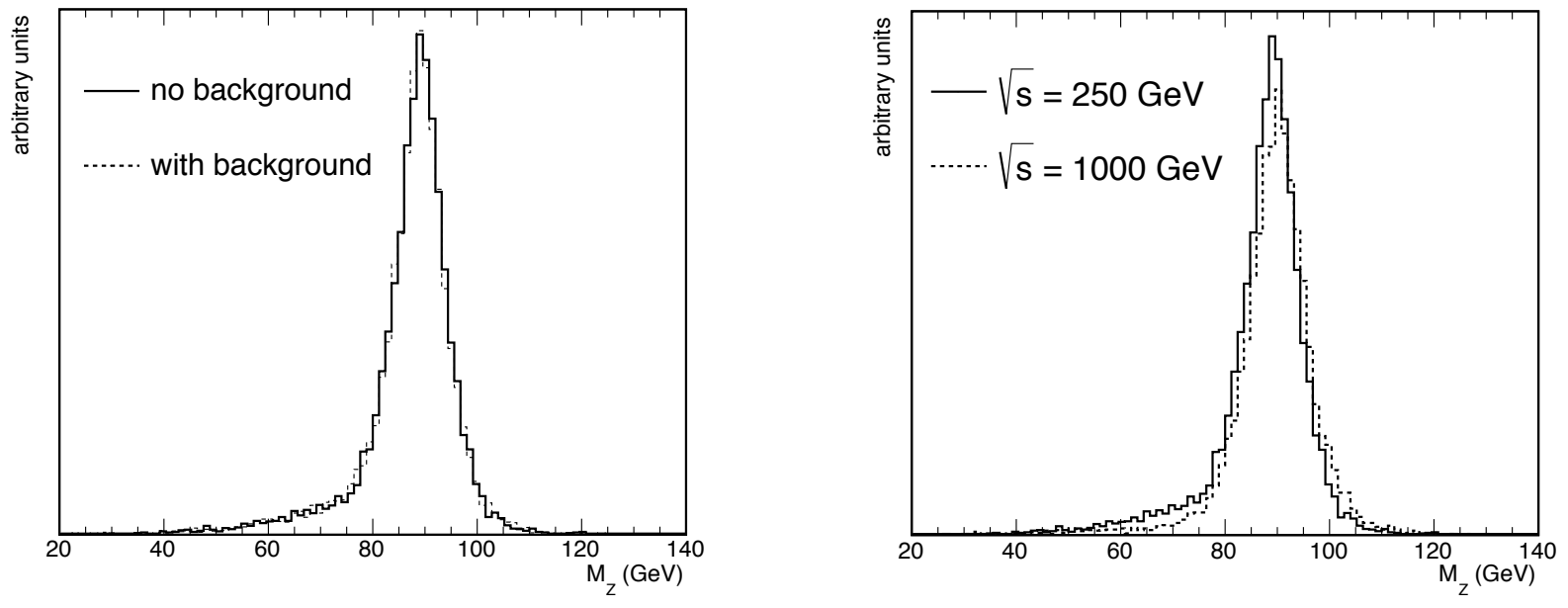

**Figure II-10.12** Comparison of the distributions of the reconstructed $\mathrm{Z}$ mass in $\mathrm{ZZ}$ events at $\sqrt{s}$ =250 GeV with and without background (left) and at two different values of $\sqrt{s}$ without background (right). In these events one $\mathrm{Z}$ decays invisibly and the other to a pair of jets.

Figure II-10.12 (left) shows a comparison of the distributions of the reconstructed $\mathrm{Z}$ mass in $\mathrm{ZZ}$ events simulated with and without background at $\sqrt{s}$ =250 GeV. The effect of the background on the reconstructed mass at these energies is partially mitigated by the jet clustering. Figure II-10.12

(right) shows a comparison of the distributions of of the reconstructed Z mass in ZZ events at $\sqrt{s}$ =250 GeV and $\sqrt{s}$ =1 TeV without background. It is expected that the relative shift of 1 GeV in the distributions can be reduced with improved calibration and calorimetric reconstruction.

## 10.7 Summary

The production of events for the benchmarking analyses described in Chapter 11 includes generation of a comprehensive set of Standard Model processes taking into account the ILC beam spectrum, full Geant4 detector simulation, generation, simulation and mixing of machine-induced background processes with the signal samples and PFA-based reconstruction of the mixed events.

To evaluate the performance of the sidloi3 detector concept and the reconstruction software as relevant for the DBD analyses kinematic properties of the machine-induced backgrounds, vertex reconstruction and flavour tagging, and performance of the particle flow algorithms have been studied in detail. While some critical work items have been identified in this process, the overall performance of the simulated detector and the simulation and reconstruction software as described in this section is adequate to carry out the DBD benchmarking analyses.

# Chapter 11
# SiD Benchmarking

Results of detailed simulation studies of the SiD detector are discussed in this chapter. First a short review of the studies performed for the SiD LOI is given. Several additional benchmark reactions were studied for the DBD. The generation of the events and common analysis tools used for several benchmark analyses are described briefly. Three new benchmark studies were performed at a centre-of-mass energy of 1 TeV. One of the LOI studies was repeated using the DBD version of the detector simulation and event reconstruction. In addition, the production of scalar tau leptons was investigated to illustrate the importance of the BeamCal detector.

## 11.1 Summary of the LOI Results

For the SiD LOI [63] several physics performance studies have been conducted to quantify the physics performance of the SiD detector concept. These studies were also used to broaden and emphasise the physics case of the ILC. From a list of physics benchmark reactions [176] six reactions were compulsory for the LOI submission. Three of the benchmark studies were conducted at a centre-of-mass energy of 250 GeV using a dataset of 250 $\text{fb}^{-1}$:

- $e^+e^- \to e^+e^-h,\ \mu^+\mu^-h$;
- $e^+e^- \to hZ,\ h \to c\bar{c},\ Z \to \nu\bar{\nu}, q\bar{q}$;
- $e^+e^- \to hZ,\ h \to \mu^+\mu^-,\ Z \to \nu\bar{\nu}, q\bar{q}$.

The remaining analyses were performed assuming $\sqrt{s}$ =500 GeV and using a dataset of 500 $\text{fb}^{-1}$:

- $e^+e^- \to \tau^+\tau^-$;
- $e^+e^- \to t\bar{t} \to 6\ \text{jets}$;
- $e^+e^- \to \tilde{\chi}_1^+\tilde{\chi}_1^-,\ \tilde{\chi}_2^0\tilde{\chi}_2^0$ assuming “Point 5” as defined in [176].

In addition to these compulsory LOI benchmark reactions SiD has also investigated the the $e^+e^- \to \tilde{b}\tilde{b},\ \tilde{b} \to b\tilde{\chi}_1^0$ process [177]. For all the LOI analyses, the SM Higgs mass was set to 120 GeV and the top quark mass was set to 174 GeV. In the following, a short summary of the results from the individual benchmark analyses will be given. All results from the LOI are summarised in Table II-11.17 as well. Compared to the DBD, the detector simulation for the LOI [63] was less detailed. In contrast to the DBD studies, beam-induced backgrounds were not considered for the LOI analyses.

For the analyses at $\sqrt{s} = 250$ GeV a dataset of 250 $\text{fb}^{-1}$ was used assuming a polarisation set of 80% right-handed $e^-$ and 30% left-handed $e^+$. This polarisation parameter set is referred to as “80eR” in this section. Additional signal samples with 80% left-handed $e^-$ and 30% right-handed $e^+$ (“80eL”) have also been investigated.

One of the key measurements is the model-independent measurement of the Higgs mass and production cross-section using the process $e^+e^- \to hZ,\ Z \to e^+e^-,\ \mu^+\mu^-,\ h \to \text{anything}$. The recoil mass against the Z can be measured very accurately using the leptonic $Z$ decays.

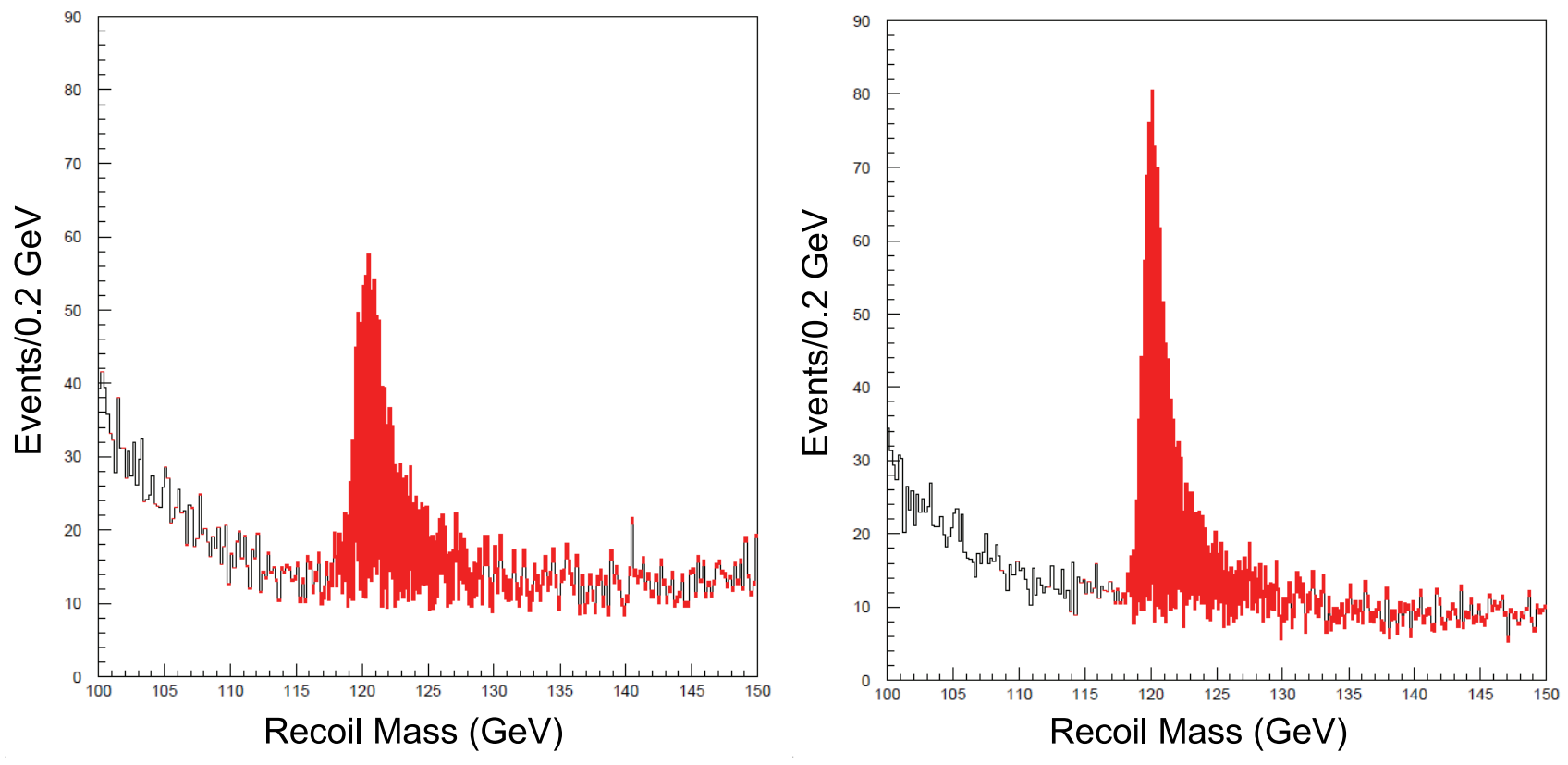

**Figure II-11.1**
Recoil mass distributions following selection cuts for $e^+e^-h$ (left) and $\mu^+\mu^-h$ (right) assuming 250 $fb^{-1}$ luminosity with 80eR initial state polarisation at $\sqrt{s} = 250$ GeV. The signal in red is added to the background in white.

The distributions for the recoil measurements in both the $e^+e^-h$ and $\mu^+\mu^-h$ channels are shown in Figure II-11.1. Main background sources include mainly di-boson production ($W^+W^-$, $ZZ$). The amount of $W^+W^-$ background can be greatly reduced by running exclusively with the 80eR configuration. A summary of the results of both leptonic $Z$ modes and using both 80eR and 80eL is given in Table II-11.1.

**Table II-11.1**
Summary of Higgs mass and $hZ$ cross-section results for different channels and the different luminosity assumptions at $\sqrt{s} = 250$ GeV. The error includes the measurement statistical error and the systematic error due to the finite statistics of the Monte Carlo training sample.

| 80eR ($fb^{-1}$) | 80eL ($fb^{-1}$) | Channel | $\Delta M_h$ (GeV) | $\Delta\sigma_{hZ}/\sigma_{hZ}$ |
|---|---|---|---|---|
| 250 | 0 | $e^+e^-h$ | 0.078 | 0.041 |
| 250 | 0 | $\mu^+\mu^-h$ | 0.046 | 0.037 |
| 250 | 0 | $e^+e^-h + \mu^+\mu^-h$ | 0.040 | 0.027 |
| 0 | 250 | $e^+e^-h$ | 0.066 | 0.067 |
| 0 | 250 | $\mu^+\mu^-h$ | 0.037 | 0.057 |
| 0 | 250 | $e^+e^-h + \mu^+\mu^-h$ | 0.032 | 0.043 |

Measuring the branching ratios of the Higgs boson is of vital importance to distinguish the SM Higgs boson from possible alternative scenarios. For the LOI the decays of the Higgs into $c\bar{c}$ and $\mu^+\mu^-$ have been studied at $\sqrt{s} = 250$ GeV using the Higgsstrahlung process, where the $Z$ decayed either in $q\bar{q}$ or $\nu\bar{\nu}$. The identification of the $h \to c\bar{c}$ decay mode took advantage of the excellent c-tagging capabilities of SiD (see [63]) and employed neural networks to separate the $c\bar{c}$ signal from the overwhelming $h \to b\bar{b}$ background. For the $c\bar{c}$ branching ratio, the finally achieved accuracies are 11% ($Z \to \nu\bar{\nu}$) and 6% ($Z \to q\bar{q}$), respectively.

For the rare Higgs decay into $\mu^+\mu^-$ the challenge is to extract the signal out of an overwhelming Standard Model background of mainly four-fermion events. While for the $Z \to \nu\bar{\nu}$ decay mode, it has been proven quite difficult to extract the signal, the LOI analysis has demonstrated sensitivity in the hadronic channel, selecting 7.6 signal events over a background event of 39.3 events with a signal selection efficiency of 62%. This yields a measurement of the cross-section for the process $e^+e^- \to hZ,\ h \to \mu^+\mu^-$ with a precision of 89%.

For the analyses at $\sqrt{s} = 500$ GeV a dataset of 500 $fb^{-1}$ was used with 80eR polarisation unless explicitly stated otherwise.

The first analysis using the 500 GeV dataset studies the process $e^+e^- \to \tau^+\tau^-$ and aims to measure the $\tau$ polarisation with high precision. The measurement of the $\tau$ polarisation allows a search for multi-TeV $Z'$ resonances. Tightly collimated jets with only a few tracks must be reconstructed to identify the underlying charged hadron and $\pi^0$ constituents. Therefore additional reconstruction algorithms were applied in a second pass of the reconstruction, which were dedicated for identifying $\tau$ decays. This leads to $\tau$ samples with purities of 85% or larger. To measure the mean $\tau$ polarisation over all $\tau$ production angles, $< P_\tau >$, the optimal observable technique [178, 179] is used. For this study two datasets with an integrated luminosity of 250 $fb^{-1}$ each were used, one with 80eR polarisation and one with 80eL polarisation. The true $< P_\tau >$ values are given by 0.528 (80eR) and -0.625 (80eL), respectively.

The results obtained for the measured polarisation are $< P_\tau >= 0.501 \pm 0.010$ (stat.) $\pm$ 0.006 (syst.) (80eR) and $< P_\tau >= -0.611 \pm 0.009$ (stat.) $\pm 0.005$ (syst.) (80eL).

The second benchmark analysis investigates top quark pair production at $\sqrt{s} = 500$ GeV, where both top quarks decay hadronically. The goal is to measure the cross-section, the top-quark mass and the forward-backward asymmetry ($A^t_{FB}$). Events are selected by requiring two b-tagged jets and the events being compatible with a six-jet configuration. The mass is then reconstructed applying a constrained kinematic fit to all pre-selected events and selecting only candidate events with a good fit probability. The top mass and cross-section is then extracted using a fit to the mass peak (see Figure II-11.2). This yields a top-quark mass of 173.918±0.053 GeV and a cross-section of 284.1±1.4 fb.

The forward-backward asymmetry measurement provides a window to new physics at the terascale [180]. A key tool for this analysis is the usage of the reconstructed vertex and jet charges. In this analysis both $A^b_{FB}$ and $A^t_{FB}$ are measured, yielding $A^b_{FB}$=$0.293 \pm 0.008$ and $A^t_{FB}$=$0.356 \pm 0.008$ assuming an integrated luminosity of 250 $\mathrm{fb}^{-1}$ with 80eR polarisation and 250 $\mathrm{fb}^{-1}$ of integrated luminosity with 80eL polarisation [181].

The last compulsory benchmark at $\sqrt{s}$ =500 GeV was a measurement of the masses of charginos and neutralinos in the processes $\mathrm{e^+e^-} \to \tilde{\chi}^+_1\tilde{\chi}^-_1 \to \mathrm{W^+W^-}\tilde{\chi}^0_1\tilde{\chi}^0_1$ and $\mathrm{e^+e^-} \to \tilde{\chi}^0_2\tilde{\chi}^0_2 \to \mathrm{ZZ}\tilde{\chi}^0_1\tilde{\chi}^0_1$. The analysis focused on the final states with four jets and missing energy and thereby focused on measuring the gaugino masses using di-jet final states from the two gauge bosons. The four reconstructed jets were then paired using a $\chi^2$ fit maximising the compatibility with the two di-jet pairs having equal masses. This then also allows separating the $\tilde{\chi}^+_1\tilde{\chi}^-_1$ from the $\tilde{\chi}^0_2\tilde{\chi}^0_2$ process by using the obtained di-jet resolutions (see Figure II-11.2).

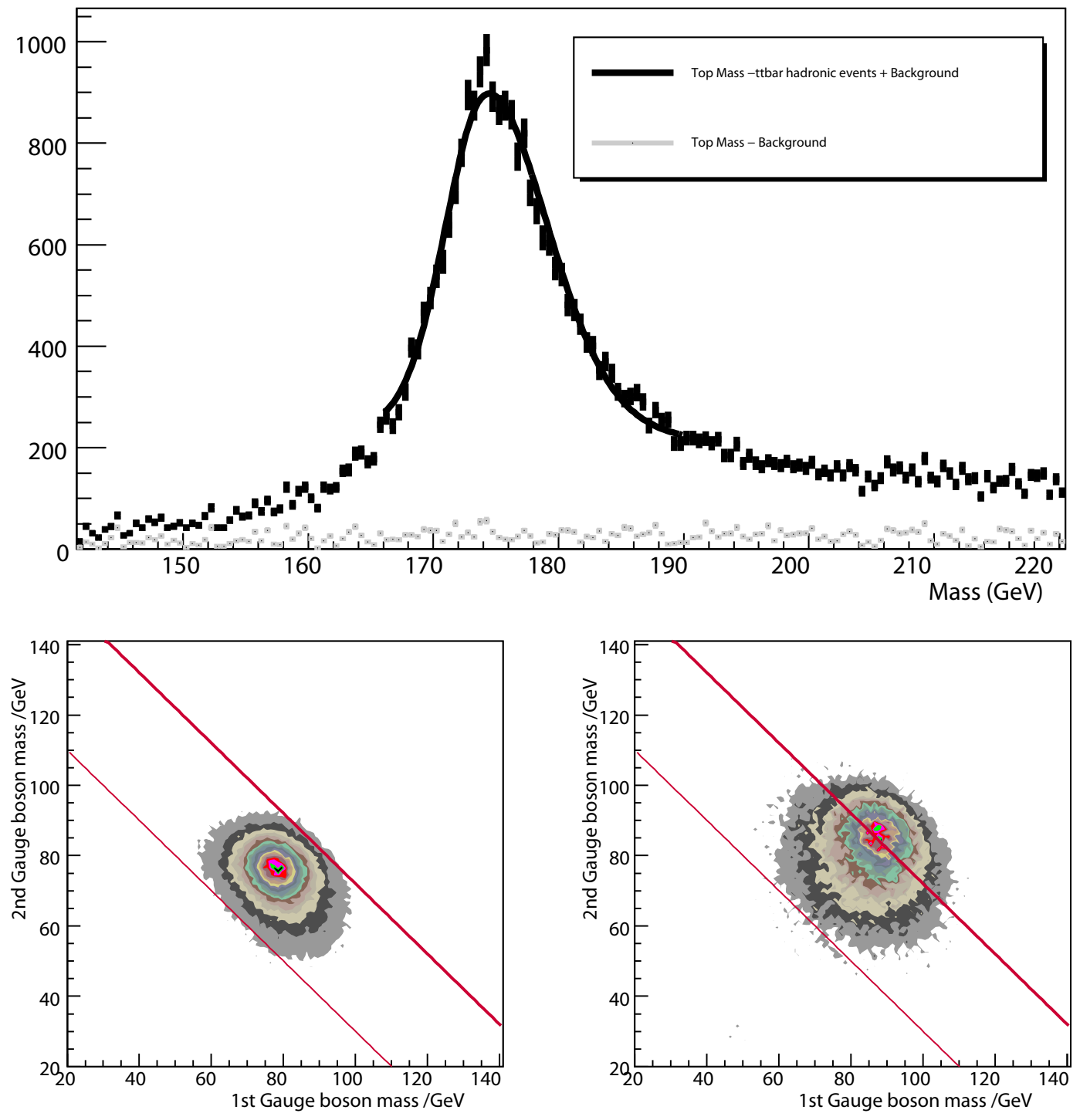

**Figure II-11.2** Top analysis: Distribution of the top invariant mass after kinematic fitting (top). Chargino/neutralino analysis: The reconstructed boson masses from the four jets, selecting chargino events with a pure chargino signal (bottom left ); and a pure neutralino signal (bottom right). The region between the two straight lines indicates the allowed chargino selection window.

The masses of the gauginos were then derived using a template fit to the energy distributions of the reconstructed bosons. For the $\tilde{\chi}^\pm_1$ and $\tilde{\chi}^0_1$ masses from the $\mathrm{e^+e^-} \to \tilde{\chi}^+_1\tilde{\chi}^-_1 \to \mathrm{W^+W^-}\tilde{\chi}^0_1\tilde{\chi}^0_1$ process the resulting mass uncertainties are 450 MeV and 160 MeV. Using the reaction $\mathrm{e^+e^-} \to \tilde{\chi}^0_2\tilde{\chi}^0_2 \to \mathrm{ZZ}\tilde{\chi}^0_1\tilde{\chi}^0_1$, uncertainties of 490 MeV and 280 MeV are obtained for the $\tilde{\chi}^0_2$ and $\tilde{\chi}^0_1$ masses, respectively, which are larger due to the smaller cross-section for $\mathrm{e^+e^-} \to \tilde{\chi}_2^0\tilde{\chi}_2^0$ and the lower sample purity.

In addition to these six compulsory benchmarks, SiD also looked into the $\mathrm{e^+e^-} \to \widetilde{\mathrm{b}}\widetilde{\mathrm{b}},\ \widetilde{\mathrm{b}} \to \mathrm{b}\tilde{\chi}_1^0$ process at $\sqrt{s} = 500$ GeV in scenarios where the sbottom is the next-to-lightest supersymmetric particle (NLSP) [182]. It has been shown, that SiD has sensitivity for this scenario up to sbottom masses close to the kinematic limit [177].

## 11.2 DBD Benchmark Reactions

For the DBD three additional benchmarks at $\sqrt{s} = 1$ TeV have been defined [183] using a Higgs mass of 125 GeV and a data sample of 1 $\mathrm{ab}^{-1}$. For half of this data sample 80% right-handed electron polarisation and 20% left-handed positron polarisation were assumed. 80% left-handed electron polarisation and 20% right-handed positron polarisation were assumed for the other half of the integrated luminosity.

The first benchmark is the process $\mathrm{e^+e^-} \to \mathrm{t\bar{t}h}$, where the Higgs decays into $\mathrm{b\bar{b}}$. The final states include eight jets (all-hadronic) and six jets, a lepton and missing energy (semi-leptonic). The aim is to measure the top Yukawa-coupling.

The second benchmark is a measurement of the Standard Model Higgs boson branching ratios into $\mathrm{b\bar{b}}$, $\mathrm{c\bar{c}}$, $\mathrm{W^+W^-}$, $\mathrm{gg}$ and $\mu^+\mu^-$ using the $\mathrm{e^+e^-} \to \nu_e\bar{\nu}_e\mathrm{h}$ production process.

The last benchmark is a measurement of the forward $\mathrm{e^+e^-} \to \mathrm{W^+W^-}$ pair cross-section considering both hadronic and leptonic decays of the $\mathrm{W^+W^-}$ pairs. The goal is to measure in situ the effective left-handed polarisation $(1 - P_{\mathrm{e}^-})(1 + P_{\mathrm{e}^+})/4$ for each of the two polarisation configurations.

Additionally, one benchmark from the LOI was repeated using the DBD detector layout and the updated simulation and reconstruction software. The $\mathrm{e^+e^-} \to \mathrm{t\bar{t}}$ process at $\sqrt{s} = 500$ GeV was chosen for this purpose.

The DBD benchmark results presented in this chapter correspond to our present level of understanding and can be refined further. This holds in particular for the measurement of the Higgs to $\mathrm{c\bar{c}}$ decay, where possibilities for further improvement of the analysis have already been identified.

### 11.2.1 Event Generation

#### 11.2.1.1 Monte Carlo event generators

The event generation was carried out in common for SiD and ILD. The WHIZARD Monte Carlo event generator was used for the generation of all $2 \to n$ processes, $n = 2...6$, where $n$ is the number final state fermions (e, $\mu$, $\tau$, u, d, s, c, b), and the two initial state particles are $\mathrm{e^+e^-}$, $e^+\gamma$, $e^-\gamma$, or $\gamma\gamma$. The $\mathrm{t\bar{t}h}$ signal process and the eight fermion backgrounds were generated using the PHYSSIM Monte Carlo program. All event samples were generated with 100% polarisation for the initial state electron and positron. The Higgs branching ratios listed in [184] were assumed for the event generation. A complete review of the event generation process is given in the chapter on common tasks and issues.

#### 11.2.1.2 Generated signal and background samples

Events from files with different 100% initial state polarisations and possibly different final states were combined at generator level to form "mixed files" with 80% electron and 20% (30%) positron polarisation for $\sqrt{s} = 1000$ (500) GeV. Only these mixed files were used as input to the full simulation and reconstruction in SiD. All mixed files are summarised in Table II-11.2.

Separate mixed background files were generated for the eight-fermion final states and for the dominant $\nu\bar{\nu}\mathrm{h}$ backgrounds, while everything else was combined together in the "all other SM processes" sample. The composition of the events in the "all other SM processes" sample is shown in Table II-11.3.

All Monte Carlo samples were processed using the full simulation of the SIDLOI3 detector. Beam-induced backgrounds from $\gamma\gamma \to \mathrm{hadrons}$ interactions and incoherent $\mathrm{e^+e^-}$ pairs were mixed with the physics events for all event samples. More details on the detector simulation and on the properties of the machine-induced backgrounds are given in Chapter 10.

**Table II-11.2**
Overview of the mixed samples used as input for the full detector simulation and reconstruction.

| Process | $\sqrt{s}$ (GeV) | $N_{Events}$ ($10^6$) | $\mathcal{L}$ (ab$^{-1}$) |
|---|---|---|---|
| $t\bar{t}h$ | 1000 | 0.4 | 52 |
| $t\bar{t}Z$, $t\bar{t}g^*$ | 1000 | 0.4 | 15 |
| $t\bar{t}$ | 1000 | 1.0 | 2.0 |
| $\nu\bar{\nu}h$, $h \to b\bar{b}$, $c\bar{c}$, $WW^*$, $gg$ | 1000 | 3.1 | 7.4 |
| $\nu\bar{\nu}h$, $h \to \mu^+\mu^-$ | 1000 | 0.5 | 6400 |
| $e\nu W$, $eeZ$, $\nu\nu Z \to e\nu qq$, $eeqq$, $\nu\nu qq$ | 1000 | 4.0 | 0.034 |
| $eeZ$, $\nu\nu Z$, $W^+W^- \to ee\mu\mu$, $\nu\nu\mu\mu$ | 1000 | 1.0 | 0.004 |
| $W^+W^-$ | 1000 | 6.0 | 2.0 |
| all other SM processes | 1000 | 6.0 | $1 \cdot 10^5 - 1.0$ |
| $t\bar{t}$ | 500 | 2.0 | 1.0 for each $m_{top}$ |
| $t\bar{t}$ background SM processes | 500 | 2.0 | varies |
| TOTAL | | 26 | |

**Table II-11.3**
Contents of the "all other SM processes" mixed sample at $\sqrt{s} = 1$ TeV. The weights of the individual processes were calculated assuming an integrated luminosity of $1$ ab$^{-1}$ for each of the two polarisations.

| Process | $\mathcal{L}$ (ab$^{-1}$) per pol. | $N_{Events}(10^5)$ P(e$^-$/e$^+$) $= -0.8/+0.2$ | $N_{Events}(10^5)$ P(e$^-$/e$^+$) $= +0.8/-0.2$ | Weight |
|---|---|---|---|---|
| $e\gamma \to e\gamma$ | $4 \cdot 10^{-5}$ | 0.5 | 0.5 | $2.5 \cdot 10^{+4}$ |
| $e^+e^- \to 2f, 4f$ | 0.034 | 3.7 | 2.0 | 29 |
| $e\gamma \to 3f$ | 0.003 | 3.5 | 3.1 | 330 |
| $e\gamma \to 5f$ | 0.25 | 3.1 | 2.1 | 4 |
| $e^+e^- \to 6f$ | 1.0 | 1.8 | 0.6 | 1 |
| $\gamma\gamma \to 2f$ | 0.001 | 5.7 | 5.7 | 7700 |
| $\gamma\gamma \to 4f$ | 0.083 | 2.5 | 2.5 | 12 |
| $\gamma\gamma \to$ minijets: | | | | |
| $4 < p_T < 40$ GeV | 0.012 | 9.2 | 9.2 | $80 - 9000$ |
| $p_T > 40$ GeV | 0.105 | 2.3 | 2.3 | 12 |

### 11.2.2 Analysis Tools

In the following, the software tools common to more than one of the DBD detector benchmark analyses are described.

#### 11.2.2.1 Jet finding

To reconstruct jets in hadronic final states, the Durham algorithm as implemented in LCFIPlus or the $k_t$ algorithm from the FastJet [185, 186] package were used. Especially for the reconstruction of jets in the forward direction, where the contribution from beam-related backgrounds is larger, the $k_t$ algorithm developed for hadron collisions is more suitable. This has already been demonstrated for the CLIC CDR [129]. More details and the parameters used for the jet finding are given in the descriptions of the individual analyses.

#### 11.2.2.2 Multivariate analysis tools

The traditional approach in high energy physics to separate a signal from backgrounds is based on a set of fixed cuts. However, for complex final states and large backgrounds this method is often not optimal. Hence multivariate analysis techniques like artificial neutral networks or boosted decision trees (BDTs) are commonly used today. The implementations of these models in the TMVA [187] software package were used for the benchmark analyses described in the following unless explicitly stated otherwise.

### 11.2.3 Measurement of the top Yukawa coupling

The measurement of the cross-section for the process $e^+e^- \rightarrow t\bar{t}h$ using two different final states is described in the following [188]. The Feynman diagrams for this process are shown in Figure II-11.3. Here $h$ is a Standard Model Higgs boson of mass 125 GeV. The diagram shown on the left represents the dominant contribution to the cross-section.

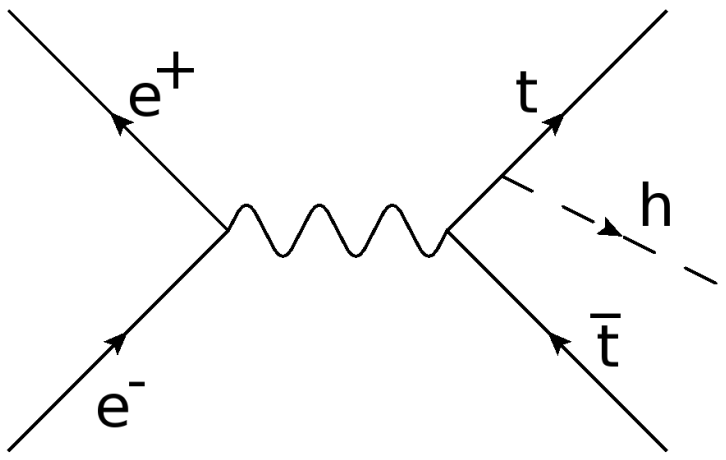

**Figure II-11.3** Diagrams for $t\bar{t}h$ production in $e^+e^-$ collisions.

Hence the measurement of the $t\bar{t}h$ cross-section at the ILC allows a direct extraction of the top Yukawa coupling, $y_t$, with good precision. The contribution to the cross-section from Higgs radiation off the intermediate $Z$ boson represents a small correction which needs to be taken into account in the extraction of $y_t$ from the measured cross-section. In the analysis presented here, the Higgs decay $h \rightarrow b\bar{b}$ is considered.

Two final states are investigated in the following:

- **8 jets:** In this case both $W$ bosons decay hadronically. Hence this final state contains eight jets out of which four originate from b-quark decays.
- **6 jets:** Here one $W$ boson decays hadronically and the other $W$ boson decays leptonically. The final state contains four b-jets, two further jets, an isolated lepton and missing energy. Only electrons and muons are considered as isolated leptons in the final state.

This study requires jet clustering in complex hadronic final states, missing energy reconstruction, flavour-tagging and reconstruction and identification of high energy leptons. Hence it represents a comprehensive check of the complete analysis chain and overall detector performance.

**Table II-11.4** Production cross-sections times branching ratios or production cross-sections for the signals and for the considered backgrounds. All samples were generated assuming a Standard Model Higgs mass of 125 GeV. The numbers for "other $t\bar{t}h$" processes in this table do not include either of the signal final states (see text). The $t\bar{t}Z$ and $t\bar{t}g^*$ samples do not contain events where both top quarks decay leptonically. The $t\bar{t}$ samples contain all possible decays of both top quarks.

| Type | Final state | P($e^-$) | P($e^+$) | Cross-section [× BR] (fb) |
|---|---|---|---|---|
| Signal | $t\bar{t}h$ (8 jets) | -80% | +20% | 0.87 |
| Signal | $t\bar{t}h$ (8 jets) | +80% | -20% | 0.44 |
| Signal | $t\bar{t}h$ (6 jets) | -80% | +20% | 0.84 |
| Signal | $t\bar{t}h$ (6 jets) | +80% | -20% | 0.42 |
| Background | other $t\bar{t}h$ | -80% | +20% | 1.59 |
| Background | other $t\bar{t}h$ | +80% | -20% | 0.80 |
| Background | $t\bar{t}Z$ | -80% | +20% | 6.92 |
| Background | $t\bar{t}Z$ | +80% | -20% | 2.61 |
| Background | $t\bar{t}g^* \rightarrow t\bar{t}b\bar{b}$ | -80% | +20% | 1.72 |
| Background | $t\bar{t}g^* \rightarrow t\bar{t}b\bar{b}$ | +80% | -20% | 0.86 |
| Background | $t\bar{t}$ | -80% | +20% | 449 |
| Background | $t\bar{t}$ | +80% | -20% | 170 |

An overview of the cross-sections for the signal final states as well as for the considered backgrounds is shown in Table II-11.4. For the measurement in the final state with six jets all other $t\bar{t}h$ events, i.e. all events where both top quarks decay leptonically or hadronically, or events where the Higgs boson does not decay into $b\bar{b}$, are treated as background. For the eight jets final state events where at least one top quark decays leptonically or where the Higgs boson does not decay into $b\bar{b}$ are considered as background.

### 11.2.3.1 Event reconstruction

As a first step of the event reconstruction chain, isolated leptons are searched for. The particle flow objects (PFOs) identified as muons or electrons are excluded from the jet reconstruction procedure. Only PFOs in the range $20^\circ < \theta < 160^\circ$ are considered in the following, because the particles originating from the signal processes are located in the central part of the detector while the beam-related backgrounds peak in the forward direction. The Durham jet clustering algorithm is used in the exclusive mode with six or eight jets. A b-tag value is obtained for each jet.

To form $\mathrm{W}$, top and Higgs candidates from the reconstructed jets, the following function is minimised for the final state with eight jets:

$$\frac{(M_{12}-M_{\mathrm{W}})^2}{\sigma_{\mathrm{W}}^2}+\frac{(M_{123}-M_{\mathrm{t}})^2}{\sigma_{\mathrm{t}}^2}+\frac{(M_{45}-M_{\mathrm{W}})^2}{\sigma_{\mathrm{W}}^2}+\frac{(M_{456}-M_{\mathrm{t}})^2}{\sigma_{\mathrm{t}}^2}+\frac{(M_{78}-M_{\mathrm{h}})^2}{\sigma_{\mathrm{h}}^2}, \quad \text{(II-11.1)}$$

where $M_{12}$ and $M_{45}$ are the invariant masses of the jet pairs used to reconstructed the $\mathrm{W}$ candidates, $M_{123}$ and $M_{456}$ are the invariant masses of the three jets used to reconstruct the top candidates and $M_{78}$ is the invariant mass of the jet pair used to reconstruct the Higgs candidate. $M_{\mathrm{W}}$, $M_t$ and $M_{\mathrm{h}}$ are the nominal $\mathrm{W}$, top and Higgs masses. The resolutions $\sigma_{\mathrm{W}}$, $\sigma_{\mathrm{t}}$ and $\sigma_{\mathrm{h}}$ were obtained from reconstructed jet combinations matched to $\mathrm{W}$, top and Higgs particles at generator level. The corresponding function minimised for the six jets final state is given by:

$$\frac{(M_{12}-M_{\mathrm{W}})^2}{\sigma_{\mathrm{W}}^2}+\frac{(M_{123}-M_{\mathrm{t}})^2}{\sigma_{\mathrm{t}}^2}+\frac{(M_{45}-M_{\mathrm{h}})^2}{\sigma_{\mathrm{h}}^2}. \quad \text{(II-11.2)}$$

### 11.2.3.2 Event selection

Events were selected using boosted decision trees as implemented in TMVA (see Section 11.2.2.2). The BDTs were trained separately for the eight and six jet final states.

The following input variables were used:

- the four highest b-tag values;
- the event thrust;
- a transition value from the Durham algorithm. For the six jet final state $Y_{6\to5}$ is used while $Y_{8\to7}$ is used for the eight jet final state;
- the number of reconstructed PFOs in the range $20^\circ < \theta < 160^\circ$;
- the number of identified isolated electrons or muons;
- the missing transverse momentum calculated from the reconstructed jets;
- the total visible energy defined as the scalar sum of all jet energies;
- the masses $M_{12}$, $M_{123}$ and $M_{45}$ as defined above.

For the eight jet final state two additional variables are included:

- $M_{456}$ and $M_{78}$ as defined above.

The output values of the BDTs for the signals and the different backgrounds are shown in Figure II-11.4 for both final states. To select events, cuts on the BDT output values are applied. The cuts were optimised by maximising the signal significance given by: $\frac{S}{\sqrt{S+B}}$, where $S$ is the number of signal events and $B$ is the number of background events. As an example, the reconstructed top and Higgs masses in six jet events after the cut on the BDT output are shown in Figure II-11.5. The selection efficiencies for signal events are 42% and 54% for the six and eight jet final states, respectively.

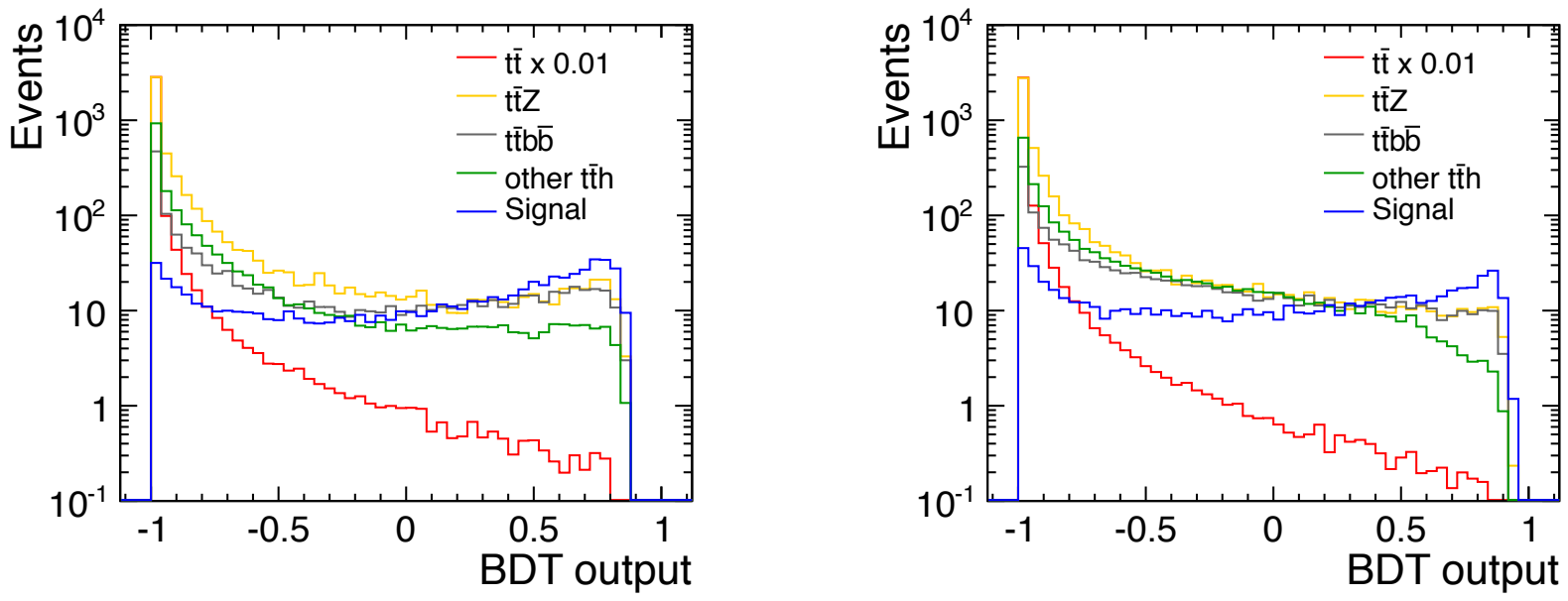

**Figure II-11.4** Output distributions of the BDTs for the eight (left) and six (right) jet final states. The signals are shown in blue while the backgrounds are shown in different colours. The distribution for $t\bar{t}$ was scaled by a factor $0.01$.

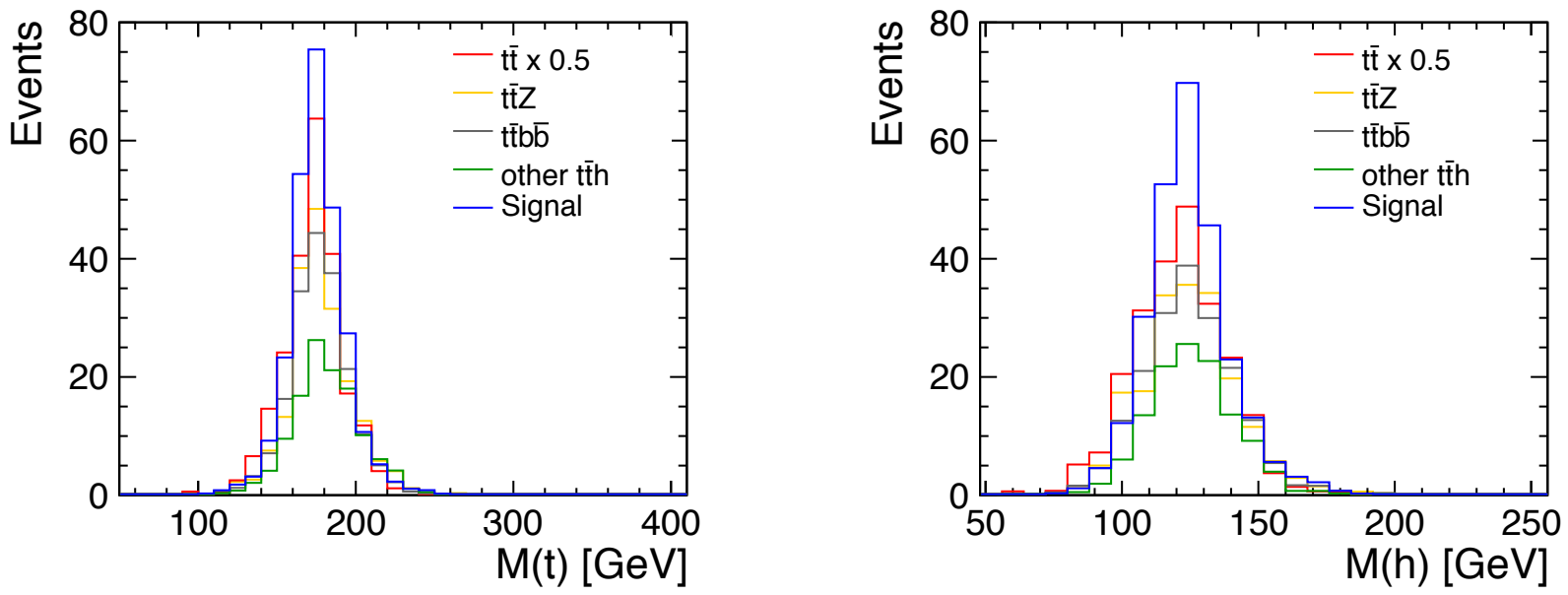

**Figure II-11.5** Reconstructed top (left) and Higgs (right) masses for selected (BDT output > 0.1978) six jet events. The signal is shown in blue while the backgrounds are shown in different colours. The distribution for $t\bar{t}$ was scaled by a factor $0.5$.

#### 11.2.3.3 Results on the cross-section and top Yukawa coupling

The cross-section can be directly obtained from the number of background-subtracted signal events after the event selection. Assuming an integrated luminosity of 1 $\mathrm{ab}^{-1}$, the cross-section can be measured with a statistical accuracy of $11.5\%$ using the eight jet final state and with a statistical accuracy of $13.2\%$ for the six jet final state. As a cross check, the analysis was repeated preselecting events with one isolated lepton for the six jet final state and events without isolated leptons for the eight jet final state. The differences in precision compared to the nominal analysis are negligible.

To extract the top Yukawa coupling from the measured cross-sections, signal Monte Carlo samples with different values of the top Yukawa coupling were generated. The dependence of the cross-section on the value of the coupling was fitted using a quadratic function. The following relation was found: $\frac{\Delta y_t}{y_t} = 0.52 \cdot \frac{\Delta\sigma}{\sigma}$. The factor between the cross-section uncertainty and the coupling uncertainty differs from $0.5$ due to the contribution from Higgsstrahlung to the $t\bar{t}h$ production cross-section. The uncertainties of the measured cross-sections translate to precisions on the top Yukawa coupling of $6.0\%$ and $6.9\%$ from the eight and six jet final states, respectively.

If both measurements are combined, the top Yukawa coupling can be extracted with a statistical accuracy of $4.5\%$. For 1 $\mathrm{ab}^{-1}$ of data with only P($e^-$) = -80%, P($e^+$) = +20% polarisation, this number would improve to $4.0\%$. The precision for the six jets final state could be improved further if $\tau$-leptons were included in the reconstruction.

### 11.2.4 Higgs branching fractions

Here the process to be studied is $e^+e^- \to \nu_e\bar{\nu}_e h$ at $\sqrt{s}$ =1 TeV, where $h$ is a SM Higgs boson of mass 125 GeV. The $\nu_e\bar{\nu}_e h$ final state occurs in the WW fusion (see Figure II-11.6) and Higgsstrahlung processes. This benchmark study provides a test of jet energy resolution, missing energy reconstruction, flavour-tagging, and reconstruction and identification of electrons and muons in the forward region.

The cross-section times branching ratio for the Standard Model Higgs boson decays into $b\bar{b}$, $c\bar{c}$, $W^+W^-$, $gg$ and $\mu^+\mu^-$ has been measured using the $e^+e^- \to \nu_e\bar{\nu}_e h$ production process. We focus on the WW fusion process which dominates at high energies since the cross-section grows as $\log(s)$. The datasets used for this analysis are shown in Table II-11.5.

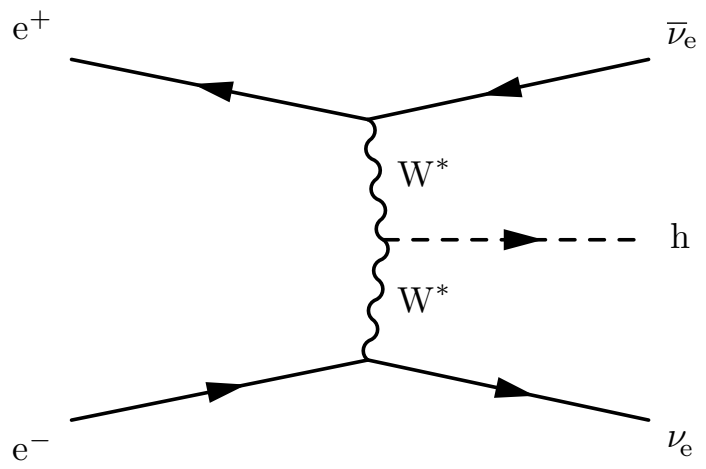

**Figure II-11.6**
Higgs production through the WW fusion process. Unlike the Higgsstrahlung interaction, the WW fusion interaction has a cross-section that grows as $\log(s)$ which results in making it the dominant interaction at $\sqrt{s}$ =1 TeV.

**Table II-11.5**
Simulated data samples used for the $\nu_e\bar{\nu}_e h$ analysis.

| Process | $P(e^-)/P(e^+)$ | $N_{Events}$ |
|---|---|---|
| Higgs | -80%/+20%<br>+80%/-20% | 1,544,378<br>1,544,398 |
| $e\nu W + eeZ + \nu\nu Z$ semileptonic | -80%/+20%<br>+80%/-20% | 6,570,292<br>5,080,159 |
| All other SM background mix | -80%/+20%<br>+80%/-20% | 3,232,672<br>2,814,719 |

The analysis of the Higgs boson decays to $b\bar{b}$, $c\bar{c}$, $W^+W^-$ and $gg$ are described first (see Section 11.2.4.1), followed by a section dedicated to $\mu^+\mu^-$ (see Section 11.2.4.4).

## 11.2.4.1 Event reconstruction for $h \to b\bar{b}, c\bar{c}, W^+W^-, gg$

To reconstruct events in the decay topology consistent with the two particle Higgs decays with no other visible event activity in the WW fusion interactions, events are clustered into two jets using the exclusive $k_t$ algorithm. This algorithm clusters particles apparently from beam activity into a beam jet thus avoiding introduction of those particles into the rest of the analysis.

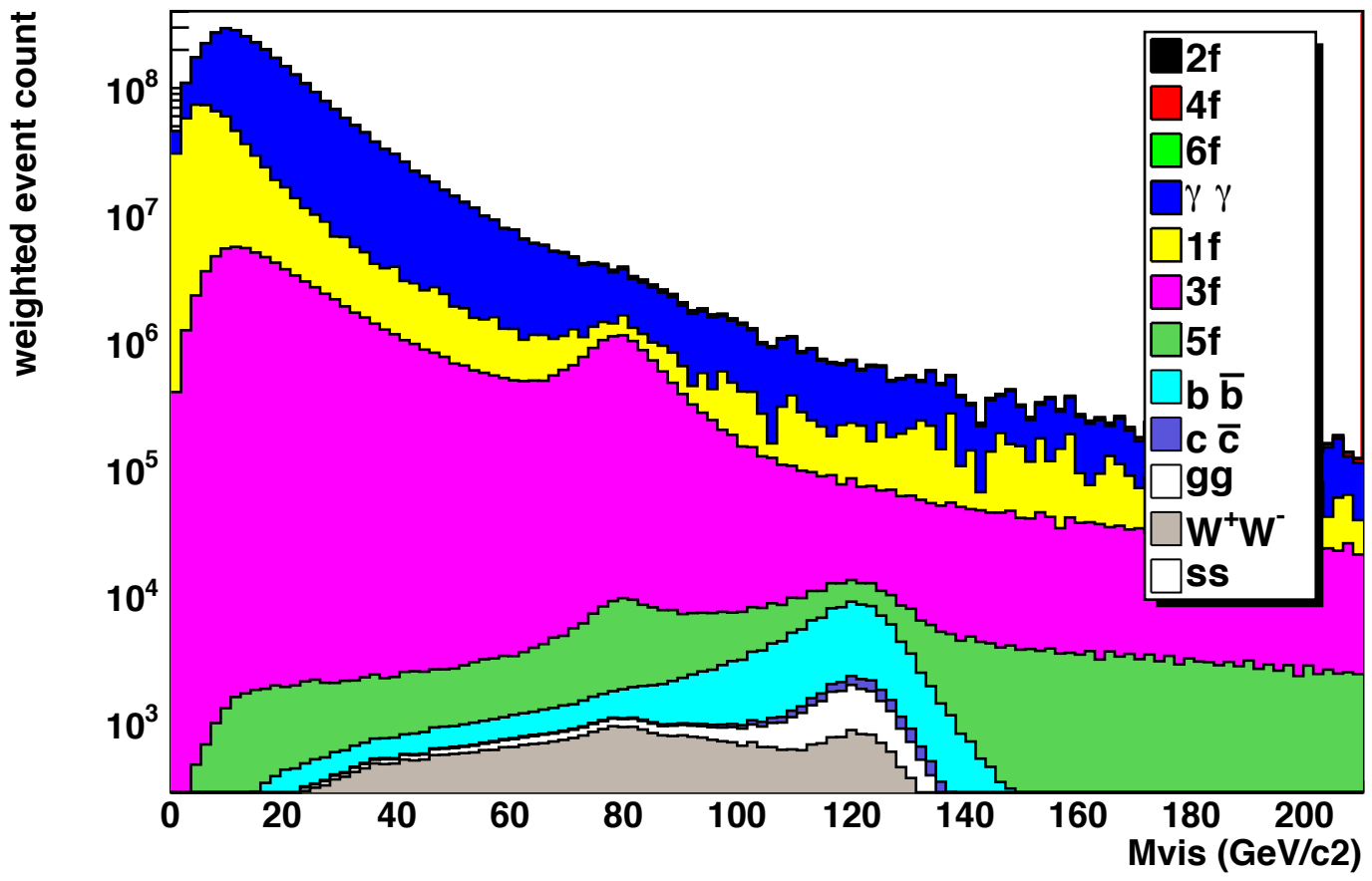

**Figure II-11.7**
Visible mass distributions for the backgrounds and $\nu_e\bar{\nu}_e h$ events for the various Higgs decay modes.

The jet clustering size parameter used for the exclusive $k_t$ algorithm was chosen to be 1.5 based on the visible mass resolutions. The visible mass distributions for the backgrounds and the $\nu_e\bar{\nu}_e h$ decay modes are shown in Figure II-11.7. From this figure one can see the poor reconstruction of the $h \to W^+W^-$ events. To address this, the PFO objects are used from the jets obtained when using the $k_t$ algorithm with a jet size parameter, R, of 0.7 and clustering into six jets. This was found to improve the rejection of beam particles.

The flavour-tagging is used as implemented in the LCFIPlus package which uses boosted decision trees on vertexing quantities to determine b-tag and c-tag probabilities for bottom and charm jets respectively. It is trained using samples of two-jet events from $Z \to b\bar{b}, c\bar{c}$ and $q\bar{q}$ at $\sqrt{s}$ = 250 GeV and the tagging is accordingly applied to all signal and background samples.

#### 11.2.4.2 Event Selection for $\mathrm{h} \to \mathrm{b\bar{b}}, \mathrm{c\bar{c}}, \mathrm{W^+W^-}, \mathrm{gg}$

Events are preselected based on the Higgs decay mode being studied using the criteria shown in Table II-11.6.

**Table II-11.6** Overview of the preselections for the different Higgs decay modes. The cuts as well as the efficiencies for signal and background events are shown.

| Higgs decay | Preselection cuts | Signal eff. | Background eff. |
|---|---|---|---|
| $\mathrm{h} \to \mathrm{b\bar{b}}$ | $50 < p^T_{\text{vis}} < 250$ GeV<br>$100 < E_{\text{vis}} < 400$ GeV<br>$110 < M_{\text{vis}} < 140$ GeV<br>$\lvert\cos(\theta_{\text{jet}})\rvert < 0.90$<br>$N_{\text{tracks}} > 15$<br>b-tag$_{1,2} > 0.2$ | 21.6% | $1.3 \times 10^{-6}$ |
| $\mathrm{h} \to \mathrm{c\bar{c}}$ | $50 < p^T_{\text{vis}} < 250$ GeV<br>$150 < E_{\text{vis}} < 400$ GeV<br>$115 < M_{\text{vis}} < 135$ GeV<br>$\lvert\cos(\theta_{\text{jet}})\rvert < 0.95$<br>$10 < N_{\text{tracks}} < 50$<br>b-tag$_{1,2} < 0.8$ | 12.2% | $1.3 \times 10^{-6}$ |
| $\mathrm{h} \to \mathrm{gg}$ | $50 < p^T_{\text{vis}} < 250$ GeV<br>$150 < E_{\text{vis}} < 400$ GeV<br>$100 < M_{\text{vis}} < 140$ GeV<br>$\lvert\cos(\theta_{\text{jet}})\rvert < 0.90$<br>$N_{\text{tracks}} > 20$<br>b-tag$_{1,2} < 0.8$<br>$M_{\text{jet},2} > 20$ GeV | 16.1% | $4.8 \times 10^{-6}$ |
| $\mathrm{h} \to \mathrm{W^+W^-}$ | $50 < p^T_{\text{vis}} < 250$ GeV<br>$150 < E_{\text{vis}} < 400$ GeV<br>$100 < M_{\text{vis}} < 140$ GeV<br>$\lvert\cos(\theta_{\text{jet}})\rvert < 0.90$<br>$N_{\text{tracks}} > 15$<br>b-tag$1, 2 < 0.8$<br>$M_{\text{jet},2} > 40$ GeV | 7.5% | $7.4 \times 10^{-6}$ |

After the preselection, Fisher discriminants, as implemented in TMVA, are then used to maximise the significance $(S/\sqrt{S+B})$ of the selection. They are trained using 10% of the signal and background events and done separately for the different polarisations and integrated luminosities. The cuts on the Fisher discriminant which maximise the significance for each decay mode are used to obtain the final results. The input variables for Fisher discriminants are given by:

- the b-tag and c-tag values of both jets;
- the masses and energies of both jets;
- the number of reconstructed PFOs;
- the number of high-momentum isolated electrons;
- the visible energy, mass and transverse momentum;
- the cosines of the polar angles of both jets;
- the angle between both jets in the plane perpendicular to the beam axis.
- the c-tag value divided by the sum of the b-tag and c-tag values for each jet (for $\mathrm{h} \to \mathrm{c\bar{c}}$ only)

The probability distributions for example Higgs decay modes are shown in Figure II-11.8. Plots showing the efficiency and significance curves vs. cuts on the Fisher discriminant output are shown in Figure II-11.9.

The composition of the samples of events passing all selections of the analysis are shown in Table II-11.7 for the polarisation P($\mathrm{e^-}$) = -80%, P($\mathrm{e^+}$) = +20% and an integrated luminosity of 500 $\mathrm{fb^{-1}}$.

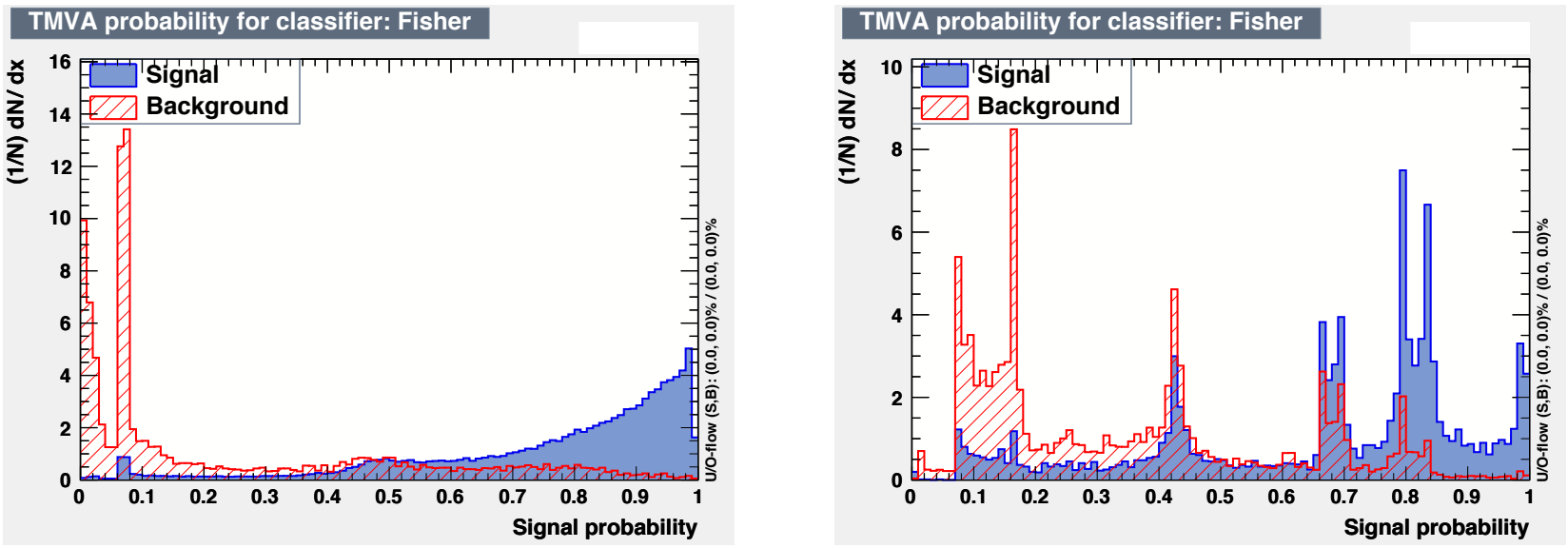

**Figure II-11.8** Probability distributions for selecting Higgs boson decays to $\mathrm{b\bar{b}}$ (left) and $\mathrm{c\bar{c}}$ (right) from the Fisher Discriminant.

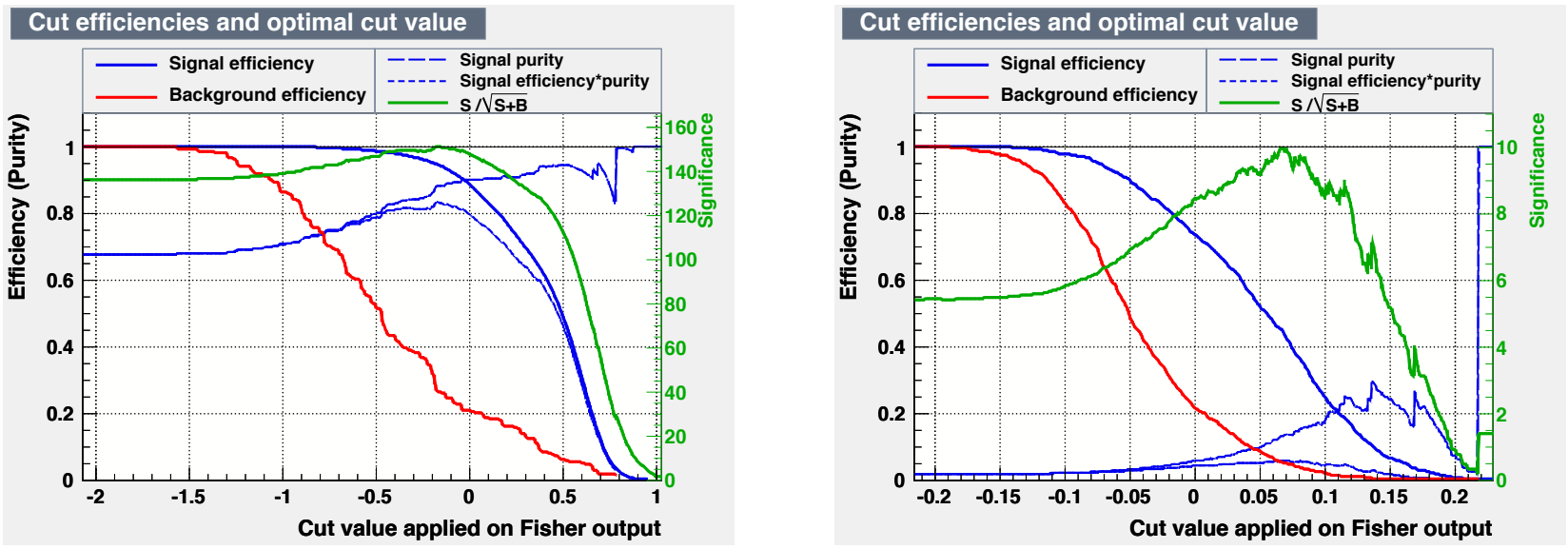

**Figure II-11.9** Efficiency and significance curves vs. cuts on the MVA Fisher discriminant output for the $\mathrm{h \to b\bar{b}}$ (left) and $\mathrm{c\bar{c}}$ selections (right).

The visible mass distribution for the $\mathrm{h \to b\bar{b}}$ selected events with the visible mass preselection cut removed is shown in Figure II-11.10 for an integrated luminosity of 500 $\mathrm{fb}^{-1}$ and the P($\mathrm{e}^-$) = -80%, P($\mathrm{e}^+$) = +20% polarisation.

**Table II-11.7** Composition of the events passing all analysis selections for each of the four Higgs decay mode studied in this analysis for the polarisations P($\mathrm{e}^-$) = -80%, P($\mathrm{e}^+$) = +20% and integrated luminosity of 500 $\mathrm{fb}^{-1}$.

| | $\mathrm{h \to b\bar{b}}$ (%) | $\mathrm{h \to c\bar{c}}$ (%) | $\mathrm{h \to gg}$ (%) | $\mathrm{h \to W^+W^-}$ (%) |
|---|---|---|---|---|
| $\mathrm{e^+e^-} \to 2$ fermions | 0.14 | 0.40 | 0.14 | 0.00 |
| $\mathrm{e^+e^-} \to 4$ fermions | 6.41 | 22.3 | 19.6 | 20.0 |
| $\mathrm{e^+e^-} \to 6$ fermions | 0.23 | 2.30 | 2.38 | 2.64 |
| $\gamma\gamma \to X$ | 1.19 | 8.11 | 11.0 | 11.9 |
| $\gamma e^+ \to X$ | 3.03 | 15.3 | 18.1 | 19.3 |
| $e^-\gamma \to X$ | 3.80 | 23.5 | 28.5 | 28.3 |
| $\mathrm{h \to b\bar{b}}$ | 83.7 | 7.00 | 0.36 | 0.96 |
| $\mathrm{h \to c\bar{c}}$ | 0.28 | 12.6 | 0.45 | 0.65 |
| $\mathrm{h \to gg}$ | 0.50 | 1.42 | 15.2 | 2.81 |
| $\mathrm{h \to WW^*}$ | 0.17 | 6.03 | 3.8 | 12.3 |

### 11.2.4.3 Results for $\mathrm{h \to b\bar{b}, c\bar{c}, W^+W^-, gg}$

The uncertainties on the cross sections times Higgs branching fractions, $\Delta(\sigma \times BR)$, are determined from the numbers of signal and background events passing each selection. The uncertainties for both polarisation configurations assuming an integrated luminosity of 500 $\mathrm{fb}^{-1}$ and for P($\mathrm{e}^-$) = -80%, P($\mathrm{e}^+$) = +20% polarisation for an integrated luminosity of 1 $\mathrm{ab}^{-1}$ at $\sqrt{s}$ =1 TeV are shown in Table II-11.8. For 1 $\mathrm{ab}^{-1}$ of integrated luminosity the precision for the $\mathrm{b\bar{b}}$ final state is about 0.5%. For the $\mathrm{W^+W^-}$ and $\mathrm{gg}$ final states precisions of about 3% can be reached while the $\mathrm{c\bar{c}}$ decay can currently be reconstructed with a precision of 7.6%. Given the results from the detector's flavour tagging performance studies, it is felt that ongoing efforts to refine the analysis of $\mathrm{h \to c\bar{c}}$ will certainly lead to significant improvements in the error on $\sigma \times BR(\mathrm{h \to c\bar{c}})$.

In Table II-11.9 the results obtained when not including the backgrounds from $\gamma\gamma$ interactions or with five fermion final states are shown to illustrate the impact of these contributions. The precision for the $\mathrm{c\bar{c}}$ decay improves by 20% while all other Higgs decays are only marginally affected. Such backgrounds can be reduced in practice by optimising the forward detector design to minimise the incoherent $\mathrm{e^+e^-}$ pair background, and through the inclusion of the Detector Integrated Dipole.

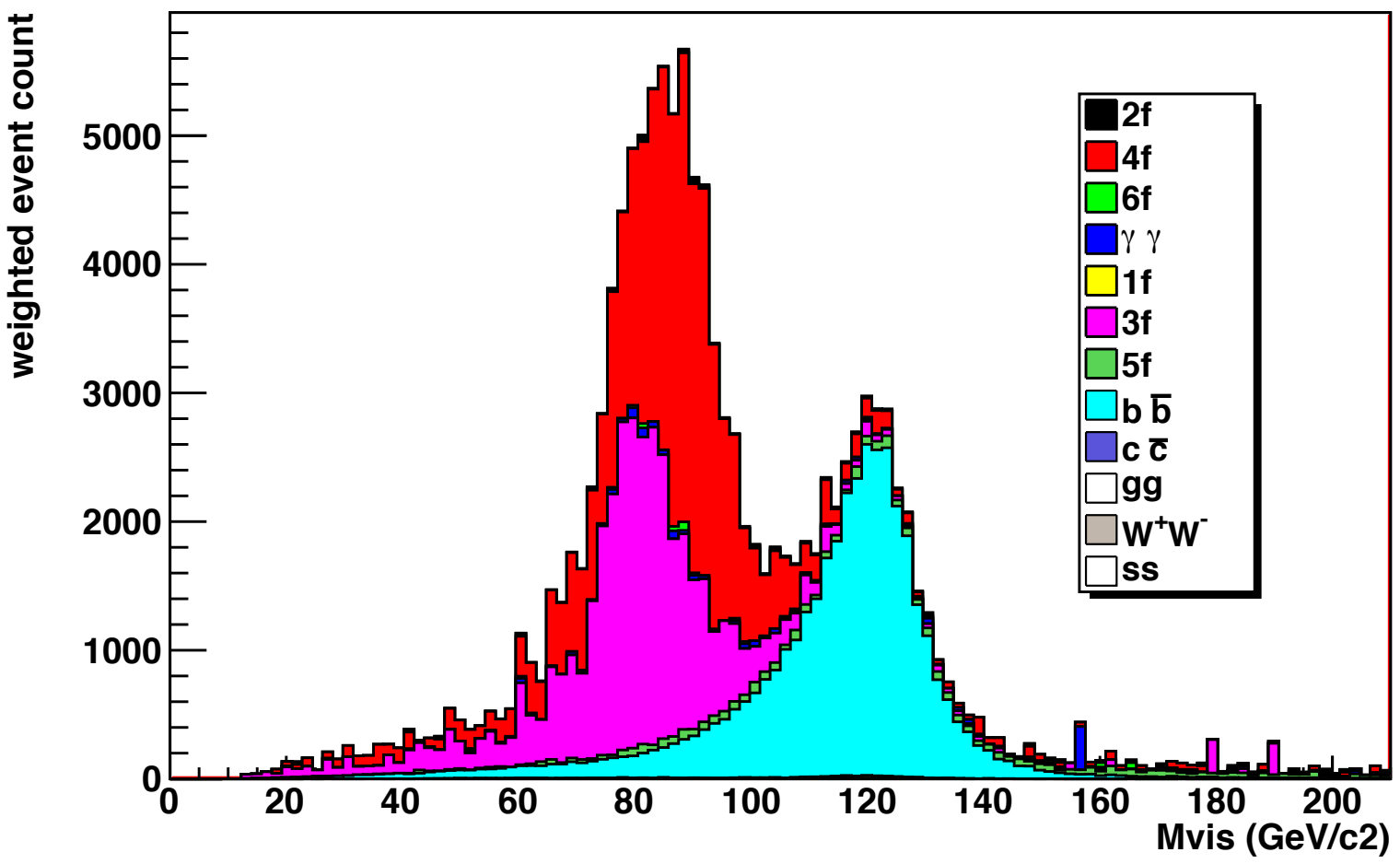

**Figure II-11.10**
The visible mass distribution for the $h \to b\bar{b}$ selected events without the visible mass preselection cut for 500 $\text{fb}^{-1}$ and the P($e^-$) = -80%, P($e^+$) = +20% polarisation configuration.

**Table II-11.8**
Relative uncertainties on the Higgs $\sigma \times BR$ expected at $\sqrt{s}$ =1 TeV using the SiD detector with integrated luminosities of 500 $\text{fb}^{-1}$ and 1 $\text{ab}^{-1}$ and polarisation sets P($e^-$) = -80%, P($e^+$) = +20% and P($e^-$) = +80%, P($e^+$) = -20%.

| | $\mathcal{L}$ = 500 $\text{fb}^{-1}$ | | $\mathcal{L}$ = 1 $\text{ab}^{-1}$ |
|---|---|---|---|
| | P($e^-$) = -80% P($e^+$) = +20% | P($e^-$) = +80% P($e^+$) = -20% | P($e^-$) = -80% P($e^+$) = +20% |
| $h \to b\bar{b}$ | 0.0067 | 0.046 | 0.0047 |
| $h \to c\bar{c}$ | 0.108 | 0.843 | 0.076 |
| $h \to gg$ | 0.044 | 0.294 | 0.031 |
| $h \to W^+W^-$ | 0.047 | 0.346 | 0.033 |

## 11.2.4.4 Event selection and results for $h \to \mu^+\mu^-$

Higgs boson decays to $\mu^+\mu^-$ are selected by requiring that there be two and only two muons in the event, that they have opposite charge, that the sum of their energies be less than 400 GeV, that the closest distance of approach of each muon to the primary event vertex be less than 7 microns in the plane transverse to the beam direction, and that the sum of the transverse momenta of the two muons be greater than 100 GeV.

In addition, the total missing energy in the event must be greater than 450 GeV, the total missing transverse momentum must be greater than 55 GeV, the number of charged particle flow objects with energy greater than 15 GeV must be less than 4, and there must not be any electrons with energy greater than 15 GeV. The muon pair invariant mass distribution following these cuts is shown for signal and background in Figure II-11.11.

The cross section times branching ratio is measured by counting the number of events with $\mu^+\mu^-$ mass in the range $124 < M_{\mu^+\mu^-} < 126$ GeV. The number of events for signal and background in this mass window is shown in Table II-11.10. For background processes the number of events in the mass window $124 < M_{\mu^+\mu^-} < 126$ GeV is calculated under the assumption that the $\mu^+\mu^-$ mass distribution is flat in the range $115 < M_{\mu^+\mu^-} < 145$ GeV. The error on the cross section times branching ratio for $h \to \mu^+\mu^-$ under different assumptions for background and luminosity is summarised in Table II-11.11.

**Table II-11.9**
Relative uncertainties on the Higgs $\sigma \times BR$ expected at $\sqrt{s}$ =1 TeV using the SiD detector with integrated luminosities of 500 $\text{fb}^{-1}$ and 1 $\text{ab}^{-1}$ and polarisation sets P($e^-$) = -80%, P($e^+$) = +20% and P($e^-$) = +80%, P($e^+$) = -20% with the five fermion and $\gamma\gamma$ backgrounds removed.

| | $\mathcal{L}$ = 500 $\text{fb}^{-1}$ | | $\mathcal{L}$ = 1 $\text{ab}^{-1}$ |
|---|---|---|---|
| | P($e^-$) = -80% P($e^+$) = +20% | P($e^-$) = +80% P($e^+$) = -20% | P($e^-$) = -80% P($e^+$) = +20% |
| $h \to b\bar{b}$ | 0.0065 | 0.026 | 0.0046 |
| $h \to c\bar{c}$ | 0.100 | 0.733 | 0.071 |
| $h \to gg$ | 0.040 | 0.234 | 0.028 |
| $h \to W^+W^-$ | 0.042 | 0.260 | 0.030 |

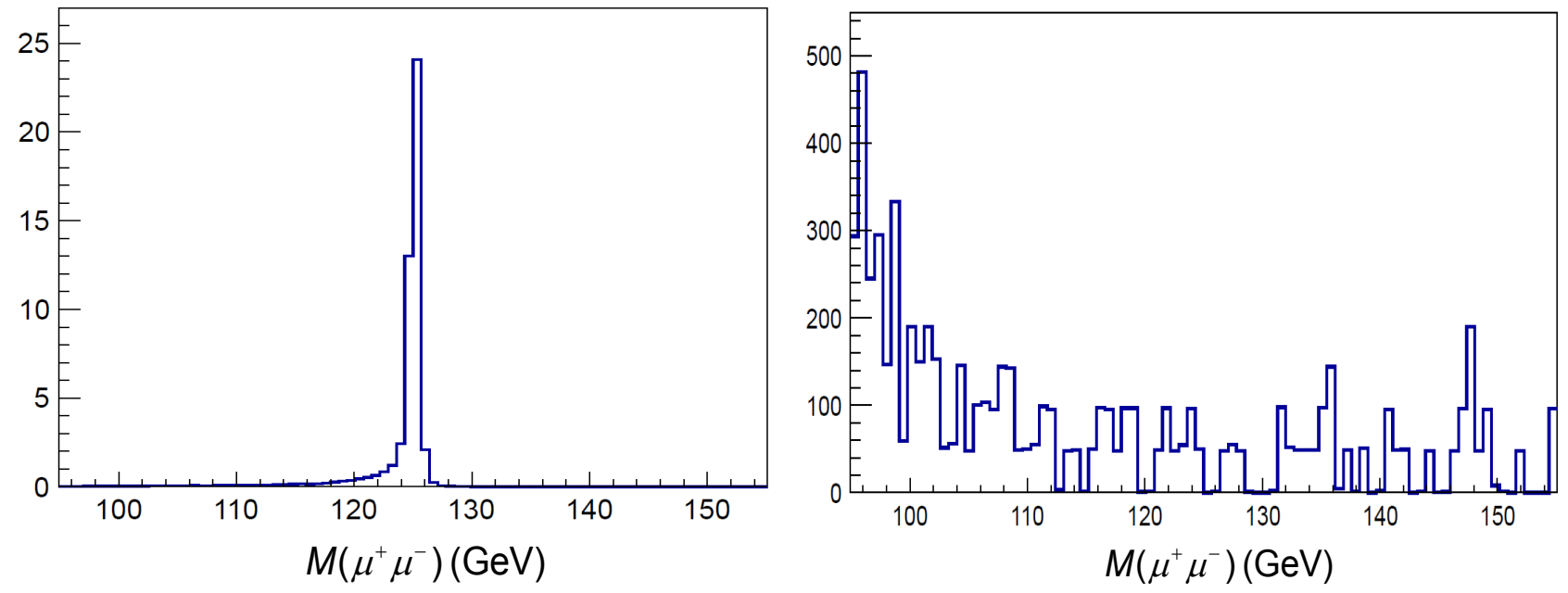

**Figure II-11.11** Muon pair mass for $h \to \mu^+\mu^-$ (left) and for all Standard Model background (right) following all cuts. The plots are normalised to $1\ ab^{-1}$ luminosity.

**Table II-11.10** Number of events passing all cuts with $124 < M_{\mu^+\mu^-} < 126$ GeV for 500 $fb^{-1}$ luminosity. The error from Monte Carlo statistics is indicated. The number of events for $e^-\gamma$ processes includes events from the $\gamma e^+$ charge conjugate process.

| Process | $N_{Events}$ |
|---|---|
| $e^+e^- \to \nu_e\bar{\nu}_e h \to \nu_e\bar{\nu}_e\mu^+\mu^-$ | $20.0 \pm 0.1$ |
| $e^+e^- \to \nu_e\bar{\nu}_e\mu^+\mu^-$ | $17.4 \pm 5.3$ |
| $e^+e^- \to \nu_\mu\bar{\nu}_\mu\mu^+\mu^-$ | $7.9 \pm 3.5$ |
| $e^+e^- \to \nu_\tau\bar{\nu}_\tau\mu^+\mu^-$ | $1.6 \pm 1.6$ |
| $e^+e^- \to t\bar{t} \to b\bar{b}\nu_\mu\bar{\nu}_\mu\mu^+\mu^-$ | $0.8 \pm 0.2$ |
| $e^+e^- \to \nu_\tau\bar{\nu}_\mu\tau^+\mu^-$ | $0.2 \pm 0.1$ |
| $\gamma\gamma \to \nu_\mu\bar{\nu}_\mu\mu^+\mu^-$ | $29.3 \pm 6.7$ |
| $e^-\gamma \to e^-\nu_\mu\bar{\nu}_\mu\mu^+\mu^-$ | $4.8 \pm 2.7$ |
| $e^-\gamma \to \nu_e\nu_\mu\bar{\nu}_\mu\mu^-\mu^+\mu^-$ | $1.6 \pm 1.6$ |
| $e^-\gamma \to \nu_e\bar{\nu}_\mu\mu^-\mu^+\mu^-$ | $0.2 \pm 0.2$ |
| $e^+e^- \to q\bar{q}\nu_\mu\bar{\nu}_\mu\mu^+\mu^-, q \neq b$ | $0.2 \pm 0.1$ |

### 11.2.5 Measurement of beam polarisation using $W^+W^-$ pairs

The baseline ILC design at $\sqrt{s}$ =1 TeV includes longitudinal electron and positron beam polarisations of 80% and 20% respectively. Polarimeters upstream and downstream of the collision point measure the average beam polarisation before and after collision, but cannot provide an estimate of the luminosity weighted beam polarisation without complex modelling of the beam collision process. Physics processes that are sensitive to beam polarisation, however, can directly measure the luminosity weighted beam polarisation.

The process $e^+e^- \to W^+W^-$ is very sensitive to beam polarisation. It is dominated by t-channel neutrino exchange in the forward region, where only left-handed electrons and right-handed positrons contribute, and continues to exhibit polarisation dependence in the central and backward regions through s-channel diagrams. The production of $W^+W^-$ in the forward region has the advantage of high statistics and negligible sensitivity to new physics.

The primary objective of this benchmark is the measurement of the effective left- and right-handed polarisations

$$P_{e^-}(L)_{\text{eff}} = \frac{(1-P_{e^-})(1+P_{e^+})}{4}, \qquad P_{e^-}(R)_{\text{eff}} = \frac{(1+P_{e^-})(1-P_{e^+})}{4}$$

**Table II-11.11** Cross section times branching ratio resolution for $h \to \mu^+\mu^-$ under different luminosity and background assumptions. The Monte Carlo statistical error for each entry in the table is shown.

| Background Processes | L ($fb^{-1}$) | $N_{\text{background}}$ | $N_{\text{signal}}$ | $\frac{\Delta(\sigma\times BR)}{\sigma\times BR}$ |
|---|---|---|---|---|
| $e^+e^- \to f\bar{f},\ \ f\bar{f}f\bar{f},\ \ t\bar{t}$ only | 500 | $28.0 \pm 6.5$ | $20.0 \pm 0.1$ | $0.35 \pm 0.02$ |
| | 1000 | $56.0 \pm 13.0$ | $40.0 \pm 0.2$ | $0.24 \pm 0.02$ |
| all background | 500 | $64.1 \pm 9.9$ | $20.0 \pm 0.1$ | $0.46 \pm 0.03$ |
| | 1000 | $128.2 \pm 19.8$ | $40.0 \pm 0.2$ | $0.32 \pm 0.02$ |

through a measurement of the W production and decay angles $\Theta$ and $\theta$ in $\mathrm{e^+e^-} \rightarrow \mathrm{W^+W^-} \rightarrow q\bar{q}q\bar{q},\ \ q\bar{q}l^-\bar{\nu}$ at $\sqrt{s}$ =1 TeV in the forward region with $0.8 < \cos\Theta$. This measurement is made independently for two different configurations, $(P_{\mathrm{e^-}}, P_{\mathrm{e^+}}) = (-0.8, 0.2)\ ,\ \ (+0.8, -0.2)$ assuming 500 $\mathrm{fb}^{-1}$ integrated luminosity for each configuration. This benchmark study provides a test of jet reconstruction of boosted W bosons in forward regions, jet energy resolution, differential luminosity measurement, and reconstruction and identification of leptons in the forward region.

Once $P_{\mathrm{e^-}}(L)_{\mathrm{eff}}$ and $P_{\mathrm{e^-}}(R)_{\mathrm{eff}}$ have been measured the actual electron and positron polarisations $P_{\mathrm{e^-}}$ and $P_{\mathrm{e^+}}$ can be extracted using the relations

$$P_{\mathrm{e^-}} = b - a \pm \sqrt{(b-a)^2 - 2(a+b) + 1}$$
$$P_{\mathrm{e^+}} = P_{\mathrm{e^-}} - 2(b-a)\ .$$

where $a = P_{\mathrm{e^-}}(L)_{\mathrm{eff}}$ and $b = P_{\mathrm{e^-}}(R)_{\mathrm{eff}}$. The sign ambiguity is resolved using knowledge of the signs of the beam polarisations.

The production angle $\Theta$ is defined to be the polar angle of the $\mathrm{W^-}$ in the $\mathrm{W^+W^-}$ rest frame. No attempt is made to measure the charges of the hadronically decaying W's in the fully hadronic topology $\mathrm{W^+W^-} \rightarrow \mathrm{q\bar{q}q\bar{q}}$, and so $|\cos\Theta|$ is used in this case. The charge of the decay lepton determines the charge of the W for the semileptonic topology $\mathrm{e^+e^-} \rightarrow \mathrm{W^+W^-} \rightarrow \mathrm{q\bar{q}}l^-\bar{\nu}$.

For the semileptonic topology the decay angle $\theta$ is defined to be the polar angle of the fermion in the $\mathrm{W^-}$ rest frame ($\mathrm{W^-} \rightarrow l^-\bar{\nu}$) or the antifermion in the $\mathrm{W^+}$ rest frame ($\mathrm{W^+} \rightarrow l^+\nu$). For the fully hadronic topology $|\cos\theta|$ is used where $\theta$ is the polar angle of either of the two jets in the $\mathrm{W^-}$ rest frame. The effective right handed polarisation $P_{\mathrm{e^-}}(R)_{\mathrm{eff}}$ is measured rather poorly in the forward region $0.8 < \cos\Theta$ due to the dominance of the polarisation configuration with left-handed electrons and right handed positrons. Therefore the effective polarisation parameters $P_{\mathrm{e^-}}(L)_{\mathrm{eff}}$ and $P_{\mathrm{e^-}}(R)_{\mathrm{eff}}$ are also measured using the entire solid angle $-1 < \cos\Theta < 1$, with the caveat that the result is only valid for Standard Model $\mathrm{W^+W^-}$ production. Only the semileptonic topology is used for measuring the polarisation outside the forward region.

Finally, it can also be assumed that the magnitudes of the beam polarisations do not change as the signs of the polarisations are changed. In this case data from the two polarisation configurations $(P_{\mathrm{e^-}}, P_{\mathrm{e^+}}) = (-0.8, 0.2)\ ,\ \ (+0.8, -0.2)$ can be combined in order to measure $|P_{\mathrm{e^+}}|$and $|P_{\mathrm{e^-}}|$. The measured parameters $P_{\mathrm{e^-}}(L)_{\mathrm{eff}}$ and $P_{\mathrm{e^-}}(R)_{\mathrm{eff}}$ are replaced by

$$\alpha = \frac{(1+|P_{\mathrm{e^-}}|)(1+|P_{\mathrm{e^+}}|)}{4}\ , \qquad \beta = \frac{(1-|P_{\mathrm{e^-}}|)(1-|P_{\mathrm{e^+}}|)}{4}$$

and the absolute polarisation values are given by

$$|P_{\mathrm{e^-}}| = \alpha - \beta \pm \sqrt{(\alpha-\beta)^2 - 2(\alpha-\beta) + 1}$$
$$|P_{\mathrm{e^+}}| = 2(\alpha-\beta) - P_{\mathrm{e^-}}\ \ .$$

#### 11.2.5.1 Event reconstruction

Particle Flow Objects (PFOs) were used as input to the analysis. Isolated electrons, muons and photons must be identified to separate W pairs from background processes and to classify a $\mathrm{W^+W^-}$ event as semileptonic or fully hadronic. The algorithm to identify isolated objects loops through electrons, muons and photons with $p_{\mathrm{T}} > 25$ GeV, removes them one at a time from the PFO list, and performs the inclusive $k_t$ jet algorithm with R=0.7 on the modified PFO list.

For each inclusive jet with $E_{\text{jet}}/E_{\text{lepton}} > 2E_{\text{lepton}}$ the variable $\rho = 2E_{\text{lepton}}(1-\cos\theta_{\text{jet}-\text{lepton}})$ is calculated where $\theta_{\text{jet}-\text{lepton}}$ is the angle between the lepton and the jet. If the minimum value of $\rho$ over all jets is greater than 2 then the object is said to be isolated. Further event reconstruction depends on whether an event contains one or zero isolated objects. If an event contains zero isolated electron, muons and photons, then the 4-jet exclusive $k_t$ jet algorithm with R=0.7 is applied to the PFO list. The three jet pair combinations are considered, and the one that minimises $(M_{12} - M_{\mathrm{W}})^2 + (M_{34} - M_{\mathrm{W}})^2$ is chosen to represent the $\mathrm{W^+W^-}$ in the fully hadronic mode. The exclusive mode of the $k_t$ jet algorithm is used because it discards beam jets and returns only central jets. In this way the $\gamma\gamma \to \mathrm{hadrons}$ background is minimised [129].

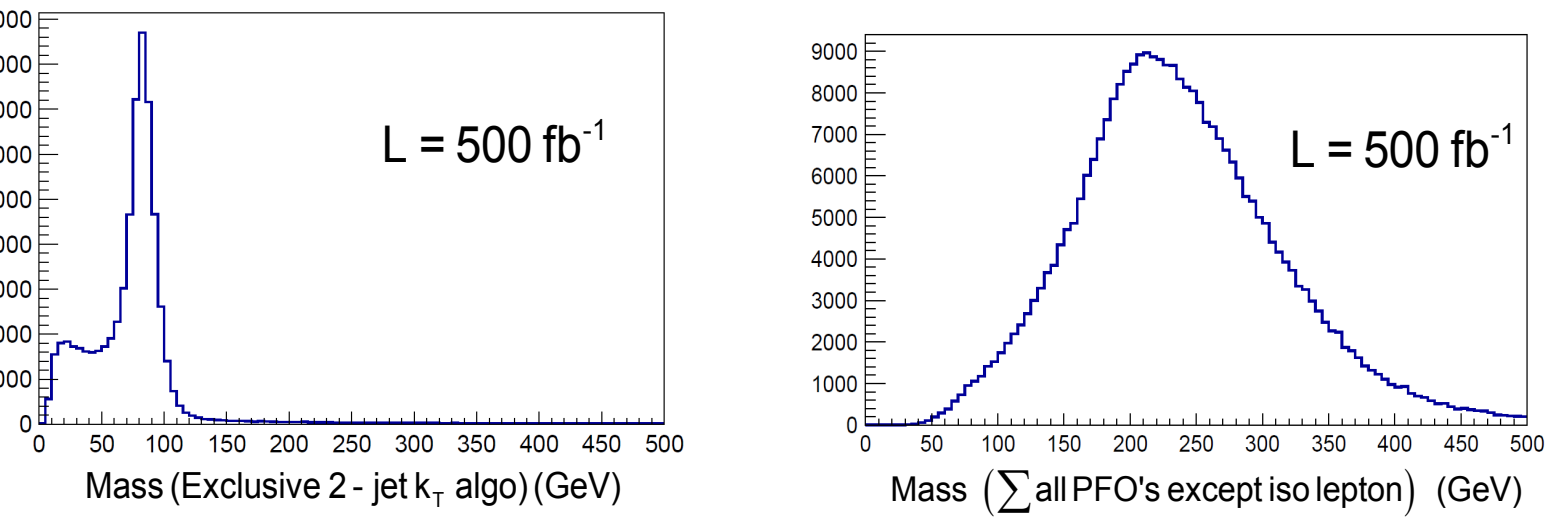

**Figure II-11.12** Mass of the hadronically decaying W in semileptonic $\mathrm{W^+W^-}$ events using the two jets from the exclusive $k_t$ jet algorithm (left) and using all PFO objects other than the isolated charged lepton (right). The broad and displaced mass distribution on the right results from including PFO objects arising from background processes.

If an event contains one isolated electron or muon and no other isolated object then the lepton is removed from the PFO list and the 2-jet exclusive $k_t$ jet algorithm with R=0.7 is performed using the modified PFO list. The di-jet system returned by the $k_t$ jet algorithm represents the hadronically decaying $\mathrm{W}$ in the semileptonic $\mathrm{W^+W^-}$ topology. The neutrino from the leptonically decaying $\mathrm{W}$ is reconstructed assuming that the $\mathrm{W^+W^-}$ is produced back-to-back: $\vec{p}_\nu = -(\vec{p}_{\text{lepton}} + \vec{p}_{\text{2jet}})$ .

### 11.2.5.2 Event selection

Semileptonic $\mathrm{W^+W^-}$ events are selected by requiring that there be exactly one isolated electron or muon, that the total number of PFO's in the hadronically decaying $\mathrm{W}$ (the two jets returned by the two-jet exclusive $k_t$ jet algorithm) be greater than 12, that the mass of the hadronically decaying $\mathrm{W}$ be greater than 60 GeV and less than 100 GeV, that the mass of the leptonically decaying $\mathrm{W}$ be less than 250 GeV, and that the energy of the hadronically decaying $\mathrm{W}$ be greater than 300 GeV. The cut on mass of the leptonically decaying $\mathrm{W}$ is required to remove $\gamma e^- \to \nu W^-$ events. Such events can provide interesting polarisation information, but are considered beyond the scope of this benchmark.

The reconstructed mass of the hadronically decaying $\mathrm{W}$ in semileptonic events is shown in Figure II-11.12 along with the mass obtained by summing together all PFO objects with the exception of the charged lepton. The effectiveness of the 2-jet exclusive $k_t$ jet algorithm in removing $\gamma\gamma \to \mathrm{hadrons}$ is clear. The number of signal and background events following these cuts is summarised in Table II-11.12.

**Table II-11.12** Number of events passing semileptonic $\mathrm{W^+W^-}$ cuts for 500 $\mathrm{fb}^{-1}$ luminosity.

| Type | Solid Angle | P($\mathrm{e}^-$) | P($\mathrm{e}^+$) | $N_{Events}$ |
|---|---|---|---|---|
| Signal | $0.8 < \cos\Theta < 1.0$ | -80% | +20% | 204031 |
| Signal | $-1 < \cos\Theta < 0.8$ | -80% | +20% | 58912 |
| Signal | $0.8 < \cos\Theta < 1.0$ | +80% | -20% | 16090 |
| Signal | $-1 < \cos\Theta < 0.8$ | +80% | -20% | 5315 |
| Background | $0.8 < \cos\Theta < 1.0$ | -80% | +20% | 7994 |
| Background | $-1 < \cos\Theta < 0.8$ | -80% | +20% | 7053 |
| Background | $0.8 < \cos\Theta < 1.0$ | +80% | -20% | 5173 |
| Background | $-1 < \cos\Theta < 0.8$ | +80% | -20% | 4962 |

Fully hadronic $\mathrm{W^+W^-}$ events are selected by requiring that there be no isolated electron, muon or photon, that the total number of PFO's in the two hadronically decaying $\mathrm{W}$'s (the four jets returned by the four-jet exclusive $k_t$ jet algorithm) be greater than 28, that the mass of each of the hadronically decaying W's be greater than 55 GeV and less than 105 GeV, and that the sum of the energies of the two hadronically decaying W be greater than 600 GeV. The number of signal and background events following these cuts is summarised in Table II-11.13.

**Table II-11.13**
Number of events passing fully hadronic $W^+W^-$ cuts for 500 $\mathrm{fb}^{-1}$ luminosity.

| Type | Solid Angle | P(e$^-$) | P(e$^+$) | $N_{Events}$ |
|---|---|---|---|---|
| Signal | $0.8 < \|\cos\Theta\| < 1.0$ | -80% | +20% | 296813 |
| Signal | $0.8 < \|\cos\Theta\| < 1.0$ | +80% | -20% | 22967 |
| Background | $0.8 < \|\cos\Theta\| < 1.0$ | -80% | +20% | 31764 |
| Background | $0.8 < \|\cos\Theta\| < 1.0$ | +80% | -20% | 12548 |

### 11.2.5.3 Beam Polarisation Measurements

The effective polarisation parameters $P_{e^-}(L)_{\rm eff}$ and $P_{e^-}(R)_{\rm eff}$ are extracted by counting events in bins of $(\cos\Theta, \cos\theta)$ and fitting for $P_{e^-}(L)_{\rm eff}$ and $P_{e^-}(R)_{\rm eff}$ with a linear least squares fit:

$$\chi^2 = \sum_i \frac{(N_i - (a\mu_i + b\nu_i)L)^2}{N_i}$$

where $N_i$ is the number of events in bin $i$, $L$ is the integrated luminosity

$$\mu_i = \int d\vec{x}_i d\vec{x}' \eta(\vec{x}')\Omega(\vec{x}_i, \vec{x}') \frac{d\sigma_{LR}}{d\vec{x}'}$$
$$\nu_i = \int d\vec{x}_i d\vec{x}' \eta(\vec{x}')\Omega(\vec{x}_i, \vec{x}') \frac{d\sigma_{RL}}{d\vec{x}'}$$

$\eta(\vec{x})$ is the detection efficiency, $\Omega(\vec{x}_i, \vec{x}')$ is the resolution function, and $d\sigma_{LR}/d\vec{x}$ and $d\sigma_{RL}/d\vec{x}$ are the true differential cross-sections for 100% polarised beams for signal *and* background. The background must be included in this way since it in general has a polarisation dependence.

There is no need to separately calculate $\eta(\vec{x})$ and $\Omega(\vec{x}_i, \vec{x}')$ since the parameters $\mu_i$ and $\nu_i$ are linearly related to the bin contents of a $(\cos\Theta, \cos\theta)$ histogram of fully simulated Monte Carlo events. Let $M_{ki}$ be the number of events in bin $i$ from a Monte Carlo sample produced with effective beam polarisations $a_k$ and $b_k$ and luminosity $L_k$. The SiD Monte Carlo samples were produced with $(P_{e^-}, P_{e^+}) = (-0.8, +0.2)$ $(k = 1)$ and $(P_{e^-}, P_{e^+}) = (+0.8, -0.2)$ $(k = 2)$. The parameters $\mu_i$ and $\nu_i$ are then given by

$$\mu_i = \frac{1}{a_1 b_2 - a_2 b_1}\left[b_2 \frac{M_{1i}}{L_1} - b_1 \frac{M_{2i}}{L_2}\right], \qquad \nu_i = \frac{1}{a_1 b_2 - a_2 b_1}\left[-a_2 \frac{M_{1i}}{L_1} + a_1 \frac{M_{2i}}{L_2}\right].$$

Ten divisions each are used for $\cos\Theta$ and $\cos\theta$ so that 100 bins are defined for each of the two event topologies, semileptonic and fully hadronic. A total of 200 bins are then used for the least squares fit of the effective polarisations $P_{e^-}(L)_{\rm eff}$ and $P_{e^-}(R)_{\rm eff}$ or $\alpha$ and $\beta$. The errors on the effective polarisations are displayed in Table II-11.14 along with the errors on the actual polarisations $P_{e^-}$ and $P_{e^+}$.

**Table II-11.14**
Polarisation errors assuming 500 $\mathrm{fb}^{-1}$ luminosity for each initial state polarisation configuration.

| $\cos\Theta$ range | $P_{e^-}, P_{e^+}$ | $\Delta P_{e^-}(L)_{\rm eff}$ | $\Delta P_{e^-}(R)_{\rm eff}$ | $\Delta P_{e^-}$ | $\Delta P_{e^+}$ |
|---|---|---|---|---|---|
| $0.8 < \cos\Theta < 1$ | -0.8,+0.2 | 0.0011 | 0.022 | 0.13 | 0.087 |
| $0.8 < \cos\Theta < 1$ | +0.8,-0.2 | 0.00036 | 0.0096 | 0.0050 | 0.024 |
| $-1 < \cos\Theta < 1$ | -0.8,+0.2 | 0.0011 | 0.0104 | 0.062 | 0.041 |
| $-1 < \cos\Theta < 1$ | +0.8,-0.2 | 0.00036 | 0.0077 | 0.0045 | 0.020 |
| $\cos\Theta$ range | $P_{e^-}, P_{e^+}$ | $\Delta\alpha$ | $\Delta\beta$ | $\Delta\|P_{e^-}\|$ | $\Delta\|P_{e^+}\|$ |
| $-1 < \cos\Theta < 1$ | sum | 0.0010 | 0.00032 | 0.0020 | 0.0029 |

## 11.2.6 Top quark cross-section and forward-backward asymmetry

### 11.2.6.1 Introduction

The top quark is the heaviest elementary particle known. The explanation for its large mass may come from beyond the Standard Model physics [189, 190]. Generally, models predict that due to its large mass, the top quark couples strongly to the particle(s) that generate the spontaneous symmetry breaking of the electroweak interaction. As such, it is important to measure the characteristics of the top quark with high precision.

In this analysis, we investigate the determination of the cross-section and the forward-backward asymmetries for both the $\mathrm{b}$ and $\overline{\mathrm{b}}$ quarks and $\mathrm{t}$ and $\overline{\mathrm{t}}$ quarks using the fully hadronic $\mathrm{t}\overline{\mathrm{t}}$ decay mode ($\mathrm{e}^+\mathrm{e}^- \rightarrow \mathrm{t}\overline{\mathrm{t}} \rightarrow \mathrm{b}\overline{\mathrm{b}}\mathrm{q}\overline{\mathrm{q}}\mathrm{q}\overline{\mathrm{q}}$). The forward-backward asymmetry is defined in Equation II-11.3.

$$A_{FB} = \frac{\sigma(\theta < 90^o) - \sigma(\theta > 90^o)}{\sigma(\theta < 90^o) + \sigma(\theta > 90^o)} \qquad \text{(II-11.3)}$$

where $\sigma(\theta < 90^o)$ is the cross-section of the events in which the $\mathrm{b}$ or $\mathrm{t}$ quark has a polar angle of less than $90^o$ in the centre-of-mass frame of reference.

### 11.2.6.2 Event Selection

The event selection presented in this study is closely based on an earlier study [181]. We require that jets be composed of at least two reconstructed particles in order to reject semi-leptonic $\mathrm{t}\overline{\mathrm{t}}$ decays and other SM backgrounds. The total energy originating from the six jets is required to be greater than 400 GeV in order to suppress events with leptons and neutrinos. We also require that each event has a particle and track multiplicity greater than 80 and 30, respectively.

The next step of the event selection is to identify the two b-jets in the signal event, which is achieved by using the LCFIPlus package [170]. The importance of correctly identifying the b-jets is twofold: to reduce the SM background and to reduce the number of combinations required to reconstruct the full signal event. As such, we require that the jet with the highest b-tag value be greater than 0.9 and the jet with the second highest b-tag value be greater than 0.4. Once the b-jets are identified, the remaining jets are assumed to be associated with the $\mathrm{W}$ boson hadronic decays. The top quark mass is determined using a kinematic fitting approach with constraints listed in Table II-11.15. The combination with the smallest $\chi^2$ is selected as the proper event configuration. The results from the kinematic fitting algorithm are presented in Figure II-11.13. Finally, we require that the reconstructed mass of the top quark candidates is between 150 GeV and 200 GeV, yielding a final signal efficiency of $27.3 \pm 0.1\%$. The signal efficiency was determined by using the Monte Carlo truth information to identify the $\mathrm{e}^+\mathrm{e}^- \rightarrow \mathrm{t}\overline{\mathrm{t}} \rightarrow \mathrm{b}\overline{\mathrm{b}}\mathrm{q}\overline{\mathrm{q}}\mathrm{q}\overline{\mathrm{q}}$ decay chain within the generic ($\mathrm{e}^+\mathrm{e}^- \rightarrow \mathrm{b}\overline{\mathrm{b}}\ 4f$).

**Table II-11.15**
The kinematic constraints used in the $\mathrm{t}\overline{\mathrm{t}}$ analysis.

| | | |
|---|---|---|
| $m(\mathrm{top}_1)$ | = | $m(\mathrm{top}_2)$ |
| $m(W_1)$ | = | 80.4 GeV |
| $m(W_2)$ | = | 80.4 GeV |
| $m(b_1)$ | = | 5.8 GeV |
| $m(b_2)$ | = | 5.8 GeV |
| $E_{tot}$ | = | $\sqrt{s}$ |
| $\vec{p}_{tot}$ | = | 0 |

After the event selection, the cross-section for $\mathrm{e}^+\mathrm{e}^- \rightarrow \mathrm{t}\overline{\mathrm{t}} \rightarrow \mathrm{b}\overline{\mathrm{b}}\mathrm{q}\overline{\mathrm{q}}\mathrm{q}\overline{\mathrm{q}}$ was calculated using Equation II-11.4.

$$\sigma = \frac{N_{tot} - N_{bkg}}{\epsilon_{sig} \int \mathcal{L} dt} \qquad \text{(II-11.4)}$$

Here $N_{tot}$ is the number of total events that survive all selection cuts, $N_{bkg}$ is the estimated background events after the selection cuts, $\epsilon_{sig}$ is the signal efficiency, and $\int \mathcal{L}dt$ is the integrated luminosity. The statistical uncertainty on the cross-section, assuming 500 $\mathrm{fb}^{-1}$ of integrated data, was calculated to be approximately 0.47% and 0.69% for the P($\mathrm{e}^-$) = +0.8, P($\mathrm{e}^+$) = -0.3 and P($\mathrm{e}^-$) = -0.8, P($\mathrm{e}^+$) = +0.3 polarisation configuration, respectively.

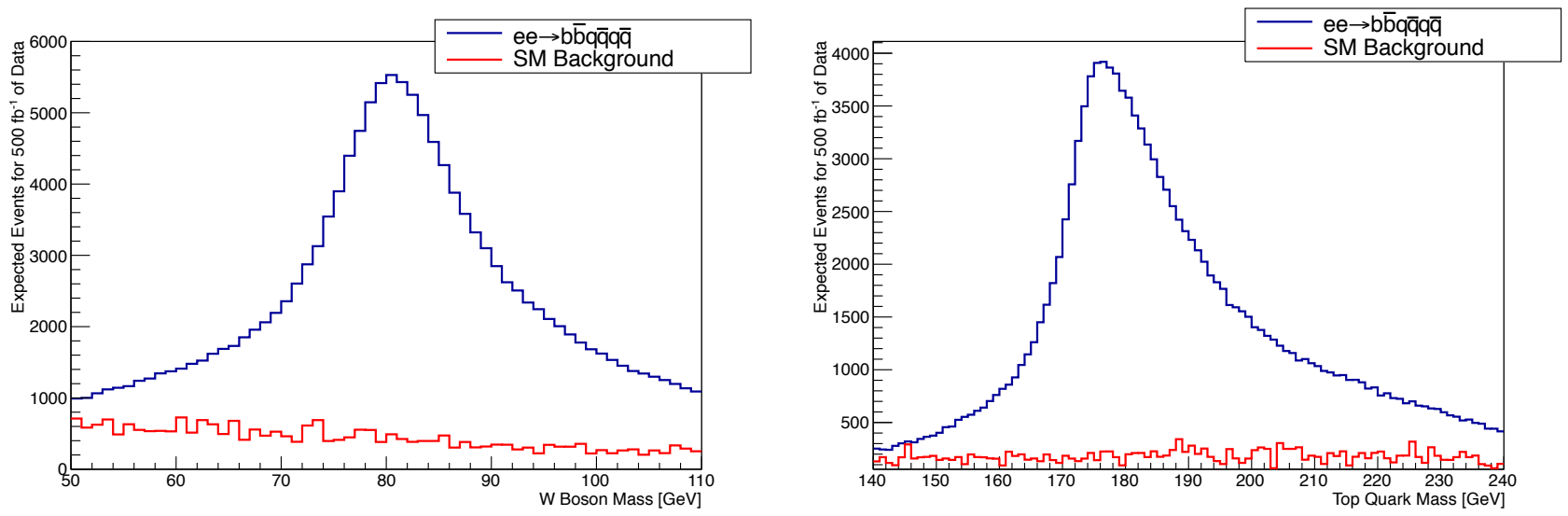

**Figure II-11.13** Mass distribution of the W boson candidates (left) and top quark candidates (right).

### 11.2.6.3 Forward-Backward Asymmetry

In order to determine the forward-backward asymmetry, the charges of the bottom and top quarks must be determined. Based on the previous studies [181], the vertex charge and the jet charge are the two variables that are used in determining the quark charge. The vertex charge is determined by calculating a momentum weighted charge using the tracks associated to the secondary vertex within the identified b-jet, as defined in Equation II-11.5.

$$Q = \frac{\sum_j p_j^k Q_j}{\sum_j p_j^k} \qquad \text{(II-11.5)}$$

Here $p_j^k$ and $Q_j$ are the momentum and the charge of the $j$-th track and $k = 0.3$ is a power weight. Similarly, the jet charge is also calculated using Equation II-11.5, however, it is determined by using all the tracks associated to a b-jet. These two variables are combined to provide a single discriminate variable, $C$, as defined in Equation II-11.6.

$$C = \frac{1-r}{1+r}; \qquad r = \prod_i \frac{f_i^{\bar{b}}(x_i)}{f_i^b(x_i)} \qquad \text{(II-11.6)}$$

Here $f_i^b(x_i)$ and $f_i^{\bar{b}}(x_i)$ is the probability density function for variable $x_i$ for the $b$ and $\bar{b}$ quarks, respectively.

The calculation of $A_{FB}$ for the top quarks can be performed by using the effective charge of the b-quark jets from Equation II-11.6 and the angle $\theta$ of the reconstructed top quark. The corrected number of top quarks in each hemisphere is calculated using Equation II-11.7.

$$N_b = (N_{tot} - N_{bkg}) \cdot \epsilon_p \cdot \epsilon_{sig} \qquad \text{(II-11.7)}$$

Here, $N_{tot}$ is the total number of events, $N_{bkg}$ is the background events, $\epsilon_p$ is the purity, and $\epsilon_{sig}$ is the signal efficiency. The purity and signal efficiency was determined by using the Monte Carlo truth information to identify the $\mathrm{e}^+\mathrm{e}^- \to \mathrm{t\bar{t}} \to \mathrm{b\bar{b}q\bar{q}q\bar{q}}$ decay chain within the generic ($\mathrm{e}^+\mathrm{e}^- \to \mathrm{b\bar{b}}\ 4f$). Finally, we require that the product of the two effective charges is negative. As a result, the expected statistical uncertainty of the top quark forward-backward asymmetry, assuming 500 $\mathrm{fb}^{-1}$ of integrated data, is approximately 2% and 2.5% for the P($\mathrm{e}^-$) = +0.8, P($\mathrm{e}^+$) = -0.3 and P($\mathrm{e}^-$) = -0.8, P($\mathrm{e}^+$) = +0.3 polarisation configuration, respectively. These numbers agree with the MC input value and previous studies [181].

#### 11.2.6.4 Conclusion

The achievable cross-section resolution for the top quark at the ILC in the $e^+e^- \to t\bar{t} \to b\bar{b}q\bar{q}q\bar{q}$ channel is less than $1\%$ for a total luminosity of 500 $fb^{-1}$. Additionally, in the case of both polarised beam configurations the achievable resolution for the top quark asymmetries is approximately $2\%$.

## 11.3 Additional Benchmarks

To illustrate the importance of the BeamCal, an additional benchmark study was performed.

### 11.3.1 Measurement of scalar tau leptons

One of the processes whose detection is very difficult at the LHC but easier at the ILC is $\tilde{\tau}$ production [191]. In this benchmark, the $\tilde{\tau}^{\pm}$ production and decay has been studied [192] at $\sqrt{s} = 500$ GeV for the kinematically accessible benchmark points B', C', D', G', and I' as proposed in [193]. At these points, the LSP is the $\tilde{\chi}^0_1$, and the NLSP is the $\tilde{\tau}^{\pm}$, with the masses shown in Table II-11.16. Having similar masses, the two particles are a candidate for the co-annihilation mechanism that can explain the WMAP relic dark-matter density. It is therefore important to identify and measure the mass of the NLSP, as well as the LSP. Several studies of the $\tilde{\tau}^{\pm}$ at the ILC have already been made. These include an analysis at D' [194], an analysis at SPS1a' [195], and a much broader analysis covering many parameter points [196].

**Table II-11.16**
SUSY Particle Masses (GeV) for the kinematically accessible benchmark points at $\sqrt{s} = 500$ GeV [193].

| Model | B' | C' | D' | G' | I' |
|---|---|---|---|---|---|
| $\tilde{\tau}^-$ | 110.6 | 170.6 | 223.9 | 158.6 | 144.6 |
| $\tilde{\chi}^0_1$ | 96.5 | 161.0 | 216.4 | 150.9 | 140.8 |

At the benchmark points analysed here, the $\tilde{\tau}^{\pm}$ has only one decay channel: $\tilde{\tau}^{\pm} \to \tau^{\pm}\tilde{\chi}^0_1$. The production of $\tau$ leptons via the two-photon process $e^+e^- \to e^+\gamma^* e^-\gamma^* \to e^+e^-\tau^+\tau^-$ is by far the most significant background process, and the BeamCal is an essential detector to veto two-photon events by detecting high energy scattered $e^+e^-$ beam particles. Figure II-11.14 shows the detection efficiency as a function of radius in the BeamCal for 5, 15, 30, 50, 100, and 150 GeV electrons. As the beamstrahlung energy has a strong radial and azimuthal dependence, the detection efficiency is calculated as a function of the distance from the outgoing beam axis at three azimuthal angles (0, 90, and 180 degrees). The inefficiency between 30 and 50 mm at $\phi = 180°$ is due to the incoming beam hole. Since the beamstrahlung background energy is the highest at $\phi \approx 90°$ (and $270°$), the detection efficiency is lower in this angular region. The efficiency to detect electrons with energy above 150 GeV is almost 100% up to 8 mrad from the beam axis.

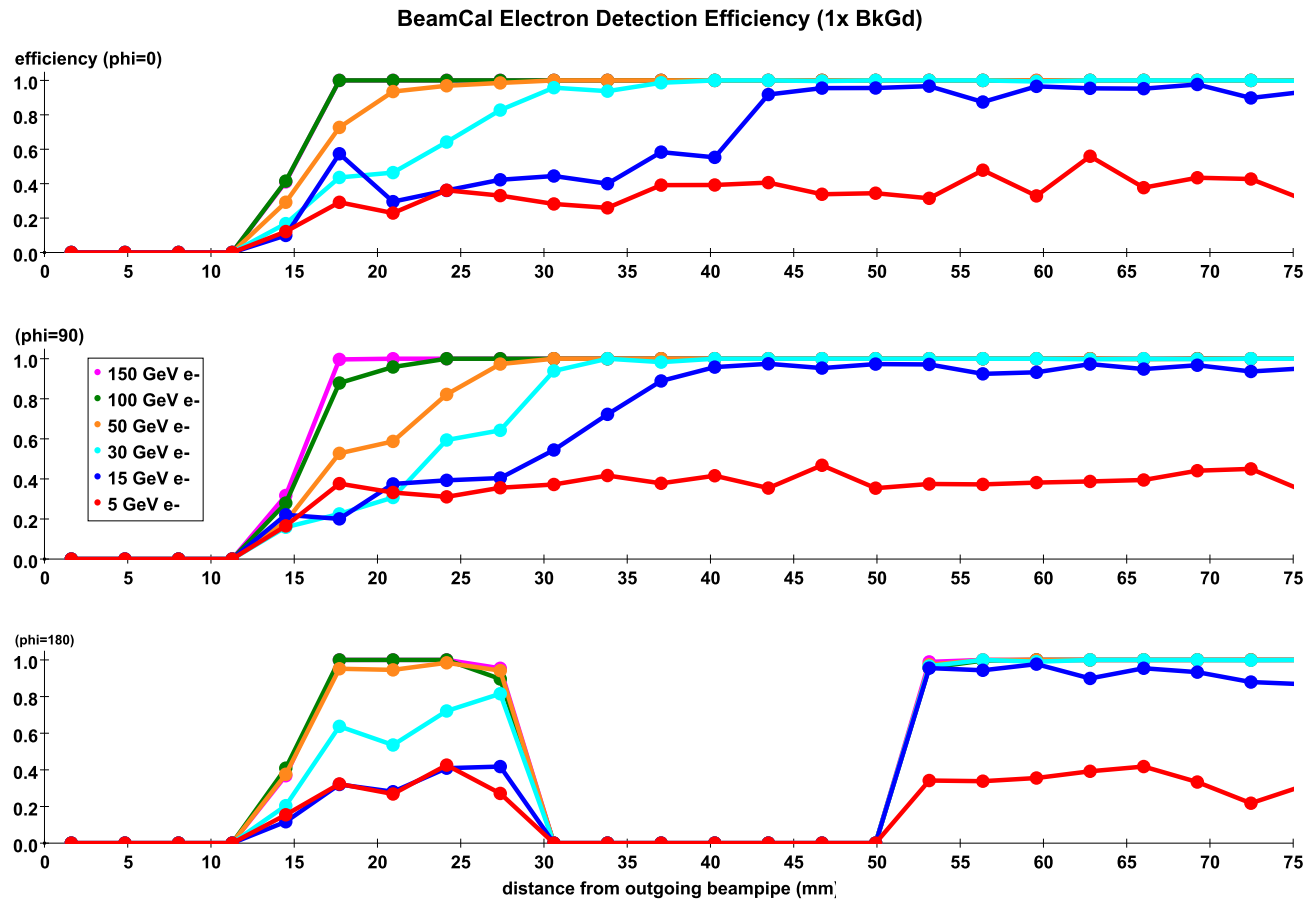

**Figure II-11.14**
The BeamCal detection efficiency of electrons, at various energies and angles, as a function of distance from the outgoing beam.

Events are grouped based on their vectorially combined transverse momentum, and it is expected that the SUSY events will often have higher vectorially combined momentum due to the momentum carried away by the $\tilde{\chi}^0_1$. Detection of the $\tilde{\tau}^\pm$ by these methods relies on a significant mass difference between the $\tilde{\tau}^\pm$ and the $\tilde{\chi}^0_1$. If the mass difference between the $\tilde{\tau}^\pm$ and $\tilde{\chi}^0_1$ is not large enough, the visible $\tau$s will not have sufficient scalar momentum to be visible above the two-photon process (even if the visible products have a preferred direction). For this reason a cut is made on the acoplanarity of the two jets in the plane perpendicular to the beampipe. The requirements that the mass of each jet is less than 1.8 GeV and the number of charged particles in each jet is either one or three, serve to select $\tau^\pm$ events and eliminate other processes such as $\mathrm{hZ} \rightarrow$ hadrons.

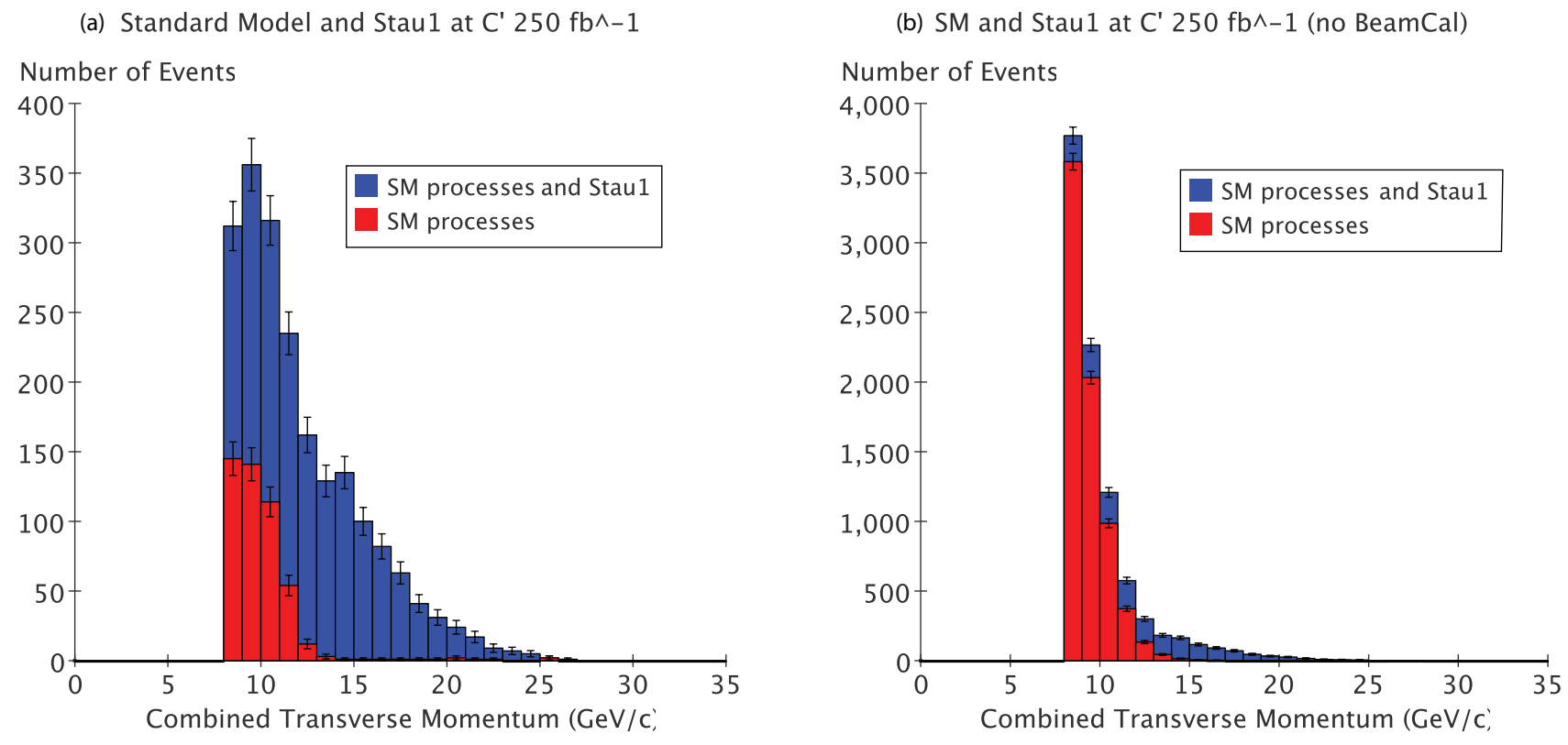

**Figure II-11.15** Fill event $p_\mathrm{T}$ distribution with (left) and without (right) the BeamCal veto at the benchmark point C'.

Figure II-11.15 shows the transverse momentum of combined SM and $\tilde{\tau}^\pm$ events (blue) and the SM backgrounds (red) at the benchmark point C' using a data sample based on an integrated luminosity of 250 $\mathrm{fb}^{-1}$ at $\sqrt{s} = 500$ GeV. Figure II-11.15 (left) shows the distribution when the BeamCal is used to veto SM backgrounds, while Figure II-11.15 (right) shows the distribution without the veto. A significant excess of $\tilde{\tau}^\pm$ signal is observable over SM backgrounds only when the BeamCal is used as a veto. The mass difference between $\tilde{\tau}^\pm$ and $\tilde{\chi}^0_1$ is 9.6 GeV at the benchmark point C', and it is possible to measure the $\tilde{\tau}^\pm$ mass with a 1 GeV uncertainty.

Figure II-11.16 shows the transverse momentum of combined SM and $\tilde{\tau}^\pm$ events (blue) and the SM backgrounds (red) at the benchmark point I' using a data sample based on an integrated luminosity of 250 $\mathrm{fb}^{-1}$ at $\sqrt{s} = 500$ GeV. Figure II-11.16 (left) shows the distribution when the BeamCal is used to veto SM backgrounds, while Figure II-11.16 (right) shows when the BeamCal is not used. Although the $\tilde{\tau}^\pm$ signal can be enhanced when the BeamCal is used for a veto, the mass difference between the $\tilde{\tau}^\pm$ and the $\tilde{\chi}^0_1$ is only 3.8 GeV at this benchmark point I' and the signal is not very visible even after applying the BeamCal veto.

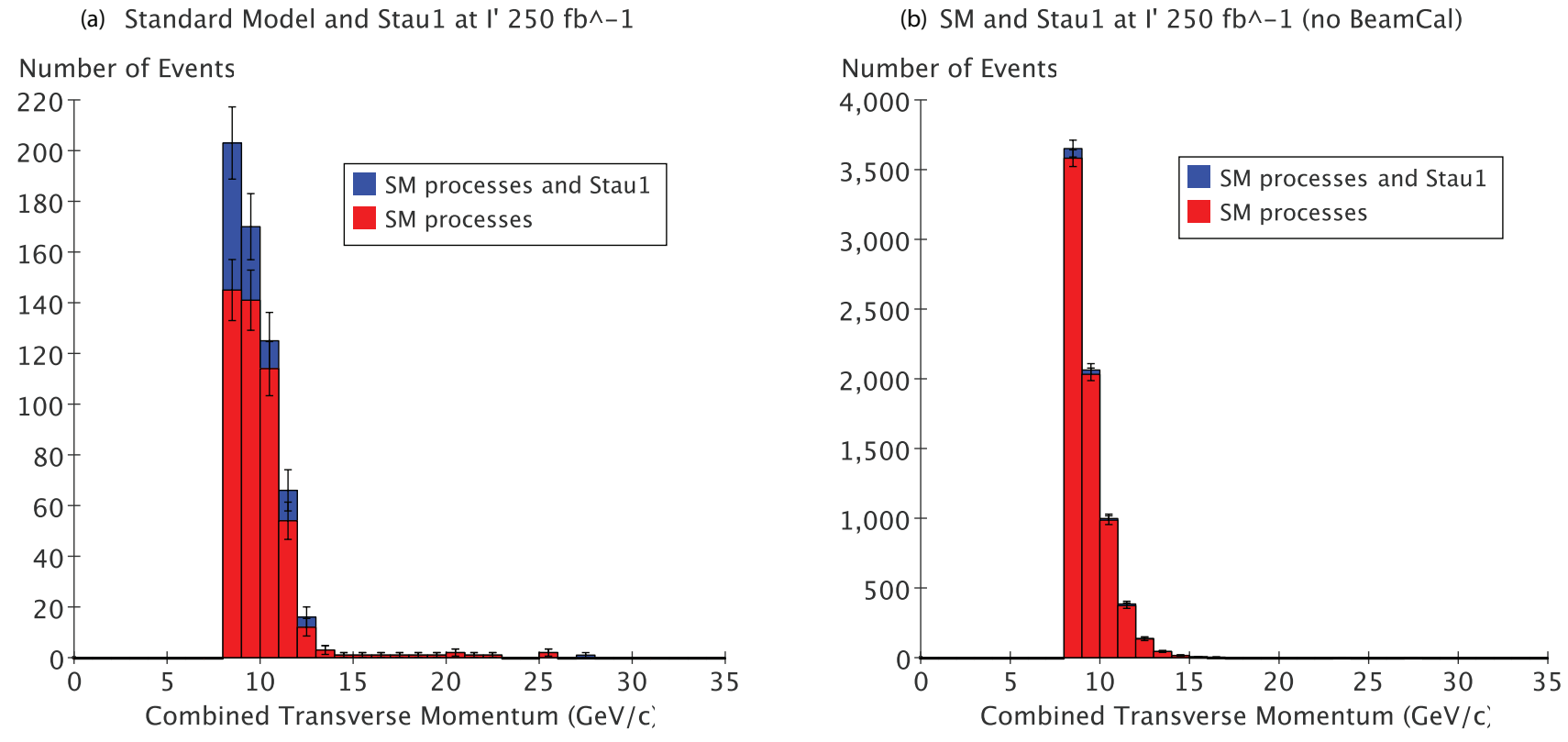

**Figure II-11.16** $p_\mathrm{T}$ distribution with BeamCal veto (left) and without (right) at the benchmark point I')

**Table II-11.17.** Summary of the SiD benchmarking results. The LOI Higgs results were obtained assuming a polarisation of $P(e^-) = +80\%$, $P(e^+) = -30\%$ while the the total luminosity was equally divided between $P(e^-) = \mp 80\%$, $P(e^+) = \pm 30\%$ for the LOI $t\bar{t}$ and $\tau^+\tau^-$ results. The $t\bar{t}h$ and $W^+W^-$ results assume $P(e^-) = -80\%$, $P(e^+) = +20\%$ for half of the integrated luminosity and $P(e^-) = +80\%$, $P(e^+) = -20\%$ for the other half. For the $\nu_e\bar{\nu}_e h$ studies only $P(e^-) = -80\%$, $P(e^+) = +20\%$ was considered. The DBD $t\bar{t}$ analysis assumes both $P(e^-) = \mp 80\%$, $P(e^+) = \pm 30\%$.

| Process $e^+e^- \to$ | $\sqrt{s}$ (GeV) | $\mathcal{L}$ ($fb^{-1}$) | SiD | Meas. Quant. | Unit | Result |
|---|---|---|---|---|---|---|
| $e^+e^-h/\mu^+\mu^-h$ | 250 | 250 | LOI | $m_H$ | GeV | ± 0.04 |
| | | | | $\sigma$ | % | ± 2.7 |
| $hZ^0 \to c\bar{c}q\bar{q}$ | 250 | 250 | LOI | BR | % | ± 6.0 |
| $hZ^0 \to c\bar{c}\nu\bar{\nu}$ | 250 | 250 | LOI | BR | % | ± 11.0 |
| $hZ^0 \to \mu^+\mu^-q\bar{q}$ | 250 | 250 | LOI | $\sigma$ | % | 89.1 |
| $\tau^+\tau^-$ | 500 | 500 | LOI | $A^\tau{}_{FB}$ | - | ± 0.0021/0.0024 |
| | | | | $< P_\tau >$ | % | ± 1.7/2.3 |
| $t\bar{t} \to 6$ jets | 500 | 500 | LOI | $m_{top}$ | GeV | 173.92 ± 0.05 |
| | | | | $\sigma$ | % | 0.49 |
| | | | | $A^t{}_{FB}$ | - | ± 0.008 |
| $\tilde{\chi}^0_2\tilde{\chi}^0_2 \to \tilde{\chi}^0_1\tilde{\chi}^0_1 Z^0Z^0$ | 500 | 500 | LOI | $m_{\tilde{\chi}^0_1}$ | GeV | ± 0.16 |
| $\tilde{\chi}^0_2\tilde{\chi}^0_2 \to \tilde{\chi}^0_1\tilde{\chi}^0_1 Z^0Z^0$ | 500 | 500 | LOI | $m_{\tilde{\chi}^+_1}$ | GeV | ± 0.45 |
| $\tilde{\chi}^+_1\tilde{\chi}^-_1 \to \tilde{\chi}^0_1\tilde{\chi}^0_1 W^+W^-$ | 500 | 500 | LOI | $m_{\tilde{\chi}^0_1}$ | GeV | ± 0.28 |
| $\tilde{\chi}^+_1\tilde{\chi}^-_1 \to \tilde{\chi}^0_1\tilde{\chi}^0_1 W^+W^-$ | 500 | 500 | LOI | $m_{\tilde{\chi}^0_2}$ | GeV | ± 0.49 |
| $t\bar{t}h$ (6 jets) | 1000 | 1000 | DBD | $\sigma$ | % | ± 13.2 |
| $t\bar{t}h$ (8 jets) | 1000 | 1000 | DBD | $\sigma$ | % | ± 11.5 |
| $t\bar{t}h$ (combined) | 1000 | 1000 | DBD | $\sigma$ | % | ± 8.7 |
| $\nu_e\bar{\nu}_e h; h \to WW^*$ | 1000 | 1000 | DBD | $\sigma \times$ BR | % | ± 3.3 |
| $\nu_e\bar{\nu}_e h; h \to gg$ | 1000 | 1000 | DBD | $\sigma \times$ BR | % | ± 3.1 |
| $\nu_e\bar{\nu}_e h; h \to c\bar{c}$ | 1000 | 1000 | DBD | $\sigma \times$ BR | % | ± 7.6 |
| $\nu_e\bar{\nu}_e h; h \to b\bar{b}$ | 1000 | 1000 | DBD | $\sigma \times$ BR | % | ± 0.47 |
| $\nu_e\bar{\nu}_e h; h \to \mu^+\mu^-$ | 1000 | 1000 | DBD | $\sigma \times$ BR | % | ± 32 |
| $W^+W^-$ | 1000 | 1000 | DBD | $P_{e^-}(L)_{\text{eff}}$ | % | ± 0.20/0.90 |
| $W^+W^-$ | 1000 | 1000 | DBD | $\|P_{e^-}\|$ | % | ± 0.25 |
| $W^+W^-$ | 1000 | 1000 | DBD | $\|P_{e^+}\|$ | % | ± 1.45 |
| $t\bar{t} \to 6$ jets | 500 | 500 | DBD | $\sigma$ | % | ± 0.47/0.69 |
| | | | | $A^t{}_{FB}$ | % | ± 2.0/2.5 |

Similarly, good signal-to-noise was achieved by the application of the BeamCal veto for the benchmark points B', C', and G', where the mass difference between $\tilde{\tau}^\pm$ and $\tilde{\chi}^0_1$ is about 10 GeV, while the $\tilde{\tau}^\pm$ signal was not strong enough for mass measurements for the benchmark points D' and I', where the mass difference is about 5 GeV.

## 11.4 Benchmarking Summary

A large set of benchmarks have been conducted with the SiD detector using both simulations of the LOI and the more detailed DBD detector variants. They illustrate the detector performance of the SiD concept for centre-of-mass energies in the range from 250 GeV up to 1 TeV. All results obtained have been summarised in Table II-11.17.

# Chapter 12
# SiD Costs

## 12.1 Introduction

The SiD cost estimate is a construction cost estimate; it does not include R&D, commissioning, operating costs, or physicist salaries.

The SiD design process has continuously monitored costs using a parametric cost model. This tool has been essential for the ongoing detector optimisation process. At various stages, detector parameters (e.g. dimensions or masses) have been transferred to a Work Breakdown Structure where it is more convenient to describe a subsystem to arbitrary levels of detail. Here we describe this method of costing for the DBD version of SiD.

**Table II-12.1**
Unit Costs agreed to by SiD, ILD, and CLIC [197].

| | agreed unit cost (US-$) | agreed error margin (US-$) |
|---|---|---|
| Tungsten for HCAL | 105/kg | 45/kg |
| Tungsten for ECAL | 180/kg | 75/ kg |
| Steel for Yoke | 1000/t | 300/t |
| Stainless Steel for HCAL | 4500/t | 1000/t |
| Silicon Detector | 6 / $cm^2$ | 2 / $cm^2$ |

The SiD baseline for the DBD has been changed from RPC's to scintillator bars for the muon system (see Chapter 5), which is also included in the costing model.

At the time of the LOI, the cost optimisation of the global SiD design was studied by using a parametric model of PFA based jet energy resolution and the parametric cost tools described here. The tracker radius, B field, and HCAL depth were varied holding the jet energy resolution fixed at 3.78% for 180 GeV jets. The cost optimal point was quite near the baseline SiD parameters of R = 1.25 m, B = 5 T, and HCAL $\lambda_I = 4.5$. This work has not been repeated.

## 12.2 Parametric cost model

The parametric model of the detector is a large set of Excel spreadsheets that first maintain a self consistent model of SiD. It is straightforward to vary parameters ranging from the most basic, such as the tracker radius and aspect ratio, to parameters such as the number of tracking layers, the number and thickness of HCAL layers, and calorimeter radiator material. The tracking layers and disks are adjusted to fit the allocated space.The calorimeter inner radii and minimal z coordinate are adjusted for the tracker size, and thicknesses are set parametrically. The solenoid model is adjusted for its radius and field, and the flux return is adjusted to roughly contain the return flux.

For each system, the cost driving component count, such as tungsten plate, silicon detectors, and readout chips for the ECAL, are calculated. The model has tables for material costs and estimates both M&S and labour costs that are associated with the actual scale of SiD.

Costs that are approximately fixed, for example, engineering, fixturing, or solenoid He plants, are imported from the separate Work Breakdown Structure program. Finally, a set of macros calculate the costs of SiD as parameters are varied. The cost process also develops a Work Breakdown

**Table II-12.2**
Summary of Costs per Subsystem.

| | M&S Base (M US-$) | M&S Contingency (M US-$) | Engineering (MY) | Technical (MY) | Admin (MY) |
|---|---|---|---|---|---|
| Beamline Systems | 3.7 | 1.4 | 4.0 | 10.0 | |
| VXD | 2.8 | 2.0 | 8.0 | 13.2 | |
| Tracker | 18.5 | 7.0 | 24.0 | 53.2 | |
| ECAL | 104.8 | 47.1 | 13.0 | 288.0 | |
| HCAL | 51.2 | 23.6 | 13.0 | 28.1 | |
| Muon System | 8.3 | 3.0 | 5.0 | 22.1 | |
| Electronics | 4.9 | 1.6 | 44.1 | 41.7 | |
| Magnet | 115.7 | 39.7 | 28.3 | 11.8 | |
| Installation | 4.1 | 1.1 | 4.5 | 46.0 | |
| Management | 0.9 | 0.2 | 42.0 | 18.0 | 30.0 |
| | 314.9 | 126.7 | 186.0 | 532.1 | 30.0 |

Structure using the SLAC program WBS. WBS facilitates the description of the costs as a hierarchical breakdown with increasing levels of detail. Separate tables describe cost estimates for purchased M&S and labour. These tables include contingencies for each item, and these contingencies are propagated by WBS. The M&S costs are estimated in 2008 US-$ except for those items described in Table II-12.1.

Labour is estimated in man-hours or man-years as convenient. The WBS had about 50 labour types, but they are condensed to engineering, technical, and clerical for this estimate. The statement of base M&S and labour in man-years by the three categories results in a cost which we believe is comparable to that used by the ILC machine, and is referred to here as the ILC cost.

Contingency is estimated for each quantity to estimate the uncertainties in the costs of the detector components. However, we do not use the ILC value system for these estimates. Items which are commodities, such as detector iron, have had costs swinging wildly over the last few years. While there is agreement on a set of important unit costs, those quantities also have "error margins". SiD, ILD, and CLIC have worked together to reach agreed values for some unit costs as shown in Table II-12.1.

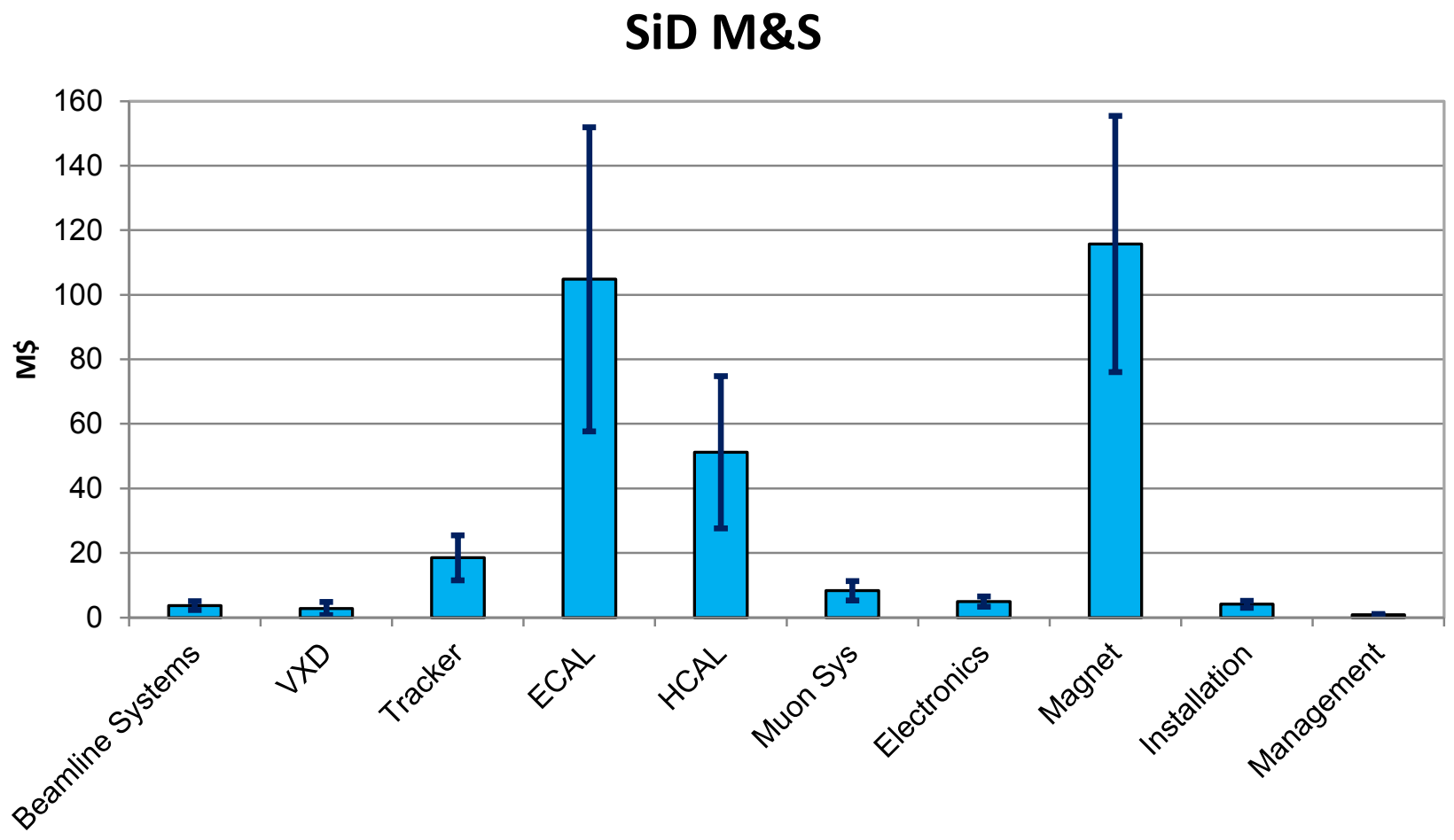

**Figure II-12.1**
Subsystem M&S Costs in million US-$, the error bars show the contingency per subsystem.

There are a substantial set of interfaces in the interaction region hall. For the purpose of this estimate, the following has been assumed:

- The hall itself, with finished surfaces, lighting, and HVAC are provided by the machine.
- Utilities, including 480 VAC power, LCW, compressed air, and Internet connections are provided.
- An external He compressor system with piping to the hall is provided. The refrigeration and associated piping is an SiD cost.
- All surface buildings, gantry cranes, and hall cranes are provided by the machine.

- Data storage systems and offline computing are provided by others.
- SiD will be assembled and will travel on a suitable platform for push-pull. This platform and its motion and alignment systems will be provided by the machine.
- QD0's and their 2K systems are provided by the machine. The beampipe is an SiD cost.

## 12.3 Results

The subsystem level summary is shown in Table II-12.2, the M&S costs are plotted in Figure II-12.1, and the labour costs are shown in Figure II-12.2. The costs are dominated by the Magnet and the ECAL. The magnet has roughly equal costs for the superconducting coil and the iron. The ECAL is dominated by the silicon detectors.

The cost estimate has several important "commodity" items whose costs have recently been fluctuating significantly. For SiD, these include most metals and processed silicon detectors. Table II-12.3 illustrated the cost sensitivity to these prices by indicating the unit cost used in the estimate and the effect on the SiD M&S cost of doubling the unit cost.

**Figure II-12.2**
Subsystem Labour

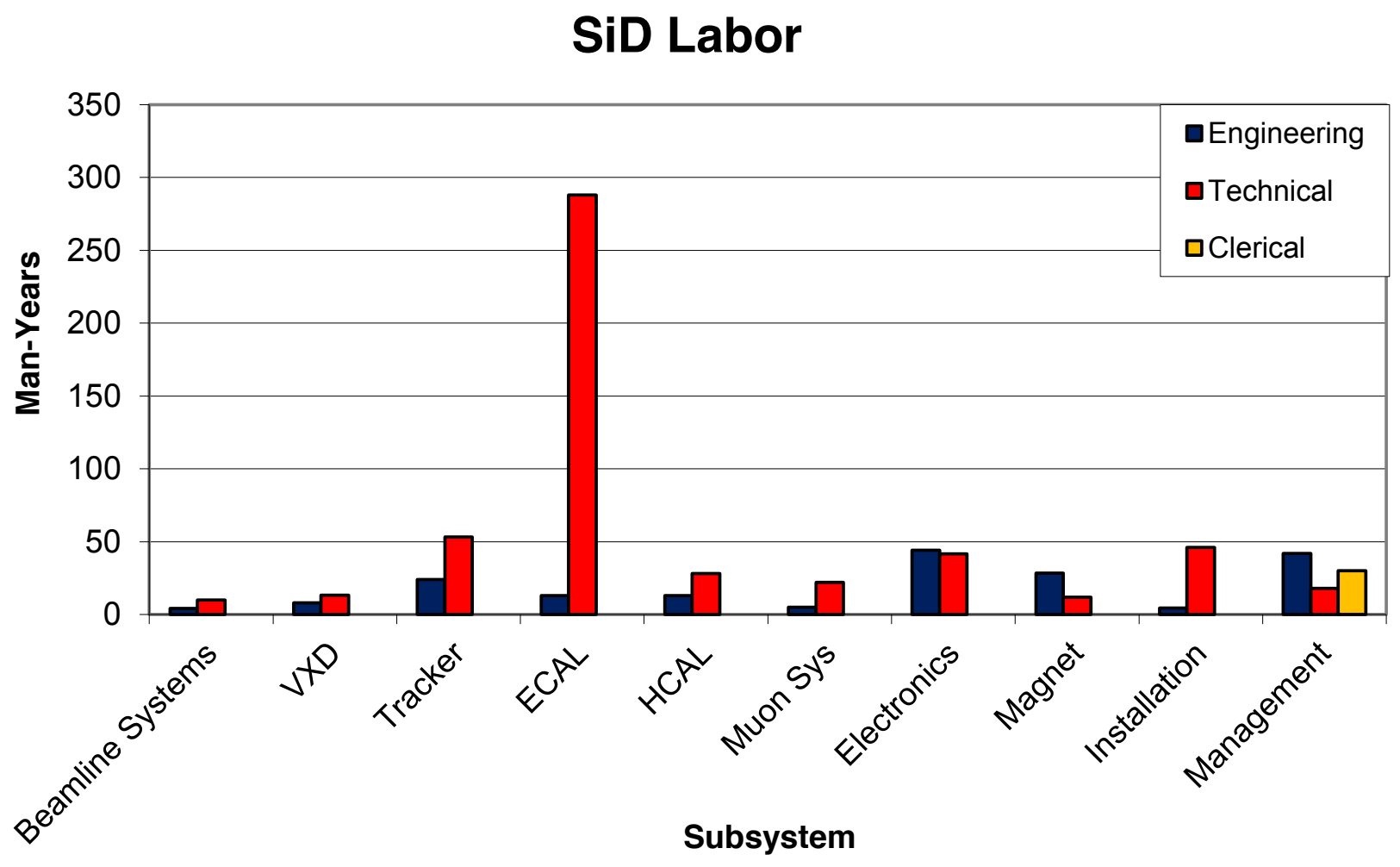

The superconducting coil cost is difficult to estimate, because there is little data and experience with coils of this size and field. An attempt was made to extract the CMS coil cost, and it is believed to be US-$ 48M for cold mass and vacuum tank.

**Table II-12.3**
Sensitivity to selected unit costs. The table shows the effect on the total M&S cost if the selected unit cost were to double.

| Material | Base Cost (US-$) | Delta Cost (M US-$) | Fractional Delta (%) |
|---|---|---|---|
| Magnet Iron | 6.00 /kg | 48 | 16 |
| Silicon Sensors | 6.00 /cm$^2$ | 79 | 26 |
| Tungsten ECAL | 180 /kg | 14 | 5 |
| Stainless | 4.5 /kg | 2 | 1 |
| HCAL Detector | 12K /m$^2$ | 42 | 14 |

A Japanese industrial estimate for the SiD coil was obtained, and it was approximately the same as CMS, but for a coil with roughly half the stored energy. Cost functions linear in the stored energy and with a 0.66 exponential dependence have been studied.

SiD has taken a conservative approach and for the parametric study has used a linear model fit to the BaBar coil at the low end and the industrial estimate at the high end. The result for the current SiD design is US-$ 55M, higher than the CMS cost, but inflation and currency exchange variations have been ignored. SiD is doing R&D on advanced conductor design, and there is some reason to expect the coil cost estimate to decrease.

**Table II-12.4**
Components of the US accounting style estimate.

| | M&S (M US-$) | Labour (M US-$) | Totals (M US-$) |
|---|---|---|---|
| Base | 315 | 81 | 396 |
| Contingency | 127 | 18 | 144 |
| Total | 442 | 99 | 540 |
| Indirect rates | 0.06 | 0.20 | |
| Indirects | 26 | 20 | 46 |
| Totals w/ indirects | 468 | 119 | 587 |
| Total in FY2016 M$ | 2008 | | 586.7 |
| Start Year | 2016 | | |
| Construction Duration | 6 years | | |
| Inflation | 3.5%/a | | |
| Factor | 1.460 | | |
| Total Escalation | | | 269.9 |
| Total | | | 856.6 |

The SiD cost in ILC value units is US-$ 315M for M&S, 186 MY engineering, 532 MY technical, and 30 MY administrative labour. The estimated M&S contingency, reflecting uncertainty in unit costs and some estimate of the maturity of this study, is US-$ 127M.

The cost in US accounting, assuming a construction start in 2016 and 3.5% per year inflation and US National Laboratory labour rates, is US-$ 857M. The components of the US accounting calculation are indicated below in

Table II-12.4.

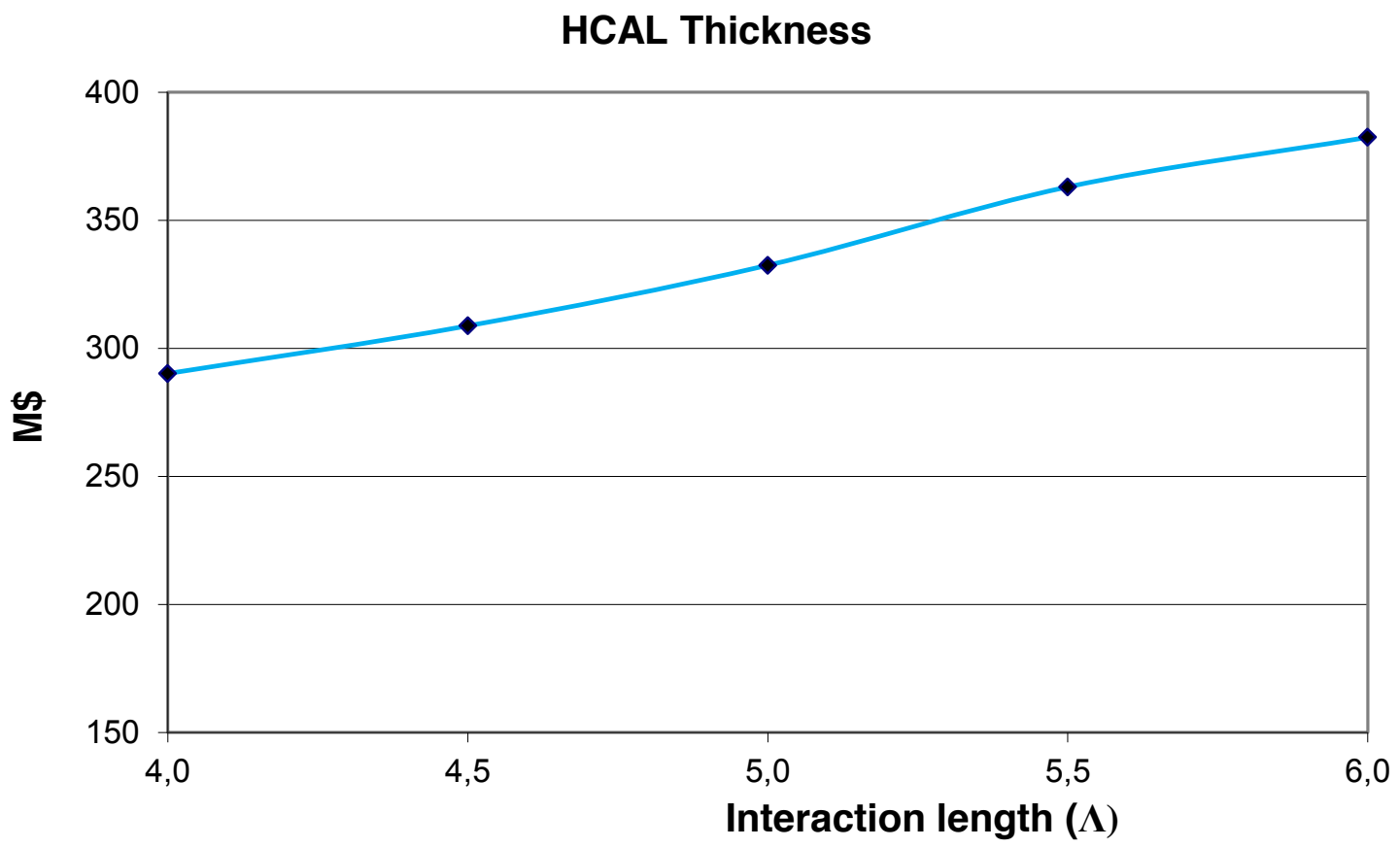

**Figure II-12.3**
Dependence of the SiD M&S base cost on the thickness of the HCAL

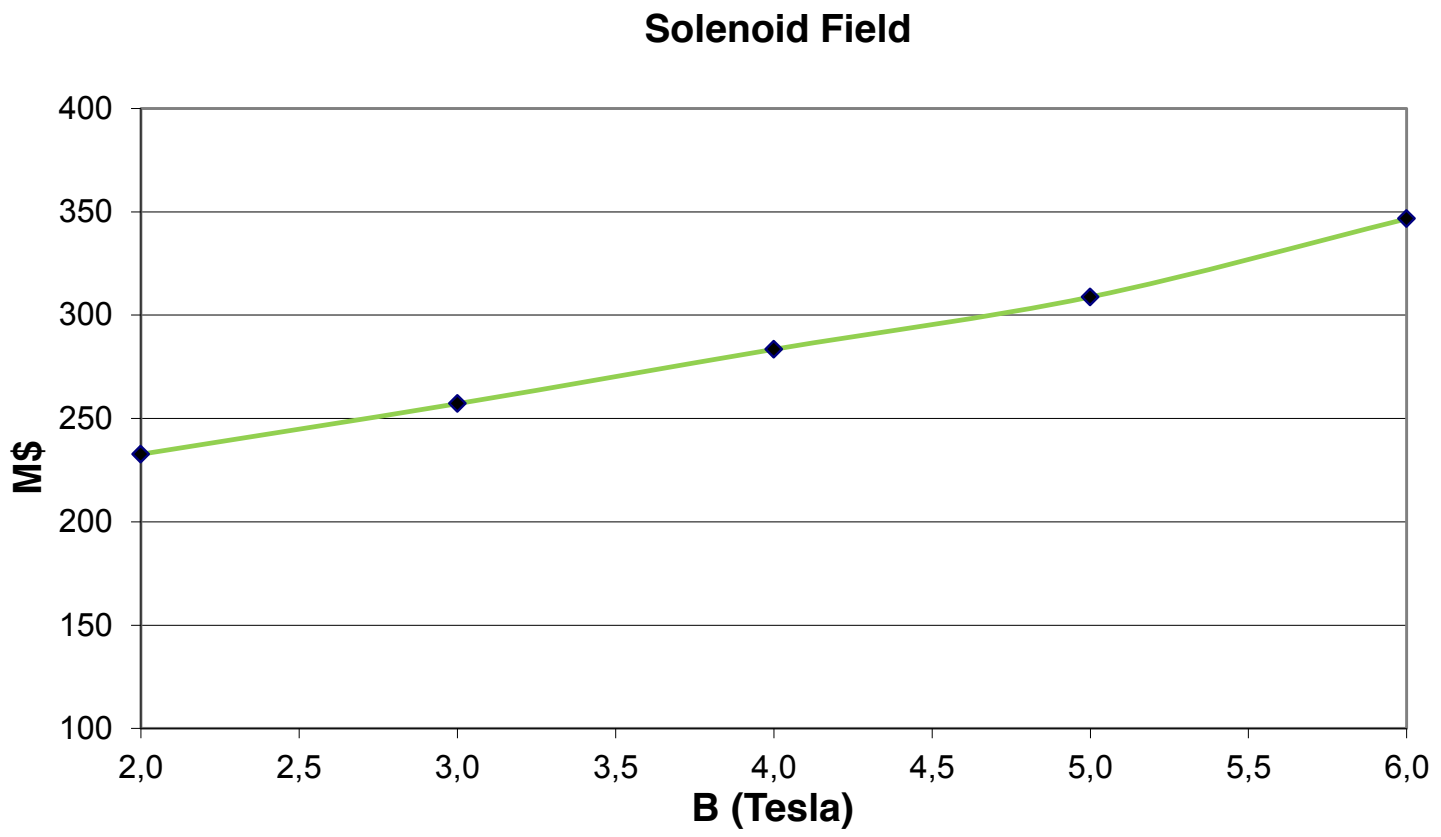

**Figure II-12.4**
Dependence of the SiD M&S base cost on the Solenoid Field

## 12.4 Parameter Dependencies

The parametric fitter enables studies of the SiD cost against the major parameters of the detector. Figure II-12.3 shows the dependence on the HCAL thickness, Figure II-12.4 on the central value of the magnetic field, and Figure II-12.5 on the tracker radius. In all case the cost is M&S base cost; contingency and labour are not included.

SiD has also examined using higher density absorber material in terms of their cost impact, especially by reducing the diameter of the coil. In this exercise, the number of layers and $\lambda_I$ has been keep constant. It has been found, that moving from an all-steel HCAL to an all-tungsten HCAL would increase the total cost of SiD by about US-$ 26M.

**Figure II-12.5**
Dependence of the SiD M&S base cost on the Tracker barrel radius

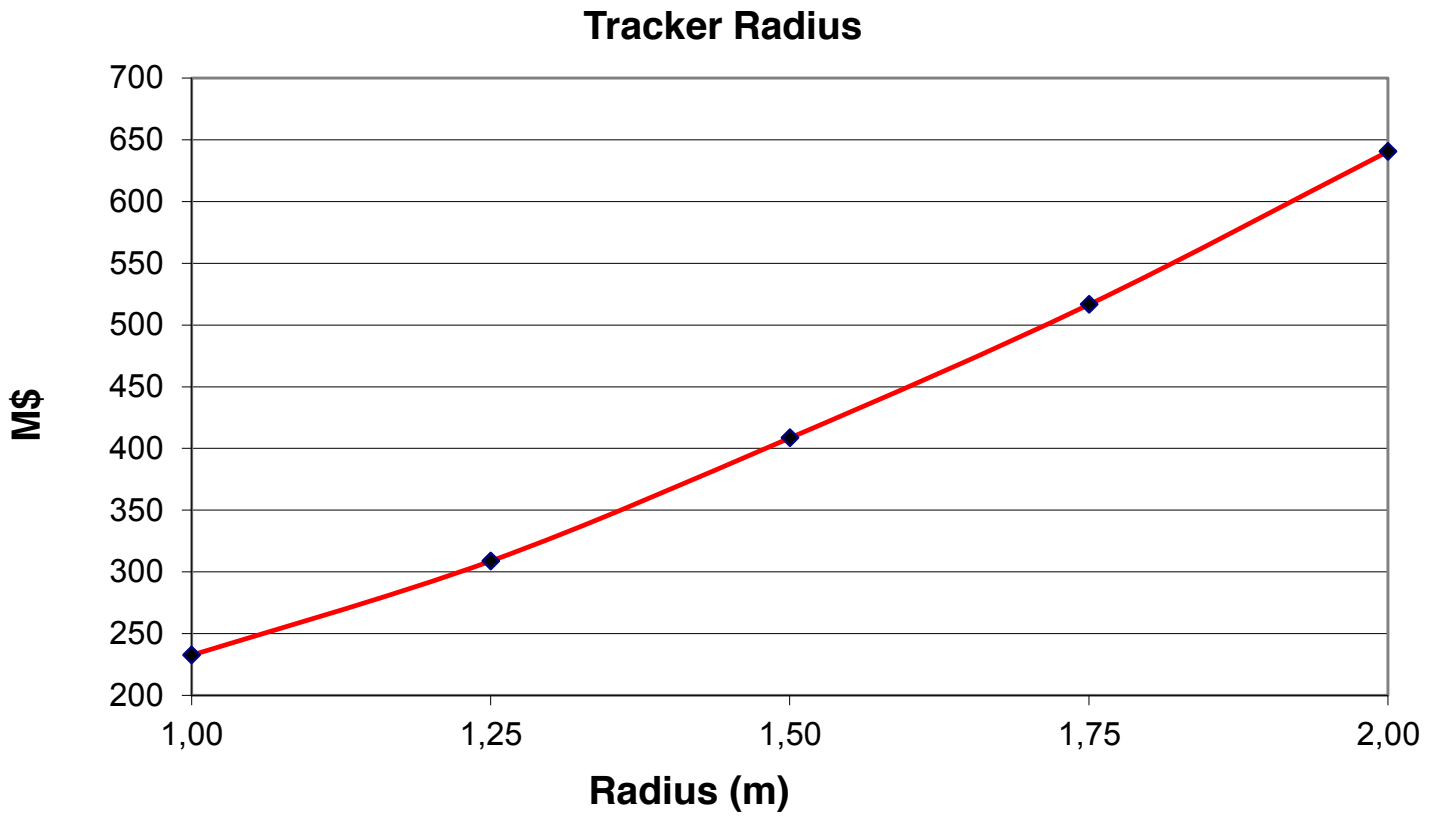

In adapting a configuration used in CLIC_SiD [129] where the transition of barrel and endcap has been optimised for cost and by using tungsten only for the barrel, the cost increase goes down to US-$ 15M. So in terms of cost optimisation, moving to a tungsten HCAL is not beneficial for SiD.

# Chapter 13
# SiD Summary

## 13.1 The Status of the SiD Detector Concept

We have presented a Detailed Baseline Design for the SiD Detector Concept for experiments at the future International Linear Collider. Our design is mature and delivers the required detector and physics performance as demonstrated so far by our simulations and benchmark physics studies.

The design of SiD represents a significant advance over the current generation of collider detectors. The baseline choices for the SiD subsystems represent our current selections in terms of level of successful R&D, measured and or simulated performance characteristics, dimensional practicality, and cost. We will continue to develop alternative technology options where they show promise for enhanced performance. The detector design presented here has been aimed specifically at a 500 GeV or 1 TeV ILC. A modified SiD design, aimed at CLIC energies up to 3 TeV, has already been described in the CLIC Conceptual Design Report (CDR) [129].

## 13.2 Further Development of the SiD Detector Concept

As a detector concept we strongly believe that, while technologies and/or their implementations may evolve over time, SiD will remain an excellent tool for exploration of physics at the International Linear Collider or CLIC.

There are several aspects to this from the detector, physics, organisational, and resource perspectives. We therefore propose to further study and develop the SiD design as new information emerges in the Higgs and possibly other new physics areas. There are areas of detector R&D that must be further developed and completed, followed by studies of specific implementations in a full technical design.

In parallel, while a limited number of physics processes have been studied for this DBD, there are other processes that should be addressed in continued studies. The sum of all these detector and physics activities points towards a lively and sustained effort on SiD as a well identified concept moving forward into the next phase of linear collider development. We therefore see SiD as a vital element of the future program and a major component of the Physics and Detectors section of the new Linear Collider Organisation.

SiD has evolved from the Letter-of-Intent stage as a largely U.S.- based activity to a more global concept with increased contributions from outside the Americas. Our aim is to expand to an even more global level of participation, and we will pursue this vigorously within the new organisation.

## 13.3 SiD and the New Global Linear Collider Organisation

The members of the SiD Detector Concept look forward to working with the new global Linear Collider Organisation. We view the new organisation as a framework within which we can advance our concept towards a full technical detector design for the ILC, and, working with our CLIC colleagues, for CLIC also. We strongly support the efforts of our Japanese colleagues to construct the ILC and will actively work to promote this project.

Organisationally, we support the creation of a group having broad representation from the concept groups and R&D collaborations to advance the physics and detector case for a linear collider. We believe that, when the time is right for the linear collider to move towards realisation, having well identified detector concepts with a substantial participation from all regions within the global organisation, will significantly benefit discussion of the funding agency contributions to the project as a whole.

# Volume 4

# Detectors

## Part III

## ILD Detailed Baseline Design

# ILD Editors

Main Editors:
T. Behnke, D. Karlen, Y. Sugimoto, H. Videau, H. Yamamoto

Tracking System:
T. Matsuda, A. Ruiz-Jimeno, A. Savoy Navarro, R. Settles, Y. Sugimoto, I. Vila, M. Vos, M. Winter

Calorimer System:
H. Abramowicz, J.C. Brient, M. Fouz, D. Jeans, I. Laktineh, W. Lohmann, R. Poeschl, F. Sefkow, F. Simon, T. Takeshita

Outer Detector System:
M. Danilov, K. Elsener, F. Kircher, V. Saveliev, U. Schneekloth

Data Acquisition:
V. Boudry

Machine Detector Interface:
K. Buesser, T. Tauchi

Integration:
K. Buesser, C. Clerc, T.Tauchi

Alignment:
M. Fernández, F. Sefkow

Software:
F. Gaede, A. Miyamoto

Performance:
M. Berggren, A. Miyamoto, T. Tanabe, M. Thomson

Costing:
C. Clerc, H. Videau

# Chapter 1
# ILD: Executive Summary

The **I**nternational **L**arge **D**etector (ILD) is a concept for a detector at the International Linear Collider, ILC [198]. In a slightly modified version, it has also been proposed for the CLIC linear collider [199].

The ILD detector concept has been optimised with a clear view on precision. In recent years the concept of particle flow has been shown to deliver the best possible overall event reconstruction. Particle flow implies that all particles in an event, charged and neutral, are individually reconstructed. This requirement has a large impact on the design of the detector, and has played a central role in the optimisation of the system. Superb tracking capabilities and outstanding detection of secondary vertices are other important aspects. Care has been taken to design a hermetic detector, both in terms of solid-angle coverage, but also in terms of avoiding cracks and non-uniformities in response. The overall detector system has undergone a vigorous optimisation procedure based on extensive simulation studies both of the performance of the subsystems, and on studies of the physics reach of the detector. Simulations are accompanied by an extensive testing program of components and prototypes in laboratory and test-beam experiments.

**Figure III-1.1**
View of the ILD detector concept.

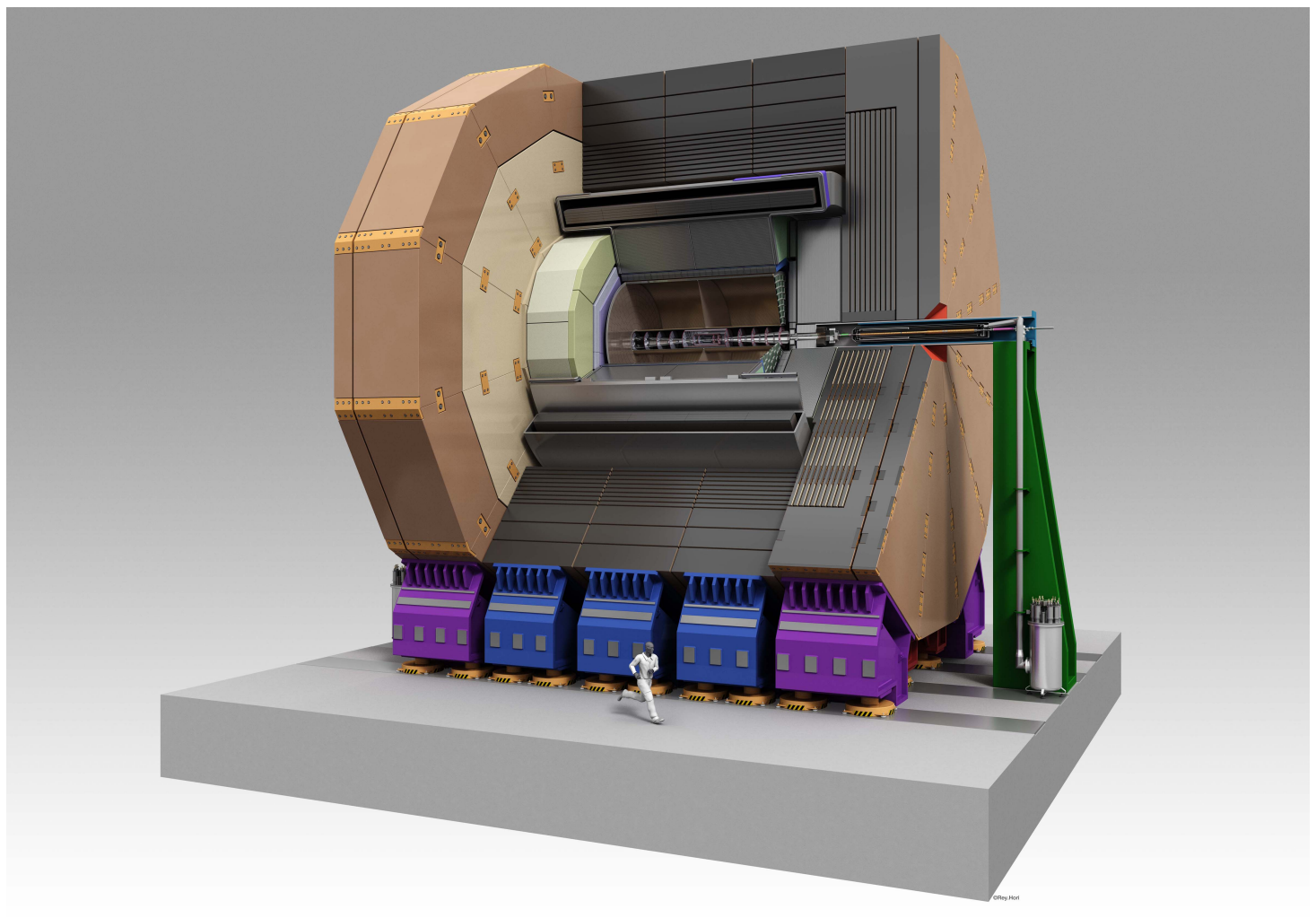

The ILD detector concept has been described in a number of documents in the past. Most recently the letter of intent [198] gave a fairly in depth description of the ILD concept. The ILD concept is based on the earlier GLD and LDC detector concepts [200, 201, 202]. Since the publication of the letter of intent, major progress has been made in the maturity of the technologies proposed for ILD, and their integration into a coherent detector concept.

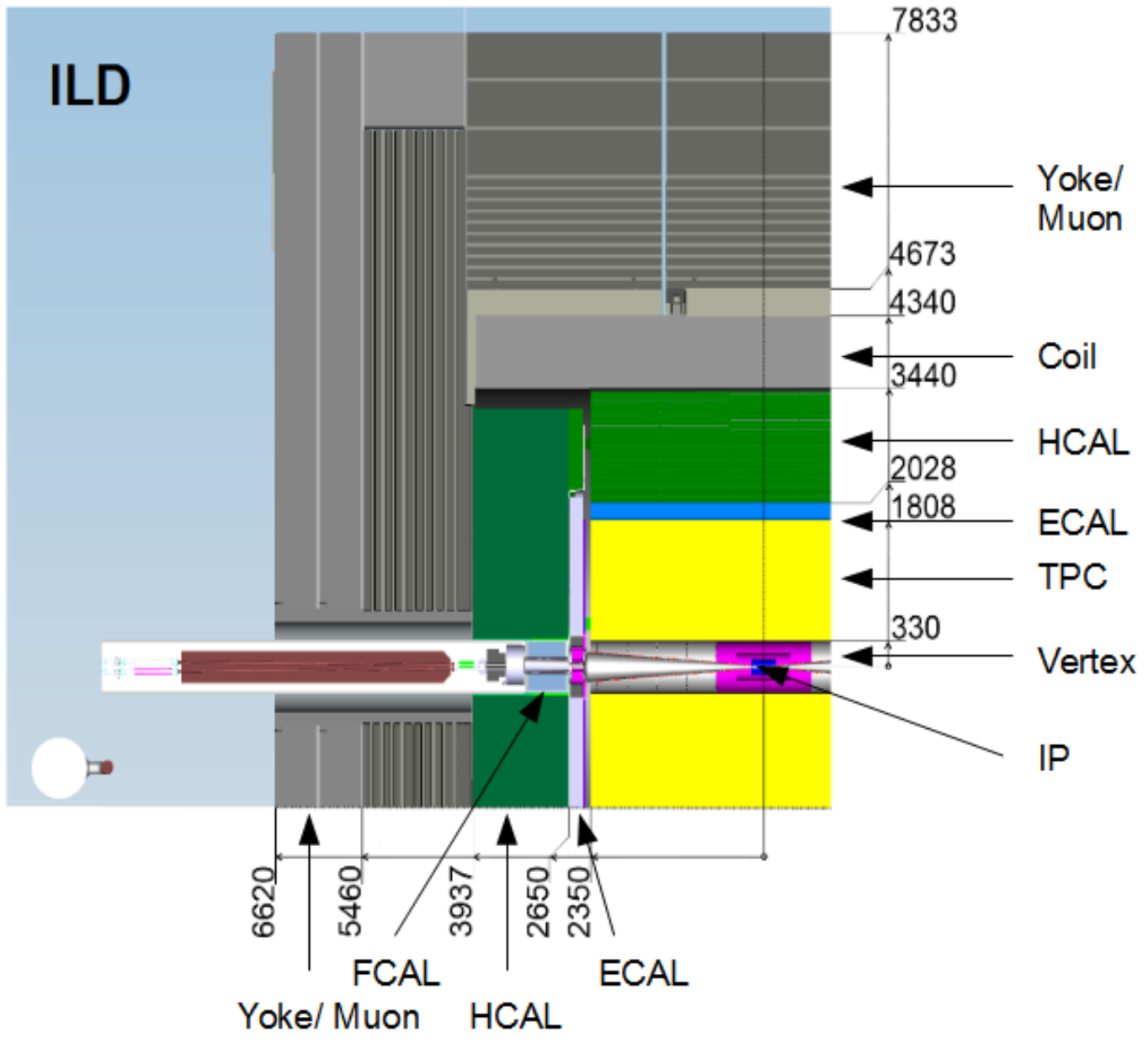

**Figure III-1.2** Quadrant view of the ILD detector concept. The interaction point is in the lower right corner of the picture. Dimensions are in mm.

## 1.1 ILD philosophy and challenges

The particle flow paradigm translates into a detector design which stresses the topological reconstruction of events. A direct consequence of this is the need for a detector system which can separate efficiently charged and neutral particles, even inside jets. This emphazises the spatial resolution for all detector systems. A highly granular calorimeter system is combined with a central tracker which stresses redundancy and efficiency. The whole system is immersed in a strong magnetic field of 3.5 T. In addition, efficient reconstruction of secondary vertices and very good momentum resolution for charged particles are essential for an ILC detector. An artistic view of the detector is shown in Figure III-1.1, a vew of a quarter of the detector is seen in Figure III-1.2.

The interaction region of the ILC is designed to host two detectors, which can be moved in and out of the beam position with a "push-pull" scheme. The mechanical design of ILD and the overall integration of subdetectors takes these operational constraints into account.

The ILC is designed to investigate the mechanism of electroweak symmetry breaking. It will allow the study of the newly found higgs-like particle at 126 GeV. It will search for and explore new physics at energy scales up to 1 TeV. In addition, the collider will provide a wealth of information on standard model (SM) physics, for example top physics, heavy flavour physics, and physics of the $\mathrm{Z}$ and $\mathrm{W}$ bosons, as discussed earlier in this document. A typical event ($t\bar{t}$ at 500 GeV) is shown in Figure III-1.3. The requirements for a detector are, therefore, that multi-jet final states, typical for many physics channels, can be reconstructed with high accuracy. The jet energy resolution should be sufficiently good that the hadronic decays of the $\mathrm{W}$ and $\mathrm{Z}$ can be separated. This translates into a jet energy resolution of $\sigma_E/E \sim 3-4\%$ (equivalent to $30\%/\sqrt{E}$ at $100\,\mathrm{GeV}$). Secondary vertices which are relevant for many studies involving heavy flavours should be reconstructable with good efficiency and purity. Highly efficient tracking is needed with large solid-angle coverage.

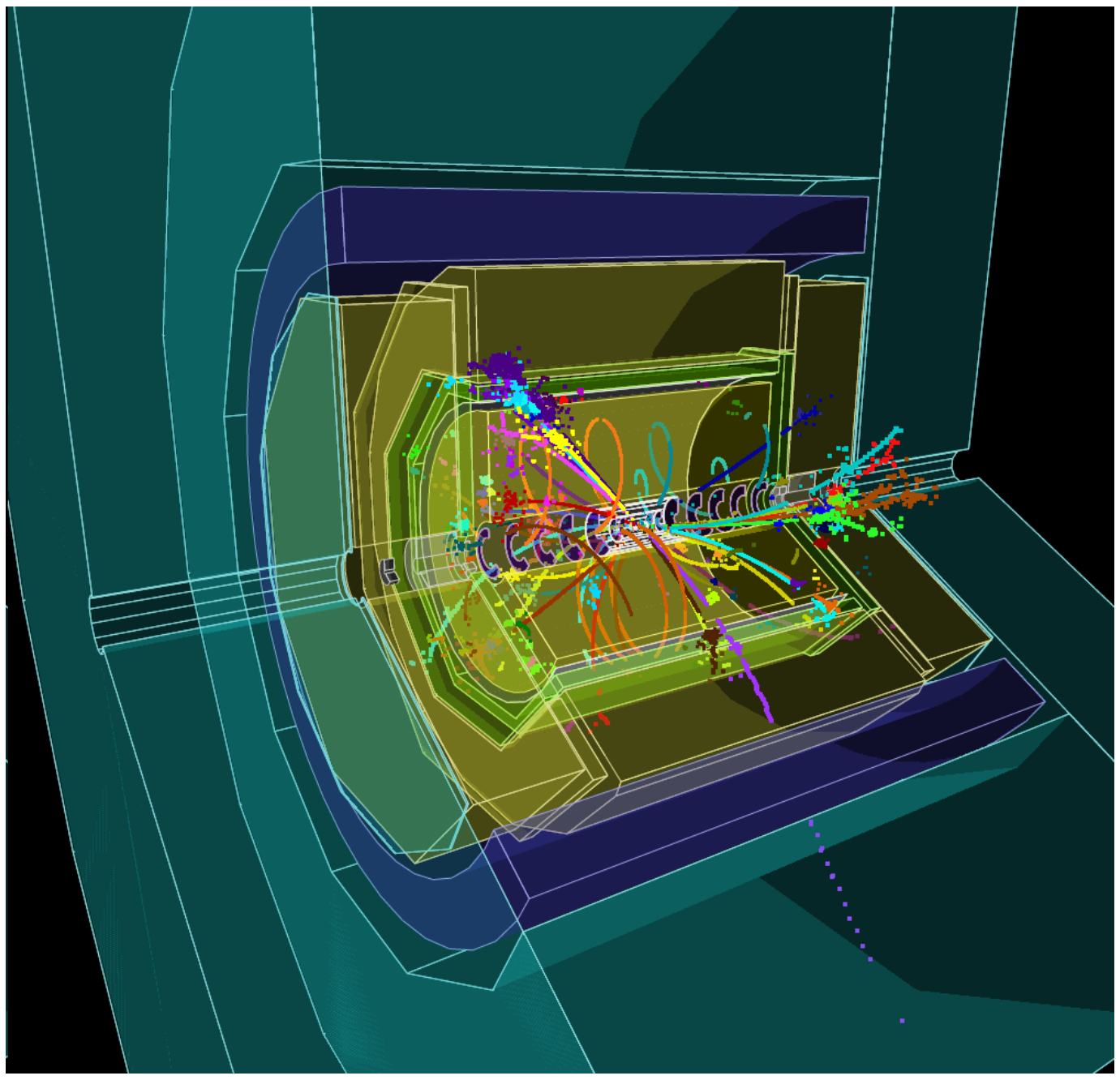

**Figure III-1.3** Three-dimensional view of a typical multi jet final state at the ILC (500 GeV $t\bar{t}$ event with multi-hadronic final state). The picture was generated by the detailed detector simulation of the ILD detector.

## 1.2 ILD layout and performance

The ILD concept has been designed as a multi-purpose detector. A high precision vertex detector is followed by a hybrid tracking layout, realised as a combination of silicon tracking with a time projection chamber, and a calorimeter system. The complete system is located inside the large solenoid. On the outside of the coil, the iron return yoke is instrumented as a muon system and as a tail catcher calorimeter.

The vertex detector is realised as a multi-layer pixel-vertex detector (VTX), with three super-layers each comprising two layers, or a 5 layer geometry. In either case the detector has a pure barrel geometry. To minimise the occupancy from background hits, the first super-layer is only half as long as the outer two. Whilst the underlying detector technology has not yet been decided, the VTX is optimised for point resolution and minimum material thickness.

A system of silicon strip and pixel detectors surrounds the VTX detector. In the barrel, two layers of silicon strip detectors (SIT) are arranged to bridge the gap between the VTX and the TPC. In the forward region, a system of two silicon-pixel disks and five silicon-strip disks (FTD) provides low angle tracking coverage.

A distinct feature of ILD is a large volume time projection chamber (TPC) with up to 224 points per track. The TPC is optimised for 3-dimensional point resolution and minimum material in the field cage and in the end-plate. It also allows $\mathrm{d}E/\mathrm{d}x$ based particle identification.

Outside the TPC a system of Si-strip detectors, one behind the end-plate of the TPC (ETD) and one in between the TPC and the ECAL (SET), provide additional high precision space points which improve the tracking performance and provide additional redundancy in the regions between the main tracking volume and the calorimeters.

A highly segmented electromagnetic calorimeter (ECAL) provides up to 30 samples in depth and small transverse cell size, split into a barrel and an end cap system. For the absorber Tungsten has been chosen, for the sensitive area silicon diodes or scintillator strips are considered.

This is followed by a highly segmented hadronic calorimeter (HCAL) with up to 48 longitudinal samples and small transverse cell size. Two options are considered, both based on a Steel-absorber structure. One option uses scintillator tiles of $3 \times 3\,\mathrm{cm}^2$, which are read out with an analogue system. The second uses a gas-based readout which allows a $1 \times 1\,\mathrm{cm}^2$ cell geometry with a binary or semi-digital readout of each cell.

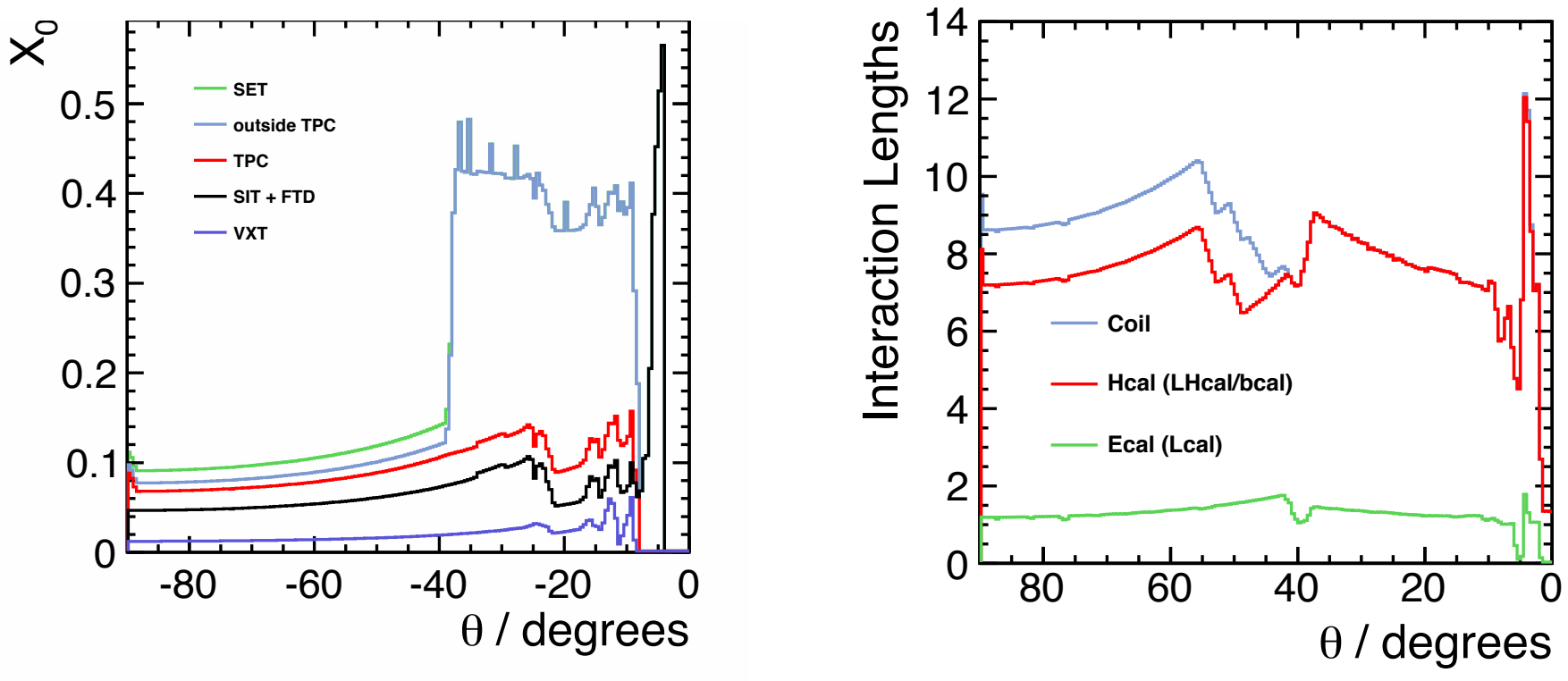

**Figure III-1.4**
Left: Average total radiation length of the material in the tracking detectors as a function of polar angle. Right: Total interaction length in the detector, up to the end of the calorimeter system, and including the coil of the detector.

At very forward angles, below the coverage provided by the ECAL and the HCAL, a system of high precision and radiation hard calorimetric detectors (LumiCAL, BeamCAL, LHCAL) is foreseen. These extend the calorimetric coverage to almost $4\pi$, measure the luminosity, and monitor the quality of the colliding beams.

A large volume superconducting coil surrounds the calorimeters, creating an axial $B$-field of nominally 3.5 Tesla.

An iron yoke, instrumented with scintillator strips or resistive plate chambers (RPCs), returns the magnetic flux of the solenoid, and, at the same time, serves as a muon filter, muon detector and tail catcher calorimeter.

To maximise the sensitivity of the detector to the physics at the ILC, the detector will be operated in a continuous readout mode, without a traditional hardware based trigger.

Precision physics at the ILC requires that the beam parameters are known with great accuracy. The beam energy and the beam polarization will be measured in small dedicated systems, which are shared by the two detectors present in the interaction region.

The ILD detector has been designed and optimised as a detector which can be used in a push-pull configuration, as described in section 5.5.

The main parameters of the ILD detector are summarised in Table III-1.1 and table III-1.2.

The performance of the ILD concept has been extensively studied using a detailed GEANT4 based simulation model and sophisticated reconstruction tools. Backgrounds have been taken into account to the best of current knowledge. A key characteristics of the detector is the amount of material in the detector. Particle flow requires a thin tracker, to minimise interactions before the calorimeters, and thick calorimeters, to fully absorb the showers. Figure III-1.4 (left) shows the material in the detector in radiation lengths, until the entry of the calorimeter. The right plot shows

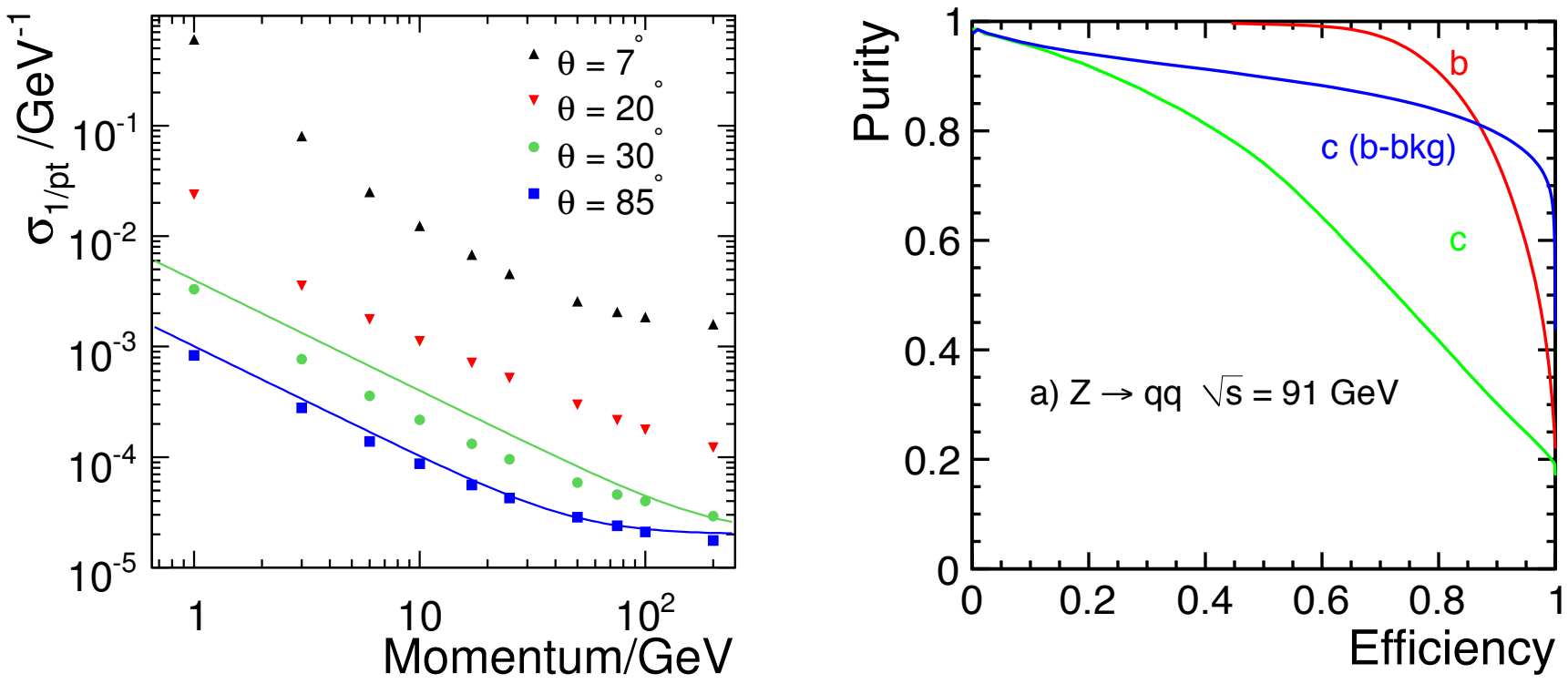

**Figure III-1.5**
Left: Momentum resolution as a function of the transverse momentum of particles, for tracks with different polar angles. Also shown is the theoretical expectation. Right: Flavour tagging performance for $Z \to q\bar{q}$ samples at different energies.

**Figure III-1.6** Fractional jet energy resolution plotted against $|\cos\theta|$ where theta is the polar angle of the thrust axis of the event.

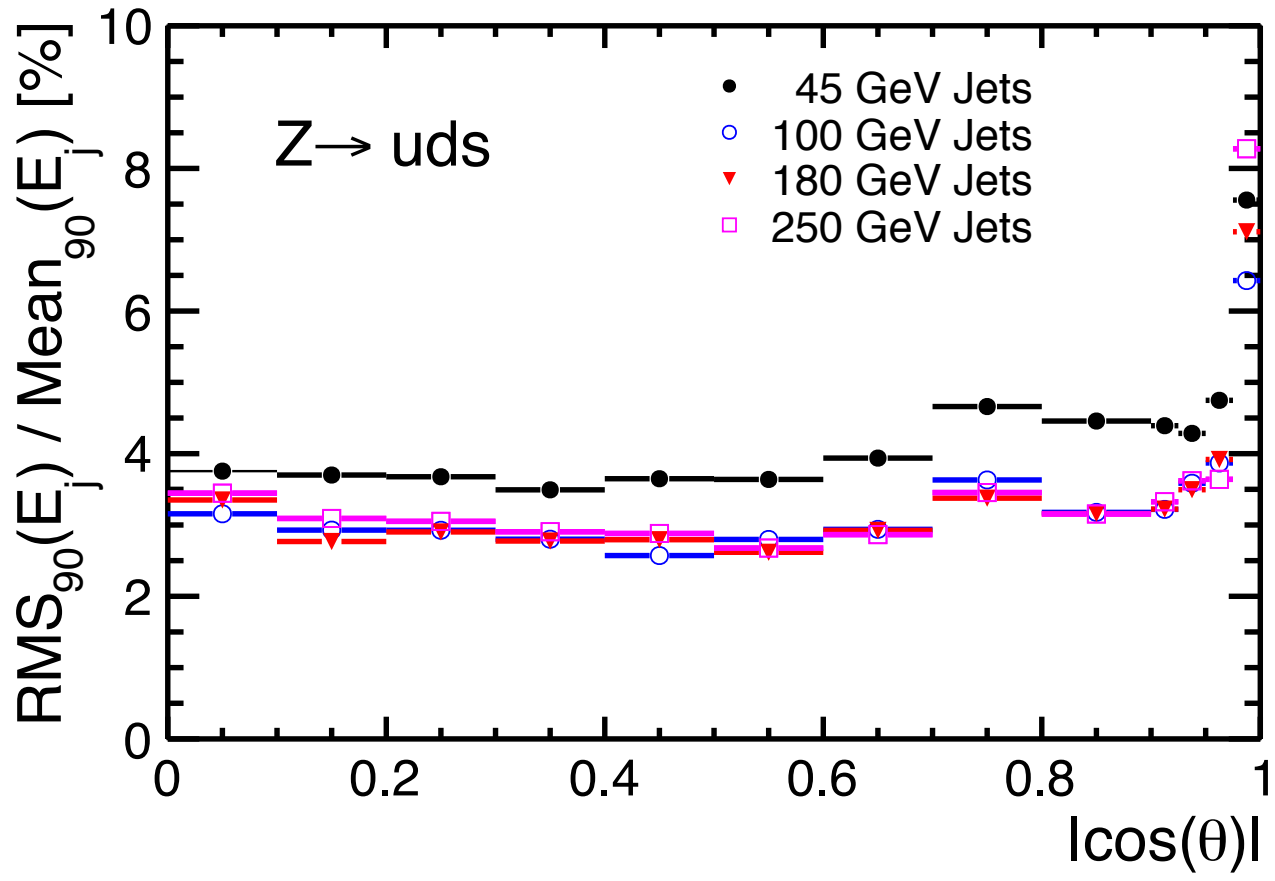

**Table III-1.1.** List of the main parameters of the ILD detector for the barrel part.

| **Barrel system** | | | | | | |
|---|---|---|---|---|---|---|
| System | R(in) | R(out) [mm] | z | comments | | |
| VTX | 16 | 60 | 125 | 3 double layers | Silicon pixel sensors, | |
| | | | | layer 1: | layer 2: | layer 3-6 |
| | | | | $\sigma < 3\mu m$ | $\sigma < 6\mu m$ | $\sigma < 4\mu m$ |
| Silicon | | | | | | |
| - SIT | 153 | 300 | 644 | 2 silicon strip layers | $\sigma = 7\mu m$ | |
| - SET | 1811 | | 2300 | 2 silicon strip layers | $\sigma = 7\mu m$ | |
| - TPC | 330 | 1808 | 2350 | MPGD readout | $1 \times 6\text{mm}^2$ pads | $\sigma = 60\mu m$ at zero drift |
| ECAL | 1843 | 2028 | 2350 | W absorber | SiECAL | 30 Silicon sensor layers, $5 \times 5$ mm$^2$ cells |
| | | | | | ScECAL | 30 Scintillator layers, $5 \times 45$ mm$^2$ strips |
| HCAL | 2058 | 3410 | 2350 | Fe absorber | AHCAL | 48 Scintillator layers, $3 \times 3\text{cm}^2$ cells, analogue |
| | | | | | SDHCAL | 48 Gas RPC layers, $1 \times 1$ cm$^2$ cells, semi-digital |
| Coil | 3440 | 4400 | 3950 | 3.5 T field | $2\lambda$ | |
| Muon | 4450 | 7755 | 2800 | 14 scintillator layers | | |

the total interaction length including the calorimeter system.

The performance of the tracking system can be summarised by its combined momentum resolution, shown in Figure III-1.5 (left). A resolution of $\sigma_{1/p_T} = 2 \times 10^{-5}$ GeV$^{-1}$ has been achieved for high momenta. For many physics studies the tagging of long lived particles is of key importance. Several layers of pixel detectors close to the IP allow the reconstruction of displaced vertices, as shown in Figure III-1.5 (right).

Calorimeter system and tracking system together enter into the particle flow performance. The performance of the ILD detector for different energies and as a function of the polar angle is shown in Figure III-1.6.

The few plots shown in this executive summary illustrate the anticipated performance of the detector and illustrate the potential for precision measurements with the ILD detector. More details

**Table III-1.2.** List of the main parameters of the ILD detector for the end cap part.

| **End cap system** | | | | | | |
|---|---|---|---|---|---|---|
| System | z(min) | z(max) [mm] | r(min), r(max) | comments | | |
| FTD | 220 | 371 | | 2 pixel disks | $\sigma = 2 - 6\mu m$ | |
| | | | | 5 strip disks | $\sigma = 7\mu m$ | |
| ETD | 2420 | 2445 | 419-1822 | 2 silicon strip layers | $\sigma = 7\mu m$ | |
| ECAL | 2450 | 2635 | | W-absorber | SiECAL | Si readout layers |
| | | | | | ScECAL | Scintillator layers |
| HCAL | 2650 | 3937 | 335-3190 | Fe absorber | AHCAL | 48 Scintillator layers $3 \times 3$cm$^2$ cells, analogue |
| | | | | | SDHCAL | 48 gas RPC layers $1 \times 1$cm$^2$ cells, semi-digital |
| BeamCal | 3595 | 3715 | 20-150 | W absorber | 30 GaAs readout layers | |
| Lumical | 2500 | 2634 | 76-280 | W absorber | 30 Silicon layers | |
| LHCAL | 2680 | 3205 | 93-331 | W absorber | | |
| Muon | 2560 | | 300-7755 | 12 scintillator layers | | |

on the performance may be found in section 6.1 of this document.

In this document the design of ILD is presented. Intense R&D has taken place over the last decade to develop the necessary technologies. This work has typically happened within dedicated R&D collaborations, which are independent but maintain very close connections to ILD. All technologies selected by ILD for one of its subsystems have been proven experimentally to meet the performance goals, or to come very close. In some cases ILD presents more than one technology for a given sub-detector. At this moment no attempt has been made by the ILD group to down-select or limit the number of different technologies. The concept group wants to remain open and flexible to be prepared to select the most modern and most powerful technology once it is necessary. However a distinction is made between options and alternatives: while options have undergone an extensive R&D program and have passed critical proof-of-concept tests, alternatives are potentially interesting and promising technologies which have not matured to a similar level at the time of writing this document.

# Chapter 2
# ILD Tracking System

Key features of the ILD detector are a very powerful and redundant tracking systems, consisting of a high precision large volume time projection chamber, surrounded by a complete Silicon based vertexing and tracking system, all contained, together with a highly granular calorimeter system, inside a 3.5 T strong solenoidal field.

## 2.1 ILD vertex system

The identification of heavy (charm and bottom) quarks and tau leptons is essential for the ILC physics programme. The reconstruction of decay vertices of short lived particles, such as $D$ or $B$ mesons, deserves therefore much attention and requires a particularly light and precise vertex detector. The vertices are tracked back by reconstructing the trajectory of the short lived particles decay products. This is achieved through the very precise measurement of the charged particles' track parameters in the vicinity of the interaction point, which are then combined with those of the other tracking detectors to reconstruct vertices. The performance of a vertex detection system may be expressed by the resolution on the impact parameter of charged particles. The main performance goal of the ILD vertexing system resumes in a resolution on the track impact parameter of $\sigma_b < 5 \oplus 10/p \sin^{3/2}\theta$ $\mu$m. In order to reach such a high performance level, the ILD vertex detector should comply with the following specifications:

- A spatial resolution near the IP better than 3 $\mu$m ;
- A material budget below 0.15% $X_0$/layer;
- A first layer located at a radius of $\sim 1.6$ cm;
- A pixel occupancy not exceeding a few %.

The power consumption should be low enough to minimise the material budget of the cooling system inside the detector sensitive volume. Power savings may be obtained by exploiting the beam time structure and power pulse the sensors equipping the detector. Alternatively, the signals may be integrated over the complete bunch train and read out in-between consecutive trains at very low frequency and thus very low power.

The required radiation tolerance follows entirely from the beam related background (i.e. beamstrahlung) (see section 5.5.6), which is expected to affect predominantly the innermost layer. The requirements for the total ionising dose and the fluence amount respectively to about $1$ kGy and $10^{11}$ $n_{eq}/cm^2$ per annum. These values assume that neutrons backscattered from the beam dump are shielded well enough to add a minor contribution to the overall radiation load.

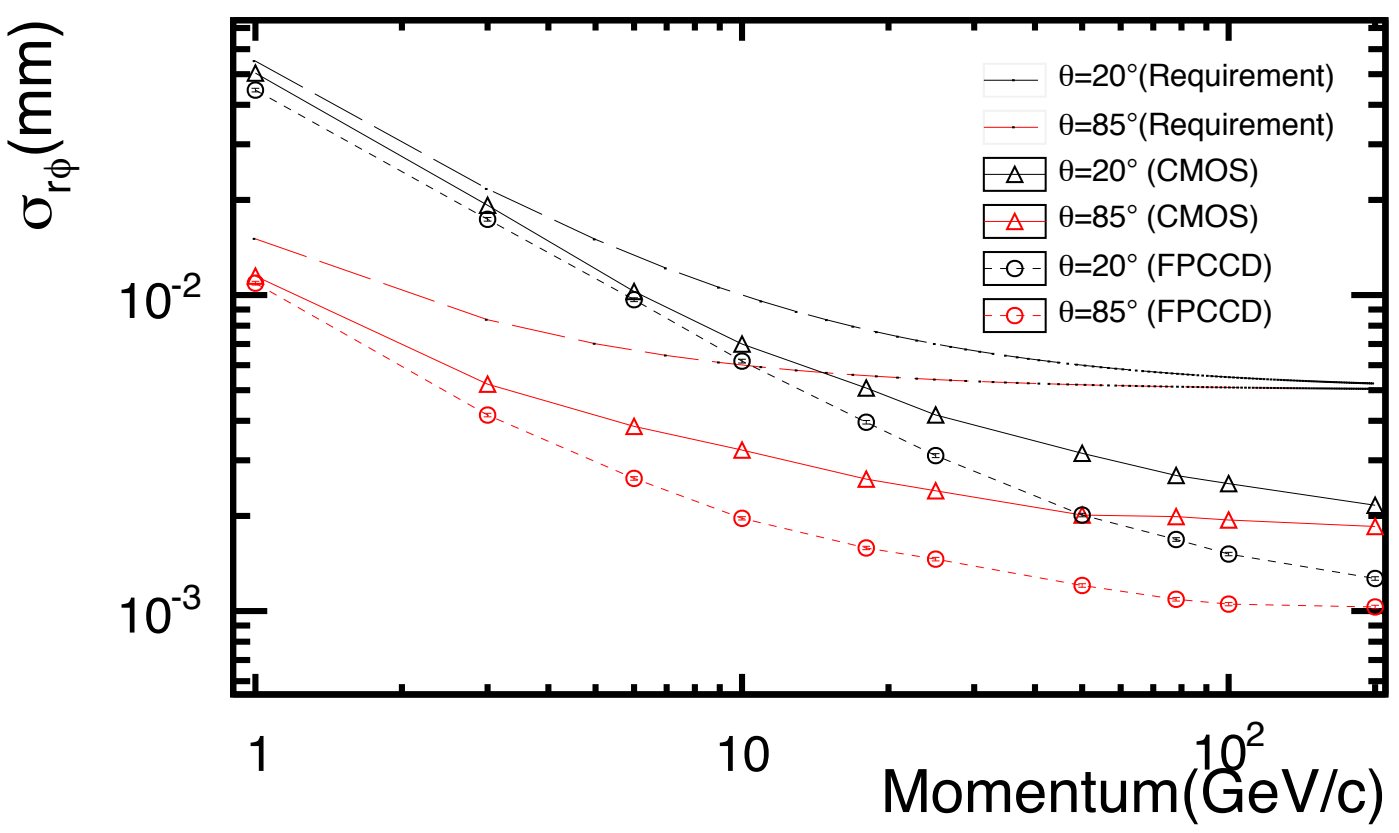

**Figure III-2.1** Impact parameter resolution of the ILD vertex detector for two different particle production angles (20° and 85°), assuming the baseline point resolution given in Table III-2.1 for the CMOS option (solid line), and the FPCCD option (dotted line). The curves with long dashes show the performance goal.

## 2.1.1 Baseline design

The baseline design of the ILD vertex detector consists of three, nearly cylindrical, concentric layers of double-sided ladders. Each ladder is equipped with pixel sensors on both sides, $\sim$ 2 mm apart, resulting in six measured impact positions for each charged particle traversing the detector. The radii covered by the detector range from $16$ mm to $60$ mm. The material budget of each ladder amounts to $\sim 0.3\%\ X_0$, equivalent to $0.15\%\ X_0$/layer.

An alternative geometry is also considered, based on five equally spaced single-sided layers, with radii ranging from $15$ to $60$ mm.

The current layout of the proposed vertex detector is summarised in Table III-2.1. It is based on extensive simulation and technical studies. The parameters are considered conservative.

**Table III-2.1** Vertex detector parameters. The spatial resolution and readout times are for the CMOS option described in section 2.1.2.1.

| | $R$ (mm) | $\lvert z\rvert$ (mm) | $\lvert\cos\theta\rvert$ | $\sigma$ ($\mu$m) | Readout time ($\mu$s) |
|---|---|---|---|---|---|
| Layer 1 | 16 | 62.5 | 0.97 | 2.8 | 50 |
| Layer 2 | 18 | 62.5 | 0.96 | 6 | 10 |
| Layer 3 | 37 | 125 | 0.96 | 4 | 100 |
| Layer 4 | 39 | 125 | 0.95 | 4 | 100 |
| Layer 5 | 58 | 125 | 0.91 | 4 | 100 |
| Layer 6 | 60 | 125 | 0.9 | 4 | 100 |

The impact parameter resolution following from the single point resolutions provided in the table is displayed in Figure III-2.1 as a function of the particle momentum, showing that the ambitious impact parameter resolution is achievable.

## 2.1.2 Pixel technologies and readout electronics

Currently three sensor technology options are actively developed for the ILD vertex detector. They have been shown to have the potential of meeting the detector requirements or to come close to them. Those technological options are CMOS Pixel Sensors (CPS) [203, 204, 205, 206], Fine Pixel CCD (FPCCD) sensors [207, 208, 209, 210], and Depleted Field Effect Transistor (DEPFET) sensors [211, 212, 213]. The development and optimisation of each technology is closely associated to a specific readout architecture. For CPS and DEPFETs a power pulsed readout is under study and offers attractive advantages. For the FPCCD, the very large number of pixels calls for a slow (low power) readout, which must be performed in between bunch trains.

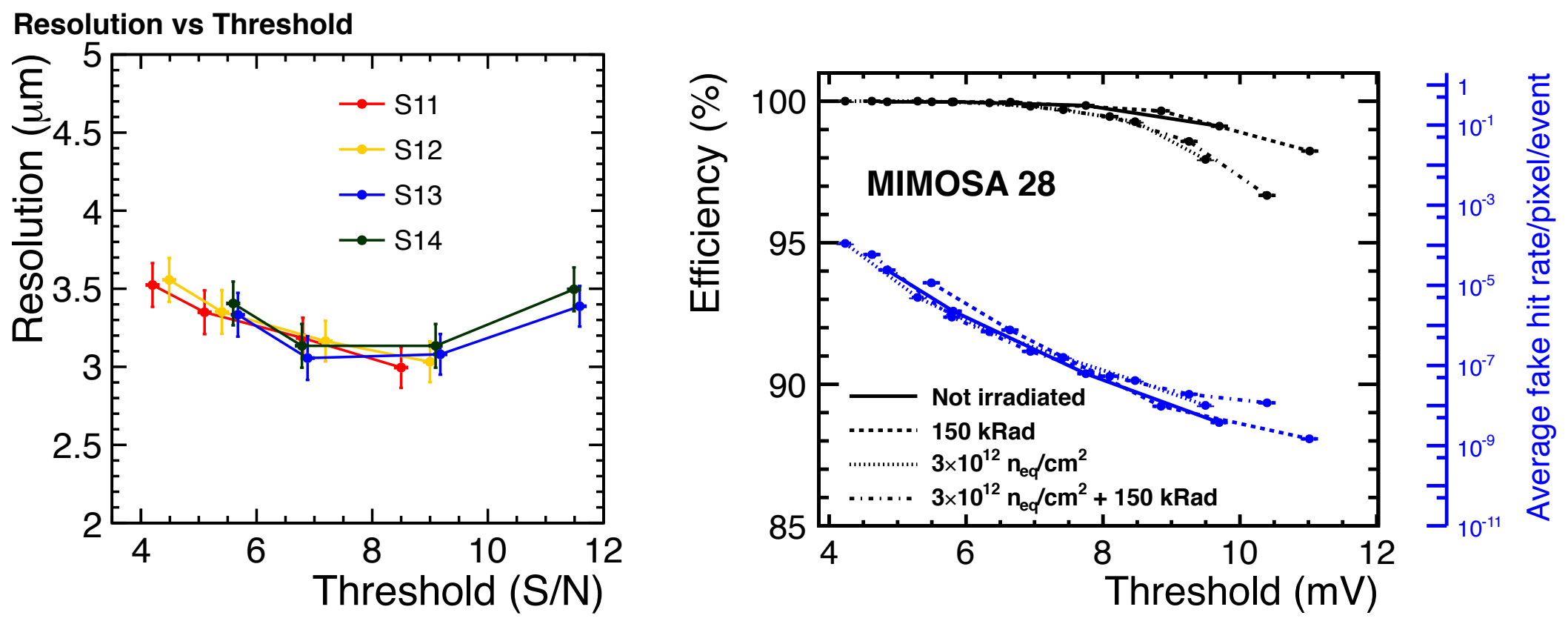

**Figure III-2.2.** Left: Single point resolution measured with a $18.4$ $\mu$m pitch as a function of the discriminator threshold (the colours refer to different in-pixel circuits). Right: Measured variations of the minimum ionising particle (m.i.p.) detection efficiency and fake hit rate (fraction of pixel noise fluctuations above threshold) of the STAR sensor as a function of the discriminator threshold, before and after irradiation ($150$ kRad, $3 \times 10^{12}$ n$_{eq}$/cm$^2$) at a coolant temperature of $30°$C.

## 2.1.2.1 CMOS pixel sensors

CMOS pixel sensors (CPS) use as a sensitive volume the $10$-$20$ $\mu$m thin, high-resistivity epitaxial layer deposited on low resistivity substrates of commercial CMOS processed chips, which makes them particularly well suited for a low-mass detector. A high density sensing node lattice is implemented on the layer, allowing for high spatial resolution. Moreover, the full signal processing circuitry may be integrated on the same substrate as the sensitive volume, which makes CPS flexible and cost effective.

Developments over the last couple of years have shown that these genuine features can be combined into a single, full scale device, fabricated by industry, which complies with the vertex detector specifications. The proof of principle was achieved with the MIMOSA-26 sensor, initially developed for the beam telescope of the European Union (EU) funded project EUDET [203].

The sensor architecture is based on a column parallel read-out with amplification and correlated double sampling inside each pixel. The chip features $1152$ columns of $576$, $18.4$ $\mu$m pitch, pixels. Each column is terminated with a precision discriminator and is read out in a rolling shutter mode at a frequency of $5$ MHz ($200$ ns/row). Due to charge sharing, the spatial resolution obtained is close to $3$ $\mu$m (see left of Figure III-2.2) despite the single bit charge encoding.

The MIMOSA-26 architecture was extended to a sensor (MIMOSA-28) [204] adapted to the (air cooled) vertex detector (PXL) of the STAR experiment at BNL [205], foreseen to start data taking in 2013. Its current assembly and upcoming operation provide valuable experience for the vertex detector at ILD. The sensor minimum ionising particle (m.i.p.) detection performances measured at $30°$C are displayed in Figure III-2.2 (right), before and after irradiation with loads well above the annual values expected at the ILC ($\lesssim 100$ kRad, $< 10^{11}$ n$_{eq}$/cm$^2$), showing the adequacy of the CPS radiation tolerance up to high temperatures.

CPS complying with the vertex detector specifications were derived from MIMOSA-26, with modified spatial resolution and read-out times, and adapted to different requirements for distinct layers.

For the inner layer, which accounts only for $\sim 10\%$ of the detector active surface, the sensors are optimised for single point resolution and short read-out time, relaxing the power consumption constraints. The conflict between high granularity and fast read-out is resolved by equipping the innermost ladders with two different types of sensors, one achieving the required spatial resolution and one providing a fast time stamp.

The high precision sensors mounted on one side of a ladder, feature square pixels with $\sim 17$ $\mu m$

pitch and provide a spatial resolution $< 3$ $\mu$m. The frame read-out time is $50$ $\mu$s, which may lead to a relatively high occupancy if the beam related background happens to be a few times higher than expected from simulations. The fast sensors are installed on the other side of the ladder. They feature rectangular pixels (e.g. $17 \times 85$ $\mu$m$^2$), which result in five times less pixels per column and therefore in a 10 $\mu$s time resolution, at the expense of an increased spatial resolution of $\sim$ 6 $\mu$m.

The combination of a very precise sensor with a much faster one in a geometry which provides a tight correlation between the two allows to achieve a spatial resolution of $< 3$ $\mu$m and a timing resolution of $\lesssim 10$ $\mu$s (see left of Figure III-2.5), from the first layer alone. This is expected to strongly suppress the perturbation of the track reconstruction due to beam related background, even if its rate is well above simulated values.

The design of these sensors is ready and has been partially validated. Further improvements might consist in integrating a low power discriminator for each pixel [206]. It reduces the read-out time to $< 5$ $\mu$s and the pixel array power consumption by at least $30\%$. This approach, which is also followed for the ALICE-ITS upgrade, is expected to be mature within $2$-$3$ years.

The sensors for the outer layers have to cover a much larger area, but see significantly lower occupancies. Therefore their design has been optimised to minimise the power consumption. Pixels of $34 \times 34$ $\mu$m$^2$ are used, organised in columns terminated with $4$-bit ADCs. They achieve a spatial resolution of $\sim 4$ $\mu$m, at a read-out time of $100$ $\mu$s. A prototype composed of $64 \times 64$ pixels is currently being tested.

The instantaneous power consumption of the full detector was evaluated to be $< 600$ W. Assuming power cycling with a conservative duty cycle value of $2\%$ (i.e. $5$ ms long periods of power dissipation encompassing the 1 ms long bunch trains), the average power dissipation is about $10$ W, an amount expected to comply with air cooling.

#### 2.1.2.2 Fine Pixel CCD

The use of FPCCD sensors allows for particularly small pixels ($\sim 5$ $\mu$m pitch), which results in a sub-micron spatial resolution and an excellent two-track separation capability. It allows simultaneously to mitigate the occupancy generated by the beam related background even when integrating the signal over many bunch-crossings.

The sensitive volume is a $\sim 15$ $\mu$m-thick epitaxial layer. It is fully depleted, resulting in a limited charge spread, which is essential to keep the number of hit pixels per hit small. The pixel occupancy is therefore expected to remain affordable even if accumulating the signals over a full bunch train without time stamping. The FPCCD instantaneous power consumption being moderate, a slow signal processing in-between consecutive bunch trains can be envisioned.

FPCCD may also be advantageous in case of intense beam-induced RF noise, to which they are intrinsically insensitive. Moreover their readout circumvents potential difficulties associated to power cycling (see section 2.1.6).

For the inner two layers, where the hit density due to pair-background is particularly high, 5 $\mu$m pitch pixels will be used, while 10 $\mu$m pitch pixels will be used for the outer four layers. As shown in Figure III-2.1, in which a single point resolution equivalent to the pixel size divided by $\sqrt{12}$ is assumed for FPCCD, a significant improvement in impact parameter resolution is achieved w.r.t. the baseline performance, reflecting the outstanding spatial resolution of the two inner layers.

The sensitive area of each FPCCD sensor is divided into $16$ areas. The horizontal registers which are embedded in the sensitive area run parallel to the detector axis, and the readout nodes are located at one end of the chip. The outputs from the sensor are connected to read-out ASICs on the ladder. The read-out ASIC consists of amplifiers, low-pass filters, correlated double samplers (CDS), and analog-to-digital converters. FPCCD sensors are operated at $\sim -40°$C in order to suppress the effect of radiation damage. Each ladder has a few tens of centimetre-long pig-tail Kapton flat cables which are connected to junction circuits outside of the cryostat using micro-connectors (see section 2.1.5). The junction circuits, which include clock drivers, data suppression circuits, optical fibre drivers, etc., are surrounding the beam pipe. The total power consumption inside the cryostat is expected to be about $35$ W.

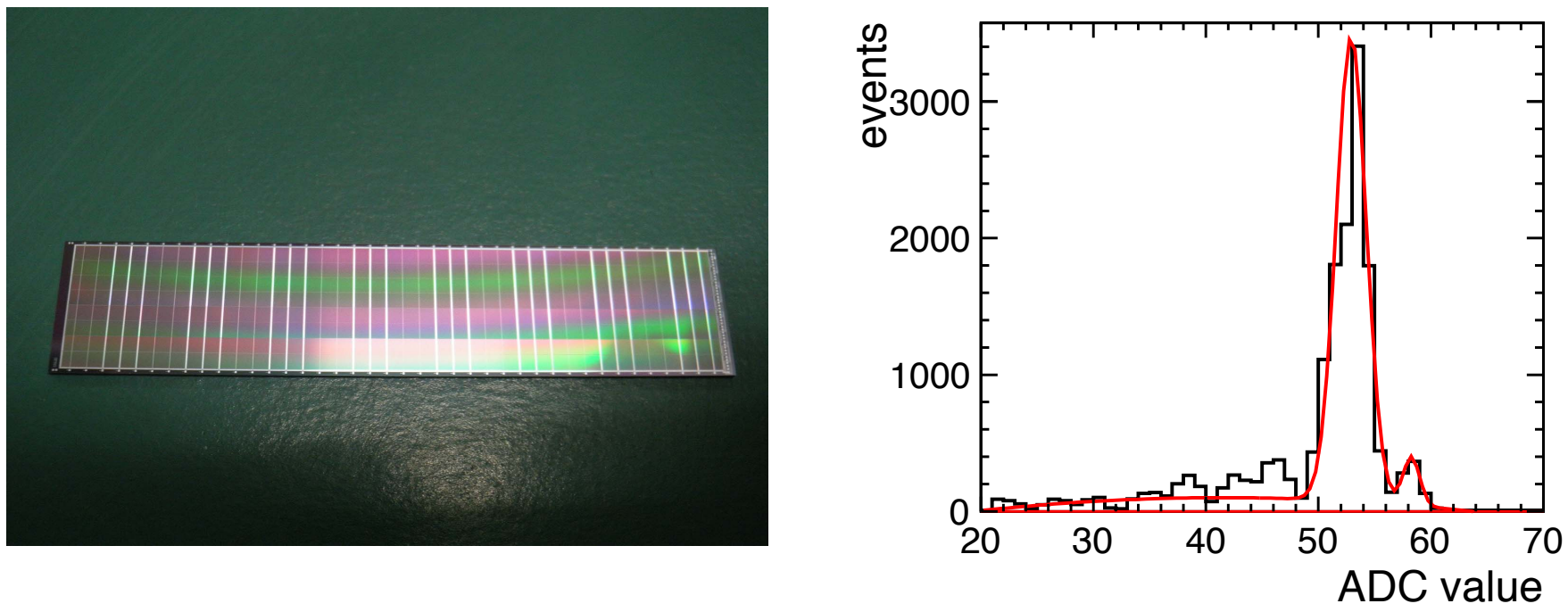

**Figure III-2.3**
Full size FPCCD prototype (left) and response to a $^{55}\mathrm{Fe}$ source (right).

Prototypes of FPCCD sensors and ASICs have been made and tested. Figure III-2.3 shows a full size prototype sensor and the response of a prototype sensor (coupled to its ASIC) to a $^{55}\mathrm{Fe}$ X-Ray source.

#### 2.1.2.3 DEPFET sensors

In the DEPFET [211, 212, 213] active pixel detector concept a field effect transistor is incorporated into each pixel of a detector-grade silicon sensor. A voltage applied over the thickness of the detector depletes the sensor of free charge carriers and ensures rapid and efficient collection of the signal on a deep implant underneath the field effect transistor. As the collected charge modulates the source-drain current, a first amplification of the signal takes place inside the pixel, that is crucial to achieve an excellent signal/noise (S/N) ratio for a thin detector.

Columns of pixels that stretch across the length of the ladder are read out by two auxiliary ASICs: the DCD [214, 215] that digitises the signal and DHP, located at both ladder ends. Both ASICs could eventually be merged. Rows of pixels are read out in a rolling shutter mode. They are addressed by an ASIC known as the SWITCHER [215], that is located on a narrow balcony on the sensor periphery. At any time during operation, only one single row of pixels is active, keeping the power consumption within the strict ILD budget.

An international collaboration [216] pursues the development of the DEPFET concept for use in the vertex detectors of future collider experiments (Belle-II, LC). Over the decade 2002-2012, realistic prototype sensors have been produced and submitted to exhaustive tests with radioactive sources and particles from beams at CERN and DESY [217, 218]. The successful production of sensors with $20 \times 20$ $\mu$m$^2$ pixels demonstrates the feasibility of the process. Sensors produced in the most recent run with a thickness of $50$ $\mu$m are found to be fully functional electrically. The response of such thin sensors to 120 GeV pions is compared to the prediction of H. Bichsel [219] in Figure III-2.4.

The internal gain of the field-effect-transistors extracted from such measurements is found to lie in the $g_q = 300\text{-}600$ pA/$e^-$ range, depending on design variations [220], sufficient to provide a S/N value of up to $40$ for a $50$ $\mu$m thick sensor.

Row read-out times of $\sim$ 80 ns have been obtained in the operation of a DEPFET sensor with the DCDv2 read-out ASIC. The R&D goal for the vertex detector is to improve the row read-out time to $\sim$ 40 ns, thus achieving a frame read-out time of 50 $\mu$s and 100 $\mu$s for the innermost and outer layers respectively. Further improvements in the read-out speed can be obtained by reading more rows in parallel (two rows are assumed for the LC estimate above, in the Belle-II design four rows are read out simultaneously), and by increasing the length of the pixels towards the end of the ladder.

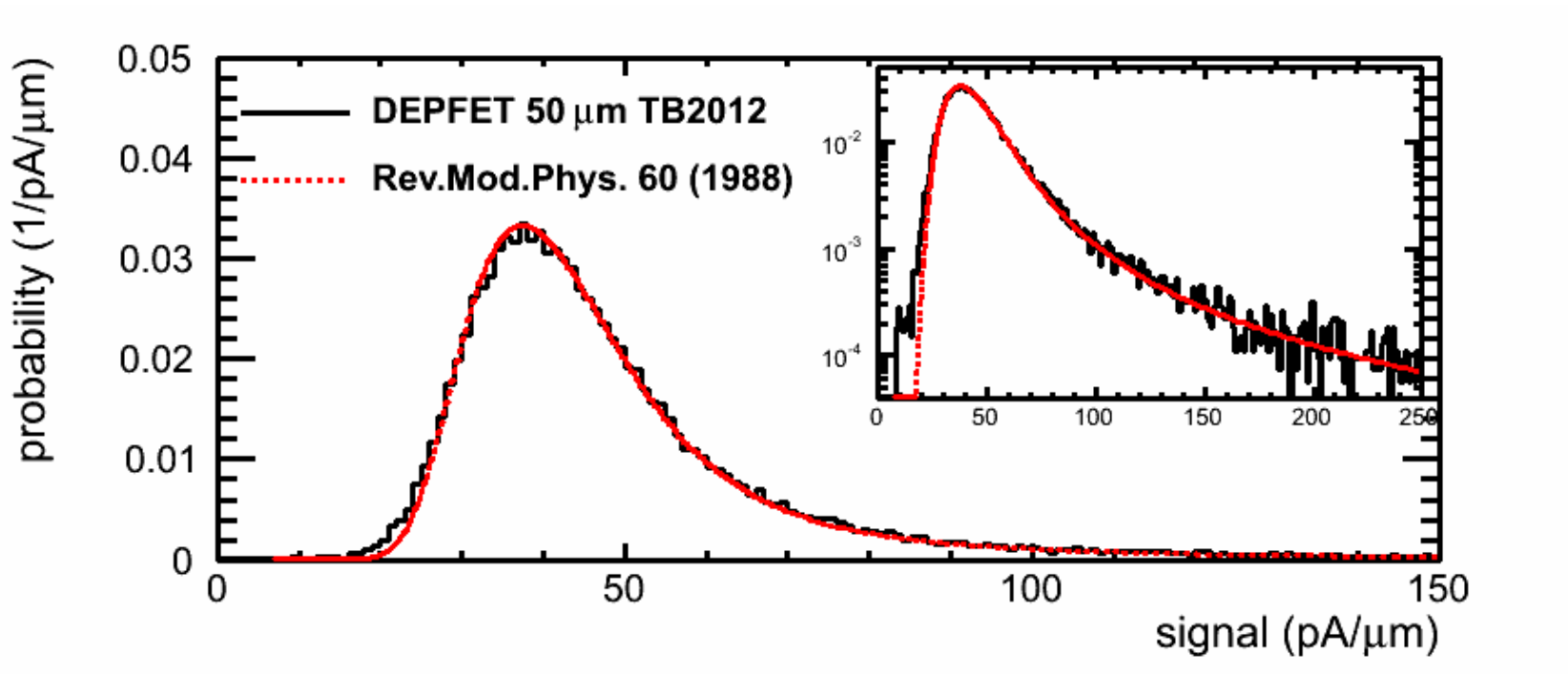

**Figure III-2.4** The response of a 50 $\mu$m thick DEPFET sensor to 120 GeV pions. The prediction of H. Bichsel [219] is compared to the measurements, the most probable value being left free to vary.

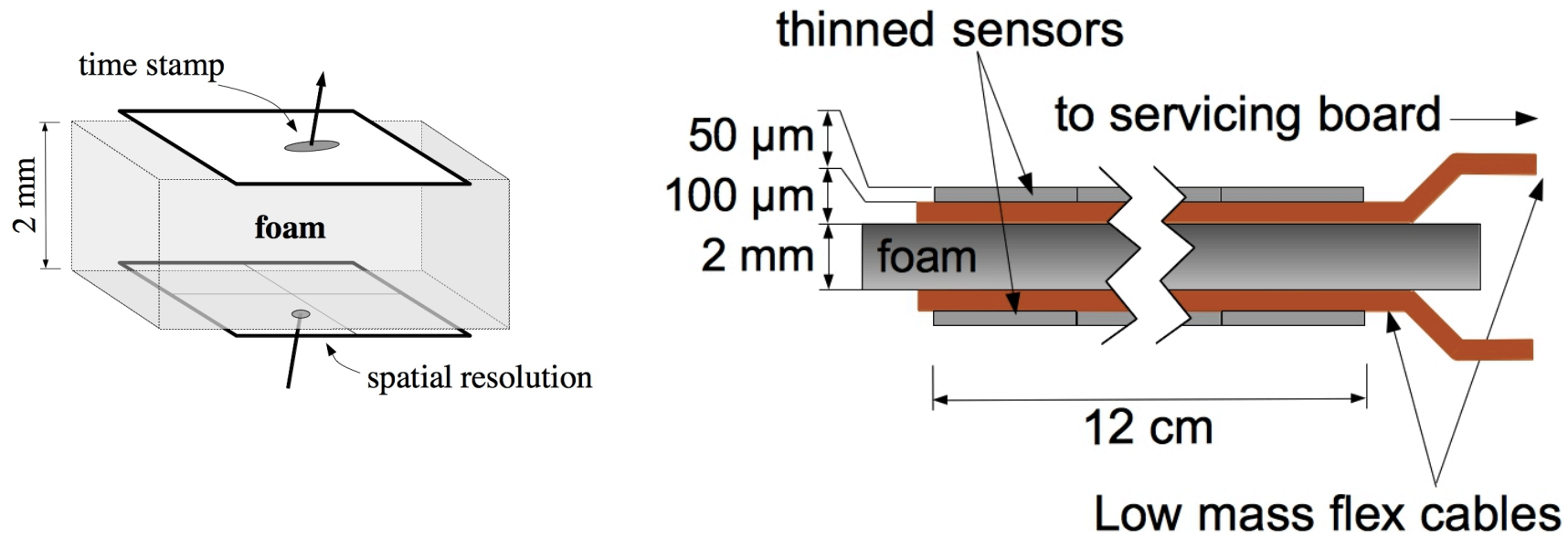

**Figure III-2.5** Left: Illustration of the double-sided ladder concept based on a high-resolution sensor on one side and a fast sensor on the other side. Right: Schematic cross-section of the double-sided ladder developed within the PLUME project.

#### 2.1.2.4 Other sensor technologies

Development of pixel detectors is a very active and fast moving area. A number of alternative technologies are under study by groups around the world, which could feature the required high granularity and low material budget. Very few of them, however, are under active development for ILD, and none have reached a degree of maturity comparable to that of the technologies described above. It is not excluded, however, that the developments undertaken for the high energy run of the ILC (see 2.1.6 ) will promote an alternative technology to those described earlier. This remark may in particular apply to high-resistivity substrate CMOS sensors and to multi-tier 3D pixel sensors.

### 2.1.3 Ladder design

The vertex detector ladders must comply with a particularly tight material budget reflecting the ambitious impact parameter resolution goals. Excellent mechanical properties are required, in particular when power pulsing is foreseen. Three options are currently under study, each related to one of the three pixel technologies introduced earlier. Two of them address the baseline design, using double-sided ladders, while the third one focusses on the alternative geometry using single-sided ladders.

The double-sided ladder design has a structure of a rigid foam core sandwiched by thin ($\sim 50\ \mu$m) silicon pixel sensors. Low density silicon carbide (SiC) and carbon foams (RVC) are considered for the core material. The number of ladders of each layer is 10, 11, and 17 for the first, second, and third layer, respectively. The width of a ladder is 11 mm in the innermost layer, and 22 mm in the outer two layers.

The hits generated by a traversing particle can be used to reconstruct a mini-vector with potential benefits in terms of resolution, alignment and reconstruction of shallow angle tracks. Moreover, as stated earlier, it allows mitigating the conflict between granularity and read-out time.

The double-sided ladder concept envisaged for CPS consists of two sensor layers mounted on a flex cable and separated by a $\sim$ 2 mm thick support layer made of very low density (few per-cent) SiC foam, as illustrated on the right of Figure III-2.5.

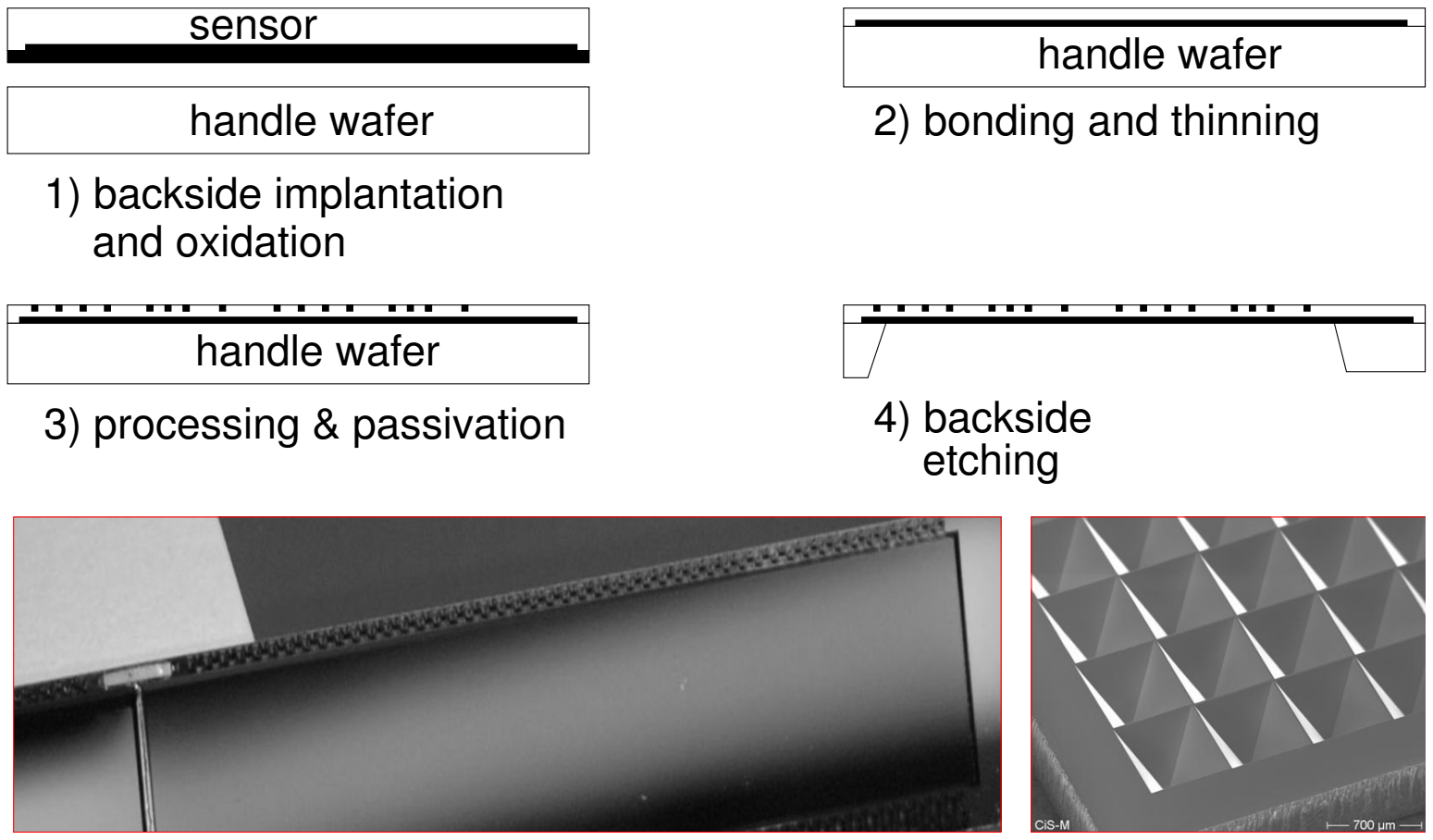

**Figure III-2.6** Illustration of the most important steps in the creation of a thin all-silicon ladder (see text). A photograph of an all-silicon ladder and a scanning electron microscope image of a detail of the support structure are also shown.

The concept is developed within the PLUME collaboration [221]. It uses $50\ \mu$m thin MIMOSA-26 sensors, but can be extended to other technologies, and may actually combine different technologies. The first complete prototype, featuring 0.6% $X_0$ total material budget, was fabricated and validated with a 100 GeV particle beam. Currently, a lighter prototype, featuring only 0.35% $X_0$, is under construction and foreseen to be tested with beam particles in 2013.

An alternative approach is investigated with the FPCCD sensors, which relies on Reticulated Vitreous Carbon foam (RVC) for the core material, with a density of 3-5% of the graphite density. RVC is somewhat less rigid than SiC foam but is expected to be sufficient because of the rigidity provided by the $\sim\ 6.5$ cm long FPCCD sensor slabs.

The single-sided, silicon-only, ladder design is pursued by the DEPFET collaboration. A fully self-supporting silicon ladder is being developed, that requires no external support structure over the full length of the ladder. The use of a single material reduces mechanical stress due to mismatching of thermal coefficients. Auxiliary detector components and power and signal lines are integrated onto an additional metal layer on the sensor, thus strongly reducing the material for this high-density interconnection on the ladder.

The process [222, 223] is schematically depicted in Figure III-2.6: (1) backside implants and oxidation of the sensor wafer; (2) bonding of the sensor and handle wafer and thinning of the former; (3) front-side processing and passivation; (4) photolithographic thinning of the handle wafer, leaving support structures around the edges. The sensor wafer is ground to a thickness of 50-75 $\mu$m. The remaining material from the handle wafer forms a support structures. The resulting self-supporting ladder has excellent mechanical properties and represents a minimal material budget. The thermo-mechanical performance of a realistic detector mock-up with thin DEPFET ladders has been characterised in the presence of a forced air flow [224, 225].

The detailed ladder design for the Belle-II vertex detector envisages 0.21% $X_0$/layer within the acceptance of the ladder [226]. This value may still be reduced for ILD by further thinning of the active material, that is one of the dominant contributions. Further reduction of the Silicon support frame may finally allow achieving the ILD goal of $\sim\ 0.15\%\ X_0$/layer.

### 2.1.4 Cooling system

Two different cooling options are considered, depending on the sensor technology.

For the CPS and DEPFET options cooling strategies which generate a minimal amount of material inside the fiducial volume, if any, are being studied. Those may be air flow cooling similar to the one used for the STAR-PXL [205] or cool nitrogen gas cooling.

For the FPCCD sensor option, in which more than 30 W is consumed inside the cryostat, two-phase $CO_2$ cooling may be used. Because of its large cooling power, typically $\sim\ 300$ J/g, a thin (OD $\sim$ 2 mm) cooling tube may be sufficient. It may be attached at the detector end plate. The increase of the material budget due to the titanium cooling tube on the end plate is only 0.3% $X_0$ if averaged over the end plate. The main heat source of a FPCCD based vertex detector (CCD on-chip amplifier and read-out ASIC) is located near the ladder ends and the end plate, so that the heat is expected to be rather easily removed. The cooling temperature is $-40$ °C. In order to prevent condensation on the cooling tube, and to avoid occupying space with a heat insulator around the tube, the inner support tube supporting the vertex detector and the inner silicon tracker should be filled with dry air.

**Figure III-2.7**
Mechanical support structure of ILD vertex detector

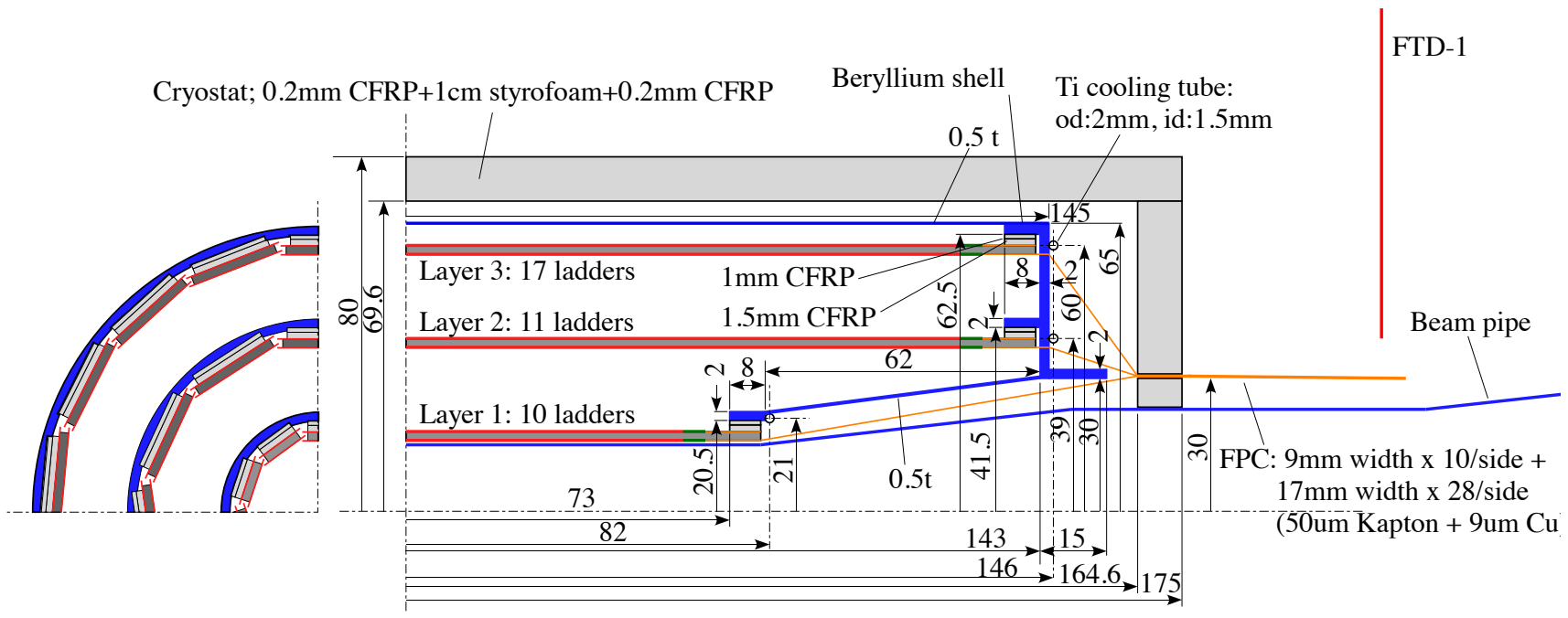

## 2.1.5 Detector mechanics

The vertex detector mechanical design implemented in the full simulation model is shown in Figure III-2.7. It is similar to the SLD vertex detector. The ladders are supported by a 2 mm thick beryllium end plate and a 0.5 mm-thick beryllium outer shell. The strength of this beryllium structure has been calculated with a finite element analysis, which showed that the largest deformation under 9.8 N compression along the beam lines is less than 2 $\mu$m. The whole detector is contained in a cryostat made of 1 cm thick styrofoam (though only mandatory for FPCCD sensors). The material budget of the cryostat including 50 $\mu$m CFRP sheets on both sides is only 0.1% $X_0$.

The vertex detector is supported by the beam pipe, the latter being supported by the inner support tube. The vertex detector is thus integrated as a part of the ILD'inner silicon trackers' inside the inner support tube.

The alignment of the vertex detector will be performed in two major steps. In the assembly phase, micrometrical pre-alignment will be performed by optical survey. After installation, a precise beam-based alignment will be achieved. The latter may proceed through two phases. The first one will consist in aligning the ladders composing a layer, using the few hundred micrometers wide overlapping bands of neighbouring ladders. The second phase will allow making the global detector alignment.

## 2.1.6 Future prospects

The vertex detector is relatively easy to upgrade or replace. The evolution of sensor technologies and performance can therefore be exploited quite efficiently, in particular to comply with the manyfold increase of the beam related background expected at a collision of $\sim$ 1 TeV. It should therefore not be an issue to introduce new sensors featuring much shorter readout times than those foreseen for the first years of data taking.

Despite the achievements described above, the detector is still premature in various aspects, and requires therefore substantial R&D.

The overall detector mechanical design is among the least advanced components. More detailed design studies, including the assembly procedure and important thermal aspects (e.g. power cycling in the experimental magnetic field) are necessary. Manufacturing real scale mechanical prototypes will be an important step of the development.

The CPS foreseen for the outer layers, equipped with ADCs, still need two years of development. For the 1 TeV run, a fast sensor achieving $\lesssim 2\ \mu s$ read-out time is under development, based on in-pixel signal discrimination and two-row simultaneous read-out. It addresses the ALICE-ITS and CBM-MVD applications and may therefore be ready for the 500 GeV run. Once the sensor development is finalised, multi-reticule sensors will be fabricated using industrial stitching, which may be used in order to suppress dead areas and improve the ladder stiffness.

The main R&D activities addressing system integration aspects focus on finalising the present double-sided ladder prototype featuring 0.35% $X_0$. The next generation of prototypes will follow, to tighten the material budget below 0.3% $X_0$.

Power cycling studies of the ladders will also be performed within the AIDA [227] project, which offers also a framework for high precision alignment studies. Finally, the integration of CPS, similar to those developed for ILD, in the STAR, ALICE, and CBM experiments is expected to generate substantial progress in most system integration aspects.

For FPCCD sensors, the radiation immunity has to be proven and the electronics downstream of the read-out ASIC needs to be developed (in particular the data suppression circuitry given the huge amount of pixels). Software developments are also needed in order to achieve efficient track finding in the presence of a large number of background hits.

For the DEPFET option, complete system integration aspects are being addressed at the occasion of the Belle-II vertex detector construction. Concerning the sensors, R&D is performed to further improve their read-out speed, motivated by the innermost vertex detector layers requirements. A more aggressive design of the all-Silicon ladder is also being investigated to meet the ILD goal of 0.15% $X_0$/layer.

R&D is also performed to develop sensors fast enough to provide bunch tagging, which may allow coming closer to the IP in order to improve the reconstruction of low momentum tracks, and will naturally be best suited to the highest energy running. VDSM CMOS processes using high-resistivity, fully depleted, substrates are being studied for this purpose, as well as 3D CPS exploiting industrial stacking techniques to interconnect multi-tier sensors at the pixel level.

## 2.2 The ILD silicon tracking system

The silicon part of the ILD tracking system is made of four components: two barrel components, the Silicon Inner Tracker (SIT) and the Silicon External Tracker (SET), one end cap component behind the endplate of the TPC (ETD), and the forward tracker (FTD). They form the Silicon Envelope [228]. The overall layout of the system is shown in Figure III-2.8.

The barrel silicon parts SIT and SET provide precise space points before and after the TPC; this improves the overall momentum resolution, helps in linking the VTX detector with the TPC, and in extrapolating from the TPC to the calorimeter. The coverage of the TPC with silicon tracking is completed by the ETD, located within the gap separating the TPC and the end-cap calorimeter. Together these systems help in calibrating the overall tracking system, in particular the TPC. The good timing resolution of the silicon detectors relative to the time between bunches in the ILC together with the high spatial precision helps in time-stamping tracks and assigning them to a given bunch within an ILC bunch train.

In the very forward region, where the TPC does not provide any coverage, a system of seven silicon disks (pixel and strips) ensures efficient and precise tracking down to very small angles. Good forward coverage is particularly important for the ILC since at $\sqrt{s} \geq 350$ GeV the relative weight of t-channel exchange processes and high-multiplicity ($2 \rightarrow 4$ and $2 \rightarrow 6$) processes increases. In a large fraction of collisions some of the outgoing particles are emitted at very small polar angle. Compared to previous experiments at $e^+e^-$ colliders, the instrumentation of the forward region of the tracking

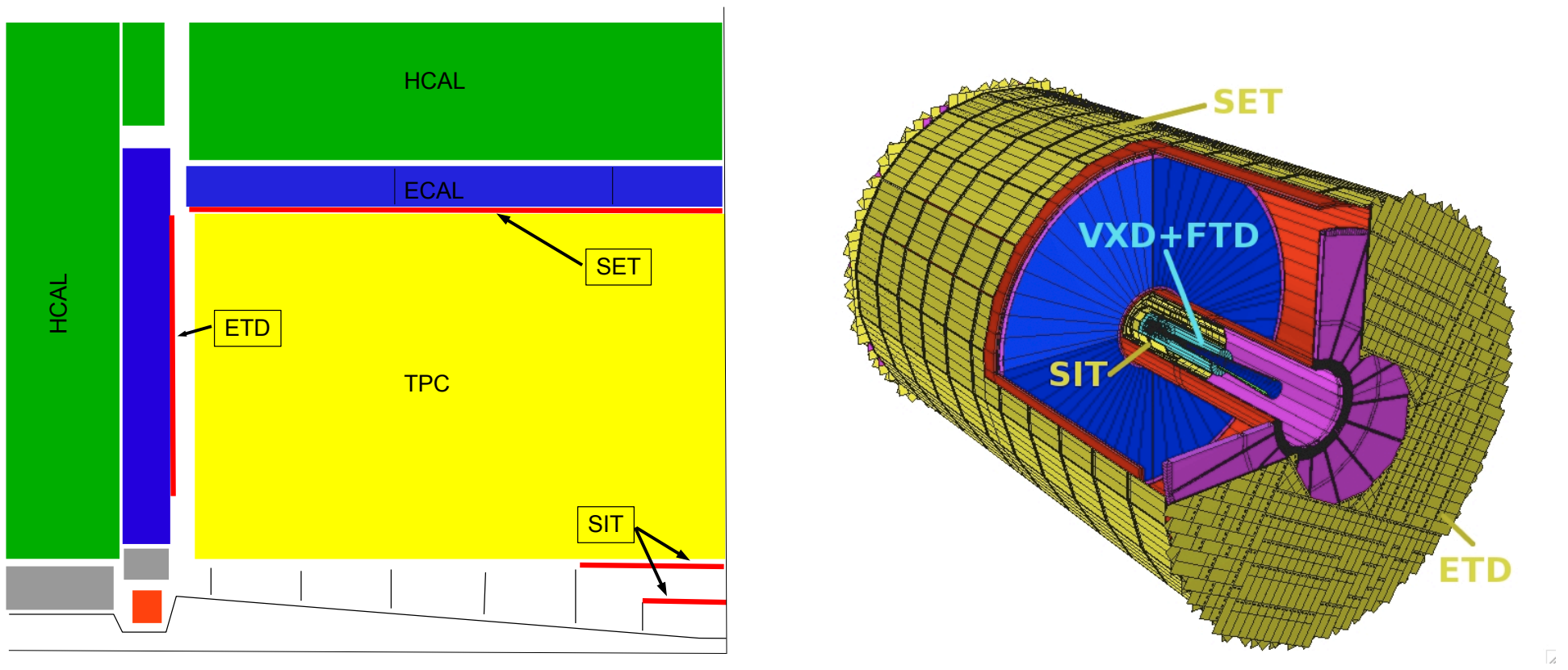

**Figure III-2.8.** Left: a quadrant view of the ILD silicon envelope system made of four components, SIT, SET, FTD, and ETD as included in MOKKA full simulation. Right: a 3D detailed GEANT 4 simulation description of the silicon system as sketched in the quadrant view on the left.

thus becomes important [229].

A special challenge to all silicon systems is the design of lightweight, thin systems that can be operated at minimum power to avoid the need for intricate cooling systems. This requires careful management of the materials for the detector support. Power consumption is minimised by power pulsing. This requirement leads to a synchronised power distribution that has to deal with large pulsed currents, which must not generate any electromagnetic interference phenomena or transients altering the front-end electronics performance during the active period. Highly integrated readout electronics moves significant processing power close to the detector, and thus reduces the number of cables needed to exit from the system. An advanced scheme is pursued to minimise the material needed to bring the necessary power to the detector. Powering schemes like DC-DC conversion or the use of super-capacitors mounted on the detector are being investigated.

The silicon tracking system of ILD has been developed by the SiLC collaboration. Detailed descriptions of the wide ranging R&D activities can be found in [230, 231, 232, 233].

## 2.2.1 The central silicon: SIT, SET, and ETD

The central silicon components SIT, SET, and ETD are realised with layers made each of two single-sided strip layers tilted by a small angle with respect to each other; this is also called ‘false’ double-sided layers. SIT includes two such layers and SET one; together they thus provide three precise space points for central tracks, the ETD adds one precise point to tracks going into the end-cap. The main parameters of the system are given in Table III-2.2.

A central design feature of the silicon envelope detectors is that the same sensor type is used throughout the system. This minimises the complexity of this large system, and will help to minimise the costs. Similarly the same mechanical design for the basic detector unit, the ladder, is used throughout. It is based on modern silicon detector technology, deep sub-micron (DSM) CMOS technology for the front-end (FE) electronics with a new on-detector electronics connection and new material technology for the support structure. Special challenges for ILD are a significant reduction in material compared to the most recent examples of large scale silicon detectors (e.g. currently running LHC detectors), operating at very low power, and reaching excellent point resolution and calibration.

The SIT is positioned in the radial gap between the vertex detector and the TPC. Its role is to improve the linking efficiency between the vertex detector and the TPC; it improves the momentum resolution and the reconstruction of low $p_T$ charged particles and improves the reconstruction of long lived stable particles. The SET is located in the barrel part between the TPC and the central barrel electromagnetic calorimeter (ECAL). The SET gives an entry point to the ECAL after the TPC field cage. It acts as the outermost silicon layer in the central barrel and also improves the overall momentum resolution. The SIT and SET, in addition to improving momentum resolution, provide time-stamping information; by combining the hits from these silicon detectors (especially the SET) with the TPC hits, a very precise time stamping is possible as explained in details in section 4.1.2.5. These two central silicon components may serve in monitoring the distortion of the TPC and for the alignment of the overall tracking.

**Table III-2.2** Main parameters of the central silicon systems SIT, SET, and ETD.

| SIT (baseline = false double-sided Si microstrips) | | | | | |
|---|---|---|---|---|---|
| Geometry | | | Characteristics | | Material |
| R [mm] | Z [mm] | cos $\theta$ | Resolution R-$\phi$ [$\mu$m] | Time [ns] | $X_0$ [%] |
| 153 | 368 | 0.910 | R: $\sigma$=7.0 | 307.7 (153.8) | 0.65 |
| 300 | 644 | 0.902 | z: $\sigma$=50.0 | $\sigma$=80.0 | 0.65 |
| **SET (baseline = false double-sided Si microstrips)** | | | | | |
| Geometry | | | Characteristics | | Material |
| R [mm] | Z [mm] | cos $\theta$ | Resolution R-$\phi$ [$\mu$m] | Time [ns] | $X_0$ [%] |
| 1811 | 2350 | 0.789 | R: $\sigma$=7.0 | 307.7 (153.8) | 0.65 |
| **ETD (baseline = single-sided Si micro-strips)** | | | | | |
| Geometry | | | Characteristics | | Material |
| R [mm] | Z [mm] | cos $\theta$ | Resolution R-$\phi$ [$\mu$m] | | $X_0$ [%] |
| 419.3-1822.7 | 2420 | 0.985-0.799 | x: $\sigma$=7.0 | | 0.65 |

The ETD is positioned between the TPC end plate and the end cap calorimeter system. The ETD provides an entry point for the calorimeter and improves the momentum resolution for charged tracks with a reduced path in the TPC. Moreover it helps reducing the effect of the material of the TPC end-plate (currently estimated to be 15% $X_0$). It thus might improve the matching efficiency between the TPC tracks and the shower clusters in the EM calorimeter. It also contributes to extending the lever arm and angular coverage of the overall tracking system at large angle. Both the ETD and the FTD ensure the full tracking hermeticity.

An intense R&D program is carried out to further develop the sensors and the overall detector concept. This work is in many areas done in close collaboration with groups from the LHC, as many of the requirements and technologies are similar.

#### 2.2.1.1 The basic silicon sensor

The microstrip sensors that will equip the SIT, SET, and ETD components are the basic element of the silicon system architecture. The baseline sensor has an area of $10 \times 10$ cm$^2$, with 50 $\mu$m pitch, 200 $\mu$m thick silicon, edgeless (i.e. with a non-active edge decreased from a few 100 $\mu$m to a few tens of $\mu$m), and an integrated pitch adapter (IPA) in order to directly connect strips with the front-end nd ASIC channels. This design allows the construction of a detector without overlapping sensors, significantly simplifying the construction and minimising the material.

The sensor is undergoing a vigorous R&D program to identify the most appropriate technology and layout. A special effort is made to find more than one vendor to produce these sensors.

Reducing the non-active edge of the sensors is an important step towards reducing the material budget and simplifying the detector mechanical construction. Edgeless sensors allow building large area seamlessly tiled detector matrices and thus getting flat tracking areas without excess of material in the overlapping region between two silicon tiles, the tiles being made of one or several unique size strip sensors. Prototype sensors have been developed and successfully tested.

The integration of the pitch adapter into the basic silicon sensor is another important step towards simplification of the sensors. This would allow the direct connection of the front-end with the sensor chip. This technology has been pioneered by the ILC community through the SiLC collaboration [232, 234, 235], and has been adopted and further developed especially by the CMS upgrade program. Recently, progress has been achieved by including Integrated Pitch Adapter (IPA) prototypes in the test structures of the silicon sensors (see Figure III-2.9). The LHC groups are testing various IPA schemes and are studying the impact on the crosstalk, noise, and other sensor characteristics. This is indeed a very important step ahead and makes it feasible to have IPA fabricated by industry.

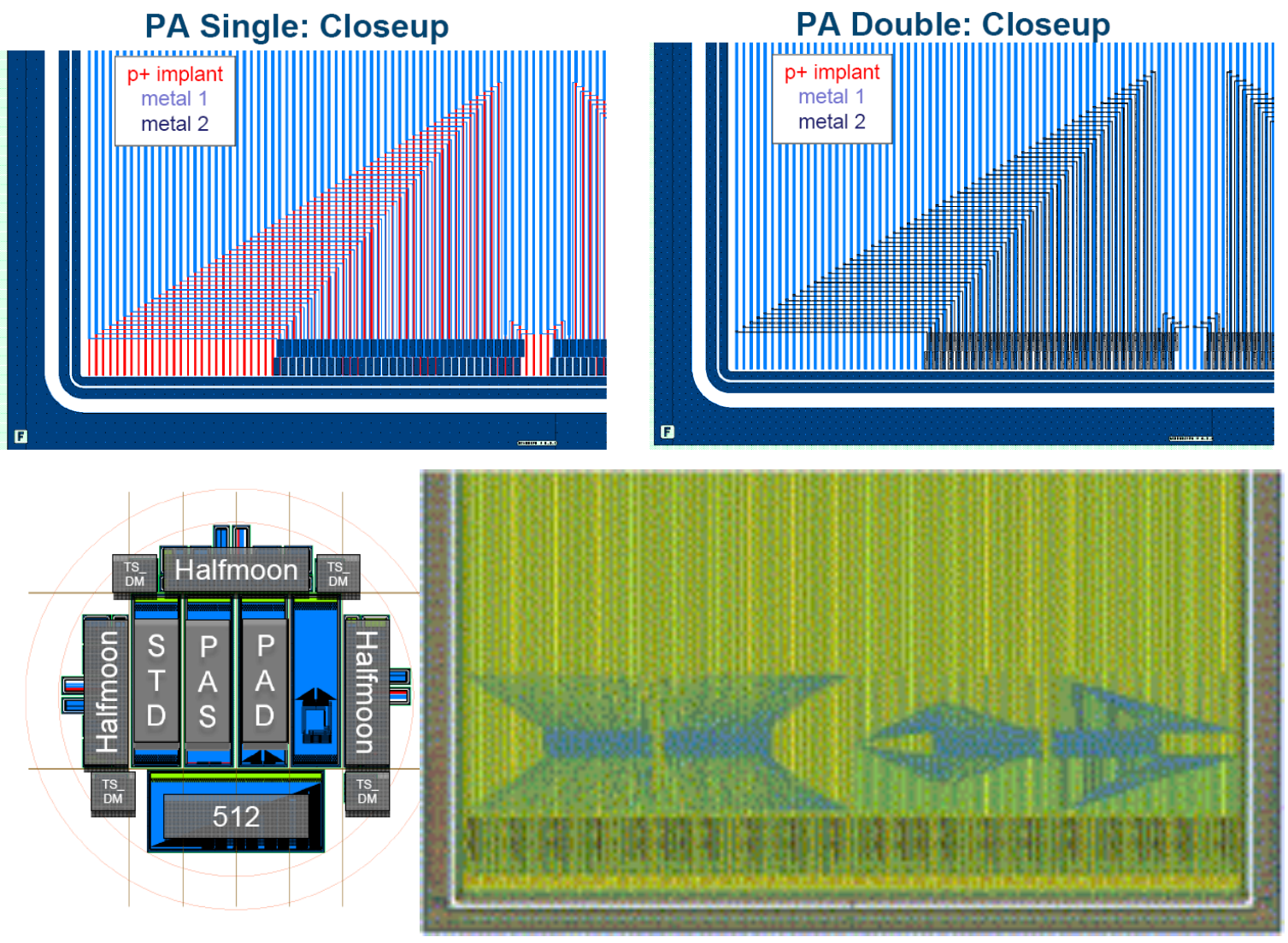

**Figure III-2.9** Picture of a silicon test-structure developed to test the integrated pitch adapter technology.

Another innovative packaging technology based on 3D-packaging using Through Silicon Vias (TSV) is currently under development in close collaboration with industry [236]. The technology has not reached the maturity needed to be included as a baseline but may well be in the future.

#### 2.2.1.2 Front-end electronics

The front-end will be based on a custom designed ASIC realised in deep sub-micron CMOS technology. It will provide the analogue to digital conversion, and a high degree of digital processing capabilities, to minimise the data volume which needs to be transferred out of the detector. The chip will be developed with low-noise, low-power consumption in mind, and will be capable of power pulsing. To arrive at a compact design, new interconnection technologies will be explored, like bump bonding or 3D vertical interconnects.

Over the past years a prototype version of such a chip, the SiTRK ASIC, has been developed in the SiLC collaboration. It pioneered this aspect by introducing high-level data processing already at the front-end thanks to an early digitisation stage [237]. It was developed in 130 nm CMOS technology and each of its components was successfully tested. It includes an analogue pipeline, low noise operational amplifier, and an 8-bit Wilkinson ADC, together with the required control circuitry [238]. The next iteration of this chip will move to the commercial 65 nm technology.

#### 2.2.1.3 Powering schemes and thermal management

A key element of the powering scheme for the silicon detectors is the use of pulsed powering. A significant part of the front-end ASIC is switched off in-between ILC bunch trains, thus reducing on average the power consumption by close to two orders of magnitude. In the development of the ASIC for silicon tracking the power pulsing capability has been taken into account.

Most of the power consumption and thus also the heat load will be concentrated in the ASICs at the ends of the ladders. A design of a cooling strategy based on forced cooled gas flow should be possible. Special care has to be taken to avoid the introduction of unwanted oscillations due to the gas flow. The complete inner tracking volume will need to be flooded with dry air to avoid condensation.

A significant reduction in the amount of material needed for the low voltage power cables can be achieved by either a DC-DC powering scheme, or by the use of super-capacitors mounted close to the front-ends. This technology is also used by the LHC experiment upgrades. First tests with ILD readout systems have delivered promising results.

#### 2.2.1.4 Mechanical design and integration

The silicon tracking system poses several challenges for the mechanical design. It should introduce a low amount of dead material, be fairly simple and modular, but at the same time stable also during external manipulations of the detector. During detector push-pull operations it should maintain its position so that a new calibration can be done quickly and efficiently. The proposed mechanical design tries to address these issues.

The inner silicon detectors are supported from a central tube inside the inner radius of the TPC. This tube is hung on either side on the TPC endplates. SIT and FTD are both connected to this tube. A challenge is the development of a lightweight but stiff structure. Given the high degree of precision required for the inner detectors it might be needed to suspend this inner tube from the TPC using remote-controlled movers. This has not yet been studied in detail, but such movers might be needed to return the system back to a good initial alignment after a push-pull operation.

Studies are under way to investigate how well different functions can be integrated into the support structure. Cooling systems, calibration systems, and possibly even cabling can become an integral part of the support system, and thus in addition to providing the needed functionality, also help in minimising the overall material. Progress in materials will also be utilised to arrive at optimised structures. An example of such an 'intelligent' system is the one proposed for the Super-B support structure [239].

Within the silicon tracker two options are studied. One uses staves, the other one is based on super-modules. Both are built with Carbonfibre Reinforced Plastic (CFRP) material [240]. These are self-supporting light, but robust, structures assembled and inserted in the corresponding barrel layers. At the very end of the detector edge, an independent part hosts the module services, the cooling, and signal cabling connections. Because of its location at the detector ends, this strategic part can be replaced rather easily.

The preferred solution for the cable routing of the SIT component is that the cables run along the inner radius of the TPC, because it reduces the amount of material around the beam pipe and in front of the FTD as well. The cabling for the SET is less critical, and will run directly inside the calorimeter inner radius in z to the end of the silicon tracker, and follow the central cabling routes from there.

#### 2.2.1.5 Calibration and alignment

The hybrid tracking system as proposed for ILD has a number of special considerations.

1. Alignment of the SIT and SET with respect to the TPC;
2. Alignment of the ETD with the FTD and/ or the TPC, but the precision required in the ETD case is not the same as for the other Si tracking components;
3. Quick-and-precise re-installation and positioning of these components after push-pull operation.

A particular challenge for the central silicon system is the precise alignment of components separated by large distances like the SIT and the SET. In principle a system of laser beams can be used to register these systems relative to each other, with sufficient precision. There is, to date, no detailed design which could be integrated with the overall ILD layout. However no fundamental problems are anticipated.

The silicon trackers are mechanically stable devices that will help to improve the absolute alignment of the overall tracking system, and of ILD as a whole. This alignment is sensitive in particular to temperature fluctuations, which will need to be understood to the 2 $\mu$m level. These alignment systematics will be very different from those for the TPC. The TPC is sensitive to ambient temperature and to atmospheric pressure variations, to non-homogeneities in electric and magnetic fields, etc. In particular the electric drift field in the TPC may depend on space charge transient effects due to variations in the machine induced backgrounds. The SIT and SET give an independent and effective means to monitor accurately such effects with real data. Experience at LEP has shown that this capability gives an invaluable redundancy during data analysis, and a unique mean to disentangle and understand anomalous behaviour. It is a necessary complement to the unique pattern recognition capabilities of the TPC.

The experience gained on LHC silicon tracking systems shows that, once the mechanical alignment is achieved with the precision of about 100 $\mu$m, the commissioning with cosmic rays allows a remarkably precise alignment. Data with colliding beams will then be used to further improve the alignment.

#### 2.2.1.6 Future R&D perspectives

Silicon tracking is a field which is developing very rapidly. The need of the LHC experiments to make major upgrades to their silicon based tracker within the next few years is driving innovation in the field. Many developments in the area of sensors, readout and mechanical construction are expected in the near future.

The baseline solutions for constructing the silicon envelope components of the ILD detector concept are well established. Beyond baseline R&D activities are pursued in parallel on new sensor technologies, new associated front-ends, and higher-level signal processing. Solutions with much higher granularity are being investigated including a full silicon pixel tracker [241], or a modest version of this proposal, where only part of the silicon envelope is made with pixels. Depending on when the ILC will be built, the silicon envelope tracking components might evolve in design and technology.

### 2.2.2 Forward silicon tracking

The forward tracking in the ILD concept contains seven tracking disks installed between the beam pipe and the inner field cage of the TPC. The first two are realised as pixel detectors to cope with the expected high occupancies in this area, the remaining five are strip detectors. The layout is given in detail in Table III-2.3. Their precise space points with a large lever arm are crucial to maintain good momentum resolution in the forward region.

The detection of charged particles emitted in the forward and backward directions faces a number of significant challenges. The magnetic field becomes less and less useful in bending charged tracks in the forward region, thus making a precise momentum determination difficult. In addition forward going jets are not opened up by the field as much as they are in the barrel, resulting in significantly larger occupancies. Finally, the disk are very close to the beam axis and are thus prone to high backgrounds from the interaction region.

**Table III-2.3**
Layout of the Forward Tracking Disks. The quoted single hit resolution for the pixel disk depends on its technological implementation which has also an effect on the material budget.

| FTD (baseline: pixels for two inner disks, microstrips for the rest) | | | | |
|---|---|---|---|---|
| | Geometry | | Characteristics | Material |
| R [mm] | Z [mm] | cos $\theta$ | Resolution R-$\phi$ [$\mu$m] | RL [%] |
| 39-164 | 220 | 0.985-0.802 | | 0.25-0.5 |
| 49.6-164 | 371.3 | 0.991-0.914 | $\sigma$=3-6 | 0.25-0.5 |
| 70.1-308 | 644.9 | 0.994-0.902 | | 0.65 |
| 100.3-309 | 1046.1 | 0.994-0.959 | | 0.65 |
| 130.4-309 | 1447.3 | 0.995-0.998 | $\sigma$=7.0 | 0.65 |
| 160.5-309 | 1848.5 | 0.996-0.986 | | 0.65 |
| 190.5-309 | 2250 | 0.996-0.990 | | 0.65 |

### 2.2.2.1 Detector optimisation

The main challenge for the forward tracker is to deliver good momentum resolution in this difficult environment. The momentum resolution scales approximately with the inverse of the B-field component orthogonal to the direction of movement of the charged particle, with the inverse of the square of the track length being measured, and with the inverse of the number of hits. Thus in the forward direction with the effective B-field approaching zero as the particle travels along the beam pipe, high precision measurements using a large lever arm are needed.

The parameters are highly constrained by the overall detector layout (and cost). Additional measurement layers would improve the momentum resolution for very high momentum tracks, but the extra material considerably reduces the reconstruction precision for the abundant low momentum tracks.

To achieve the best momentum measurement within the constraints described above, the FTD is instrumented with those detectors that yield the most precise $r\phi$-measurement within the tight material budget. Micro-strip detectors have proven to be capable of resolutions of several microns with minimum channel and power density. The orthogonal radial measurement is only relevant in the pattern recognition stage and little is gained by improving the radial resolution beyond the several hundred microns that is readily obtained with pairs of micro-strip detectors under a small stereo angle.

The extrapolation of the trajectories of charged particles emitted at very shallow angle to the interaction point is crucial for flavour tagging of very forward jets. The optimal segmentation and placement has been studied in the context of the CLIC study [242].

To achieve a precise measurement of the longitudinal impact parameter the radial segmentation of the innermost disks is crucial. An optimisation of the vertexing performance for very shallow tracks requires a first precise measurement at minimal distance from the interaction point. Even a small amount of material before the first measurement severely degrades the measurement. The best performance in this respect is obtained by minimising the gap between the $z$-position corresponding to the end-of-stave of the vertex detector and that of the innermost disk. Services of the vertex detector barrel must avoid as much as possible the line of sight between the interaction point and the innermost disk.

### 2.2.2.2 Pixel disk implementation

Due to their higher occupancy, the first two disks are implemented using highly granular pixel detectors. As in section 2.1, three technologies are under consideration: CPS, CCD and DEPFET.

For the CMOS based technology (CPS) each of the two stations closest to the interaction point (IP) (FTD1 and FTD2) would be equipped with 50 $\mu$m thin CPS sensors on their front and back sides, thus providing four high resolution space points per track traversing the end-cap. Each station side is composed of 16, slightly overlapping, petals featuring a structure similar to the one of the double-sided ladders equipping the vertex detector. With a mechanical support including a 2 mm thick, low density, SiC foam layer, the total material budget of a station is expected to be $\sim$ 0.5% $X_0$, averaged over the azimuthal angle.

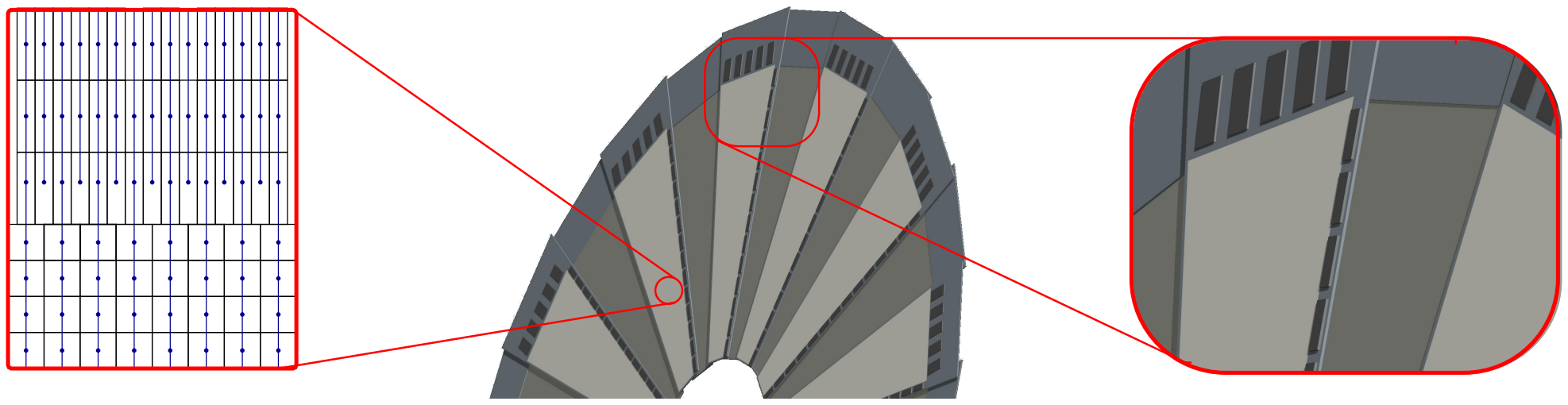

**Figure III-2.10.** A half-disk for the FTD in the DEPFET petal concept. The rightmost zoom image shows a detail of the end-of-petal area that houses the read-out electronics and the balcony with the steering chips. The leftmost image shows the region at $R = 8$ cm where both the column width and the $R$-dimension of the pixels changes.

The petals are equipped with four sensors, which intercept different 3 cm wide radius ranges. Each sensor is equipped with the same number of pixels, so that the size of the pixels increases with increasing radius. Each sensor type is thus composed of an active area of 480 $\times$ 1152 pixels. The dimensions of the pixels vary from 26 $\times$ 29 $\mu\mathrm{m}^2$ for the innermost sensor to 26 $\times$ 67 $\mu\mathrm{m}^2$ for the outermost one. The single point resolution varies accordingly from $\lesssim$ 6 to 9 $\mu\mathrm{m}$, depending on the sensor.

The CPS architecture reproduces the one used for the vertex detector, based on a continuous read-out in rolling shutter mode allowing efficient power saving. The sensitive area is subdivided in columns of 480 pixels read out continuously and in parallel. Within each column, the pixels are read out in pairs to accelerate the read-out, resulting in a frame read-out time of $\lesssim$ 50 $\mu$s. Each column ends with a discriminator, whose outputs are processed through a sparsification circuitry integrated in the sensor periphery.

Each petal dissipates nearly 10 W, resulting in a total instantaneous power dissipation of $\sim$ 600 W per end-cap. Assuming a powering duty cycle of $\lesssim$ 2 %, the average power dissipation per end-cap is in the order of 10 W, compatible with air flow cooling.

Industrial stitching may allow combining several of the four chips composing a petal in one single silicon slab, for the benefit of material budget, system integration and alignment. This possibility remains however to be assessed in view of the fabrication yield and handling issues.

Fine pixel CCDs (FPCCDs) can be used as sensors for FTD1 and FTD2. Each disk is divided into sectors and each sector is covered by one trapezoidal FPCCD sensor. FPCCD sensors can be as thin as 50 $\mu$m and the material budget can be $0.15\%X_0$ per layer including the support structure. Electrical connection (clock input and signal output) is made at the outer edge of the sensors using Kapton flexible cables.

From the consideration of beam background and pixel occupancy, a pixel size of 10 $\mu$m can be used with the signal accumulation in one train and read out between trains. Spatial resolution better than 3 $\mu$m is expected with this pixel size. The total number of readout channels is 1400/disk and the power dissipation is 21 W/disk (15 mW/ch) including on-chip source followers and the front-end ASICs.

The DEPFET-based all silicon ladder concept is adapted to the geometry of the Forward Tracking Disks, as shown in Figure III-2.10. The solution is optimised to yield excellent $r - \phi$ resolution of 3-5 $\mu$m, with narrow read-out columns oriented radially. The read-out electronics is located at the end-of-petal area on the upper rim of the petal. The steering chips are located on the balcony that is visible in the zoom image. The active area of the petal is divided in two sections that meet at $R = 8$ cm. A zoom image is shown in the leftmost panel of Figure III-2.10. In each of these the column pitch fans out gradually from 25 $\mu$m to 50 $\mu$m with increasing radial distance to the beam line. Along the column direction (i.e. radially) three pixel dimensions are moreover used: 25 $\mu$m in the innermost region, 50 $\mu$m for $6 < R < 8$ cm and 100 $\mu$m for $R > 8$ cm. The increased granularity at small radius ensures good vertexing performance for very shallow tracks that leave no hits on the barrel vertex detector and helps to cope with the strongly increasing background levels towards the smallest $R$ values.

The material budget of this arrangement is comparable to that of the barrel vertex detector based on the same all silicon concept (0.15% $X_0$/layer). The power consumption per unit area is slightly less due to the slightly smaller column density, allowing for cooling through a forced air flow.

#### 2.2.2.3 Strip disk implementation

In the following paragraphs the key design aspects of the microstrip-based FTD disks are briefly presented.

Given the relatively low hit occupancy expected for the disks three to seven, the detector technology based on AC coupled p-on-n fine-pitch microstrip silicon sensors is proposed as the detector baseline. Two sensors will be glued on opposite sides of the same petal frame with a stereo angle configuration allowing for a two dimensional determination of the particle's impact point on the petal; furthermore, for minimising the multiple scattering, the sensors will be manufactured on wafers 200 $\mu$m thick and a second metal layer will be used to fan-in the signals into the FEE input pads, avoiding the material burden introduced by an external pitch adaptor.

As an option beyond the above introduced baseline, with somewhat reduced material budget, real double-side microstrip sensors are being considered. Similar sensors have been already employed successfully [243].

The strip FTD requires around 4000 front-end readout chips with 256 channels and a 50 $\mu$m pitch. The FE readout chip will have the typical architecture used for Si-strip readout in high energy physics experiments [244], adapted to the particular conditions of the experiment and the sensors. A channel composed of preamplifier and shaper followed by analogue pipelines, sparsification and analogue to digital conversion stages and simple slow control and processing digital electronics is planned. Special care has to be taken to maintain an acceptable ratio between noise and power consumption. A limit of 700 $\mu$W per channel and a figure of noise of 400 $e^-$ for a detector capacitance of 20 pF with 2 $\mu$s shaping time are established as specifications. Power reduction techniques, including switching off analogue modules during defined periods, and the use of a deep-submicron technology are essential to manage the constrains. Prototypes of the constituent modules, the main channel and several multichannel chips are being planned [245].

The outer FTD disks are segmented in 16 petals mounted in two half rings manufactured in composite material. The petal consists in a trapezoidal shaped frame made of monolithic high modulus carbon fibre material laminate (M55J) while the ring is designed as a sandwich structure with two skins of high modulus carbon fibre laminate with a foam or honeycomb core. The petals must have a good face finishing (planarity) allowing for the proper gluing of the sensors. Four sensors will be glued in each petal, two per petal face (false double-sided sensor). The electronic hybrids will be located on the frame edge. Adjacent petals are staged along the z coordinate allowing for a petal's edge overlapping along the azimuthal coordinate ($\varphi$).

#### 2.2.2.4 Power distribution system

With the current design one disk of the FTD strip system will use less than 200 W of power during electronics-on time, or less than 40 W on average. This estimate has been computed taking into account the front-end and power distribution electronics dissipation; a conservative effective duty cycle of 20% has been assumed. For the complete strip detector less than 400 W of power needs to be dissipated [246].

Two different power distribution system topologies are under study for the FTD sub-detector. One is based on DC-DC power converters; the other is based on super-capacitors and low voltage regulators. DC-DC based power distribution system has been the preferred option for the latest generation of HEP experiments; however, this topology presents some limitations due to stability and EMI issues [247] that have to be analysed in detail.

A supercapacitor-based power distribution system has been selected as an alternative option to power up the strip FTD system [248]. The FTD will be powered remotely by a current source that supplies low current to the periphery of each disk. At that location a set of super-capacitors will supply the high peak current to the periphery of each petal, where LV regulators stabilise the voltage at the input of each hybrid electronics. The most important element of this option are the super-capacitors. Supercapacitors for power applications are emerging devices [249]. The high power density of these capacitors makes them a very suitable solution for the characteristics required for a power pulsing system. The use of this type of device is new for HEP experiments but not for industrial applications. Key issues which need to be studied are to understand the radiation hardness of these devices, and to optimise the number of power cycles before failure.

At the modest heat loads of the of strip-based FTD disks, about 6-9 mW/cm$^2$, an air-forced cooling system seems a feasible and reliable solution. Cooling implementations have been studied which include air conducts in the CFRP supporting cylinder for blowing cool air into the detector, and extract the heated air. The actual challenge for this cooling system is to probe its validity to extract the heat dissipated by the two inner pixelated disks which have a much higher heat load. Finite element simulations are being carried out to optimise the layout and get a first estimate of the required air flow and temperature. Moreover, the mechanical instabilities introduced by the blowing air should be studied in dedicated mockups.

#### 2.2.2.5 Integration and System Aspects

The FTD system is to be installed inside the boundaries defined by the beam pipe outer radius and the inner surface of the support cylinder which encloses all the inner volume delimited by the TPC inner radius. All the FTD disks are supported by this cylinder. The vertex detector and the SIT are directly supported from the beam pipe, which is turn is hung from the support cylinder through the third disk of the FTD. To comply with this extra requirement, the FTD ring three will be reinforced.

For each petal a dimensional metrology measurement of predefined fiducial marks will be carried out at each step of the assembly procedure, including the final mounting of the eight petals in its corresponding half supporting ring. During system assembly, the lower half ring will be mounted directly on the lower half of the inner support cylinder. After the installation of all the lower half rings, VTX, SIT, and beam pipe, the upper half ring will be connected to the already mounted lower half ring. Finally, the upper part of the support cylinder will complete the assembly of the FTD systems inside the support cylinder envelope. The fully equipped support cylinder will then be finally inserted as a whole into the inner volume bounded by the TPC inner radius.

Structural (deformation and displacements) and environmental (temperature and humidity) real-time monitoring of the FTD supporting structure is instrumental to achieve the design accuracy under the major detector movements required by the push-pull operating mode. Structural monitoring should be focused on the monitoring of those overall deformation modes called 'weak modes' to which the conventional track-based off-line alignment algorithms are blind. As monitoring technology in-fibre Bragg grating sensors (FBG) will be used [250]; FBG sensors have become a very attractive solution for strain and temperature monitoring in hostile or hazardous environments. In particular, their small weight, size and intrinsic immunity to EMI combined with the absence of electrical signals and cables have recently encouraged the use of FBG sensors in high energy physics experiments [251, 252]. Moreover, FBG sensors can be embedded in composite materials, allowing the fabrication of the so-called smart structures where the actual sensors are just part of the CFRP laminate, allowing for a straight forward integration into the supporting structure with a negligible interference from the point of view of the material budget. This is the technological solution adopted for the real-time monitoring of the FTD supporting structure [253].

A careful electrical grounding design is necessary to preserve the performance of the front-end electronics. For this reason, it will be necessary to develop an electromagnetic compatibility (EMC) plan [254] that systematically approaches the grounding design and quantifies the immunity/ emission of the electronic systems to integrate safely the FTD system. The EMC plan comprises two basic steps; grounding topology definition and EMC test.

## 2.3 The ILD TPC system

The central tracker of ILD is a Time Projection Chamber (TPC). A TPC tracker in a linear collider experiment offers several advantages. Tracks can be measured with a large number of three-dimensional $(r, \phi, z)$ space points. The point resolution, $\sigma_{\mathrm{point}}$, and double-hit resolution, which are moderate when compared to silicon detectors, are compensated by continuous tracking. The TPC presents a minimum amount of material as required for the best calorimeter and PFA performance. A low material budget also minimises the effects due to the $\simeq 10^3$ beamstrahlung photons per bunch-crossing which traverse the barrel region [255]. Topological time-stamping in conjunction with inner silicon detectors is an important tool that is explained in section 6.1.2.5. To obtain good momentum resolution and to suppress backgrounds, the detector will be situated in a strong magnetic field of $3.5$ T. Under this condition a point resolution of better than $100$ $\mu$m for the complete drift and a double hit resolution of $< 2$ mm are possible.

Continuous tracking facilitates the reconstruction of non-pointing tracks which are significant for the particle-flow measurement and for the reconstruction of physics signatures in many scenarios. The TPC yields particle identification via the specific energy loss $dE/dx$ which is valuable for many physics analyses.

Over the past years systematic R&D work to develop the design of a high-resolution TPC for a linear collider detector has been pursued in the context of the LCTPC collaboration [227, 256, 257, 258].

### 2.3.1 Design of the TPC

The main parameters for the TPC are summarised in Table III-2.4. The overall dimensions of the ILD detector and the TPC have been optimised to obtain the best physics performance, as described in the ILD Letter of Intent (LOI) [198]. The design goal has been to maintain a very low material budget and to achieve the required single and double-point resolution. The mechanical structure of the TPC consists of an endplate, where the readout of the amplified signals takes place using custom-designed electronics, and a fieldcage, made from advanced composite materials. Two options for the gas amplification systems are Micromegas [259] and Gas Electron Multipliers (GEM) [260]. At present either option would use pads of size $\approx 1 \times\ 6\ \mathrm{mm}^2$, resulting in about $10^6$ pads per endplate. An alternative technology of a pixelated readout with much smaller pitch is being investigated [261].

The readout endplate (Figure III-2.11) is a concentric assembly of modules. The modules are self-contained and integrate the gas amplification, readout electronics, supply voltages, and cooling [262].

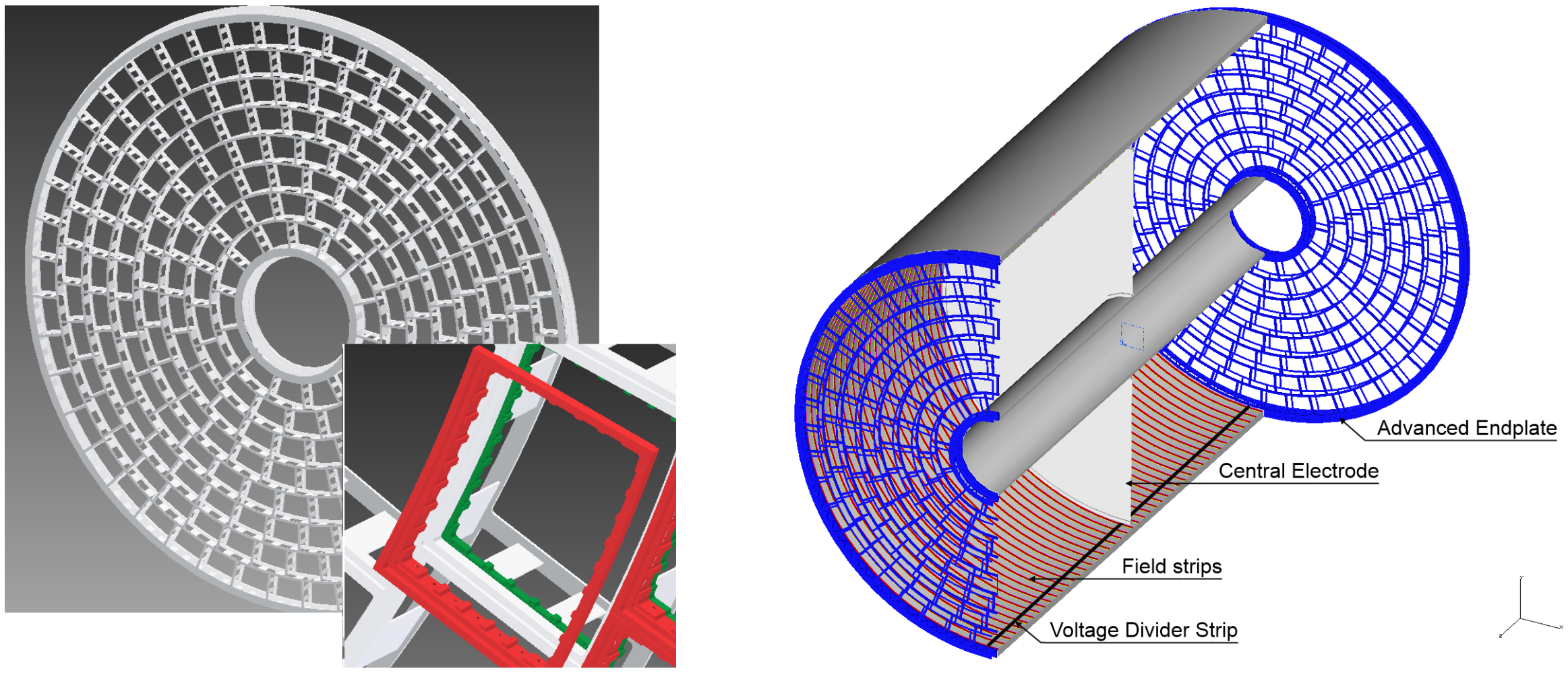

**Figure III-2.11.** Left: Drawing of the proposed end-plate for the TPC. In the insert a backframe which is supporting the actual readout module, is shown. Right: Conceptual sketch of the TPC system showing the main parts of the TPC (not to scale).

**Table III-2.4** Performance and design parameters for the TPC with standard electronics and pad readout.

| Parameter | | | |
|---|---|---|---|
| Geometrical parameters | $\mathrm{r_{in}}$<br>329 mm | $\mathrm{r_{out}}$<br>1808 mm | z<br>$\pm$ 2350 mm |
| Solid angle coverage | up to $\cos\theta \simeq 0.98$ (10 pad rows) | | |
| TPC material budget | $\simeq 0.05\ \mathrm{X_0}$ including outer fieldcage in $r$ | | |
| | $< 0.25\ \mathrm{X_0}$ for readout endcaps in $z$ | | |
| Number of pads/timebuckets | $\simeq$ 1-2 $\times$ $10^6$/1000 per endcap | | |
| Pad pitch/ no.padrows | $\simeq 1\times$ 6 mm$^2$ for 220 padrows | | |
| $\sigma_{\mathrm{point}}$ in $r\phi$ | $\simeq$ 60 $\mu$m for zero drift, $<$ 100 $\mu$m overall | | |
| $\sigma_{\mathrm{point}}$ in $rz$ | $\simeq 0.4 - 1.4$ mm (for zero – full drift) | | |
| 2-hit resolution in $r\phi$ | $\simeq 2$ mm | | |
| 2-hit resolution in $rz$ | $\simeq 6$ mm | | |
| dE/dx resolution | $\simeq 5$ % | | |
| Momentum resolution at B=3.5 T | $\delta(1/p_t) \simeq 10^{-4}$/GeV/c (TPC only) | | |

#### 2.3.1.1 Gas amplification system

The gas amplification system for a pad-based TPC will be either GEM or Micromegas (see [263] and [264] for examples of results using small prototypes). It has been demonstrated that both amplification technologies combined with pad readout can be built as modules which cover large areas with little dead space.

The use of Multi-Wire Proportional Chamber (MWPC) technology has been ruled out [264], because it does not meet the ambitious performance goals.

Two or three GEM foils are stacked together to achieve sufficient charge amplification. For a GEM readout the transverse diffusion within the GEM stack itself is enough to spread the charge over **Figure III-2.12** Measured point resolution using data taken with a triple GEM stack at a magnetic field of 4 T, recorded in a small TPC prototype with T2K gas.

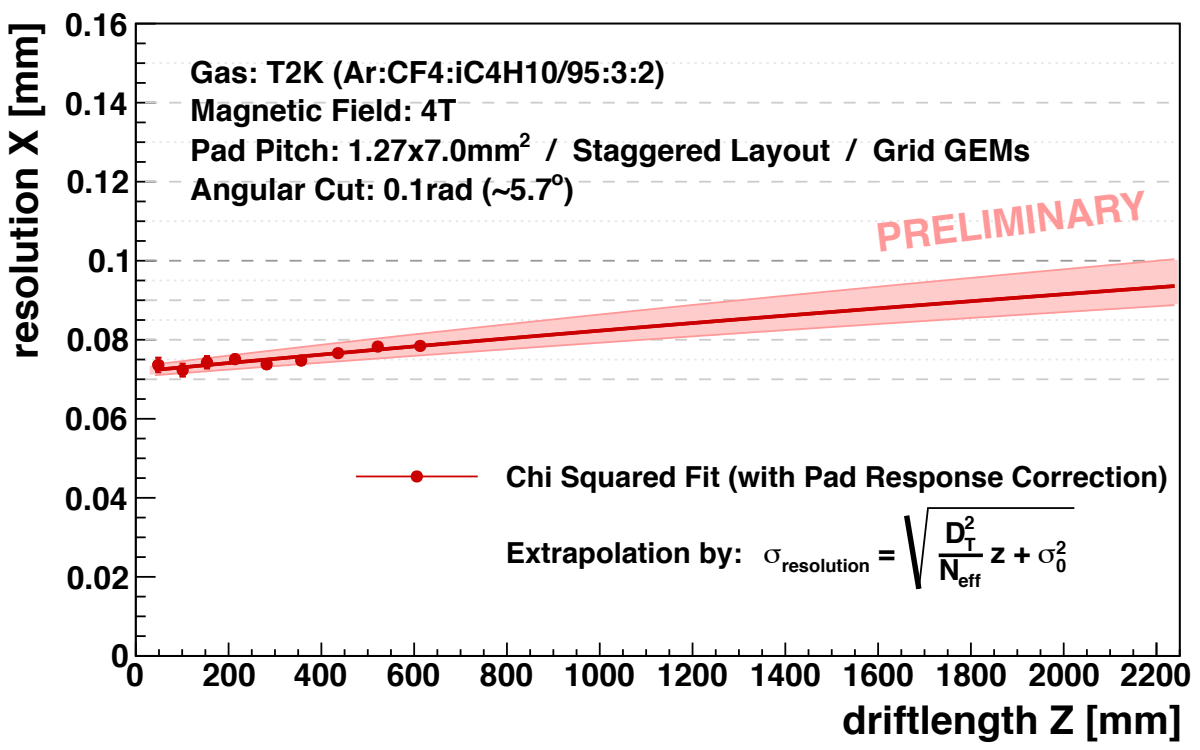

several 1 mm wide pads, which enables a good point reconstruction.

Micromegas have enough amplification in a single structure, but the spatial extent of the signals is very small on the readout plane. In order to spread the charge signal out over several pads, Micromegas use a resistive coating on the anode surface with resistivity of order 2 to 3 M$\Omega$ per square [258] [265].

The properties of the gas determine the drift velocity and the diffusion constant [266]. The parameters are chosen to minimise the diffusion in the transverse and longitudinal directions, to preserve an intrinsically excellent resolution.

For a drift length of more than 2 m and a high field of 3.5 T, the so-called T2K gas mixture (Ar-CF4(3%)-isobutane(2%) [267]) is a promising candidate, as was demonstrated using a small prototype chamber [268] and equipped with Grid GEMs [269, 270]. Data taken with that chamber are presented in Figure III-2.12: the single point resolution versus the drift length as measured in a 4 T magnetic field. The length of the chamber was 600 mm, and the extrapolation (error indicated by pink shaded area [271]) to full drift for the ILD TPC agrees with the goal of a resolution of better than $100\ \mu$m, as predicted by simulation [198].

#### 2.3.1.2 Endplate

The modules are integrated on an endplate (Figure III-2.11) which closes the TPC gas volume and supports the modules. It is important that he endplate is designed to have low mass, while retaining the required mechanical and thermal stability [262].

The material of the endplate in front of the endcap calorimeter can potentially disturb the particle flow performance. In recent studies [258, 272, 273] the particle flow performance was evaluated using the PandoraPFA program [274] and the full ILD simulation for a range of endplate thicknesses $X_0$. Increasing from 15% to 60% $X_0$ degrades the jet energy resolution from $4.2\%$ to $4.8\%$ for 45 GeV jets and $3.2\%$ to $3.3\%$ for 100 GeV jets (and about the same for 250 GeV jets). From this a material budget of up to $25\% X_0$ for the endplate seems acceptable.

A prototype for a low mass endplate has been designed and built for use with the Large Prototype (LP) TPC. It meets the requirements for the ILD TPC [262]. It is based on a thinned aluminium structure stabilised by a system of adjustable struts (Figure III-2.13 left). Finite element methods, which were validated by measurements, show that this system provides adequate stability and precision (an example is pictured in Figure III-2.13 right). The current design of the endplate foresees 240 modules of approximately $17 \times 21$ cm$^2$, as they are used in the large prototype. Depending on the final choice of technology the size will be optimised.

### 2.3.1.3 Readout module

Each readout module consists of the gas amplification system (GEM or Micromegas), the pad plane, the readout electronics, and cooling.

A design issue for the multi-GEM systems is the provision of a support system that keeps the GEM surfaces both flat and parallel without introducing dead space or adding too much material to the detector. Several options have been developed and successfully operated [258]. With a recently demonstrated system [269, 270] based on ceramic spacers, good flatness and mechanical stability could be demonstrated while introducing only about 2% of dead space.

The Micromegas system with one stage amplification has a fine wire mesh mounted in front of the readout pad plane. A system of small pillars maintains a constant distance between the mesh and the pad plane. This system has been shown to operate very reliably over long periods. The pillars introduce a dead area of a few %.

### 2.3.1.4 Readout electronics

Small pads of $1 \times 6$ mm$^2$ area require that the electronics per channel does not exceed this footprint; the most modern readout system for a TPC, the AFTER system developed for the T2K experiment [275], has a footprint per channel of about three times this area. A picture of Micromegas-based modules mounted on the LP is seen in Figure III-2.14 left and an event in figure III-2.14 right, taken during a testbeam run at DESY [276]. They demonstrate that, for this technology and a pad size of between $2.7 - 3.2 \times 6.1$ mm$^2$, a solution exists which fits inside the current module boundaries.

Efforts are underway to develop more compact, fast, low noise, and power pulsed systems [277, 278]. The fundamental layout consists of a charge sensitive preamplifier, a fast ADC, and a digital signal processing unit which is used to analyse the data online, find pulses, determine time and charge, and, where applicable, reduce the total amount of data.

The power management relies critically on the ability of the system to use power pulsing. Power pulsing has been demonstrated for the S-ALTRO16 system [278]. Even with power pulsing, however, an active cooling of the endplates will be needed, for which two-phase $CO_2$ cooling is planned. The power-pulsing goal is to reduce the power consumption to less than 100 W/m$^2$ (1 kW per endplate).

An alternative readout concept relies on the coupling of a gas amplification system and a pixelated silicon chip [261]. The Timepix chip, derived from the Medipix family of chips, has been used in a series of proof-of-principle experiments. The pixel sizes are about $50 \times 50$ $\mu$m and thus are small compared to the contribution from diffusion. The Timepix chip allows both time and charge to be measured per pixel, providing potentially a very detailed view of the charge pattern on the end plate. Challenges of this system are the large number of pixels, the readout speed, and the robust and safe integration of the silicon pixel chip with the gas amplification system. For the moment this system is considered to be an interesting variant to the more traditional pad-based readout systems, but is not yet far enough advanced to be proposed as an alternative to the baseline.

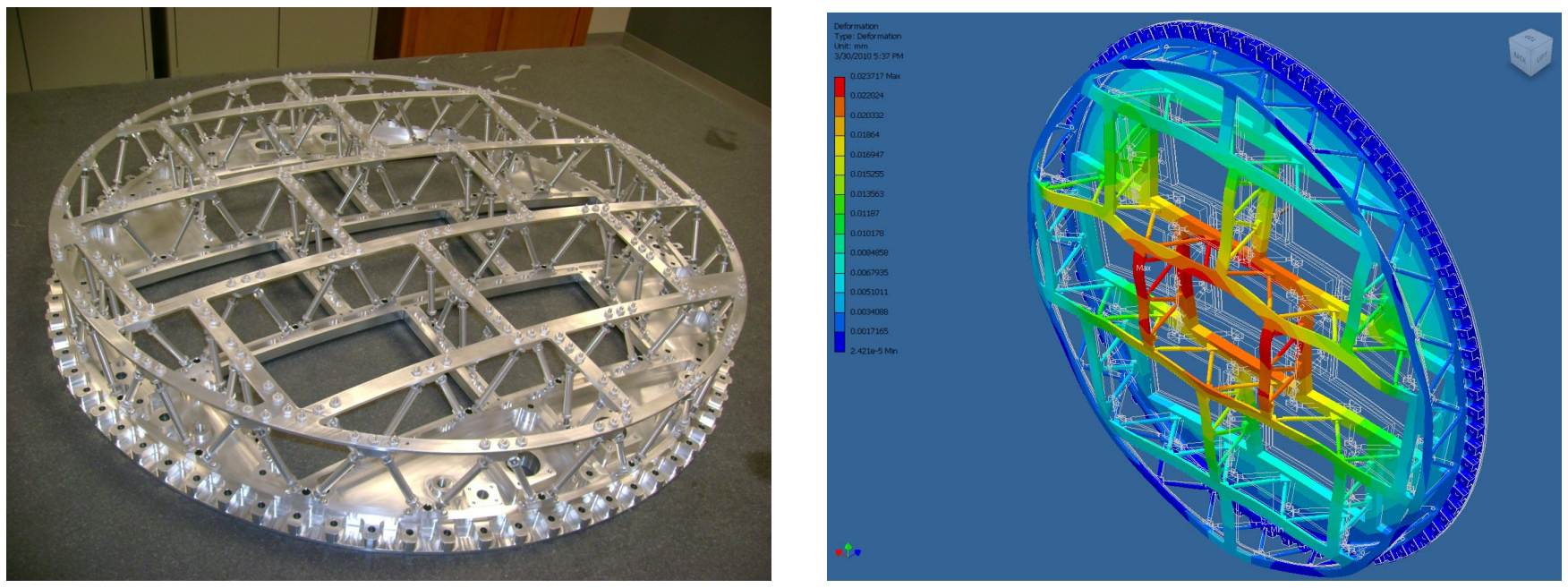

**Figure III-2.13**
Left: A low mass endplate for the LP. Right: Study of deflection of the LP endplate due to a load on the central module: load = 100 N, deflection = 23 $\mu$m.

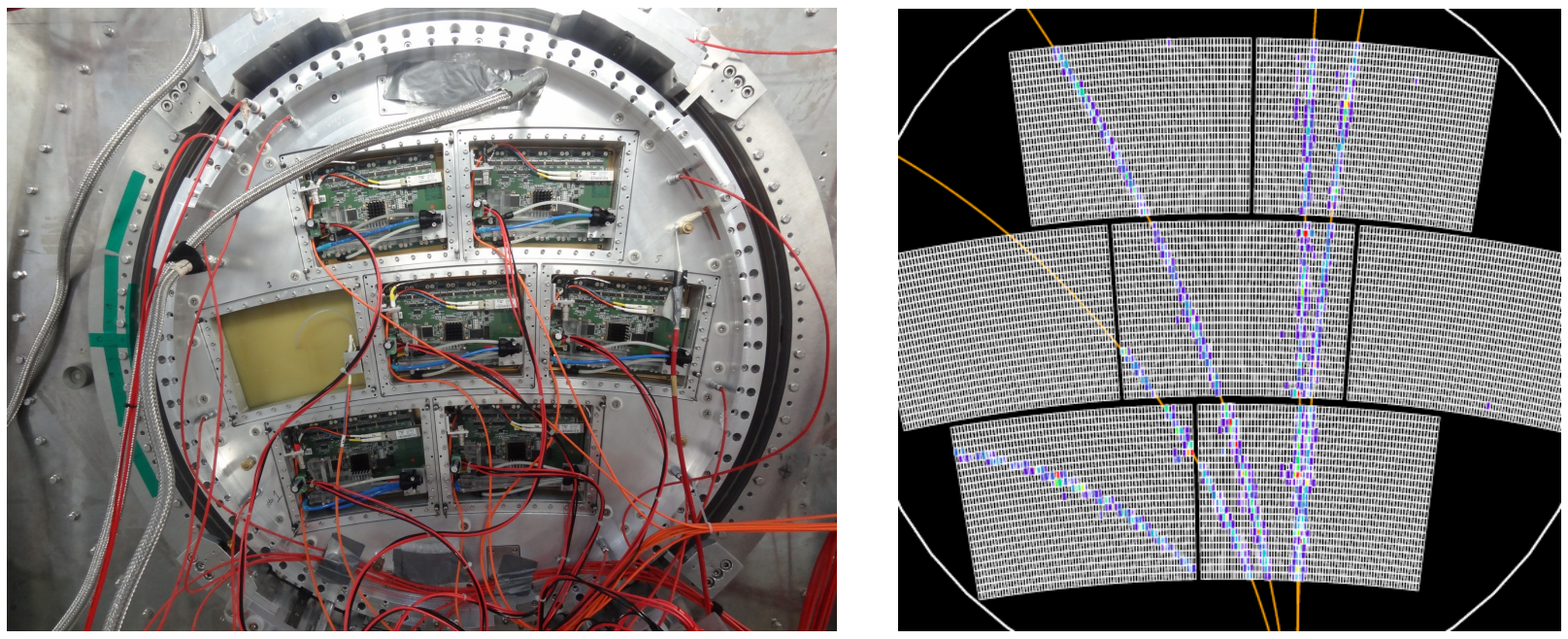

**Figure III-2.14** Left: Six micromegas modules (with resistive anode) mounted on the endplate of the large prototype, equipped with highly integrated electronics. Right: An event recorded using a micromegas equipped readout at teh DESY testbeam facility.

### 2.3.1.5 Fieldcage

The inner and the outer fieldcages will be built using composite materials [279]. A core made of honeycomb is covered on the inside and outside with a layer of glass-fibre reinforced epoxy. On the drift-volume side of the inner and outer cylinders, Kapton sheets with metallised potential strips provide insulation and field-shaping electrodes. The potential of the strips is defined by a resistive divider mounted inside the gas volume. Mirror strips on the back of the Kapton sheets shield the field against the grounds on the outside of the TPC, where each cylinder will be covered with grounding sheets.

The conceptual design of the fieldcage has been tested and demonstrated with the LP [279]. Based on this a prototype a material budget of 1% $X_0$ for the inner and 3% $X_0$ for the outer fieldcage seems to be feasible. The fieldcage will provide a homogeneous electric field. Simulations show that field distortions due to the electrical properties of the field cage alone should stay below $50\ \mu$m [280]. Designs exist for the transition from the fieldcage to the endcap, which will add only minimal material in the corner region. Experience from the large prototype shows that the mechanical tolerances of the system, in particular the parallelism of the cathode and the endcap, are difficult to achieve. A careful survey of the cathode and the endcap is mandatory to measure possible deviations so that they can be corrected later on. The design of the central cathode membrane is less well developed. At the moment a thin membrane is stretched between a light-weight inner and an outer ring, at $z = 0$, similar to the central cathode design of the ALICE TPC [281].

### 2.3.1.6 Support structure

The TPC support structure will be non-magnetic, have a low thermal expansion coefficient, be robust in all directions (x,y,z), maintain accuracy and stability over long time periods, absorb vibrations, and provide a position accuracy of 100 $\mu$m or better.

In the present design the TPC endplates are suspended from the solenoid. A number of spokes run radially along the faces of the calorimeter to the TPC endplates (Figure III-2.11 right). With the total mass of the TPC estimated to be around 2 t, the weight is not a problem. A mechanism must be developed which prevents the TPC to move in the longitudinal direction to ensure that the system is not damaged in case of earthquakes and simplifies the recovery of the alignment of the TPC after a push-pull cycle. In the present design, supports using double-T beams made of lightweight carbon, carbon fibre reinforced composite (CFRP) or by a system of flat CFRP ribbons are being studied. The ribbon system needs less space in the endcap-barrel transition region, but requires an additional fixation of the TPC in longitudinal direction.

The TPC fieldcage will support the inner and outer silicon trackers. While there are no conceptual issues, this additional load on the fieldcage might require a stiffer system and more material than anticipated.

**Figure III-2.15.** (Left) Effects of the ion disks on the electrons inside the TPC volume. The effects without (top half of the figure) and with (botton half of the figure) are compared. (Right) Expected distortions in $r-\phi$ as simulated for a disk of charge located inside the TPC drift volume at different z-values, for a range of radii. For more details see the text.

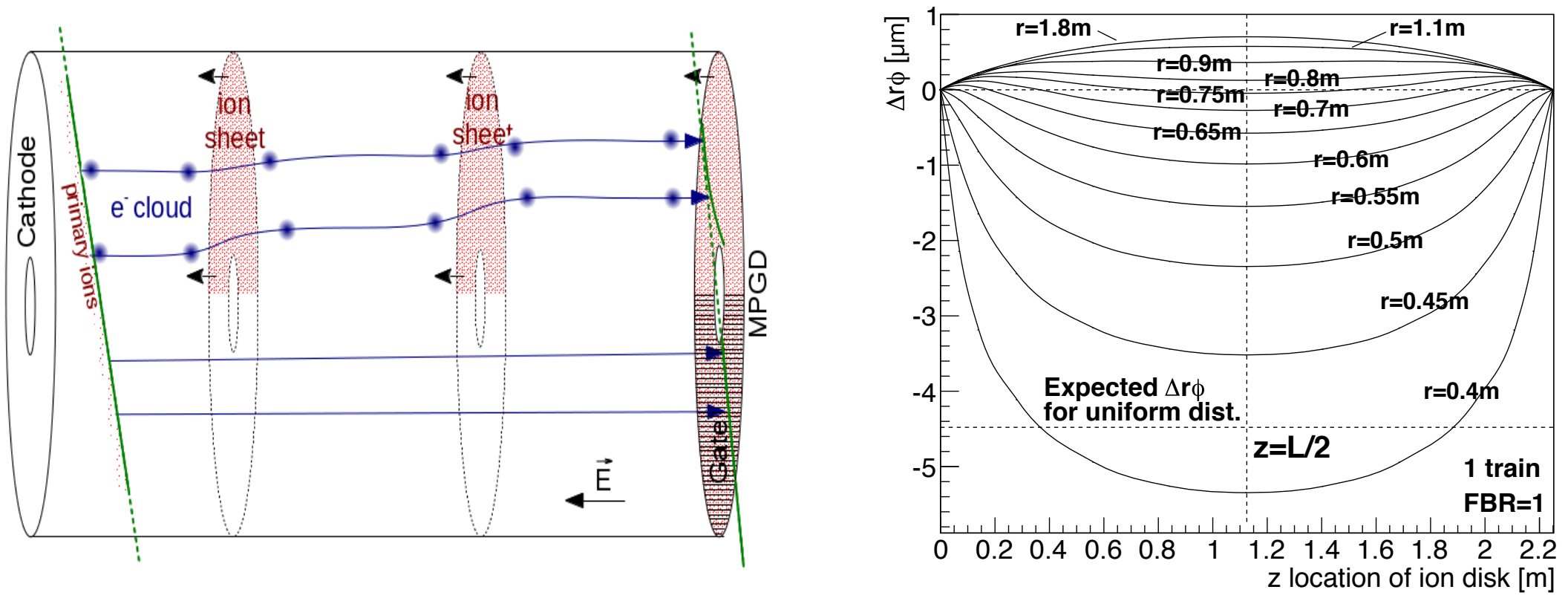

## 2.3.1.7 System performance

The design of the TPC system has been studied extensively to understand its impact on the overall performance. The electric field is covered in section 2.3.1.5. The main distortions in the TPC will originate from the magnetic field, which will need to be measured with high precision, as described in section 2.3.2.

Another source of deviations is the accumulation of charge inside the TPC drift volume. Studies and detailed simulations have been performed to understand the impact of ions on detector performance which were produced during the primary ionisation and during the amplification phase [282, 283]. The requirement of continuous operation during an ILC bunch train implies that no gating of primary ions will be possible during a bunch train.

The studies [282, 283] at a 500 GeV ILC have shown that the effects of primary ions are manageable and that effects from ions between a gate and the amplification region are negligible. Distortions arising from the so-called ‘ion discs’ due to the secondary-ion backflow into the drift region from the micro pattern gas detector (MPGD) gas-amplification region can result inup to 60 $\mu$m of transverse displacement of the drifting electrons. The ion discs arise because the TPC is active during the 1 ms bunch train followed by a 199 ms pause, while the backflow ions from the amplification region take about 1 s to drift out of the TPC. An ion gate can eliminate the discs by gating before the ions can enter the drift region, as seen in Figure III-2.15. In the upper half of this sketch of a TPC the drift of electrons in the TPC volume in the presence of ion feedback is shown. In the lower half the same situation but this time with an ion gating grip, is illustrated. [282, 283]. The track near the cathode gives rise to electron clouds that drift to the anode. The track registered by the MPGD at the anode in the x, y plane is indicated by solid lines, distorted (above) and undistorted (below). Present gas candidates are compatible with there being 3 discs in the chamber (without gating) and with a feedback ratio (the product of gas amplification and intrinsic ion suppression by the MPGD) of about 3; a distortion of $\approx$ 60 $\mu$m would occur without gating, at the inner fieldcage. The right plot shows the results from a simulation, for a single ion disk. Shown is the expected distortion as a function of the disk position in z, and at different radial positions within the TPC. The charge assumed for the disk is that from the normal operation. To obtain the total distortion this needs to be multiplied by the number of trains over which the TPC integrates (3) and the ratio between primary and secondary ions (1:3). With a maximum distortion of $6.4\mu m$ as seen in the plot, this results in a total distortion of around $60\mu m$, as quoted above. While it is in theory possible to correct for these distortions, it is planned to include an ion gate [284] into the design of all MPGD options; prototype gates have been manufactured.

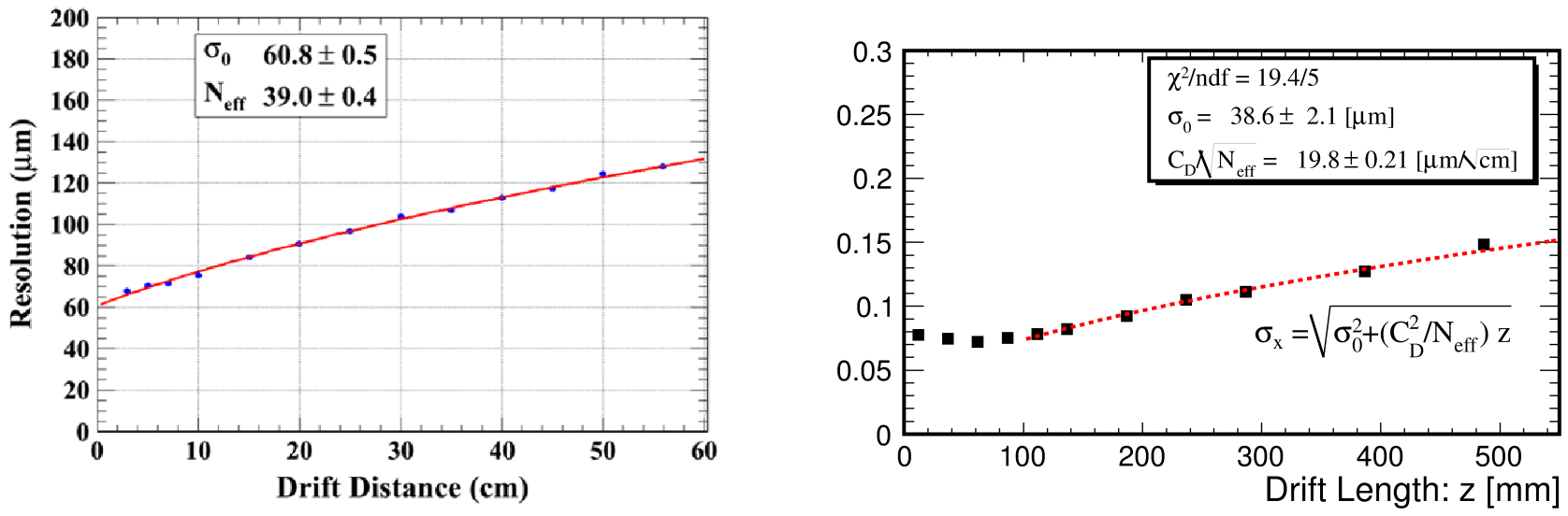

**Figure III-2.16** Preliminary results from the large prototype running at 1 Tesla magnetic field, for (left) a micromegas readout [258] and (right) a GEM based readout [276]

## 2.3.2 Calibration and internal alignment of the TPC

As described in the LOI [198], achieving a momentum resolution an order of magnitude better than collider detectors to date will require significant effort. The systematics of the internal alignment of the TPC must be well understood to guarantee its performance. Redundant tools [285, 286] for solving this issue are $Z$ peak running, laser system (described below), a good B-field map, possibly complemented by a matrix of Hall-plates/ NMR-probes outside the TPC, and use of the SIT and SET-layers inside the inner fieldcage and outside the outer fieldcage. In general based on experience at LEP, about 10 pb$^{-1}$ of data at the $Z$ peak during commissioning could be sufficient for the alignment of the different subdetectors, and typically 1 pb$^{-1}$ during the year may be needed depending on the background and operation of the linear collider (e.g., after push-pull operations). For detector calibration, the accelerator is requested to deliver about $10^{32}$/cm$^2$/s at the $Z$ peak.

For alignment purposes a laser system is foreseen and may be integrated into the fieldcage [281, 287]. The laser system could either be used to create calibration tracks inside the drift volume, or to illuminate calibration spots on the cathode. Electrons will be released from these spots via the photo effect. These electrons then drift into the drift volume at well defined places and at well defined times, under the influence of all field components along its path. Such a system is being tested at the LP [258], and is being used at the T2K experiment [275, 288].

## 2.3.3 Status of R&D for the ILD TPC

To date at the LP, the GEM and Micromegas-based readout systems have been tested, both equipped with either pad-based or pixelated readout. GEM-equipped LP endplates have been tested with two and three modules. Figure III-2.16 shows preliminary resolution results from earlier [258] and recent [276] LP running. For both technologies the basic system goals have been reached.

The pixel-based readout scheme for a TPC [261] has so far only been used with small systems with up to eight readout chips. It has been shown to work with both GEM-based and Micromegas-based systems. Missing is the proof that large area readouts can be realised in this technology.

In addition to the issues described above, the following tasks are important:

- continue electron beam tests to perfect correction and alignment procedures;
- future tests for momentum resolution, for two-track resolution, and for performance in a jet environment;
- further reduction of the pad size is a topic for the far future.

# Chapter 3
# ILD Calorimeter System

Particle flow as a basis for event reconstruction has striking and far-reaching consequences in the design of the detector. This is particularly true for the calorimeter system. Particle flow implies that each particle be reconstructed individually in the detector. This requires unprecedented granularity for the electromagnetic and hadronic calorimeters, ECAL, HCAL and FCAL. To determine the properties of the particles precisely the system should be thick and hermetic, to minimise energy leackage. For a realistic detector a compact calorimeter design is favorable [289]. The imaging capabilities are thus emphasised more than the intrinsic single particle energy resolution, although the latter is still an important ingredient to the particle flow performance for jets [274].

## 3.1 Calorimeter overview

The calorimeter system for ILD consists of a nearly cylindrical barrel system and two large end caps. At very small angles dedicated calorimeter systems provide hermeticity and sensitity to measure the luminosity and monitor beam parameters.

The barrel and end cap calorimeter is divided in depth into an electromagnetic and hadronic section. The principal role of the ECAL is to identify photons and measure their energy. The capability to separate photons from each other and from near-by particles is of prime importance. The ECAL forms the first section for hadron showers and, with its fine segmentation, makes important contributions to the hadron hadron separation. The HCAL is optimized to measure neutral hadrons well and thus has to provide the topological resolution power for separating them from the showers of the much more abundant charged hadrons which must be matched with tracks.

In the very forward region, three systems, LumiCal, BeamCal and LHCAL, are proposed. These system serve as luminosity monitor (LCAL) and beamstrahlungsmonitor (BeamCal), and they close the coverage down to very small angles, also for neutral hadrons (LHCAL).

The transverse and longitudinal segmentation of both calorimeters has been optimised based on detailed simulation and test beam data. It has been shown that the granularity must be of the order of $X_0$ in all three dimensions. This implies that a sampling calorimeter is the best option for both ECAL and HCAL. For the ECAL the most compact design can be realised with tungsten as absorber material. For the HCAL iron is chosen as this allows an excellent energy resolution for hadrons at manageable granularity.

For the ECAL, silicon pad diodes lead to the highest possible compactness (and effective Molière radius) and exhibit excellent stability of calibration. As an option scintillating strips with silicon photo-sensor readout are studied, which provide a similar effective segmentation. The two technologies can be combined in order to reach a cost-performance optimum.

For the hadronic calorimeter, two options have been developed: one based on scintillator tiles with silicon photo-sensors and analogue read-out electronics, and one based on gaseous devices with two-bit, so-called semi-digital readout but finer transverse segmentation. The main development for gaseous readout planes uses glass resistive plate chambers (RPCs), but structures based on micromegas are being considered as alternatives. For the HCAL barrel, two different geometrical concepts of routing the read-out are being proposed, independently of the read-out technology.

### 3.1.1 The challenges of high granularity

The ILD calorimeter system will have about $10^8$ channels in total. Handling such a large number of channels presents a significant challenge. A compact and hermetic design can only be realised if the readout system is as much as possible integrated into the sensitive layers. Special care needs to be taken to minimise the power consumption of the readout. The ILD concept has adopted a scheme to power pulse the detector in between bunch trains, which will be an important ingredient of the power management strategy of the calorimeter as well. The large number of channels will also impact the calibration strategies for the detector. In particular in view of the anticipated push-pull operation a fast and efficient calibration will be needed.

The main challenges of the calorimeter system are the development of optimised and cost effective sensor systems, the design of a low power integrated readout electronics, the development of an effective thermal management and calibration strategy, and a mechanical concept which combines large stability with minimal dead zones.

A central component to the design is the readout electronics. Is is based for all calorimeter systems on a family of custom developed ASICs which are derived from a common base system. The main features are

- Auto trigger and zero-suppression to reduce the data volume;
- Fully digital output;
- Power-pulsing capability to reduce the power dissipation by a factor 100 down to values of about 25 $\mu\mathrm{W/channel}$.

The ASIC family is based on a track and hold scheme with a pre-amplifier shaper sequence allowing for low noise and large dynamic range, and integrating analogue storage pipelines and digitisation. Slow control parameters ensure various configurations and therefore versatility with respect to sensor properties. The large number of channels requires minimising the data lines and the power. The digital readout integrated in the ASICs is therefore common to all the calorimeters. It has been designed to be daisy chained using a token ring mode, without any external components.

The following ASICs have been developed and are used for the different systems:

- HaRDROC [290] to read out the RPCs of the Semi-Digital Hadronic Calorimeter (SDHCAL).
- MICROROC [291] to readout the micromegas alternative of the SDHCAL.
- SPIROC [292] to readout the silicon photomultipliers, SiPM, of the analogue hadron calorimeter and of the electromagnetic calorimeter based on scintillators.
- SKIROC [293] to readout the Si pin diodes of the silicon tungsten electromagnetic calorimeter.

A generic data acquisition system has been developed which is used to read out all different systems [294]. This DAQ includes many features which are proposed for a full DAQ for the ILD detector, but is also suitable for use at a test beam setup.

### 3.1.2 Beam tests

A key role in the development of a calorimeter suitable for particle flow is played by extensive test beam experiments. A large international effort has been ongoing for a number of years, within the CALICE collaboration, to organise common test beam experiments for the different technologies considered. Given the complexity and the scale of the setups common infrastructure such as mechanical devices, readout systems, data acquisition and software were essential to successfully organise this effort. CALICE has been able to expose all major technologies to test beams and collect large amounts of data. Power pulsing has been demonstrated by the SDHCAL beam test which supports the approach for all proposed technologies. It has also been verified that interactions of shower particles do not affect the stable operation of the embedded ASICs [295].

Detailed comparisons have been made between the test beam data and different simulation models. With recent results it is possible to match the simulation and data within typically 5% [296, 297, 298], sufficiently good to reliably model the performance of the detector. In the future further analysis of the data and more test beam campaigns will be needed to continually improve the understanding of data and simulation.

## 3.2 The electromagnetic calorimeter system

The particle flow paradigm has a large impact on the design of the electromagnetic calorimeter system. A key requirement is the capability of the system to separate overlapping showers from each other. A calorimeter for particle flow thus needs to be able to do pattern recognition in the shower. The electromagnetic section has a number of tasks to fulfill. It should be able to reconstruct photons in the presence of close-by particles. It should be able to reconstruct the detailed properties of the shower, such as shower shape, starting point and energy to distinguish early starting electromagnetic showers from hadronic ones. It should be noted that about half of the hadronic showers start inside the electromagnetic calorimeter. Thus an excellent three-dimensional granularity of the device is of utmost importance.

Earlier studies [198] have shown that the separation and reconstruction continues to improve, even with pixel size smaller than the Molière Radius. The study was done with square pixels between 1 mm and 2 cm size. For ILD a pixel size of $5{\times}5\,\mathrm{mm}^2$ has been chosen.

In order to have a better separation of close-by showers in the calorimeter, a system with a small Molière radius is advantageous. Further help in the separation between electromagnetic and hadronic showers can come from a large ratio between interaction length and radiation length. A small radiation length will move the start of the electromagnetic shower earlier in the calorimeter, while a large interaction length will reduce the fraction of hadronic showers starting in the ECAL.

The particle flow approach requires that the calorimeters are placed inside the magnetic coil, see Sec. 1.2. This has a major impact on the layout of the detector, and on the cost. Therefore, a compact calorimeter is preferred in order to minimise the overall physical thickness, which in turn reduces the size of the coil. For the ECAL tungsten is a good choice for the radiator as it is dense, and has a large ratio of interaction length to radiation length. The final system layout is a compromise between performance and cost. The energy resolution scales with $\sqrt{T}$, where $T$ is the individual absorber plate thickness, while the cost scales linearly with the surface area of the readout layers. For ILD a solution with 30 readout layers and a thickness of the ECAL of 24 $\mathrm{X}_0$ has been chosen as the baseline. The optimisation of the layout is ongoing.

For a chosen pad size of $5 \times 5\,\mathrm{mm}^2$ silicon pin diodes are a good choice. They can cover large areas, are reliable and simple to operate, allow for a thin readout layer and can operate in the 3.5 T strong central magnetic field. While the very thin silicon layers offer excellent performance for the tracking capabilities of the calorimeter, the energy resolution is somewhat degraded. Here a less compact device, with a thicker readout layer, will show better performance.

As an alternative option a sensitive layer based on scintillator strips could be used. With scintillators individual tiles of size $5 \times 5\,\mathrm{mm}^2$ are difficult to realise. By using strips of $5 \times 45\,\mathrm{mm}^2$arranged in alternative directions an effective granularity approaching $5 \times 5\,\mathrm{mm}^2$ can be achieved. However, the reconstruction becomes more complicated, in particular in dense jets.

An alternative to silicon could be micromegas chambers. This technology however is significantly less advanced than either the silicon or the scintillator option.

In the following sections the detailed design of the ECAL will be presented. In addition industrial aspects will be addressed, with variables such as numbers of producers, time of production, etc..

The requirements on granularity, compactness and particle separation lead to the choice of a sampling calorimeter with tungsten (radiation length $X_0 = 3.5$ mm, Moliere Radius $R_M = 9$ mm and interaction length = 99 mm) as absorber material. This allows for a compact design with a depth of roughly 24 $X_0$ within 20 cm and, compared to e.g. lead, a better separation of electromagnetic showers generated by near-by particles. To achieve an adequate energy resolution, the ECAL is longitudinally segmented into 30 layers, possibly with varying tungsten thicknesses. In order to optimise the pattern recognition performance, the active layers (either silicon diodes or scintillator) are segmented into cells with a lateral size of 5 mm.

### 3.2.1 Detector implementation

Figure III-3.1 shows the position of the electromagnetic calorimeter in the ILD detector, the trapezoidal form of the modules and how it is envisaged to be interfaced mechanically with the hadron calorimeter.

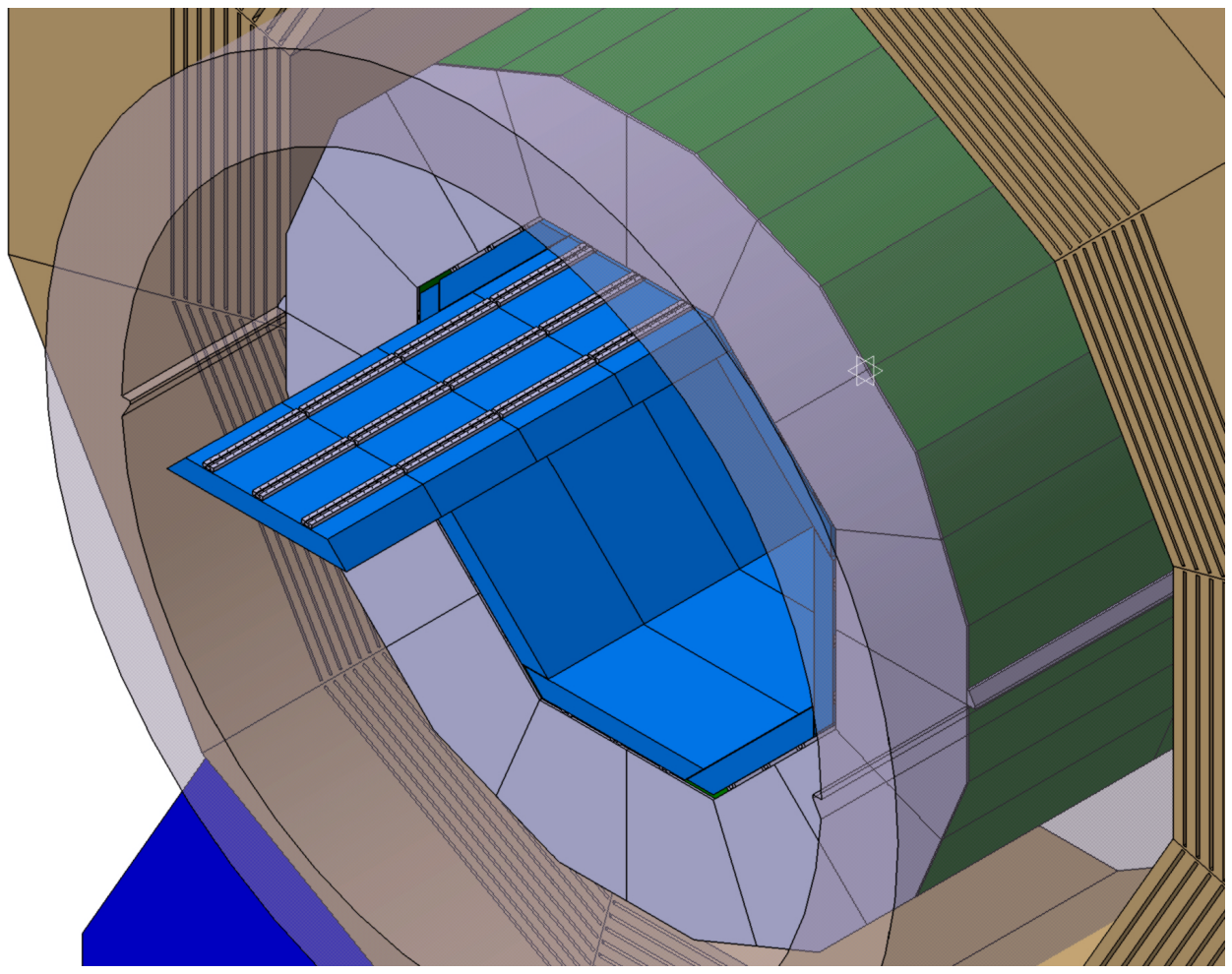

**Figure III-3.1** The electromagnetic calorimeter (in blue) within the ILD Detector.

After several years of successful operation of small so called *physics prototypes* the focus of the work turns to the realisation of *technological prototypes*, see e.g. [299]. These prototypes address the engineering challenges which come along with the realisation of highly granular calorimeters.

#### 3.2.1.1 Alveolar structure and general integration issues

The mechanical structure consists of a carbon reinforced epoxy (CRP) composite structure, which supports every second tungsten absorber plate. The carbon fibre structure ensures that the tungsten plates are at a well defined distance, and provide the overall mechanical integrity of the system (the so-called alveolar structure). Into the space between two tungsten plates another tungsten plate is inserted, which supports on both sides the active elements, the readout structure and necessary services. This results in a very compact structure with minimal dead space. The mechanical structure is equally well suited for both proposed technologies. Figure III-3.2 shows a prototype which is 3/5 of the size of a final structure for the barrel. For the end-cap region alveolar layers of up to 2.5 m length have been fabricated. While in the barrel the shape of all alveolar structures is the same, three different shapes of alveolar structures are needed in the end-caps. Recent studies revealed that in the end-caps considerable forces are exerted onto the thin carbon fibre walls, which enclose the alveolar structure. This issue has to be addressed in the coming R&D phase.

Figure III-3.3 shows a cross section through a calorimeter layer for the electromagnetic calorimeter with silicon (SiECAL), and one layer for the electromagnetic calorimeter with scintillator (ScECAL). The two readout layers of the SiECAL will be mounted on two sides of a tungsten slab, which is

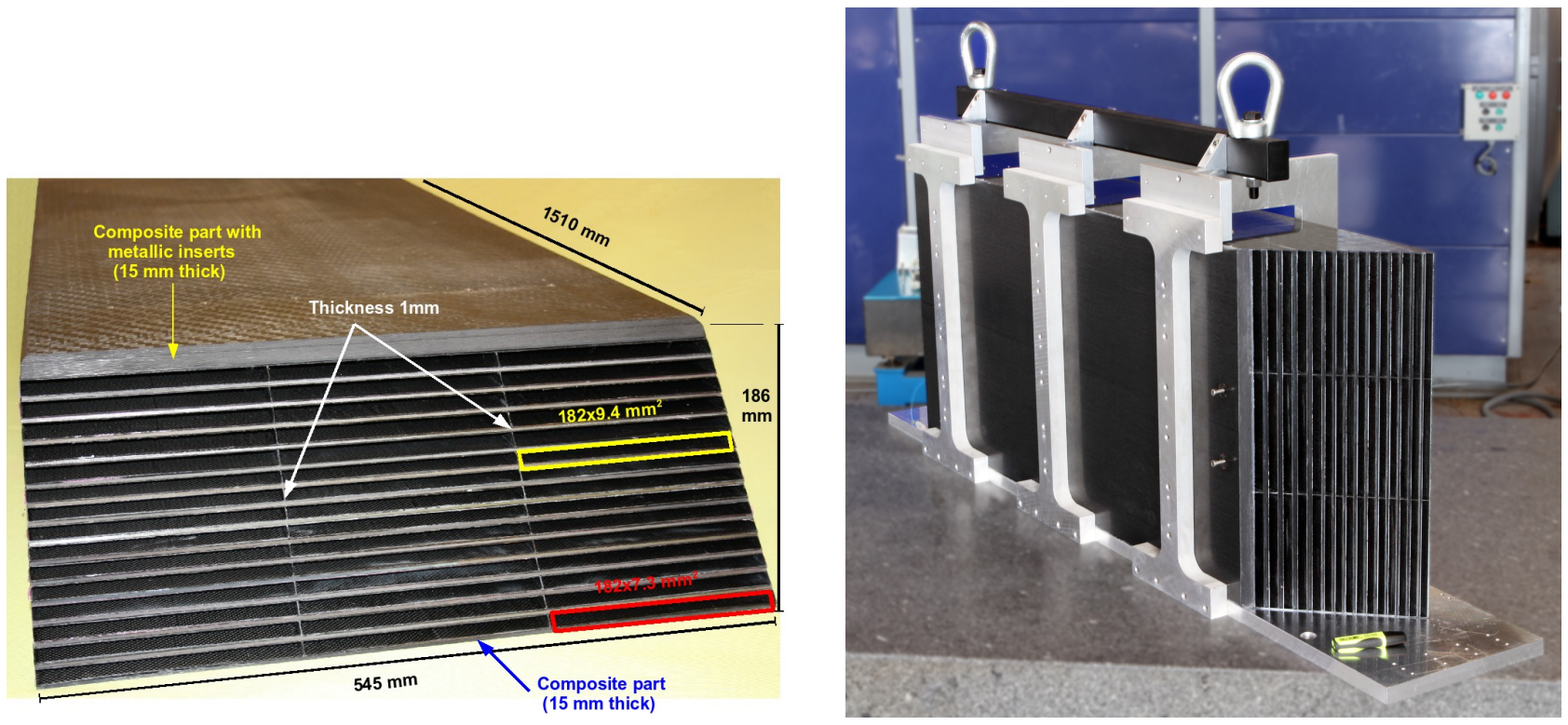

**Figure III-3.2**
Left: Front view with dimensions on the alveolar structure which houses the sensitive parts of the electromagnetic calorimeter prototype. Right: Side view on the completed structure and its mechanical protection.

inserted into the alveoli of the mechanical stricture. The insertion process has been successfully implemented for the physics prototype with short layers and has in addition been demonstrated with layers of up to 1.3 m length for a mechanical demonstrator of the technological prototype. In case of the ScECAL one side of the tungsten board will be equipped with scintillating strips for the $x$ direction and the other side with strips for the $y$ direction.

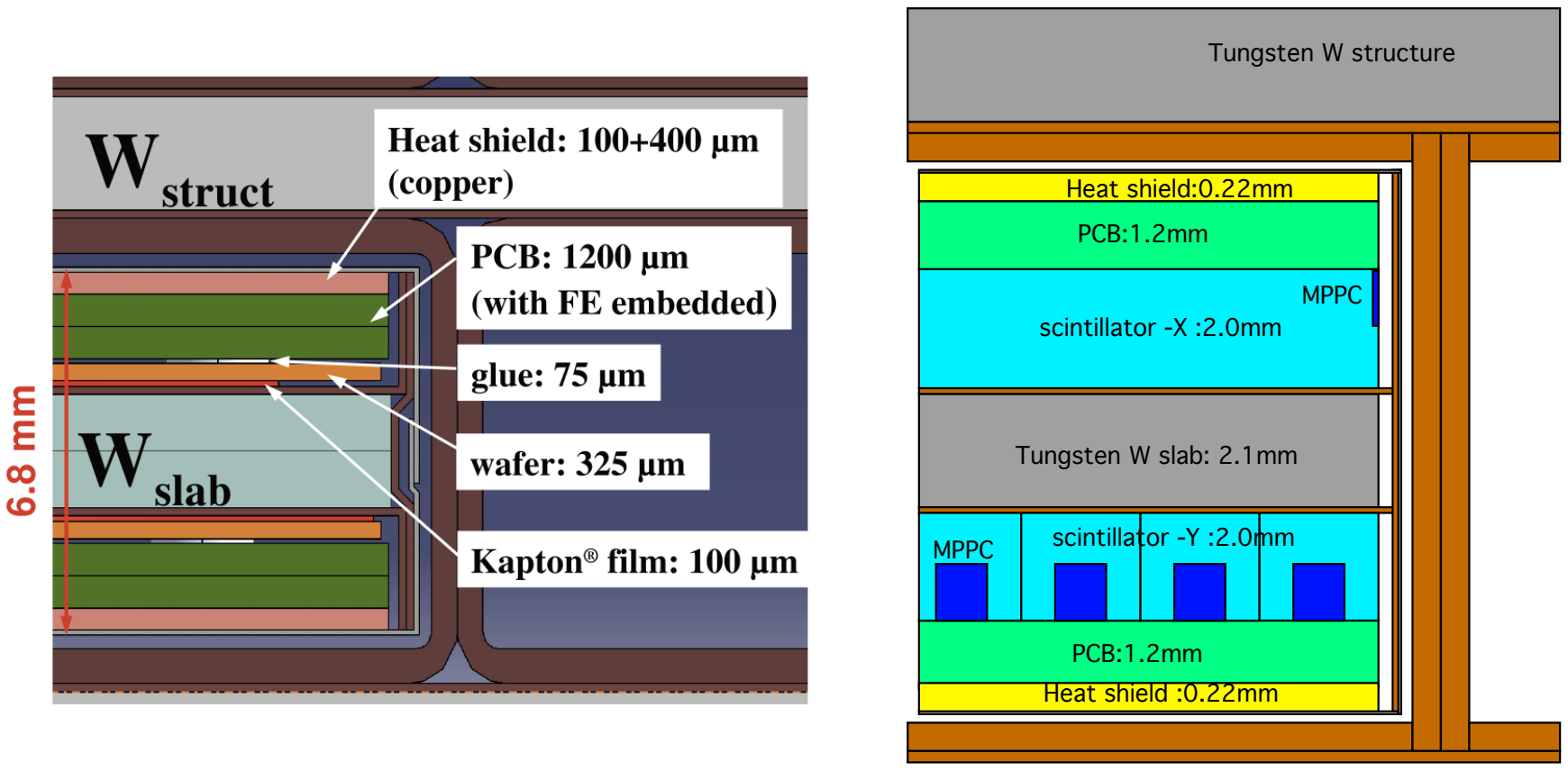

**Figure III-3.3**
Cross sections through electromagnetic calorimeter layers for the silicon option (left) and for the scintillator option (right).

A notable difference between the two options is the thickness of the active sensors. While silicon wafers can be produced easily with a thickness of a few 100 microns, the scintillator thickness needs to be at least 1 mm. This is due to the size of the photo-sensor MPPC and number of photons detected by this sensor.

#### 3.2.1.2 Silicon wafers:

An example of a silicon wafer matrix as employed in the current R&D phase is shown in Fig. III-3.5. Silicon allows for a thin and easily segmented readout detection system suited for high granularity. The proposed technology is shown to deliver an excellent signal to noise ratio, which will allow to detect also small energy deposits, thus facilitating the two particle separation. The R&D goal for the S/N ratio is 10:1 at 1 MIP level.

The wafers are composed of silicon with a typical resistivity of $5\,\mathrm{k\Omega\cdot cm}$. To achieve full depletion, the bias voltage to be applied to the wafers is between 100 V and 200 V. While the manufacturing of these wafers is a well known technique, a key challenge is to produce these wafers at a low cost in order to reduce the cost since a surface of about 3000 $\mathrm{m}^2$ will be needed for ILD. Contacts and discussions with industry are being developed.

The measurements with the physics prototype revealed cross talk between the guard ring which surrounds the silicon wafers and neighbouring silicon pads resulting in so-called square events, the frequency of which increased with the energy of primary electrons [300]. Currently an R&D effort is under way to understand this problem and to optimise the system. These "square events" however show a very distinct signature so that during the offline reconstruction they could be removed. The final performance of the system was shown to be essentially unchanged by this cross-talk effect.

### 3.2.1.3 Scintillator strips

Optimal particle flow performance requires a cell size of about $5 \times 5\text{mm}^2$. The ScECAL has a strip width of 5 mm, while the length of the strip may depend on the performance of the details of the Particle Flow Algorithm. Current results indicate that 45 mm long strips are sufficient. Good hermeticity is also relevant for the calorimeter, which inspired a design in which the photosensor is embedded into the strip resulting in a dead area of 1.9% of the total surface. In order to reduce the passive area different ideas to extract the scintillation light at the bottom of the scintillator are under study.

### 3.2.1.4 Front-end electronics

Figure III-3.4 shows the performance of the SKIROC circuit. The curve shows the threshold of the 50% trigger efficiency as a function of the injected charge, in units of MIPs. The signal over noise ratio is about ten, which meets the goal.

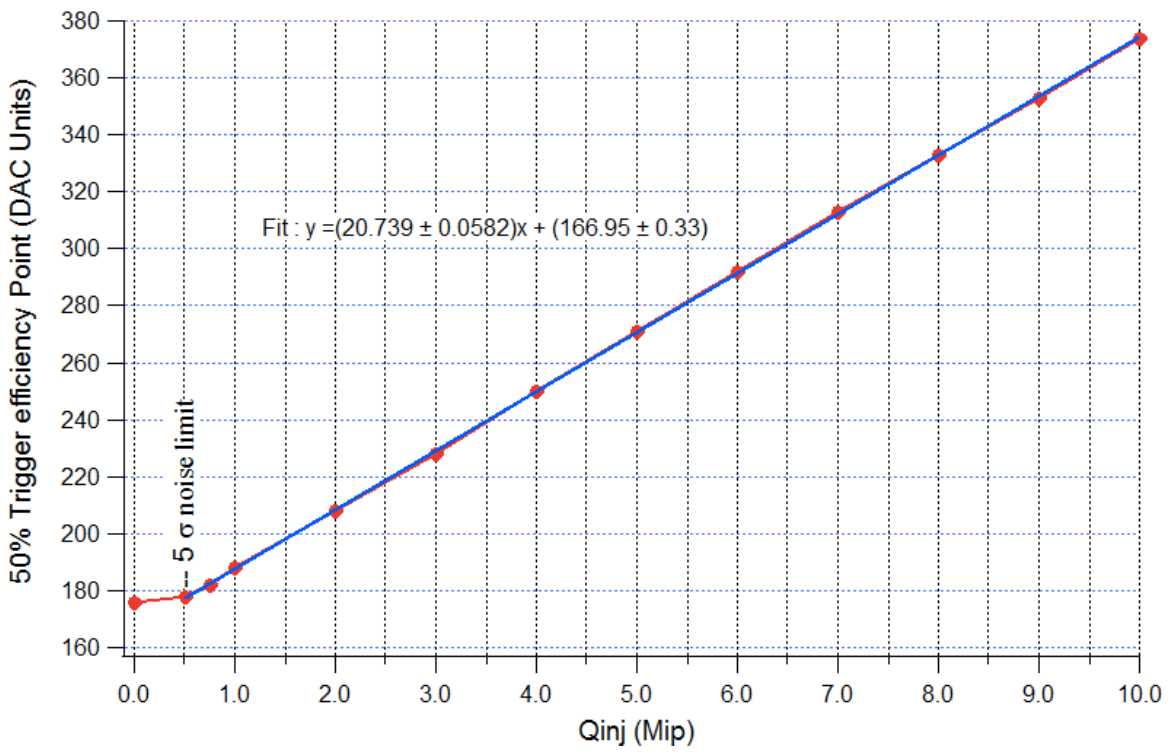

**Figure III-3.4** Validation of the SKIROC circuit; shown is the threshold for 50% trigger efficiency as a function of the injected charge, in units of MIPs.

The front-end electronics has to be integrated into the calorimeter layers as illustrated in Figure III-3.3. This is a major challenge for the construction of the calorimeter. In the current design the room available for the readout circuits (ASICs) and the interface boards between the ASICs and the sensitive material is about 1.2 mm. Figure III-3.5 shows a picture of four ASICs bonded onto a PCB. The PCB is a very thin multi-layer board. PCBs for 16 ASICs are now available and will be equipped soon with ASICs. It should be noted that one of the major challenges to be solved in the near future is the planarity of the PCB. This issue is currently addressed in collaboration with industrial partners as well as by revising the entire assembly process of the detector. For protection purposes the ASICs will be encapsulated. For this encapsulation standard industrial processes can be applied.

### 3.2.1.5 Module assembly and cooling system:

A calorimeter layer will have a length of up to 1.5 m in the barrel and up to 2.5 m in the end-caps and will be composed of several units which carry the sensitive devices as well as the front end electronics. Great care is taken in the development of the technique to interconnect the individual units. The signal transfer along the slab to the interconnection pad must be very reliable and at the same time should not exert mechanical or thermal stress e.g. to the silicon wafers which are very close to the interconnection pads. Good progress has been made in the past years and a viable solution is currently applied to the first layers of the technological prototype of the SiECAL. The sensitive ensemble is then to be inserted into the alveolar structure which houses the calorimeter layers. The integration cradles are under development and a first integration test with a demonstrator has been successfully conducted.

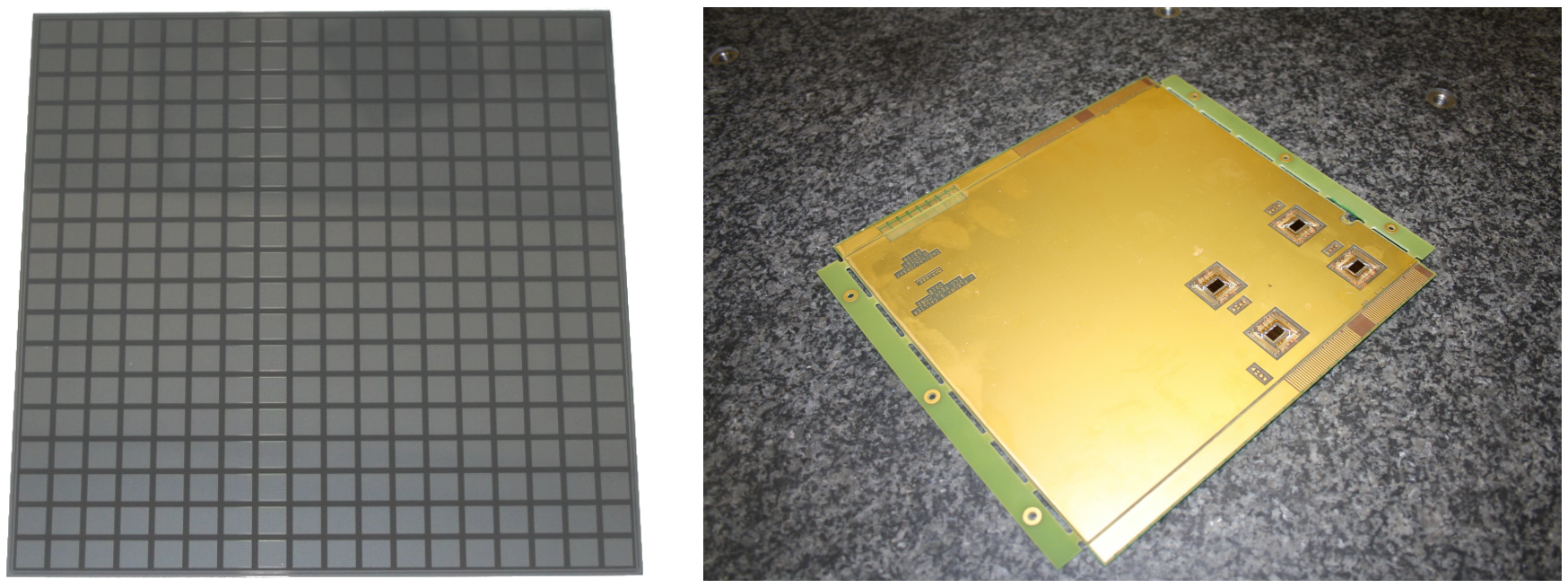

**Figure III-3.5** Left: Example of a silicon wafer matrix as studied for the large scale prototype. Right: The wafers are mounted onto interface boards, PCB. The photo shows a PCB for the CALICE SiECAL prototype with wire bonded readout ASICs.

For this demonstrator a leak-less water system for cooling has been developed [301]. A heat exchanger will be coupled to a copper drain at the outer part of the ECAL layers. The FPGAs mounted at the end of the modules are a major source of heat and are directly connected to the cooling pipes. Thin copper plates will ensure heat evacuation of residual heat from the inner parts of the detector layers. Earlier studies for SiECAL have shown that the temperature gradient along an ECAL layer is about $6^o$ C in the detector end-caps and only $2.2^o$ C in the barrel region. Due to this comparatively small temperature gradient the concept of applying cooling only at the detector ends seems possible. The cooling pipes routed from outside the detector to the ECAL module will be passed in the 3 cm wide radial space between the HCAL and the ECAL.

#### 3.2.1.6 Signal and power cable routing:

The gap between the electromagnetic and hadronic calorimeter is also used for the power and signal cables. Low voltage of about 5 V is needed for the front-end electronics while high voltage between 200 V or 50 V is needed to saturate the silicon wafers or to operate the silicon photomultipliers. Although no detailed concept exists today it is likely that the power will be brought close to a detector module and then fanned out to the individual layers. The data coming from the detector layers will be concentrated in a device on top of the module and routed out via one cable per module. This cable can also be used to transmit slow control and configuration commands to the individual detector elements. The challenge is to reduce the number of cables to an absolute minimum. In the ideal case signal propagation and power delivery would share the same cable.

### 3.2.2 Detector optimisation

The main parameters of the ECAL to be optimised are the inner radius of the detector, its thickness (in $X_0$), the number of detection layers, and the segmentation within the layers. These parameters determine the detector performance as well as its cost. A full cost-performance optimisation has not yet been performed, however several aspects required to perform such an optimization have been studied. The cost of the detector option with silicon sensors scales mainly with the surface area of the wafers, while the cost of the scintillator option scales with the number of channels. Many aspects of the optimization are, at least to first order, independent of the particular technological solution.

The inner radius of the ECAL is chosen to be relatively large, allowing particles within a hadronic jet to spread, thereby increasing particle separation in the calorimeters. An optimization of the inner radius and the magnetic field has been performed in the context of the Letter of Intent [198], where the jet energy resolution was found to scale approximately as $\sigma_E/E \sim R^{-1}B^{-0.3}$. The inner radius of the ECAL is strongly correlated with the outer radius of the TPC. The actual outer radius of the TPC has an effect on the momentum resolution of the tracking system. This dependence on the variation in momentum resolution is however relatively weak.

#### 3.2.2.1 Optimization of the silicon-only ECAL option

It has been shown in the ILD LOI and validated in beam test studies (see section 3.2.3 and [302]) that the silicon diodes are perfectly suited to meet the precision requirements of an ECAL at a future linear collider. The cost of such a high precision system has been estimated based on current best knowledge and careful extrapolations, and is presented in chapter 7. Several options are under study to understand the scaling of costs and to find ways to optimise the cost-performance ratio.

In the baseline model, the ECAL consists of 30 silicon (Si) and 29 tungsten (W) layers. Some details of the design are given in Table III-3.1. Five alternative ECAL models (26, 20, 16, 12 and 10 layers) have been studied. Their parameters are also summarised in Table III-3.1. Other configuration parameters such as the total tungsten thickness, a $1:2$ ratio of W thickness between inner and outer absorber layers, carbon fibre, cooling layers, Si thickness etc. remain the same for the six models.

**Table III-3.1** ECAL models with different numbers of layers, $N_{lay.}$ and layer thicknesses, $d$, corresponding to different layer sets.

| $N_{lay.}$ | 10 | | 16 | | 20 | | 26 | | 30 | |
|---|---|---|---|---|---|---|---|---|---|---|
| W layers | 6 | 3 | 10 | 5 | 13 | 6 | 17 | 8 | 20 | 9 |
| $d$ [mm] | 6.7 | 13.3 | 4.0 | 8.0 | 3.2 | 6.3 | 2.4 | 4.8 | 2.1 | 4.2 |

The detector performance is studied using the example of the jet energy resolution measured as $\mathrm{rms}_{90}$ of the jet invariant mass distribution. The jets are reconstructed by the PandoraPFA algorithm [274] using $Z \to q\bar{q}$ events generated at $\sqrt{s} = 91$, 200, 360 and 500 GeV. Defined in [198], the $\mathrm{rms}_{90}$ is the root-mean-squared deviation from the mean of the jet invariant mass distribution, in the region around the mean, which contains 90% of the reconstructed events. To avoid the barrel/end-cap overlap region, a cut on the polar angle of the generated $q\bar{q}$ system of $|\cos\theta_{q\bar{q}}| < 0.7$ is applied.

The results in Figure III-3.6 and Table III-3.2 show the relative jet energy resolution for a single jet. A degradation of 10% in jet-energy resolution is observed going from 30 to 20 layers for events at 91 GeV, a smaller deterioration for higher energies. Going below 20 layers, the resolution starts to degrade significantly, though again less so at higher energies.

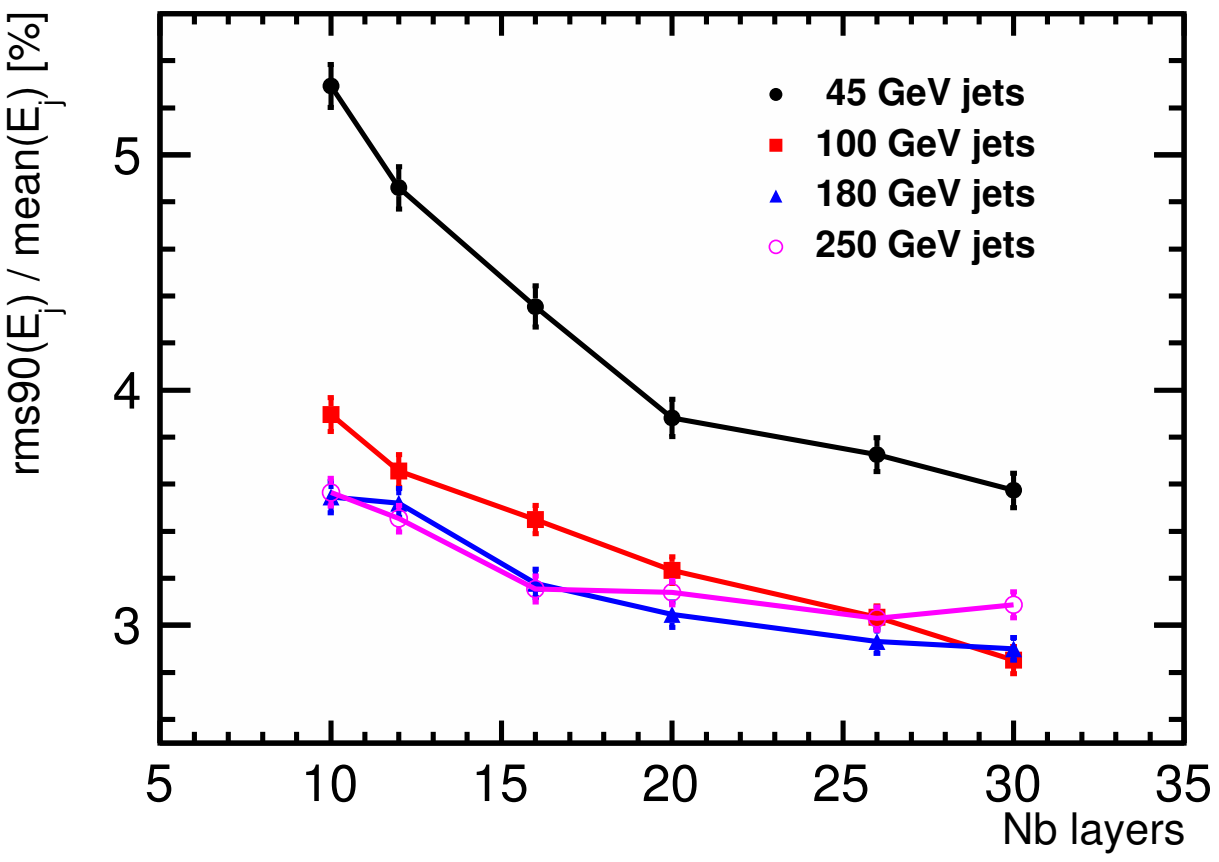

**Figure III-3.6** Dependence of the relative jet energy resolution ($\mathrm{rms}_{90}/E_j$) for single jets on the number of ECAL layers for events with $|\cos\theta_{q\bar{q}}| < 0.7$, for the SiECAL option. The resolutions are shown for $e^+e^- \to Z \to u\bar{u}, d\bar{d}, s\bar{s}$ events at $\sqrt{s} = 91$, 200, 360 and 500 GeV.

**Table III-3.2**
Jet energy resolution for $Z \to uds$ events with $|\cos\theta_{q\bar{q}}| < 0.7$ expressed as $\mathrm{rms}_{90}/E_j$ for SiECALs with different number of layers.

| Jet energy | 10 layers | 16 layers | 20 layers | 26 layers | 30 layers |
|---|---|---|---|---|---|
| 45 GeV | 5.3 ± 0.1 | 4.4 ± 0.1 | 3.9 ± 0.1 | 3.7 ± 0.1 | 3.6 ± 0.1 |
| 100 GeV | 3.9 ± 0.1 | 3.5 ± 0.1 | 3.2 ± 0.1 | 3.0 ± 0.1 | 2.9 ± 0.1 |
| 180 GeV | 3.6 ± 0.1 | 3.2 ± 0.1 | 3.1 ± 0.1 | 2.9 ± 0.1 | 2.9 ± 0.1 |
| 250 GeV | 3.6 ± 0.1 | 3.2 ± 0.1 | 3.1 ± 0.1 | 3.0 ± 0.1 | 3.1 ± 0.1 |

#### 3.2.2.2 Optimzation of the scintillator strip ECAL option

Two-jet events at 200 GeV centre of mass energy have been simulated with a scintillator-based ECAL and different strip lengths, and have been analysed using PandoraPFA. The jet energy resolution achieved as a function of different strip lengths is shown in Figure III-3.7. To resolve the ambiguities introduced by crossing strips of 5 mm width and different length a strip splitting algorithm (SSA) [303] has been developed, which is run in addition to PandoraPFA. For the purpose of this particular analysis, the strip thickness was set to the same as the SiECAL sensor thickness (0.5 mm). Results with and without the use of SSA are shown. At least for this type of event, no strong deterioration with increasing the strip length is found after applying SSA. A strip length of 45 mm is chosen for the baseline.

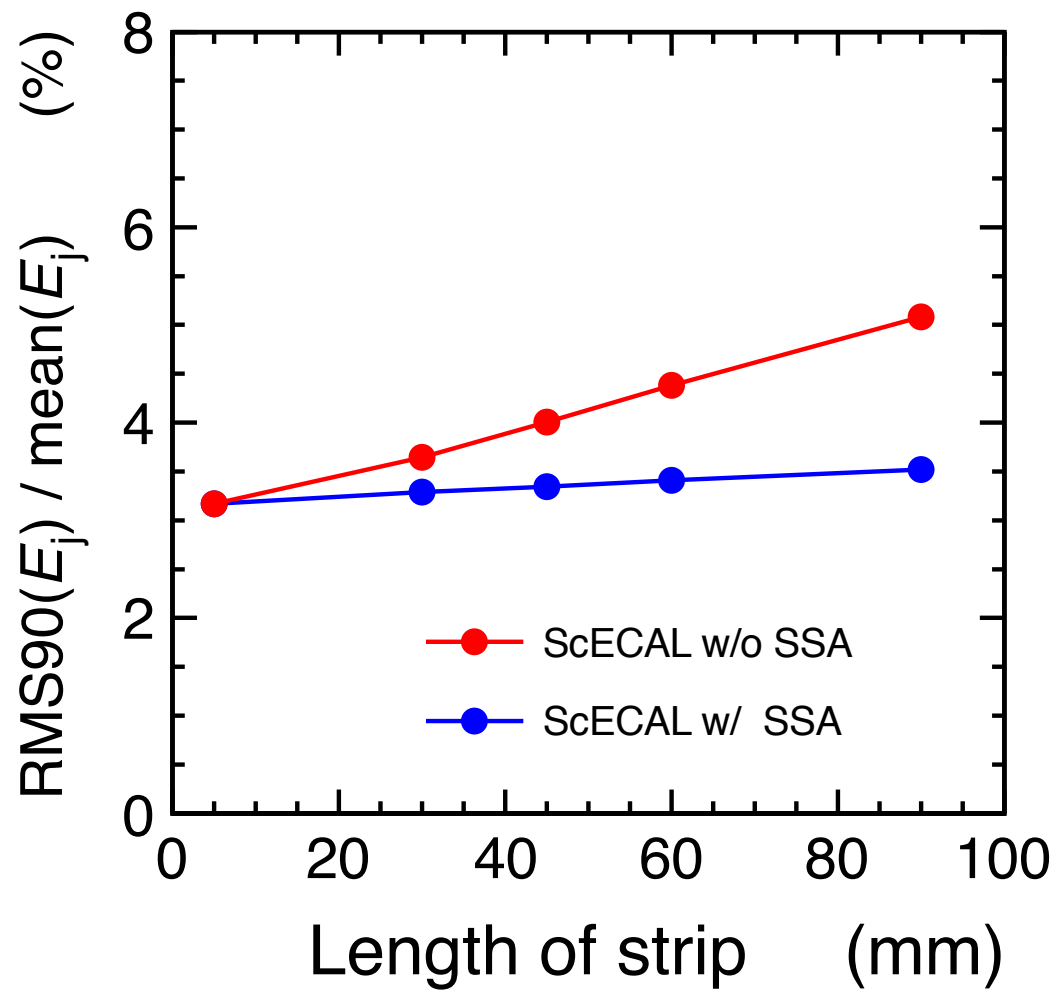

**Figure III-3.7**
Jet energy resolution for $e^+e^- \to Z \to u\bar{u}, d\bar{d}, s\bar{s}$ event at 200 GeV centre of mass energy as a function of strip length, for the ScECAL option, shown without (red) and with (blue) the strip splitting algorithm.

#### 3.2.2.3 The hybrid ECAL option

Another option could be a mixture of silicon layers and scintillator-strip layers (Hybrid ECAL). Such an design, restricting the use of silicon sensors to the more critical areas of the detector (up to around the position of the maximum of electromagnetic showers), may be more cost-effective than a silicon-only ECAL. A number of configurations were studied, all using the same 27 layer tungsten absorber structure, the first 20 layers with a thickness of 2.1 mm, the remaining 7 of 3.5 mm. Note that this is a slightly different structure to the default ECAL, so results cannot be directly compared.

Different arrangements of the sensitive layers were studied. The inner layers were instrumented with silicon sensors and the outer layers with scintillator, with the following arrangements: (20 silicon + 8 scintillator); (14 silicon + 14 scintillator); and (8 silicon + 20 scintillator). Models with scintillator-only and silicon-only readout were also studied. The thickness of the sensitive detectors was 0.5 mm for silicon and 2.0 mm for scintillator. Note that the total thicknesses of these ECAL models are different, models with more scintillator layers having a larger thickness.

Events were reconstructed using the standard reconstruction chain, including PandoraPFA including the strip splitting algorithm. The jet energy resolution in $e^+e^- \to q\bar{q}(q = uds)$ events generated at centre of mass energies of 91, 200, 360, and 500 GeV was measured for each of these ECAL models. Figure III-3.8 shows this jet energy resolution as a function of the fraction of ECAL layers which are scintillator, for each of the jet samples. For 45 GeV jets, the jet energy resolution does not degrade with increasing scintillator layers. At higher energies, degradation in performance is seen for an increasing number of scintillator layers, particularly when the scintillator fraction is above 50%.

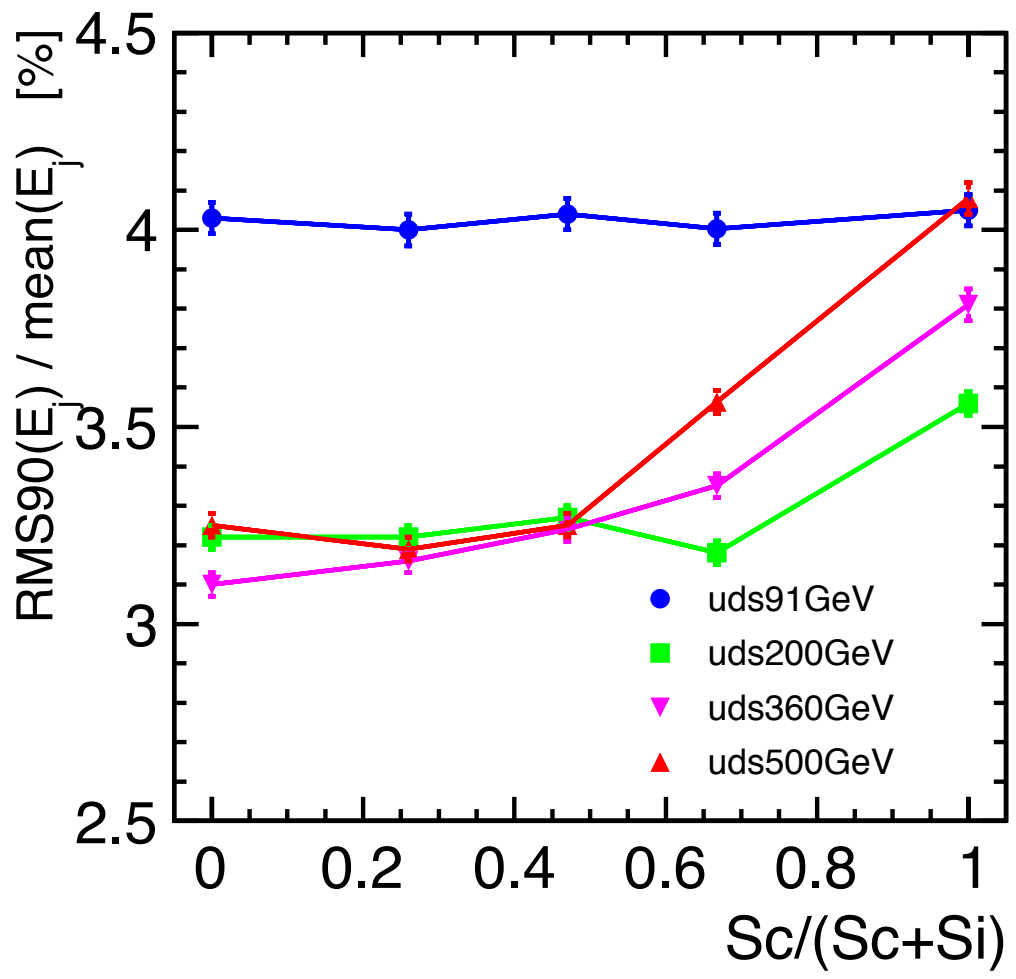

**Figure III-3.8** Jet energy resolution in $q\bar{q}(q = uds)$ events at different centre of mass energies, using a hybrid ECAL with silicon (Si) and scintillator (Sc) layers. The jet energy resolution is shown as a function of the fraction of scintillator layers in the ECAL, (Sc/(Sc+Si)). The total number of ECAL layers (Sc+Si) is 28.

## 3.2.3 Test beam validation

Within the framework of the CALICE collaboration physics and technological prototypes for both options of the electromagnetic calorimeters have been built and tested in beam tests since 2004. They were exposed to a wide variety of particle beams (electrons/ positrons, pions, protons and muons) over a wide range of momenta, between 2 and 180 GeV.

### 3.2.3.1 SiECAL test beam validation

The CALICE SiECAL group has designed and built a so-called "physics prototype" [304], shown in Figure III-3.9, whose aim was to demonstrate the ability of this ECAL to meet the performance requirements. It had an active area of $18 \times 18\,\mathrm{cm}^2$ and 30 sampling layers. The active sensors had a granularity of $1 \times 1\,\mathrm{cm}^2$, giving a total of nearly 10k readout channels. These data have been used to calibrate the detector, to measure its performance, and to tune and validate the simulation of the SiECAL and particle interactions within it. Since these tests have been carried out over a number of years, they have also given important information about the long term stability of the detectors and associated systems.

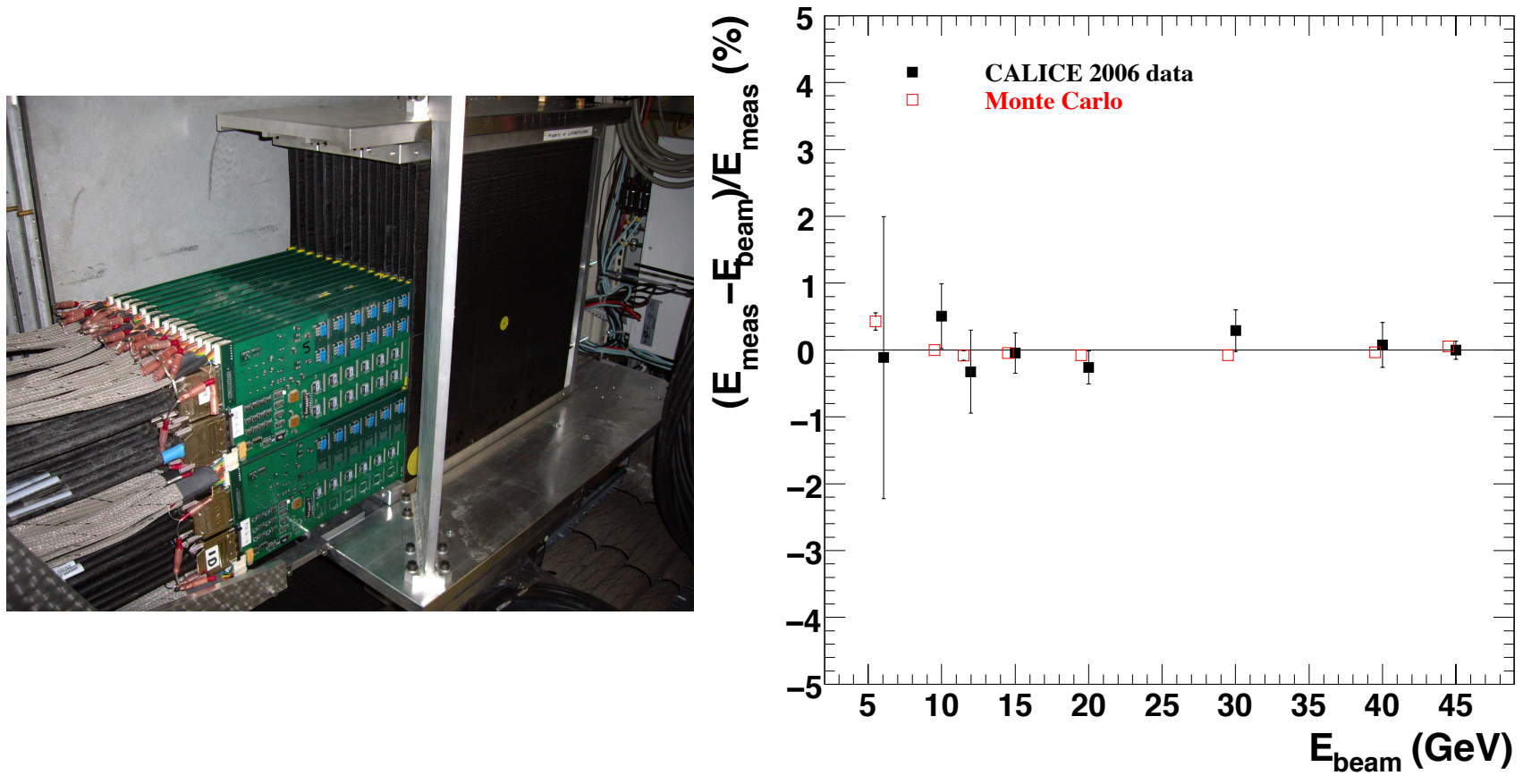

**Figure III-3.9** (Left) the SiECAL physics prototype. (Right) linearity of the energy response, as measured in both real data and simulation [296].

The SiECAL physics prototype has been successfully and stably operated over a period of five years. No major systematic problems were identified with the concept of this detector or with its technical design. The detector could be operated reliably and will minimal intervention also for lobger periods of time. The signal-over-noise ratio in the physics prototype was measured to be 7.5:1 [304]. This value was confirmed over the years. First measurements with the technological prototype show similar of even better performance.

The response of the detector to electrons is presented in [296]. The energy response is found to be linear to within 1% in the energy range between 1 and 45 GeV, as shown in Figure III-3.9. The energy resolution for electrons was measured to be $16.6/\sqrt{E(\mathrm{GeV})} \oplus 1.1\%$. Both the energy response and the longitudinal shower profiles of electron showers are well described in the simulation, as is the effective Molière radius. The results are compatible with the values assumed for the full detector simulation of the ILD detector as indicated above.

The data collected with hadron beams have been used to constrain the models for hadronic showers implemented in GEANT4, with the FTFB_BERT physics list giving the best description of the data [305, 306]. Analysis of overlaid "MIP"-like and EM shower events shows that the efficiency to distinguish them (in the ECAL alone) begins to decrease at a separation of 3 cm, to a minimum of around 50% for overlapping particles [307].

The position resolution of the physics prototype was found to be about 0.6 mm for electrons with energy above 20 GeV. The angular resolution is found to be $(106 \pm 2)/\sqrt{E} \oplus (4 \pm 1)\,\mathrm{mrad}$ along the $x$ direction and $(100 \pm 2)/\sqrt{E} \oplus (14 \pm 1)\,\mathrm{mrad}$ along the $y$ direction [308]. The differences between the two directions can be explained by the different arrangement of the detector layers in $x$ and $y$ direction.

In the coming years the technological prototype will be progressively equipped. The finer granularity will allow for a more precise tomography of hadronic showers. Currently, data recorded with first layers are analysed for the electromagnetic response of the new prototype.

#### 3.2.3.2 ScECAL test beam validation

A physics prototype of the scintillator ECAL has been built and exposed to test beam. The prototype consists of 30 active layers, each of which includes 72 scintillator strips readout by photosensors. The minimum detecting unit has a 45 mm long and 10 mm wide plastic scintillator and a MPPC (Multi Pixel Photon Counter) semiconductor photosensor packaged in a 4.2 $\times$ 3.0 $\times$ 1.3 mm$^3$ housing. There is an one mm hole for a wave length shifting fibre to absorb the scintillation light generated by the charged particles and guide it to the end where the MPPC is located. The thickness of the unit is 3 mm [309].

The basic performance of the calorimeter has been tested in a hadron beam at Fermilab. The linearity of the system is shown in Figure III-3.10(left), the energy resolution is shown in figure III-3.10(right). The results include a temperature correction calibrated using data for an temperature range between 19 to 28°C. The deviation from a linear behaviour is determined to be less than 2% and the energy resolution is found to be $12.9/\sqrt{E\mathrm{GeV}} \oplus 1.2\%$ for 2 - 32 GeV electron beams [310].

A first layer of the technological prototype has been constructed and tested at DESY in autumn 2012. This layer combines absorber material, scintillator and read out electronics. The layer is equipped with 144 scintillator strips each 5 mm wide and MPPCs. MIP like signals have been observed. The efficiency and S/N ratio will be studied with these data. The power pulsing mode will be tested in 2013.

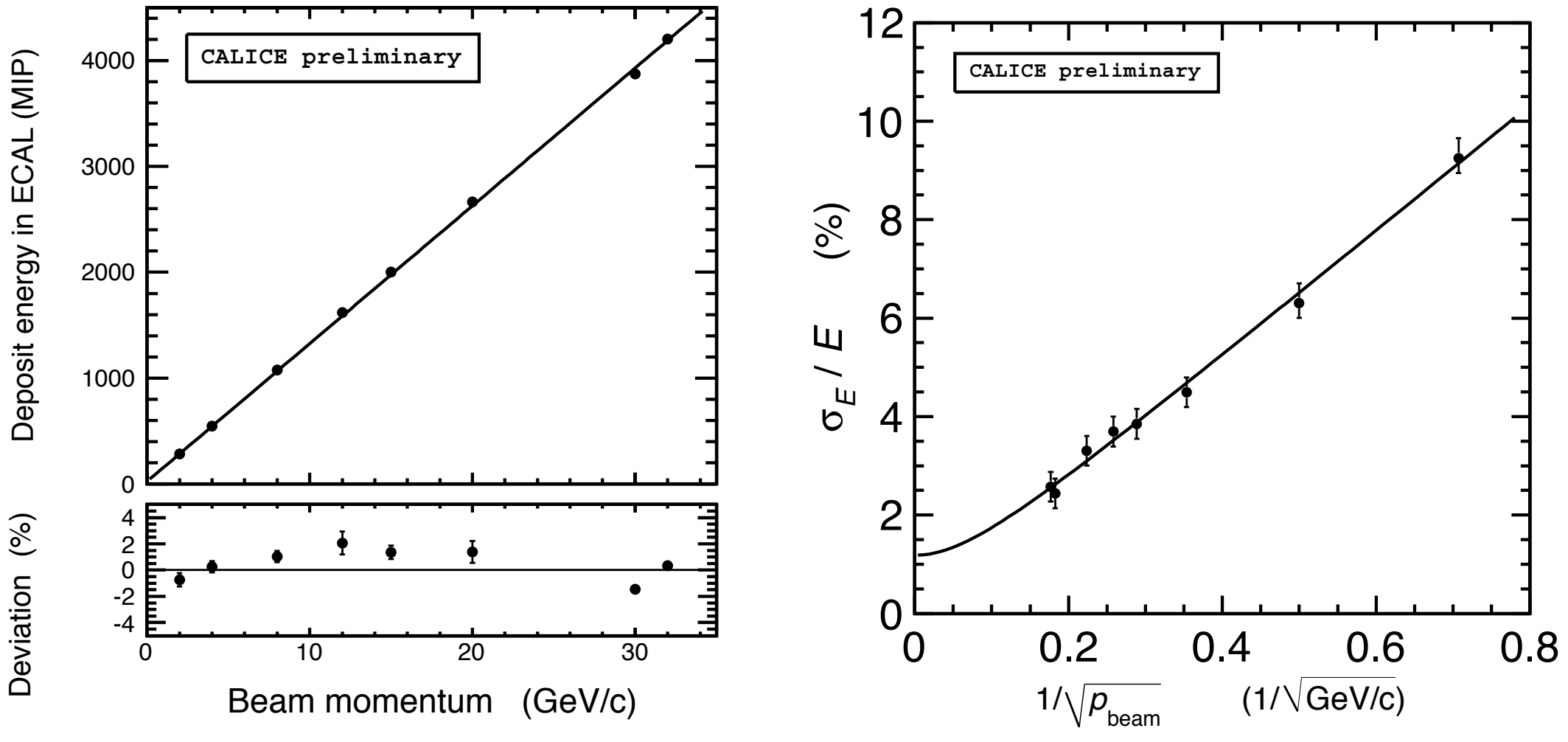

**Figure III-3.10.** Left: Response curve for the physics prototype of the ScECAL prototype vs. beam energy, with the deviation from a linear behaviour shown in the bottom part of the plot. The linearity is better than 2%. Right: Measured energy resolution of the physics prototype of the ScECAL prototype after temperature correction, for electrons in the energy range between 2 and 32 GeV. [310]

**Figure III-3.11** Left: Correlation between calibration constants obtained on a cosmics test bench and in beam test. Right: Comparison between calibration constants obtained in two different data taking periods in 2006. Results are taken from [304].

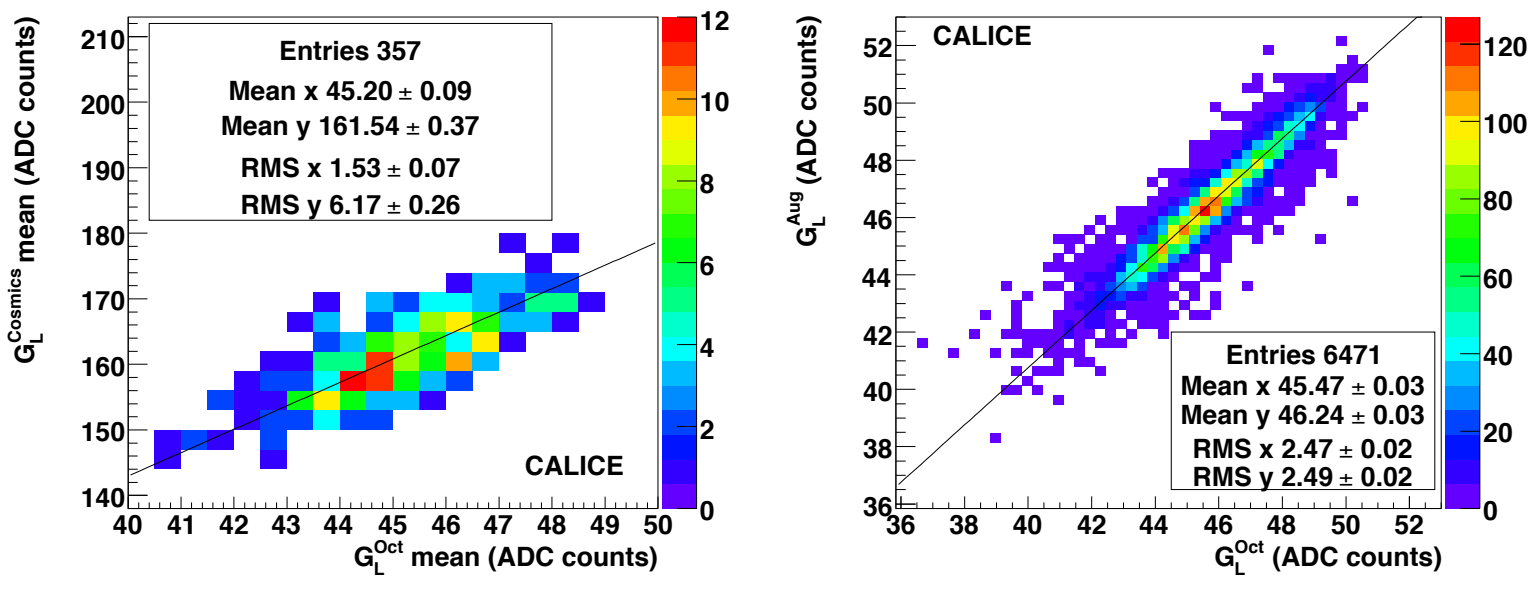

## 3.2.4 Calibration and alignment

As shown above the energy resolution of the electromagnetic calorimeters is of order $15\%/\sqrt{E}$. Therefore a calibration procedure at the percent level seems to be sufficient. In the following the experience from beam test campaigns and the resulting projection to a full calorimeter system will be outlined for the two options of the electromagnetic calorimeter.

### 3.2.4.1 SiECAL calibration

The calibration factors were stable over long time periods to the % level, where the variations are mainly due to different experimental conditions (e.g. cable length) at the beam test sites. The calibration constants showed no influence from external factors like temperature. The correlation of calibration constants obtained for different periods of data taking 2006 are published in [304] (see Figure III-3.11). In [311, 312] it is shown that the correlation coefficient is 83.8% between the calibration constants obtained at FNAL in 2008 and at CERN in 2006. Considering that many operations like mounting, un-mounting, and shipping occurred between 2006 and 2008, this high correlation coefficient demonstrates the stability with time of the SiECAL prototype. The same level of correlation exists between calibration constants derived for the beam tests in 2008 and 2011 at FNAL.

The test beam experience gives confidence that the calibration can be well controlled for a full SiECAL. In this case all detector modules will have to be scanned by a muon beam in a test beam experiment prior to the installation. This operation would take several months but it should be possible to complete this in time for the start of detector operation. Afterwards it should be possible to monitor the calibration constants with muons. An alternative are charged pions which pass through the detector as a MIP or track segments of secondaries after a hadronic interaction. For the latter no study exists for the electromagnetic calorimeter so far but the studies performed for the analogue hadronic calorimeter give confidence that the monitoring of the calibration constants is feasible using this method.

The detailed alignment procedure has not been worked out. However, given the position resolution of about 1 mm, see section 3.2.3, the alignment of the detector has to be precise to about 100 $\mu$m. This value should be easily achievable with electron and muon tracks measured in the TPC and other tracking detectors, provided that these are correctly aligned.

#### 3.2.4.2 ScECAL calibration

There are more than 10 million channels of small scintillator strip units in this calorimeter option. The stability of the light output has to be controlled and monitored. Three calibration schemes are under investigation.

In the first system light from an LED is guided with a clear optical fibre to the strips and coupled into the strip through notches in the fibre. This system will be used to monitor the stability of the system. Experience from test beams show that this is possible to within a few %. The other systems use particles from the beam halo or pions within a jet that traverse the scintillators. These particles behave as minimum ionising particles for the scintillator strips. The arrival of these particles is synchronous with the beam and thus the calibration schemes can be applied in power pulsing mode. For this dedicated track finding methods have been developed. Simulation studies show that with the proposed segmentation 50 hits/cell/day will be recorded using muons from a dedicated run of the accelerator at the $Z$-pole. Therefore a couple of days of running at the $Z$ pole will be sufficient for calibration. The in situ calibration can also be done using beam-halo muons. They are distributed isotropically over the tunnel diameter. Their density is estimated to be about 4.1 muons/$\mathrm{cm}^2$/s [313] without a muon spoiler. The energies are high enough to pass through the detector. Hence they can be used for the MIP calibration of the calorimeter endcap. With the current accelerator parameters and tunnel design about 500 seconds are needed to collect enough halo muons in each cell. However it should be noted that this procedure depends critically on the beam line design. For example, if muon spoilers are introduced to suppress the halo-muon flux the rate can easily go down by 2 orders of magnitude.

### 3.2.5 Future directions

CALICE has completed a series of full-size proof-of-principle tests with physics prototypes of both silicon and scintillator ECAL technologies. Large data sets have been collected and demonstrate the performance at ILD. The emphasis in the more realistic second generation technological prototypes is shifted towards a demonstration of the feasibility of a compact integrated detector design fulfilling the ambitious demands on compactness and hermeticity. Operational challenges not yet addressed with prototypes are the power-pulsed front-end electronics and the on-detector zero suppression in auto-triggered mode, which requires continuous and precise on-line controls of thresholds. Beam campaigns with technological prototypes started in 2012, with the focus rather on calibration and stability than on shower physics, and will continue for a couple of years. The campaigns, together with the design studies given in this document, will make it possible to construct an EM calorimeter system for a real ILC detector.

Until actual construction of the detector many aspects of the system will be continued to see improvements as a result of further R&D. Some of the issues to be addressed are:

- silicon technology: sensor guard rings, AC coupling, chip bonding, PCB thickness;
- scintillator technology: developments of MPPC with more pixels and photon readout system;
- further hybrid simulation study;
- development of mass production and mass test system of sensor;
- alternative sensor technologies: e.g. MAPS;
- further studies of power pulsing;
- possible reduced scope (for cost reasons): reduced layers, radius. Estimate of cost scaling.

## 3.3 The Hadronic calorimeter system

The role of the HCAL is to separate the deposits of charged and neutral hadrons and to precisely measure the energy of the neutrals. Their contribution to the jet energy, around 10% on average, fluctuates over a wide range from event to event, and the accuracy of the measurement is the dominant contribution to the particle flow resolution for jet energies up to about 100 GeV. For higher energies, the performance is dominated by confusion, and both topological pattern recognition and energy information are important for correct track cluster assignment.

The HCAL is conceived as a sampling calorimeter with steel absorber and scintillator tiles (analogue HCAL) or gaseous devices (semi-digital HCAL) as active medium. Due to the rigidity of stainless steel, a self-supporting structure without auxiliary supports (dead regions) can be realised. Moreover, in contrast to heavier materials, iron with its moderate ratio of hadronic interaction length ($\lambda_I = 17$ cm) to electromagnetic radiation length ($X_0 = 1.8$ cm) allows a fine longitudinal sampling in terms of $X_0$ with a reasonable number of layers in a given total hadronic absorption length, thus keeping the detector volume and readout channel count at an acceptable level. This fine sampling is beneficial both for the measurement of the sizeable electromagnetic energy part in hadronic showers and for the topological resolution of shower substructure, needed for particle separation and weighting. Two baseline technology options have been developed, the scintillator-tile based AHCAL and the Glass Resistive Plate Chamber (GRPC) based SDHCAL.

With the advent of novel, multi-pixel Geiger mode silicon photo-diodes, so-called SiPMs, high granularities as required for a particle flow detector can be realised with the well-established and robust scintillator technology at reasonable cost. The scintillator tiles provide both energy and position measurement and thus allow to optimise amplitude and spatial resolution together. They exhibit a very homogenous response and with 3 mm thickness allow for a compact design with high MIP efficiency for tracking inside showers and calibration purposes. The transverse segmentation suggested by simulations is about $3 \times 3$ cm$^2$ and leads to a number of read-out channels an order of magnitude smaller than in the digital case with $1 \times 1$ cm$^2$ cells. The CALICE AHCAL [314] was the first device that used the novel SiPM technology on a large scale, and its robustness and reliability has encouraged other experiments, e.g. T2K, CMS and Belle, to apply it in their detector upgrades.

Gaseous detectors are good candidates for the active layers of a sampling calorimeter of high granularity. In addition to their excellent efficiency, gaseous detectors provide very good homogeneity. Another important advantage of the gaseous detectors is the possibility to have very fine segmentation. Indeed the segmentation is to a large extent driven by the electronics readout granularity used to read them. The thickness of gaseous detectors is also of importance for an ILD hadronic calorimeter to be placed inside the magnetic field. Highly efficient gaseous detectors can be built with a thickness of less than 3 mm. Other gaseous detectors such as micromegas and GEMs could also be alternatives to GRPC once the technology of producing large areas of such detectors cost effectively becomes available.

For the barrel calorimeter, two different absorber geometries are being proposed. The first version is separated longitudinally into 2 rings and azimuthally into 16 modules. The signal readout is guided along the z axis towards the barrel/ end-cap gap. Alternatively, the 2nd version is segmented into 5 rings in z and 8 modules in azimuth. The signals are guided towards the outer perimeter in a similar way as in the ECAL. The main advantage of the first is the accessibility of the module level electronics and connections for maintenance and repair. On the other hand, the second provides superior rigidity and less deformation in the transverse plane. In principle, both geometries can be combined with all proposed technologies. However, the detailed engineering is presently being worked out for scintillator in the first, and for gaseous readout in the second approach.

## 3.3.1 Detector optimisation

### 3.3.1.1 AHCAL design optimisation

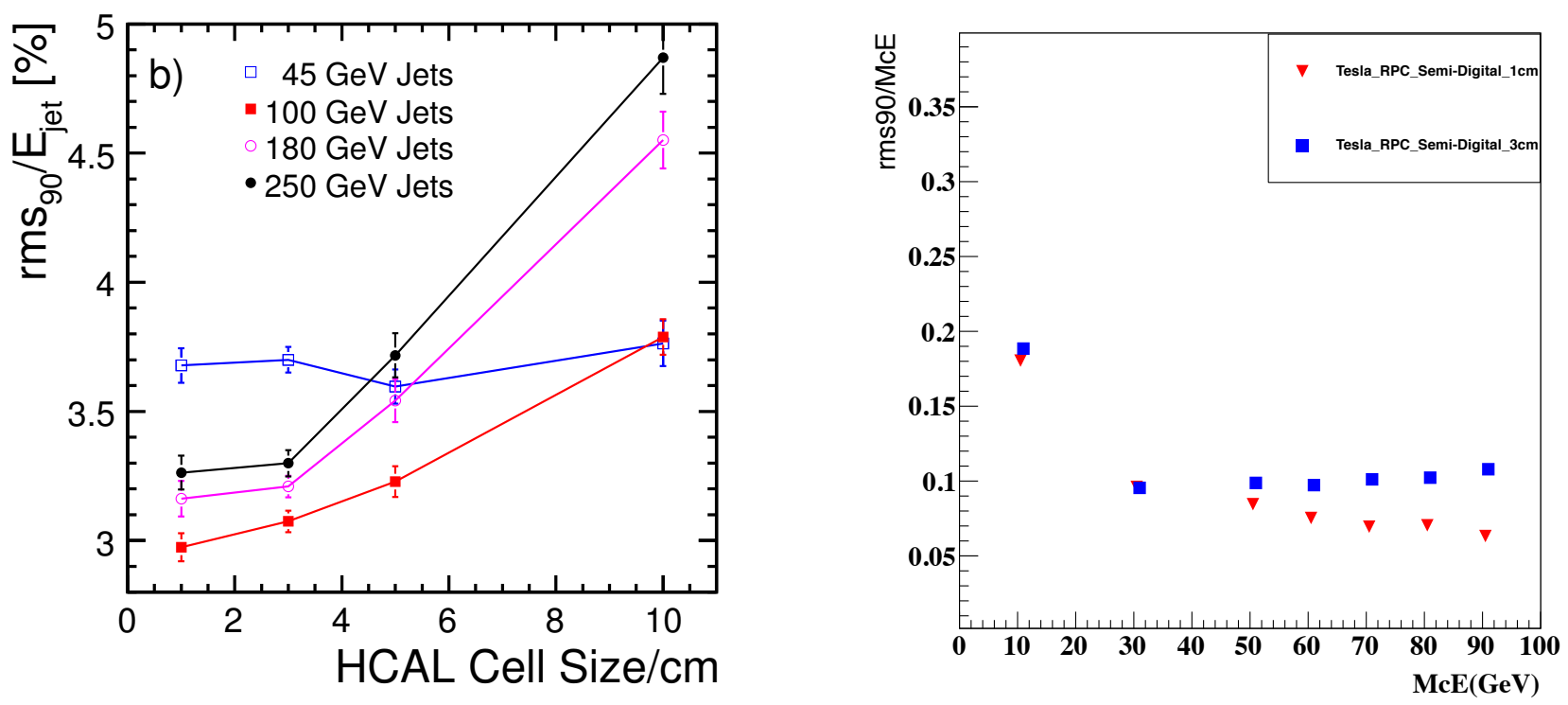

**Figure III-3.12** Optimization of the hadron calorimeter cell sizes. Left: Particle flow jet energy resolution as a function of the AHCAL cell size. Right: Single $K_0$ energy resolution for particles showering in the SDHCAL for two different cell sizes.

The parameters of the AHCAL have been optimised using full detector simulations with particle flow reconstruction, comparing the performance for different design parameters. Of particular relevance are the thickness of the calorimeter and the cell size. The former strongly influences the energy resolution at higher jet energies due to potential leakage out of the back of the detector, while also driving the size of the solenoid, and the latter is crucial for the two-particle separation but also affects the overall system cost and complexity due to the impact on the channel count.

Figure III-3.12 *left* shows the particle flow performance as a function of the lateral segmentation of the AHCAL readout layers. It is apparent that going below a size of $3 \times 3$ cm$^2$ does not provide substantial advantages, while larger cells lead to reduced performance, resulting in the choice of $3 \times 3$ cm$^2$ for the size of the AHCAL scintillator tiles. With the same studies, the depth of the calorimeter was optimized. In order to not reduce the performance at 1 TeV, where typical jet energies are up to 250 GeV, a depth of 48 layers, corresponding to 6 $\lambda_I$ was chosen.

### 3.3.1.2 SDHCAL design optimisation

The fine granularity of the hadronic calorimeter is an important asset to provide an excellent tracking capability needed for PFA but this is not the only element in favor of high granularity in the case of the SDHCAL. The energy measurement performance of the SDHCAL depends essentially on its capability to account for the particles produced within the hadronic shower. It is then necessary to find the best cell size which allows one to account for the many tracks produced in the hadronic shower. The first optimization studies indicated that a few mm cell size is the one which leads to the best energy resolution using a simple binary readout. However, this leads to a huge number of electronics channels (more than 200 million) making the technical realization of such a detector extremely complicated. In order to reduce this number without deteriorating the physics performance a compromise was found. It consists of choosing larger cell size while going from a simple binary to a three-threshold electronics (2 bit) readout. The role of the different thresholds is to help separating among one, few and many particles crossing the same cell. A detailed study using a full ILD detector model with 48 active layers and 6 $\lambda_I$ shows that a 1 cm size cell achieves better precision than a one of 3 cm size, as shown in figure III-3.12 right. This is the option that was selected for the SDHCAL base line.

### 3.3.2 Detector implementation

The first version of the mechanical design for the calorimeter barrel is based on two rings with 16 modules each. One module has a weight of almost 20 tons, which is manageable with standard installation techniques. The modules are constructed independently of the active layers, which can be inserted before or after installation of the modules. There are 48 absorber plates, 16 mm thick each, held together by 5 mm thick side panels in the $rz$ planes; no additional spacers are foreseen. The active layers will contribute 4 mm of steel to each absorption layer, and require 5.5 $\mathrm{mm}$ for instrumentation (3 mm thick scintillator plus readout and calibration devices). The structure has been extensively simulated using finite element methods, including the integration of the heavy ECAL structure. Maximum deformations are found to be less than 3 mm, if the barrel structure is supported by two rails in the cryostat.

Presently the boundaries between modules are pointing in $\varphi$ and in $z$. Variants with non-pointing boundaries have been validated in finite element calculations as well, but are disfavoured to ease the mechanical construction. The pointing geometry does not degrade the performance as long as the cracks between modules are filled with absorber material, and if the active instrumentation extends up to the boundary within tolerances, which is the case in the present scintillator layer design.

This mechanical concept has been fully developed, with horizontal and vertical prototypes successfully assembled and tested (see Figure III-3.13.) The measurements done on the horizontal prototype demonstrate that the required tolerances (flatness better than 1 mm over the full area of about 2 m x 1 m) and mechanical stability can be achieved with realistic stainless steel structures using roller-levelled plates. This avoids a cost-intensive machining of the delivered rolled steel sheets. In addition a mechanical design for the end caps has been completed and is also shown in Figure III-3.13.

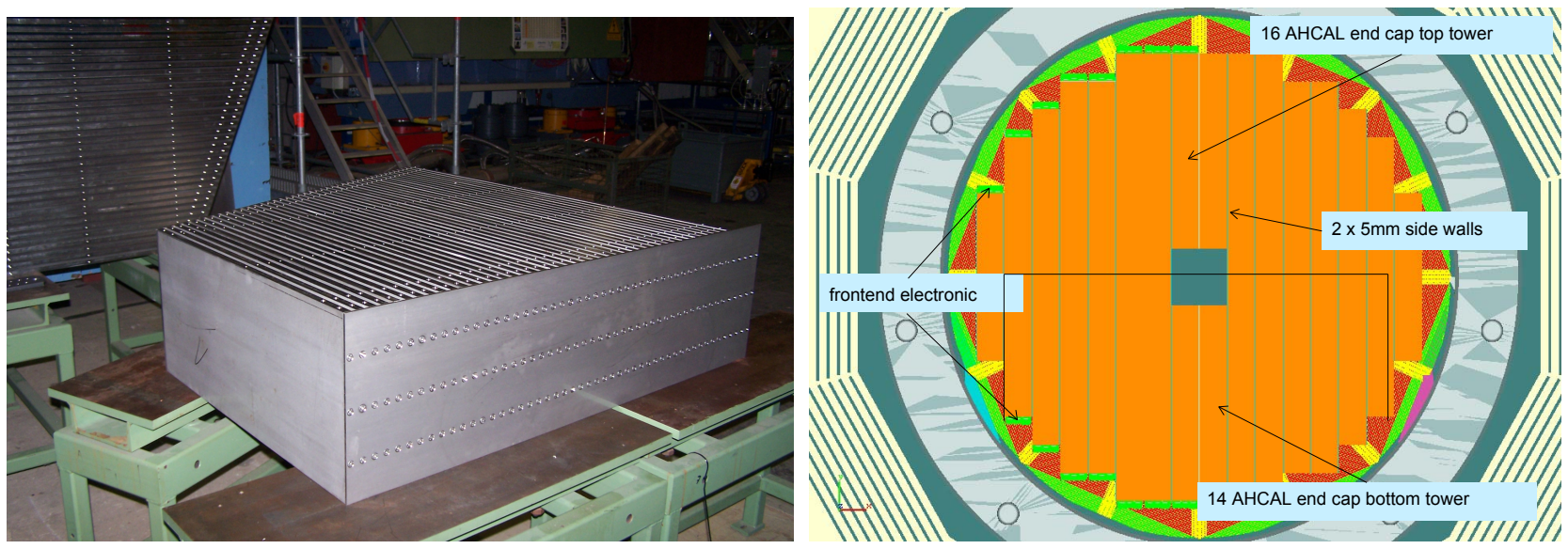

**Figure III-3.13** Left: the mechanical prototype for the first version of the barrel structure. Right: mechanical design of the HCAL endcap.

The second mechanical structure is a self-supporting mechanical structure called the V structure. The structure has been designed to eliminate the projective holes and cracks so none of the particles produced close to the detector centre could escape detection. The V structure has additional advantages. It eliminates in principle the space between the barrel and the Endcaps avoiding the shower deformation which results not only because of this space but also of the different cables and services needed in CMS-like mechanical structures. In this structure the different services such as the gas tubes, data collection and electric cables of both the barrel and the Endcaps are taken out from the outer radius side. Detailed studies have shown that the deformation of this structure is extremely low and its robustness was verified experimentally with the SDHCAL technological prototype built with a self-supporting structure following the design of the V structure.

### 3.3.2.1 AHCAL readout technology and implementation

The arrangement of the active layers with internal and external electronics components is sketched in Figure III-3.14. The layer consists, from bottom to top, of a 0.5 mm thick steel support plate covered with reflector foil, the scintillator tiles (3 mm), the printed circuit board with electronics components (2 mm), covered with reflector foil from underneath, and a polymide foil for insulation. The PCB carries the SPIROC readout ASICs, introduced in section 3.1.1, and auxiliary components as well as a LED based optical calibration system. Interfaces for data acquisition, clock and control, for power distribution and for calibration system steering are accessible at the end face. Since the ASICs are operated in power-pulsed mode, no cooling is needed inside the detector volume.

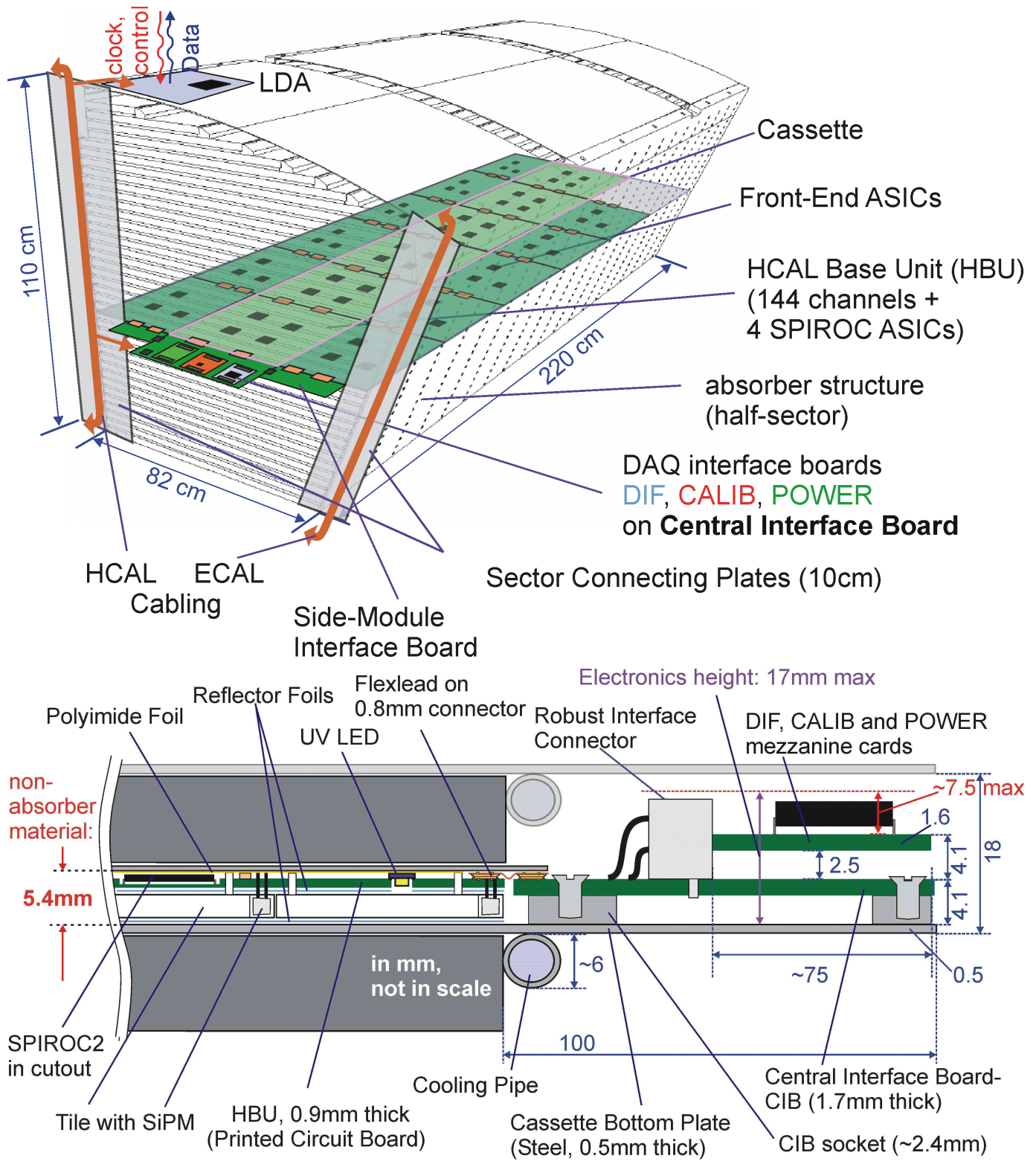

**Figure III-3.14** Arrangement of AHCAL layers with electronic components (left), cross section of an active layer (right).

The PCB is subdivided into units (HCAL base units, HBUs) of smaller size, manageable for automated mounting and soldering techniques. The standard unit is 12 by 12 tiles, $36 \times 36$ cm$^2$ , so six units are aligned along $z$ to fill a half barrel. In order to accommodate the variation in layer width with increasing radius, 4 different HBUs, 8 to 12 tiles wide, are needed. At the layer edges, tiles with smaller size, e.g. $2 \times 3$ cm$^2$, are placed such that the width of the uninstrumented region near the sector boundary is on average 2.5 mm, but never larger than 5 mm. The electronics at the end face will require cooling, mainly due to the use of FPGAs in the DIF (Detector InterFace board introduced in section 3.1.1). The boards will extend 5 to 10 cm in $z$, but occupy only a fraction of the full width in $\varphi$, thus leaving space for ECAL and main tracker services as well as for the TPC support along radial directions.

Figure III-3.15 shows the details of the scintillator tiles, with a thickness of 3 mm and embedded wavelength shifting fiber which couples the light to an embedded SiPM. This new design is based on the experience with the physics prototype, and has been adapted for easier manufacturing. In extensive laboratory and beam tests, the tiles together with the SiPMs have been proven to deliver the expected performance in terms of signal yield and uniformity.

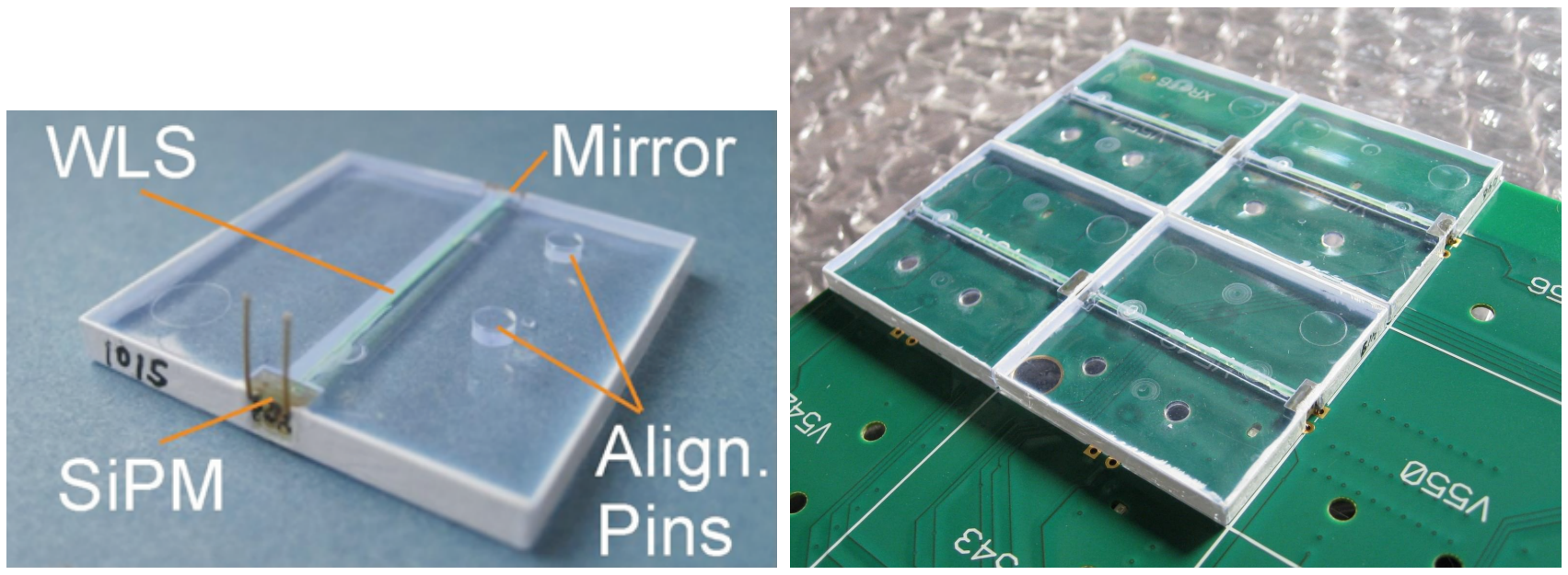

**Figure III-3.15**
The AHCAL scintillator tiles with embedded SiPMs, mounted on the readout PCB.

### 3.3.2.2 SDHCAL readout technology and implementation

The basic unit of the SDHCAL is a cassette that contains the active layer. The cassettes whose walls are made of 2.5 mm thick stainless steel are inserted into the mechanical structure. The structure itself is made of 1.5 cm thick plates of the same material. The cassette walls together with the structure plates play the role of the absorber. In total 2 cm of stainless steel is separating two active layers. The active layer itself is composed of a GRPC detector and its embedded readout electronics. The former is made of two glass plates. The anode plate has a thickness of 0.7 mm, the cathode plate of 1.1 mm. The two plates are separated by 1.2 mm space which is maintained constant by a special spacers (see Figure III-3.16). The distance and the size of these spacers were optimized to eliminate dead zones in the detector while providing an uniform electric field between the two plates. The two glass plates are covered on their outer side by a conductive painting. A high voltage is applied to these layers to create an electric field between the plates. The gap between the two plates is filled with a gas mixture of TFE(93%), CO2 (5%) , SF6(2%). The first gas provides the primary electrons when ionized by a charged particle (8 electrons/mm) while the second and the third are photon and electron quencher respectively. Their role is to limit the size of the avalanche that follows the creation of primary electrons. Gas tightness is provided by a frame made of robust insulating material. The frame is 3 mm wide resulting in a the dead area of less than 1.3%. A gas distribution system was developed. It allows to renew the gas content of the chamber in an efficient way taking into consideration the fact that gas inlets and outlets are to be on one side of the chamber. The system is designed to reduce the gas consumption. This and the recycling progress achieved by the RPC-gas group at CERN are important elements to reduce the cost of the gas consumption.

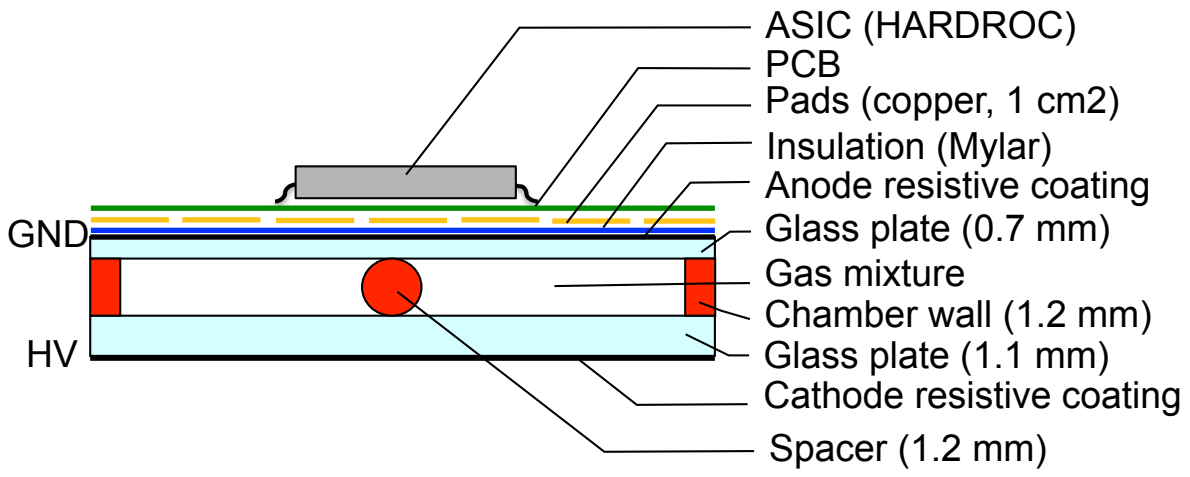

**Figure III-3.16**
Cross sectional drawing of the GRPC used in the SDHCAL.

The avalanche signal created between the two glas plates is read out inductivly by a pad plane. A very thin (0.8 mm), 8-layer printed circuit board (PCB) has been designed. One side of the PCB hosts the readout ASICs called HARDROC. The other side carries the signal pick-up pads with a pad area of $1$ cm$^2$. The PCB is designed to connect the ASICs to each other (DAISY chain). The PCB size chosen for the technological prototype was $33 \times 50$ cm$^2$. To read out large GRPC the PCB were conceived to be connected to each other using tiny connectors which are capable to transmit the signal as well as the different electric powers from one PCB to another. For the technological prototype boards of 1 m$^2$ were constructed by connecting four PCBs to form so-called slabs (see Figure III-3.17). Each slab is then connected to the data acquisition through a detector interface board (DIF) which hosts an FPGA responsible for the communication with the 48 ASICs of one slab. Three slabs were soldered together in an appropriate way to ensure the same grounding for the three of them and to have a flat surface on the pads side. The boards are then fixed on the top cover of the cassette acquiring in this way a better rigidity. This is then fixed on the cassette that contains the GRPC. The total active layer thickness is less than 6 mm. The total thickness of one cassette is less than 11 mm.

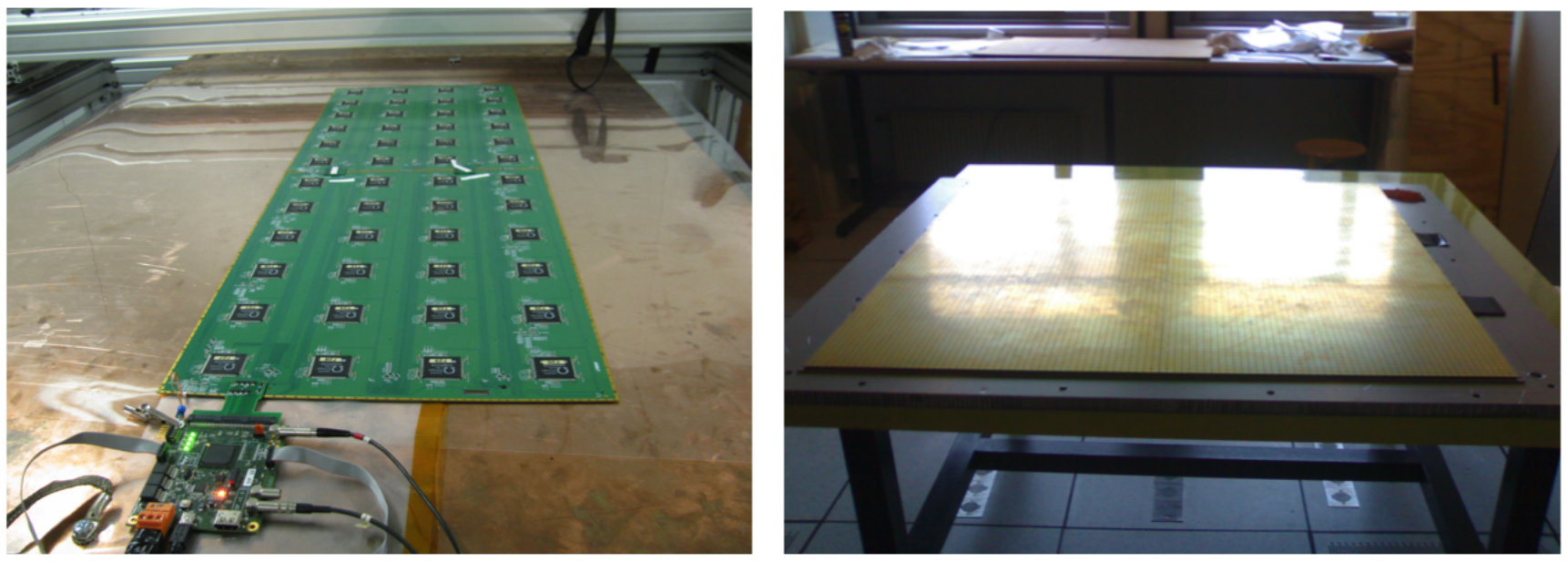

**Figure III-3.17** An electronic slab made of two boards hosting each 24 ASIC and connected with a tiny connector. The interface DAQ board(DIF) is also shown (left) and the final 1 $m^2$ board.

## 3.3.3 Test beam results

### 3.3.3.1 AHCAL test beam results and operational experience

From the extensive CALICE test beam program, in which the AHCAL physics prototype [314] modules were used from 2006 until 2011, a wealth of results on detector performance, simulation validation and operational experience are available. The long-term operation of the AHCAL physics prototype, together with a large number of assembly and disassembly procedures, often coupled with long-distance shipping of the detector, has provided substantial information of the stability and reliability of the AHCAL technology. The number of observed non-working channels is very moderate at roughly 2%, most of which are due to broken solder points at the connection of the SiPMs to the PCB leading to the front-end electronics that were caused by deformations of the board during detector movements [314].

The linearity and the energy resolution – a key performance parameter even for a particle flow detector — of the AHCAL have been studied using pion beam at different energies [315]. In Figure III-3.18(left) the reconstructed single particle energy is shown as a function of the beam energy. The deviations from a linear response are within $\pm 1\%$. The uncorrected energy resolution is shown in Figure III-3.18(right). In addition the energy resolution after applying a software compensation technique is shown. The AHCAL has an $\mathrm{e}/\pi$ ratio of approximately 1.2. The compensation algorithm makes use of the fact that electromagnetic sub-showers have different spatial characteristics compared to purely hadronic energy deposits. Using the high granularity of the calorimeter fluctuations between the electromagnetic and hadronic component of the shower can be corrected on an event by event basis. This improves the resolution by close to 20%, reaching a stochastic term of $45\%$.

In the analysis, events from different data taking periods with operating temperatures ranging between 15 and 25 $^o$C were combined. The overall good performance is demonstrate that a reliable temperature corrections can be applied. It shows that the temperature sensitivity of the photon sensor does not limit intrinsically the performance.

Electromagnetic showers are used to validate the simulation of the detector as well as to assess possible intrinsic performance limits. Both the linearity of the response and the energy resolution for electrons are very well reproduced by simulations once saturation effects of the photon sensor are taken into account in the event reconstruction [316]. Since the simulations do not include a modeling of response non-uniformities and gaps between tiles, this good agreement demonstrates that the non-uniformities present in the detector do not affect the electromagnetic performance, and thus are irrelevant for the performance for hadrons.

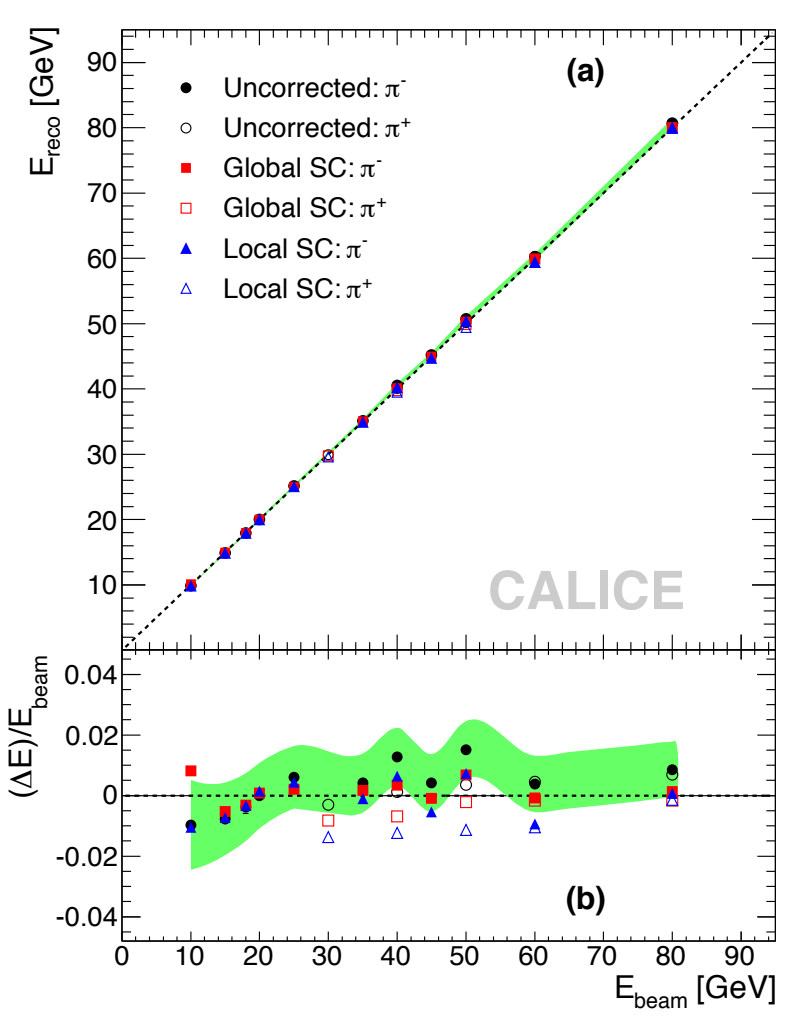

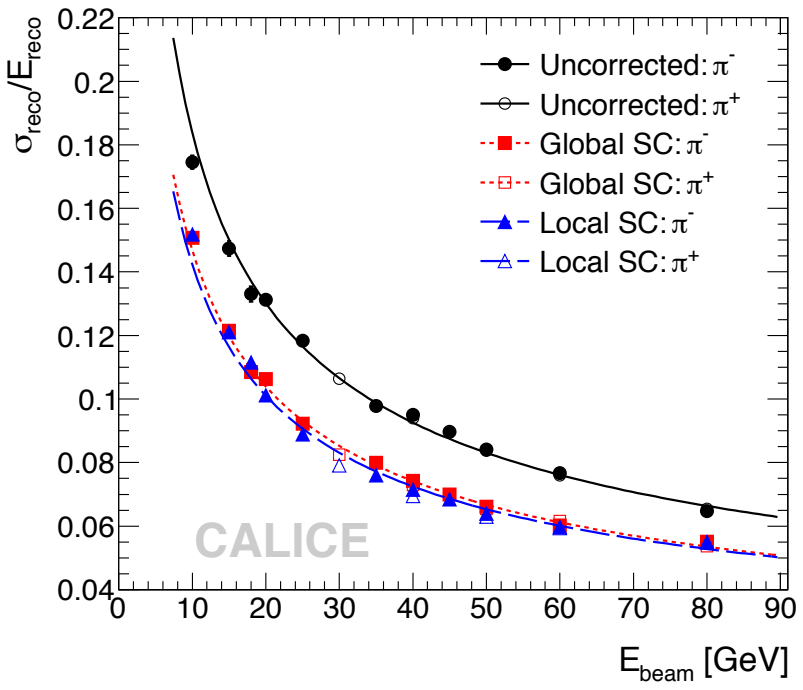

| fit results | | |
|---|---|---|
| | stochastic | constant |
| initial | 57.6% | 1.6% |
| global SC | 45.8% | 1.6% |
| local SC | 44.3% | 1.8% |

**Figure III-3.18** Reconstructed energy (left) and energy resolution (right) of the AHCAL for pion showers starting in the first five calorimeter layers. Shown are results obtained with a simple energy sum and with a local and a global software compensation (SC) technique, respectively. The green band indicates the systematic error of the calibration, and is shown around the results with with initial energy reconstruction. Figure taken from [315].

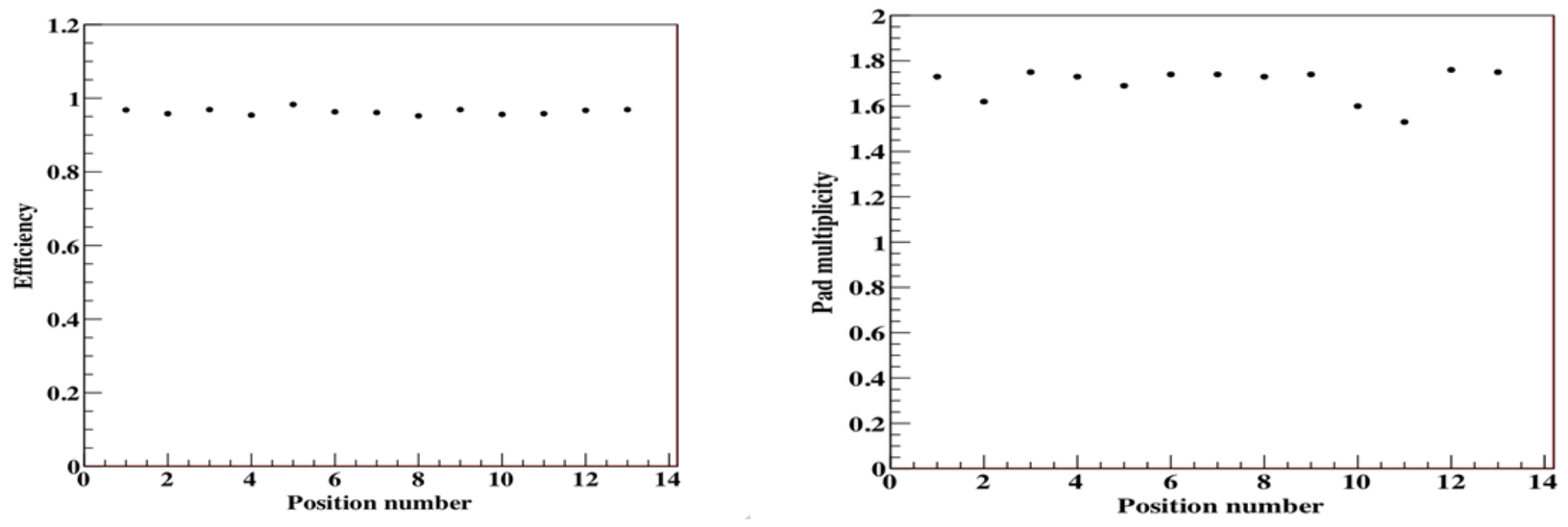

**Figure III-3.19** Efficiency (left) and multiplicity (right) measured for different points of the SDHCAL prototype cassette. The measurement points include the critical area where readout boards join, and where potentially areas of lower efficiency are introduced.

Beyond this evaluation of the performance of an imaging analogue scintillator HCAL, the high granularity of the detector has also been used for detailed investigations of the substructure of hadronic showers to study the realism of various Geant4 shower models. These studies include the measurement of shower profiles [317] and of secondary high-energy particle production within hadronic showers, accessible via minimum-ionizing tracks identified within the showers [318]. While older Geant4 physics lists often disagree with data, state-of-the-art physics lists are in general able to provide a good description of the measurements. Overall, these results give additional confidence in the realism of the AHCAL simulation in full detector performance predictions for ILD.

#### 3.3.3.2 SDHCAL test beam results and operational experience

A technological prototype for the SDHCAL was built. The mechanical structure of this prototype is constructed using 1.5 cm thick stainless steel plates. The flatness of the plates was measured using a laser-based interferometer system and was found to be better than 500 $\mu$m. This result guarantees that for the V structure proposed for the SDHCAL, a tolerance of less than 1 mm is achievable. This mechanical structure can host up to 50 cassettes described above.

The first cassettes were extensively tested using a cosmic-ray test bench and particle beam at CERN. Both the efficiency and the multiplicity of the GRPC cassettes were studied. These studies (see Figure III-3.19) showed high efficiency and good homogeneity and validated the cassette concept.

In addition a single cassette was tested in a magnetic field of 3 Tesla (H2 line at CERN) applying the power-pulsed mode [319]. The results indicated clearly that the use of the power-pulsed mode in such a magnetic field is possible. The behavior of the detector (efficiency, multiplicity etc.) was found to be similar to those obtained in the absence of both the magnetic field and the power-pulsed mode (see Figure III-3.20).

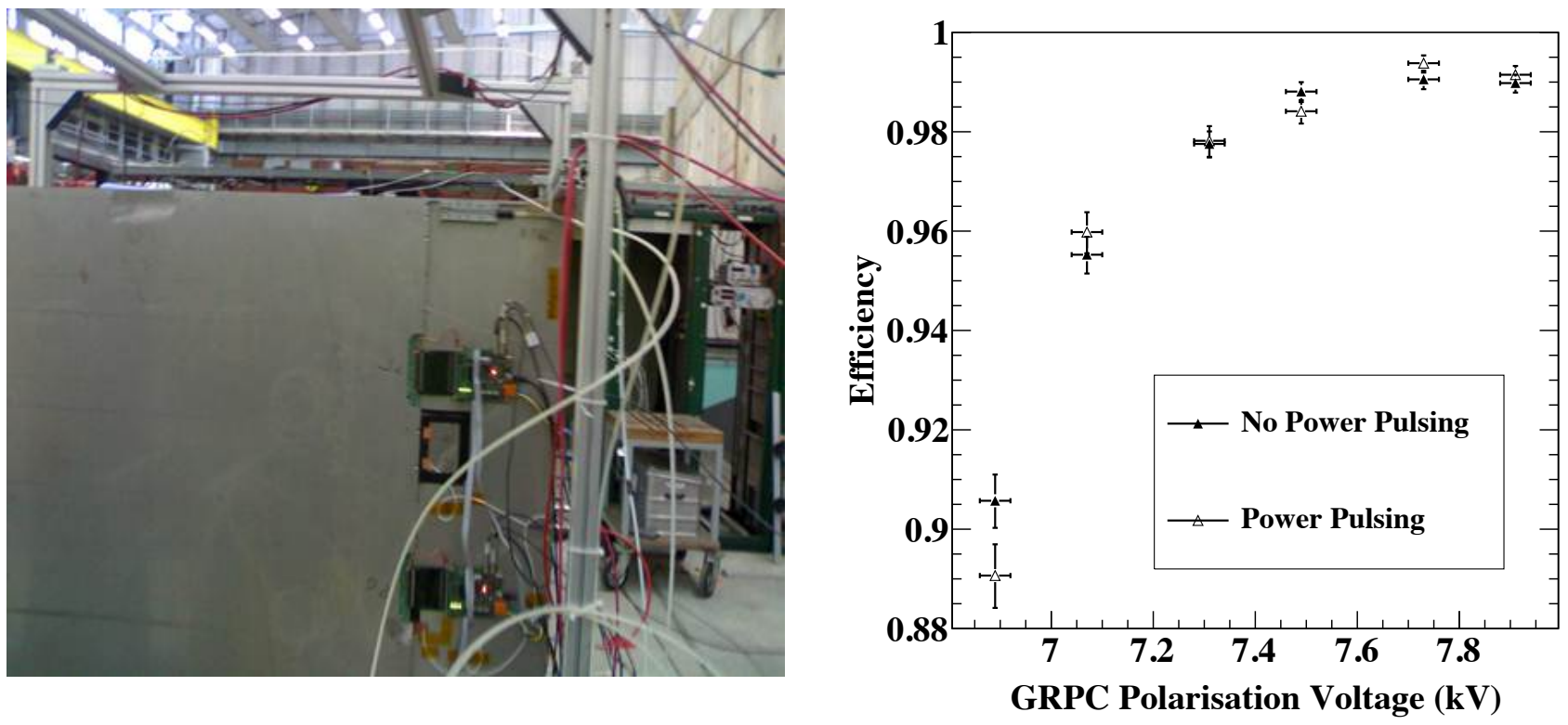

**Figure III-3.20** (left) Photograph of a prototype cassette of the SDHCAL. (right) Efficiency of the SDHCAL module measured with the power-pulsing in a 3-Tesla magnetic field at CERN.

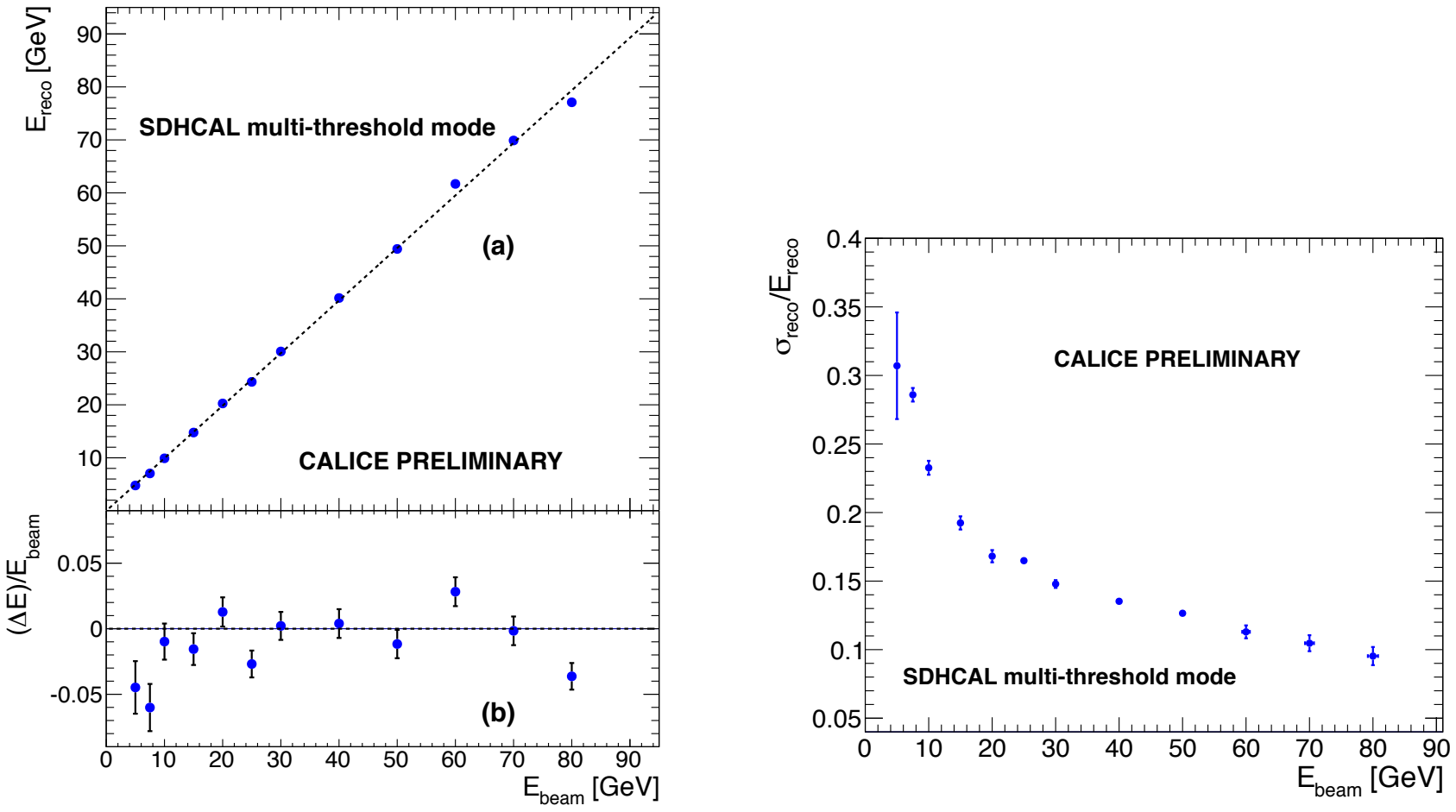

**Figure III-3.21** Reconstructed energy linearity (left) and energy resolution (right) for Pions using a weighted sum of the three-threshold number of hits and with no data correction, for the SDHCAL prototype [320].

The prototype construction lasted less than 6 months. A commissioning test at CERN in 2011 allowed to understand the behavior of the complete system.

In April 2012 the prototype was exposed to pion, muon and electron beam at both the PS and the SPS at CERN. The power-pulsed mode was applied to all electronic channels of the prototype using the beam time structure (0.3 ms on-time duration for the PS beam and 9 s for the SPS beam every 45 s). A basic water-based cooling system was used to control the temperature particularly in the case of the SPS where the power consumption reduction is only 5 (to be compared with a factor of more than 100 in the ILC case). Data were collected continuously in a triggerless mode. The DAQ stops when the memory of one ASIC is full. Data are then transferred to a storage station and then the acquisition starts again.

Preliminary results [320] obtained from this short test beam confirm the excellent results of the binary-readout DHCAL physics prototype which uses the same active medium (GRPC) (DHCAL). The SDHCAL prototype results obtained with a minimum data treatment (no corrections) show clearly that excellent linearity and good resolution could be achieved on large energy scale as can be seen in Figure III-3.21. In future analyses the data from the tests will be used to study thoroughly the hadronic showers topology and to improve the energy resolution by, among others, separating the electromagnetic and the hadronic contribution as was done in the case of the AHCAL option. The separation between close-by showers is expected to benefit from the high granularity on the one hand and from the very low noise of the detector ($< 1$ Hz/cm$^2$) on the other hand.

## 3.3.4 Technical validation

### 3.3.4.1 AHCAL technical validation

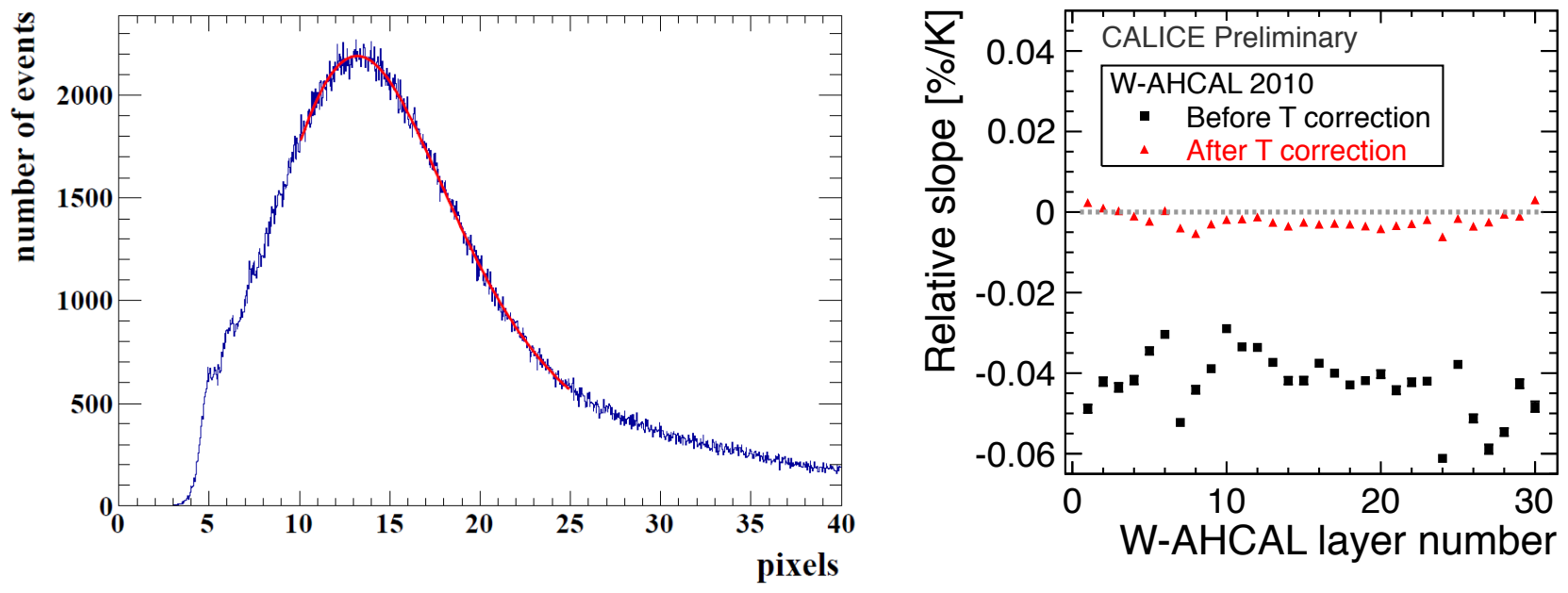

**Figure III-3.22** (Left) Response of a sensitive layer of the AHCAL technical prototype to 2 GeV electrons.(Right) Layer wise distribution of the relative response variation per degree change in temperature for minimum-ionizing particles with (red) and without (black) temperature correction [298].

To scale the technology of the analogue HCAL up to a full collider detector, particular care has been taken to minimize dead space and to maximize the depth of the calorimeter inside of the magnetic coil. The new generation of front-end electronics, based on the SPIROC2 ASIC, have been fully designed, with first boards successfully taking data in beam. These HCAL base units (HBU) each take 144 scintillator tiles, 3 mm thick with embedded wave length shifting (WLS) fiber and improved SiPMs. These SiPMs have considerably reduced noise rates compared to those installed in the physics prototype, resulting in a significant reduction of the noise occupancy. The relative impact of the thinner tiles (3 mm thick compared to the previously used 5 mm thick strip) on the energy resolution was simulated to be 2–7%. The scintillator tiles and electronics perform as expected, with the response to minimum ionizing particle shown in figure III-3.22 (left), giving a light yield of approximately 15 photo-electrons/ MIP. The electronics also provides the capabilities for self-triggering and precise time-stamping with a resolution of approximately 300 ps. The ASICs have channel by channel voltage control, and an LED calibration system is interfaced into the read-out boards .

In addition to the use of scintillator tiles with embedded WLS fibers, directly coupled scintillators are being considered for the AHCAL. Two designs have been established [321, 322], with the second one directly compatible with the current HBU design. For these scintillator tiles, promising first results with a molding procedure compatible with mass production have been achieved, demonstrating that large scale production of the required tiles is possible.

### 3.3.4.2 SDHCAL technical validation

The quality of data obtained during three weeks of data taking validates the SDHCAL concept as proposed in the LOI. This is especially encouraging since no gain correction was applied to the electronics channels to equalize their response. However a gain correction mode is elaborated and tested during the test beam. It will be applied in the future to assess the effect of such correction on the energy resolution.

Another important aspect is the full success of the power-pulsing mode applied to the more than 460000 channels of this prototype. The performance of the 48 cassettes during the whole test beam period remained stable and identical to that observed for single cassettes operated with a permanent powering. Power-pulsing is used to reduce the power consumption, which also significantly reduces the heat load and thus the temperature variations of the GRPC chambers. This largely simplifies the high voltage system.

## 3.3.5 Calibration and alignment

Key issues, such as the capability to fully calibrate the HCAL detector with minimum-ionizing particles and the ability to reliably correct the temperature dependence of the response of the photo-sensors for temperature variations far outside of the range expected in ILD have already been demonstrated with the physics prototype. The performance of the temperature correction is illustrated in figure III-3.22 (right) for minimum-ionizing particles recorded in the AHCAL [298]. The capability of the SiPM to detect single photons is used in addition to internally calibrate the gain of the system [323].

## 3.3.6 Future R&D

### 3.3.6.1 AHCAL future R&D

The large data sets taken with the AHCAL physics prototype hold the potential for further analysis, in particular in the area of detailed validation of GEANT4 hadronic shower models. A particularly interesting field has recently been opened here with the addition of data taken with tungsten absorbers.

The R&D plans for the analogue HCAL mainly go into the direction of fully demonstrating the concepts for a real detector, further improving the production and performance of components and exploiting the capabilities of the new electronics in test beams. The time scale of the R&D, in particular involving larger prototypes, will be driven by the available funding.

In the November 2012 test beam at CERN, one HCAL layer with 4 HBUs has been successfully tested with hadrons. The data will allow to further expand the investigations of the time structure of hadronic showers in steel and tungsten begun by earlier studies at CERN. A laboratory test of one full readout slab consisting of 6 HBUs is foreseen in the existing mechanical prototype with absorber layers of the same size as in the ILD HCAL barrel. Beyond 2012, the construction of a vertical stack with a minimum of 10 to 12 HBUs is planned. This stack will use the existing wedge-shaped mechanical prototype of a barrel module as absorber. This structure will be tested in electron beams at DESY in 2013. It is planned to be expanded to a full hadronic system for tests at CERN in 2014 or beyond.

On the basic technological front, new types of photo-sensors are being explored, in close cooperation with developers in research and industry. The goal is to push the limits in dynamic range, noise and device uniformity. The electronics and integration concept is versatile enough to accommodate advances on the sensor and tile side, integrate them into existing test structures and combine different types in the same beam tests. In this way, sensor technology and system integration can be optimised together.

### 3.3.6.2 SDHCAL future R&D

Large GRPC of 1 m$^2$ were developed and built for the technological prototype. However, larger GRPC are needed in the future DHCAL with the largest one being $290 \times 91$ cm$^2$. These large chambers with gas inlet and outlet on one side need a dedicated study to guarantee a uniform gas gap everywhere, independent of the mounting angle of the plate. It is also necessary to ensure an efficient gas distribution as it was done for the 1 m$^2$ chambers. The readout of such chambers needs to be as efficient as the one of the technological prototype. An upgrade of the readout ASIC is under way. The new ASIC will be directly addressable and can be easily bypassed in case of failure. Although no major difficulties are expected the R&D needed to validate the feasibility of the SDHCAL with the V mechanical structure will start soon. In addition it is needed to improve the interface boards (DIF) for the control of the ASICs synchronization and data transfer. Indeed, the space left between the active layer of one module and the cryostat is only 5 cm. This means that the DIF components should be optimized to cope with the volume availability.

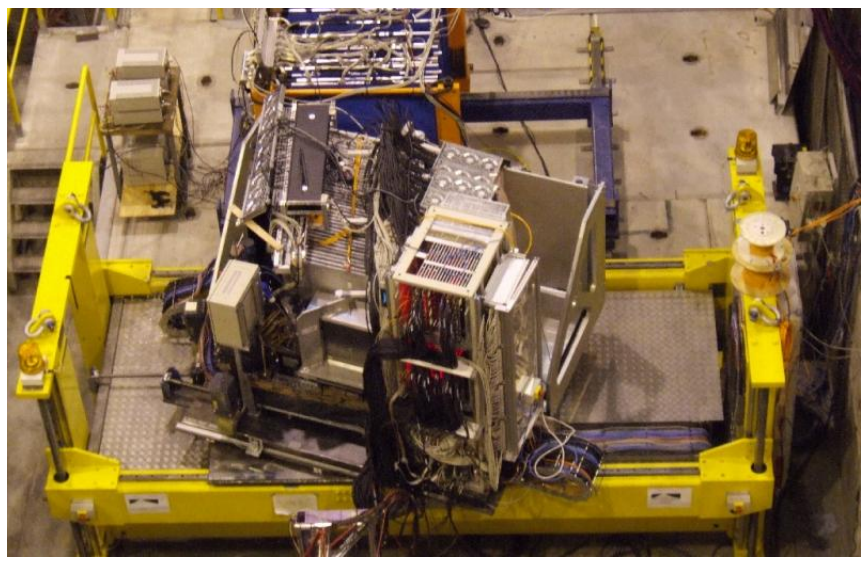

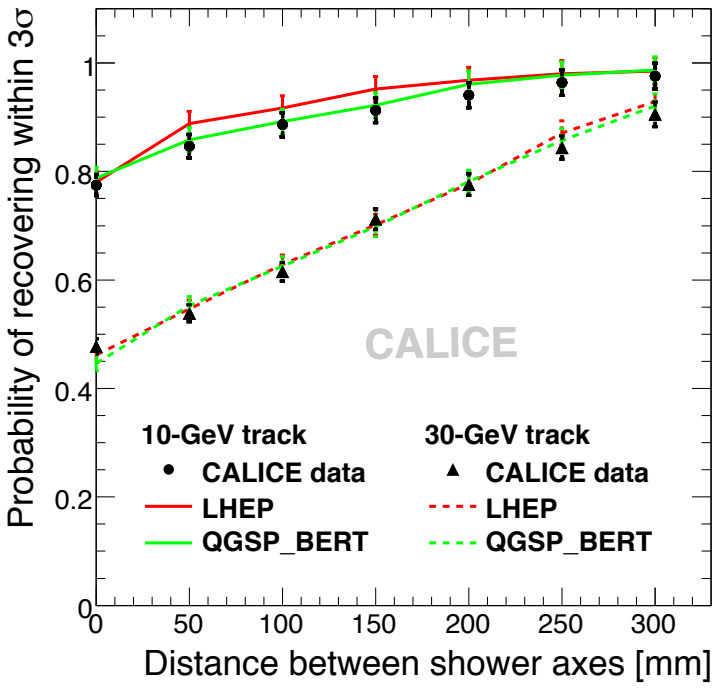

**Figure III-3.23**
Left: CALICE test beam set-up at CERN. Right: Probability to recover the energy of a 10 GeV neutral hadron within three sigma of the detector resolution as a function of the distance from a 10 GeV and 30 GeV charged hadron, respectively, using the Pandora PFA for test beam showers mapped into the ILD detector [302].

## 3.4 Particle flow performance of the ILD calorimeter system

Based on data taken with the physics prototypes of the SiECAL and the AHCAL the particle flow performance of the ILD calorimeter concept has been studied [302]. Two displaced showers measured in CALICE prototypes of an analogue hadron and an electromagnetic calorimeter were mapped into the ILD detector geometry and processed by the Pandora particle flow algorithm for event reconstruction. Figure III-3.23 (left) shows the setup used. The right part of this figure shows the probability to recover the energy of a 10 GeV neutral hadron within three sigma of the detector resolution as a function of the distance to a 10 GeV and 30 GeV charged pion, compared with simulations using different physics lists in GEANT4. The good agreement of data and simulations, in particular for the QGSP_BERT physics list, underlines the reliability of full detector simulations in predicting the particle flow performance of the detector system.

## 3.5 Forward calorimetry

Two special calorimeters are foreseen in the very forward regions of the detector [324], denoted hereafter as LumiCal and BeamCal. LumiCal will measure the luminosity with a precision of better than $10^{-3}$ at 500 GeV centre-of-mass energy[1], and BeamCal will perform a bunch-by-bunch estimate of the luminosity and, supplemented by a pair monitor, assist beam tuning when included in a fast feedback system [325]. Both calorimeters extend the detector coverage to low polar angles, important e.g. for new particle searches with missing energy signature [326]. The additional low angle hadron calorimeter LHCAL extends the coverage of the hadron calorimeter to the polar angle range of LumiCal. A sketch of the design is shown in Figure III-3.24.

LumiCal is positioned in a circular hole of the end-cap electromagnetic calorimeter ECAL. BeamCal is placed just in front of the final focus quadrupole. LumiCal covers polar angles between 31 and 77 mrad and BeamCal between 5 and 40 mrad.

Due to the high occupancy originating from beamstrahlung and two-photon processes, both calorimeters need a fast readout. In addition, the lower polar angle range of BeamCal is exposed to a large flux of low energy electrons, resulting in radiation depositions up to one MGy per year. Hence, radiation hard sensors are needed.

[1] At 1 TeV centre-of-mass energy this requirement is relaxed to $3 \times 10^{-3}$ due to the expected lower statistics of the relevant physics processes.

## 3.5.1 Mechanical concept

Monte Carlo simulations have been performed to optimise the design. In both calorimeters a robust electron and photon shower measurement is essential, making a small Molière radius preferable. Compact, cylindrical sandwich calorimeters using tungsten absorber disks of one radiation length thickness, interspersed with finely segmented silicon (LumiCal) or GaAs (BeamCal) sensor planes, as sketched in Figure III-3.24, are found to match the requirements [324]. For the innermost part of BeamCal, adjacent to the beam-pipes, also Chemical Vapour Deposition (CVD) diamond sensors are considered. Since LumiCal is used to measure precisely the polar angle of scattered electrons[2], it must be centred around the outgoing beam.

Both calorimeters consist of two half-cylinders. The tungsten absorber disks are embedded in a mechanical frame stabilised by steel rods. Finite element calculations were done for the support structure to ensure the necessary precision and stability. The sensors are fixed on the tungsten half-disks and connected via a flexible PCB to the front-end readout. The gap between the absorber disks is minimised to about 1 mm to achieve the smallest possible Molière radius.

The distance between the two calorimeters of LumiCal and the position of the beam with respect to the calorimeter axis must be known to about 1 mm and 500 $\mu$m, respectively. A laser based position monitoring system has been developed [327] to control the position of LumiCal e.g. with respect to QD0 with the necessary precision.

## 3.5.2 LumiCal

Bhabha scattering will be used as the gauge process for the luminosity measurement. The cross section can be calculated precisely from theory [328], and the luminosity, L, is obtained as $\mathrm{L} = \mathrm{N_B}/\sigma_\mathrm{B}$, where $\sigma_\mathrm{B}$ is the integral of the differential cross section over the considered polar angle range, and $\mathrm{N_B}$ the number of counted events in the same range. Bhabha scattering events were generated using the BHWIDE generator [329]. Electromagnetic showers were simulated and reconstructed using the standard ILD software tools. The sensor pad size was chosen to obtain sufficient polar angle resolution and to keep the polar angle measurement bias small for fully contained electron showers [324]. The energy resolution is $\sigma_\mathrm{E}/\mathrm{E} = \mathrm{a_{res}}/\sqrt{\mathrm{E_{beam}}\ (\mathrm{GeV})}$, where E and $\sigma_\mathrm{E}$ are, respectively, the central value and the standard deviation of the distribution of the energy deposited in the sensors for a beam of electrons with energy $\mathrm{E_{beam}}$ and $\mathrm{a_{res}} = (0.21 \pm 0.02)\sqrt{\mathrm{GeV}}$, as shown in Figure III-3.25. From the energy depositions in the pads for the passage of minimum ionising particles and for showers of

[2]'Electrons' is used here to describe equally electrons and positrons originating from Bhabha scattering.

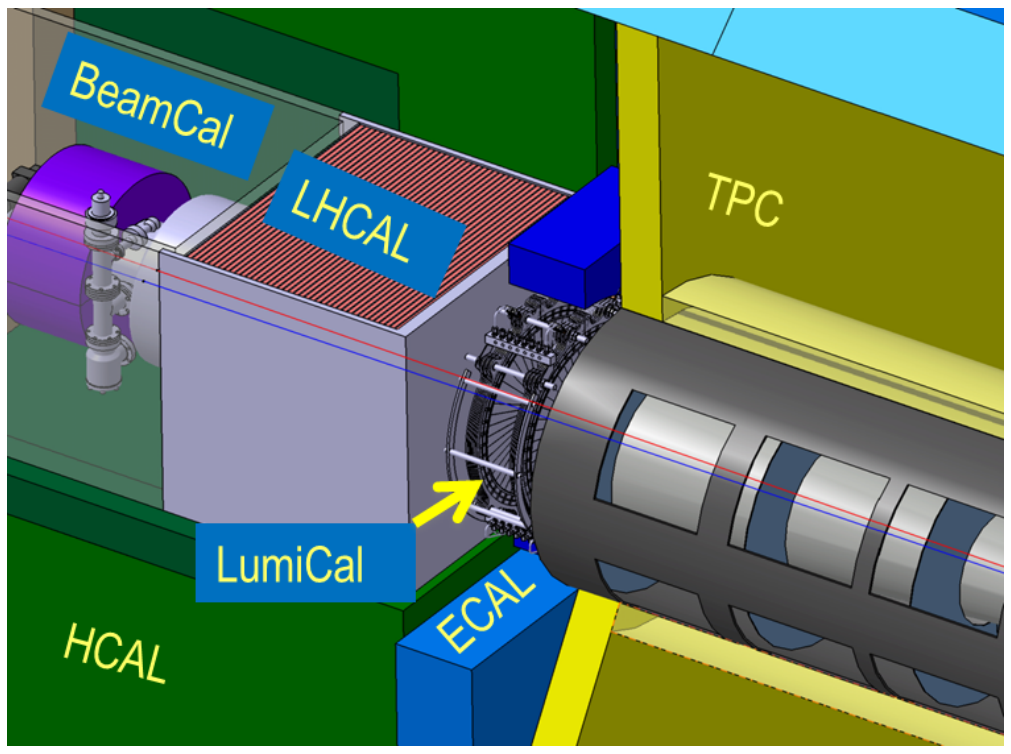

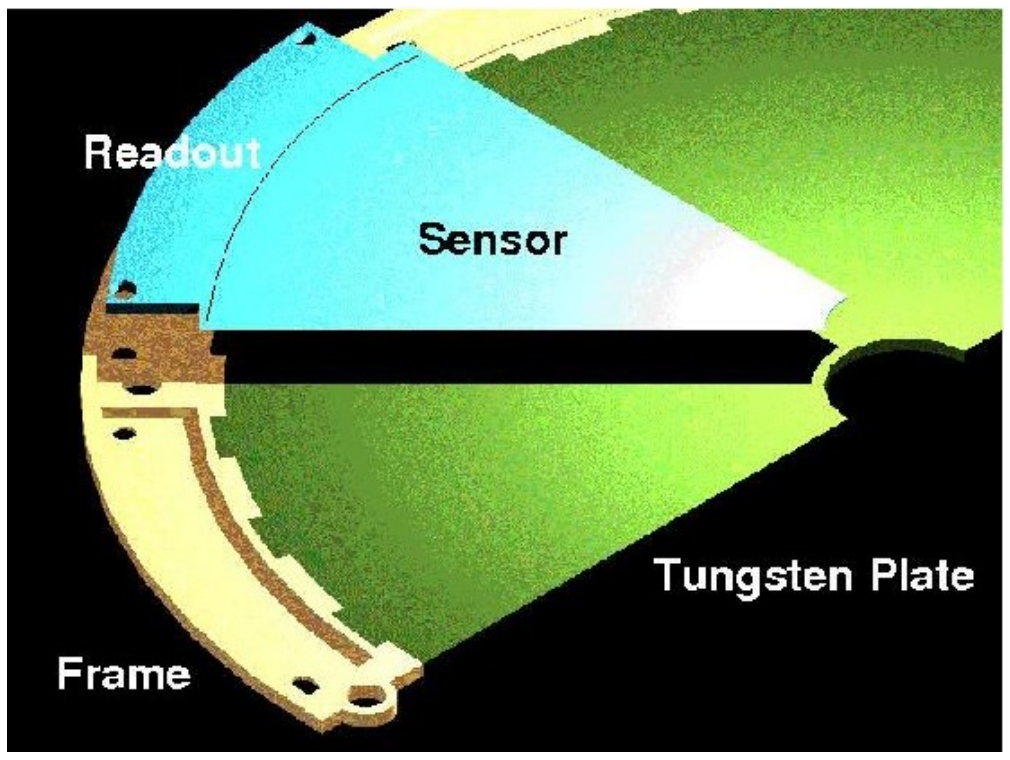

**Figure III-3.24.** Left: The very forward region of the ILD detector. LumiCal, BeamCal and LHCAL are carried by the support tube for the final focusing quadrupole QD0 and the beam-pipe. TPC denotes the central track chamber, ECAL the electromagnetic and HCAL the hadron calorimeter. Right: A half layer of an absorber disk with a sensor sector and front-end electronics.

250 GeV electrons [330], the distribution of the charge deposited in a single pad, $Q_{pad}$, was estimated to range between $4 < Q_{pad} < 6000$ fC. Signal digitisation with a 10-bit ADC preserves the energy measurement.

Prototypes of LumiCal sensors have been designed and manufactured by Hamamatsu Photonics. Their shape is a ring segment of 30°. The thickness of the n-type silicon bulk is 0.320 mm. The pitch of the concentric $p^+$ pads is 1.8 mm and the gap between two pads is 0.1 mm. The bias voltage for full depletion ranges between 39 and 45 V, and the leakage currents per pad are below 5 nA. Pad capacitances between 8 pF for the smallest pads and 25 pF for the largest pads were measured [331].

### 3.5.3 BeamCal

BeamCal will be hit after each bunch-crossing by a large amount of beamstrahlung pairs. For the current ILC beam-parameter set [332], beamstrahlung pairs were generated with the GUINEA-PIG program [333]. Inside the ILD detector an anti-DID field [334] was assumed. The energy deposited in the sensors of BeamCal per bunch crossing allow a bunch-by-bunch luminosity estimate and the determination of beam parameters with a precision of better than 10% [325]. Applying a shower-finding algorithm, single high energy electrons, as illustrated in Figure III-3.25. can be detected with high efficiency even at low polar angles.

The signals expected on the pads range up to 40 pC. Digitising with a 10-bit ADC has no impact on the performance of the calorimeter [335]. The dose and the neutron fluence in the sensors after one year of operation with nominal beam parameters are estimated for a sensor layer at the depths of the shower maximum to be about 1 MGy and $0.4 \times 10^{12}$ neutrons per $mm^2$ and year, respectively, near the beam-pipe.

CVD diamond sensors were obtained from Element6 and IAP Freiburg. Large area GaAs sensors, as shown in Figure III-3.26, were produced by means of the Liquid Encapsulated Czochralski method, doped by a shallow donor (Sn or Te), and then compensated with Chromium. This results in a semi-insulating GaAs material with a resistivity of about $10^7$ $\Omega$m.

Sensors were exposed to a 10 MeV electron beam at the S-DALINAC accelerator [336]. The diamond sensors were found to keep good performance under irradiation of up to 7 MGy [337]. The GaAs shows a significant drop in charge collection efficiency as shown in Figure III-3.26, but even

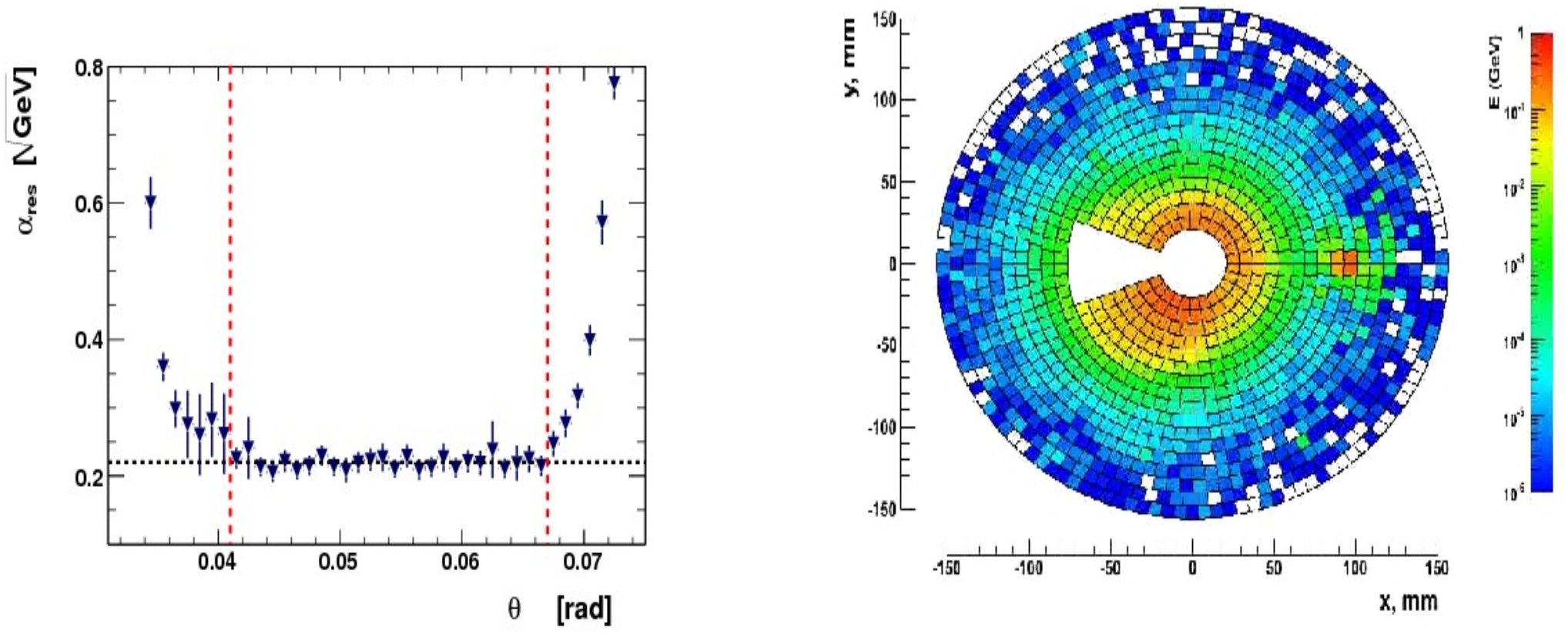

**Figure III-3.25.** Left: The energy resolution, $a_{res}$, for electrons as a function of the polar angle, covering the range of LumiCal. Right: The distribution of the energy deposited by beamstrahlung pairs after one bunch crossing in the sensors of BeamCal at a depth of 5 radiation lengths. Superimposed is the deposition of a single high energy electron, seen as red spot on the right side.

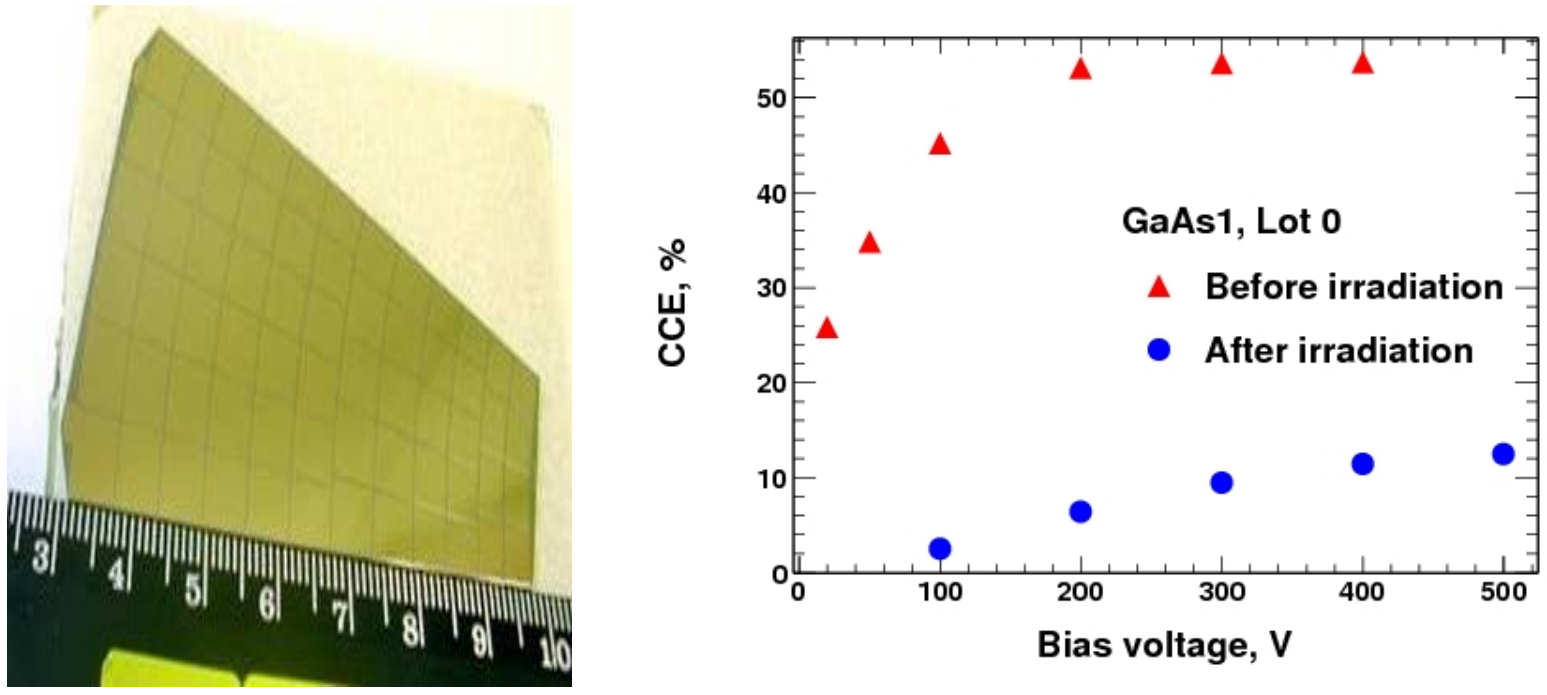

**Figure III-3.26**
Left: A prototype of a GaAs sensor sector for BeamCal with pads of about 30 mm$^2$ area. Right: The charge collection efficiency (CCE) as a function of the applied voltage for a GaAs sensor before and after irradiation.

after irradiation with 1.2 MGy a signal from a MIP is still visible [338].

## 3.5.4 ASIC developments

The readout comprises a physics mode and a calibration mode. In the physics mode signals from electromagnetic showers are recorded. In the calibration mode smaller signals from relativistic muons, considered here as minimum ionising particles, must be detected to be used for alignment and calibration. Signals from the subsequent bunch crossings, separated in time by about 300 ns, must be resolved. To reduce power dissipation switching off the power between bunch trains is implemented. An architecture [339, 340] comprising a charge sensitive amplifier and a shaper was chosen for the LumiCal ASIC. A variable gain in both the charge amplifier and the shaper is implemented by a mode switch. The peaking time of the shaper output signal is 60 ns. ASICs, containing 8 front–end channels, were designed and fabricated in 0.35 $\mu$m CMOS technology. A micrograph of the prototype, glued and bonded on the PCB, is shown Figure III-3.27. Measurements of the performance are published elsewhere [341]. A dedicated low power, small area, multichannel ADC is designed and produced. It comprises eight 10-bit power and frequency (up to 24 MS/s) scalable pipeline ADCs and the necessary auxiliary components. The active size of the ASIC is $3.17\ \mathrm{mm} \times 2.59\ \mathrm{mm}$. Eight ADC channels are placed in parallel with 200 $\mu$m pitch and are followed by the serialiser and LVDS pads, while the analog and digital peripheral circuits are on the ASIC sides. Measurements of the static and dynamic parameters, power scaling, and cross-talk are performed and published elsewhere [342].

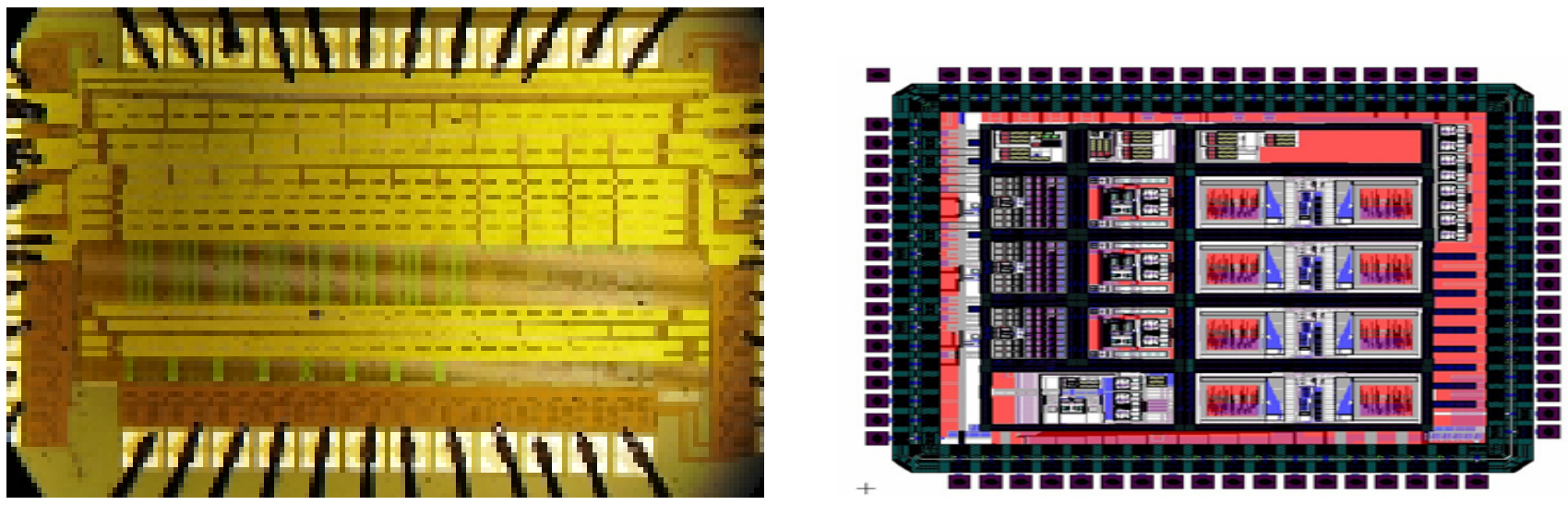

**Figure III-3.27**
Left: Micrograph of the front–end ASIC. Right: Micrograph of the Bean ASIC.

The Bean (BeamCal Instrumentation IC), shown in Figure III-3.27, is designed and produced in a 180-nm CMOS process. Each channel has a dual-gain charge amplifier, a filter, and a successive approximation register ADC. Groups of channels can be put into an adder that combines the outputs and provide a fast feedback signal which will be used for beam tuning and diagnostics. Two different gains can be selected for physics and calibration modes of operation. Both the signal and the adder output are digitised using a custom 10-bit successive approximation register ADC. The full conversion takes less than 250 ns, the adder output is available in less than 1 $\mu$s. Tests with prototype chips have confirmed the performance [343].

### 3.5.5 Beam tests

Prototypes of sensor planes assembled with FE and ADC ASICs, as shown in Figure III-3.28, were built using LumiCal and BeamCal sensors [344]. The detector plane prototypes were installed in an electron beam and the trajectories of beam particles were measured by four planes of a silicon strip telescope. The front-end electronics outputs were sampled synchronously with the beam clock, a mode to be used at the ILC. Data were taken for different pads and also for regions covering pad boundaries. Signal-to-noise ratios of better than 20 are measured for beam particles both for LumiCal and BeamCal sensors, as illustrated in Figure III-3.29. The impact point on the sensor is reconstructed from the telescope information. Using a colour code for the signals on the pads the structure of the sensor becomes nicely visible, as also seen in Figure III-3.29. The sensor response was found to be uniform over the pad area and to drop by about 10% in the area between pads.

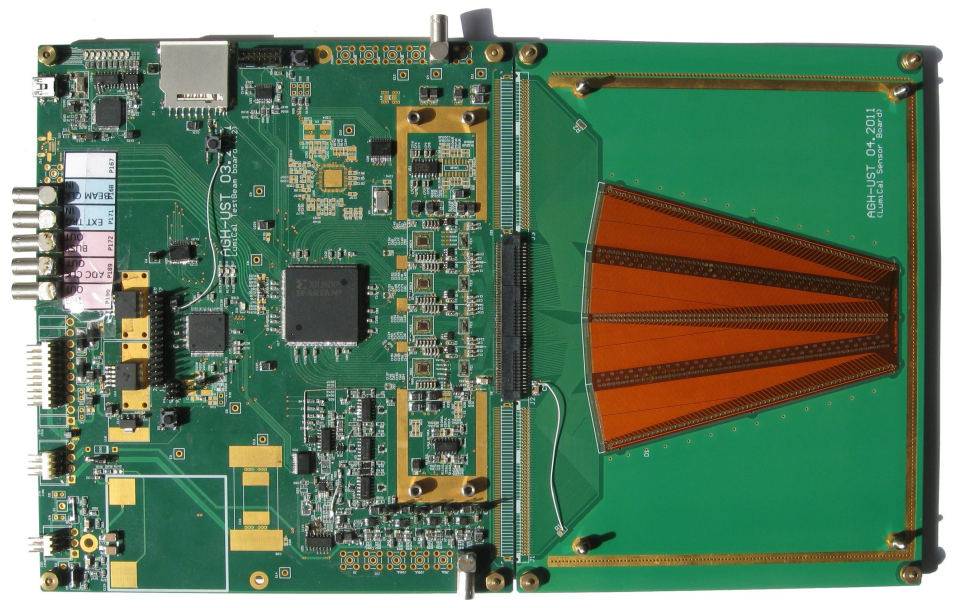

**Figure III-3.28**
Photograph of LumiCal readout module with sensor connected.

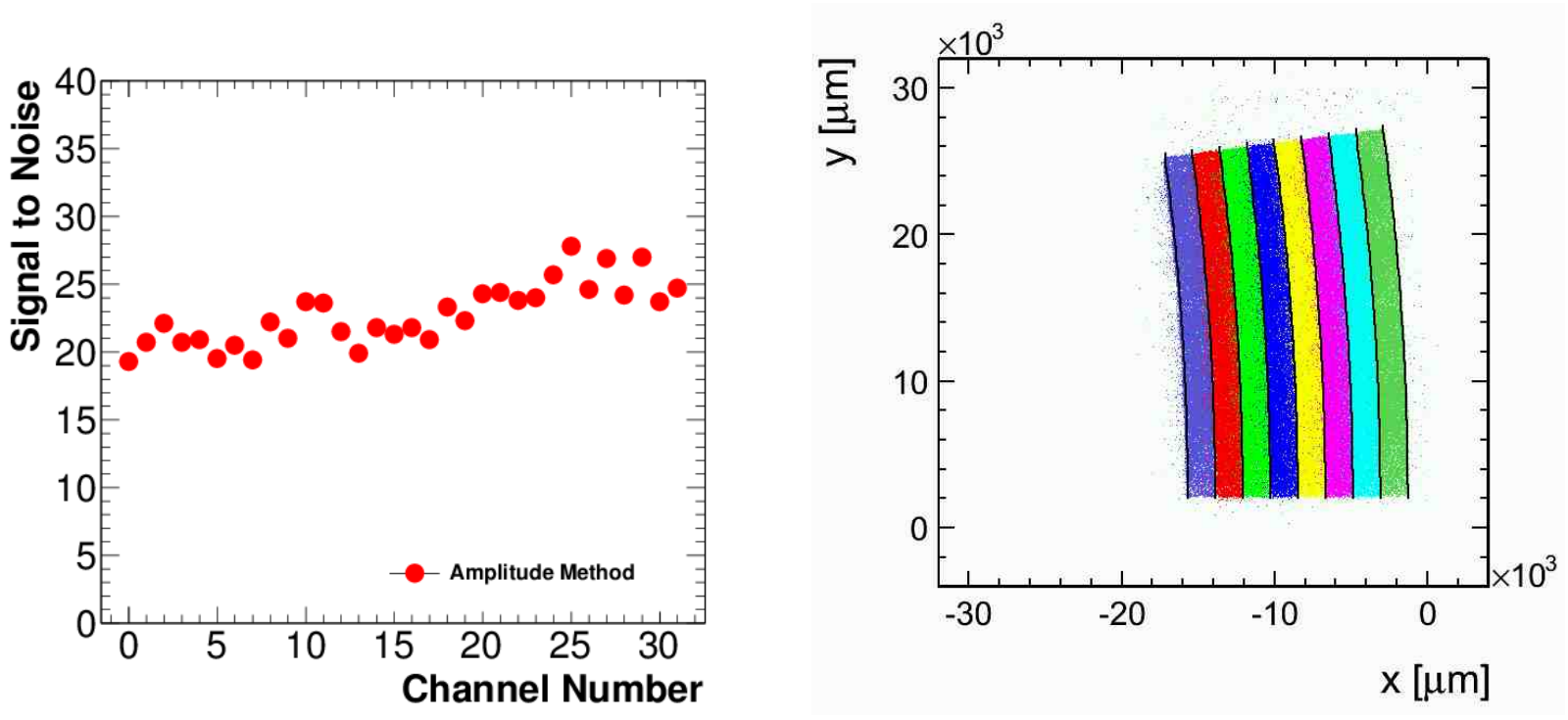

**Figure III-3.29**
Left: The signal-to-noise ratio of all read-out channels before calibration. Right: Distribution of the predicted impact points on pads with a colour coded signal.

### 3.5.6 Summary and future plans

The design of the forward calorimeters for ILD has been optimised for a precise luminosity measurement, and to assist beam tuning to optimise the accelerator operation. Dedicated sensors and ASICs have been produced and tested. A fully assembled sensor plane segment was studied in the beam. The functionality was demonstrated with excellent performance. A concept has been developed how these detectors can be integrated into the ILD detector. In the future, studies of a calorimeter prototype are needed to fully establish the design of the system. A new generation of ASICs for LumiCal using 130 nm CMOS technology is under development to reduce power dissipation and space for the on board electronics. The Bean-ASIC will be extended to a multi-channel version with a digital memory array on chip. Also effort will be invested in the development of a beamstrahlung photon calorimeter, GamCal, important for beam diagnostics [325].

# Chapter 4
# ILD Outer Detector

## 4.1 The ILD muon system/ tail catcher

A stable, highly efficient muon identification system with excellent hadron rejection is an important requirement to meet the physics goals of the ILD detector. The ILD muon system provides a number of measurement stations outside the solenoid coil, which supplement the measurements taken with the calorimeter system and the tracker. It is used to identify the muons and to act as a tail catcher, to recover energy which is leaking out of the back of the calorimeter. However, the barrel part location behind the coil limits its role to fairly high momentum particles.

The muon system/ tail catcher instruments the iron return yoke in the barrel and in the forward region. The yoke barrel part is equipped with one sensitive layer in front of the iron yoke, 10 layers spaced 14 cm apart, followed by three sensitive layers spaced by 60 cm apart. The forward part of the yoke is equipped with 10 layers spaced by 14 cm, followed by two sensitive layers spaced by 60 cm. The overall layout of the muon system/ tail catcher is shown in Figure III-4.1.

Two main options are investigated for the sensitive layers, scintillator strips equipped with wavelength shifting fibres and read out with silicon photomultipliers (SiPM), or resistive plate chambers (RPC). The main parameters of the system are summarised in Table III-4.1.

**Figure III-4.1**
Sensitive Layers of ILD Muon System/Tail Catcher

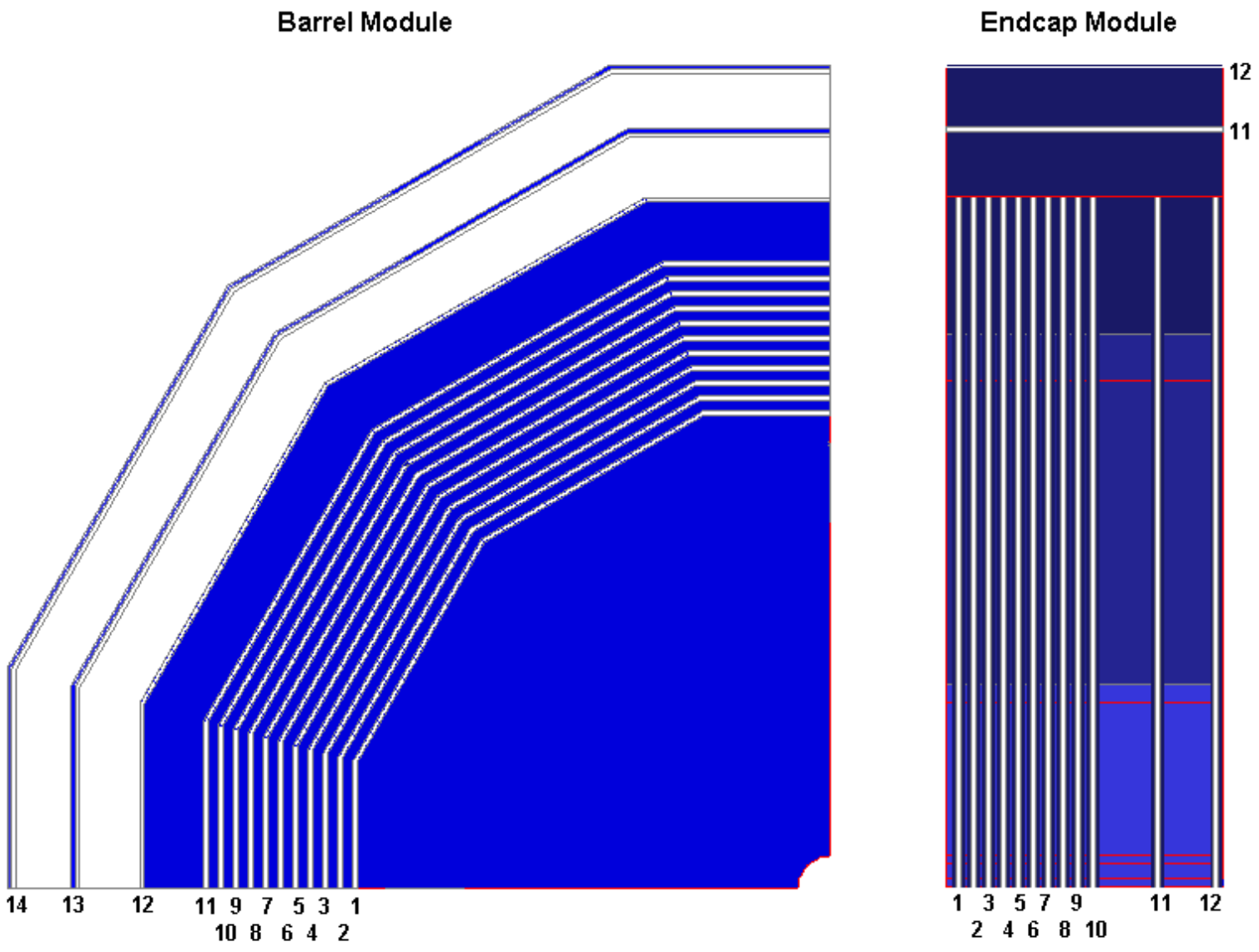

**Table III-4.1**
Table of parameters of the ILD muon system.

| Modules: | Barrel: 3 | Endcap: 2 |
|---|---|---|
| Rmin, Rmax, length [mm] | 4450, 7760, 2800 | 300, 7760, 2560 |
| No. of sens. layers | 14 | 12 |
| Scintillation strips: | total 125000 | |
| thickness, width, length [mm] | 10, 30, 2800 | |

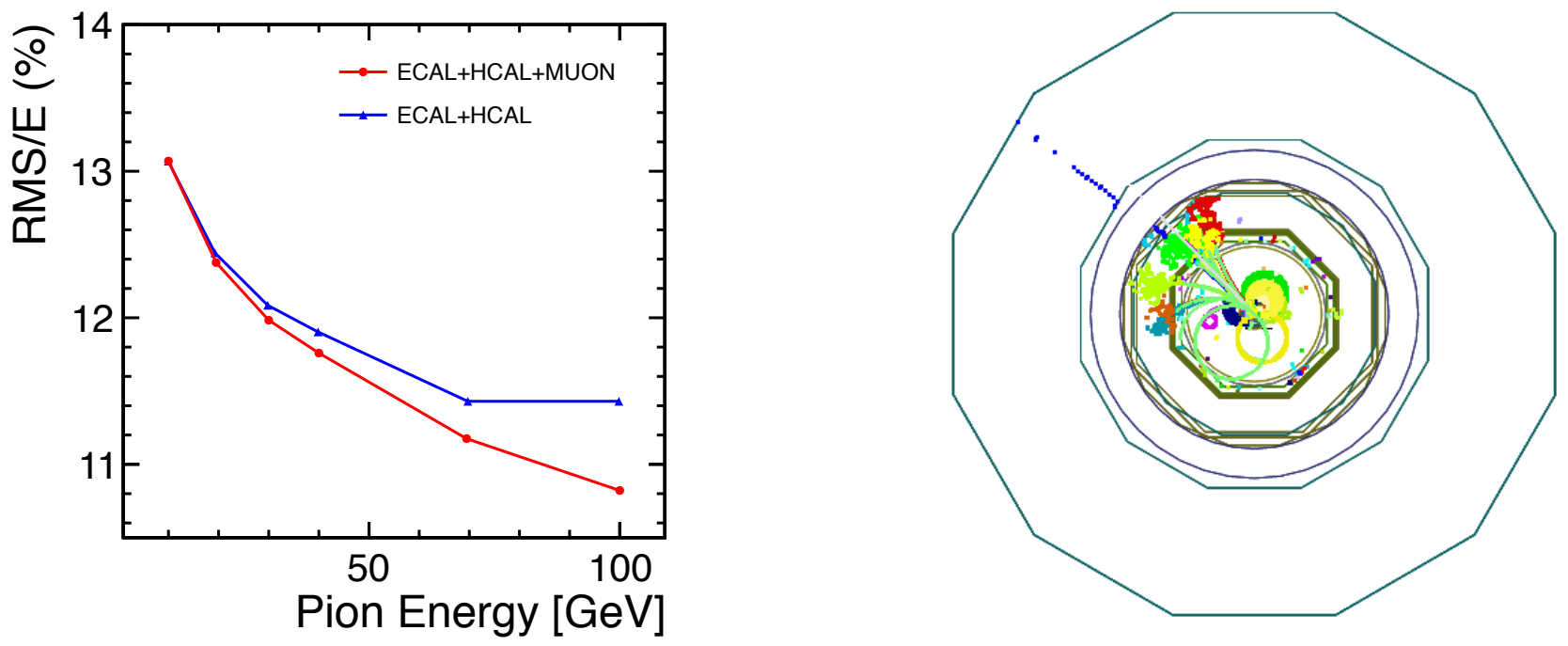

**Figure III-4.2**
Left: Energy resolution of pions without and with tail catcher. Right: Event display of the 50 GeV $b$-jet with muon track in muon system.

## 4.1.1 Muon system layout

The requirement that the muon system/tail catcher serves both as a muon identifier and as a tail catcher impacts its design. The first section of the system provides ten relatively closely spaced layers, to act as a calorimeter. Mechanical constraints limit the iron thickness between readout stations to be at least 10 cm. At the rear of the muon system the distance between stations in much increased, since they only need to act as a muon tracker. Three layers in the barrel, two in the endcap are spaced 60 cm apart [345].

The potential improvement of the jet energy resolution with a perfect tail catcher, as estimated from simulation, is shown in Figure III-4.2 (left). The fact that the coil adds about two interaction lengths of material in front of the muon system limits the effect of the tail catcher. To maximise its impact a sensitive layer is placed in front of the iron yoke, directly behind the coil and the first 10 layers are spaced more closely to improve the calorimetric performance of the system.

With the anticipated point resolution of about 1 cm and the current design, the achievable momentum resolution for muons is limited by multiple scattering up to momenta of 7 GeV. However in particular for muons inside jets the addition of the information from the muon system/tail catcher can significantly improve the purity of the muon sample, as shown in Figure III-4.2 (right).

## 4.1.2 Technologies

The main option for the sensitive layers will use extruded scintillation strips with a thickness of 7-10 mm and a width of 25-30 mm. A 1 mm wide extruded groove running along the center of the strip will take a commercially available wave length shifting (WLS) fibre. The scintillator strips will be covered on the outside by a layer of $TiO_2$, that is co-extruded alongside the scintillator during the extrusion process. The maximal length of strips required for ILD is 270 cm. The technology was successfully tested in ITEP [346].

The signals will be readout from both sides of the strips by silicon photo multipliers, coupled to the wave length shifting (WLS) fibres. Reading out both sides of a strip offers the possibility to define the position of the hits along the strip, which will help in reducing the fake rate in the muon system.

Fig. III-4.3 (left) shows the design of the scintillator strip. The right picture presents the signal (number of photons) of the scintillation strip with WLS and SiPM readout from both sides.

Resistive plate chambers (RPC) are considered as alternative sensitive layers. Main features are excellent granularity up to $1 \times 1$ cm$^2$ pads and one threshold (1-bit) digital readout. Several types of RPCs have been successfully constructed and tested in the HEP community and within the ILC R&D program [347].

**Figure III-4.3**
Left: Technology of muon system sensitive elements: schematic view of scintillator strip with SiPM readout. Right: Signal from both sides of the 2 m length scintillator strip with WLS and SiPM readout.

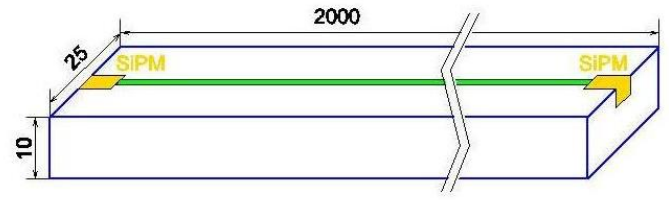

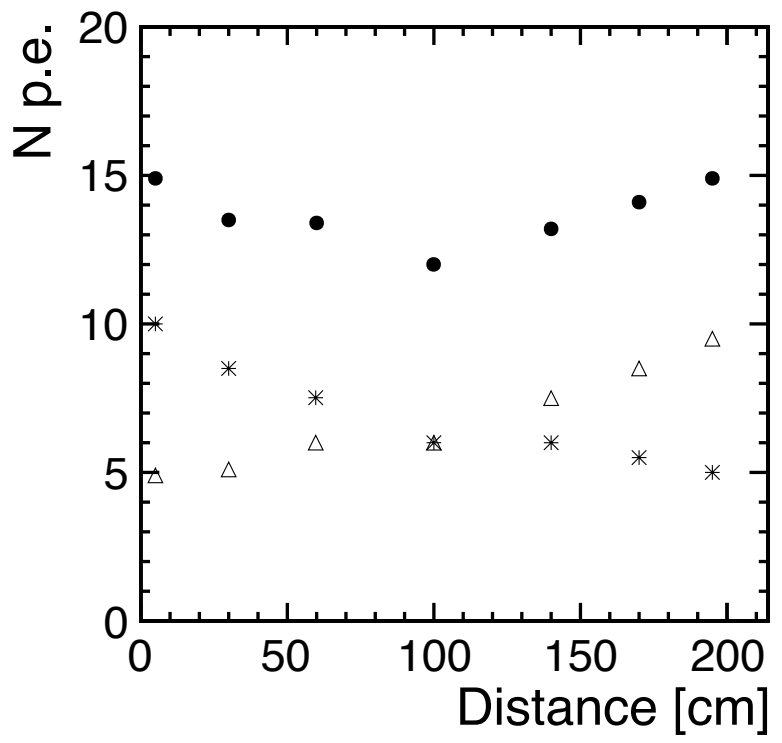

## 4.1.3 Performance

The performance of muon system and tail catcher was studied with the full ILD Monte Carlo simulation and reconstruction chain. In the model the sensitive elements are implemented as square tiles of 30 x 30 mm$^2$ and a thickness of 10 mm, with SiPM readout, similar to the AHCAL tiles. This is a simplification compared to the proposed system, which will reduce significantly the fake-rate problem present in a readout with long strips.

### 4.1.3.1 Muon Identification

One of the main tasks of the muon system/tail catcher is the identification of isolated muons. The main source for wrongly identified muons are pions. A number of scenarios have been identified how this can happen:

- pions can decay into emitting a muon, which is then detected in the muon system;
- pions may pass the calorimeter system without interaction ('sail through') and are detected in the muon system;
- particles from the shower in the calorimeter may pass to the muon system and are detected there.

The muon identification for single particles is based on the analysis of the hits in the sensitive stereo layers of the muon system in coincidence with a region of interest defined as a cone extrapolated along the direction of the track measured in the tracking system and calorimeter system. The angle of the cone is defined as a function of the multiple scattering angle.

Figure III-4.4 left shows the efficiency of the muon identification and the contamination with pions as a function of the energy of the particles. The colour of the lines corresponds to the layers of the muon system which are used for the muon identification: blue are the first 10 layers, green are more widely spaced layers. In jets the muon is accompanied by hadronic background in the same region of interest. Study of the identification of muons in jets and the contamination by hadrons was performed using semileptonic decays of $b$-quarks. The results are shown in Figure III-4.4 (right). The blue lines correspond to the identification of muons in the first 11 layers of the barrel part, normalised to 5 GeV muons inside jet. The green lines represents data recorded beyond layer 11, normalised to 7 GeV muons due to the fact that low energetic muons do not reach this layers. The results show that an identification efficiency of more than 97% can be reached for energies higher than 7 GeV, with a hadronic contamination at the few percent level.

The muon identification power at low energies, below 4 GeV, is affected by dead material in front of the muon system and by the deflecting of particles in the magnetic field, which do not allow the particles to reach the sensitive layers of the muon system. The muon identification at low energy is possible using the highly granular structure of the ILD calorimeter system which allows the identification of muons as minimum ionising particles (mip) like tracks in the calorimeter system.

#### 4.1.3.2 Muon system as tail catcher

The first layers of the muon system serve as a tail catcher, measuring the energy which leaks through the end of the calorimeter system. Figure III-4.5 shows the effect of an ideal tail catcher (no dead material between the calorimeter and the tail catcher) and the realistic scenario at ILD, with two interaction lengths of material in front of the tail catcher, as a function of the total depth of the calorimeter system. For 6 $\lambda$, the value for the ILD calorimeter system, a roughly 10% improvement is possible with the tail catcher [348].

A prototype of the muon system/tail catcher was successfully tested during the 2007-2012 CALICE test beam campaign with ECAL and analogue HCAL. A tail catcher was placed behind the HCAL instrumented with scintillator strips and readout with SiPMs [348]. Results from the tests show that the proposed system delivers the anticipated performance and thus validates the technology needed to built a muon system for ILD.

## 4.2 The ILD coil and yoke system

The ILD detector design asks for a nominal 3.5 T and maximum 4 T solenoidal central field in a warm aperture of 6.88 m in diameter and 7.35 m in length. In addition, in order to suppress background from incoherent pairs from beamstrahlung, an anti-DID (Detector-Integrated-Dipole) is needed. In order to achieve high precision tracking with the TPC, accurate field mapping after construction is requested.

The iron yoke will be instrumented to be used for the detection of muons and for measuring showers escaping the hadron calorimeter (tail catcher). In addition, the yoke serves as the main mechanical structure of the ILD detector and, combined with the calorimeters, should make the detector self-shielding in terms of radiation protection. To allow work in the vicinity of the detector while its magnet is powered, the fringe field should be less than 50 G at 15 m from the IP, in the radial direction.

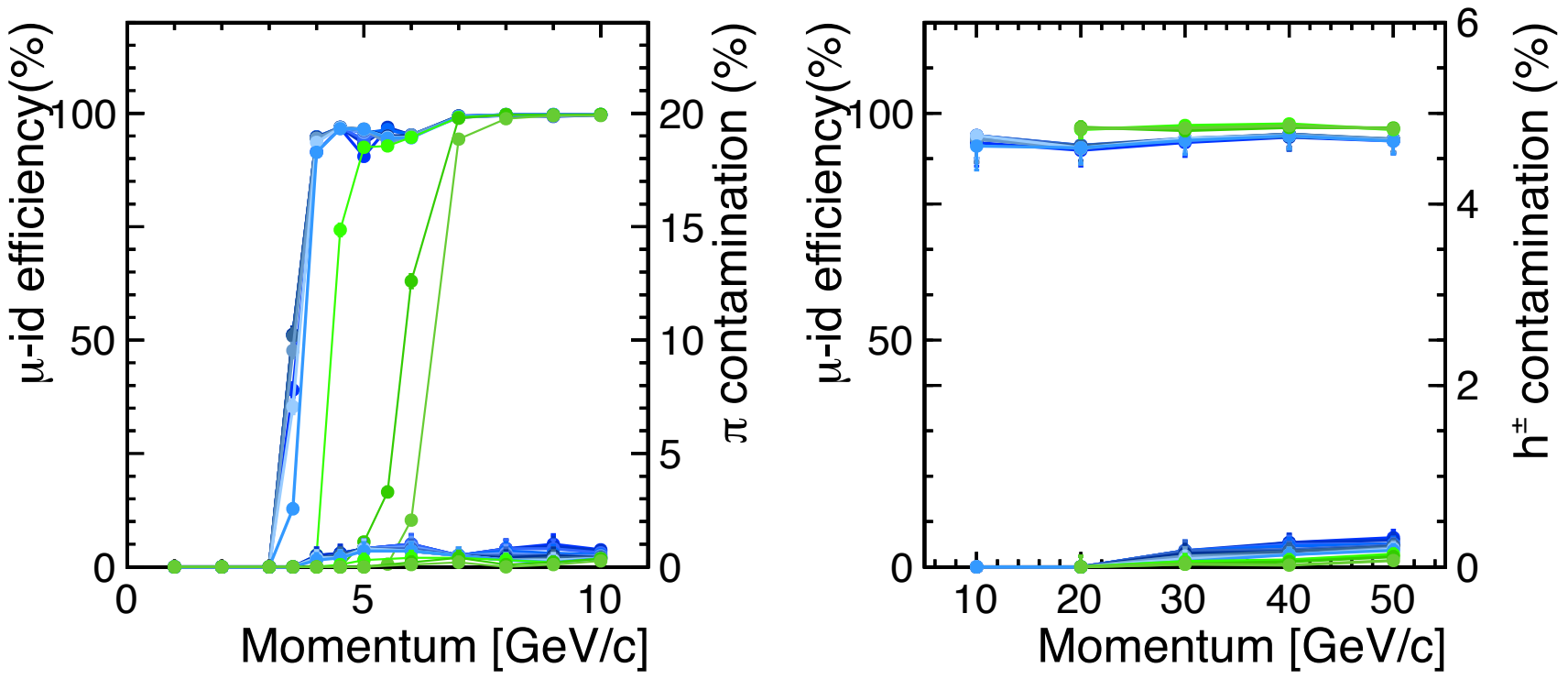

**Figure III-4.4** Simulated muon efficiencies and contamination, left: muon efficiency and contamination as function of energy for single particles, right: muon efficiency and hadron contamination as function of energy for $b$-jet.

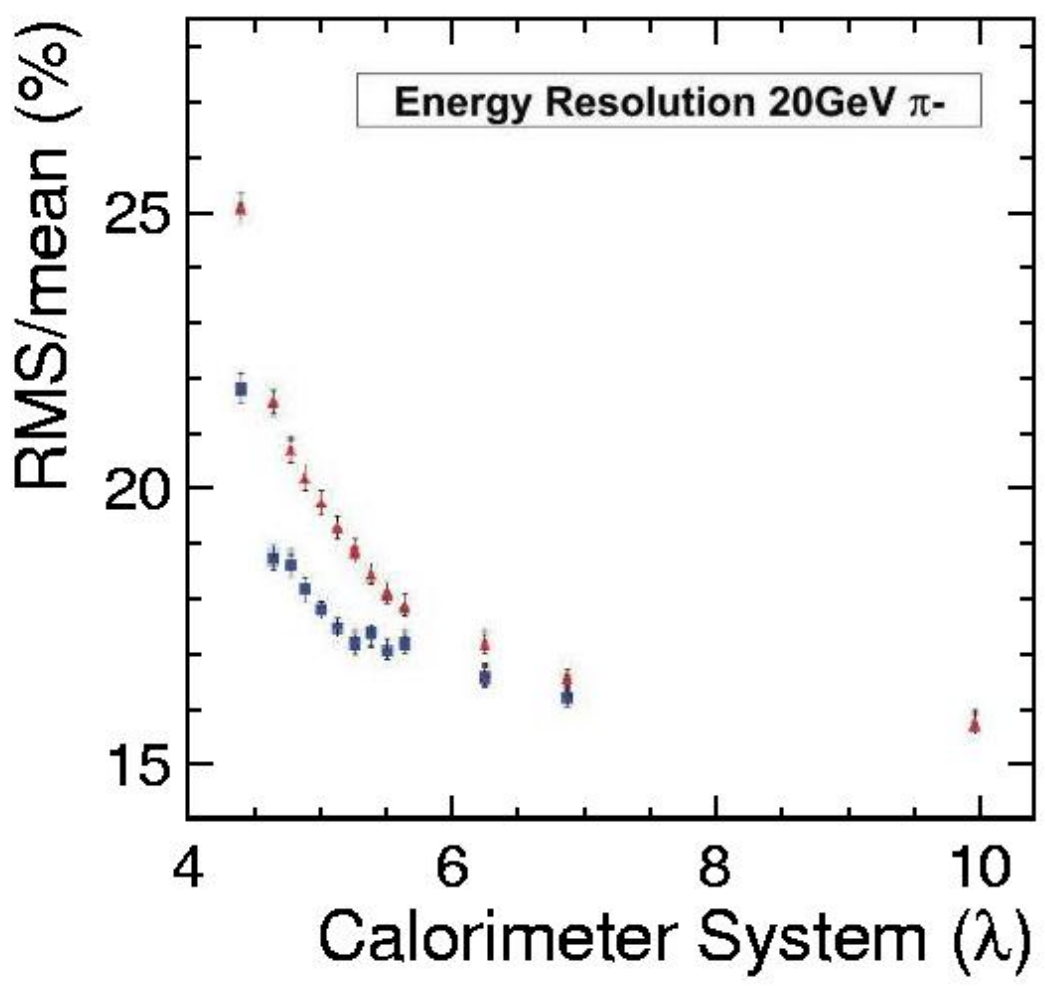

**Figure III-4.5**
Muon system as tail catcher: comparison of energy resolution of a calorimeter system with a tail catcher without a coil separating the two, in blue, with a system including a simulated coil (about 2 $\lambda$ of dead material) in front of the tail catcher, in red, for 20 $GeV$ $\pi$s.

## 4.2.1 Magnet design

The ILD magnet design is very similar to the CMS one, except for its geometrical dimensions, and the presence of the anti-DID. Consequently, many technical solutions successfully used for CMS [349] are proposed for ILD. The magnet consists of three main parts:

- the superconducting solenoid coil, made of three modules, mechanically and electrically connected. With its thermal shields, it makes up the cold mass, supported inside the vacuum tank by several sets of tie-rods;
- the anti-DID, located on the outer radius of the main solenoid, the dipolar magnetic field of which enables to reduce the beam background in the vertex and tracking volume;
- the iron yoke, consisting of the barrel yoke and the two end-cap yokes, of dodecagonal shape. The yokes are laminated to house muon detectors.

A detailed description of the conceptual design of the ILD magnet system is given in [350]. The main parameters and characteristics are summarised in this section. A schematic cross section of the magnet is given in Figure III-4.6. The main geometrical parameters of the ILD magnet are summarised in Table III-4.2.

**Table III-4.2.** ILD magnet main parameters

| | | | |
|---|---|---|---|
| Cryostat inner radius [mm] | 3440 | Barrel yoke outer radius [mm] | 7755 |
| Cryostat outer radius [mm] | 4400 | Yoke overall length [mm] | 13240 |
| Cryostat length [mm] | 7810 | Barrel weight [t] | 6900 |
| Cold mass weight [t] | 168 | End cap weight [t] | 6500 |
| Barrel yoke inner radius [mm] | 4595 | Total yoke weight [t] | 13400 |

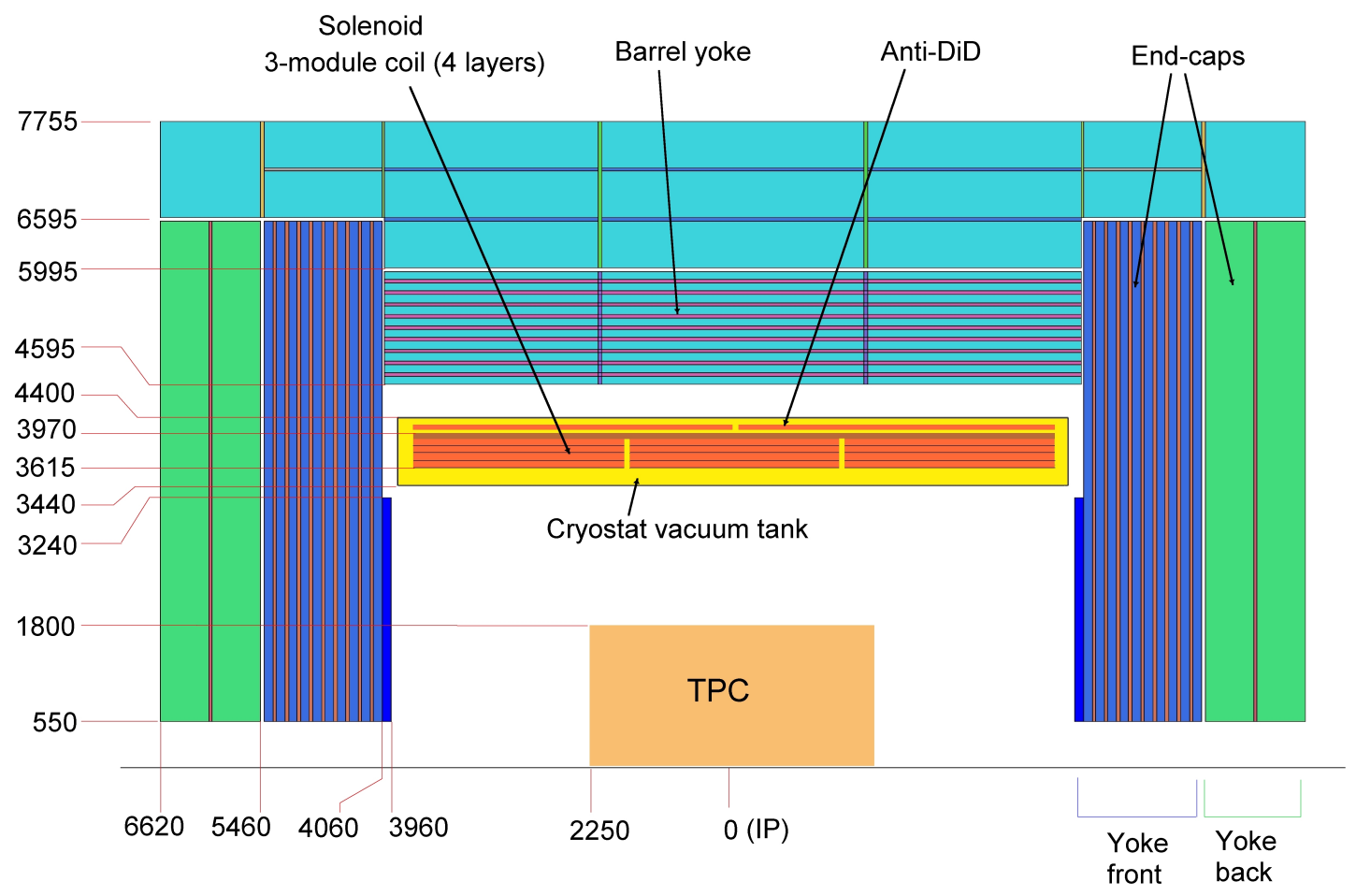

**Figure III-4.6**
ILD magnet cross section, dimensions are in mm (half upper part, cylindrical symmetry)

## 4.2.2 Solenoid design

The ILD solenoid main parameters are given in Table III-4.3. The 7.35 m length of the ILD coil enables to make it in three modules, each 2.45 m long. The reasons of this choice of three modules, rather than two or one, are linked to the fabrication of the external mandrel, to winding and impregnation as well as to transport and handling. Moreover, this enables to have shorter unit lengths of conductor, of about 2.6 km, and to join the units in known positions and in low field regions, on the outer radius of the solenoid. Each module consists of four layers, with 105 turns per layer.

**Table III-4.3**
ILD solenoid main parameters

| Design maximum solenoid central field [T] | 4.0 | Nominal current [kA] | 22.5 |
|---|---|---|---|
| Maximum field on conductor [T] | 4.77 | Total ampere-turns solenoid [MAt] | 27.65 |
| Field integral [T*m] | 32.65 | Inductance [H] | 9.26 |
| Coil inner radius [mm] | 3615 | Stored energy [GJ] | 2.27 |
| Coil outer radius [mm] | 3970 | Stored energy per unit of cold mass [kJ/kg] | 13.5 |
| Coil length [mm] | 7350 | | |

The conductor design uses a superconducting cable, electrically stabilised and mechanically reinforced. The temperature safety margin is around 1.93 K, assuming a maximum operating temperature in the coil of 4.5 K.

The winding will be done inside the coil mandrel, using the inner winding technique, similarly to CMS [351]. This Al-alloy mandrel, about 50 mm thick, has several important other roles, as it will also be used as a mechanical support, a path for the indirect cooling of the coil (done with cooling tubes where liquid helium circulates welded on the outer radius of the mandrel), and a quench back tube (induced currents in this mandrel in case of quench or fast discharge enable a uniform quench of the coil and a limited radial temperature gradient). The anti-DID and the tie rods supporting the whole cold mass will be attached to the mandrel. The cold mass will be indirectly cooled by saturated liquid helium at 4.5 K, circulating in a thermosiphon mode.

The coil protection in case of quench uses an external dump circuit. With a dump voltage of 500 V, the maximum temperature within the coil does not exceed 82 K.

### 4.2.3 Anti-DID design

The magnetic dipole field $B_x$ generated by the anti-DID should reach 0.035 T at z=3 m from the IP, and should extend up to z=5 m. The anti-DID coil is formed with two dipoles centred on the beam axis with their magnetic field in opposite direction. The anti-DiD design parameters are given in Table III-4.4. Details of the design can be found in [350].

**Table III-4.4.** ILD anti-DID main parameters

| | | | |
|---|---|---|---|
| Design dipole central field on beam axis [T] | 0.035 | Nominal current [A] | 1075 |
| Position of max dipole field in z [m] | 3 | Maximum field on conductor [T] | 2.0 |
| Anti-DID total length in z [mm] | 6820 | Anti-DID inner radius [mm] | 4160 |

The anti-DID is located within the same cryostat as the main solenoid, and benefits from the cryogenics of the main coil. The preferred superconductor is NbTi to tolerate some deformation of the winding pack but other superconductors (like $Nb_3Sn$ and $MgB_2$) will be evaluated at a more advance stage of the design.

The manufacturing of the four poles constituting the anti-DID is independent from the main solenoid. It is proposed to do the winding inside a coil casing, similarly to the ATLAS barrel toroids [352]. The winding procedure and tooling will be validated with a winding test using a dummy conductor.

### 4.2.4 Assembly of the solenoid

The proposed assembly of the solenoid is similar to CMS [353]. The three modules of the main solenoid will be assembled on the ILC experimental site in a surface hall. They will be stacked vertically for the mechanical coupling. After the completion of the solenoid assembly, the anti-DID poles will be fixed on the main solenoid in the same vertical position, and all their connections (mechanical, electrical and cryogenic) done.

After the installation of the thermal screens in vertical position, the cold mass is swivelled to the horizontal position on its supporting platform, and brought to the position where it can be inserted into the outer cylinder of the vacuum tank which is fixed in cantilever to the central yoke barrel.

### 4.2.5 Ancillaries

The classical power circuit will consist of a two-quadrant converter (25 kA, ±20 V), a dump resistance allowing both fast and slow discharges, and redundant current breakers. A superconducting high critical temperature (HTS) link is the preferred option for the flexible power lines. The current leads will be built as well with HTS superconductor. The anti-DID will have its own power circuit with similar characteristics as the one described for the main coil, connected through the same chimney across the yoke as the solenoid.

The magnet control and safety systems consist of (a) controls for all operation phases, (b) a system to safely discharge the energy of the magnet and (c) redundant quench detectors (QDs) on coil modules, anti-DiD poles and on the superconducting busbars connected to the HTS power lines.

A common refrigerator will be used to cool down the main solenoid and the anti-DID. It is also able to extract the dynamic losses during the various magnet ramps or discharges. An estimate for the cryogenic losses is 400 W at 4.5 K.

### 4.2.6 Final tests and field mapping

A full test of the magnet at its nominal current is mandatory before the inner detectors are installed. A complete field map of the magnet, to an accuracy of about 1 G in an overall field of 4 T, i.e. with a relative accuracy of around 2 $10^{-5}$, is needed. Possibilities to reach such a measurement accuracy could be to use a differential method, or to aim for a very large number of measurement points during the field mapping.

### 4.2.7 Iron yoke design

The yoke has several functions. It provides the flux return of the solenoidal field and reduces the outside stray fields to an acceptable level. It is instrumented with detectors for muon identification and tail catching of hadronic showers. In addition, the yoke is the main mechanical structure of the detector. The ability for access and work in the interaction region (IR) hall during beam operation requires the detector to be self-shielding. The design allows for a fast opening in order to get access to the inner detector components.

For the inner part of the yoke a fine segmentation of the iron was chosen, 10 layers of 100 mm thick plates with 40 mm gaps for detectors to be inserted for good muon reconstruction, rejection of hadron background and good performance of the tail catcher (see section 4.1). This segmentation is in particular useful for the tail catcher, whereas a similar performance of the muon system could be achieved by arranging the detectors in groups of layers. In addition to the inner fine segmentation, some 560 mm steel plates are added on the outer part mainly to reduce the stray field.

During beam operation the IR hall has to be accessible due to the push-pull concept. Since all activities in a high magnetic field are very cumbersome and potentially dangerous, a field limit of 50 G at 15 m radial distance from the beam line was agreed upon [354]. Two- and three-dimensional FEM field calculations were done using the CST EM Studio program, varying the thickness and geometry of the iron in the barrel and end-caps until the goal of less than 50 G at 15 m radial distance was achieved. This was obtained with three 560 mm thick steel plates in the barrel and two 560 mm plates in each end-cap in addition to the ten 100 mm thick inner layers. This results in a total thickness of the iron of 2.68 m in the barrel and 2.12 m in the end-caps, respectively. In order to obtain the desired limit, all gaps between the steel plates on the outer radius have to be closed with iron. The only exception are the gaps between the barrel rings and between barrel and end-caps. This space will be needed for cables, cooling pipes and other services.

It should be noted, that the field calculations assume no additional magnetic material outside the yoke and that the results are at the limit of the accuracy of the FEM calculations.

The strong magnetic field, maximum of 4 T, introduces large magnetic forces on the end-caps, which were calculated using different FEM programs (CST EM Studio and ANSYS). The largest force, an inward pulling force in the z-direction of about 180 MN, acts on each end-cap, which has to be taken into account in the mechanical design.

### 4.2.8 Barrel yoke design

The solenoid with the central subdetectors is supported by the central barrel ring, the only stationary part around the interaction point. Both outer rings can be moved independently along the z-direction to allow access to muon chambers and services. A dodecagonal shape was chosen in order to reduce the weight and size of the sections. The twelve segments come in two slightly different sizes to avoid segment edges pointing towards the beam line. The average weight of a segment is about 190 t. Fig. III-4.7 gives an overview of the design.

The 10 plates of an inner segment and the three outer plates are welded together with 30 x 40 mm spacers between the plates along the segment edges. Segments are then bolted together on all sides using M36 bolts. (bigger on outside). Shear keys between the segments prevent radial displacement, whereas shear pins on the inner and outer edges are used to prevent movements along the z-direction.

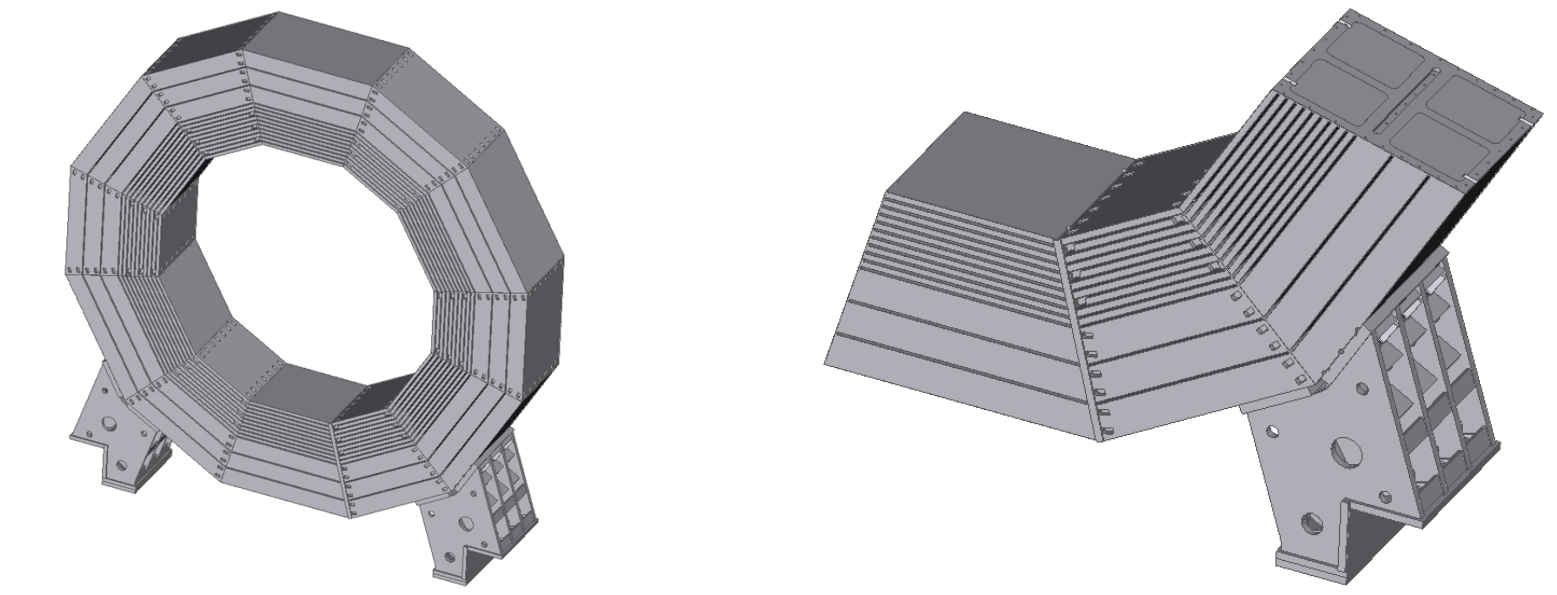

**Figure III-4.7**
The yoke barrel design: general view of one barrel ring (left) and detailed view of a sector with one supporting foot (right)

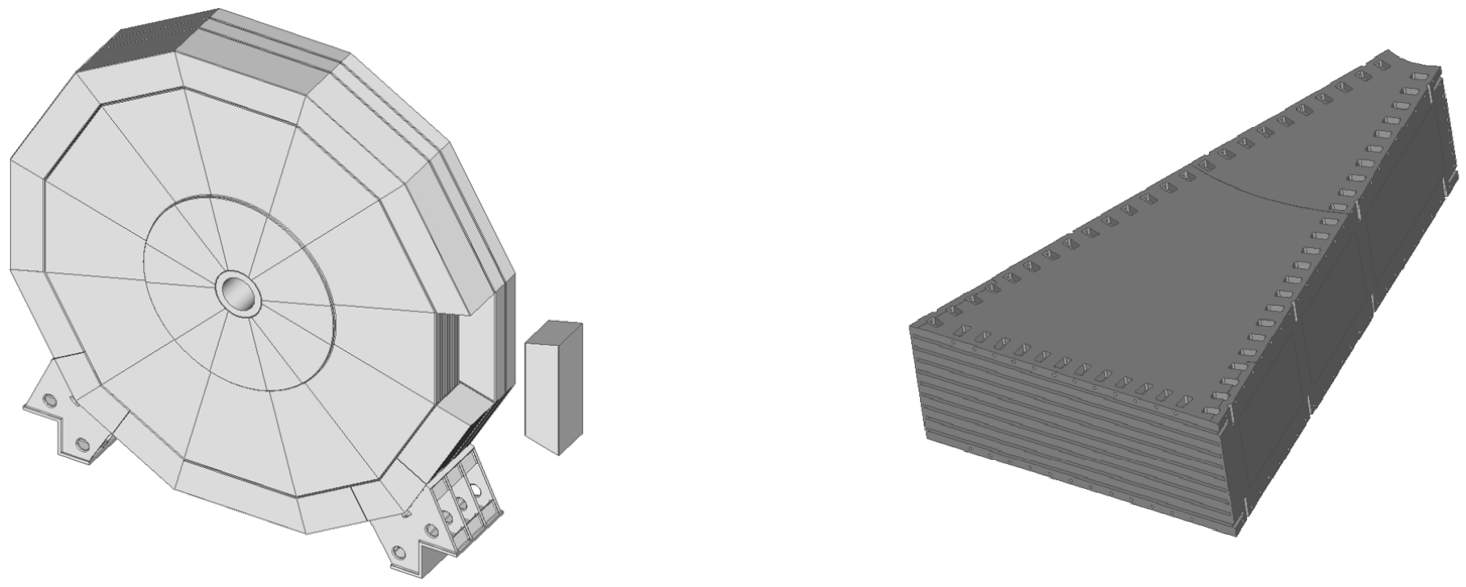

**Figure III-4.8**
The yoke end-cap design: overview (left) and detailed view of one sector (right)

The fully assembled barrel ring is a very stiff structure. The maximum vertical deformation of an outer ring is 1.6 mm, which is due to the gravitational load. At the very end of the coil there is a radial magnetic field component acting on the inner plate of the outer ring, which introduces a force of about 1.3 MN. This leeds to a 1.5 mm radial deformation of the plate.

Each barrel ring has a mass of about 2300 t, including the support feed. The central barrel ring has to carry an additional weight of almost 1000 t, the mass the cryostat with the coil, barrel calorimeters and central tracking detectors. For the calculation of deformation and stress the cryostat was approximated by a single 50 mm thick steel cylinder attached to the barrel at 12 points. The additional gravitational load was introduced by increasing the density of the cylinder. The maximum vertical deformation is 4 mm.

### 4.2.9 End-cap yoke design

The design of the end-cap is more challenging compared to the barrel due to the large magnetic forces, about 180 MN acting in the z-direction. Several geometries were considered. A design with radial supports instead of horizontal supports was chosen due to the larger second moment of area, better transfer of force to the barrel, symmetric iron distribution and a minimum of dead material. This design minimises the end-cap deformation and stress. An overview of the design as shown in Fig. III-4.8 The end-cap is made out of twelve wedge-shaped segments, extending from the inner hole to the outside of the yoke, consisting of 10 inner 100 mm thick plates, and two outer plates 560 mm thick. In addition, a 100 mm thick steel plate was introduced to improve the self-shielding of the detector.

Similar to the barrel, the 10 plates of an inner segment are welded together with spacers along the segment edges. Thus forming rigid structures, with the spacers acting as supports. Segments are then bolted together on the front and back sides using M36 bolts. A central cylindrical support tube of 1.0 m (1.2 m) inner (outer) diameter is bolted to the individual inner and outer plates, making a rigid connection of the inner and outer parts.

The maximum deformation of the end-cap due to the magnetic force of 180 MN is about 3 mm.

The force are transmitted to the barrel through z-stops the resulting stress is less than 200 MPa. The total weight of one end-cap is about 3250 t.

### 4.2.10 Yoke assembly

After a full trial assembly at the manufacturer, the barrel end-cap segments with a maximum weight of 200 and 90 t, respectively, are transported to the experimental site. In case of vertical access shaft, the assembly of the barrel rings and end-caps is done in the surface building above the IR region. Complete barrel rings and the end-caps are then lowered into the IR hall, similar to the CMS assembly.

The design does not have to be changed for a mountain site with horizontal access tunnels. Barrel and end-cap segments have to be transported into the IR hall, where the rings and end-caps are then assembled. This requires more work and time spent in the IR hall and requires a 250 t crane in the IR hall.

# Chapter 5
# The ILD Detector System

A central part of the activities of the ILD group has been the integration of the different sub-systems into a coherent detector, and the coordination between the detector and the machine. In this section a coherent integration scheme is presented, with a first realistic estimate about space and extra material this requires. Also described are systems which concern the complete detector as overall calibration scenarios, data acquisition, and central software and tools. The chapter closes with a description of the assembly procedure which is planned for ILD, and a discussion of the impact the different sites discussed for ILC will have on this procedure.

## 5.1 ILD integration

The integration of the different sub-detectors into a coherent and functioning ILD detector concept is an important aspect of the ILD work. Not only the mechanical integration, but also the coordination of the services, cabling, cooling strategies, thermal stabilisation and alignment of the various sub-detectors is an on-going task, which evolves with the better knowledge about the respective detector technologies. Moreover, the envisaged push-pull scenario at the ILC imposes additional requirements.

### 5.1.1 Mechanical concept

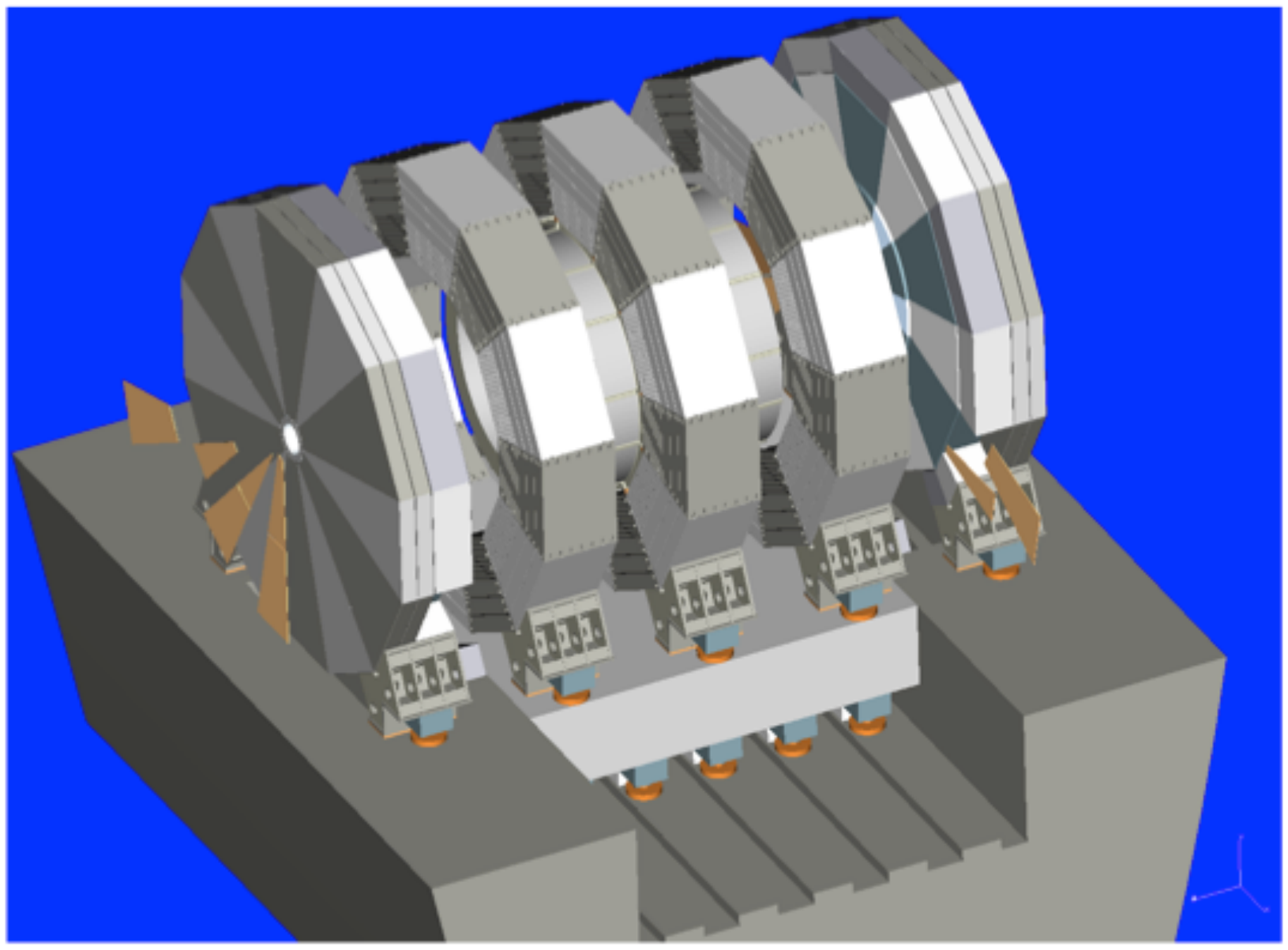

**Figure III-5.1**
The mechanical design of ILD.

The mechanical design of the ILD detector is shown in figures III-5.1 and III-1.1. The major components are the five parts of the iron return yoke: three barrel rings and two endcaps. The central barrel ring carries the cryostat with the solenoid coil in which the barrel calorimeters are installed. The TPC and the outer silicon envelope detectors are also suspended from the cryostat using tie rods. The endcap calorimeters are supported by the endcap yoke sections which can be moved independently from the barrel sections. The beam pipe, the vertex detector and the other inner silicon detectors are supported from a structure of carbon fibre reinforced plastic (CFRP), which hangs at the flanges of the TPC field cage. The whole structure can be aligned with respect to the beam axis using actuators and a laser alignment system. The final focus quadrupole (QD0) magnets are mounted independently of the yoke endcaps in a support structure that carries the magnets and the forward calorimeters. This structure is supported from a pillar outside of the detector and is suspended from the solenoid cryostat using tie rods. The QD0 magnets are also monitored by an alignment system and can be moved using actuators.

### 5.1.1.1 Yoke and magnet

The mechanical design of the return yoke and the solenoid is described in section 4.2. The central barrel yoke ring supports the detector magnet solenoid. The magnet cryostat has been designed to carry the load of all central detectors, i.e. barrel calorimeters, TPC, inner tracking. The cryostat itself is bolted to 24 double-brackets that are welded to the inner support structure of the yoke barrel. Figure III-5.2 shows how the outer cryostat shell is fixed to the barrel yoke (left) and the simulated deformation of the cryostat under its own load and the load of the barrel calorimeters (right). Under the assumption that the loads are distributed evenly over the cryostat flanges, maximum deformations of less than 1.3 mm are expected. Simulations for a more realistic support system, where the barrel calorimeters are supported by rails in the cryostat, yield maximum deformations of $\approx 2.5$ mm. Details of the magnet cryostat integration are described in [355].

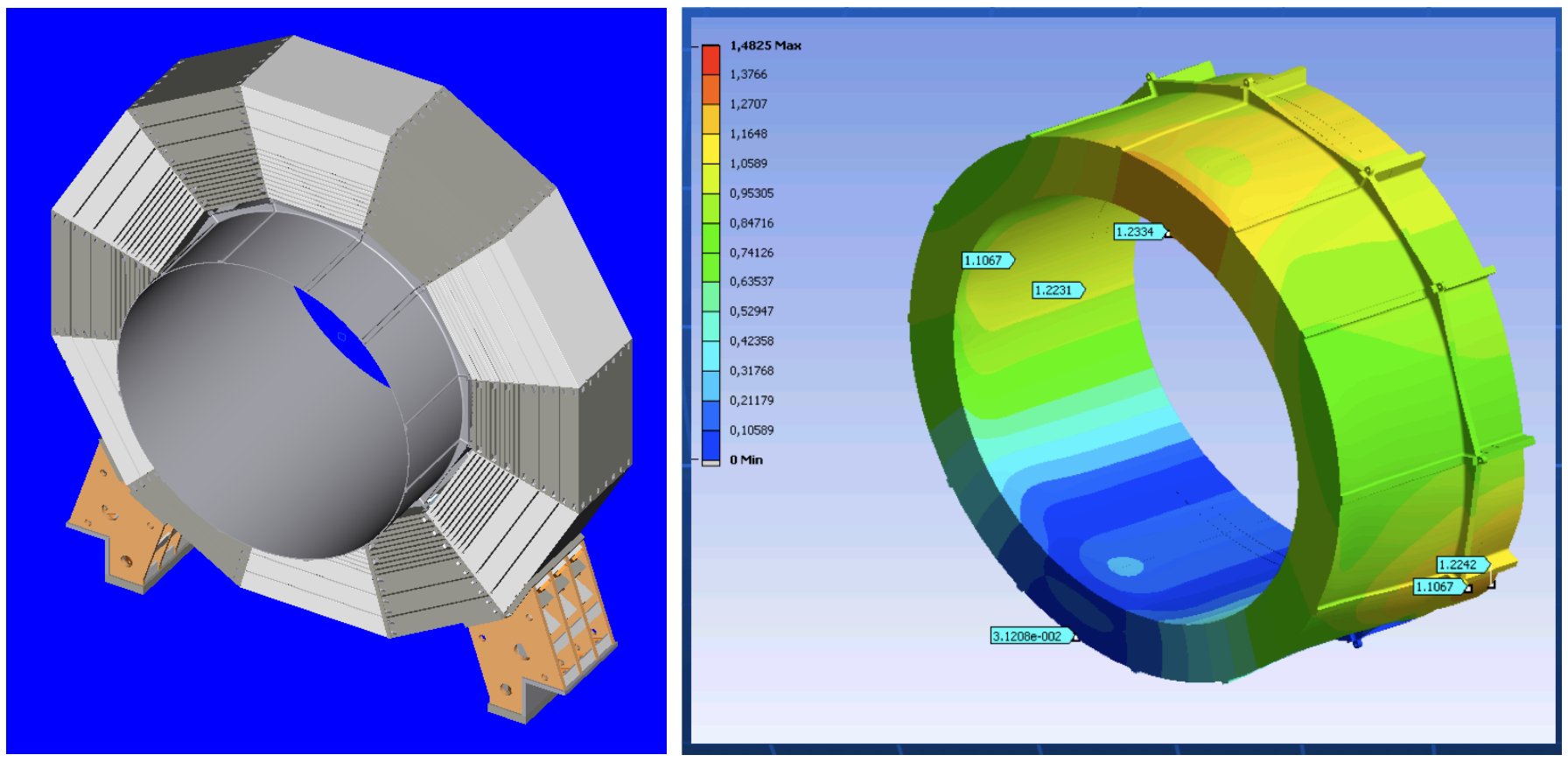

**Figure III-5.2** Integration of the solenoid cryostat and the central yoke ring (left). FEM simulation of the cryostat deformations under its own mass and the mass of the barrel detectors (right).

### 5.1.1.2 Hadronic barrel calorimeters

The hadronic barrel calorimeters are installed in the cryostat of the detector solenoid. The calorimeter modules are assembled in rings and are supported inside the cryostat. Two different mechanical absorber structures are under investigation. The structure of the analogue hadronic calorimeter (AHCAL) is shown in Figure III-5.3, the structure of the semi-digital hadronic calorimeters (SDHCAL) is shown in Figure III-5.4. Two rings of eight AHCAL modules form the barrel that is installed in the cryostat. All services for the modules are accessible from the open ends of the barrel. In the SDHCAL case, the barrel consist of five rings that are assembled from eight wedge-shaped modules each. The services for the calorimeter run in this case on the outside of the barrel. In both cases, the barrels are supported from rails in the cryostat. The mass of the HCAL steel absorber structures is of the order of 600 t. The expected distortions are in the order of a few millimetres.

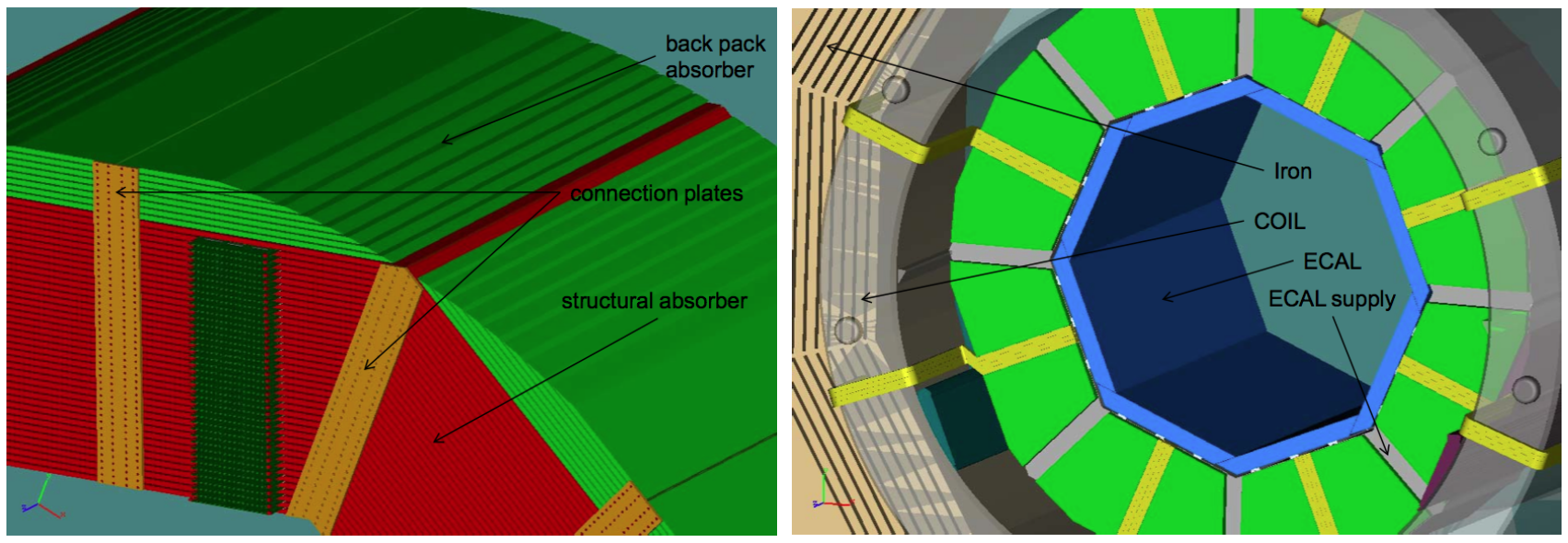

**Figure III-5.3**
Integration of the AHCAL structures into the cryostat.

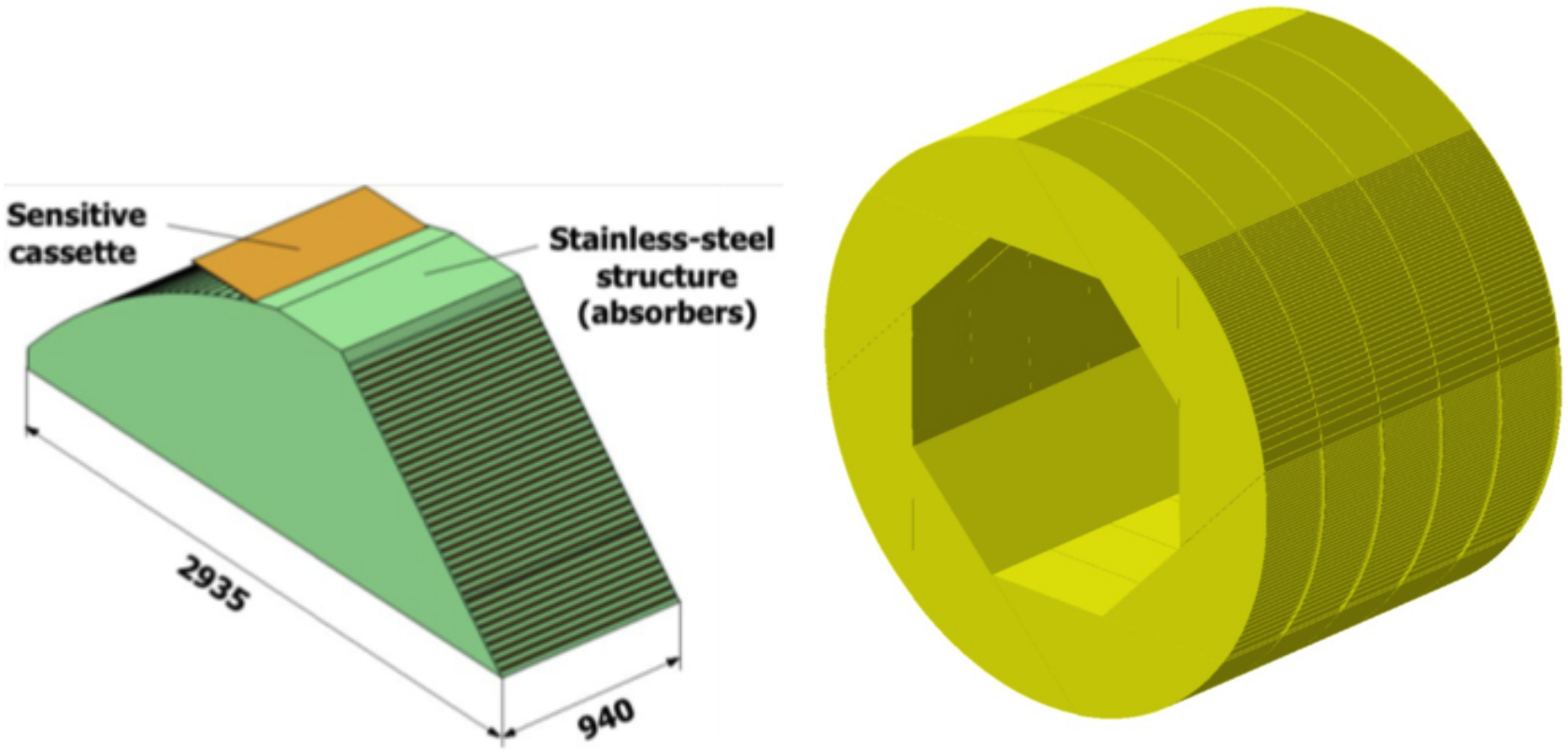

**Figure III-5.4**
Integration of the SDHCAL structures.

## 5.1.1.3 Electromagnetic barrel calorimeters

The electromagnetic calorimeter (ECAL) modules are supported by rails from the HCAL barrel modules. Figure III-5.5 shows the installation procedure of the ECAL modules. An external cradle that holds a rotatable support cage will be used during the installation phase.

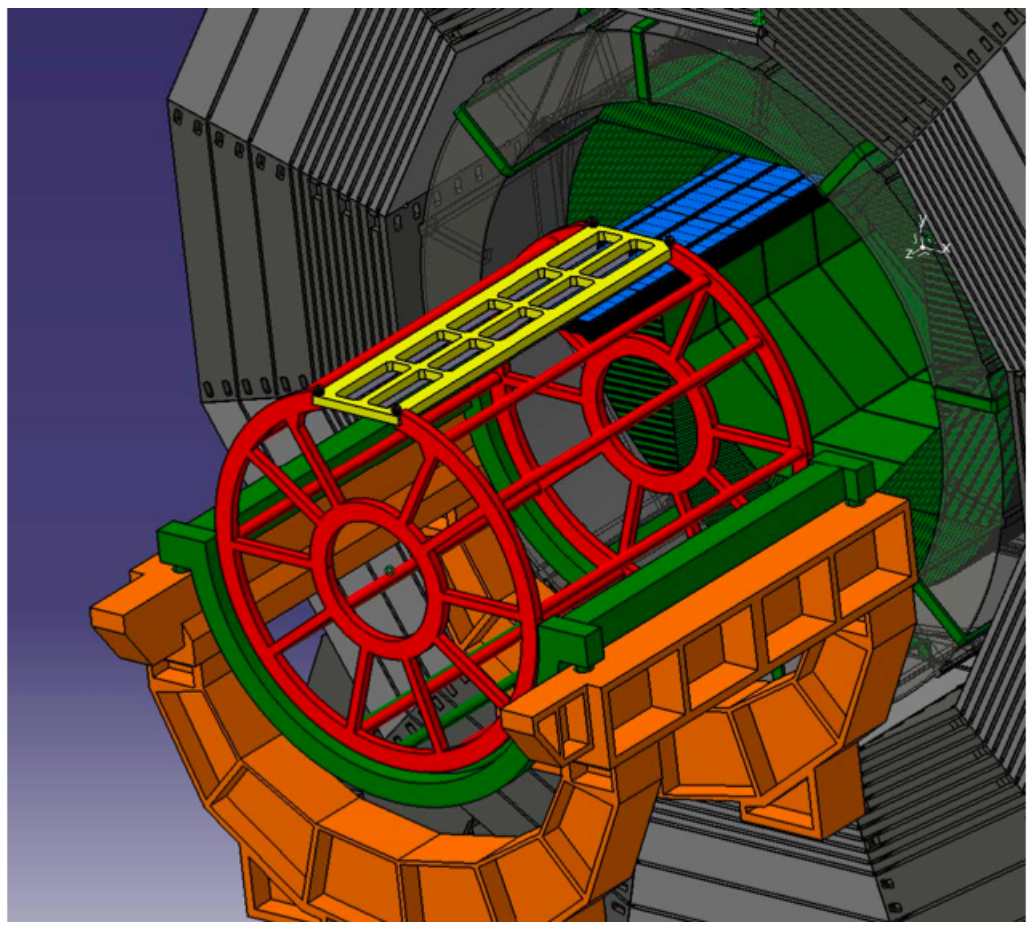

**Figure III-5.5**
ECAL installation.

## 5.1.1.4 Endcap calorimeters

The endcap calorimeters are supported from the endcap iron yoke. Figure III-5.6 shows the endcap assembly where the ECAL and HCAL endcap detectors are supported from the iron yoke endcap. The support for the HCAL endcap from the yoke needs to balance the bending of the iron yoke in the strong magnetic field (c.f. section 4.2).

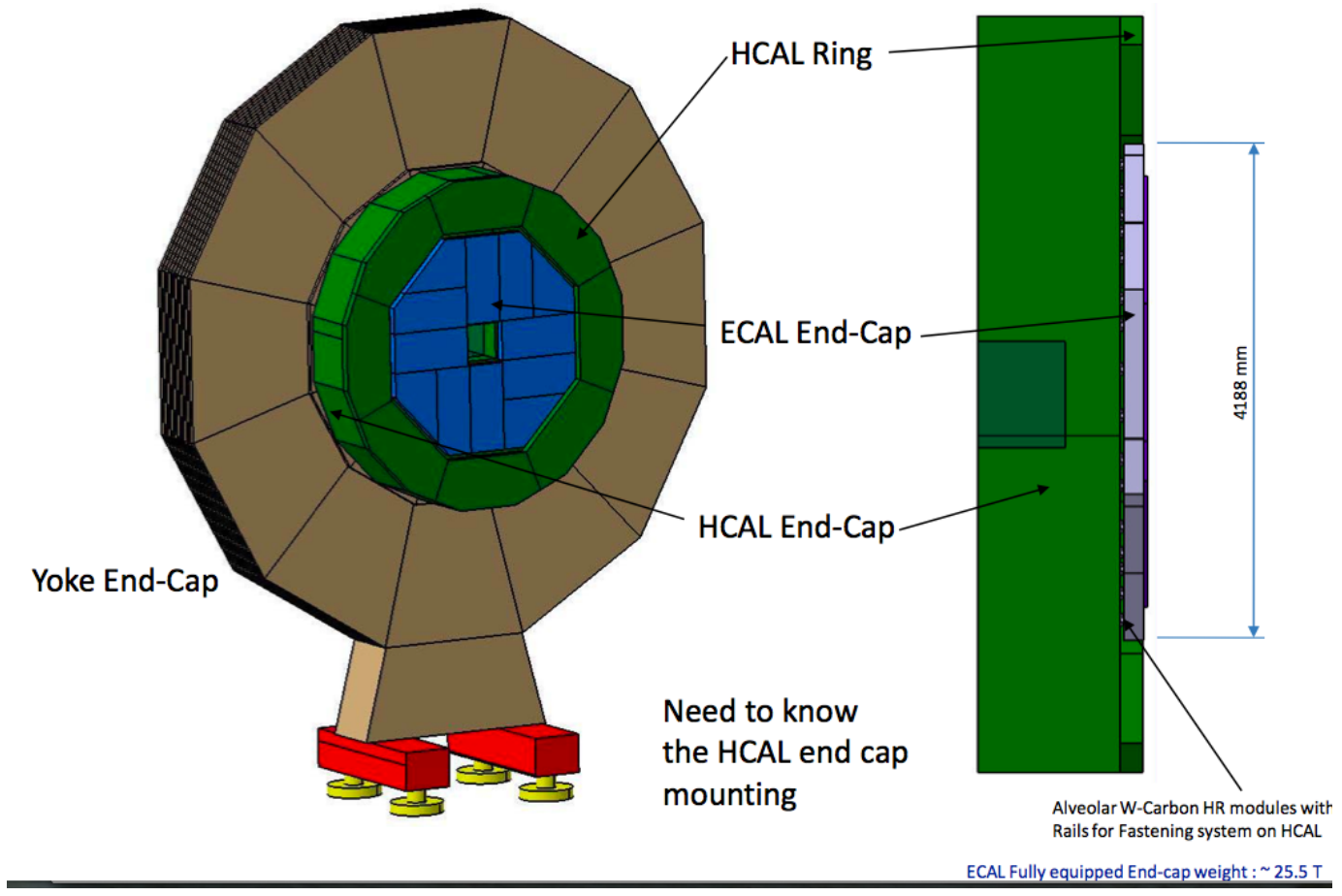

**Figure III-5.6**
The endcap detectors of the HCAL and the ECAL are supported by the yoke endcap.

### 5.1.1.5 TPC

The Time Projection Chamber (TPC) will be supported from the solenoid cryostat by a system of either double-T beams from lightweight CFRP or by a system of flat CFRP ribbons. Both systems will run along the front face of the HCAL barrel (see Figure III-5.7). The ribbon system needs less space in the endcap-barrel transition region, but requires an additional fixation of the TPC in longitudinal direction. A U-bracket with a spring based suspension would fix the TPC w.r.t. the ECAL barrel calorimeter.

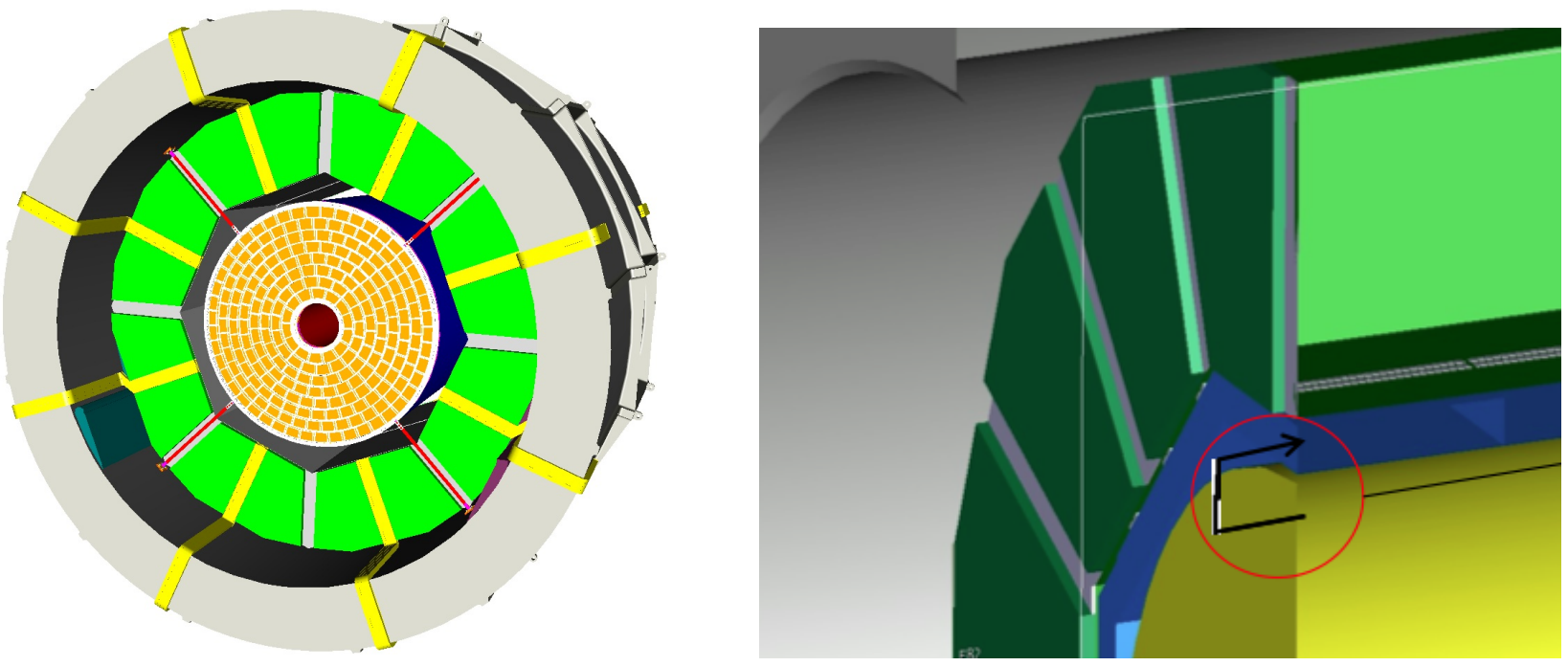

**Figure III-5.7**
TPC support from the cryostat.

### 5.1.1.6 Inner tracking system

The inner tracking system consists of the Silicon Inner Tracker (SIT), the Forward Tracking Disks (FTD) and the Vertex Detector (VTX). These detectors will be mounted together with the Beryllium beam pipe in the Inner Support Structure (ISS), as indicated in Figure III-5.8. The ISS consists of a CRFP tube that is fixed to the end plates of the TPC. This support system needs to be remotely adjustable to allow for alignment of the inner trackers and the beam pipe after a push-pull operation. As the push-pull system will align the overall ILD detector axis only to $\pm 1$ mm, a re-adjustment of the beam pipe might be necessary to keep the stay-clear margin between the beam pipe and the cone of background radiation at safe levels. Details of the inner detector system are described in [356].

### 5.1.1.7 Forward region

The forward detectors (c.f. section 3.5) LumiCal, BeamCal and LHCal are supported by the same structure that supports also the QD0 magnet (c.f. section 5.5.2). A support tube with a square cross section extends from the external pillar and is suspended from the coil cryostat with a tie-rod system. The support structure is a double-tube structure where the inner tube supports the QD0 magnet and the outer tube supports the forward detectors. This decouples the heavy masses of the calorimeters from the QD0 and eases the alignment procedures of the final focus magnets. A detailed view of the forward region is shown in Figure III-3.24.

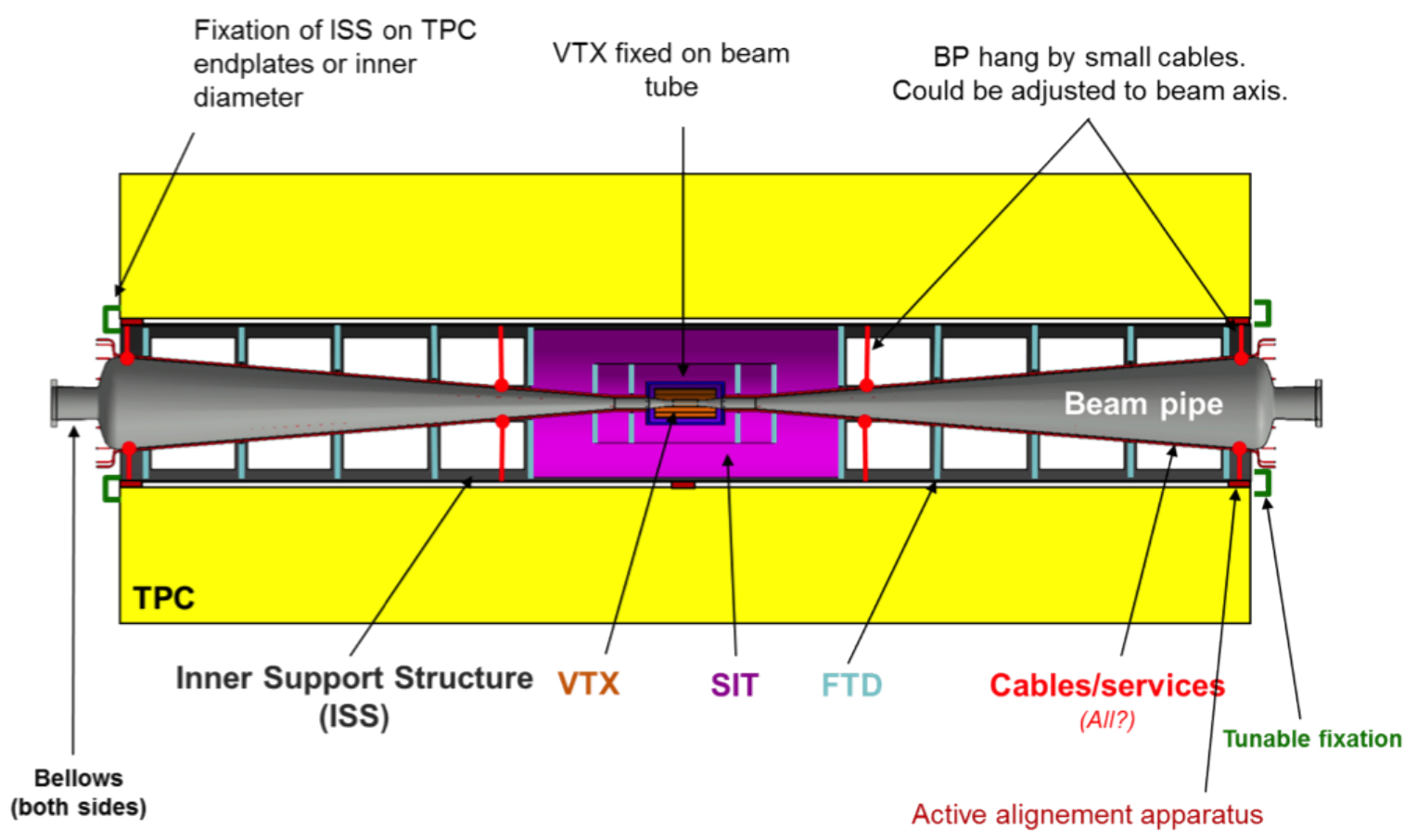

**Figure III-5.8**
Top: support of the inner tracking detectors. Bottom: schematic representation of the cable distribution along the beam pipe (from IP to the position of TPC endplate).

## 5.1.2 Detector assembly

### 5.1.2.1 Non-mountain sites

The main assemblies of ILD are the five rings of the iron yoke, three in the barrel part and two end caps. The assembly scenario will be similar to the CMS experiment at the LHC:

The detector will be pre-assembled and tested in a surface building. The large sub-assemblies will then be lowered into the experimental hall through a large vertical access shaft. The dimensions of the shaft and of the (temporary) crane for these operations are given by the masses and dimensions of the biggest assembly piece. In the case of ILD this would be the central yoke ring, which carries the solenoid coil. The size and mass of this biggest piece drive the requirements for the central shaft diameter (18 m) and the capacity of the hoist crane (3500 t).

The five yoke rings are mounted on air pads and can therefore be moved easily within the underground experimental hall. In the beam position and during the push-pull movement, the detector is mounted on the transport platform. In the maintenance position, the detector can be opened and the yoke rings can move independently away from the platform. Figure III-5.9 shows the detector opened for maintenance, in the beam position and in the maintenance area. The hall layout needs to foresee enough space in the maintenance position to allow the complete opening of the detector rings. Access to the inner detector parts and, in the maintenance area, the removal of large detector components (e.g. the time projection chamber) needs to be possible.

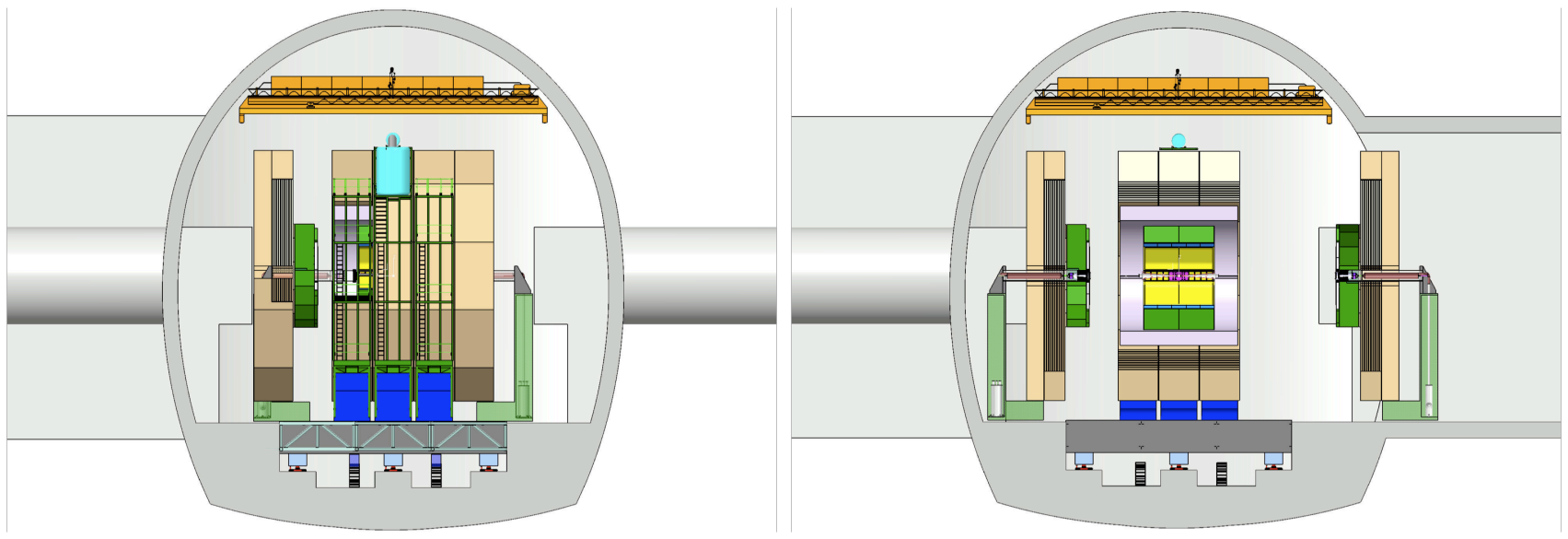

**Figure III-5.9**
ILD detector opened on the beam line (left) and in the maintenance area (right) [198].

#### 5.1.2.2 Mountain sites

Possible ILC sites in Japan are different to the other reference sites as they are situated in mountainous regions where a vertical access to the experimental hall might not be possible. Instead, horizontal tunnels of $\approx$ 1 km length will serve as access ways into the underground experimental area. As the tunnel diameters and the transport capacities are limited for technical and economic reasons, a modified assembly scheme for the ILD detector is applied for these sites: In these cases, it is foreseen to also pre-assemble most detector parts on the surface. However, the yoke rings are too big and heavy and can only be assembled in the underground hall. The yoke would be transported in segments into the hall where enough space for the yoke assembly and the necessary tools need to be provided. The largest part of the ILD detector, which should not be divided and therefore needs to be transported in one piece, is the superconducting solenoid coil. Its outer diameter of $\approx$ 8.7 m puts stringent limits on the diameter of the access tunnel.

The detector assembly procedures in mountain sites are part of an on-going optimisation process that needs to balance the requirement for space - linked to the time needed for the detector assembly - and the cost of the underground caverns.

### 5.1.3 Service paths and interfaces

A number of services (cables, cooling, gases) are needed for the operation of the ILD detector. The understanding of the needs and the analysis of their distribution inside the detector are major issues of the integration and mechanical design studies. Figure III-5.10 shows the main service paths

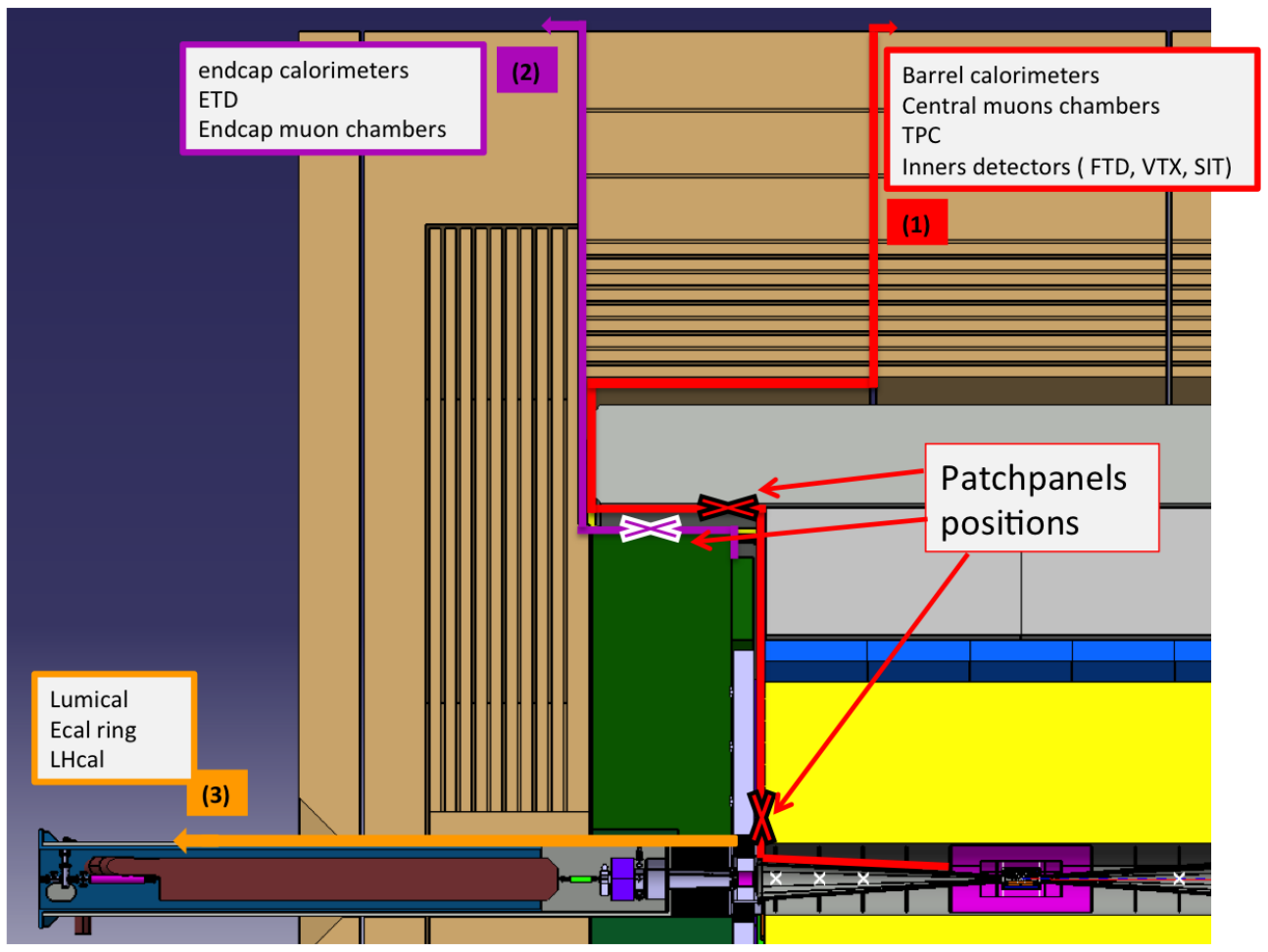

**Figure III-5.10**
Illustration of the main service paths in the ILD detector.

within the ILD detector. The routing of services is foreseen as follows:

1. All the services of the barrel detectors will be routed outwards via the endcap/barrel gaps, then along the outer radius of the coil, and finally between the central yoke rings. The assembly procedures of the inner detectors (SIT, FTD, VTX) and the volume of cables associated to each, imply that all the inner detector services will follow the same way.
2. The endcap detector services will run in the same gap, up to the coil outer diameter and will then be fixed on the return yoke endcap.
3. The forward components (forward calorimeters, QD0 magnets, support structures) will be built as one unit, and the required services will be distributed along the QD0 support structure.

The locations of the cable patch panels are under study, taking into account:

- assembly and maintenance procedures;
- the power distribution considerations as power convertor positions need to be chosen to limit the voltage loss in the cables;
- optimisation of the overall volume of services;

- material budget and impact on physics performance.

The requirements for both service lines and patch panels define the space that needs to be reserved in the detector for gaps. The geometry of these gaps is a major issue to be optimised as these gaps are dead zones in the detector. In addition to the space that is needed for services, additional dead zones are needed for construction tolerances, mechanical deformation of detector parts under loads (gravitational and magnetic), space needed for integration and assembly, alignment tools. etc..

Each sub-detector group has begun to define the amount of cables they foresee for power distribution and signal transmission as well as the power consumption. The later is particularly critical as it is needed to define the cooling method and the distribution of fluids. Currently, the amount of cables for the barrel calorimeters and the TPC is estimated to be small with respect to the numbers of channels in these sub-detectors, less than 3000 cables per side of the detector. For the inners part (SIT, FTD, VTX) up to 600 cables per side might be needed. This represents an average are of about 2000 $cm^2$ occupied by cables and piping at the radius of the HCAL. Figure III-5.11 shows the cable paths and their occupancies in the gap between barrel and endcap detectors.

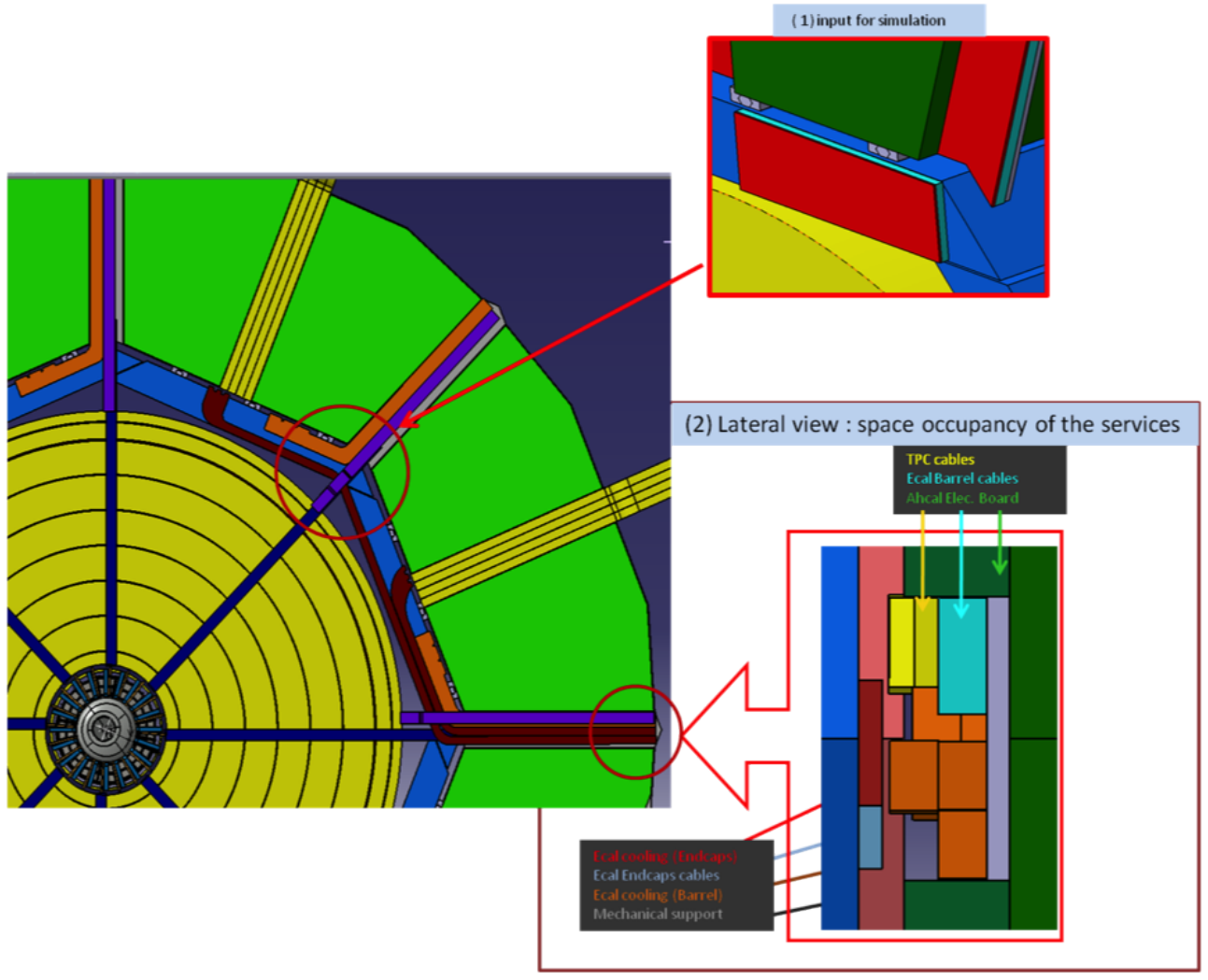

**Figure III-5.11** Front view of the barrel calorimeters (green/blue), TPC (yellow), inner part and of their associated services. (1) The volume occupied by cables and services in each way-out has been translated into equivalent thickness of conductor and insulator to be implemented in the simulation model. (2) Lateral view of one way-out, with representation of the space needed per sub-detectors services.

The inner detector layout is very challenging as the amount of cables from the inner silicon detectors (VTX, FTD, SIT) represent dead material immediately around the beam pipe and may become a source of background. It is presently estimated to be some few percent of $X_0$ at some positions of the beam pipe. In addition a mass of some few kg of material needs to be supported by the light structure of the beam pipe. Specific R&D on the definition of the cables according to the nature of the conductor and the optimisation of the insulator is mandatory in order to minimise the effects of the services on the physics performance of the detector. A schematic view of the cable routing in the inner detector is shown in Figure III-5.8.

### 5.1.4 General Safety Issues

The final ILD installation needs to follow the applicable safety rules, given by the collaborative aspect of the project and requested by the site retained. Some general rules have been established to allow the co-existance of the two detectors, ILD and SiD, in one underground experimental hall (c.f. section 2.3). Other safety aspects need to be respected in the integration scheme for ILD. Among others, this covers:

- Mixing of flammable gases will be done in protected areas and only non-flammable mixtures will be sent to the experimental cavern;
- Fire prevention will be an important feature of the integrated design - a particular worry is to prevent fire from propagating inside the vaccum vessel of the coil, e.g. caused by the explosion of a super-capacitor;
- General protection against earthquakes will be developed, to protect the ILD detector during assembly, maintenance and operation.

## 5.1.5 ILD modelling

Three different types of models are being used for the design of the ILD detector. While engnieering and placeholder models are needed for the mechanical design of the detector, physics simulation models are used to study the detector performance:

- Placeholder models are used for global integration purposes. They describe the boundaries and volumes of the sub-elements and enable fast integration, checks for conflicts and compliance of the interfacing components. They also include reserved space that is needed for assembly purposes and tolerances. Different technology options for sub-detectors need to fit into the global sub-detector placeholder to enable and check plug compatibility;
- Detailed engineering models of the sub-detectors form the basis of the construction. They define how to assemble a component from parts and provide exact geometry and material description. Detailed models exist for each sub-detector option and are the basis of the cost evaluations;
- Physics simulation models are used in the Monte-Carlo simulations of the detector performance. They describe the segmentation, shape, and physics behaviour of the active and passive components.

While the placeholder and the detailed engineering models are usually derived from CAD systems, the ILD physics model is part of the Geant4 based full detector simulation MOKKA. Figure III-5.12 shows, for the example of the ECAL barrel detector, the three model types.

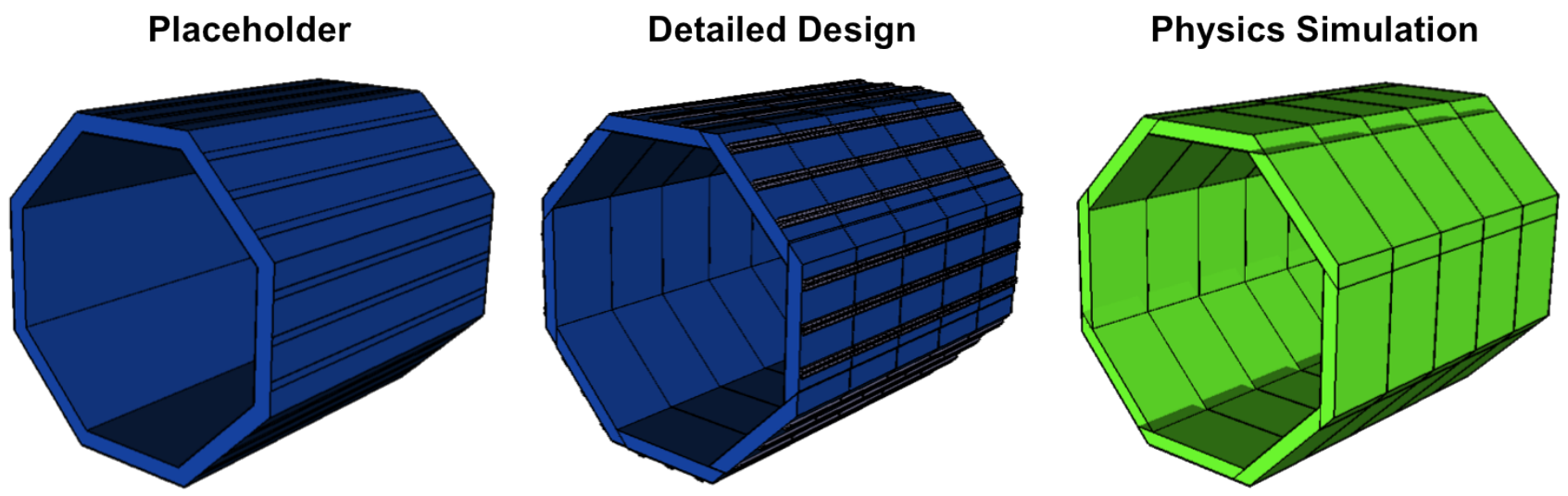

**Figure III-5.12** Different models describe the ILD detector (this example: the ECAL barrel).

While the CAD models (placeholders and engineering) are by default stored in the ILC Engineering Data Management System (ILC-EDMS) [357], a process has been set up to convert the geometric information from the MOKKA model into a 3D format that allows comparison with the engineering models using the design analysis tools. Figure III-5.13 shows an overlay of the simulation and the engineering model of the ECAL barrel. Differences and overlaps are colour-coded so that the compatibility of the models can be checked quickly.

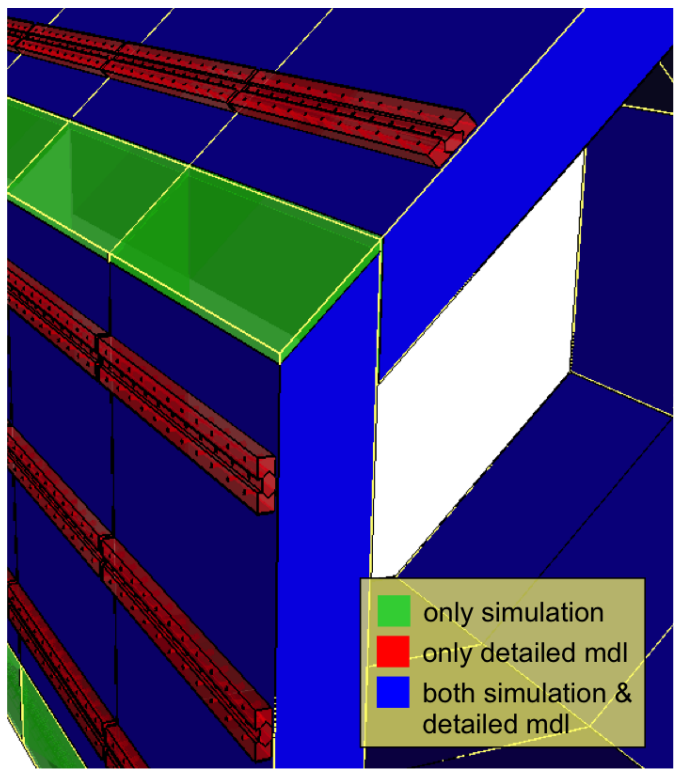

**Figure III-5.13** Geometry comparison of the simulation and the detailed engineering model of the ECAL barrel detector. The blue parts are in both models, while the red ones are only in the engineering model (labelled "mdl") and the green ones are only in the Geant4 description.

## 5.2 ILD alignment and calibration

### 5.2.1 Alignment of the tracking system

Achieving one order of magnitude improvement on momentum resolution over state of the art detectors imposes stringent requirements on the performance of the tracking devices and on the stability of their support structures. The list of structural properties needed for mechanical supports is long: low mass material, stiff and lightweight, stable to electric and magnetic fields, robust against temperature and humidity gradients. The particular conditions at the ILC add two new sources of instability: power pulsing of the electronics and push-pull operation. Power pulsing induces temperature changes around front-end electronics and vibrations that can be propagated as oscillation modes to higher order structures. Opening and closing of the detectors, and moving detectors between maintenance and on-beam position will demand a quick re-alignment of the full experiment.

The quest for spatial precision spans the life time of the detector. During construction, accurate positioning of individual active elements is granted by tight assembly tolerances and then measured using Coordinate Measuring Machines (CMM). Elements are arranged together into higher order structures (modules, supermodules and sub-detectors). Once the full sub-detector is assembled, global measurements of its degrees of freedom are obtained by standard survey techniques. During operation, hardware alignment systems can monitor internal stability of sub-detectors and even movements of sub-detectors with respect to each other. The ultimate micrometer-level geometrical description of the experiment is achieved by means of exhaustive but time consuming track alignment algorithms. Tracks have very little sensitivity to deformations of the support structures that make $\chi^2$ invariant, the so-called weak modes. Some of these will be measured using an innovative monitoring of structural stability in ILD: Fiber optical sensors laid inside the support structures are sensitive to changes that induce strain in the fiber and, therefore, deformations of the supports can be recorded and taken into account in the analysis.

#### 5.2.1.1 Silicon sensor hardware alignment system

The internal hardware alignment of the Si-microstrip tracker uses infrared (IR) laser tracks to align consecutive layers of Si detectors. This system exploits the partial absorption/transmission of Si to infrared light, generating signals in consecutive sensors that are measurable by the readout electronics. The aluminum back-metallization of the sensor is removed locally to allow downstream propagation of the IR light, using the so-called alignment passages. The transmittance to IR light is maximized using the top and bottom passivation layers as an anti-reflection coating. IR optimized sensors produced at CNM-IMB (Barcelona, Spain) showed maximum transmittance values of 50% (30% increase with respect to untreated sensors) for sensors of 50 $\mu$m pitch.

The IR laser system uses optical fibers and collimators inside the tracker volume while the corresponding laser heads remain outside. It provides quick reconstruction ($<$ 1 min) of the positions of the measured modules with a relative resolution (between consecutive measurements) of 10 $\mu$m. Therefore, it can monitor shifts and rotations on a short time scale. It does not, however, distinguish global movements of the monitored structure from global movements of the laser beams.

#### 5.2.1.2 Structural and environmental monitors using fiber optical sensors

A fiber Bragg grating (FBG) is a type of distributed Bragg reflector embedded in a short segment of optical fiber that reflects particular wavelengths of light and transmits all others. Any strain (temperature, pressure, vibration etc.) at the grating will cause a shift and magnitude change of the reflections. This change allows for very accurate measurements of the magnitude of strain.

Distributed sensing is achieved by recording gratings for different wavelengths in the same fiber. Deformation, displacement, temperature, humidity etc. can then be sampled at different locations along the fiber. The light transmitted in the fiber is immune to electromagnetic disturbances, does not dissipate power and works in a wide range of temperatures. Due to the negligible loss of light in the fiber, all readout electronics for this system will be placed outside of the experiment.

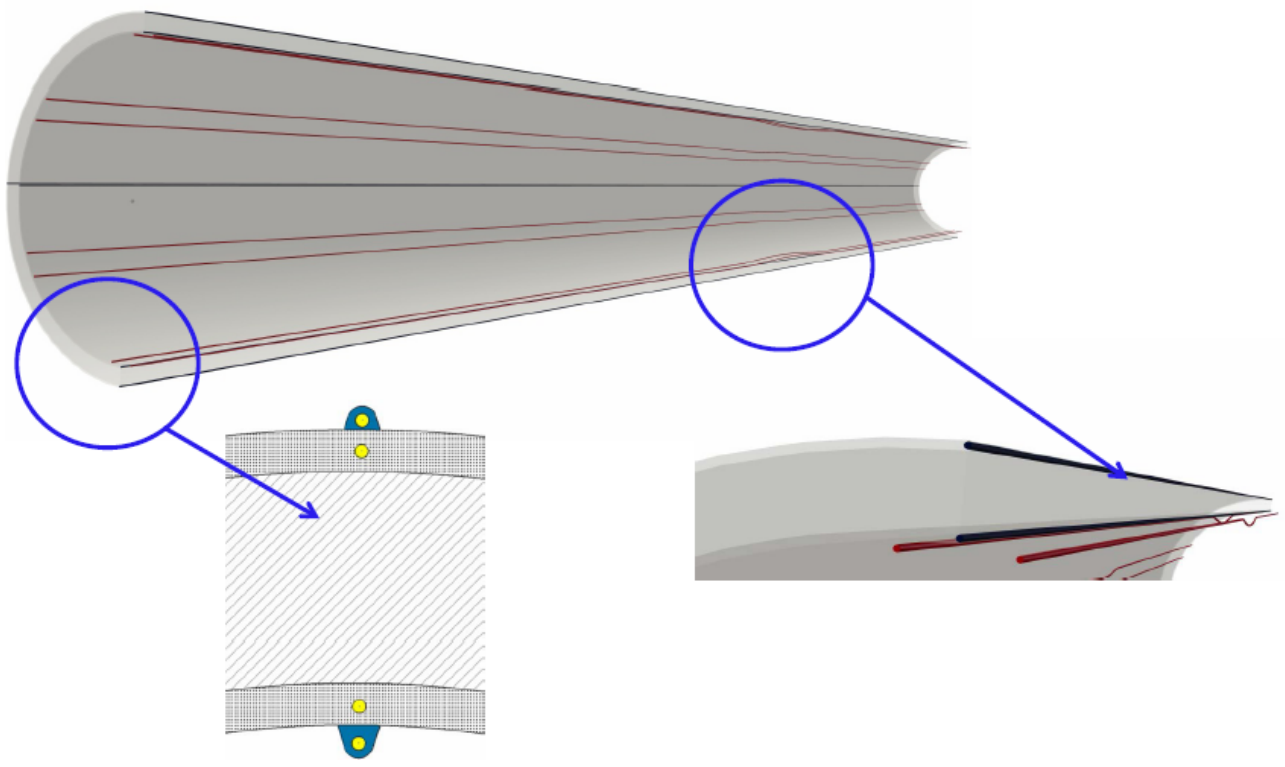

**Figure III-5.14** Embedding of fibers along the tracker support structure for structural monitoring of distortion, temperature, humidity, etc..

In ILD, fiber optical sensors will be embedded in carbon fiber reinforced plates. The measured deformation will help to determine the shape of the support structure, measure its displacements with respect to locking points, and calculate temperature distribution and gradients. In particular, these smart support structures will be able to detect and diagnose weak modes. Figure III-5.14 shows a cylindrical support structure with fibers inserted near the inner and outer boundaries. From the difference in traction and compression from top and bottom fibers, the shape of the structure can be obtained.

### 5.2.1.3 Track-based Alignment

Individual tracking sensors have an excellent intrinsic spatial resolution. During construction, the position of sensors inside each module can be measured with a precision of $\sim$5 $\mu$m. Modules are then assembled into higher order structures, positioned and surveyed with uncertainties in the range of 200-500 $\mu$m. Hardware alignment procedures reduce this uncertainty to a level of about 100 $\mu$m, such that tracks can be reconstructed by software alignment algorithms. Track alignment takes the task of reducing alignment uncertainties below the intrinsic sensor resolution. To reach the required precision reliably the track based alignment must include the constraints from construction, survey and hardware alignment (laser system and fiber Bragg system) into the global alignment procedure. This will dramatically reduce the number of degrees of freedom and speed up alignment after push-pull.

Typically, the alignment sample is composed of a mixture of collision data and tracks from cosmics and beam halo. These tracks are useful as they allow to relate different parts of the detector (upper and lower half, both end-caps). Tracks with known momentum are extremely valuable, both as a means to determine some of the weakly constrained alignment parameters and as a monitoring tool to validate the alignment. This role has traditionally been played by tracks from resonances with a well-known mass (Z, J/$\Psi$ and $\Upsilon$ resonances).

The alignment precision should be such that the momentum resolution of the tracker is degraded by less than 5%, which leads to alignment constraints of $\sim$2 $\mu$m for the vertex detector, $\sim$4 $\mu$m for the inner silicon tracking, $\sim$ 6 $\mu$m for the outer silicon tracking and $\sim$ 20$\mu$m for the TPC.

### 5.2.1.4 Pixel alignment

The pixel system will be aligned in two steps. Within each layer, alignment will be achieved using hadronic tracks crossing the overlap region between ladders (a few thousands per day at nominal beam parameters). Across different layers and with respect to the rest of the detector, tracks from muon pairs at the Z-pole (estimated several thousands in a day) will be used.

#### 5.2.1.5 Alignment of the silicon strip tracking system

The total number of degrees of freedom of the ILD tracker is of the order of $10^5$. The time required to accumulate sufficient tracks to resolve all of them will be of the order of months, comparable to the push-pull period.

After a push-pull operation the goal of ILD is to obtain a quick re-alignment. If the relative sensor positions inside a module are known to 5 $\mu$m precision (from construction data) the number of degrees of freedom will be reduced by a factor 2-10, depending on the sub-detector. Another sizeable reduction will be obtained if the internal degrees of freedom of the subsystems (ladders and rings, cylinders and disks) are stable over time. This requires a careful design of the mechanical support of the detector. Relative movements of these rigid bodies can be monitored using the laser alignment system and the embedded fiber optical sensors. In total, 26 degrees of freedom will be present if only the sub-detectors need to be re-aligned against each other.

#### 5.2.1.6 Alignment after Push-Pull

Push-pull of the full detector between the interaction point and the garage position and opening and closing of the experiment for maintenance are evident sources of misalignment. To facilitate maintenance and accessibility to different detector regions, a modular detector design will allow to open the detector by pulling it on a combined system of air and grease pads. A system able to control the relative distance from bulky elements to delicate objects like the beam pipe has been implemented in the CMS experiment with a distance measuring system, using about 200 sensors (contact and non-contact distance meters) to accomplish this task. This system will ensure precise and safe handling and overall repositioning of the experiment. Internal alignment of each sub-detector has to be obtained using hardware alignment and cosmics in the maintenance area.

A special task is the alignment of the final focus magnets (QD0). An optical alignment system based on RASNIK sensors has been proposed for CLIC [358]. The alignment requirements at CLIC are about 5-10 times more stringent than at the ILC. An adaptation of this system to ILD is under investigation.

### 5.2.2 Calorimeter calibration

This section discusses general calibration issues for particle flow calorimeters. Aspects specific to each technology are summarised in the subsystem sections. The discussion is limited here to the single particle energy calibration; corrections at jet level are part of the particle flow reconstruction and need to take tracking information on an event-by-event basis into account.

Calibration as a general term is used for several aspects of the calorimeter reconstruction. For the channel-to-channel normalisation we use the term equalisation, to be distinguished from the corrections of time-dependent effects, induced for example by temperature or pressure variations. Tracing such variations is called monitoring. Establishing an absolute scale in units of GeV is again a separate task, and different scales, electromagnetic, hadronic or weighted scales need to be distinguished. If applied at particle level, they may depend on the clustering definition. Other corrections, like for dead materials, may be applied at particle level, too.

A common feature of all particle flow calorimeter technologies is their relative insensitivity to any sort of stochastic calibration or alignment uncertainty. The large number of cells required for the topological resolution is an asset rather than a burden, since the precision with which these effects need to be known scale with $\sqrt{N}$, where $N$ is the number of channels.

In contrast, coherent systematic effects must be corrected with higher precision, depending on the fraction of the detector affected. If it is the entire calorimeter, the precision required is given by the constant term aimed at, about 1% for the ECAL and 2-3% for the HCAL. The challenge of the high granularity is that time-dependent corrections cannot be applied at cell level, and cell-wise corrections require stability over time to reach statistical precision. On the other hand, since every cell is individually read out, one is free to form averages over space or time according to the specific problem, but finding the optimal averaging procedure and identifying the leading effects is often an involved analysis and intimately related to understanding the detector and its systematics. The procedures needed in practice can only be developed from real data. Such studies form an important part of the test beam data analysis, and they are also the reason why the next generation of prototypes must undergo beam tests at system level again to obtain realistic performance figures.

#### 5.2.2.1 Calibration scheme

The calibration of the electromagnetic and hadronic response of the calorimeter proceeds in the following general steps:

1. Test bench characterization of sensor parameters at cell level
2. Inter-calibration of the electronic response of all individual cells using muon test beams, and conversion to the MIP scale
3. Verification of the electromagnetic scale and linearity using electron beams impinging directly on the detector modules
4. Determination of the hadronic response using hadron test beams
5. Determination of combined ECAL and HCAL hadronic response, including weighting procedures
6. Verification of dead material corrections at inter-module connections using hadron test beams
7. In-situ validation and monitoring using kinematic constraints, tracker information and track segments in hadronic showers

#### 5.2.2.2 Channel equalisation

On the test bench, one measures parameters like, for example, gain, efficiency, dependence on operating conditions, and the non-linear response function of the photo-sensors. Also time-consuming threshold scans can be done to equalise zero-suppression DACs at the input stage of the front end electronics to compensate for channel-to-channel gain variations. Measurements are done at different production stages, for example before and after integration into the scintillator tiles. Already for the beam tests this was done for ECAL and HCAL prototypes using semi-automatic procedures, and studies are underway to automatise this further.

The inter-calibration with muon beams must be done for all cells and all detector layers. Thanks to the modular design, this can be done with the bare active layers before insertion into the absorber, or with the assembled modules. In the CERN test beam 12 hours were needed for the AHCAL to acquire sufficient statistics on a stack with a square meter front face and 38 layers. This would translate into about two months for the entire ILC detector, or less, if more layers are aligned after each other in the beam. Likewise, the analysis of the calibration data and the fits to the pulse height spectra can be massively parallelised.

For the silicon ECAL with its excellent stability and linearity, only one electronic conversion factor per channel is needed. For scintillator options, due to the non-linearity of the photo-sensors, two inter-calibration constants per channel are required, the MIP and the single photon response. For gaseous detectors, the response is proportional to the product of efficiency and pad multiplicity which are also determined with muon beams (for the three thresholds).

#### 5.2.2.3 Energy scales

The response to electromagnetic showers on the MIP scale can be uniquely predicted by simulations and verified in test beams with known energy. The electromagnetic (em) scale is thus related to the MIP scale by a simple conversion factor MIP, or hit per GeV. In practice, the usefulness of the em scale depends on the linear range for electrons. While for the silicon ECAL, no significant non-linearities were found, the scintillator detectors start to deviate at the 2% level around 50 GeV.

The hadronic response for the AHCAL shows small deviations from linearity (less than 2% up to 80 GeV), mainly due to the non-compensating features of the structure ($e/\pi \sim 1.19$) [315] and to leakage. It can be predicted by simulations and was verified with test beam data. For the gaseous calorimeters, the hit multiplicity noticeably deviates from a linear behaviour for energies above 30 GeV.

In the semi-digital option, (2 bits per cell), weighted combinations of the hit multiplicity for each threshold are used. With this procedure, a linearity at the percent level was achieved up to 80 GeV. The combined ECAL and HCAL measurement requires the application of weighting factors in any case. Making use of the fine granularity, or even of resolved shower sub-structure, allows to significantly improve the resolution. The weighted energy scale depends on the applied algorithm.

Additional corrections will be necessary to account for uninstrumented regions or additional material from support structures, electronics and service lines, at the ECAL-HCAL transition and at inter-module boundaries. This must be extracted from simulations which need to be benchmarked in test beams with realistical ECAL and HCAL prototypes combined.

Apart from the inter-calibration, which must be done for every individual active detector element, we assume that all studies addressing the absolute em and weighted scalers can be done with single representative sample structures. One such sample structure should be immersed into a magnetic field to verify the modeling of magnetic effects.

#### 5.2.2.4 Monitoring techniques

The above calibration scheme needs to be complemented by monitoring techniques in order to take time-dependent variations into account when the test beam based calibrations are applied to collider data. The general approach is that if the MIP scale - or the MIP hit multiplicity - is maintained and under control, all derived scales are stabilised as well.

Test beam experience has demonstrated that the MIP scale of the silicon-based ECAL is intrinsically stable. Variations of the MIP scale of scintillation detectors are mainly due to changes of the electronic response of the photo-sensor, induced by changed thermal conditions, whereas the hit multiplicity of gaseous detectors mainly varies as a consequence of temperature or pressure-induced gain variations.

Apart from slow-control recording of bias voltages, temperatures and pressures, the monitoring is based on mostly innovative techniques, namely in-situ MIP calibration using track segments in hadron showers, and, for photo-detectors, auto-calibration of the photo-sensor gain using LED light. This can be extracted from the spacing between peaks in the pulse-height spectrum attributed to small, discrete numbers of registered photo-electrons, and does not require LED light stability.

In principle it is also possible to adjust the voltage in order to compensate the temperature variation, and use the gain to watch the stability. This will be tried with the second generation prototypes now under construction. The use of radioactive sources is not necessary according to present understanding.

Changes in the amplification of the read-out chain were checked independently and found to be much smaller than those of the sensors. We therefore do not discuss them further here; they are absorbed in the other corrections. The pedestals of the read-out electronics are regularly monitored using random trigger events; this also detects and monitors dead or noisy channels.

#### 5.2.2.5 In-situ calibration

The absolute calibration of the ECAL can be verified and adjusted by comparison with the tracker or using electrons and photons kinematically constrained like Bhabha's or return to the Z. This does not require any running at the Z peak.

Due to the underground location, the orientation of the detector layers, the power pulsing, and due to the high granularity, cosmic rays might not be sufficient for monitoring the MIP scale in-situ. However, thanks to the excellent imaging capabilities of the calorimeters, MIP-like track segments can be identified in hadronic showers and used for calibration purposes. This has been demonstrated using the CALICE AHCAL test beam data, and the potential for in-situ calibration of the ILD detector was studied in simulations - for details see [318]. Although typically two tracks are found in each shower which are used for the calibration of 20 cells, it is even at the Z resonance not possible to obtain a channel-by-channel calibration within realistic running times. However, the method is well suited for the determination of average corrections for a sub-section of the detector, e.g. a layer in a module.

At the Z pole, 1 $\mathrm{pb}^{-1}$ is sufficient to provide at least 1000 identified tracks per layer module out to AHCAL layer 20, while 20 $\mathrm{pb}^{-1}$ are necessary to reach out to the last AHCAL layer, layer 48. For the last layers in the calorimeter, also $Z^0 \rightarrow \mu^+\mu^-$ events contribute significantly to the overall statistics, reducing the required integrated luminosity to 10 $\mathrm{pb}^{-1}$.

At 500 GeV, significantly larger integrated luminosities are necessary to achieve the same precision due to the much lower cross section. Less than 2 $\mathrm{fb}^{-1}$ will allow for a 3% calibration for each layer-module out to layer 20, so even at full energy running a monitoring of the calibration on the layer-module level will be possible. Also here, muons contribute to the calibration of the last layers in the detector.

Similar luminosity are expected to be necessary for the semi-digital gaseous detectors, the decreased statistics due to small cells (by a factor of 9), being compensated by the binomial statistics governing the efficicency determination (with a average 95%) and the relative uniformity of the sensors. The relative weights of the hit population above the 3 thresholds can be monitored from Z Z channels with one of the Z decaying into hadronic channels; charged hadronic with energy measured in the tracker can also be used to control the previous population ratios for different energies. For reference, the possibilities with cosmic muons were also studied. At the surface, the rate of cosmic muons with energies above 10 GeV (necessary to ensure penetration through the complete detector) is approximately 20 $\mathrm{Hz/m}^2$. Taking the duty cycle of the electronics of 0.5% into account, the detectable muon rate reduces to 0.1 $\mathrm{Hz/m}^2$. The area of one layer module is around 2.5 $\mathrm{m}^2$, so for horizontal calorimeter layers about 70 minutes would be sufficient to acquire 1000 tracks. In underground locations, this needed time will increase with increasing depth. For non-horizontal layers, in particular also in the endcaps, the needed time is significantly higher, and in deep underground, this will presumably be prohibitive. However, also cosmic muons will be a valuable calibration and monitoring tool for parts of the detector. For the endcaps, there is the additional possibility of using muons from the beam halo. Their rate depends strongly on the shielding of the detector, but is expected to be between 100 $\mathrm{Hz/m}^2$ and 10 $\mathrm{kHz/m}^2$ at full energy. The fact that these muons arrive in time with the beam, and thus don't suffer an effective rate reduction due to power pulsing, make them well suited for the calibration of the end-cap calorimeter. Even a cell-by-cell calibration using these muons might be possible.

#### 5.2.2.6 Required accuracy

Using fully detailed simulations of the ILD detector and reconstruction based on the Pandora particle flow algorithm, we have modeled different scenarios of statistically independent as well as coherent mis-calibration effects, affecting the entire HCAL or parts (module layers) of it. Purely statistical variations, like those arising from calibration errors or random aging effects, hardly affect the energy resolution at all. However, they may degrade the in-situ MIP calibration capability. From this, a moderate requirement of the inter-calibration stability to be ensured by hardware design of $\pm 10\%$ is derived.

Coherent effects which could for example arise from uncorrected temperature variation induced changes of the response are potentially more harmful, as they directly show up in the constant term, if they affect the entire detector. However, these are easy to detect, and even a 5% variation only mildly propagates into the jet energy resolution. Systematic effects shifting sub-sections like layers are unnoticeable unless they exceed about 15%, comfortably in range of the in-situ calibration method accuracies.

We have demonstrated the validity of these simulation based estimates by treating our AHCAL test beam experiment like a collider detector, using cell-by-cell inter-calibrations only from data taking at a different site, under different conditions and after having it exposed to disassembly, transport and re-assembly influences. Applying only in-situ monitoring techniques, we re-established the scale and reproduced the resolution. Imperfections absent in any simulation showed up, but were successfully compensated.

#### 5.2.2.7 Conclusion on calibration

All in all, we conclude that the high granularity and channel count is a blessing rather than a curse. On one hand, due to the law-of-large-numbers suppression of statistical effects, the requirements on individual cell precision are very relaxed. Coherent effects, on the other hand, can be studied with any desired combination of channels, be it layers, longitudinal sections, electronics units or according to any other supposed hypothesis of systematic effects. The high degree of redundancy and the full information for each channel provide maximum freedom, without having to rely on intrinsic homogeneity as in the case of internal, in-transparent optical or analog summing in less finely segmented readout. Testing the second generation prototypes under beam conditions will be an important step towards working out the procedures for the full detector in detail and demonstrate their performance.

## 5.3 ILD data acquisition and computing

The DAQ system for the ILD concept has to fulfill the needs of a high luminosity, high precision experiment without compromising on rare or yet unknown physics processes. Although the average collision rate of the order of a few kHz is small compared to the LHC, peak rates within a bunch train will reach several MHz due to the bunched operation. In addition, the ILC physics goals require higher precision than has ever been achieved in a colliding beam experiment. This increased precision is to a large extent achieved through increased granularity and thus leads to a substantially bigger number of readout channels than that used in previous detectors

Taking advantage of the bunch train operation at the ILC, event building without a central trigger, followed by a software-based event selection was proposed in [359] and has been adopted for ILD. This will assure the needed flexibility and will be able to cope with the expected complexity of the physics and detector data without compromising on efficiency or performance. The only foreseeable drawback from bunched operation is the reduction of cosmics event recording and their use for calibration and alignment.

The LHC experiments have up to $10^8$ front-end readout channels and a maximum event building rate of 100 $\mathrm{kHz}$, moving data with up to 300 $\mathrm{GB/s}$ (with an average throughput of $\lesssim 200\,\mathrm{GB/s}$ required [360, 361, 362]). The proposed ILD DAQ system will be less demanding in terms of data throughput but the number of readout channels is likely to be a factor of 10 or more larger. The computing requirements for event processing at the ILC, in terms of storage and CPU, are also going to be less demanding than those of the LHC experiments. The details of the DAQ and computing system depend to a large extent on the developments in microprocessors and electronics and the final design of the different sub-detector electronic components. Therefore the DAQ and computing system presented here has to be rather conceptual, highlighting some key points to be addressed in the coming years.

In contrast to past and recent colliders such as HERA, Tevatron or LHC, which have a continuous rate of equidistant bunch crossings, the ILC has a pulsed operation mode. The nominal parameter set [332] of the ILC at 500 $\mathrm{GeV}$ ($1\,\mathrm{TeV}$) with

- 1312 bunch crossings in a train about $1\,\mathrm{ms}$ long,
- 366 $\mathrm{ns}$ between bunch crossings inside a bunch train and
- a bunch train repetition rate of 5 $\mathrm{Hz}$

results in a burst of collisions at a rate of 2.7 $\mathrm{MHz}$ over $\lesssim 1\,\mathrm{ms}$ followed by 199 $\mathrm{ms}$ without any interaction. The overall collision rate of 13 $\mathrm{kHz}$ is significantly smaller than the event building rate at the LHC experiments

The very large number of readout channels for ILD will require signal processing and data compression already at the detector electronics level as well as high bandwidth for the event building network to cope with the data.

The traditionally deployed front-end electronics (VFE), will have to be fully integrated in the detectors. This is achievable by dedicated Readout-Chips (ROC) that are able to treat a large number of channels ($\sim 100$) and perform amplification, auto-triggering, time-stamping, signal (and eventually time) digitisation and local storage. A local zero-suppression is mandatory as only a very small fraction of the channels are hit. It can be achieved ROC by ROC or channel by channel, the optimum depending on the occupancy per ROC.

The data volume will be dominated by machine background, mainly from pair production from beam-beam interactions, as described in section 5.5.6. The rate of hadronic $e^+e^-$ events at design luminosity is expected to be of the order of 0.1 per bunch train and will contribute less then $1\,\%$ to the data recorded.

The particle flow analysis approach requires, as much as possible, to follow all particles individually through the detector. At the electronics level, this implies the capacity to be able to trigger with a high level of efficiency on a mip signal. If this seems obvious for trackers, it is more challenging for calorimeter systems, as the noise level will have to be kept at a low level.

The memory size that is needed for a given ROC depends on the expected machine background rate in its position in the detector, the internal noise which is expected to be uniform and the actual physics rates.

A large security margin should be kept since a "RAM full" event implies a local loss of information and, depending on the reconstruction procedure, either a loss of luminosity or modification of the calibration. Although the very high number of channels allows for software restoration of information, the tracking of individual, unavailable cells will add complexity to the reconstruction.

Table III-5.1 lists for the major ILD detector components and options the number of channels, occupancy, noise frequency and the expected data volume per train.

**Table III-5.1**
Data Volume in MB per bunch train for the major ILD detector components, for nominal ILC operation at 500 GeV.
The noise frequency reflects the noise rate taken during an acquisition;
The occupancy is the fraction of occupied channel during one Bunch crossing.
∗ Numbers to be updated.
∗∗ Raw output; after online treatment, a maximum of 5% of this data should be kept on tape.

| Sub-detector | Channels $[10^6]$ | Beam induced [Hits/BX] | Noise [Hits/BX] | Data volume per train [MB] |
|---|---|---|---|---|
| VTX (CPS) | 300 | 1700 | 1.2 | $< 100$ |
| VTX (FPCCD) | 4200 | 1700 | 1200 | 135 |
| TPC | 2 | 216 | 2000 | 12 |
| FTD | 1 | 260 | 0.3 | 2 |
| SIT | 1 | 11 | 0.3 | 6 |
| SET | 5 | 1 | | 1 |
| ETD | 4 | | | 7 |
| SiECAL | 100 | 444 | 29 | 3 |
| ScECAL | 10 | 44 | 40 | |
| AHCAL | 8 | 18000 | 640 | 1 |
| SDHCAL | 70 | 28000 | 70 | |
| MUON | 0.1 | | 8 | $\leq 1$ |
| LumiCal | 0.2 | | | 4 |
| BeamCal | 0.04 | | | 126** |

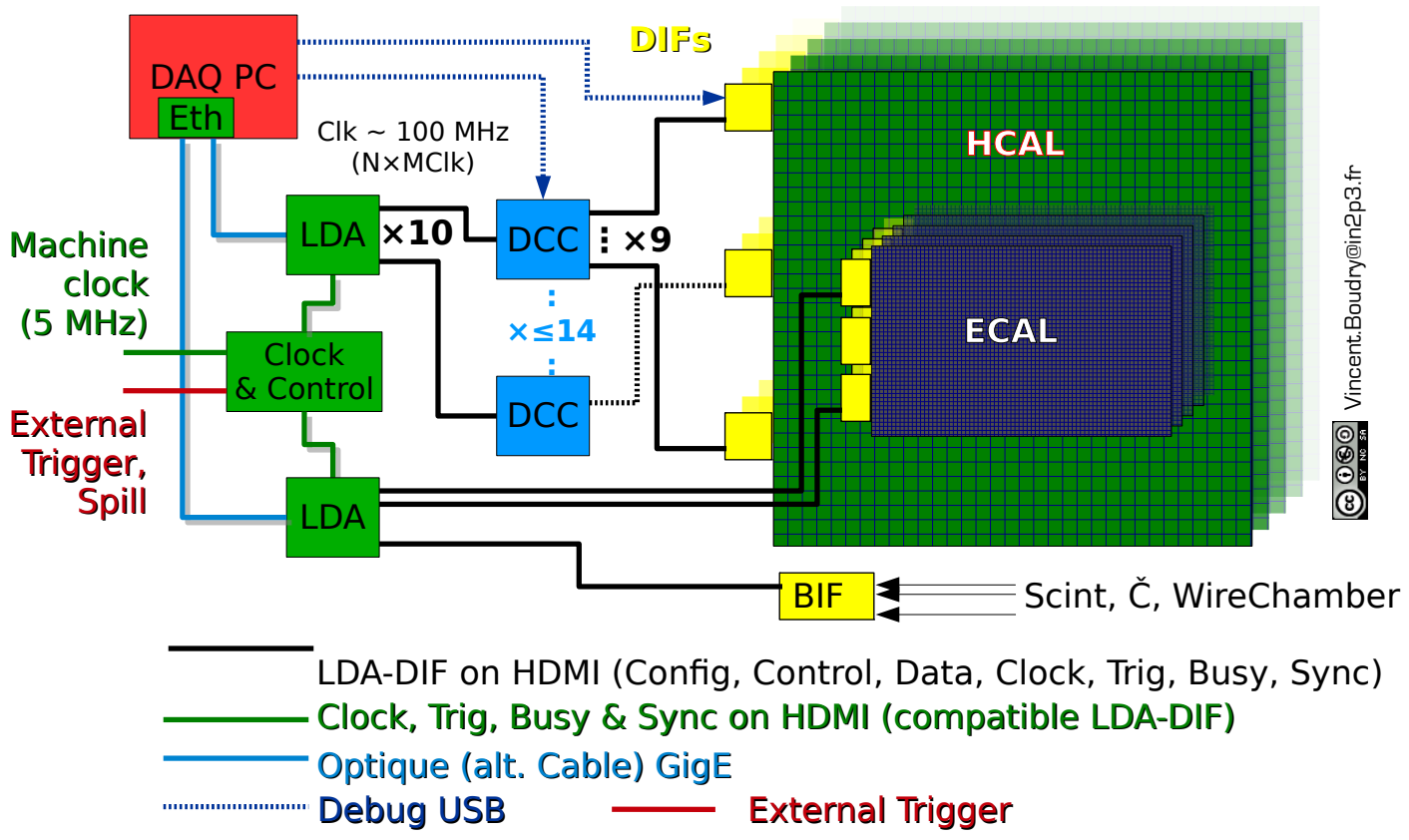

**Figure III-5.15**
General layout of the calorimeters DAQ system.

## 5.3.1 DAQ structure

The general layout of the proposed DAQ system for calorimeters is shown in Figure III-5.15. It is representative of the general structure of the complete system, including the tracking detectors: one or several PCB's equipped with ROC's (up to a few hundred for the calorimeters) are interfaced to the DAQ by specific front-end boards; these ensure the control of the ROC, dispatching and collection of the signals, the readout sequencing and mode of operation. The FE boards are linked with single cables to concentrator boards (LDA's) which ensure the separation of data (in and out) based on standard protocols from the DAQ specific fast signals (clock, busy, spill). The fast signals are handled by dedicated cards to ensure synchronous operation of the complete detector. The size of the boards and the location of functions on each board is still subject of intense R&D.

### 5.3.1.1 Front-end electronics:

The front-end electronics on the sub-detector or sensor level has to be detector specific; it will digitise and store the data of ∼2600 bunch crossings.

In contrast to the central DAQ system, the front-end readout electronics for the different sub-detector prototype designs has started, with the realisation of technological prototypes (AHCAL, SDHCAL, SiECAL, TPC, FCAL, Silicon Trackers and VTX) taking into account realistic engineering as well as detector performance tests. Several approaches are underway for the calorimeters, TPC, silicon trackers and vertex detectors. Common to all the designs is a highly integrated front-end electronics with signal shaping, amplifying, digitising, hit detection, data storage and highly multiplexed data transfer to reduce the number of cables. Some designs foresee data processing such as noise detection or cluster finding already at this stage to further reduce cables.

For a highly granular detector like the ILD with the resulting large channel counts both the material budget as well as the power consumption are areas of concern. Minimising the number of cables by data processing and multiplexing already on the sensor level is required as well as high density electronics with low power consumption. A common approach to reduce the power consumption is to turn the front-end electronics off in the gaps between bunch trains. First systems have been designed and built with this power pulsing capability. For a recent overview of such a system see [363].

The VFE chip will be powered a few 10$\mu$s before the arrival of a spill. Recent studies [319, 363] have shown that this preparation time is enough to get the stability of the amplifier and the ADC threshold to a level of one percent. The data is stored locally in the VFE chips. The pipeline length has to be adapted to data flux plus expected noise to avoid saturation of the local memory. Online procedures are of prime importance to control noisy cells. One noisy cell will effectively blind the entire region corresponding to all cells that are covered by its ROC.

Fast, hence online, correction procedures are mandatory to either suppress the noisy channels or correct their gain. Given the number of channels, the overall gains and mask configurations might change rather often, and the possibility of time evolving calibration parameters should be studied carefully as well as the failure rate of electronics and the stability of detectors w.r.t. the noise level.

#### 5.3.1.2 Middle level DAQ:

The middle level includes all the intermediate cards dispatched in the detector. It will have to be designed to use standard tools as much as possible and be fault tolerant, in particular by the use of signal path redundancy.

Standard protocols for data exchange should be used in order to profit from the development of commercial products. The usage of distribution networks [361] in the first running phase of the LHC has fulfilled all requirements.

The xTCA (ATCA modified for acquisition) is likely to become the new standard for DAQ in the coming years. Its use in current and upcoming experiments (PANDA, LHC detector upgrade) should be followed closely.

Since machine parameters and beam conditions such as the beam energy or the polarisation will be a vital input for the high precision physics analyses, they should be stored together with the data. The time structure and data volume are similar, hence a common DAQ and data storage model is envisaged.

In addition, the ILD will be operated in a truly international collaboration, with partners all over the world. Similarly to the global accelerator network (GAN), a global detector network (GDN) is proposed to operate the ILD detector remotely from the participating institutes. First experience with (not so) small set-up was gained with the CALICE remote control room at DESY during test beams at CERN and FNAL. For large setups the experience from the CMS remote operation centres at CERN, FNAL and DESY will be taken into account. The design of the DAQ and control system should have remote operation features built in from the start.

## 5.3.2 Data processing

The details of the DAQ and computing system depend to a large extent on the developments in microprocessors and electronics and the final design of the different sub-detector electronic components. The key points that will have to be addressed in the coming years are highlighted below.

The reconstructed events are written to the storage systems in an object oriented data format that is suited for further analysis with appropriate pointers into the raw data file containing the bunch train data. LCIO [364] provides a first version of such an event data and is in use for several test beam efforts and for the offline analyses within ILD.

### 5.3.2.1 Event building and prompt reconstruction

The purpose of the online event processing will mainly be event classification, calibration, alignment and data quality monitoring. Although no event rejection is foreseen, a scheme of event finders may be used to identify "bunches of interest" which could then be used for the physics analysis or for fast analysis streams.

Event building and prompt reconstruction will be performed on the *Online Filter Farm* – a sufficiently large farm of processing units near the detector, connected to the front-end electronics via the Common Event Building Network. The raw data of a complete bunch train is kept in the raw data file after compression. This is essential as many detectors will integrate over several bunch crossings or even the full bunch train. The event reconstruction will be an iterative process where in a first step a preliminary reconstruction will be done on the data from every sub-detector. Bunches of interest are then identified by exploiting correlations in time and space between the data from all sub-detectors. After calibration and alignment finally a full event reconstruction is performed on the event data.

An event filter mechanism run at prompt reconstruction will provide the necessary meta data for fast event selection at the physics analysis level. One processor per Bunch Train Event building of all data from the bunch train will be done in a single processing unit. Hence all data of the complete train will be available for the event processing without further data transfer which is essential since many detectors will integrate over several bunch crossings.

Each processing unit of the Online Filter Farm will process the data of one complete bunch train at a time.

### 5.3.2.2 Offline computing

The further offline data processing will exploit a Grid infrastructure for distributed computing using a multi-tier like approach following closely what is done for the LHC-experiments [365, 366]. The offline computing tasks such as the production of more condensed files with derived physics quantities (DST/AOD), Monte Carlo simulations and re-processing of the data will be distributed to the various tiers of the ILC-computing Grid. Setting up a data Grid and suitable data catalogues will allow the physicists to efficiently access the data needed for their analyses.

## 5.3.3 Outlook and R&D

Key elements of the DAQ systems have to be defined to guide the R&D of the sub-detector front-end electronics especially when entering the technical prototype stage. The effort started in the European FP6 project EUDET program is pursued in the FP7 project AIDA, with emphasis on the integration of the various DAQ systems (Pixel EUDAQ, CALICE, FCAL, Silicon trackers) for common test beams by 2014-15. These tests will serve as a first test-stand for an ILD data acquisition system. Large bandwidth systems such as xTCA are also being evaluated, e.g. for the TPC.

Due to the timescales involved and the rapid changing computing and network market, a decision on the DAQ hardware will be made as late as possible to profit from the developments in this area. A performance enhancement by a factor of 20 is expected in bandwidth, storage and processing in the coming 10 years.

## 5.4 ILD software and tools

The ILD detector concept uses the iLCSoft software framework which provides the core tools LCIO [364], Gear [367], Mokka [368] and Marlin [369] as well as reconstruction and analysis tools for LC detector R&D. The framework, which has already been used for a massive Monte Carlo production for ILD's Letter Of Intent [198], has since been extended and improved with the focus on enabling a more realistic simulation and reconstruction of physics events in the ILD detector.

LCIO provides a hierarchical event data model and persistency and is used by all detector concepts for linear colliders, providing a basis for common software developments and the exchange of algorithms and tools. A recent major release of LCIO (2.0) comprises many new features, the most important of which are:

- direct access to runs and events in a file allowing for efficient overlay of background events;
- introduction of 1D and 2D TrackerHit classes enabling a more detailed description of Si-strip and pixel detectors;
- extension of the Track class to hold many TrackStates per track: typically one at the IP, the first and last hit and at the face of the calorimeter are stored;
- introduction of a ROOT [370] dictionary to facilitate analysis of LCIO data;
- a number of extensions to the Event Data Model.

Gear provides an API for querying the detector geometry and material distribution at the reconstruction stage, including a detailed description of the measurement surfaces of tracking detectors including their sensitive and insensitive materials.

The Geant4 [371] based full simulation application Mokka's handling of the generator information has been enhanced with respect to the treatment of long lived and exotic particles and their decay vertices. The description of the ILD sub-detectors has been made more realistic by introducing gaps, imperfections as well as support and service materials.

Marlin is the C++ application framework that is used for further processing of the simulated detector response and is based on LCIO and Gear. Marlin contains a plug-in mechanism that supports the modular development of user software packages. The per-module and global configuration of the application is performed via an XML steering file with optional overwrite through command line arguments. A logging mechanism ensures that the actual configuration is stored for future reference and reproducibility.

### 5.4.1 Detector models in Mokka

All ILD sub-detectors in Mokka have been implemented including a significant amount of engineering detail such as mechanical support structures, electronics and cabling as well as dead material and cracks. Some sub-detectors, for which this level of realism had not been reached at the time of the LOI, have been completely re-written for this report.

Where possible, sub-detectors have been implemented in a way that is agnostic to the actual readout technology, for others different implementations exist. The following three Mokka models have been created for comparison of the different technology options with Monte Carlo simulations:

- **ILD_o1_v05:** ILD model with analogue HCAL and Si-ECAL
- **ILD_o2_v05:** ILD model with semi-digital HCAL and Si-ECAL
- **ILD_o3_v05:** ILD model with analogue HCAL and Scintillator-Strip-ECAL

Figure III-5.16 shows a 3D view of one of these simulation models. The simulation models comprise the following sub-detectors:

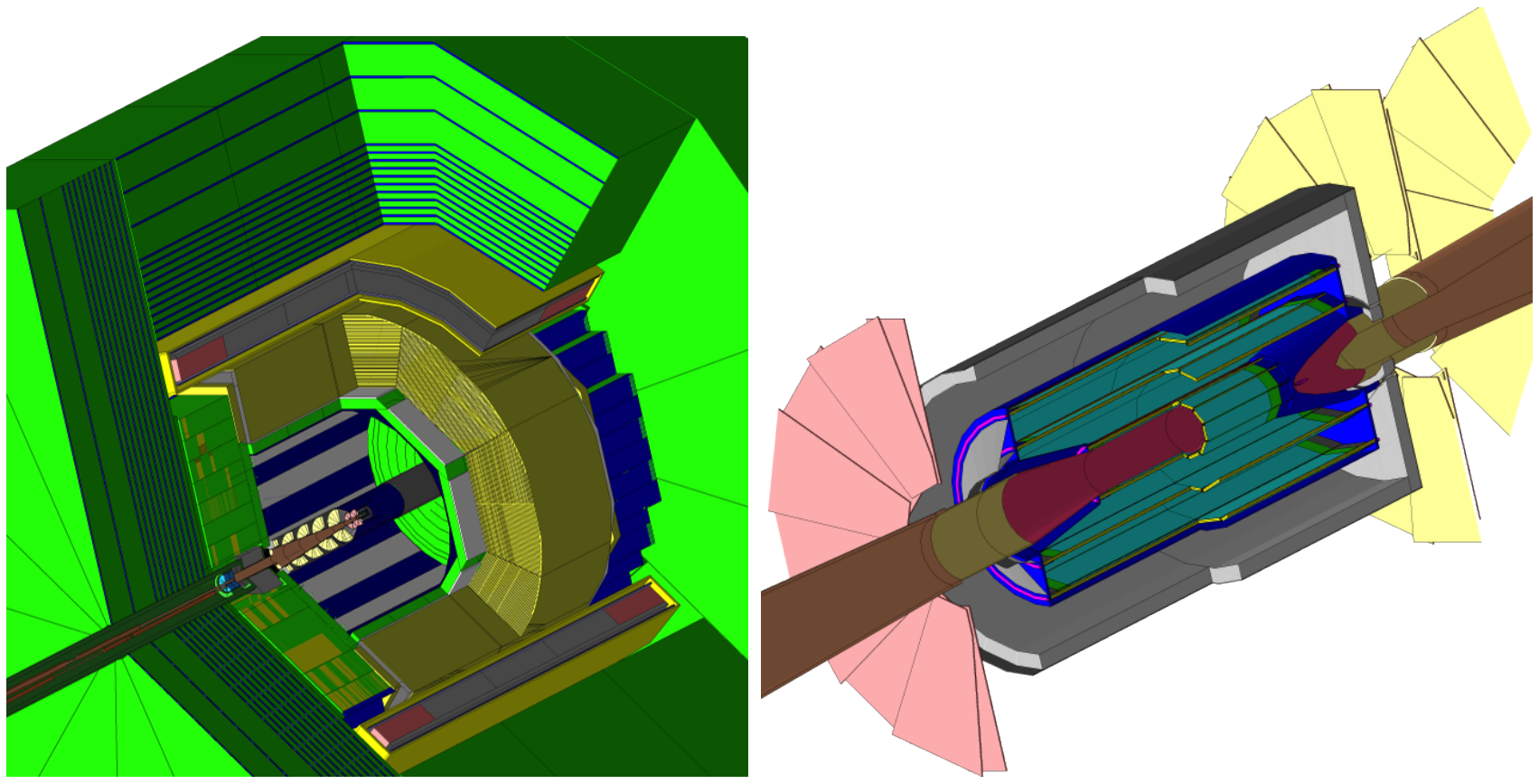

**Figure III-5.16** The ILD simulation model. *Left:* 3D view of the ILD simulation model ILD_o1_v05, from inside to outside: VTX, SIT, TPC, SET, ECAL, HCAL, Coil, Yoke and in the forward direction: FTD, LumiCal, LHCal, BeamCal. *Right:* Blowup of the inner tracking detectors in the simulation model.

- **VTX:** detailed description of the sensitive and support part of the ladders in the three double layers including a surrounding cryostat;
- **SIT/SET:** Si-Strip detectors with planar ladders of sensitive and realistically averaged support material;
- **FTD:** Si-Pixel and Si-Strip disc detectors, built from sensitive Si-petals on a space frame support structure – a realistic estimate of the material budget from power and readout cables for the inner tracking detectors VTX, SIT and FTD has been averaged into an $Al$ cylinder running just inside the TPC field cage and a cone just around the beam pipe in order to correctly account for the effect of multiple scattering;
- **TPC:** cylindrical volume filled with correct gas mixture, surrounded by a realistic field cage and a conservatively estimated back plane comprised of material for the space frame structure, electronics and cooling pipes;
- **ECAL:** detailed description of the alveolar layer structure with $W$ as absorber material and a readout part that is either based on Si-wafers with $5 \times 5\ mm^2$ cell size or on $5 \times 45\ mm^2$ scintillator strips – gaps between modules are properly modeled;
- **HCAL:** realistic models for the analogue and semi-digital HCAL options with a different layout of the absorber and readout structure; the gaps and electronics are properly modeled in both cases; Birk's law is taken into account for the analogue case;
- **Muon:** the iron Yoke has been instrumented with scintillator based active layers. At the moment tiles with $3 \times 3\ cm^2$ granularity are used, for muon detection and serving as a tail catcher for the HCAL; this is different than the detector baseline which uses 3 cm wide and 1 m long strips;
- **LumiCal, LHCal, BeamCal:** the forward calorimeters are also modeled realistically with their corresponding sandwich structure consisting of $W$ absorbers and $Si$ ( LumiCal, LHCal) and diamond (BeamCal) readout, respectively.

## 5.4.2 Marlin: Reconstruction and analysis system

More than 20 million events have been fully simulated with the Mokka detector models described above and then reconstructed with the following Marlin modules :

#### 5.4.2.1 Digitisation

The digitisation of hits in the tracking detectors is performed by a parameterisation of the point resolution as established by the R&D groups and shown in Table III-5.2. In the case of the Si-Strip detectors in SIT/SET and FTD, 1D TrackerHits are created at the digitisation stage and then combined into 3D space points that are used for pattern recognition in order to correctly account for ghost hits. Calorimeter hits are scaled with a calibration factor according to the sampling fraction, where in the case of the semi-digital HCAL effects of cross talk between neighboring cells are included.

**Table III-5.2.** Effective point resolution as used in the digitisation of the tracking detectors' response.

| Detector | Point Resolution | | |
|---|---|---|---|
| VTX | $\sigma_{r\phi,z}$ | $=$ | $2.8\mu\mathrm{m}$ (layer 1) |
| | $\sigma_{r\phi,z}$ | $=$ | $6.0\mu\mathrm{m}$ (layer 2) |
| | $\sigma_{r\phi,z}$ | $=$ | $4.0\mu\mathrm{m}$ (layers 3-6) |
| SIT | $\sigma_{\alpha_z}$ | $=$ | $7.0\mu\mathrm{m}$ |
| | $\alpha_z$ | $=$ | $\pm 7.0^\circ$ (angle with z-axis) |
| SET | $\sigma_{\alpha_z}$ | $=$ | $7.0\mu\mathrm{m}$ |
| | $\alpha_z$ | $=$ | $\pm 7.0^\circ$ (angle with z-axis) |
| FTD | $\sigma_r$ | $=$ | $3.0\mu\mathrm{m}$ |
| Pixel | $\sigma_{r_\perp}$ | $=$ | $3.0\mu\mathrm{m}$ |
| FTD | $\sigma_{\alpha_r}$ | $=$ | $7.0\mu\mathrm{m}$ |
| Strip | $\alpha_r$ | $=$ | $\pm 5.0^\circ$ (angle with radial direction) |
| TPC | $\sigma^2_{r\phi}$ | $=$ | $(50^2 + 900^2 \sin^2\phi + \left((25^2/22) \times (4T/B)^2 \sin\theta\right)(z/\mathrm{cm}))\,\mu\mathrm{m}^2$ |
| | $\sigma^2_z$ | $=$ | $(400^2 + 80^2 \times (z/\mathrm{cm}))\,\mu\mathrm{m}^2$ |
| | where $\phi$ and $\theta$ are the azimuthal and polar angle of the track direction | | |

#### 5.4.2.2 Track reconstruction

The reconstruction of charged particles is done with a set of new C++ packages recently added to iLCSoft: Kaltest, IMarlinTrK, Clupatra and FwdTracking, replacing the previously used FORTRAN tracking code that dated back to LEP. Clupatra is a TPC pattern recognition algorithm that combines topological clustering methods for seed finding with Kalman Filter based extrapolations for picking up hits. Optionally the hit search can be extended inwards to include the Si-tracking detectors. FwdTracking [372] is a newly developed pattern recognition for the FTD that is based on cellular automatons and hopfield networks. The IMarlinTrK package provides the interface to the track fitter based on a Kalman Filter implemented using KalTest [373]. SiliconTracking is a package for standalone tracking in VTX, SIT/SET and FTD. FullLDCTracking finally combines the track segments from all sub-detectors into a consistent final list of tracks which is then used as input to particle flow.

#### 5.4.2.3 Particle flow

PandoraPFANew [274] is an implementation of the *particle flow algorithm (PFA)*, which recently has been re-written to be detector and framework independent. MarlinPandora is a Marlin package that converts the calorimeter hit and track objects from LCIO objects into corresponding data structures used in PandoraPFA, augmented with relevant information from the detector geometry and with suitable track quality cuts applied. The resulting list of particle flow objects is then converted back into a list of ReconstructedParticles which is used for further analysis. PandoraPFANew uses sophisticated clustering algorithms and track-cluster matching as an initial step. The application of re-clustering methods, based on cluster energy to track momentum comparisons, is crucial to eventually achieve the optimal jet energy resolution based on single particle reconstruction.

#### 5.4.2.4 Vertex finding and jet flavour tagging

LCFIVertex [374] is a package for vertex finding, based on the ZVTop algorithm and for jet flavour tagging using Artificial Neural Networks (ANNs). LCFIPlus is a new package that provides improved flavour tagging, introducing additional input variables and replacing the ANNs with Mutli Variate Analysis techniques, as well as a new jet clustering algorithm [375]. The secondary vertex finding is run centrally as part of the standard reconstruction, whereas jet finding and flavour tagging are run by the users, individually tuned according to the specific needs of their physics analysis.

#### 5.4.2.5 Background overlay

The main source of background are hits from $e^+e^-$ pairs, resulting in considerable hit densities, predominantly in the VTX detector (see section 5.5.6). While these hits pose a non-trivial problem to the pattern recognition, eventually the effect of ghost tracks can be reduced very efficiently by requiring hits in the SIT which provides an exact time stamp per bunch crossing [198]. Thus pair background is not overlayed for the Monte Carlo events analysed here, with the exception of the BeamCal, where an averaged energy weighted hit density from pair background is taken into account in reconstruction.

The other source of background are multi-peripheral $\gamma\gamma \rightarrow$ hadrons events. These give raise to much lower occupancies – on average one expects 1.7 (4.1) low multiplicity events per bunch crossing at nominal beam conditions for 500 GeV (1 TeV). As these events of course result in real tracks and clusters, fully simulated $\gamma\gamma \rightarrow$ hadrons events are overlaid statistically before the reconstruction. As these events come from a different vertex than the physics event, their z-position is smeared with a spread of $\sigma_{vertex,z} = 300\mu\text{m}\ (225\mu\text{m})$ at 500 GeV (1TeV) reflecting the resulting difference of the z-position of the vertices of independent events originating from a Gaussian beam profile.

### 5.4.3 Monte Carlo productions

Around 20 million Monte Carlo events (see Table III-5.3) have been produced on the Grid with the help of a newly developed production system: GridProd. It is based on a MySQL data base and a set of python scripts with a web interface, that provides an up-to-date view of the requested and already processed events. It allows to query the file catalog based on a set of meta-data tags, such as center of mass energy, process name and type.

**Table III-5.3**
Number of Monte Carlo Events produced for the DBD.

| Process | $E_{CMS}$ | Detector model | Events $[10^6]$ |
|---|---|---|---|
| $e^+e^- \rightarrow \nu\nu h$ | | | 0.3 |
| $e^+e^- \rightarrow t\bar{t}h$ | | | 0.04 |
| $e^+e^- \rightarrow W^+W^-$ | 1 TeV | ILD_o1_v05 | 1.4 |
| SM background (2-8 fermion) | | | 11.4 |
| $e^+e^- \rightarrow t\bar{t}h$ | 1TeV | ILD_o2_v05 | 0.04 |
| $e^+e^- \rightarrow t\bar{t}$ | 500 GeV | ILD_o1_v05 | 0.8 |
| SM background (2-8 fermion) | | | 6.8 |
| total | | | 20.9 |

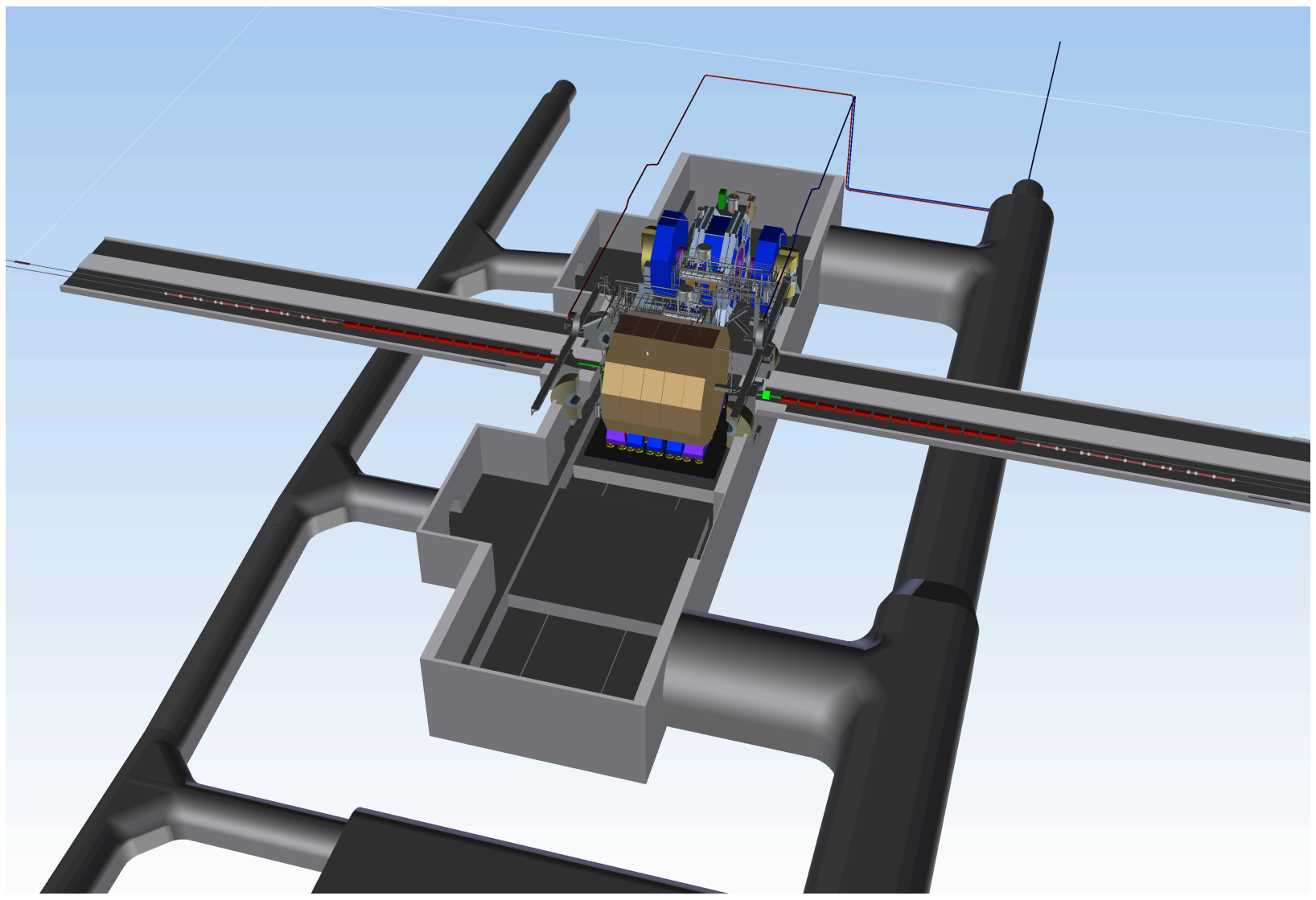

**Figure III-5.17** The push-pull system at the ILC: ILD (front) and SiD (back) in the experimental hall (example of a mountainous ILC site.

## 5.5 ILD - machine-detector interface and experimental area

This section deals with the ILD specific issues of the Machine-Detector Interface and the experimental areas [376].

### 5.5.1 ILD push-pull issues

The ILC push-pull system is described in section 2.3. Adapting this system to the ILD detector is a challenge, taking into account the dimensions and masses of the detector on the on hand and the required precision on the other hand. Figure III-5.17 shows a design of a detector hall with the push-pull system.

The ILD detector is somewhat larger than SiD and is designed to be assembled from slices in a similar way as the CMS detector. The detector placement on the push-pull platform preserves detector alignment and distributes the load evenly onto the floor. The platform will carry also some of the detector services like electronic racks. The ILD yoke slices each have their own motion system that is based on air pads and grease pads. In the parking position, the detector can be opened for maintenance by moving the yoke slices on air pads from the platform.

### 5.5.2 Final focus magnets

The interaction region of ILD is designed to fulfil at the same time the requirements from the ILC machine as well as the needs of the detector. As the allowed focal length range of the inner final focus quadrupoles (QD0) for ILC ($3.5$ m $\leq L^* \leq 4.5$ m) is smaller than the detector size, the QD0 magnet of the final lens needs to be supported by the detector itself. As a consequence, SiD and ILD will have their own pair of QD0 magnets that move together with the respective detector during push-pull operations. In contrast, the QF1 magnets of the final lenses with a focal length of $L^* = 9.5$ m are not supported by the detectors and stay on the beam line during detector movements. A set of vacuum valves between the QD0 and the QF1 magnets defines the break point for the push-pull operations. The biggest concerns for the QD0 support system are the alignment and the protection against ground motion vibrations. The limit on vibrations is given at 50 nm above 1 Hz [354].

Due to these tight requirements, the support of the magnets in the detector is of special importance. ILD has chosen a design where the magnets are supported from pillars that are standing directly on the transport platform (see Figure III-5.18). In the detector, the magnets are supported by a system of tie rods from the cryostat of the solenoid coil. This design de-couples the detector end caps from the QD0 magnets and allows a limited opening of the end caps also in the beam position, without the need to break the machine vacuum (see Figure III-5.9). In addition, the QD0 magnets are coupled via the pillar directly to the platform and limit in that way the number of other vibration sources. Simulations taking into account realistic ground motion spectra for different sample sites have been done to understand the vibration amplification in the QD0 support system [377]. These studies show that, with the exception of very noisy sites, the requirements for the QD0 magnets are fulfilled with large safety margins. Even if the additional amplification characteristics of the platform (c.f. [378]) are taken into account, the total integrated vibration amplitudes are expected to be in the order of less than 10 nm for frequencies above 1 Hz.

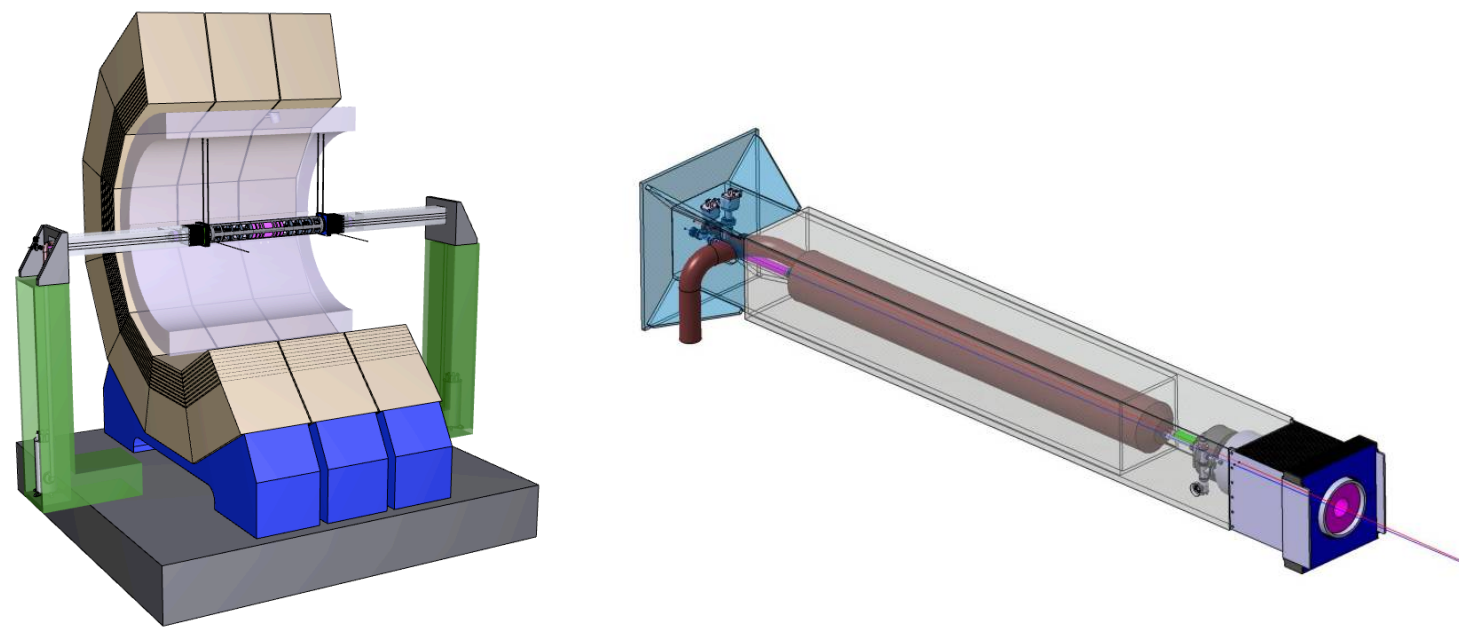

**Figure III-5.18** Left: support system of the QD0 magnets in ILD; the inner parts of the detector and the end caps are not shown [198]. Right: double-tube support structure for the QD0 magnet and the forward calorimeters.

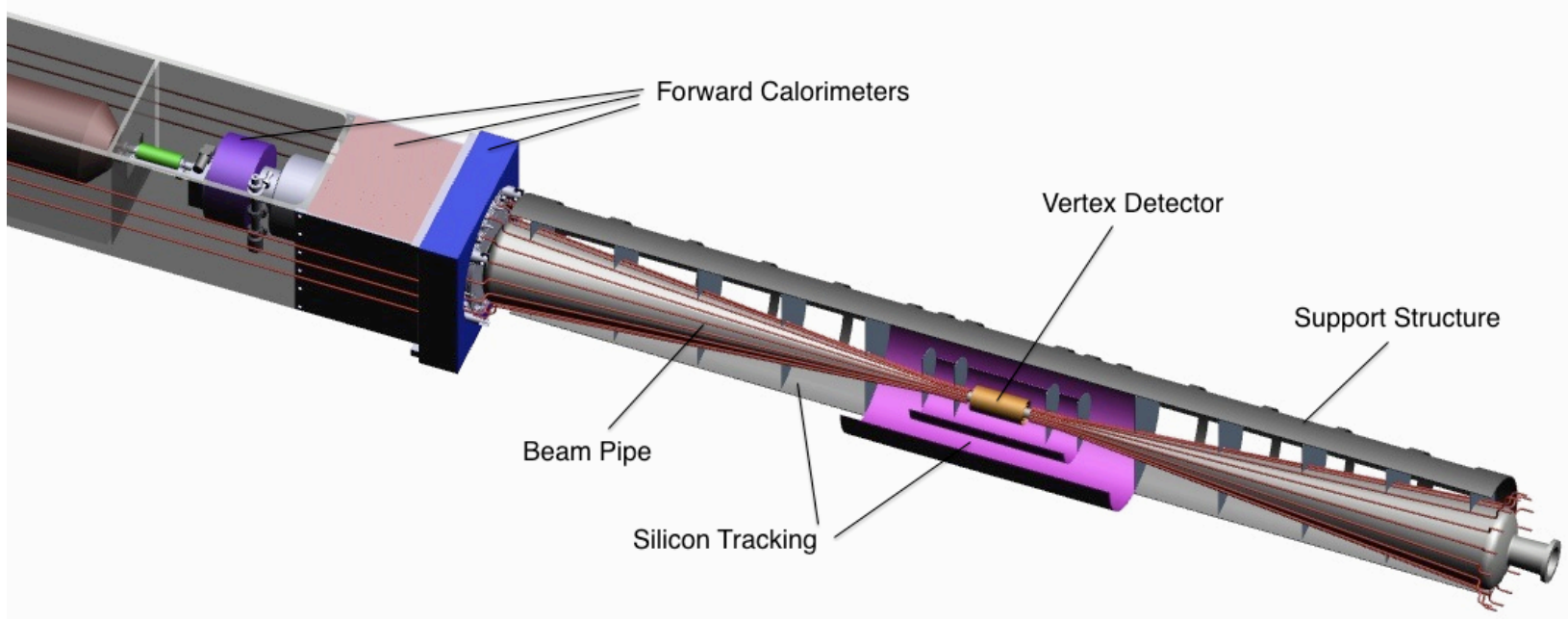

**Figure III-5.19** The interaction region of ILD.

The proper alignment of the QD0 magnets with respect to the axis that is defined by the QF1 quadrupoles is also of crucial importance. While the alignment accuracy of the detector axis after the movement into the beamline is moderate (horizontal: $\pm$ 1 mm and $\pm$ 100 $\mu$rad), the requirements for the initial alignment of the quadrupoles are much tighter: $\pm$ 50 $\mu$m and $\pm$ 20 $\mu$rad. An alignment system that comprises an independent mover system for the magnets and frequency scanning interferometers is part of the detector design.

### 5.5.3 Beam pipe and interaction region

The central interaction region of ILD comprises the beam pipe, the surrounding silicon detectors, the forward calorimeters and the interface to the QD0 magnets (see Figure III-5.19).

The most delicate component of this region is the very light beam pipe made from Beryllium, that is surrounded by the vertex detector and the intermediate silicon tracking devices. A carbon fibre reinforced cylindrical structure will form the mechanical support for these elements. This tube is attached to the inner field cage of the surrounding time projection chamber (not shown in the figure). As the horizontal alignment tolerance of the detector axis after push-pull operations is $\pm$1 mm, an adjustment system is needed to eventually re-align the tube structure with the beam pipe and the inner tracking detectors. This is especially important to keep the stay-clear distances to the tracks of the beam induced background particles within the beam pipe.

The beam pipe opens conically away from the interaction point to allow enough space for the beam induced background, most importantly the electron-positron pairs from beamstrahlung. The shape of the beam pipe results in a rather large volume that needs to be kept evacuated by means of vacuum pumps that are on both sides 3.3 m away from the interaction point. Simulations show that the vacuum requirements for the ILC can be met with this configuration [198].

The forward calorimeters are discussed in detail in section 3.5.

### 5.5.4 Experimental area for flat surface ILC sites

The design of the underground experimental cavern for the non-mountainous sites follows a z-shape floor layout. The common interaction point is in the middle of the hall, the detectors move in and out of the beam position on their transport platforms. Alcoves in the maintenance positions allow for lateral space that is needed to open the detectors. Figure III-5.20 shows the layout in the parking position for ILD. The detector is shown in fully opened position that allows for the removal of the large detector parts. The biggest element that might need to be removed from the detector (though not in routine maintenance periods) is the superconducting solenoid. Enough space is foreseen to manoeuvre the parts of the detector in the hall and bring them safely to the vertical access shafts. In addition, space for the detector services (c.f. section 2.3) is available in this design.

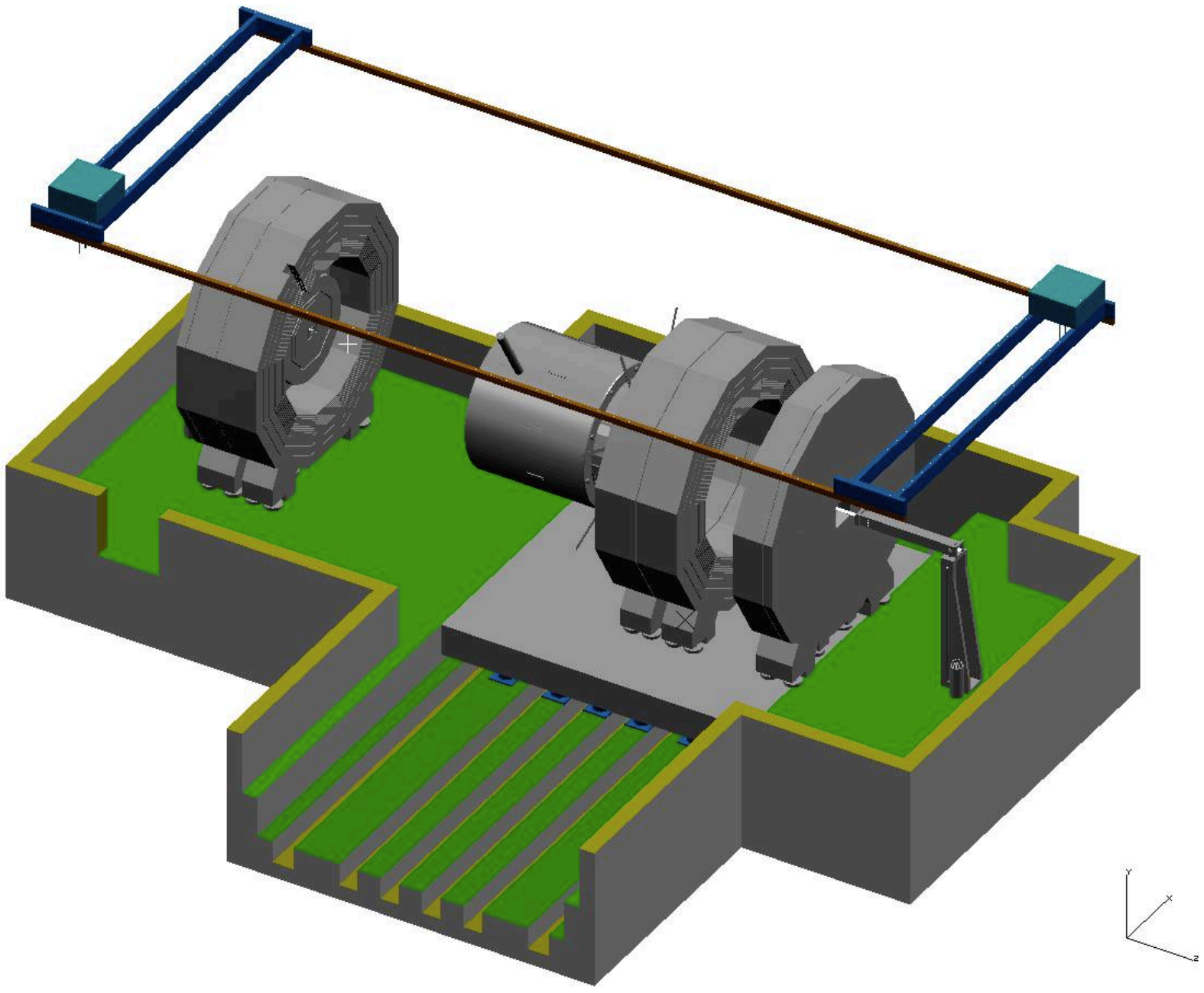

**Figure III-5.20** Conceptual design of the underground facilities for ILD. The detector is opened in the maintenance position, the crane coverage is shown [379].

Access to the underground cavern is enabled via five vertical shafts. One central big shaft is located directly above the interaction point with a diameter of 18 m. This shaft would be used during the assembly of both detectors where the big parts are pre-assembled on the surface and then lowered through the big shaft directly onto the respective transport platform. Two smaller diameter shafts (8 m for SiD and 10 m for ILD) are needed in the maintenance positions to allow access from the surface while one detector is at the beam position and blocks the access to the big shaft. Two additional smaller shafts of $\approx$5 m diameter are foreseen for elevators and services.

As the yoke rings will be moved on air pads within the hall, the crane covering the maintenance area needs to have a modest capacity of preferably $2 \times 40$ t. However, a temporary hoist with a capacity of up to 3500 t is needed on the surface over the main access shaft to lower the big detector parts during the primary assembly.

## 5.5.5 Experimental area for mountainous ILC sites

The design of the underground facilities for the mountainous ILC sites is discussed in section 2.3. The current design has been optimised to the needs of ILD as well. Figure III-5.21 shows the space that is needed to open ILD for maintenance in the parking position. Alcoves in the hall provide enough lateral space to move the endcaps away from the platform. Large detector parts, e.g. the beam pipe or the QD0 magnets, can be removed in that area.

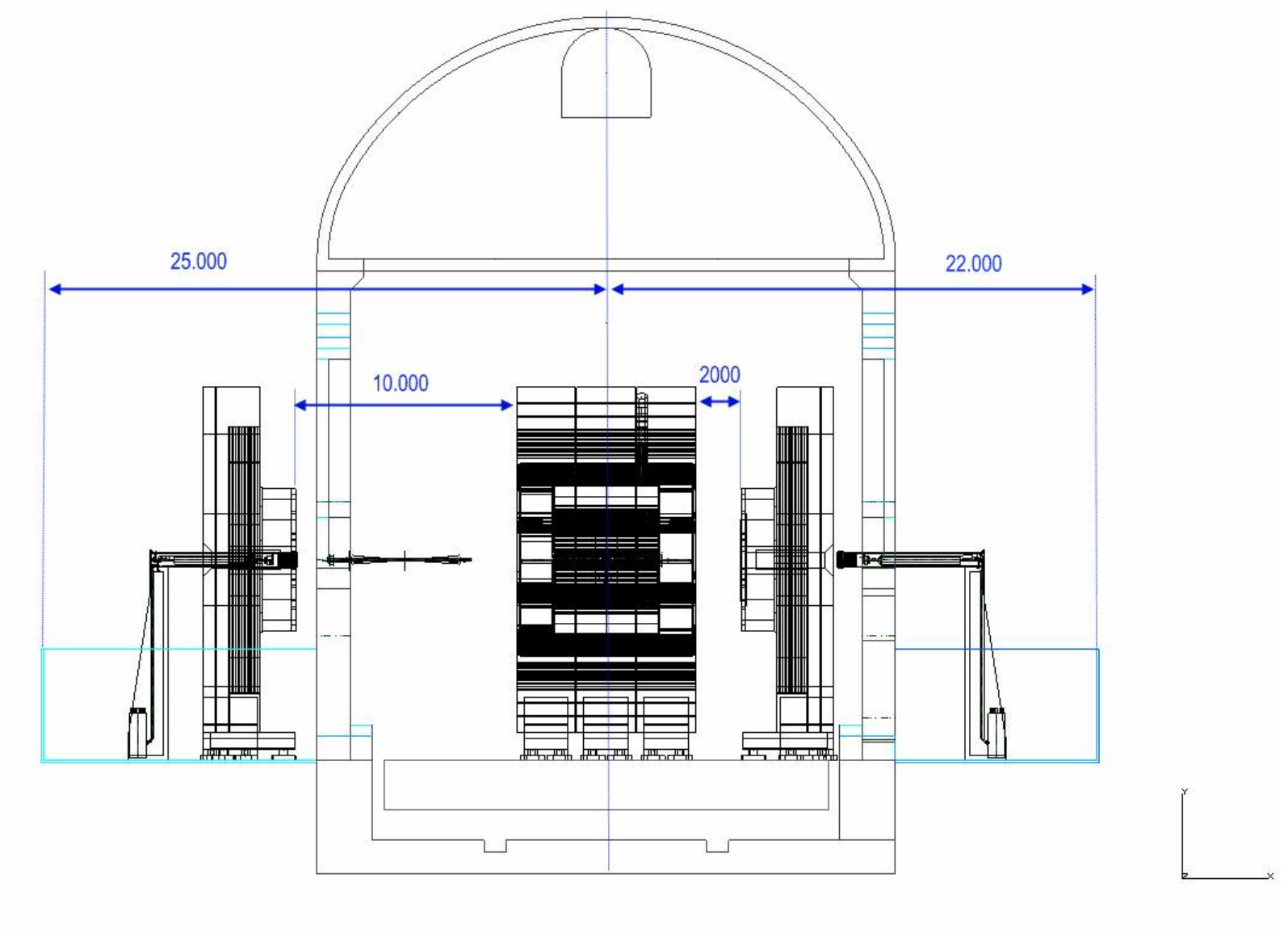

**Figure III-5.21** Opening ILD in the maintenance alcove of the mountain site hall.

## 5.5.6 Machine induced backgrounds

Machine-induced backgrounds have been studied in detail for the ILD detector [198, 255] and have recently been updated for the latest machine parameter sets at 500 GeV and 1000 GeV collision energies [380]. The main relevant background stems from $e^+e^-$ pairs from beamstrahlung which are produced in the highly charged environment of the beam-beam interaction. The background levels found are well below the critical limit for most sub-detectors. The sub-detector most sensitive to beam-related backgrounds is the vertex detector, which features an inner radius value dictated by the maximum affordable beamstrahlung hit rate.

Table III-5.4 summarises the expected background levels in the ILD sub-detectors for the nominal ILC beam parameter sets at 500 and 1000 GeV cms energy. The background occupancies have been simulated with the ILD full detector simulation with the nominal detector geometries and 3.5 T magnetic field in the solenoid that is overlaid with an 'anti-DID' dipole component [334]. The double-layer option has been chosen for the vertex detector in these simulations.

### 5.5.6.1 Background uncertainties

As the vertex detector is most critical with respect to beam induced backgrounds, detailed studies have been performed to understand the impact of different detector geometries and simulation parameters like the choice of range cut parameters in Geant4. The number of hits on the vertex detector change up to 30% when changing parameters in the simulation, which indicates the magnitude of the uncertainties for these simulation results. Another study of the uncertainties of the background simulations has been done in [255]. From these studies and using a rather conservatice approach a safety factor for backgrounds between 5 and 10 is used in ILD.

**Table III-5.4.** Pair induced backgrounds in the subdetectors for nominal 500 GeV and 1 TeV collision energy beam parameters [380]. The numbers for the ECAL and the HCAL are summed over barrel and endcaps. For the vertex detecor, the double-layer option has been chosen for this simulation. The TPC hits are the digitised hits that would be written to the data acquisition system. The errors represent the RMS of the hit number fluctuations of $\approx$ 100 bunch crossing (BX) simulations.

| Sub-detector | Units | Layer | 500 GeV | 1000 GeV |
|---|---|---|---|---|
| VTX-DL | hits/cm$^2$/BX | 1 | $6.320 \pm 1.763$ | $11.774 \pm 0.992$ |
| | | 2 | $4.009 \pm 1.176$ | $7.479 \pm 0.747$ |
| | | 3 | $0.250 \pm 0.109$ | $0.431 \pm 0.128$ |
| | | 4 | $0.212 \pm 0.094$ | $0.360 \pm 0.108$ |
| | | 5 | $0.048 \pm 0.031$ | $0.091 \pm 0.044$ |
| | | 6 | $0.041 \pm 0.026$ | $0.082 \pm 0.042$ |
| SIT | hits/cm$^2$/BX | 1 | $0.0009 \pm 0.0013$ | $0.0016 \pm 0.0016$ |
| | | 2 | $0.0002 \pm 0.0003$ | $0.0004 \pm 0.0005$ |
| FTD | hits/cm$^2$/BX | 1 | $0.072 \pm 0.024$ | $0.145 \pm 0.024$ |
| | | 2 | $0.046 \pm 0.017$ | $0.102 \pm 0.016$ |
| | | 3 | $0.025 \pm 0.009$ | $0.070 \pm 0.009$ |
| | | 4 | $0.016 \pm 0.005$ | $0.046 \pm 0.007$ |
| | | 5 | $0.011 \pm 0.004$ | $0.034 \pm 0.005$ |
| | | 6 | $0.007 \pm 0.004$ | $0.024 \pm 0.006$ |
| | | 7 | $0.006 \pm 0.003$ | $0.022 \pm 0.006$ |
| SET | hits/BX | 1 | $0.196 \pm 0.924$ | $0.588 \pm 2.406$ |
| | | 2 | $0.239 \pm 1.036$ | $0.670 \pm 2.616$ |
| TPC | hits/BX | - | $216 \pm 302$ | $465 \pm 356$ |
| ECAL | hits/BX | - | $444 \pm 118$ | $1487 \pm 166$ |
| HCAL | hits/BX | - | $18049 \pm 729$ | $54507 \pm 923$ |

# Chapter 6
# ILD Performance

The performance of the ILD detector has been studied in detail both in terms of technical performance criteria, and in terms of selected physics processes. In this chapter the system performance of the ILD detector is discussed. Then a series of different physics studies done using full Monte Carlo at different center of mass energies from 250 to 1000 GeV are presented. These analyses have not been selected to demonstrate the physics reach of the ILC facility, but rather to stress the detector and its performance. For completness results from earlier studies done in the context of the letter of intent, with a slightly different detector model, are also summarised.

## 6.1 ILD performance

The overall performance of ILD is established using a detailed GEANT4 model [368] and full reconstruction of the simulated events. Using full simulation and a realistic reconstruction ensures that the performance is as realistic as possible and takes into account the detailed knowledge on detector mechanics, dead areas, and non-perfect response.

### 6.1.1 Software for performance studies

Three distinct detector models have been implemented in the Mokka GEANT4 detector simulation program. The only differences are in the technology choices for the ECAL and HCAL. The first model (ILD_o1_v5) simulates a SiW ECAL and a scintillator tile analogue HCAL, the second (ILD_o2_v5) simulates a SiW ECAL and semi-digital RPC-based HCAL, and the final model (ILD_o3_v5) includes a scintillator strip ECAL and the analogue tile scintillator HCAL. The different detector models are treated equally and provide a demonstration of the performance of the different technology options within ILD. Because of the relative maturity of the reconstruction software, the majority of the physics studies are performed using the SiW ECAL, which assumes a $5 \times 5$ mm$^2$ transverse cell size, and the steel-scintillator HCAL option with $3 \times 3\,\text{cm}^2$ tiles; unless otherwise stated, the detector model used in the performance studies is ILD_o1_v05

The level of detail included in the detector simulation represents a significant step forward compared to the ILD LoI. Most of the sub-detectors in the ILD models have been implemented with a significant amount of engineering detail such as mechanical support structures, electronics and cabling as well as dead material and cracks. The material budget associated with the support structures and services are based on the best current estimates from the detector R&D groups. In addition, the material associated with the delivery of power and cooling to the sub-detectors have been implemented in the simulation so as to provide a reasonable description of the mean material budget. The improvements in the simulation are crucial for a realistic demonstration of particle flow and tracking performance. A description of the detector parameters and the reconstruction software can be found in Section 5.4.

All events are reconstructed based on a sophisticated reconstruction chain, including using a

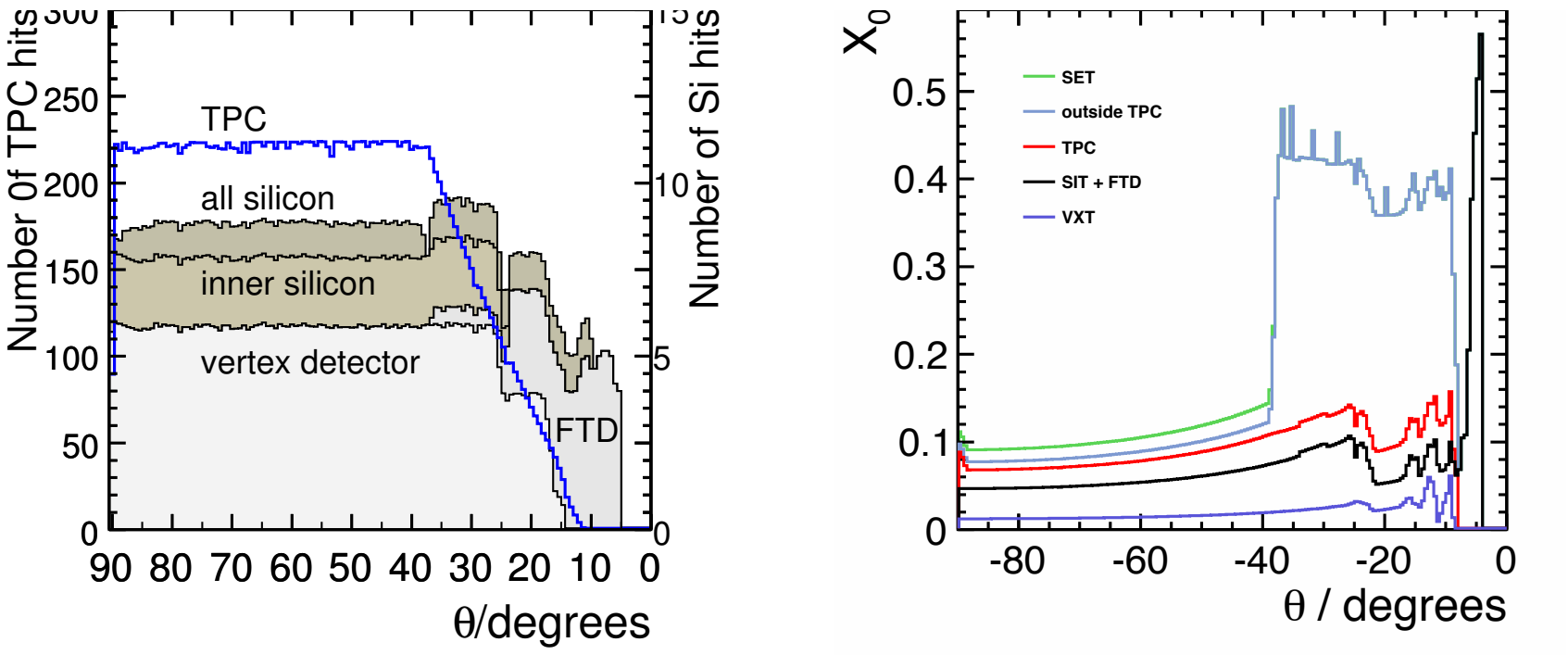

**Figure III-6.1** (Left) Average number of hits for simulated charged particle tracks as a function of polar angle. (Right) Average total radiation length of the material in the tracking detectors as a function of polar angle.

Kalman-filter based track reconstruction, MarlinTrk, the PandoraPFA particle flow algorithm and the LCFIPlus flavour tagging package.

### 6.1.2 ILD tracking performance

ILD tracking is designed around three subsystems capable of standalone tracking: VTX, FTD and the TPC. These are augmented by three auxiliary tracking systems, the SIT, SET and ETD, which provide additional high resolution measurement points. The momentum resolution goal [381] is

$$\sigma_{1/p_T} \approx 2 \times 10^{-5}\ \text{GeV}^{-1}.$$

This level of performance ensures that the model-independent selection of the higgsstrahlung events from the recoil against leptonic $Z \to \mu^+\mu^-$ decays is dominated by beam energy spread rather than the detector resolution. The performance goal for the impact parameter resolution is

$$\sigma_{r\phi} = 5\ \mu\text{m} \oplus \frac{10}{p(\text{GeV})\sin^{3/2}\theta}\ \mu\text{m}. \tag{III-6.1}$$

Meeting this gaol is crucial for the flavour tagging performance, and in particular the efficient separation of charm and bottom quark decays of the higgs boson.

#### 6.1.2.1 Coverage and material budget

Figure III-6.1a shows, as a function of polar angle, $\theta$, the average number of reconstructed hits associated with simulated 100 GeV muons. The TPC provides full coverage down to $\theta = 37°$. Beyond this the number of measurement points decreases. The last measurement point provided by the TPC corresponds to $\theta \approx 10°$. The central inner tracking system, consisting of the six layer VTX and the two layer SIT, provides eight precise measurements down to $\theta = 26°$. The innermost and middle double layer of the VTX extend the coverage down to $\theta \sim 16°$. The FTD provides up to a maximum of five measurement points for tracks at small polar angles. The SET and ETD provide a single high precision measurement point with large lever arm outside of the TPC volume down to a $\theta \sim 10°$. The different tracking system contributions to the detector material budget, including support structures, is shown in Figure III-6.1b. The spikes at small polar angles correspond to the support structures, electronics and cabling in the around the TPC endcap region. The bump at around $90°$ for the TPC corresponds to the central cathode membrane. Compared to the letter of intent the material has overall increased slightly due to the more detailed and realistic simulation, except for the TPC endplate where it has grown by close to 50%. This is explained in more detail in the TPC section 2.3.

### 6.1.2.2 Tracking efficiency

With over 200 contiguous readout layers, pattern recognition and track reconstruction in a TPC is relatively straightforward, even in an environment with a large number of background hits. In addition, the standalone tracking capability of the VTX enables the reconstruction of low transverse momentum tracks which do not reach the TPC. Hermetic tracking down to low angles is important at the ILC [229] and the FTD coverage enables tracks to be reconstructed to polar angles below $\theta = 7°$.

Figure III-6.2 shows, as a function of momentum and polar angle, the track reconstruction efficiency in simulated (high multiplicity) $t\bar{t} \to 6$ jet events at $\sqrt{s}$ =500 GeV and 1 TeV respectively. Efficiencies are plotted with respect to MC tracks that stem from a region of 10 cm around the IP with $p_t > 100$ MeV and $cos(\theta) < 0.99$, excluding decays in flight and requiring at least 90 % purity. For the combined tracking system, the track reconstruction efficiency is on average 99.7 % for tracks with momenta greater than 1 GeV across the entire polar angle range, and it is larger than 99.8 % for $cos(\theta) < 0.95$.

The effects of background from coherent pair background and from multi-peripheral $\gamma\gamma \to \mathrm{hadrons}$ events are taken into account by overlaying the corresponding number of events. For the pair background the correct number of bunch crossings resulting form the foreseen readout times are overlayed.

**Figure III-6.2**
Tracking Efficiency for $t\bar{t} \to 6$ jets at 500GeV and 1 TeV plotted against (left) momentum and (right) $\cos\theta$.

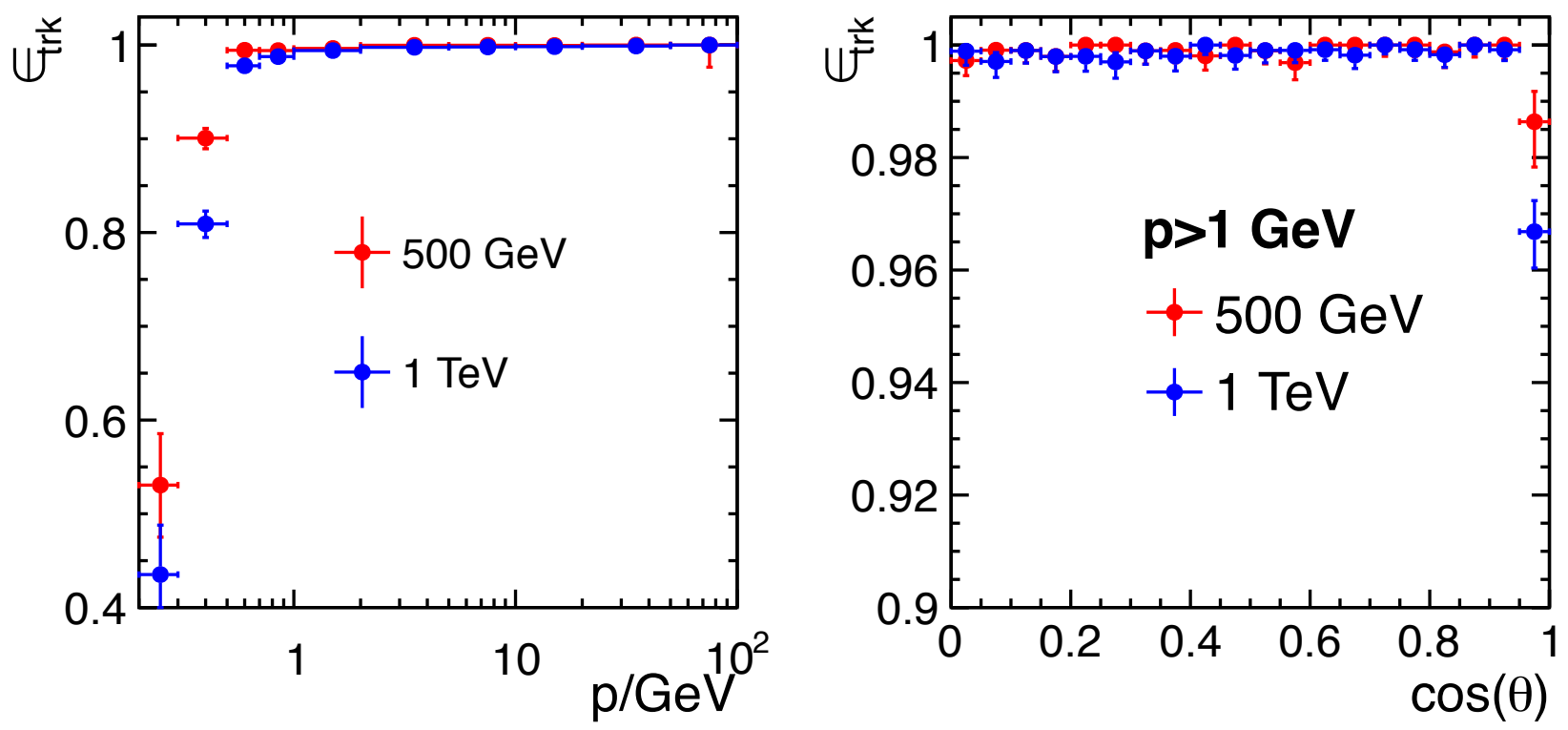

### 6.1.2.3 Momentum resolution for the overall tracking system

The momentum resolution with the ILD simulation and full reconstruction is shown in Figure III-6.3a. The study was performed using muons generated at fixed polar angles of $\theta = 7°, 20°, 30°$ and $85°$, and the momentum was varied over the range $1 - 200$ GeV. For two polar angles, this is compared to the expected parametric form of, $\sigma_{1/p_T} = a \oplus b/(p_T \sin\theta)$, with $a = 2 \times 10^{-5}$ GeV$^{-1}$ and $b = 1 \times 10^{-3}$. As can be seen, at a polar angle of $85°$, the required momentum resolution is attainable over the full momentum range from 1 GeV upwards. This remains true over the full length of the barrel region of the detector, where the TPC in conjunction with the SET is able to provide the longest possible radial lever arm for the track fit. For high momentum tracks, the asymptotic value of the momentum resolution is $\sigma_{1/p_T} = 2 \times 10^{-5}$ GeV$^{-1}$. At $\theta = 30°$, the SET no longer contributes, the effective lever-arm of the tracking system is reduced by 25 %. Nevertheless, the momentum resolution is still within the required level of performance. In the very forward region, the momentum resolution is inevitably worse due to the relatively small angle between the $B$-field and the track momentum.

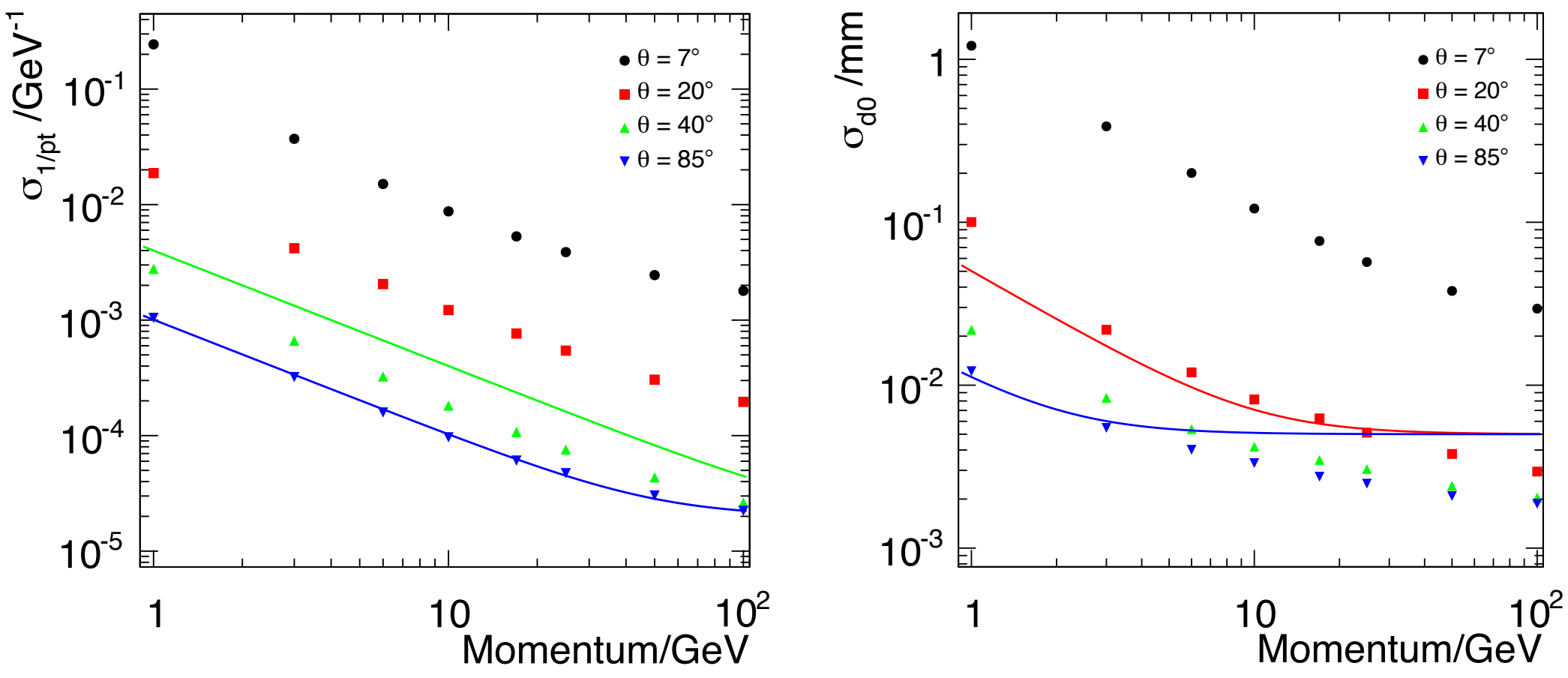

**Figure III-6.3.** (Left) Transverse momentum resolution for single muon events as a function of the transverse momentum for different polar angles. The lines show $\sigma_{1/p_T} = 2 \times 10^{-5} \oplus 1 \times 10^{-3}/(p_T \sin\theta)$ for $\theta = 30°$ (green) and $\theta = 85°$ (blue). (Right) Impact parameter resolution for single muon events as a function of the transverse momentum for different polar angles. The lines show $\sigma_{r\phi} = 5\ \mu\text{m} \oplus \frac{10}{p(\text{GeV})\sin^{3/2}\theta}\ \mu\text{m}$ for $\theta = 20°$ (red) and $\theta = 85°$ (blue).

### 6.1.2.4 Impact parameter resolution

Figure III-6.3b shows $r\phi$ impact parameter resolution as a function of the transverse track momentum. The required performance is achieved down to a track momentum of 1 GeV, whilst it is exceeded for high momentum tracks where the asymptotic resolution is 2 $\mu$m. The $rz$ impact parameter resolution (not shown) is better than $\sim 10\,\mu$m down to momenta of 3 GeV and reaches an asymptotic value of $< 5\,\mu$m for the whole barrel region. Because of the relatively large distance of the innermost FTD disk to the interaction point, the impact parameter resolution degrades for very shallow tracks, $\theta < 15°$. The impact parameter resolution here assumes perfect alignment of the tracking systems.

### 6.1.2.5 Topological time-stamping

The hybrid tracking concept, combining a TPC with silicon tracking devices, is quite powerful also in terms of time-stamping performance. Since the TPC drifts the tracks while the silicon pixels are fixed in space, the silicon can act as an external $z$ detector ($T_0$ device). Drifting TPC tracks are well-measured in $r\phi$ and angle; extrapolating a TPC track to match related silicon hits establishes where the track was in the $z$ direction. An detailed description of this technique for a TPC and a similar one for a standard drift chamber is found in [382]. The time-stamping in ILD is found to be precise to $\simeq 2$ ns (to be compared to $\simeq 300$ ns between BXs at the ILC) so that the bunch crossing which produced the track (the T0) can be uniquely identified. Cosmic background tracks can be eliminated with this tool. It is also viable in the CLIC environment [383].

## 6.1.3 ILD particle flow performance

Many important physics channels at the ILC will consist of final states with at least six fermions, setting a "typical" energy scale for ILC jets as approximately 85 GeV and 170 GeV at $\sqrt{s} = 500$ GeV and $\sqrt{s}$ =1 TeV respectively. Meeting the performance goal of a jet energy resolution of $< 3.5\,\%$ ensures an efficient separation of hadronic decays from W, Z and H bosons. The current performance of the PandoraPFA algorithm applied to ILD Monte Carlo simulated data is summarised in Table III-6.1.

The observed jet energy resolution ($\text{rms}_{90}$) is not described by the expression $\sigma_E/E = \alpha/\sqrt{E/\text{GeV}}$. This is not surprising, as the particle density increases it becomes harder to cor-rectly associate the calorimetric energy deposits to the particles and the confusion term increases. The single jet energy resolution is also listed. The jet energy resolution $(\mathrm{rms}_{90})$ is better than 3.7 % for jets of energy greater than $40$ GeV. The resolutions quoted in terms of $\mathrm{rms}_{90}$ should be multiplied by a factor of approximately 1.1 to obtain an equivalent Gaussian analysing power[274]. Despite, the inclusion of dead material in the Monte Carlo simulation, the resolutions achieved are between 2 % and 7 % better than for the previous detector model described in [198]. In part this reflects a number of improvements to the particle flow reconstruction software. Nevertheless, it can be concluded that the additional dead material associated with services does not significantly degrade the jet energy resolution.

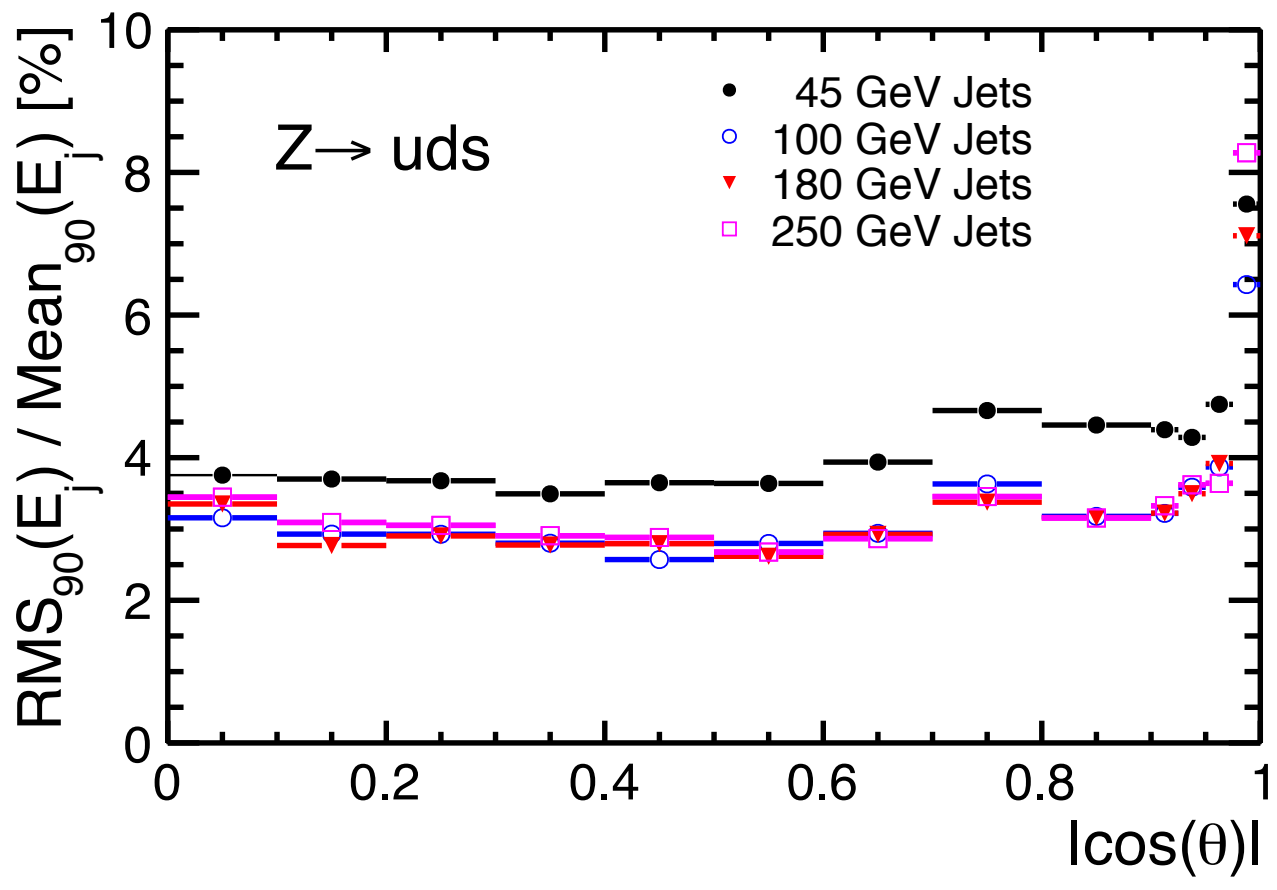

**Figure III-6.4** Fractional jet energy resolution plotted against $|\cos\theta|$ where theta is the thrust axis of the event.

Figure III-6.4 shows the jet energy resolution for $\mathrm{Z} \rightarrow$uds events plotted against the cosine of the polar angle of the generated $\mathrm{q\bar{q}}$ pair, $\cos\theta_{\mathrm{q\bar{q}}}$, for four different values of $\sqrt{s}$. Due to the calorimetric coverage in the forward region, the jet energy resolution remains good down to $\theta = 13°$ $(\cos\theta = 0.975)$.

### 6.1.4 Flavour tagging performance

Identification of $\mathrm{b}$-quark and $\mathrm{c}$-quark jets plays an important role within the ILC physics programme. The vertex detector design and the impact parameter resolution are of particular importance for flavour tagging. The LCFIPlus flavour tagging software uses boosted decision trees to discriminate $\mathrm{b}$ jets from $\mathrm{udsc}$ jets ($\mathrm{b}$-tag), $\mathrm{c}$ jets from $\mathrm{udsb}$ jets ($\mathrm{c}$-tag), and $\mathrm{c}$ jets from $\mathrm{b}$ jets ($\mathrm{bc}$-tag).

The flavour tagging performance [384] of ILD was previously studied for the two vertex detector geometries considered, three double-sided ladders (VTX-DL) and five single-sided (VTX-SL) ladders. No significant differences in the input variables for the multivariate analysis were seen. Here results are presented only for the double-layer layout. The flavour tagging performance is studied using simulated and fully reconstructed samples for $\mathrm{Z} \rightarrow q\bar{q}$ reactions, shown in Figure III-6.5a, and

**Table III-6.1.** Jet energy resolution for $\mathrm{Z} \rightarrow$uds events with $|\cos\theta_{\mathrm{q\bar{q}}}| < 0.7$, expressed as, $\mathrm{rms}_{90}$ for the di-jet energy distribution, the effective constant $\alpha$ in $\mathrm{rms}_{90}/E = \alpha(E_{jj})/\sqrt{E_{jj}/\mathrm{GeV}}$, and the fractional jet energy resolution for a single jets, $\sigma_{E_j}/E_j$. The jet energy resolution is calculated from $\mathrm{rms}_{90}$.

| Jet Energy | $\mathrm{rms}_{90}$ | $\mathrm{rms}_{90}/\sqrt{E_{jj}/\mathrm{GeV}}$ | $\sigma_{E_j}/E_j$ |
|---|---|---|---|
| 45 GeV | 2.4 GeV | 24.7 % | $(3.66 \pm 0.05)$ % |
| 100 GeV | 4.0 GeV | 28.3 % | $(2.83 \pm 0.04)$ % |
| 180 GeV | 7.3 GeV | 38.5 % | $(2.86 \pm 0.04)$ % |
| 250 GeV | 10.4 GeV | 46.6 % | $(2.95 \pm 0.04)$ % |

$ZZZ \to q\bar{q}q\bar{q}q\bar{q}$ reactions, shown in Figure III-6.5b. The latter process is forced to decay into the same quark pairs for all three Z decays. The $\gamma\gamma \to \text{hadrons}$ backgrounds are not overlaid for this study. The boosted decision trees are retrained for the different energies and different final states. A slight performance degradation is seen by increasing the jet energy. The performance also degrades by increasing the number of jets in the final state, which can be attributed to reconstruction effects in busy environments.

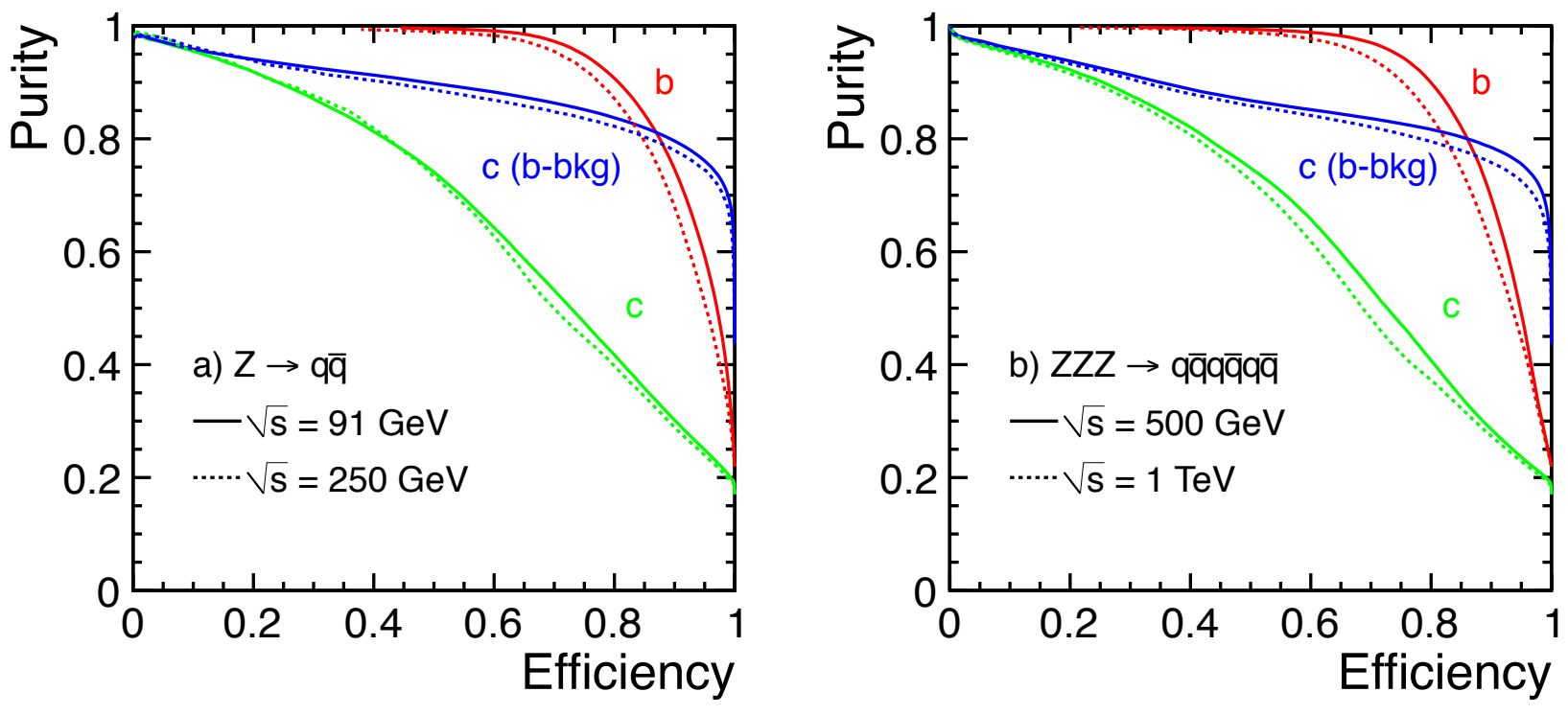

**Figure III-6.5** Flavour tagging performance plots for (a) $Z \to q\bar{q}$ samples at $\sqrt{s} = 91$ GeV and 250 GeV, and (b) $ZZZ \to q\bar{q}q\bar{q}q\bar{q}$ samples at $\sqrt{s} = 500$ GeV and 1 TeV.

### 6.1.5 Comparison of detector models

To compare the AHCAL and SDHCAL options, the $e^+e^- \to t\bar{t}h$ benchmark signal samples were simulated and fully reconstructed using dedicated detector models (ILD_o1_v05 and ILD_o2_v05, respectively) and reconstruction software for each option, and analyzed as described in Sec. 6.3.4. It was found that the there were no significant differences in the mass resolutions of the top and higgs candidates.

## 6.2 ILD physics performance at 250 and 500 GeV

In this section the performance of ILD is described for $\sqrt{s} = 250$ GeV and $\sqrt{s} = 500$ GeV. More details may be found in [198]. The results are summarised in Table III-6.2. These measurements demonstrate the excellent performance of the ILD detector for many different final states. In this chapter three topics are reviewed in more detail, which stress in particular the detector performance.

### 6.2.1 Higgs recoil mass reconstruction

The precise determination of the properties of the higgs boson is one of the main goals of the ILC. In particular, the model independent determination of the higgs boson branching ratios is central to the physics goals of the ILC. Here the measurement of the $e^+e^- \to hZ$ cross section from the recoil mass distribution in $Zh \to e^+e^-X$ and $Zh \to \mu^+\mu^-X$ events, determines the absolute $g_{hZZ}$ coupling. In $Zh \to \mu^+\mu^-X$ events the recoil mass resolution is determined by the beam-energy spread and the muon momentum resolution, whereas for $Zh \to e^+e^-X$ events Bremsstrahlung and final-state radiation (FSR) dominate. The reconstructed recoil mass distributions for simulated events is shown in Figure III-6.6. Measurement precisions on the $hZ$ production cross section of 3.6 % and 4.3 % were obtained from the respective $\mu^+\mu^-$ and $e^+e^-(n\gamma)$ final states. In the $\mu^+\mu^-$ final state, the resolution is limited by the beam energy spread rather than by the momentum resolution of the detector.

**Table III-6.2.** A summary of the main physics benchmark measurements presented in the ILD LOI and [385].

| $\sqrt{s}$ | Observable | Precision | Comments |
|---|---|---|---|
| 250 GeV | $\sigma(\mathrm{e^+e^-} \to \mathrm{Z}h)$ | $\pm 0.30$ fb (2.5 %) | Model Independent |
| | $m_h$ | 32 MeV | Model Independent |
| | $m_h$ | 27 MeV | Model Dependent |
| 250 GeV | $Br(h \to \mathrm{b\bar{b}})$ | 2.7 % | includes 2.5 % |
| | $Br(h \to \mathrm{c\bar{c}})$ | 7.3 % | from |
| | $Br(h \to gg)$ | 8.9 % | $\sigma(\mathrm{e^+e^-} \to \mathrm{Z}h)$ |
| 500 GeV | $\sigma(\mathrm{e^+e^-} \to \tau^+\tau^-)$ | 0.29 % | $\theta_{\tau^+\tau^-} > 178°$ |
| | $A_{FB}$ | $\pm 0.0025$ | $\theta_{\tau^+\tau^-} > 178°$ |
| | $P_\tau$ | $\pm 0.007$ | excluding $\tau \to a_1\nu$ |
| 500 GeV | $\sigma(\mathrm{e^+e^-} \to \tilde{\chi}_1^+\tilde{\chi}_1^-)$ | 0.6 % | |
| | $\sigma(\mathrm{e^+e^-} \to \tilde{\chi}_2^0\tilde{\chi}_2^0)$ | 2.1 % | |
| | $m(\tilde{\chi}_1^\pm)$ | 2.4 GeV | from kin. edges |
| | $m(\tilde{\chi}_2^0)$ | 0.9 GeV | from kin. edges |
| | $m(\tilde{\chi}_1^0)$ | 0.8 GeV | from kin. edges |
| 500 GeV | $\sigma(\mathrm{e^+e^-} \to \mathrm{t\bar{t}})$ | 0.4 % | $(\mathrm{bq\bar{q}})$ $(\mathrm{\bar{b}q\bar{q}})$ only |
| | $m_t$ | 40 MeV | fully-hadronic only |
| | $m_t$ | 30 MeV | + semi-leptonic |
| | $\Gamma_t$ | 27 MeV | fully-hadronic only |
| | $\Gamma_t$ | 22 MeV | + semi-leptonic |
| | $A_{\mathrm{F}B}^t$ | $\pm 0.0079$ | fully-hadronic only |
| 500 GeV | $\sigma(\mathrm{e^+e^-} \to \tilde{\mu}_L^+\tilde{\mu}_L^-)$ | 2.5 % | |
| | $m(\tilde{\mu}_L)$ | 0.5 GeV | |
| 500 GeV | $m(\tilde{\tau}_1)$ | $0.1\,\mathrm{GeV} \oplus 1.3\sigma_{\mathrm{LSP}}$ | SPS1a' |
| 1 TeV | $\alpha_4$ | $-1.4 < \alpha_4 < 1.1$ | SPS1a' |
| | $\alpha_5$ | $-0.9 < \alpha_5 < +0.8$ | WW Scattering |

## 6.2.2 Tau reconstruction

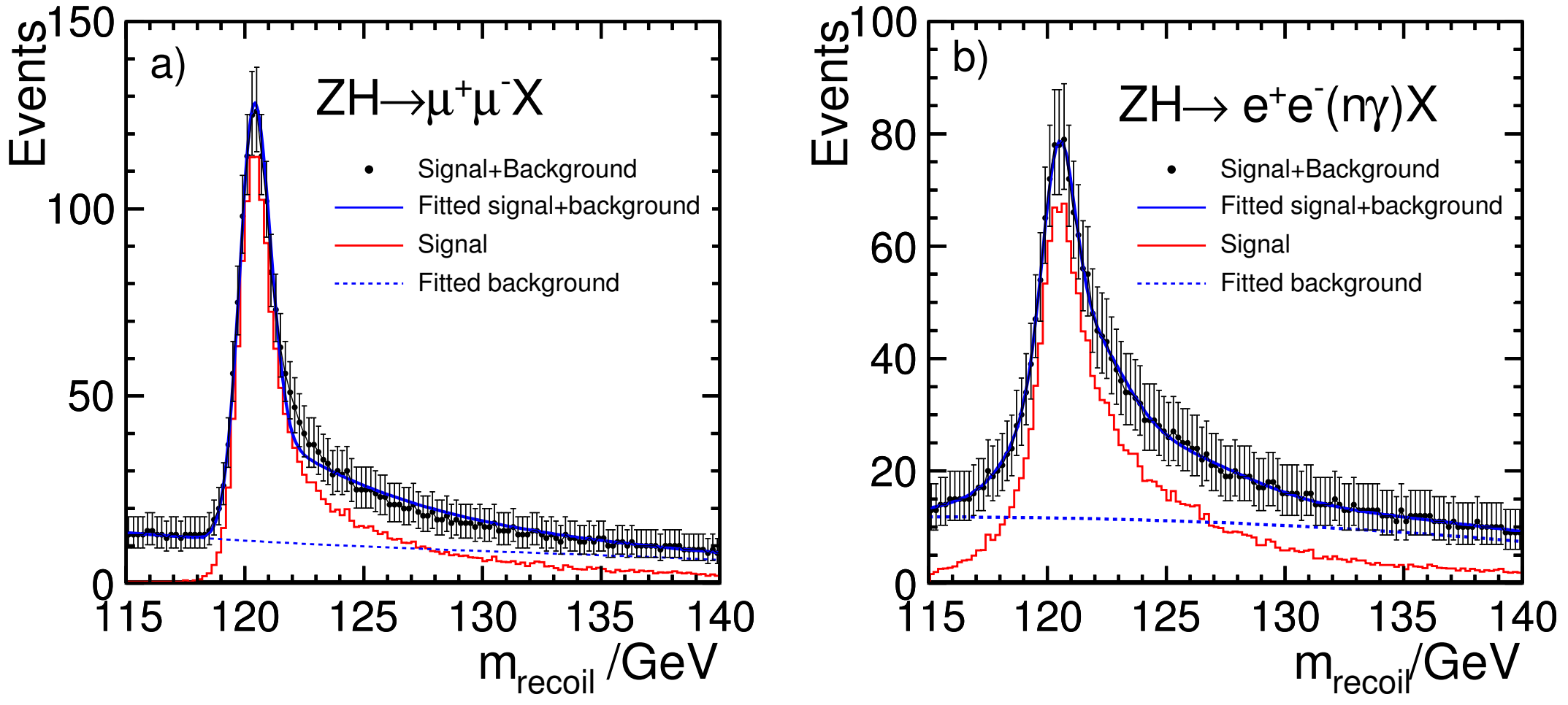

**Figure III-6.6.** Results of the model independent analysis of the Higgs-strahlung process $\mathrm{e^+e^-} \to h\mathrm{Z}$ in which a) $\mathrm{Z} \to \mu^+\mu^-$ and b) $\mathrm{Z} \to \mathrm{e^+e^-}$ (including the reconstruction of bremsstrahlung and FSR photons). The results are shown are for the $P(e^+, e^-) = (+30\,\%, -80\,\%)$ beam polarisation.

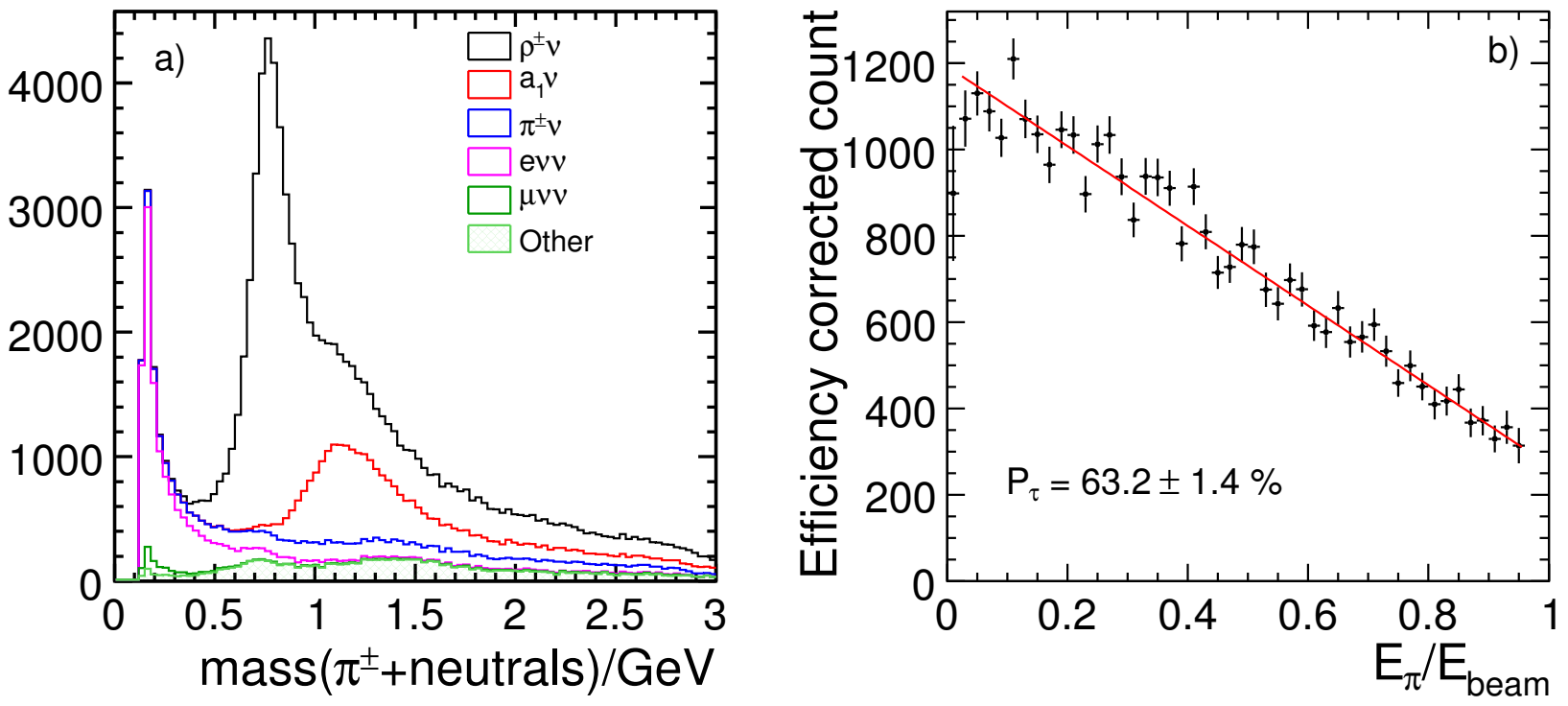

**Figure III-6.7**
a) The invariant mass distribution for selected 1-prong tau-candidates and b) The efficiency corrected reconstructed pion energy distribution for selected $\tau \to \pi\nu$ candidates.

The higgs recoil mass analysis provides a clear demonstration of excellent tracking performance of the ILD detector concept. The reconstruction of $\tau^+\tau^-$ events at $\sqrt{s} = 500$ GeV provides a challenging test of the detector performance in terms of separating nearby tracks and photons. The reconstruction of the tau final states enables the mean tau polarisation $P_\tau$ to be determined. For the tau polarisation measurement, the $\tau \to \pi\nu$ and $\tau \to \rho\nu$ decays have the highest sensitivity. The separation of the 1-prong decay modes relies on lepton identification and the ability to separate the neutral energy deposits from $\pi^0$ decays from the hadronic shower. The invariant mass distribution for 1-prong events is shown in Figure III-6.7a.

A neural network approach based on nine input variables is used to identify the tau decays modes. The variables include: the total energy of the identified photons, the invariant mass of the track and all identified photons (Figure III-6.7a); and electron and muon particle identification variables based on calorimetric information and track momentum.

**Table III-6.3**
Purity and efficiency of the main tau decay mode selections.

| Mode | Efficiency | Purity |
|---|---|---|
| $e\nu\nu$ | 98.9 % | 98.9 % |
| $\mu\nu\nu$ | 98.8 % | 99.3 % |
| $\pi\nu$ | 96.0 % | 89.5 % |
| $\rho\nu$ | 91.6 % | 88.6 % |
| $a_1\nu$ (1-prong) | 67.5 % | 73.4 % |
| $a_1\nu$ (3-prong) | 91.1 % | 88.9 % |

Table III-6.3 shows the efficiency and purity achieved for the six main tau decay modes. The selection efficiency is calculated with respect to the sample of $\tau^+\tau^-$ after the requirement that the two tau candidates are almost back-to-back. The purity only includes the contamination from other $\tau^+\tau^-$ decays. The high granularity and the large detector radius of ILD results in excellent separation.

### 6.2.3 Strong EWSB

If strong electroweak symmetry breaking (EWSB) is realised in nature, the study of the WW-scattering processes is particularly important. At the ILC, the $\mathrm{W^+W^-} \to \mathrm{W^+W^-}$ and $\mathrm{W^+W^-} \to \mathrm{ZZ}$ vertices can be probed via the processes $\mathrm{e^+e^-} \to \nu_e\bar{\nu}_e\mathrm{q\bar{q}q\bar{q}}$ where the final state di-jet masses are from the decays of two W-bosons or two Z-bosons. Separating the two processes through the reconstruction of the di-jet masses provides a test of the jet energy resolution of the ILD detector. Strong EWSB can be described by an effective Lagrangian approach in which there are two anomalous quartic gauge couplings, $\alpha_4$ and $\alpha_5$ which are identically zero in the SM. Figure III-6.8 shows, for $\nu_e\bar{\nu}_e\mathrm{WW}$ and $\nu_e\bar{\nu}_e\mathrm{ZZ}$ events at $\sqrt{s} = 1$ TeV, a) the reconstructed di-jet mass distribution, and b) the distribution of average reconstructed mass, $(m_{ij} + m_{kl})/2.0$. Clear separation between the W and Z peaks is obtained, demonstrating that the ILD jet energy resolution is sufficient to separate the hadronic decays of gauge bosons.

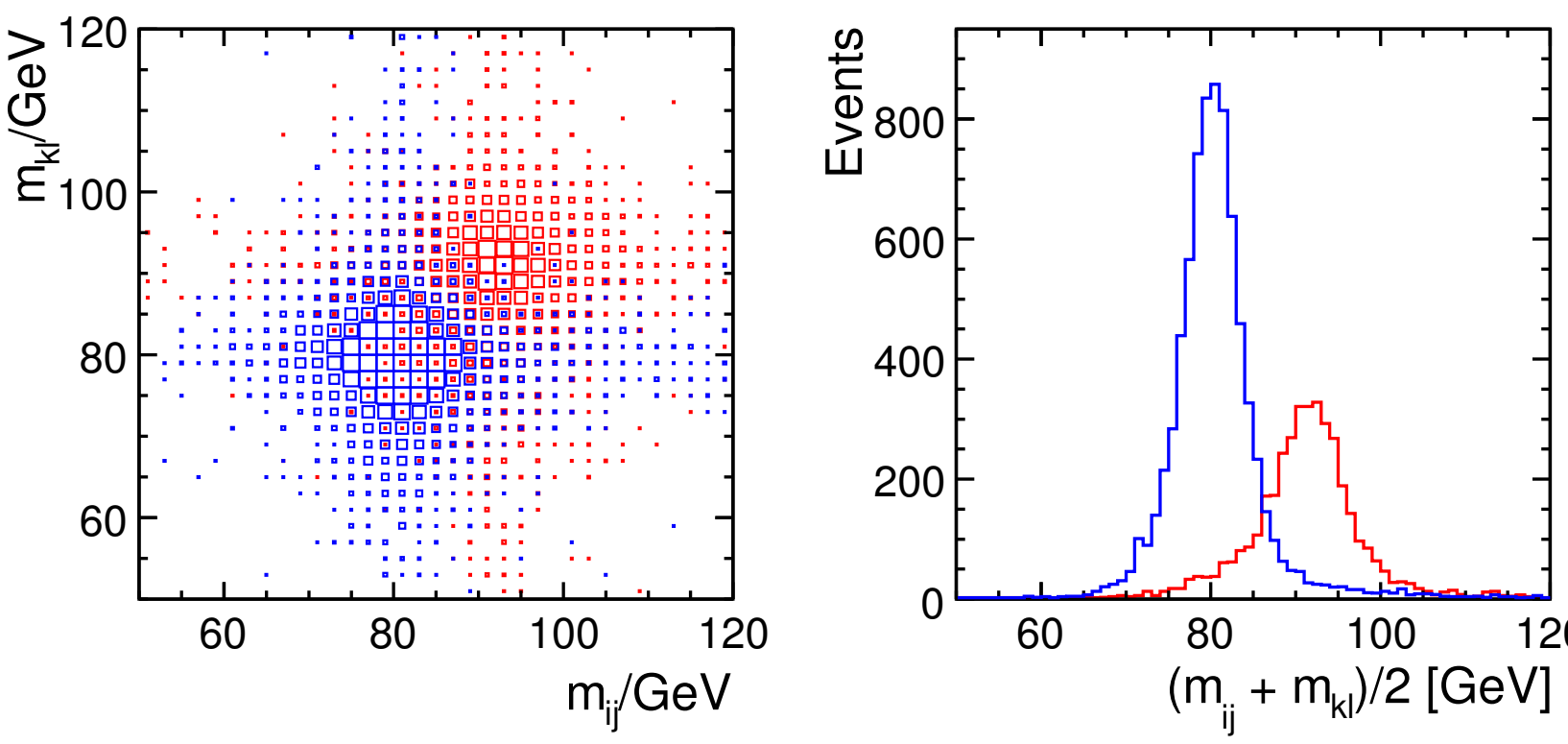

**Figure III-6.8**
a) The reconstructed di-jet mass distributions for the best jet-pairing in selected $\nu_e\bar{\nu}_e\mathrm{WW}$ (blue) and $\nu_e\bar{\nu}_e\mathrm{ZZ}$ (red) events at $\sqrt{s} = 1\,TeV$. b) Distributions of the average reconstructed di-jet mass, $(m_{ij} + m^B_{kl})/2.0$, for the best jet-pairing for $\nu_e\bar{\nu}_e\mathrm{WW}$ (blue) and $\nu_e\bar{\nu}_e\mathrm{ZZ}$ (red) events.

## 6.3 ILD benchmarking

In chapter 1.4, the list of benchmark reactions is described which have been studied by the detector groups (for more detail see [386]). The result of the analyses of these benchmarks are briefly presented in this section. The generation of both signal, physics background, and machine background was done as a common effort between ILD and SiD and is described in detail in chapter 2.2. The detector simulation software and detector model used are described in chapter 5.4. Events for the analyses were generated and simulated with the detailed GEANT4 based ILD model, and centrally reconstructed. The PandoraPFA and LCFIPlus algorithms (described in chapter 2.2) were used.

The first three benchmark processes presented are at $\sqrt{s}$=1 TeV. They were chosen partly to demonstrate the capability of the detectors under the conditions of the ILC operating at 1 TeV, partly to exploit the opportunities that this higher energy would bring. More specifically:

$e^+e^- \to \nu\bar{\nu}h$ is intended to test the detector capabilities in simple topologies.

$e^+e^- \to W^+W^-$ is complementing the first benchmark by topologies with jets at higher energies and at lower angles.

$e^+e^- \to t\bar{t}h$ is intended to demonstrate the capability of the detector to disentangle very complicated final states.

These processes were studied assuming an integrated luminosity ($L$) of 1 ab$^{-1}$, and with polarised beams. Using the convention that $\mathcal{P}_{p^-,p^+}$ denotes a configuration of $p-\%$ degree of polarisation for the electrons, $p+\%$ for the positrons, the full sample was evenly divided in two samples with $\mathcal{P}_{-80,+20}$ and $\mathcal{P}_{+80,-20}$. The full sample is referred to as the full DBD sample in the following, while the two sub-samples are called the DBD $\mathcal{P}_{-80,+20}$ and $\mathcal{P}_{+80,-20}$ samples.

The last of the benchmark processes was the analysis of $\mathrm{e^+e^-} \to \mathrm{t\bar{t}}$ at $\sqrt{s}$ = 500 GeV. The integrated luminosity was assumed to be 500 fb$^{-1}$, evenly divided in a $\mathcal{P}_{-80,+30}$ sample and a $\mathcal{P}_{+80,-30}$ one. This particular reaction was chosen to compare the current more detailed ILD model to the one used in earlier studies to understand the impact the improved simulation model has on the physics reach.

## 6.3.1 Common reconstruction tools

### 6.3.1.1 Isolated lepton finding

In several analyses the task is to identify an isolated lepton within a jet. The strategy proposed in [387] is to force the jet clustering algorithm to form a given number of jets, e.g. four in the case of semi-leptonic $t\bar{t}$ decays. The searched lepton has distinct features with respect to other particles in the jet. A lepton is called "isolated" if it is either the particle with the highest momentum (the "leading" particle) in the jet or if it has a large transverse momentum with respect to the jet axis. The two variables $x_T$ and $z$ are defined to take these two configurations into account:

$$x_T = \frac{p_{T,\text{lepton}}}{m_{\text{jet}}}$$

where $p_{T,\text{lepton}}$ is the transverse momentum of the identified lepton with respect to its jet axis and $m_{\text{jet}}$ is the mass of the jet ($m^2 = E^2 - \mathbf{p}^2$), and

$$z = \frac{E_{\text{lepton}}}{E_{\text{jet}}},$$

which corresponds to the fraction of jet-energy from the lepton. The distribution for leptons in semi-leptonic and fully hadronic $t\bar{t}$ events can be seen in Figure III-6.9, left. The fraction $z$ is restricted to values smaller than 1. The variable $x_T$ must be less than 1/2 which is the kinematic limit of a jet taken at rest where the lepton and the other particles are almost back-to-back and share the same energy $m_{jet}/2$.

### 6.3.1.2 Jet clustering

At lepton colliders, exclusive jet algorithms – in which every particle is assigned to a jet – have been favoured. An example of this algoritm-type is the Durham algorithm [388]. However, at the ILC such algorithms work poorly: while it is still true that all particles from the main interaction can be assigned to jets in a unique fashion, and that this interaction does not contain an "underlying event", the large cross section for $\gamma\gamma \rightarrow \text{hadrons}$ implies that most interesting events will be accompanied by several unrelated $\gamma\gamma \rightarrow \text{hadrons}$ events ("pile-up events") in the same bunch crossing. Exclusive jet algorithms will inevitably include particles from the pile-up events into the jets.

This problem was studied at CLIC, where the pile-up conditions are much more challenging than at ILC [199]. It was concluded that the use of inclusive algorithms, developed for hadron colliders, was well-suited to mitigate this problem. The algorithm used was the $k_t$ algorithm [389], as implemented in the FASTJET package[390]. In the $k_t$ algorithm, the measure of distance between two objects $i$ and $j$ (particles or proto-jets) is $d_{ij} = \min(p_{Ti}^2, p_{Tj}^2)[\Delta^2_{\eta ij} + \Delta^2_{\phi ij}]/R^2$. In each iteration, $d_{ij}$ is calculated for all $ij$. The smallest of all $d_{ij}$ (=$d_{min}$) is compared with the smallest $p^2_{Tk}$ (=$p^2_{T,min}$) of all objects $k$. If $p^2_{T,min} < d_{min}$ the corresponding object $k$ is removed from the event. In the opposite case, the corresponding objects $i$ and $j$ are merged. The procedure is then repeated until an end condition is fulfilled, which might be that only $N_{jet}$ objects still are left to consider, or that the lowest measure in the iteration was above a pre-defined limit. As the polar distance is measured in pseudo-rapidity ($\eta$), rather than polar angle, particles close to the beam axis - where $\eta$ tends to $\pm\infty$ - are less likely to be considered close enough to be merged, and are more likely to be removed as particles not belonging to any jet. Hence, the algorithm will remove low $p_T$, low polar angle particles - typical for pile-up events - from the jets. By choosing appropriate values of the parameters $R$ and $N_{jet}$, an optimal performance can be found. This optimum will in general not be the same for different benchmark reactions, due to the differences in number of expected jets, amount of invisible energy, or the angular distributions of the signal reaction.

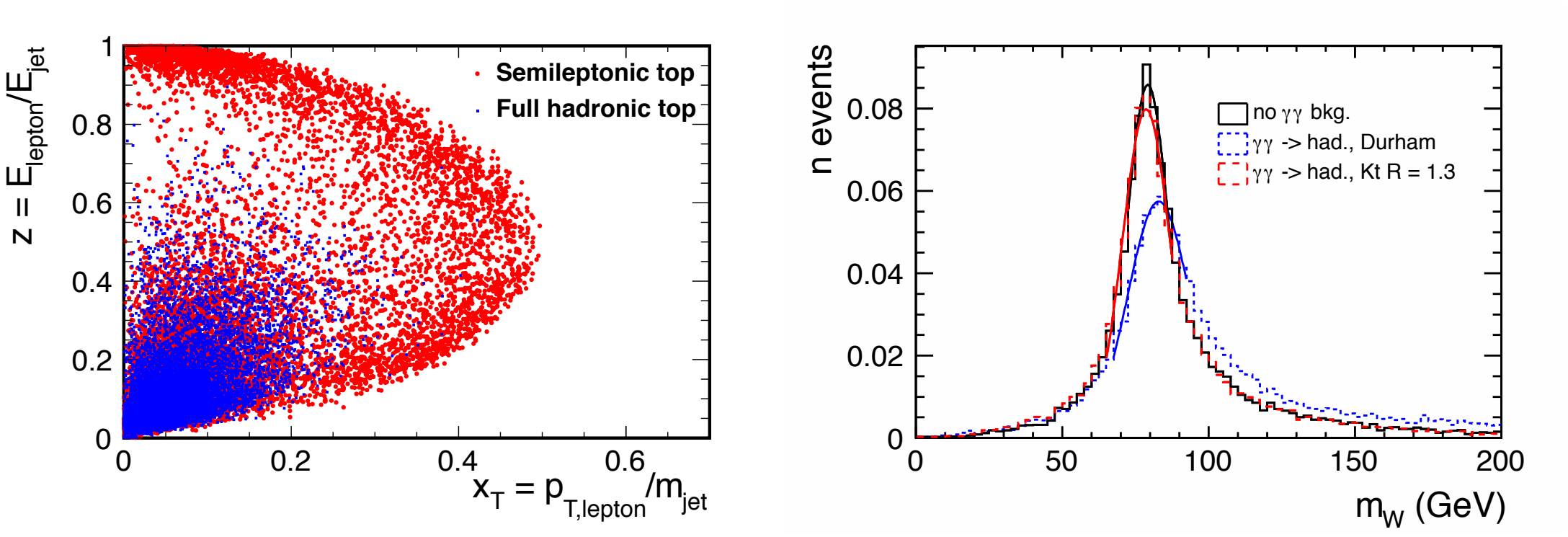

**Figure III-6.9.** Left: Distribution of variables $x_T$ and $z$ used to identify isolated leptons in semi-leptonic (red) and full hadronic (blue) top events. Right: Reconstructed di-jet mass distribution in $e^+e^- \rightarrow \mathrm{WW} \rightarrow \ell\nu qq$. Black solid: Durham algorithm, no pile-up; Blue dotted: Durham algorithm, with pile-up; Red dashed : $k_t$ algorithm, with pile-up.

In Figure III-6.9 (right) these features are demonstrated. The figure shows the di-jet mass distribution for the $\mathrm{WW} \rightarrow \ell\nu qq$ reaction. The isolated lepton was removed before each jet algorithm was applied. It can be clearly seen that the result without pile-up is almost restored.

### 6.3.2 $\mathrm{e^+e^-} \rightarrow \nu\bar{\nu}\mathrm{h}$

At $\sqrt{s} = 1$ TeV, the Higgs boson is mainly produced via the WW-fusion process ($\mathrm{e^+e^-} \rightarrow \nu_e\bar{\nu}_e h$), with a cross section exceeding the maximum cross section close to threshold (at around 250 GeV), where production is dominated by the Higgs-strahlung ($\mathrm{e^+e^-} \rightarrow Zh$) process, as shown in Figure III-6.10.

For this analysis [391, 392], Higgs boson production and decay were fully simulated, as was relevant background processes. In the simulation, the SM Higgs boson BRs for $m_h$= 125 GeV were used, see table III-6.4. The aim of the study is to determine to which accuracy the cross section weighted branching ratios ($\sigma$·BR) can be determined from the data for the decay modes $\mathrm{b\bar{b}}$, $\mathrm{c\bar{c}}$, $gg$, $\mathrm{WW}^*$, and $\mu^+\mu^-$.

**Table III-6.4**
Higgs branching ratios, from [184].

| Mode | $\mathrm{b\bar{b}}$ | $\mathrm{c\bar{c}}$ | $gg$ | $\mathrm{WW}^*$ | $\mu^+\mu^-$ | $\tau^+\tau^-$ | $\mathrm{ZZ}^*$ | $\gamma\gamma$ | $\mathrm{Z}\gamma$ |
|---|---|---|---|---|---|---|---|---|---|
| BR (%) | 57.8 | 2.7 | 8.6 | 21.6 | 0.02 | 6.4 | 2.7 | 0.23 | 0.16 |

**Figure III-6.10**
Higgs production cross section as a function of $\sqrt{s}$ for $\mathcal{P}_{-80,+20}$.

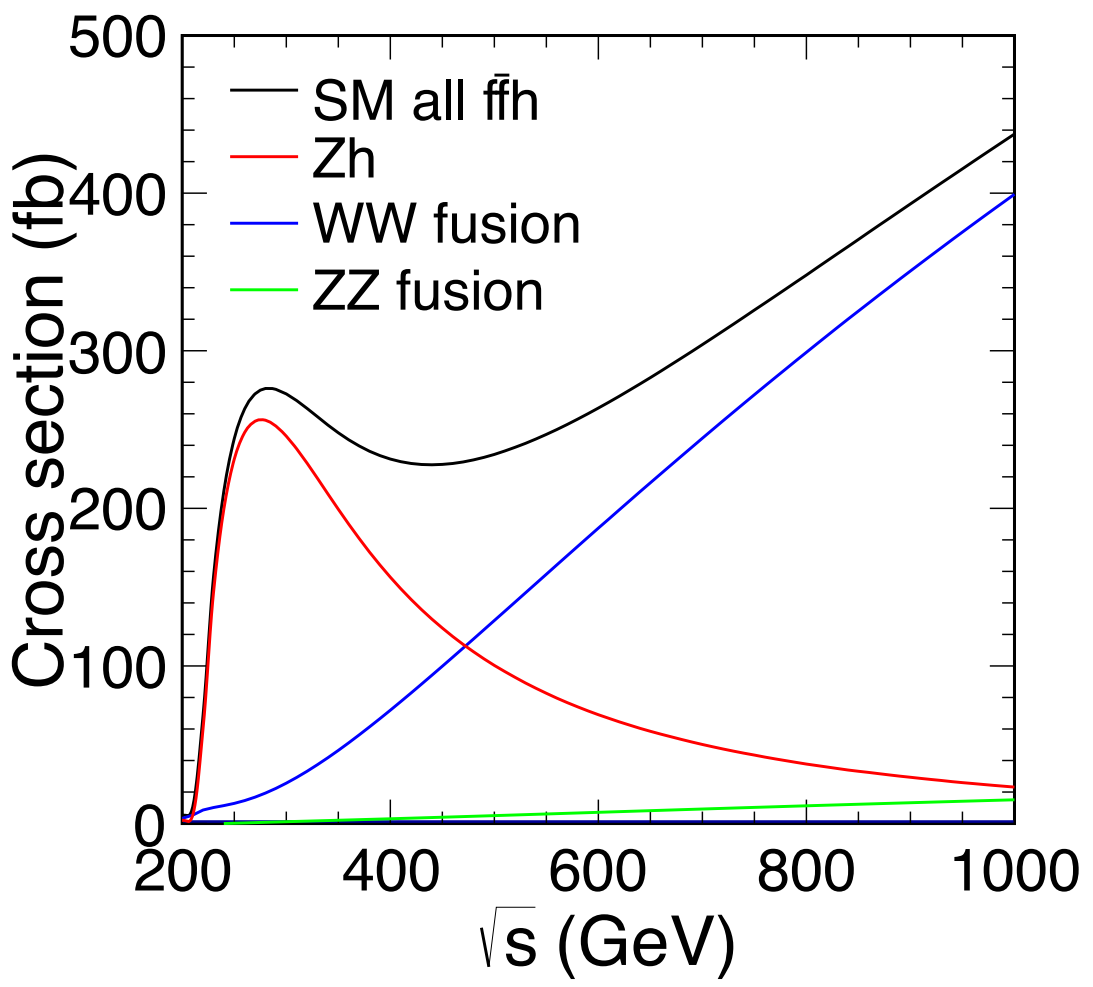

In the hadronic decay channels of the Higgs, the final state forms two jets and flavour tagging performance is crucial to measure the BRs. In the $h \rightarrow \mathrm{WW}^*$ channel, only the fully hadronic decay mode, $h \to \mathrm{WW}^* \to \mathrm{q\overline{q}q\overline{q}}$, was considered. At $\sqrt{s} = 1$ TeV, higher instantaneous luminosity is expected than at 250 or 500 GeV. This, together with the rising Higgs production cross section, implies that one can accumulate observable amounts of $h \to \mu^+\mu^-$ events ($\sigma$·BR$= 0.089$ fb for $\mathcal{P}_{-80,+20}$).

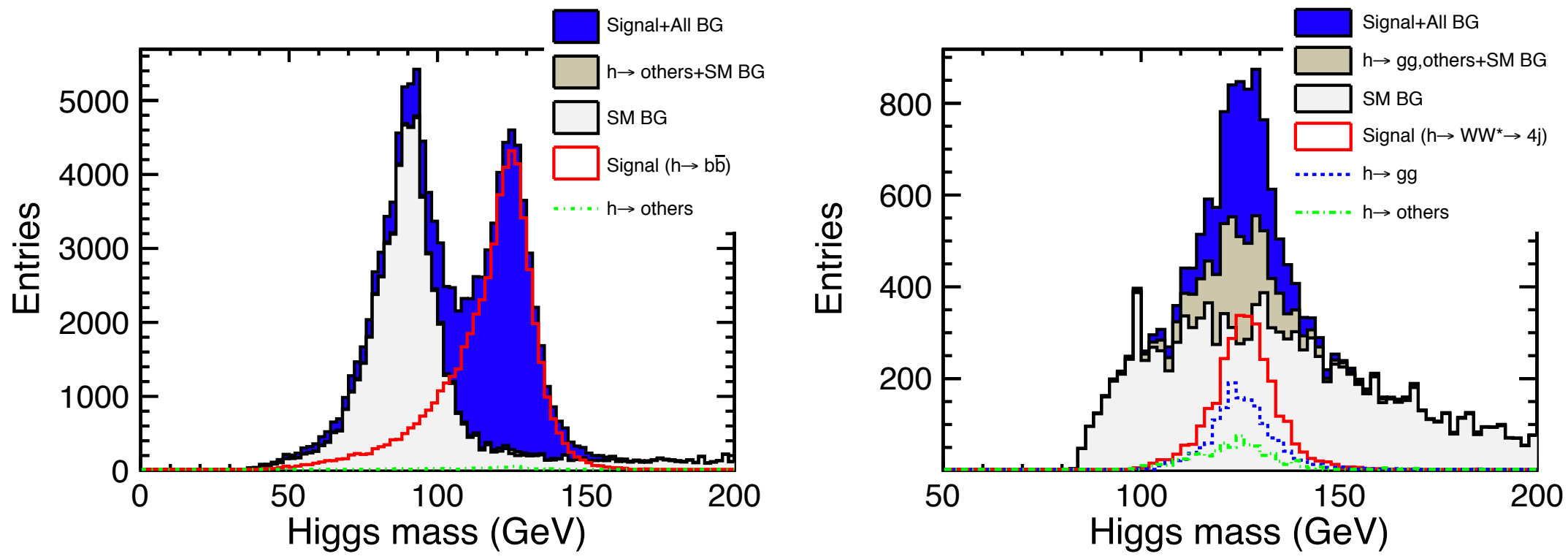

**Figure III-6.11.** Left:Reconstructed $h \to \mathrm{b\overline{b}}$ di-jet mass distribution after the b-tagging selection. Right: Reconstructed Higgs mass distribution in $h \to \mathrm{WW}^*$ fully hadronic decay channel. Both figures correspond to the DBD $\mathcal{P}_{-80,+20}$ sample.

In the $\boldsymbol{h \to \mathrm{b\overline{b}}, \mathrm{c\overline{c}},}$ **and** $\boldsymbol{gg}$ **channels**, the events have in common that they contain two jets with a di-jet mass consistent with the Higgs mass and that they have large missing energy due to the neutrinos. Flavour tagging is crucial to distinguish the decay channels.

Jets were reconstructed by first employing the $k_t$ jet clustering algorithm with $R = 1.1$ and $N_{jet} = 2$ to remove particles from pile-up events, and then the Durham algorithm on the remaining particles. In order to reduce the background, it was required that the visible energy and longitudinal momentum should be small, while the transverse momentum should be high. Cuts based on the total particle-multiplicity and the polar angle of the jets were applied to reduce the 2-fermion background. Finally, the Higgs candidate events for flavour tagging were selected by requiring the mass of the di-jet to be in $[110,\ 150]$ GeV. The efficiency to select $h \to \mathrm{b\overline{b}}$, $\mathrm{c\overline{c}}$ and $gg$ at this stage were 35.0%, 37.3% and 35.9%, respectively, while the major background was the $\nu\bar{\nu}q\bar{q}$ (non-Higgs) final state.

A flavour tagging template fitting was performed to extract $\sigma$·BR for the different channels. The flavour templates of $h \to \mathrm{b\overline{b}}$, $\mathrm{c\overline{c}}$, $gg$, and backgrounds were obtained from the flavour tagging boosted-decision tree output of LCFIPlus. Figure III-6.11 (left) shows the reconstructed $h \to \mathrm{b\overline{b}}$ di-jet mass distribution after applying a b-tagging cut for the DBD $\mathcal{P}_{-80,+20}$ sample. By repeating the template fit 5000 times on distributions generated by a toy Monte Carlo, the measurement expected accuracies on $\sigma$·BR could be evaluated.

In the fully hadronic $\boldsymbol{h \to \mathrm{WW}^*}$ **channel**, the expected final state is four jets consistent with $\mathrm{WW}^*$, with total mass consistent with the Higgs mass, while having large missing energy and missing transverse momentum. Background from pile-up events was removed by employing the $k_t$ jet clustering algorithm with $R = 0.9$ and $N_{jet} = 4$. The remaining particles were forced to into a four-jet configuration using the Durham algorithm. From the reconstructed four jets, the jet pairing yielding the di-jet mass closest to $m_{\mathrm{W}}$ was assumed to be the $\mathrm{W}$. The other di-jet should have a mass between 15 and 60 GeV. In the jet clustering, it was demanded that the Durham algorithm should show a preference for the four-jet configuration. Subsequently, pre-selections similar to those of the two-jet channel were applied. In this channel, $h \to \mathrm{b\overline{b}}$ could be a major background, therefore the $\mathrm{b}$-likeness from LCFIPlus was required to be low.

The distribution of the reconstructed Higgs mass in the $h \to \mathrm{WW}^*$ hadronic decay channel is shown in Figure III-6.11 (right) for the DBD $\mathcal{P}_{-80,+20}$ sample. Signal selection efficiency of $h \to \mathrm{WW}^*$ was 12.4% and remaining major backgrounds are 4-fermions ($e^+e^- \to \nu\bar{\nu}q\bar{q}$), 3-fermions ($e\gamma \to \nu q\bar{q}$) and other decay channels of the Higgs. The relative measurement error on $\sigma$·BR was evaluated by $\frac{\sqrt{N_s+N_{BG}}}{N_s}$, where $N_s(N_{BG})$ is the number of signal (background) events in the signal region.

The $\boldsymbol{h \to \mu^+\mu^-}$ **channel**, due to its very low branching-ratio was only studied in the DBD $\mathcal{P}_{-80,+20}$ sample where the Higgs production cross section is larger. The main backgrounds are $\mathrm{e^-e^+} \to \nu\bar{\nu}\mu^-\mu^+$ and $\gamma\gamma \to \nu\bar{\nu}\ell^-\ell^+$. Events with two reconstructed high momentum isolated tracks were selected, provided that the two tracks were identified as muons. The invariant mass of the di-muon system was required to be between 95 and 155 GeV, and its energy to be lower than 400 GeV. Fully leptonic events were selected by requiring low multiplicity and high missing energy. The di-$\tau$ background was reduced by requiring that the significance of the impact parameters should be small. The signal efficiency at this pre-selection stage was found to be 81.1 %. Further cuts on missing energy and transverse momentum, the minimum angle to the beam-axis of the muons and on energy detected in the very forward calorimeter were applied. The final signal efficiency after all cuts was 37.0 %.

Figure III-6.12 shows the reconstructed di-muon mass of $h \to \mu^+\mu^-$. After the final selection was applied, the resulting invariant mass distributions for the background and the signal were fitted individually. Those fits were used to generate mass-distributions for 5000 pseudo-experiments, assuming $L$ = 500 or 1000 fb$^{-1}$ with $\mathcal{P}_{-80,+20}$. The signal and background was fitted to each of the pseudo-experiments, and the distribution of the fit-results was used to evaluate the statistical accuracy of $\sigma$·BR.

The statistical uncertainties for all studied decay-modes are summarised in Table III-6.5 separately for the $\mathcal{P}_{-80,+20}$ and $\mathcal{P}_{+80,-20}$ DBD samples. In addition, the obtainable precisions assuming the full 1 ab$^{-1}$ sample was collected with $\mathcal{P}_{-80,+20}$ are given.

**Table III-6.5**
Summary of the accuracies of ($\sigma \cdot Br$) at $\sqrt{s} = 1$ TeV. The shown values correspond to statistical errors only.

| $L$ | 500 fb$^{-1}$ | | 1 ab$^{-1}$ |
|---|---|---|---|
| Beam polarisation | $\mathcal{P}_{-80,+20}$ | $\mathcal{P}_{+80,-20}$ | $\mathcal{P}_{-80,+20}$ |
| $\Delta\sigma\mathrm{BR}/\sigma\mathrm{BR}(\mathrm{h} \to \mathrm{b\bar{b}})$ | 0.54% | 2.1% | 0.39% |
| $\Delta\sigma\mathrm{BR}/\sigma\mathrm{BR}(\mathrm{h} \to \mathrm{c\bar{c}})$ | 5.7% | 36.8% | 3.9% |
| $\Delta\sigma\mathrm{BR}/\sigma\mathrm{BR}(\mathrm{h} \to \mathrm{gg})$ | 3.9% | 25.7% | 2.8% |
| $\Delta\sigma\mathrm{BR}/\sigma\mathrm{BR}(\mathrm{h} \to \mathrm{WW^*} \to 4\mathrm{j})$ | 3.6% | 23.7% | 2.5% |
| $\Delta\sigma\mathrm{BR}/\sigma\mathrm{BR}(\mathrm{h} \to \mu^+\mu^-)$ | 41% | - | 31% |

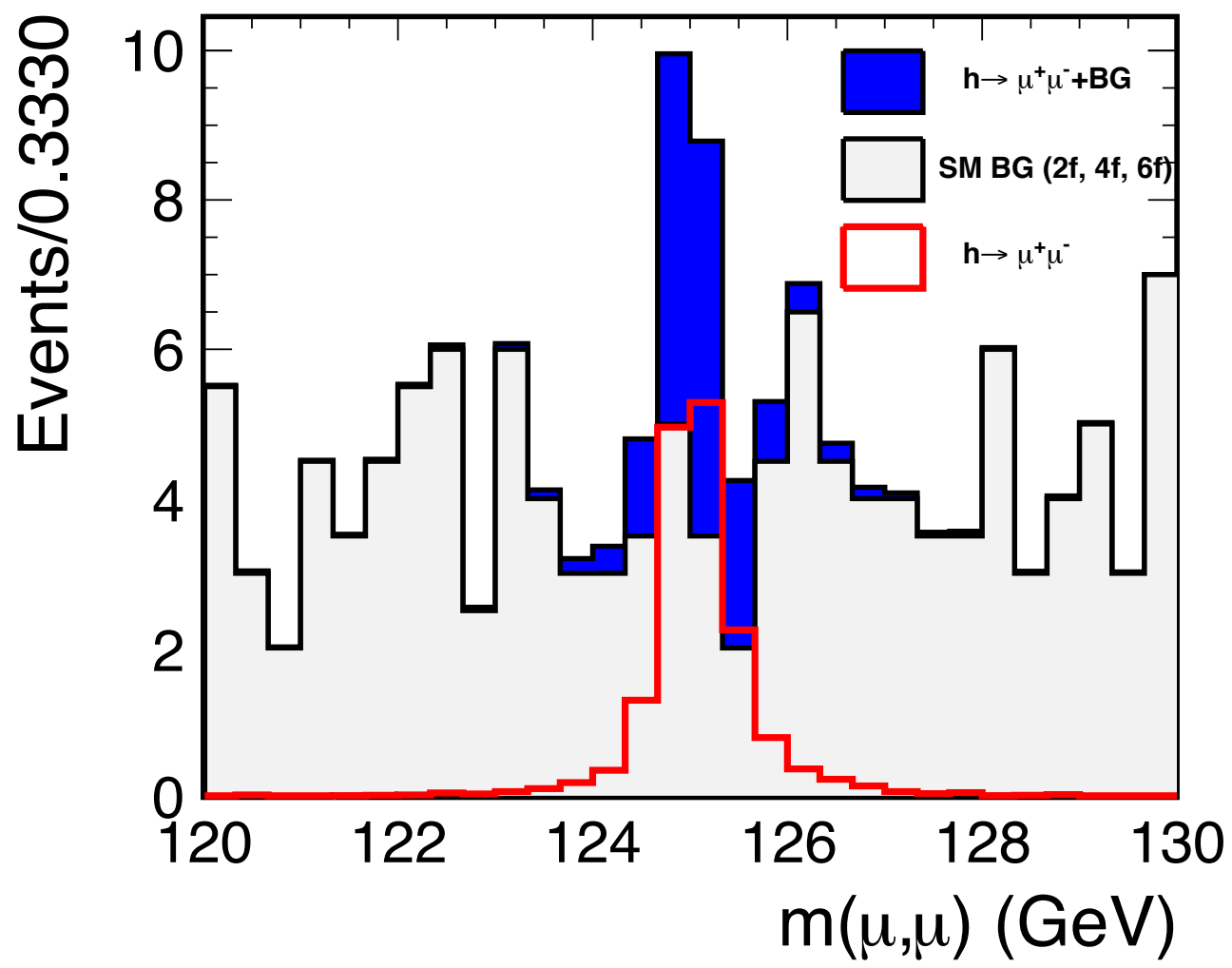

**Figure III-6.12**
Reconstructed di-muon mass distribution of $h \to \mu^+\mu^-$ in the DBD $\mathcal{P}_{-80,+20}$ sample.

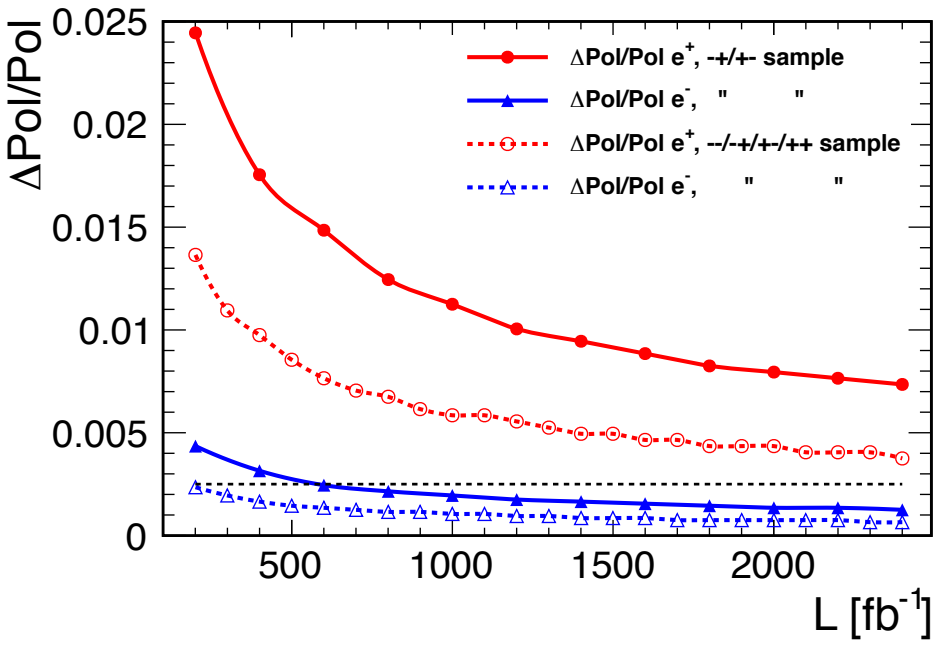

**Figure III-6.13** Relative error of the polarisation measured from the study of the WW scattering cross section. The different curves correspond to different beam polarisations (see legend). The black dashed line indicates the error from the direct measurement of the polarisation using the polarimeters.

### 6.3.3 $e^+e^- \to W^+W^-$

The use of beam polarisation is very beneficial for the physics programme at the ILC. Many examples from the Standard Model as well as beyond the Standard Model [393] demonstrate that having simultaneously polarised electron and positron beams will be very useful for the discovery of new particles, analysing signals in a model independent manner or resolving precisely the underlying model.

At the ILC operated at $\sqrt{s}$=1 TeV, the expected nominal beam-polarisations are $\mathcal{P}_{\pm 80,\mp 20}$. For both beams, the polarisation measurement will be done with Compton polarimetry (see 2.4), with an expected precision of 0.25%. This can limit the usefulness of the beam polarisation: For instance, in the measurement of the Triple Gauge Couplings (TGCs) of the $\mathrm{W}$, the uncertainty due to a 0.25% error on the beam polarisation is of a similar size as the expected statistical error.

To get a higher precision in the polarisation measurement it is necessary to use annihilation data. The benchmark $\mathrm{WW}$ process is ideally suited for this purpose due to its high cross section and to the pure left(right) handedness of the $\mathrm{W}^{-(+)}$ couplings to fermions. In the method used here, information on the angular distribution of the production angle $\cos\theta_\mathrm{W}$ of the $\mathrm{W}^-$ with respect to the direction of the incoming electron beam is exploited. Even though this method would profit from having data at the $++$ and $--$ helicity combinations, it does not depend crucially on the existence of such samples, in contrast to the well-known Blondel scheme [394, 395].

To estimate the possible precision of a polarisation measurement using $\mathrm{W}$-pair production a detailed study has been done and is described in [396, 397]. Only semi-leptonic $\mathrm{WW}$ events were used in this analysis, since they allow to unambiguously determine the charge of the $\mathrm{W}$ bosons. Semileptonic decays of the $\mathrm{W}$ bosons were selected using the isolated lepton-finder. The $k_T$ algorithm with R=1.3 and $N_{jet} = 2$, applied to all particles except the isolated lepton, was used to reject particles from pile-up events. Cuts on number of reconstructed particles, missing mass, missing energy and missing transverse momentum were used to further reduce background. Finally, a 2C kinematic fit was performed, with - apart from energy and momentum conservation - the constraint that the two reconstructed bosons should have equal mass. In order to perform the fit, it was necessary to remove $\mathrm{W}^+\mathrm{W}^- \to \tau\nu qq$ events, which was done by cutting on the discriminant $\tau_{disc}$, described in [398]. Events where the fitted mass was between 40 and 120 GeV were accepted for further analysis. The efficiency to select $\mathrm{e}^+\mathrm{e}^- \to \mathrm{W}^+\mathrm{W}^- \to \mu(e)\nu qq$ was 36%, yielding a sample with 82% purity.

Templates of the $\cos\theta_\mathrm{W}$ distributions were created for different polarisations and the data was fitted to the templates in order to determine the polarisation. It was found that the uncertainty on the electron (positron) beam polarisation would be 0.0016 (0.0023) using the DBD sample. The correspondig fractional uncertainties are 0.0019 and 0.0113, respectively. How the relative precission depends on integrated luminosity is shown by the solid curves in Figure III-6.13.

As mentioned above, the method profits from having samples with the beams having the same-sign polarisation. If the same $L$ of 1 ab$^{-1}$ is divided between $\mathcal{P}_{-80,+20}$, $\mathcal{P}_{-80,-20}$, $\mathcal{P}_{+80,+20}$ and $\mathcal{P}_{+80,-20}$ in the proportions 4:1:1:4, the uncertainty of the electron (positron) beam polarisation was found to be further reduced to 0.00084 (0.0012), yielding a relative uncertaitny of 0.0011 (0.0060). The integrated luminosity dependence is shown by the dashed curves in Figure III-6.13.

It should be pointed out that the precision of the angular fit method does not depend on assuming that the TGCs have their SM values. Indeed, in [399], it was shown that simultaneously fitting the polarisation and the TGCs is possible, and that changes greater than 0.02% to the polarisation stemming from TGC contributions could be disentangled from the beam-polarisation in the angular fit method.

### 6.3.4 $e^+e^- \to t\bar{t}h$

The precision measurement of the top–higgs Yuwaka coupling ($g_{t\bar{t}h}$) is an important benchmark for the ILC, in particular to assess the capabilities of the detectors to analyse complicated event topologies.

This study [400] investigates the semi-leptonic and fully hadronic decay modes of the $\mathrm{t\bar{t}}$ system with the higgs boson decaying via the $\mathrm{b\bar{b}}$ mode. The semi-leptonic decay mode leads to a signal of six jets, an isolated lepton and missing energy. The fully hadronic decay mode results in a signal of eight jets. Both decay modes include four $\mathrm{b}$ jets. The signal was reconstructed by locating isolated leptons in the event. The number of isolated leptons was used to divide the analysis samples into the semi-leptonic and hadronic decay modes to ensure no overlap. In the semi-leptonic sample, the number of isolated leptons is required to be exactly one; this isolated lepton candidate was then set aside whilst forcing the rest of the event into six jets. In the hadronic sample, the number of isolated leptons was required to be zero; these events were then forced into eight jets. Particles from the $\gamma\gamma \to \mathrm{hadrons}$ events were discarded by using the $k_t$ algorithm with R=1.2. The subsequent jet–finding and flavour–tagging procedures were performed using LCFIPlus. The top quarks were reconstructed using a $\mathrm{b}$ jet plus a $\mathrm{W}$ boson, where each $\mathrm{W}$ boson was formed either from two jets not tagged as $\mathrm{b}$ jets or, in the case of the semi-leptonic sample, from the isolated lepton and missing momentum. The Higgs was reconstructed using the two remaining $\mathrm{b}$ jets. The optimal combination of jets in the event was chosen so that the top and Higgs candidates have the most consistent mass.

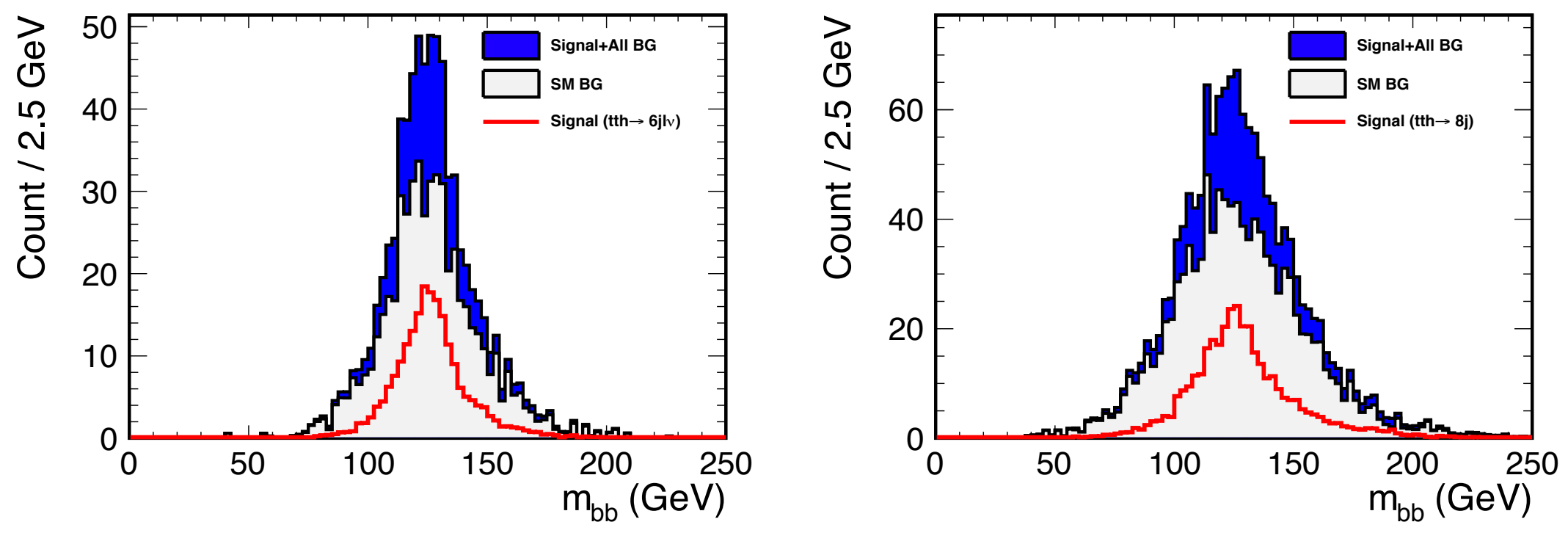

**Figure III-6.14.** Reconstructed $h$ mass for the optimal jet combination of events passed the isolated lepton and Boosted Decision Tree cuts. Left: Semi-leptonic channel. Right: Fully hadronic channel. Both figures show the full DBD sample.

The main backgrounds to this process are $\mathrm{t\bar{t}b\bar{b}}$ and $\mathrm{t\bar{t}Z}$ as these can easily mimic the signal, and $\mathrm{t\bar{t}}$ due to the huge relative cross section compared to the signal. The backgrounds were reduced by a multivariate analysis technique employing boosted decision trees [401] and the results were cross checked by a cut-based analysis. The input variables include the total visible energy and number of particles in the event, the $\mathrm{b}$-likeness of the jets, event shape variables such as the thrust and those from the jet-clustering algorithm, the reconstructed masses of the top, $\mathrm{W}$ and higgs candidates and their consistency, and the helicity angle of the Higgs decay. The final discriminant was optimised to yield the maximum statistical significance. At this point, the efficiency of selecting $\mathrm{t\bar{t}}h$ events was 50.0% with a purity of the selected sample of 29.1%. The reconstructed di-jet masses for the pair most likely to be the $h$ are shown in Figure III-6.14 for the two channels separately.

Using the full DBD sample, the statistical precision of $g_{\mathrm{t\bar{t}}h}$ was found to be 6.9% for the semi-leptonic mode and 5.4% for the hadronic mode. The combined statistical precision was found to be

$$\left(\frac{\Delta g_{\mathrm{t\bar{t}}h}}{g_{\mathrm{t\bar{t}}h}}\right) = 4.3\%$$

.

The quoted error is statistical only. It is expected that the theoretical uncertainties on background cross sections and higgs branching ratios will be at the sub-percent level at the time the 1 TeV phase of the ILC starts. Also the relevant machine parameters – luminosity, energy and polarisation – are expected to be controlled to a similar precision. Therefore, the main systematic uncertainty is expected to be the estimate of signal and background efficiencies. These uncertainties could, for instance, be evaluated from the data itself: the theoretically well-understood $\mathrm{t\bar{t}Z}$ channel is quite similar to the $\mathrm{t\bar{t}}h$ channel, and can serve as a proxy to determine the signal efficiency; similarly, the $\mathrm{t\bar{t}}$ background could be estimated by selecting events with similar topology as the signal, but with very low probabilities that there are more then two $b$ jets. The best procedure to follow is currently under study.

### 6.3.5 $e^+e^- \to t\bar{t}$ at $E_{CMS}$ = 500 GeV

The ILC provides an ideal environment to measure the couplings at the $\mathrm{t\bar{t}Z}$ and $\mathrm{t\bar{t}}\gamma$ vertex. The produced $\mathrm{t(\bar{t})}$ quark decays almost exclusively into a $\mathrm{bW}$ pair. The $\mathrm{b}$ quark hadronises giving rise to a jet, while the $\mathrm{W}$ can either decay hadronically into light quarks, which turn into jets, or leptonically into a charged lepton and a neutrino. The semi-leptonic process is defined to be case where one $\mathrm{W}$ decays hadronically while the other one decays leptonically.

Analyses of both the semi-leptonic and the fully hadronic mode have been done[402]. The latter is the benchmark reaction from the LOI, and is presented in this section, while the former is discussed in Section 6.3.6.2.

#### 6.3.5.1 Analysis of the fully hadronic decay

The top quark forward-backward asymmetry, $A^{\mathrm{t}}_{\mathrm{FB}}$, provides an important test of the SM. In the fully-hadronic channel the $\mathrm{t}$ and $\mathrm{\bar{t}}$ can be identified by tagging the $\mathrm{b/\bar{b}}$ from the charge of the secondary vertex from charged $\mathrm{b}$-hadron decays. This measurement provides a test of the vertex reconstruction capability of ILD.

The six-jet final state is reconstructed using the Durham jet finder, and the jets subsequently are analysed with LCFIPlus to assign b-tag values. The two jets with the highest b-tag are considered to be the jets from the $\mathrm{b}$ quarks, while events for which one of the $\mathrm{b}$-tag values is smaller than 0.3 were rejected. The two $\mathrm{W}$ bosons were reconstructed from the remaining four jets. For each possible combination of $\mathrm{W}$ bosons and $\mathrm{b}$-quarks, a $\chi^2$ was formed comparing the reconstructed mass, energy and $\mathrm{b}$-quark momentum with the expected values. The combination yielding the best $\chi^2$ was selected. Events where the best $\chi^2$ was too large, the di-jet mass of either $\mathrm{W}$-candidate was far from $m_{\mathrm{W}}$, or tri-jet mass of either $\mathrm{t}$-candidate was far from $m_{\mathrm{t}}$ were rejected. For each of the two identified b-jets, the charge of the secondary vertices were reconstructed. Events with like-sign combinations were rejected as were events with two neutral secondary vertices. The efficiency to select fully hadronic $\mathrm{t\bar{t}}$ events was 13 %. Of the selected events, 60 % had the correctly identified top quark charge. Figure III-6.15 shows the distribution of the cosine of the reconstructed polar angle of the tagged top-quark, showing a clear forward-backward asymmetry. The relative numbers of events in the forward and backward hemispheres, accounting for the charge identification/mis-identification probabilities, were used to determine $A^{\rm t}_{\rm FB} = 0.34$ for a left handed polarised electron beam and $A^{\rm t}_{\rm FB} = 0.44$ for a right handed polarised one. The errors of these quantities, corrected for the statistics expected at $\mathcal{P}_{-80,+30}$ and $\mathcal{P}_{+80,-30}$ beam polarisation, are 3.0% and 3.2%, respectively. The statistical and systematic errors have been added in quadrature; however the statistical ones largely dominates.

The measured asymmetry for left-handed electron beam polarisation, $A^{\rm t}_{\rm FB} = 0.344 \pm 0.011$, agrees well with the result presented in the LOI: $A^{\rm t}_{\rm FB} = 0.334 \pm 0.008$, taking into account that the LOI analysis was assuming twice the integrated luminosity. Further improvement of the result can be expected in the future since the charge determination is not yet optimised in the new LCFIPlus package.

### 6.3.6 Other physics processes

In this section, we present studies of physics channels that are not benchmarks, but nevertheless depend on the details of detector hardware, software and analysis and potentially might have changed substantially with respect to the LOI.

Many new analyses have indeed been performed since the LOI, and are presented in [403]. What is presented here are only those done with the updated event generator, with all backgrounds taken into account and with the full detector simulation model.

A new analysis of Higgs self-coupling was done and is presented in this section. It contains both a new analysis at $\sqrt{s} = 500$ GeV, and also an extended study what an ILC operating at 1 TeV would bring to our knowledge of the properties of the Higgs.

The $\rm t\bar{t}$ channel has been studied beyond the benchmark-measurement of $A^{\rm t}_{FB}$ from fully hadronic decays. The more precise measurement that can be done in the semi-leptonic channel has been carried out.

#### 6.3.6.1 Higgs self-coupling.

The ILC running at 500 GeV and 1 TeV offers the opportunity to measure the Higgs trilinear self-coupling, which is very difficult to do at LHC if the Higgs mass is around 125 GeV [404, 405, 406, 407, 408]. It would be the first non-trivial test of the Higgs potential, crucial to understand the nature of Higgs' mechanism and the spontaneous breaking of electroweak symmetry. Many physics models beyond the Standard Model have been studied that show significant deviations of the Higgs self-coupling [409, 410, 411, 412, 413, 414, 415, 416, 417, 418, 419, 420, 421, 422]. Depending on

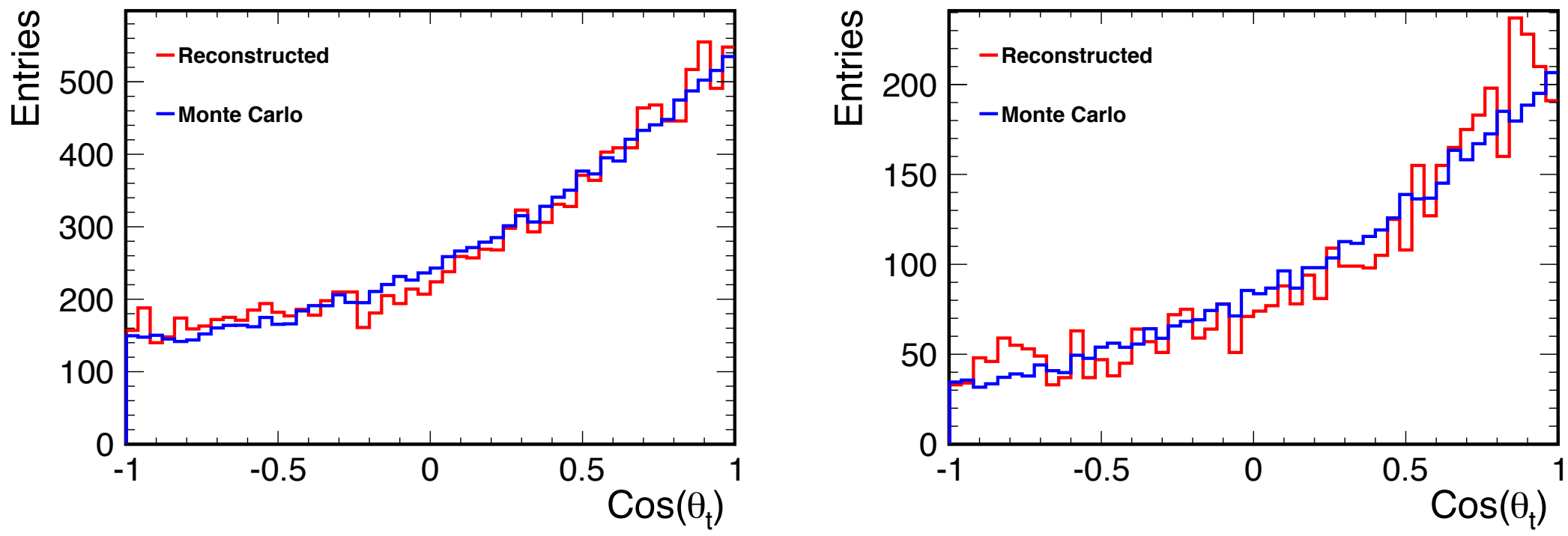

**Figure III-6.15.** Distribution of the true (blue curve) and reconstructed (red curve) polar angles of the identified top quark in fully-hadronic $\rm t\bar{t}$ events. The expected contributions from events with the wrong charge have been subtracted from the observed distribution. Left plot: $\mathcal{P}_{-100,+100}$; right plot: $\mathcal{P}_{+100,-100}$. Both plots assume $L =$ 250 fb$^{-1}$.

the model, the deviations could be as large as 100%, but also as low as 10%. It is a great challenge to measure the Higgs self-coupling at the ILC, and it has been investigated by many groups over the years [423, 424, 425, 426, 427, 428, 429].

At ILC, the measurement of the trilinear Higgs self-coupling can be carried out through two leading processes shown in the Figure III-6.16: the Higgs-strahlung process $\mathrm{e^+e^-} \to Zhh$ and the WW fusion process $\mathrm{e^+e^-} \to \nu\bar{\nu}hh$ [430, 431, 432, 433, 434]. cross sections of these two processes are also shown in Figure III-6.16. The $\mathrm{e^+e^-} \to Zhh$ process has its maximum cross section at around $\sqrt{s} = 500$ GeV and the WW fusion process becomes important at around $\sqrt{s} = 1$ TeV.

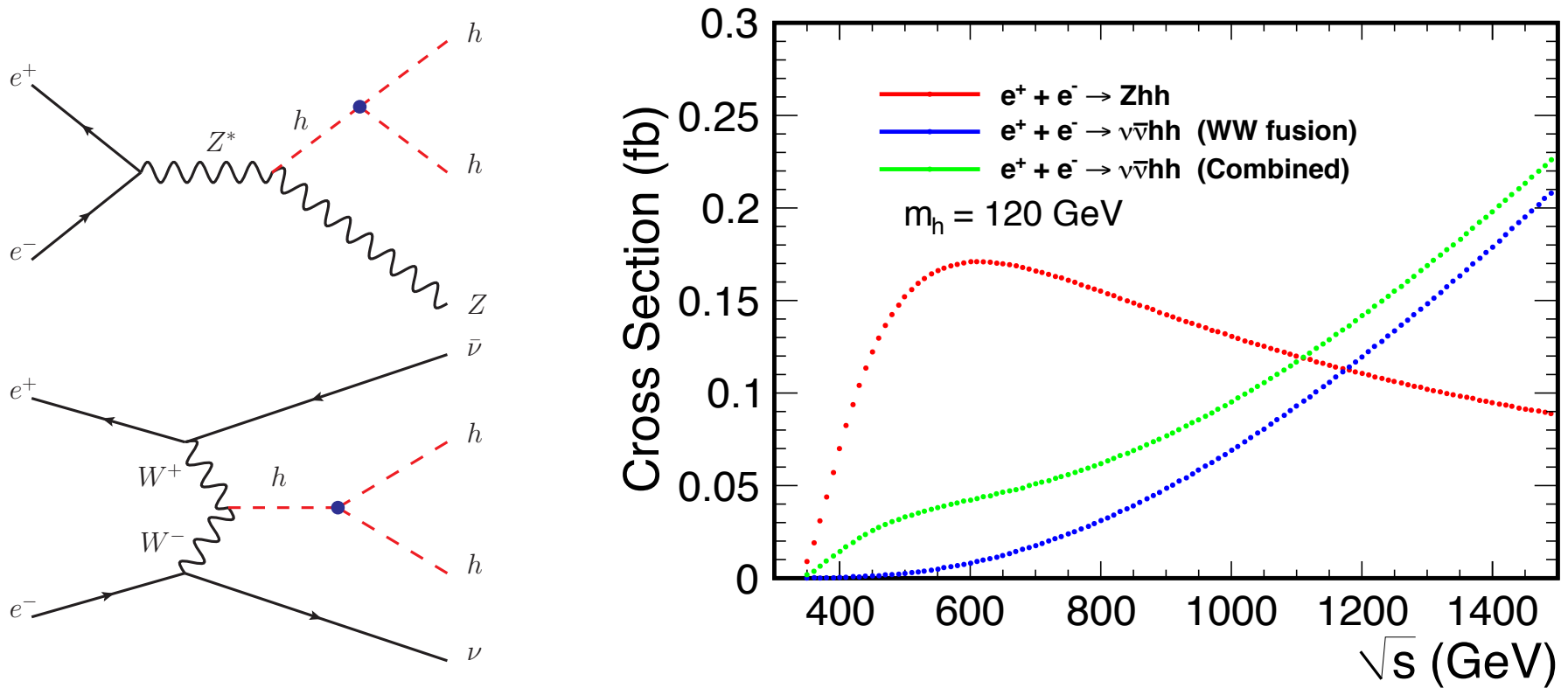

**Figure III-6.16.** Left: The Feynman diagrams involving the trilinear Higgs self-coupling for the two processes: $\mathrm{e^+e^-} \to Zhh$ (top) and $\mathrm{e^+e^-} \to \nu\bar{\nu}hh$ (bottom); Right: Cross section for these two processes as a function of $\sqrt{s}$ for $m_h$=120 GeV. The blue dotted line shows the cross section for $\mathrm{e^+e^-} \to \nu\bar{\nu}hh$ from the WW fusion process alone, while the green dotted line shows the sum of the WW fusion contribution and the contribution from $\mathrm{e^+e^-} \to Zhh \to \nu\bar{\nu}hh$.

In the absence of interfering diagrams, the relative uncertainty of the coupling of a given diagram is half the relative uncertainty of the measured cross section. However, in both the Higgs-strahlung and the WW fusion processes, there exist Feynman diagrams which have the same final state but that are not related to the Higgs self-coupling. These diagrams largely degrade the sensitivity of Higgs self-coupling to the cross section: For $\mathrm{e^+e^-} \to Zhh$ at 500 GeV, the relation becomes $\delta\lambda/\lambda = 1.8\,\delta\sigma/\sigma$, while for $\mathrm{e^+e^-} \to \nu\bar{\nu}hh$ at 1 TeV, it becomes $\delta\lambda/\lambda = 0.85\,\delta\sigma/\sigma$. This is illustrated in the Figure III-6.17 were the relation between $\sigma$ and and $\lambda$ is shown for the two cases. Recently, a weighting method has been developed [435]. It gives events where the observed invariant mass of the two Higgses is in the region where the self-coupling process is more important a higher weight and events in a region depleted of the self-coupling process a lower one. As can be seen comparing the slopes of the red and blue curves in Figure III-6.17, this method enhances the sensitivity of Higgs self-coupling, so that the factors become 1.66 and 0.76, respectively.

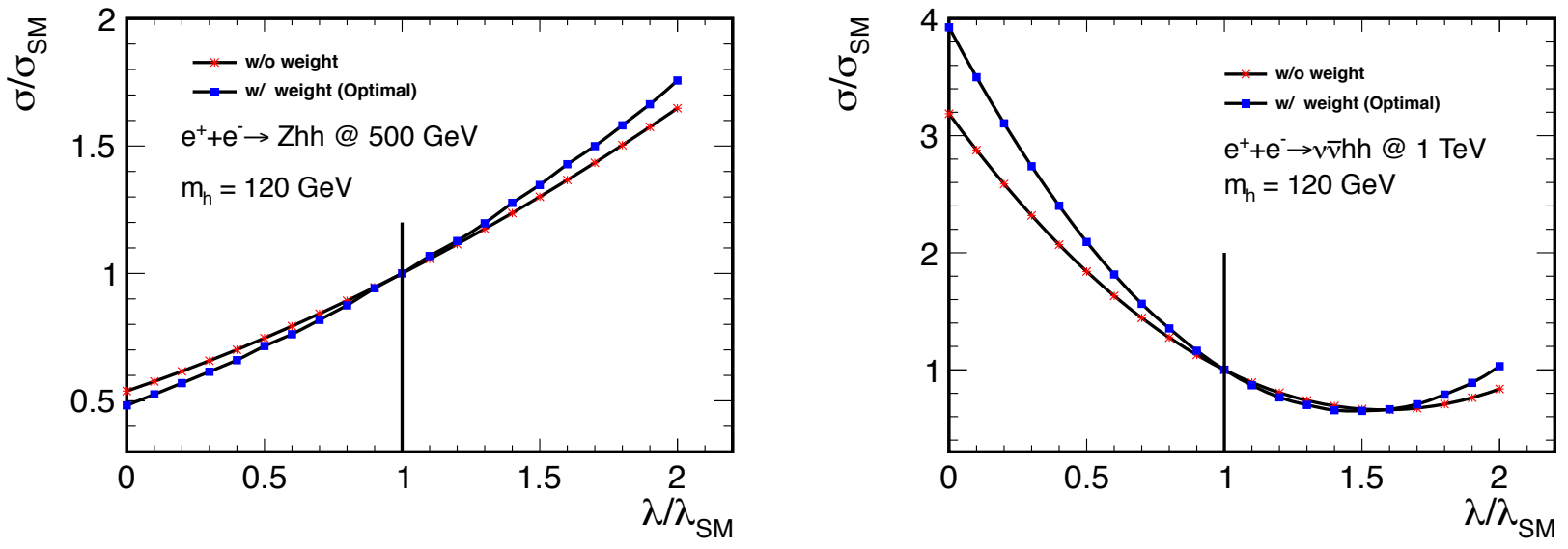

**Figure III-6.17** The sensitivity of the Higgs self-coupling for the two processes: $\mathrm{e^+e^-} \to Zhh$ (left) and $\mathrm{e^+e^-} \to \nu\bar{\nu}hh$ (right). The red ones are without weighting and the blue ones are with the optimal weighting described in the reference [435].

Based on the full detector simulation of ILD (see sect. 5.4), a new analysis of $\mathrm{e^+e^-} \to Zhh$ at 500 GeV was performed considering all the decay modes of Z ($\ell^+\ell^-$, $\nu\bar{\nu}$, and $\mathrm{q\bar{q}}$) and with both Higgses decaying to $\mathrm{b\bar{b}}$. The analysis strategy is fully described in [436] and in [437]. Even with $\mathcal{P}_{-80,+30}$, the cross section of the signal process is still very small, $\sim$ 0.22 fb, so $L$= 2 $\mathrm{ab}^{-1}$ is assumed. The remaining numbers of signal and background events are summarised in Table III-6.6 for different search modes. The study shows that by combining all the modes with a likelihood ratio test, the expected $\mathrm{Z}hh$ excess significance is 5.0 $\sigma$ and the cross section of $\mathrm{e^+e^-} \to \mathrm{Z}hh$ can be measured to the accuracy of 27%, corresponding to the precision of Higgs self-coupling of 44% [438], applying the weighting method in reference [435].

**Table III-6.6.** The numbers of the remaining signal and background events in each search mode of the $\mathrm{Z}hh$ analysis based on the full detector simulation at 500 GeV, with $\mathcal{P}_{-80,+30}$. The last two columns are $\mathrm{Z}hh$ excess significance (i) and cross section measurement significance (ii). The $\mathrm{q\bar{q}}hh$ mode is separated to two categories: (a) $\mathrm{b\bar{b}}hh$ dominant, (b) light $\mathrm{q\bar{q}}hh$ dominant.

| Search Mode | Signal | Background | Significance (i) | Significance (ii) |
|---|---|---|---|---|
| $\mathrm{q\bar{q}}hh$ $(a)$ | 13.6 | 30.7 | $2.2\sigma$ | $2.0\sigma$ |
| $\mathrm{q\bar{q}}hh$ $(b)$ | 18.8 | 90.6 | $1.9\sigma$ | $1.8\sigma$ |
| $\nu\bar{\nu}hh$ | 8.5 | 7.9 | $2.5\sigma$ | $2.1\sigma$ |
| $\mathrm{e^+e^-}hh$ | 3.7 | 4.3 | $1.5\sigma$ | $1.1\sigma$ |
| $\mu^+\mu^-hh$ | 4.5 | 6.0 | $1.5\sigma$ | $1.2\sigma$ |

At 1 TeV, one expects a clearer signal, due to the larger contribution of $\mathrm{WW}$ fusion process, which has lower background, and has lower amount of interference from other double-Higgs diagrams compared to the Higgs-strahlung process. The process $\mathrm{e^+e^-} \to \nu\bar{\nu}hh$ at 1 TeV was studied with both Higgses decaying to $\mathrm{b\bar{b}}$. An initial study was based on the fast simulation framework SGV [439]. The analysis followed a strategy quite similar to the analysis at 500 GeV, and showed that indeed a precision on the Higgs self-coupling of $\sim 17$ % is achievable with L $= 2$ $\mathrm{ab}^{-1}$ and $\mathcal{P}_{-80,+20}$[435]. Using the same strategy, a preliminary analysis using fully simulated ILD events confirms these results: It was found that 35.7 signal events were expected, with a background of 33.7 events. This yields expected uncertainties $\delta\sigma/\sigma = 23$ % and $\delta\lambda/\lambda = 18$ %, ie. a $5\sigma$ observation of Higgs self-coupling. It is also found that the statistical significance of the double-Higgs production excess is expected to be $7.2\sigma$.

### 6.3.6.2 Further $\mathrm{t\bar{t}}$ studies

In the semi-leptonic mode, $\mathrm{t\bar{t}} \to (\mathrm{bW})(\mathrm{bW}) \to (\mathrm{bqq'})(\mathrm{b}\ell\nu)$, the charged lepton allows the determination of the $\mathrm{W}$ charge, and hence to separate $\mathrm{t}$ and $\mathrm{\bar{t}}$. At the same time, the hadronically decaying $\mathrm{t}$ allows to determine the direction of the $\mathrm{t}$ or $\mathrm{\bar{t}}$.

The isolated decay lepton was identified, and it was found that the decay lepton could be identified with an efficiency of 85%. The $\mathrm{b}$ jets among the remaining four jets were identified as those with the highest value of the $\mathrm{b}$-tag from LCFIPlus, while the two remaining jets were associated with the $\mathrm{W}$. The $\mathrm{b}$ jet to combine with the two jets from the $\mathrm{W}$ to form the $\mathrm{t}$ system was decided by the choice giving the total 3-jet mass closest to $m_{\mathrm{t}}$. It has to be noted however that the final state gives rise to ambiguities in the correct association of the $b$ quarks to the $W$ bosons, see [440] for an explanation. These ambiguities affect mainly the reconstruction in case of a left handed electron beam. The ambiguities can be nearly eliminated by requiring a high quality of the event reconstruction. The control of the ambiguities however requires an excellent detector performance and event reconstruction. The signal selection efficiency was 27.6% in case of a left handed electron beam and 56.5% in case of a right handed electron beam The resulting spectrum of the polar angle of the $\mathrm{t}$ quark is shown in Figure III-6.18 (left). From this spectrum, one could determine forward-backward asymmetry: $A^{\mathrm{t}}_{\mathrm{FB}} = 0.36$ for a left handed polarised electron beam and $A^{\mathrm{t}}_{\mathrm{FB}} = 0.41$ for a right handed polarised electron beam. The statistical precision on these quantities is 1.7% and 1.3%, respectively.

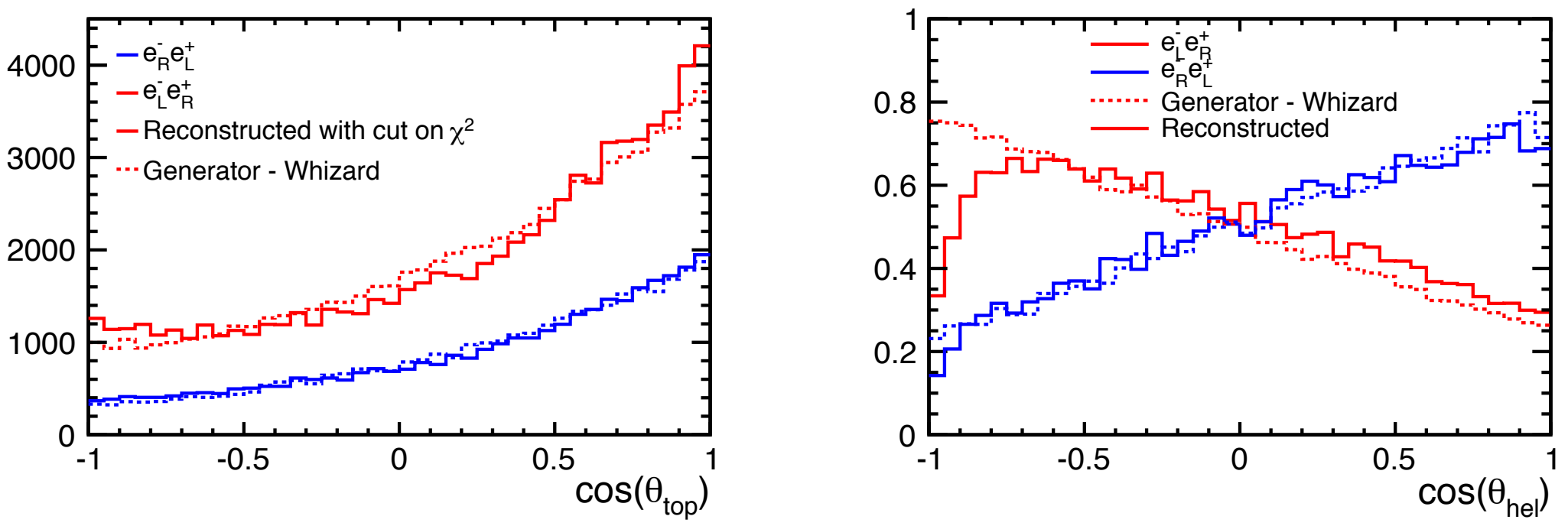

**Figure III-6.18.** Left: Reconstruction of the direction of the $\mathrm{t}$ quark for two different beam polarisations. The plot shown is an update of the one presented in [440]. Note that the figure does not include background, however, it is known from the studies in [387] that the background is negligible. Right: Generated and reconstructed distribution of the helicity angle $\cos\theta_{hel}$.

Measurements using optimised observables are investigated in [441]. These observables are the top pair production cross section for left and right-handed polarised beams and the fraction of right-handed ($\mathrm{t}_R$) and left handed top quarks ($\mathrm{t}_L$). Following [442], the fraction of $\mathrm{t}_L$ and $\mathrm{t}_R$ can be determined with the slope of the helicity asymmetry ($\lambda_\mathrm{t}$). In the top quark rest frame the distribution of the polar angle $\theta_{hel}$ of a decay lepton is

$$\frac{1}{\Gamma}\frac{d\Gamma}{d\cos\theta_{hel}} = \frac{1+\lambda_\mathrm{t}\cos\theta_{hel}}{2}$$

where $\lambda_\mathrm{t}$ varies between $+1$ and $-1$ depending on the fraction of $\mathrm{t}_R$ and $\mathrm{t}_L$. The observable $\cos\theta_{hel}$ can easily be measured at the ILC, and is less sensitive to ambiguities in the event reconstruction than eg. $A^\mathrm{t}_\mathrm{FB}$. The slope of the differential cross section wrt. $\cos\theta_{hel}$ directly measures $\lambda_\mathrm{t}$, and hence the net polarisation of a top quark sample. The result of a full simulation study is shown in Figure III-6.18 (right), where it can be seen that parton-level spectrum is only slightly distorted by hadronisation and detector effects. The remaining discrepancies in case of left handed electron beams can be explained by reconstruction inefficiencies for low energetic final state leptons. By fitting the slopes in the interval $-0.6 < \cos\theta_{hel} < 0,9$, the helicity asymmetry could be determined: $\lambda_\mathrm{t} = -0.48$ (left-hand polarised electron beam) and $\lambda_\mathrm{t} = +0.51$ (right-hand polarised electron beam). The errors of these quantities are 3.3% and 3.7%, respectively. Statistical and systematic contributions have been added in quadrature. Note that for $\lambda_\mathrm{t}$ and $A^\mathrm{t}_\mathrm{FB}$, the dominant systematic error is expected to come from the ambiguities discussed above. The role of theory errors will have to be evaluated in the future.

# Chapter 7
# ILD Costs

In this chapter an estimate of the cost of the ILD detector concept is presented. The costs shown are essentially the construction costs. Person-power needed has not been studied with the same level of detail, and only a very rough estimate is presented in section 7.3.7. The costing is an evolution of the one presented in the letter of intent for ILD [198] but has been significantly further developed and detailed. A major difference is that many of the costs are now based on experience gained with actual prototypes. The basis of the costing form the work breakdown structures for the different sub-detectors (WBS) which have been developed to include the materials, the fabrication process, the assembly and the commissioning. In this chapter abbreviated versions of the WBS are shown, to describe the main cost components.

What has been costed is the baseline ILD detector including the different options. Where costs of the different options are very different a mean price and a range is quoted. In the second part of the chapter the scaling of the main component is discussed, to provide the material for a cost - performance optimization. It can be expected that many of the costs quoted will change significantly once serious industrialisation studies are undertaken.

## 7.1 Methodology of costing

The method used by ILD is based on the methodology developed for the accelerator parts of the TDR, and is very similar to the one which was used for the RDR [443]. An attempt has been made to use for major components unit costs common with SiD and CLIC [444] detectors, as shown in Table III-7.1.

**Table III-7.1**
Unit costs agreed to by SiD, ILD, and CLIC [444].

| | agreed unit cost [ILCU] | agreed error margin [ILCU] |
|---|---|---|
| Tungsten for HCAL | 105/kg | 45/kg |
| Tungsten for ECAL | 180/kg | 75/ kg |
| Steel for Yoke (raw material) | 1000/t | 300/t |
| Stainless Steel for HCAL | 4500/t | 1000/t |
| Silicon Detector | 6/cm$^2$ | 3/cm$^2$ |

It should be noted that these common costs can only serve as a guideline, as the detailed costs depend on many factors. The ILD estimate has started from the costs in the table, and has adjusted them if needed to take into account specific ILD circumstances. A particularly important example is the cost for the silicon sensor. For the ECAL a careful study with industry has revealed scope for significant savings, as will be discussed below, considering the very special application in the ECAL. This has been taken into account.

It is obvious that the costs quoted have a large uncertainty. In particular raw material costs - which for some part like e.g. the yoke are quite important - vary widely and might change significantly with time.

A major difference compared to the previous cost estimate is that many numbers are based on actual prototyping work. From this work a detailed description of the fabrication and assembly sequences including tests and tooling is known, which can be extrapolated to the full detector, and a first approximation of the item prices are available, in some instances with quantities still far from the final numbers needed but nevertheless significant (for example 400k channels for the sDHCAL). Where possible, estimates of the testing and commissioning costs are included.

An engineering study of the integration of the ILD detector has been done, as it is documented in chapter 5.1. Proper integration will also require tools and special setups which have been wherever possible included in the estimate. It should be noted that these costs might be very site-dependent.

Costs for the ILD detector estimated in different currencies are converted into ILCU using the Purchase Power Parity (PPP) system. The conversion factor is based on the actual costs of a defined set of items, rather than the currency conversion rate at some point in time. For the purpose of this document the PPP rates shown in Table III-7.2 have been used.

**Table III-7.2** Conversion rate based on purchase power parity used in the cost estimate.

| currency | Dollar | Euro | Yen |
|---|---|---|---|
| ILCU | 1 | 0.9732 | 127.3 |

The results should nevertheless be treated with care. In some cases where the cost estimates are based on concrete offers, or have been obtained based on previous experiments, the input numbers might already include implicit currency conversions. Wherever possible or known, these effects have been unfolded, to obtain consistent results.

No attempt was made to guess the impact of future escalation. Contingencies are currently not taken explicitly into account, but some estimates based on real fabrications include them implicitly, for example for the coil. Spares were accounted for only for construction, not for maintenance. No R&D costs are included, except in some cases costs for industrialisation. No maintenance and operations costs has been estimated.

## 7.2 ILD work breakdown structure

A condensed WBS for the different sub-systems can be found in the following section together with comments and remarks on the way costs were determined. The following items were estimated for the sub-systems:

- the procurements of materials including costs for testing,
- the procurements of the sensors including costs for testing,
- the procurements of the front-end electronics,
- the cost of needed assembly and the needed tooling,
- the cost of the local data-acquisition,
- some estimate of the transportation costs, though this is very imprecise given that the location of the experiment is not yet known,
- the costs for assembly on site, including costs for tooling,
- the costs for services.

## 7.3 ILD cost evaluation

The different sub-detectors for ILD have reached different levels of maturity, which is clearly reflected in the cost estimate. Not in all cases extensive production and industrialisation studies have been done, nor are for all system tooling costs etc well known. However for the most expensive parts of ILD, in particular the calorimeter and the yoke, such studies are available at least at a preliminary level.

The building of prototypes has often provided a starting value for the procurements, though of course for smaller numbers than what will eventually be needed. Thus scaling factors have been applied to reduce the actually quoted prices, after discussions with suppliers, which reflect the current best knowledge about costs at the time of ordering large quantities.

To provide the cost of operations and related tooling, an estimate of the fabrication is needed. Again the prototype construction provides valuable input to this.

The descriptions accompanying each sub-detector should provide enough detail to the reader to understand the limits of the relevant cost estimate. A summary table at the end will put all this into a global perspective.

### 7.3.1 Vertex detector

The vertex detector exists in 3 versions. They differ essentially by the sensors and the read-out electronics. The cost of the mechanical installation and services has been taken to be identical. It is based on the cost of the STAR vertex detector, which is constructed using the same technology as proposed for the ILD CMOS option.

For the CMOS version the sensor price comes from the STAR experiment and the electronics from the Mimosa prototypes, for the FPCCD the information comes from prototypes, and for the DEPFET version it comes from the Belle II experiment. For the different options the cost vary between 3.2 MILCU and 4.2 MILCU. The value used for the ILD cost estimate is 3.4 MILCU.

### 7.3.2 Silicon tracking

The silicon tracking contains four disks with pixels, close to the vertex detector, 12 forward disks with strips, two cylinders of strip detectors of the SIT and the outer tracking, all made with the same strip technology. Experience from the LHC experiments has played an important role in the cost estimate. The cost of the readout ASICS is based on current 130 nm technology; after the conversion to the new 65 nm technology these costs might change. The cost for the inner Silicon system (SIT and FTD) together is estimates to be 2.3 MILCU, for the outer Silicon susyem (SET and ETD) 21 MILCU.

### 7.3.3 Time Projection Chamber

The estimate of the TPC price comes largely from the prices found in the construction of the STAR and ALICE TPCs. It has been updated for inflation but does not contain any added contingency. The cost of the field cage includes the experience from the construction of the large prototype, which was built in industry, using technology similar to the one to be used for a full scale field cage. A significant part of the TPC cost will be in the readout electronics, estimated to be around 30% of the total cost. The field cage - the iner cylinder, the outer cylinder, and the endplates, will account for around 20% of the cost, the rest being in tooling, anciliary systems and control systems. The total cost of the detector is estimated to be 35.9 MILCU.

### 7.3.4 Calorimeter System

The calorimeter system is a major part of the ILD detector, and one of the largest single cost items. In Table III-7.3 costs for major components are shown, with their current (prototype) costs, and the anticipated costs for the full detector construction. The cost of the tungsten used is based on the agreed value but translated back into the original currency (EUR) and the re-converted into ILCU using the PPP scheme.

**Table III-7.3**
Expected prices for major components in the calorimeters, see text for further comments. Except where explicitly stated all current costs are based on actual costs of components procured for the prototypes.

| Material | Cost [ILCU] | System | Comment |
|---|---|---|---|
| Tungsten | 123/kg | SiECAL, ScECAL, AHCAL, FCAL | quote from manufacturer (130 EUR/kg) |
| Stainless Steel | 5/kg | AHCAL, SDHCAL | processing costs to be added (1-4 EUR/ kg) |
| Si sensors | 3/cm$^2$ | SiECAL | based on extrapolation of current quotation of 5 EUR/ cm$^2$ |
| SiPM | 1/pc | ScECAL, AHCAL, muon | based on manufacturer extrapolation, current price 7-10 EUR/piece |
| ASIC | 0.22-0.25/ch | SiECAL, ScECAL, AHCAL | current price 0.5 EUR/ch |
| ASIC | 0.1/ch | SDHCAL | current price of 0.18 EUR/ch |
| PCB | 7900/m$^2$ | SiECAL | prototype |
| PCB | 2600/m$^2$ | ScECAL | extrapolated from prototype price of 10800/m$^2$ |
| PCB | 1800/m$^2$ | SDHCAL, AHCAL | for AHCAL extrapolated from prototype price of 10800/m$^2$ |

For the Silicon Tungsten ECAL a very complete and careful study has been performed to build and understand the WBS. Studies with industry have been undertaken to understand the costs of the major components: Tungsten plates, silicon sensors, and readout boards. The costing table for the electromagnetic calorimeter is shown in Table III-7.4. It should be noted however that this represents a snapshot, and that significant room for further optimization of these costs exist.

**Table III-7.4**
Cost table of the electromagnetic calorimeter.

| SiECAL | | ScECAL | |
|---|---|---|---|
| Item | Cost [kILCU] | Item | Cost [kILCU] |
| Tungsten | 16310 | Tungsten + carbon parts | 18500 |
| Carbon fiber structure | 2130 | Module realisation | 1700 |
| Silicon sensors | 75000 | Scintillators | 1030 |
| Readout ASIC | 16500 | Photo Detectors | 10200 |
| Readout Board | 21000 | Readout ASIC | 2500 |
| Materials | 1300 | Readout Board | 25000 |
| Cables, connectors | 2220 | Readout System | 6200 |
| Tooling | 9300 | Cables, connectors | 1000 |
| Assembly | 13500 | Power supplies | 4100 |
| Integration | 500 | Tooling | 3800 |
| Sum SiECAL | 157760 | Sum ScECAL | 74000 |

A major cost item for the SiECAL is the cost of the silicon wafers. The quoted number corresponds to a cost of $3$ ILCU/cm$^2$. This estimate is based on a current cost of $5-6$ ILCU /cm$^2$ in 2011. Since then careful studies have resulted in a much increased production efficiency, which led to a reduction of this price by about a factor of 2, to $3$ ILCU/cm$^2$. This price can be so low because the structures for the SiECAl are comparativly simple, and the requirements on the rate of dead pixels and the acceptable leakage current can be relaxed compared to other silicon based detectors. A further reduction of the price is not excluded.

For the Scintillator based option of the electromagnetic calorimeter the silicon based photon detectors are a major expense. Quotes have been obtained from industrial suppliers for the large number of detectors needed for the complete system. Current small scale production runs result in prices per detector of around 10 EUR, but it seems realistic to expect that a reduction to a level of 1 EUR/ channel can be realised. The assembly procedures for the scintillator ECAL are not yet as well understood as for the Si based ECAL. At the moment no estimate of the assembly cost for the scintillator planes is included in the cost estimate.

**Figure III-7.1** Beginning of the full silicon tungsten electromagnetic calorimeter WBS.

| | | | | | | | | | | | | |
|---|---|---|---|---|---|---|---|---|---|---|---|---|
| 3 | | 2.1 | Electromagnetic Calorimeter | | | | | | | | | |
| 4 | | 2.1.1 | | General procurement (Barrel + endcaps ?) | | | | | | | | |
| 5 | 1 | | 2.1.1.1 | Mechanical structures | | | | | | | | |
| 6 | 2 | | | 2.1.1.1.1 | W | | | | | | | |
| 7 | 3 | | | | | barrel | | | | | | |
| 8 | 4 | | | | | | Material | | 90.3 | ton | | Industry Several supplier |
| 9 | 5 | | | | | | Dimensional inspection | | 24000 | plates | 3D measurement system | HOME/Industry |
| 10 | 6 | | | | | endcaps | | | | | | |
| 11 | 7 | | | | | | Material | | 42 | ton | | IndustrySeveral suppliers |
| 12 | 8 | | | | | | Dimensional inspection | | 12000 | plates | 3D measurement system | HOME/Industry |
| 13 | 9 | | | 2.1.1.1.2 | Carbon fibers prepeg | | | | | | | |
| 14 | 10 | | | | | K ( H structure) | | | 8600 | m$^2$ | | Industry |
| 15 | 11 | | | | | 3K ( Modules) | | | 19000 | m$^2$ | | Industry |
| 16 | 12 | | | | | Carbon plates( modules) 12 K | | | | | | |
| 17 | 13 | | | | | | Th. :2mm | BARREL | 40 | plates | | Industry |
| 18 | 14 | | | | | | Th. :15 mm | BARREL | 40 | plates | | Industry |
| 19 | 15 | | | | | | Th. :2mm | ENDCAPS | 24 | m2 | | Industry |
| 20 | 16 | | | | | | Th. :15 mm | Endcaps | 24 | m2 | | Industry |
| 21 | 17 | | | 2.1.1.1.3 | Metal inserts | | | | 1728 | inserts | | Industry |
| 22 | 18 | | | 2.1.1.1.4 | Rails | | | | 80 | rails | | Industry |
| 23 | 19 | | 2.1.1.2 | Detector | | | | | | | | |
| 24 | 20 | | | 2.1.1.2.1 | Wafers | | processed | | 308800 | | | Industry several suppliers |
| 25 | 21 | | | 2.1.1.2.2 | ASIC | | | | 1235200 | | | Industry |
| 26 | 22 | | | 2.1.1.2.3 | PCB | | | | 75500 | | | Industry |
| 27 | 23 | | | 2.1.1.2.4 | Boards | | | | | | | |
| 28 | 24 | | | | | DC/DC converters | | | 400 | | | |
| 29 | 25 | | | | | Data Concentrator | | | 440 | | | |
| 30 | 26 | | | | | GDCC | | | 140 | | | |
| 31 | 27 | | | 2.1.1.2.5 | other elect. Components | | | | | | | |
| 32 | 28 | | | | | Active | | | | | | |
| 33 | 29 | | | | | passive | | | | | | |

The hadronic calorimeter has been costed in both options, the analogue(AHCAL) and the semi-digital option (SDHCAL). The main cost items for both versions are shown in Table III-7.5. For both AHCAL and SDHCAL significant prototypes have been built, which provide important information for the cost estimate. For the AHCAL the same cost of 1 ILCU/ piece is used for the SiPM as for the ScECAL version discussed above. More detailed work has been done for both options for the barrel part of the calorimeter. The cost of the end-caps has been estimated based on the sensitive area and the total system weight. A significant part of the cost is the readout boards, which are for both options complex large multi layer printed circuit boards. The quoted prices are based on several independent quotes and on actual experience with the prototypes.

In the very forward region two small calorimeter systems close the coverage, LumiCal, BeamCal. LumiCal and BeamCal have been carefully studied and costed. A major cost item are the sensors, which are based on silicon and diamond technology. In total a cost of 8.05 MILCU is estimated. Note that ILD discusses the possibilty to add a third system in the forward direction, LHCAL, for which however no detailed design and thus no cost estimate exists at the moment.

### 7.3.5 Magnet

The magnet system has three major components, the coil, the iron return yoke, and the ancillaries. For the coil, CMS has been used as a "prototype", the complete actual CMS fabrication chart has been revisited for ILD, taking into account the change in dimensions, the variations in technology and assembly, and the cost escalation since the building of CMS. The cost of the ancillaries is also derived directly from CMS.

For the yoke, the weight has been estimated and the agreed upon price for machined and assembled iron (see Table III-7.1) has been used. The estimate is based on a fairly detailed engineering model, including assembly procedures. The yoke iron is a large item of the total cost. This is driven not by requirements from physics but results from the request to control the stray field outside of the ILD detector to a level of 50 G at 15 m distance from the detector [354].

**Table III-7.5** Cost breakdown for the two HCAL options AHCAL and SDHCAL.

| AHCAL | | SDHCAL | |
|---|---|---|---|
| Item | Cost [kILCU] | Item | Cost [kILCU] |
| Absorber | 5200 | Absorber | 6500 |
| Module production | 3400 | Module mechanics | 2300 |
| | | | |
| Cassettes | 2100 | | |
| Scintillators | 1500 | RPC incl cassettes | 6800 |
| Reflective Foil | 1200 | | |
| Photo sensors | 7700 | | |
| | | | |
| ASIC | 1800 | ASIC | 6600 |
| Readout Board | 13200 | Readout Board | 13000 |
| Readout | 2300 | Readout | 2000 |
| Cabling, connections | 1000 | Elec Integration | 1600 |
| HV/ LV supplies | 1000 | Services incl. HV/ LV | 200 |
| | | | |
| Cooling system | 1000 | Cooling System | 1000 |
| | | Gas System | 900 |
| Tooling, testing | 500 | Testing | 200 |
| Assembly, installation | 2800 | Assembly, tooling | 3900 |
| DAQ | 200 | | incl. |
| Sum AHCAL | 44900 | Sum SDHCAL | 44800 |

The cost of the coil is based on the information from CMS obtained from CERN. Since this information is given in CHF, most of the components however are sourced in the EUR area, the cost estimate has been converted into EUR based on a sensible currency exchange rate of 1.5, before translated into ILCU using the PPP scheme.

**Table III-7.6** Cost table of the coil and the iron yoke.

| Item | Cost [kILCU] | Item | Cost [kILCU] |
|---|---|---|---|
| **Coil** | | **Yoke** | |
| Conductor and winding | 12900 | Steel, including machining | 80400 |
| Internal Cryogenics | 1000 | Support | 1700 |
| Suspension system | 560 | Moving System | 3500 |
| tooling, assembly | 10000 | Assembly | 6700 |
| Qualification, testing | 1100 | Survey | 500 |
| | | | |
| **Ancillaries** | | | |
| Cryogenics, vacuum | 6800 | Integration | 933 |
| Electrical installation | 1700 | Field Mapping | 560 |
| Control and Safety system | 350 | Engineering | 2200 |
| Sum | 131000 | | |

**Figure III-7.2**
Summary plot of the relative contribution by the different sub-components to the total cost of the ILD detector.

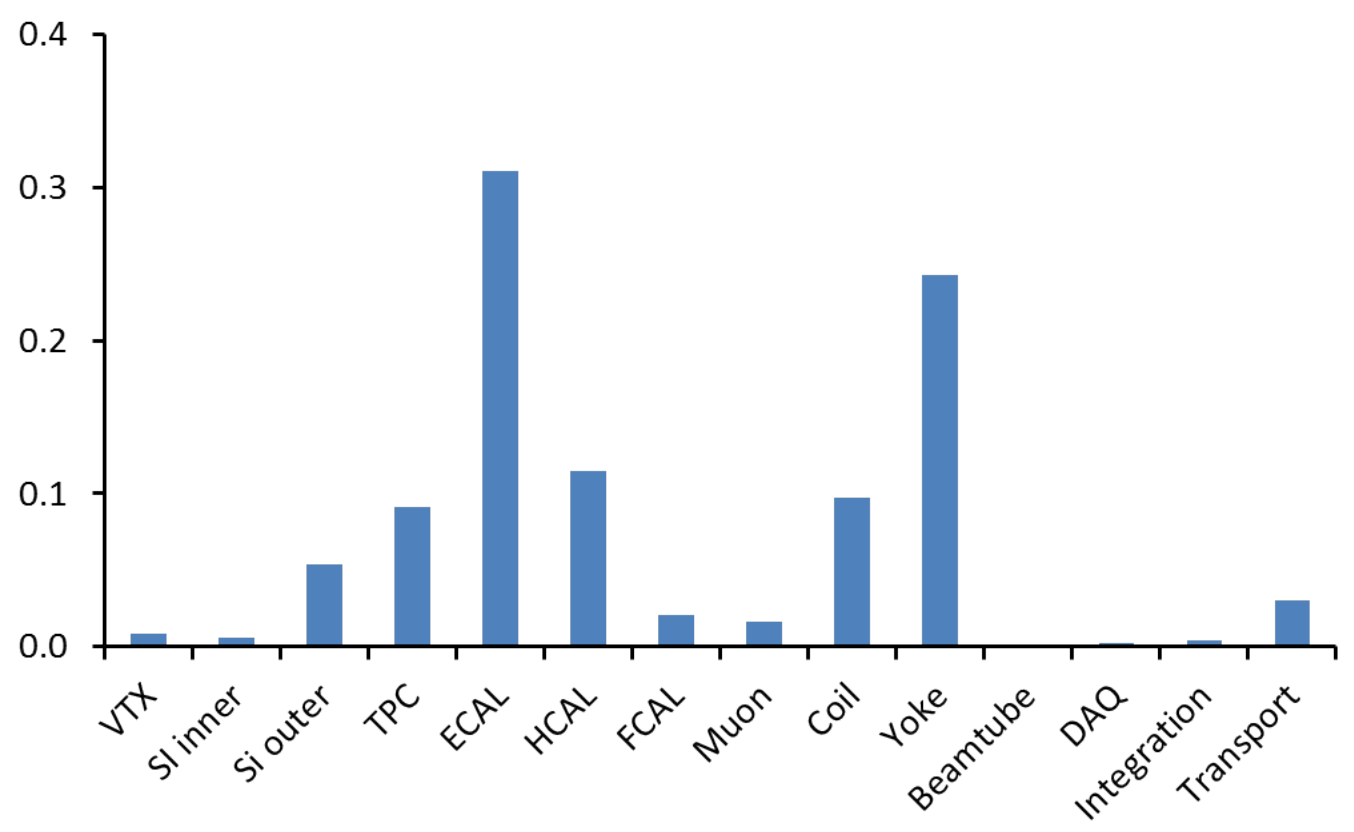

### 7.3.6 Muon system

The muon system being made of scintillator read out with SiPM like the AHCAL, the costs have been derived from there. It corresponds mostly to the procurements of materials without assembly and tooling. The cost is dominated by the costs if the sensor system. In total 6.5 MILCU is estimated.

### 7.3.7 Cost summary

The total cost of the ILD detector is summarised in Table III-7.7. The distribution of the costs

**Table III-7.7**
Summary table of the cost estimate of the ILD detector. Depending on the options used the cost range is between 336 Mio ILCU and 421 Mio ILCU.

| System | Option | Cost [MILCU] | Mean Cost [MILCU] |
|---|---|---|---|
| Vertex | | | 3.4 |
| Silicon tracking | inner | 2.3 | 2.3 |
| Silicon tracking | outer | 21.0 | 21.0 |
| TPC | | 35.9 | 35.9 |
| ECAL | | | 116.9 |
| | SiECAL | 157.7 | |
| | ScECAL | 74.0 | |
| HCAL | | | 44.9 |
| | AHCAL | 44.9 | |
| | SDHCAL | 44.8 | |
| FCAL | | 8.1 | 8.1 |
| Muon | | 6.5 | 6.5 |
| Coil, incl anciliaries | | 38.0 | 38.0 |
| Yoke | | 95.0 | 95.0 |
| Beamtube | | 0.5 | 0.5 |
| Global DAQ | | 1.1 | 1.1 |
| Integration | | 1.5 | 1.5 |
| Global Transportation | | 12.0 | 12.0 |
| Sum ILD | | | 391.8 |

among the different systems is shown in Figure III-7.2.

The cost driving items are the yoke, and the calorimeter system. The cost for the integration is an estimate of the scenario described in section 5.1, and might vary significantly with different scenarios. It includes the extra cost for the large platform (see chapter 5.5.1) on which the detectors moves, as well as the extra costs of the cryogenics needed to allow a cold move of the detector. The offline computing represents a significant cost. Owing to the continued large advances in computing technology, we have estimated this at $20\%$ of the equivalent cost for a LHC detector.

A first estimate of the person-power needed has been done. For each calorimeter it is estimate to be around 200 MY, for the coil, 500 MY. From this the total person-power needed is extrapolated to be around 1400 MY. The average cost per MY has been taken to be 93 kILCU including overheads. This value is typical for the mix of qualifications needed for a sophisticated project like the ILD. The estimate only includes the person-power needed to build the detector, and does not include needs to finish the R&D or work out a detailed design of the detector. The person-power is then estimated 130 MILCU.

The study has been carried out assuming that the detector is in a push-pull configuration. Most of the sub-system costs are only marginally affected by this assumption, with the exception of the yoke cost and the integration costs, as discussed above. It has been estimated that without these requirements the total cost of the detector might be reduced by some $10\%$.

## 7.4 Detector cost dependencies

The ILD detector as presented in this document has been strongly driven by the physics requirements. At this moment no complete cost - performance optimisation has taken place. With the information known now and available based on real prototyping experience such an overall optimisation can be performed. In this section the dependence of the main cost items on input assumptions are discussed.

The parameters which have been considered for possible scalings are the following:

- a characteristic transverse size of the detector chosen as the inner radius of the ECAL barrel;
- a characteristic longitudinal size of the detector chosen as the length of the ECAL barrel or TPC;
- the number of samples for the ECAL (for a given number of radiation lengths);
- the calorimeter cell sizes.

The study was done under the assumption that the technologies remain the same. This then results in typical cost changes below 25% of the system cost. More significant cost changes imply changes in the technologies.

### 7.4.1 Scaling with the field

The nominal magnetic field is 3.5 T, but the magnet is designed to withstand 4 T. Reducing the field below 3.5 T might offer some cost savings, but also results in a loss of the physics potential of the detector, inparticular its upgradability to higher energies. ILD therefore does not consider this option of de-scoping.

### 7.4.2 Scaling the detector size

The dimensions of the detector parts inside the TPC are dictated by considerations of background and assembly procedures. They are not very relevant for costing. Most relevant for the costing is the inner radius of the coil, and the inner radius of the calorimeter system. Another important consideration is the length of the TPC, as this drives the length of the calorimeter system and the coil and the Yoke.

The cost scaling has been studied under two scenarios: the aspect ratio of the detector remains constant, with corresponding correlated changes of radius and length, or the radius alone is changed. In Figure III-7.3 the cost impact when only changing the outer radius of the TPC is shown, on the left are the effects on sub-components, on the right the global effect is shown.

### 7.4.3 Changing the ECAL

The scaling of the number of ECAL readout layers has been done under the assumption that the total number of radiation lengths in the ECAL is kept constant. The area of sensitive medium and the number of readout channels then scale proportional to the number of samplings. On the other hand as the total amount of radiator does not change, the thickness of the absorber plates changes and the cost for manufacturing the plates varies. Reducing the number of samples will reduce the overall thickness of the ECAL even when the total amount of absorber material stays constant. For example, going from 30 to 20 samples will reduce the radial thickness by 20 mm. This space in turn can be used to reduce the constraints on the sensitive part in particular on the PCB. All this together makes the scaling essentially proportional to the area and then to the sampling.

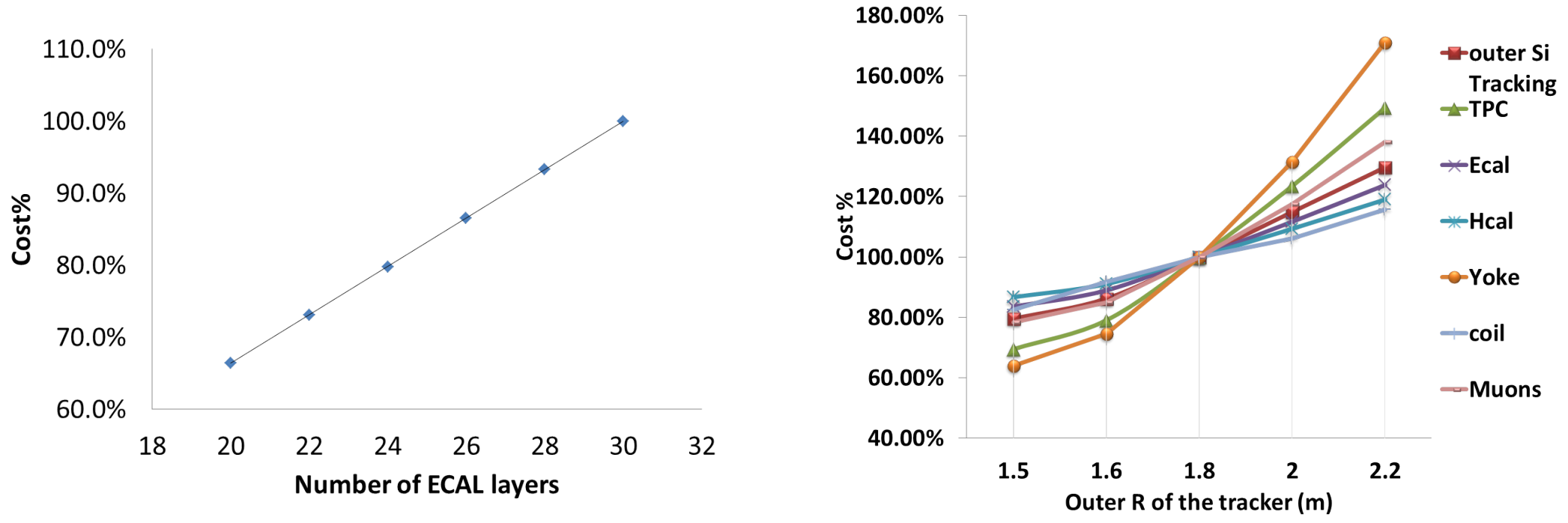

**Figure III-7.3.** Left: Dependence of the SiECAL cost on the number of sensitive layers. (right) Scaling of the cost of the ILD dector if the transverse size is changed, split into relative cost changes for the different sub-detectors.

The cell sizes of the electromagnetic calorimeter cannot be easily reduced any further with the current technological solution. To go below, a new design, may be a totally different approach will be needed. Increasing the cell sizes within the same technology will have only a minor impact on the cost, as the cost roughly scales with the area of silicon, not so much the number of readout channels. There is some effect due to a different cost of the printed circuit boards and other ancillary equipment. We estimate that reducing the number of cells by an order of magnitude reduces the cost of the ECAL by less than 10%, or 3% of the total detector cost. The impact on the cost for the scintillator version may be larger but it is unlikely that scaling up the size in this version would be considered.

### 7.4.4 Scaling the hadronic cell size

For the hadronic calorimeter changing the cell sizes will result in a changed number of FE chips, calibration devices etc. We estimate that a reduction of the number of readout channels by an order of magnitude reduces the cost of the digital HCAL by about 20%, of the analogue HCAL by about 10%. This has to be balanced with a significant performance loss.

## 7.5 Conclusion

The cost of the ILD detector has been estimated to be about 400 MILCU. It includes the material to build the detector, but does not include cost escalation and contingencies. Person-power is with few exceptions not included. The dependence of the cost on the main detector parameters has been studied, and effects of order $10\%$ or less per item on the total detector cost have been found. To illustrate the possibilities, a cost reduction of 20% can be reached by reducing the inner radius of the ECAL to 150 cm, without changing the length. The quoted cost of the ILD detector is comparable to the total cost of the large LHC detectors.

# Chapter 8
# ILD Summary

The ILD detector concept as presented in this document has been developed over a number of years. It is the result of the work of a large group of people from around the world. The design has matured over the last few years to a point where a realistic proposal for a detector has been made. The concept has undergone a careful optimization and validation using detailed simulation studies.

A key difference between this document and the Letter of Intent, published in 2009, is that essentially all technologies proposed as part of the ILD concept have been experimentally validated. ILD has worked very closely with R&D groups on different topics to proposed, develop and validate technologies. As elaborated in this document, significant progress has been achieved in many areas, and the ILD concept is technologically now on a very sound basis.

The evaluation of the anticipated performance of the ILD concept has been done using full and detailed simulation. Great care has been taken to include to the best of the current knowledge imperfections in the detectors, dead zones, dead material and the like. With this the level of realism in the simulation was significantly increased. In key areas - for example the particle flow performance - some key experiments have been done which demonstrate that the simulation indeed correctly describes the data. Thus not only the realism but also the reliability of the ILD simulation could be improved and demonstrated.

In the progress of the experimental validation of the different technologies significant prototypes of the main detector components have been designed and built. This is in particular the case for the calorimeters and key parts of the tracker. In doing so not only the performance of these components was studied but also significant experience was gained on the cost of these modules. Some of these prototypes included many 10s of thousands of channels thus approaching a level where true mass production of components is needed. The experience gained in this process entered into the way a cost estimate of the final detector has been developed. Nevertheless caution is advised when interpreting these numbers, as many of the cost factors are difficult to extrapolate many years into the future, and thus the final price is subject to significant uncertainties.

It has been the policy of the ILD concept group to propose a detector which is technically feasible, and which includes only minimal extrapolations from current levels of technology. In many cases more than one technology meets these requirements and are proposed as options for different subsystems. The ILD concept group intentionally does not want to take technological decisions at this stage, at a time where the ILC project is still not approved, to not block the way for more modern technologies yet to come, or to support future improvements of current technologies. While this makes the definition of a baseline ILD detector at this moment difficult, it keeps the options for ILD open to either improve performance even further, or to reduce cost.

Once the ILC project becomes reality ILD is prepared to rather quickly initiate a process to finalise a technology selection. Based on the large body of experience and a well functioning cooperation with the major R&D groups this selection should be driven by the scientific needs of the project, and the realities of funding as they exist at the time of the decision taking.

Until then the ILD concept group intends to continue to improve the detector design, push further on the development of technologies, and study the integration of the different subdetectors into one coherent detector.

# Volume 4

## Detectors

# Part IV

## Summary and Future Plans

On July 4, 2012, the ATLAS and CMS Collaborations announced the discovery of a particle resulting from the search at the LHC for the Standard Model Higgs Boson. This newly discovered boson has a mass of 125-126 GeV and the early measurements, albeit of limited precision, are consistent with the Standard Model Higgs Boson. The clearest evidence comes from the gamma-gamma and Z-Z channels; after the 2012 data is fully analysed the nature will be better known. The discovery immediately increases the relevance and interest in the ILC, since the ILC offers precision measurements of the new boson's properties when operating near the Higgs-strahlung threshold, between 215 and 250 GeV.

In addition to the important studies of this new Higgs Boson candidate an ILC capable of also operating at the top threshold at 350 GeV provides crucial precision in the study of top quark properties. ILC top quark measurements would far surpass the precision of those of the LHC. Among the top measurements the ILC would make is the precise top mass at threshold, of relevance to the interest in the question of vacuum stability. The ILC and the ILC detectors have been designed specifically to exploit this opportunity and they are now ready to do so.

Over the past few years many studies have been conducted toward the realisation of an experimental program at the International Linear Collider (ILC). These have ranged from phenomenological studies on the physics potential of the ILC to the pre-engineering designs of detectors. Beginning with a call for Letters of Intent (LOIs) in 2007, a large global community has marshalled resources and effort to achieve significant progress, which is described in the preceding parts.

The physics requirements for the ILC detectors are demanding. Two detector design groups, ILD and SiD, have independently developed concepts based on complementary strategies and technologies. They have demonstrated the capability to realise the unprecedented resolutions in impact parameter, charged particle momentum, and jet energies demanded by the ILC physics program. Advances in the state of the art of detector technology operating in the clean environment offered by the electron-positron collisions make these capabilities possible. Both detector designs employ calorimetry based on the particle flow algorithm (PFA), with full detector information utilised to reconstruct quark jets. The two detector designs adopt different approaches to achieve excellent PFA performance: SiD chooses a compact design enabling finer granularity of calorimeters in a cost-constrained environment, while ILD chooses to enable larger particle separation with a larger detector.

The two detector concepts have been studied through detailed and realistic simulation codes. While collaborating closely with theorists, the capabilities of the ILC detectors to study many physics topics have been evaluated using these simulation tools. The results of these studies are summarised in the physics and detector sections, clearly demonstrating that the detector designs can realise the physics potential of the ILC.

The detector R&Ds have been conducted by international collaborative groups, working closely with the detector design teams to formulate the two detector designs. In some cases, these detector R&D groups extend their applications beyond the ILC, e.g. to CLIC and the B-factory upgrade. In particular, close collaborations exist between the ILC and CLIC detector efforts both in hardware and software. These collaborations allowed effective use of limited resources.

The detector groups also have worked in close collaboration with the GDE accelerator team, especially on MDI issues and in guiding the choice of machine parameters to ensure good physics performance for the ILC. The ILC Research Directorate has fostered and coordinated these collaborations, such as with the design of the push-pull system and the SB2009 working group. The design of the push-pull system required closely cooperation of the relevant detector and accelerator experts. The SB2009 common task group worked closely with the accelerator team to optimise ILC machine parameters for lowered cost and power consumption while protecting the physics performance.

The details contained here are meant to provide a reference on the progress that has been achieved, working in parallel with the GDE's effort toward a TDR on the collider. It should be a valuable resource for future project planning, demonstrating that efforts can move forward with confidence that the designs can successfully pursue the important physics goals of the ILC.

When the ILC project is realised, these detectors can be built with the technologies that have been developed and described in this document. Their performances have been verified with realistic and detailed detector configurations. However, engineering design for construction remains to be done. The call for LOIs did not require commitment of detector concept groups to actually build the detectors. Even though many advances have been achieved, the detector designs are still mostly conceptual. As the project approaches reality, detailed engineering designs will be needed and, once it is possible, these must be adapted to the selected site.

While optimisation possibilities remain, calling for future R&D efforts, the maturity of the detector R&D justifies increased engineering studies at this point. Each subsystem of each detector design can be further optimised for performance and cost through such engineering studies. In any case, significant work remains to bring the technical designs to the level of construction readiness of the collider.

An era in the preparation for the ILC has passed. The 2012 discovery of a Higgs Boson candidate at the LHC makes the project even more compelling. The detector R&D and detector design efforts have reached a significant level of maturity. Everyone involved wants the ILC to be realised in the near term so its scientific program can commence with studies of the 125-126 GeV Higgs-like boson.

# Volume 4

# Detectors

## Part V

## Appendices

# SiD Bibliography

# ILD Bibliography

# List of Signatories

The following list of signatories represents a comprehensive list of those people who have contributed to the R&D and design work, for both the accelerator and the detectors, which is summarised in this report. The list also includes those people who wish to indicate their support for the next phases of the worldwide ILC effort.

It should be noted that inclusion in this list does not indicate any formal commitment by the signatories. It does not indicate commitment to the specific detector designs presented, nor exclusive support for ILC over other collider programs.

A. Abada[171], T. Abe[24], T. Abe[236], J. M. Abernathy[379], H. Abramowicz[265], A. Abusleme[231], S. Aderhold[47], O. Adeyemi[333], E. Adli[357,251], C. Adloff[164], C. Adolphsen[251], K. Afanaciev[209], M. Aguilar[31], S. Ahmad[93], A. Ahmed[382], H. Aihara[375], R. Ainsworth[237,139], S. Airi[154], M. Aizatskyi[208], T. Akagi[73], M. Akemoto[71], A. Akeroyd[367], J. Alabau-Gonzalvo[108], C. Albajar[46], J. E. Albert[379], C. Albertus[281], J. Alcaraz Maestre[31], D. Alesini[174], B. Alessandro[128], G. Alexander[265], J. P. Alexander[43], A. Alhaidari[243], N. Alipour Tehrani[33], B. C. Allanach[323], O. Alonso[311], J. Alwall[210], J. W. Amann[251], Y. Amhis[167], M. S. Amjad[167], B. Ananthanarayan[83], A. Andreazza[386,122], N. Andreev[58], L. Andricek[186], M. Anduze[172], D. Angal-Kalinin[258], N. Anh Ky[106,394], K. A. Aniol[18], K. I. Aoki[148], M. Aoki[148], H. Aoyagi[137], S. Aoyama[250], S. J. Aplin[47], R. B. Appleby[343,40], J. Arafune[96], Y. Arai[71], S. Araki[71], L. Arazi[404], A. Arbey[50], D. Ariza[47], T. Arkan[58], N. D. Arnold[7], D. Arogancia[194], F. Arteche[113], A. Aryshev[71], S. Asai[375], T. Asaka[137], T. Asaka[212], E. Asakawa[220], M. Asano[333], F. B. Asiri[58], D. Asner[225], M. Asorey[286], D. Attié[21], J. E. Augustin[169], D. B. Augustine[58], C. S. Aulakh[226], E. Avetisyan[47], V. Ayvazyan[47], N. Azaryan[142], F. Azfar[358], T. Azuma[246], O. Bachynska[47], H. Baer[355], J. Bagger[141], A. Baghdasaryan[407], S. Bai[102], Y. Bai[384], I. Bailey[40,176], V. Balagura[172,107], R. D. Ball[329], C. Baltay[405], K. Bamba[198], P. S. Bambade[167], Y. Ban[9], E. Banas[268], H. Band[384], K. Bane[251], M. Barbi[362], V. Barger[384], B. Barish[17,65], T. Barklow[251], R. J. Barlow[392], M. Barone[58,65], I. Bars[368], S. Barsuk[167], P. Bartalini[210], R. Bartoldus[251], R. Bates[332], M. Battaglia[322,33], J. Baudot[94], M. Baylac[170], P. Bechtle[295], U. Becker[185,33], M. Beckmann[47], F. Bedeschi[126], C. F. Bedoya[31], S. Behari[58], O. Behnke[47], T. Behnke[47], G. Belanger[165], S. Belforte[129], I. Belikov[94], K. Belkadhi[172], A. Bellerive[19], C. Belver Aguilar[108], A. Belyaev[367,259], D. Benchekroun[184], M. Beneke[57], M. Benoit[33], A. Benot-Morell[33,108], S. Bentvelsen[213], L. Benucci[331], J. Berenguer[31], T. Bergauer[224], S. Berge[138], E. Berger[7], J. Berger[251], C. M. U. Berggren[47], Z. Bern[318], J. Bernabeu[108], N. Bernal[295], G. Bernardi[169], W. Bernreuther[235], M. Bertucci[119], M. Besancon[21], M. Bessner[47], A. Besson[94,306], D. R. Bett[358,140], A. J. Bevan[234], A. Bhardwaj[326], A. Bharucha[333], G. Bhattacharyya[241], B. Bhattacherjee[150], B. Bhuyan[86], M. E. Biagini[174], L. Bian[102], F. Bianchi[128], O. Biebel[182], T. R. Bieler[190], C. Biino[128], B. Bilki[7,337], S. S. Biswal[221], V. Blackmore[358,140], J. J. Blaising[164], N. Blaskovic Kraljevic[358,140], G. Blazey[217], I. Bloch[48], J. Bluemlein[48], B. Bobchenko[107], T. Boccali[126], J. R. Bogart[251], V. Boisvert[237], M. Bonesini[121], R. Boni[174], J. Bonnard[168], G. Bonneaud[169], S. T. Boogert[237,139], L. Boon[233,7],

G. Boorman[237,139], E. Boos[180], M. Boronat[108], K. Borras[47], L. Bortko[48], F. Borzumati[272],
M. Bosman[294], A. Bosotti[119], F. J. Botella[108], S. Bou Habib[401], P. Boucaud[171], J. Boudagov[142],
G. Boudoul[89], V. Boudry[172], D. Boumediene[168], C. Bourgeois[167], A. Boveia[52], A. Brachmann[251],
J. Bracinik[315], J. Branlard[47,58], B. Brau[346], J. E. Brau[356], R. Breedon[320], M. Breidenbach[251],
A. Breskin[404], S. Bressler[404], V. Breton[168], H. Breuker[33], C. Brezina[295], C. Briegel[58],
J. C. Brient[172], T. M. Bristow[329], D. Britton[332], I. C. Brock[295], S. J. Brodsky[251], F. Broggi[119],
G. Brooijmans[42], J. Brooke[316], E. Brost[356], T. E. Browder[334], E. Brücken[69], G. Buchalla[182],
P. Buchholz[301], W. Buchmuller[47], P. Bueno[111], V. Buescher[138], K. Buesser[47], E. Bulyak[208],
D. L. Burke[251], C. Burkhart[251], P. N. Burrows[358,140], G. Burt[40], E. Busato[168], L. Butkowski[47],
S. Cabrera[108], E. Cabruja[32], M. Caccia[389,122], Y. Cai[251], S. S. Caiazza[47,333], O. Cakir[6],
P. Calabria[89], C. Calancha[71], G. Calderini[169], A. Calderon Tazon[110], S. Callier[93], L. Calligaris[47],
D. Calvet[168], E. Calvo Alamillo[31], A. Campbell[47], G. I. E. Cancelo[58], J. Cao[102], L. Caponetto[89],
R. Carcagno[58], M. Cardaci[201], C. Carloganu[168], S. Caron[97,213], C. A. Carrillo Montoya[123],
K. Carvalho Akiba[291], J. Carwardine[7], R. Casanova Mohr[311], M. V. Castillo Gimenez[108],
N. Castro[175], A. Cattai[33], M. Cavalli-Sforza[294], D. G. Cerdeno[111], L. Cerrito[234], G. Chachamis[108],
M. Chadeeva[107], J. S. Chai[260], D. Chakraborty[217], M. Champion[254], C. P. Chang[201], A. Chao[251],
Y. Chao[210], J. Charles[28], M. Charles[358], B. E. Chase[58], U. Chattopadhyay[81], J. Chauveau[169],
M. Chefdeville[164], R. Chehab[89], A. Chen[201], C. H. Chen[203], J. Chen[102], J. W. Chen[210],
K. F. Chen[210], M. Chen[330,102], S. Chen[199], Y. Chen[1], Y. Chen[102], J. Cheng[102], T. P. Cheng[351],
B. Cheon[66], M. Chera[47], Y. Chi[102], P. Chiappetta[28], M. Chiba[276], T. Chikamatsu[193],
I. H. Chiu[210,210], G. C. Cho[220], V. Chobanova[186], J. B. Choi[36,36], K. Choi[157], S. Y. Choi[37],
W. Choi[375,248], Y. I. Choi[260], S. Choroba[47], D. Choudhury[326], D. Chowdhury[83],
G. Christian[358,140], M. Church[58], J. Chyla[105], W. Cichalewski[263,47], R. Cimino[174], D. Cinca[337],
J. Clark[58], J. Clarke[258,40], G. Claus[94], E. Clement[316,259], C. Clerc[172], J. Cline[187], C. Coca[206],
T. Cohen[251], P. Colas[21], A. Colijn[213], N. Colino[31], C. Collard[94], C. Colledani[94], N. Collomb[258],
J. Collot[170], C. Combaret[89], B. Constance[33], C. A. Cooper[58], W. E. Cooper[58], G. Corcella[174],
E. Cormier[29], R. Cornat[172], P. Cornebise[167], F. Cornet[281], G. Corrado[123], F. Corriveau[187],
J. Cortes[286], E. Cortina Gil[303], S. Costa[308], F. Couchot[167], F. Couderc[21], L. Cousin[94],
R. Cowan[185], W. Craddock[251], A. C. Crawford[58], J. A. Crittenden[43], J. Cuevas[283], D. Cuisy[167],
F. Cullinan[237], B. Cure[33], E. Currás Rivera[110], D. Cussans[316], J. Cvach[105], M. Czakon[235],
K. Czuba[401], H. Czyz[365], J. D'Hondt[399], W. Da Silva[169], O. Dadoun[167], M. Dahiya[327], J. Dai[102],
C. Dallapiccola[346], C. Damerell[259], M. Danilov[107], D. Dannheim[33], N. Dascenzo[47,238], S. Dasu[384],
A. K. Datta[67], T. S. Datta[115], P. Dauncey[80], T. Davenne[259], J. David[169], M. Davier[167],
W. De Boer[90], S. De Cecco[169], S. De Curtis[120], N. De Groot[97,213], P. De Jong[213],
S. De Jong[97,213], C. De La Taille[93], G. De Lentdecker[307], S. De Santis[177],
J. B. De Vivie De Regie[167], A. Deandrea[89], P. P. Dechant[49], D. Decotigny[172], K. Dehmelt[257],
J. P. Delahaye[251,33], N. Delerue[167], O. Delferriere[21], F. Deliot[21], G. Della Ricca[388],
P. A. Delsart[170], M. Demarteau[7], D. Demin[142], R. Dermisek[87], F. Derue[169], K. Desch[295],
S. Descotes-Genon[171], A. Deshpande[252], A. Dexter[40], A. Dey[81], S. Dhawan[405], N. Dhingra[226],
V. Di Benedetto[58,123], B. Di Girolamo[33], M. A. Diaz[231], A. Dieguez[311], M. Diehl[47], R. Diener[47],
S. Dildick[331], M. O. Dima[206], P. Dinaucourt[167,93], M. S. Dixit[19], T. Dixit[252], L. Dixon[251],
A. Djouadi[171], S. Doebert[33], M. Dohlus[47], Z. Dolezal[34], H. Dong[102], L. Dong[102], A. Dorokhov[94],
A. Dosil[112], A. Dovbnya[208], T. Doyle[332], G. Doziere[94], M. Dragicevic[224], A. Drago[174],
A. J. Dragt[345], Z. Drasal[34], I. Dremin[179], V. Drugakov[209], J. Duarte Campderros[110],
F. Duarte Ramos[33], A. Dubey[272], A. Dudarev[142], E. Dudas[171,171], L. Dudko[180], C. Duerig[47],
G. Dugan[43], W. Dulinski[94], F. Dulucq[93], L. Dumitru[206], P. J. Dunne[80], A. Duperrin[27],
M. Düren[147], D. Dzahini[170], H. Eberl[224], G. Eckerlin[47], P. Eckert[297], N. R. Eddy[58],

W. Ehrenfeld[295], G. Eigen[314], S. Eisenhardt[329], L. Eklund[332], L. Elementi[58], U. Ellwanger[171], E. Elsen[47], I. Emeliantchik[209], L. Emery[7], K. Enami[71], K. Endo[71], M. Endo[375], J. Engels[47], C. Englert[49], S. Eno[345], A. Enomoto[71], S. Enomoto[197], F. Eozenou[21], R. Erbacher[320], G. Eremeev[269], J. Erler[287], R. Escribano[294], D. Esperante Pereira[108], D. Espriu[311], E. Etzion[265,33], S. Eucker[47], A. Evdokimov[336,107], E. Ezura[71], B. Faatz[47], G. Faisel[201], L. Fano[125], A. Faraggi[340], A. Fasso[251], A. Faus-Golfe[108], L. Favart[307], N. Feege[257], J. L. Feng[321], T. Ferber[47], J. Ferguson[33], J. Fernández[283], P. Fernández Martínez[108], E. Fernandez[294,293], M. Fernandez Garcia[110], J. L. Fernandez-Hernando[130], P. Fernandez-Martinez[32], J. Fernandez-Melgarejo[33], A. Ferrer[108], F. Ferri[21], S. Fichet[117], T. Fifield[347], L. Filkov[179], F. Filthaut[97,213], A. Finch[176], H. E. Fisk[58], T. Fiutowski[2], H. Flaecher[316], J. W. Flanagan[71], I. Fleck[301], M. Fleischer[47], C. Fleta[32], J. Fleury[93], D. Flores[32], M. Foley[58], M. Fontannaz[171], K. Foraz[33], N. Fornengo[128], F. Forti[126,391], B. Foster[47,140], M. C. Fouz[31], P. H. Frampton[353], K. Francis[7], S. Frank[224], A. Freitas[361], A. Frey[64], R. Frey[356], M. Friedl[224], C. Friedrich[48], M. Frigerio[163], T. Frisson[167], M. Frotin[172], R. Frühwirth[224], R. Fuchi[378], E. Fuchs[47], K. Fujii[71], J. Fujimoto[71], H. Fuke[134], B. Fuks[94,33], M. Fukuda[71], S. Fukuda[71], K. Furukawa[71], T. Furuya[71], T. Fusayasu[195], J. Fuster[108], N. Fuster[108], Y. Fuwa[95,159], A. Gaddi[33], K. Gadow[47], F. Gaede[47], R. Gaglione[164], S. Galeotti[126], C. Gallagher[356], A. A. Gallas Torreira[112], L. Gallin-Martel[170], A. Gallo[174], D. Gamba[140,33], D. Gamba[128], J. Gao[102], Y. Gao[24], P. H. Garbincius[58], F. Garcia[69], C. Garcia Canal[288], J. E. Garcia Navarro[108], P. Garcia-Abia[31], J. J. Garcia-Garrigos[108], L. Garcia-Tabares[31], C. García[108], J. V. García Esteve[286], I. García García[108], S. K. Garg[408], L. Garrido[311], E. Garutti[333], T. Garvey[167,261], M. Gastal[33], F. Gastaldi[172], C. Gatto[58,123], N. Gaur[326], D. Gavela Pérez[31], P. Gay[168], M. B. Gay Ducati[109], L. Ge[251], R. Ge[102], A. Geiser[47], A. Gektin[100], A. Gellrich[47], M. H. Genest[170], R. L. Geng[269], S. Gentile[387,127], A. Gerbershagen[33,358], R. Gerig[7], S. German[111], H. Gerwig[33], S. Ghazaryan[47], P. Ghislain[169], D. K. Ghosh[81], S. Ghosh[115], S. Giagu[387,127], L. Gibbons[43], S. Gibson[139,33], V. Gilewsky[143], A. Gillespie[328], F. Gilman[20], B. Gimeno Martínez[108], D. M. Gingrich[312,279], C. M. Ginsburg[58], D. Girard[164], J. Giraud[170], G. F. Giudice[33], L. Gladilin[180], P. Gladkikh[208], C. J. Glasman[343,40], R. Glattauer[224], N. Glover[49], J. Gluza[365], K. Gnidzinska[263], R. Godbole[83], S. Godfrey[19], F. Goertz[54], M. Goffe[94], N. Gogitidze[179,47], J. Goldstein[316], B. Golob[144,341], G. Gomez[110], V. Goncalves[290], R. J. Gonsalves[256], I. González[283], S. González De La Hoz[108], F. J. González Sánchez[110], G. Gonzalez Parra[294], S. Gopalakrishna[103], I. Gorelov[352], D. Goswami[86], S. Goswami[229], T. Goto[71], K. Gotow[398], P. Göttlicher[47], M. Götze[385], A. Goudelis[165], P. Goudket[258], S. Gowdy[33], O. A. Grachov[309], N. A. Graf[251], M. Graham[251], A. Gramolin[15], R. Granier De Cassagnac[172], P. Grannis[257], P. Gras[21], M. Grecki[47], T. Greenshaw[339], D. Greenwood[181], C. Grefe[33], M. Grefe[111], I. M. Gregor[47], D. Grellscheid[49], G. Grenier[89], M. Grimes[316], C. Grimm[58], O. Grimm[53], B. Grinyov[100], B. Gripaios[323], K. Grizzard[141], A. Grohsjean[47], C. Grojean[294,33], J. Gronberg[178], D. Grondin[170], S. Groote[370], P. Gros[240], M. Grunewald[310], B. Grzadkowski[382], J. Gu[102], M. Guchait[262], S. Guiducci[174], E. Guliyev[172], J. Gunion[320], C. Günter[47], C. Gwilliam[339], N. Haba[75], H. Haber[322], M. Hachimine[197], Y. Haddad[172], L. Hagge[47], M. Hagihara[378], K. Hagiwara[71,158], J. Haley[216], G. Haller[251], J. Haller[333], K. Hamaguchi[375], R. Hamatsu[276], G. Hamel De Monchenault[21], L. L. Hammond[58], P. Hamnett[47], L. Han[364], T. Han[361], K. Hanagaki[223], J. D. Hansen[211], K. Hansen[47], P. H. Hansen[211], X. Q. Hao[70], K. Hara[71], K. Hara[378], T. Hara[71], D. Harada[83], K. Harada[161], K. Harder[259], T. Harion[297], R. V. Harlander[385], E. Harms[58], M. Harrison[13], O. Hartbrich[47,385], A. Hartin[47], T. Hartmann[300], J. Harz[47], S. Hasegawa[197], T. Hasegawa[71], Y. Hasegawa[248], M. Hashimoto[38], T. Hashimoto[62], C. Hast[251], S. Hatakeyama[135], J. M. Hauptman[118], M. Hauschild[33], M. Havranek[105], C. Hawkes[315], T. Hayakawa[197], H. Hayano[71], K. Hayasaka[198], M. Hazumi[71,253], H. J. He[24],

C. Hearty[317,104], H. F. Heath[316], T. Hebbeker[235], M. Heck[90], V. Hedberg[183], D. Hedin[217],
S. M. Heindl[90], S. Heinemeyer[110], I. Heinze[47], A. Hektor[205], S. Henrot-Versille[167], O. Hensler[47],
A. Heo[23], J. Herbert[258], G. Herdoiza[138], B. Hermberg[47], J. J. Hernández-Rey[108], M. J. Herrero[111],
B. Herrmann[165], A. Hervé[384], J. Hewett[251], S. Hidalgo[32], B. Hidding[333,318], N. Higashi[375],
N. Higashi[71], T. Higo[71], E. Higón Rodríguez[108], T. Higuchi[150], M. Hildreth[354], C. T. Hill[58],
S. Hillert[295], S. Hillier[315], T. Himel[251], A. Himmi[94], S. Himori[272], Z. Hioki[374], B. Hippolyte[94],
T. Hiraki[159], K. Hirano[136], S. Hirano[197], K. Hirata[71], T. Hirose[276], M. Hirsch[108], J. Hisano[197],
P. M. Ho[210], A. Hoang[302], A. Hocker[58], A. Hoecker[33], M. Hoeferkamp[352], M. Hoffmann[47],
W. Hollik[186], K. Homma[72], Y. Homma[154], S. Honda[378], T. Honda[71], Y. Honda[71],
N. T. Hong Van[106], K. Honkavaara[47], T. Honma[71], T. Hori[236], T. Horiguchi[272], Y. Horii[197],
A. Horio[196], R. Hosaka[377], Y. Hoshi[271], H. Hoshino[197], K. Hosoyama[71], J. Y. Hostachy[170],
G. W. Hou[210], M. Hou[102], A. Hoummada[184], M. S. Hronek[58], T. Hu[102], C. Hu-Guo[94],
M. Huang[24], T. Huang[102], E. Huedem[58], F. Hügging[295], J. L. Hugon[330], C. Hugonie[173],
K. Huitu[335], P. Q. Hung[380], C. Hunt[80], U. Husemann[90], G. Hussain[24], D. Hutchcroft[339],
Y. Hyakutake[79], J. C. Ianigro[89], L. E. Ibanez[111], M. Ibe[96], M. Idzik[2], H. Igarashi[78], Y. Igarashi[71],
K. Igi[236], A. Ignatenko[209], O. Igonkina[213], T. Iijima[198,197], M. Iinuma[73], Y. Iiyama[20], H. Ikeda[134],
K. Ikeda[71], K. Ikematsu[301], J. I. Illana[281], V. A. Ilyin[207,180], A. Imhof[333], T. Inagaki[236],
T. Inagaki[197], K. Inami[197], S. Inayoshi[248], K. Inoue[161], A. Irles[108], S. Isagawa[71], N. Ishibashi[378],
A. Ishida[375], K. Ishida[212], N. Ishihara[71], S. Ishihara[77], K. Ishii[71], A. Ishikawa[272], K. Ishikawa[375],
K. I. Ishikawa[72], K. Ishikawa[75], T. Ishikawa[71], M. Ishitsuka[275], K. Ishiwata[17], G. Isidori[174],
A. Ismail[251], S. Iso[71], T. Isogai[197], C. Issever[358], K. Itagaki[272], T. Itahashi[223], A. Ito[275], F. Ito[378],
S. Ito[272], R. Itoh[71], E. Itou[71], M. I. Ivanyan[26], G. Iwai[71], S. Iwamoto[375], T. Iwamoto[116],
H. Iwasaki[71], M. Iwasaki[71], Y. Iwashita[95], S. Iwata[71], S. Iwata[276], T. Izubuchi[13,236], Y. Izumiya[272],
S. Jablonski[401], F. Jackson[258], J. A. Jacob[316], M. Jacquet[167], P. Jain[40], P. Jaiswal[59],
W. Jalmuzna[263], E. Janas[401], R. Jaramillo Echeverría[110], J. Jaros[251], D. Jatkar[67], D. Jeans[375],
R. Jedziniak[58], M. J. Jenkins[176,40], K. Jensch[47], C. P. Jessop[354], T. Jezynski[47], M. Jimbo[35],
S. Jin[102], O. Jinnouchi[275], M. D. Joergensen[211], A. S. Johnson[251], S. Jolly[309], D. T. Jones[340],
J. Jones[258,40], R. M. Jones[343,40], L. Jönsson[183], N. Joshi[237], C. K. Jung[257,150], N. Juntong[343,40],
A. Juste[88,294], W. Kaabi[167], M. Kadastik[205], M. Kado[167,33], K. Kadota[197], E. Kajfasz[27],
R. Kajikawa[197], Y. Kajiura[197], M. Kakizaki[377], E. Kako[71], H. Kakuhata[377], H. Kakuno[276],
A. Kalinin[258], J. Kalinowski[382], G. E. Kalmus[259], K. Kamada[47], J. Kaminski[295], T. Kamitani[71],
Y. Kamiya[116], Y. Kamiya[71], R. Kammering[47], T. Kamon[266], J. I. Kamoshita[60], T. Kanai[275],
S. Kananov[265], K. Kanaya[378], M. Kaneda[33], T. Kaneko[71,253], S. Kanemura[377], K. Kaneta[75],
W. Kang[102], D. Kanjilal[115], K. Kannike[205], F. Kapusta[169], D. Kar[332], P. Karataev[237,139],
P. E. Karchin[403], D. Karlen[379,279], S. Karstensen[47], Y. Karyotakis[164], M. Kasemann[47],
V. S. Kashikhin[58], S. Kashiwagi[273], A. Kataev[98], V. Katalev[48], Y. Kataoka[116], N. Katayama[150],
R. Katayama[375], E. Kato[272], K. Kato[155], S. Kato[71], Y. Kato[153], T. Katoh[71], A. Kaukher[47],
S. Kawabata[71], S. I. Kawada[73], K. Kawagoe[161], M. Kawai[71], T. Kawamoto[116], H. Kawamura[71],
M. Kawamura[71], Y. Kawamura[248], S. Kawasaki[71], T. Kawasaki[212], H. Kay[47], S. Kazama[375],
L. Keegan[111], J. Kehayias[150], L. Keller[251], M. A. Kemp[251], J. J. Kempster[237], C. Kenney[251],
I. Kenyon[315], R. Kephart[58], J. Kerby[7], K. Kershaw[33], J. Kersten[333], K. Kezzar[151],
V. G. Khachatryan[26], M. A. Khan[23], S. A. Khan[242], Y. Khoulaki[184], V. Khoze[49], H. Kichimi[71],
R. Kieffer[33], C. Kiesling[186], M. Kikuchi[377], Y. Kikuta[71], M. Killenberg[47], C. S. Kim[408],
D. W. Kim[63], D. Kim[23], E. J. Kim[36,36], E. S. Kim[23], G. Kim[23], H. S. Kim[23], H. D. Kim[245],
J. Kim[63], S. H. Kim[378], S. K. Kim[245], S. G. Kim[87], Y. I. Kim[358,140], Y. Kimura[71], R. E. Kirby[251],
F. Kircher[21], Y. Kishimoto[96], L. Kisslinger[20], T. Kitahara[375], R. Kitano[272], Y. Kiyo[146,71],
C. Kleinwort[47], W. Klempt[33], P. M. Kluit[213], V. Klyukhin[180,33], M. Knecht[28], J. L. Kneur[163],

B. A. Kniehl[333], K. Ko[251], P. Ko[158], D. Kobayashi[275], M. Kobayashi[71], N. Kobayashi[71], T. Kobayashi[116], M. Koch[295], P. Kodys[34], U. Koetz[47], E. N. Koffeman[213], M. Kohda[210], S. Koike[71], Y. Kojima[71], K. Kolodziej[365], Y. Kolomensky[319,177], S. Komamiya[375], T. Kon[244], P. Konar[229], Y. Kondou[71], D. Kong[23], K. Kong[338], O. C. Kong[201], T. Konno[275,275], V. Korbel[47], J. G. Körner[138], S. Korpar[344,144], S. R. Koscielniak[279], D. Kostin[47], K. Kotera[248], W. Kotlarski[382], J. Kotula[268], E. Kou[167], V. Kovalenko[333], S. V. H. Kox[170], K. Koyama[75], M. Krämer[235], S. Kraml[170], M. Krammer[224], M. W. Krasny[169], F. Krauss[49], T. Krautscheid[295], M. Krawczyk[382], K. Krempetz[58], P. Križan[341,144], B. E. Krikler[80], A. Kronfeld[58], K. Kruchinin[237,139], D. Krücker[47], K. Krüger[47], B. Krupa[268], Y. P. Kuang[24], K. Kubo[71], T. Kubo[71], T. Kubota[347], T. Kubota[275], Y. Kubyshin[298,180], V. Kuchler[58], I. M. Kudla[202], D. Kuehn[47], J. H. Kuehn[92], C. Kuhn[94], S. Kulis[2], S. Kulkarni[170], A. Kumar[10], S. Kumar[86], T. Kumita[276], A. Kundu[16], Y. Kuno[223], C. M. Kuo[201], M. Kurachi[198], A. Kuramoto[253], M. Kurata[375], Y. Kurihara[71], M. Kuriki[73,71], T. Kurimoto[377], S. Kuroda[71], K. Kurokawa[71], S. I. Kurokawa[71], H. Kuwabara[276], M. Kuze[275], J. Kvasnicka[105], P. Kvasnicka[34], Y. Kwon[408], L. Labun[210], C. Lacasta[108], T. Lackowski[58], D. Lacour[169], V. Lacuesta[108], R. Lafaye[164], G. Lafferty[343], B. Laforge[169], I. Laktineh[89], R. L. Lander[320], K. Landsteiner[111], S. Laplace[169], K. J. Larsen[213], R. S. Larsen[251], T. Lastovicka[105], J. I. Latorre[311], S. Laurien[333], L. Lavergne[169], S. Lavignac[22], R. E. Laxdal[279], A. C. Le Bihan[94], F. R. Le Diberder[167], A. Le-Yaouanc[171], A. Lebedev[13], P. Lebrun[33], T. Lecompte[7], T. Leddig[300], F. Ledroit[170], B. Lee[25], K. Lee[158], M. Lee[177], S. H. Lee[260], S. W. Lee[267], Y. H. Lee[210], J. Leibfritz[58], K. Lekomtsev[71], L. Lellouch[28], M. Lemke[47], F. R. Lenkszus[7], A. Lenz[49,33], O. Leroy[27], C. Lester[323], L. Levchuk[208], J. Leveque[164], E. Levichev[15], A. Levy[265], I. Levy[265], J. R. Lewandowski[251], B. Li[24], C. Li[364], C. Li[102], D. Li[102], H. Li[380], L. Li[195], L. Li[247], L. Li[364], S. Li[102], W. Li[102], X. Li[102], Y. Li[24], Y. Li[24], Y. Li[24], Z. Li[251], Z. Li[102], J. J. Liau[210], V. Libov[47], L. Lilje[47], J. G. Lima[217], C. J. D. Lin[204], C. M. Lin[154], C. Y. Lin[201], H. Lin[102], H. H. Lin[210], F. L. Linde[213], R. A. Lineros[108], L. Linssen[33], R. Lipton[58], M. Lisovyi[47], B. List[47], J. List[47], B. Liu[24], J. Liu[364], R. Liu[102], S. Liu[167], S. Liu[247], W. Liu[7], Y. Liu[102], Y. Liu[337,58], Z. Liu[361], Z. Liu[102], Z. Liu[102], A. Lleres[170], N. S. Lockyer[279,317], W. Lohmann[48,12], E. Lohrmann[333], T. Lohse[76], F. Long[102], D. Lontkovskyi[47], M. A. Lopez Virto[110], X. Lou[102,372], A. Lounis[167], M. Lozano Fantoba[32], J. Lozano-Bahilo[281], C. Lu[232], R. S. Lu[210], S. Lu[47], A. Lucotte[170], F. Ludwig[47], S. Lukic[396], O. Lukina[180], N. Lumb[89], B. Lundberg[183], A. Lunin[58], M. Lupberger[295], B. Lutz[47], P. Lutz[21], T. Lux[294], K. Lv[102], M. Lyablin[142], A. Lyapin[237,139], J. Lykken[58], A. T. Lytle[262], L. Ma[258], Q. Ma[102], R. Ma[312], X. Ma[102], F. Machefert[167], N. Machida[377], J. Maeda[276], Y. Maeda[159], K. Maeshima[58], F. Magniette[172], N. Mahajan[229], F. Mahmoudi[168,33], S. H. Mai[201], C. Maiano[119], H. Mainaud Durand[33], S. Majewski[356], S. K. Majhi[81], N. Majumdar[241], G. Majumder[262], I. Makarenko[47], V. Makarenko[209], A. Maki[71], Y. Makida[71], D. Makowski[263], B. Malaescu[169], J. Malcles[21], U. Mallik[337], S. Malvezzi[121], O. B. Malyshev[258,40], Y. Mambrini[171], A. Manabe[71], G. Mancinelli[27], S. K. Mandal[150], S. Mandry[309,186], S. Manen[168], R. Mankel[47], S. Manly[363], S. Mannai[303], Y. Maravin[149], G. Marchiori[169], M. Marcisovsky[105,45], J. Marco[110], D. Marfatia[338], J. Marin[31], E. Marin Lacoma[251], C. Marinas[295], T. W. Markiewicz[251], O. Markin[107], J. Marshall[323], S. Martí-García[108], A. D. Martin[49], V. J. Martin[329], G. Martin-Chassard[93], T. Martinez De Alvaro[31], C. Martinez Rivero[110], F. Martinez-Vidal[108], H. U. Martyn[235,47], T. Maruyama[251], A. Masaike[159], T. Mashimo[116], T. Masubuchi[116], T. Masuda[159], M. Masuzawa[71], Z. Mateusz[401], A. Matheisen[47], H. Mathez[89], J. Matias[293], H. Matis[177], T. Matsubara[276], T. Matsuda[71], T. Matsui[377], S. Matsumoto[161], S. Matsumoto[150], Y. Matsumoto[220], H. Matsunaga[71], T. Matsushita[154], T. S. Mattison[317], V. A. Matveev[142], U. Mavric[47], G. Mavromanolakis[33], K. Mawatari[399], S. J. Maxfield[339], K. Mazumdar[262],

A. Mazzacane[58,123], R. L. Mccarthy[257], D. J. Mccormick[251], J. Mccormick[251],
K. T. Mcdonald[232], R. Mcduffee[324], P. Mcintosh[258], B. Mckee[251], M. Medinnis[47], S. Mehlhase[211],
T. Mehrling[47,333], A. Mehta[339], B. Mele[127], R. E. Meller[43], I. A. Melzer-Pellmann[47], L. Men[102],
G. Mendiratta[83], Z. Meng[316], M. H. Merk[213,400], M. Merkin[180], A. Merlos[32], L. Merminga[279],
A. B. Meyer[47], A. Meyer[235], N. Meyners[47], Z. Mi[102], P. Michelato[119], S. Michizono[71],
S. Mihara[71], A. Mikhailichenko[43], D. J. Miller[309], C. Milstene[403], Y. Mimura[210], D. Minic[398],
L. Mirabito[89], S. Mishima[387], T. Misumi[13], W. A. Mitaroff[224], T. Mitsuhashi[71], S. Mitsuru[71],
K. Miuchi[154], K. Miyabayashi[200], A. Miyamoto[71], H. Miyata[212], Y. Miyazaki[161], T. Miyoshi[71],
R. Mizuk[107], K. Mizuno[3], U. Mjörnmark[183], J. Mnich[47], G. Moeller[47], W. D. Moeller[47],
K. Moenig[48], K. C. Moffeit[251], P. Mohanmurthy[269], G. Mohanty[262], L. Monaco[119], S. Mondal[81],
C. Monini[170], H. Monjushiro[71], G. Montagna[359,124], S. Monteil[168], G. Montoro[298], I. Montvay[47],
F. Moortgat[53], G. Moortgat-Pick[333,47], P. Mora De Freitas[172], C. Mora Herrera[30], G. Moreau[171],
F. Morel[94], A. Morelos-Pineda[280], M. Moreno Llacer[108], S. Moretti[367,259], V. Morgunov[47,107],
T. Mori[71], T. Mori[272], T. Mori[116], Y. Morita[71], S. Moriyama[96,150], L. Moroni[121], Y. Morozumi[71],
H. G. Moser[186], A. Moszczynski[268], K. Motohashi[275], T. Moulik[249], G. Moultaka[163],
D. Moya Martin[110], S. K. Mtingwa[215], G. S. Muanza[27], M. Mühlleitner[91], A. Mukherjee[85],
S. Mukhopadhyay[241], M. Mulders[33], D. Muller[251], F. Müller[47], T. Müller[90], V. S. Mummidi[83],
A. Münnich[47], C. Munoz[111], F. J. Muñoz Sánchez[110], H. Murayama[375,319], R. Murphy[7],
G. Musat[172], A. Mussgiller[47], R. Muto[71], T. Nabeshima[377], K. Nagai[71], K. Nagai[378],
S. Nagaitsev[58], T. Nagamine[272], K. Nagano[71], K. Nagao[71], Y. Nagashima[223], S. C. Nahn[185],
P. Naik[316], D. Naito[159], T. Naito[71], H. Nakai[71], K. Nakai[71,223], Y. Nakai[161], Y. Nakajima[177],
E. Nakamura[71], H. Nakamura[71], I. Nakamura[71], S. Nakamura[159], S. Nakamura[71],
T. Nakamura[116], K. Nakanishi[71], E. Nakano[222], H. Nakano[212], H. Nakano[272], Y. Namito[71],
W. Namkung[230], H. Nanjo[159], C. D. Nantista[251], O. Napoly[21], Y. Nara[4], T. Narazaki[272],
S. Narita[131], U. Nauenberg[324], T. Naumann[48], S. Naumann-Emme[47], J. Navarro[108], A. Navitski[47],
H. Neal[251], K. Negishi[272], K. Neichi[270], C. A. Nelson[255], T. K. Nelson[251], S. Nemecek[105],
M. Neubert[138], R. Neuhaus[218], L. J. Nevay[237], D. M. Newbold[316,259], O. Nezhevenko[58],
F. Nguyen[175], M. Nguyen[172], M. N. Nguyen[251], T. T. Nguyen[106], R. B. Nickerson[358],
O. Nicrosini[124], C. Niebuhr[47], J. Niehoff[58], M. Niemack[43], U. Nierste[92], H. Niinomi[159],
I. Nikolic[169], H. P. Nilles[295], S. Nishida[71], H. Nishiguchi[71], K. Nishiwaki[67], O. Nitoh[277], L. Niu[24],
R. Noble[251], M. Noji[71], M. Nojiri[71,150], S. Nojiri[197,198], D. Nölle[47], A. Nomerotski[358],
M. Nomura[135], T. Nomura[201], Y. Nomura[319,177], C. Nonaka[198], J. Noonan[7], E. Norbeck[337],
Y. Nosochkov[251], D. Notz[47], O. Novgorodova[48,12], A. Novokhatski[251], J. A. Nowak[349],
M. Nozaki[71], K. Ocalan[192], J. Ocariz[169], S. Oda[161], A. Ogata[71], T. Ogawa[248], T. Ogura[248],
A. Oh[343], S. K. Oh[156], Y. Oh[23], K. Ohkuma[61], T. Ohl[145], Y. Ohnishi[71], K. Ohta[189],
M. Ohta[71,253], S. Ohta[71,253], N. Ohuchi[71], K. Oishi[161], R. Okada[71], Y. Okada[71,253],
T. Okamura[71], H. Okawa[13], T. Okugi[71], T. Okui[59], K. I. Okumura[161], Y. Okumura[52,58],
L. Okun[107], H. Okuno[236], C. Oleari[390], C. Oliver[31], B. Olivier[186], S. L. Olsen[245], M. Omet[253,71],
T. Omori[71], Y. Onel[337], H. Ono[214], Y. Ono[272], D. Onoprienko[251], Y. Onuki[116,150], P. Onyisi[371],
T. Oogoe[71], Y. Ookouchi[159], W. Ootani[116], M. Oreglia[52], M. Oriunno[251], M. C. Orlandea[206],
J. Orloff[168], M. Oroku[375,71], R. S. Orr[376], J. Osborne[33], A. Oskarsson[183], P. Osland[314],
A. Osorio Oliveros[282], L. Österman[183], H. Otono[223], M. Owen[343], Y. Oyama[71], A. Oyanguren[108],
K. Ozawa[71,375], J. P. Ozelis[58,191], D. Ozerov[47], G. Pásztor[304,99], H. Padamsee[43], C. Padilla[294],
C. Pagani[119,386], R. Page[316], R. Pain[169], S. Paktinat Mehdiabadi[101], A. D. Palczewski[269],
S. Palestini[33], F. Palla[126], M. Palmer[58], F. Palomo Pinto[285], W. Pan[102], G. Pancheri[174],
M. Pandurovic[396], O. Panella[125], A. Pankov[227], Y. Papaphilippou[33], R. Paparella[119],
A. Paramonov[7], E. K. Park[75], H. Park[23], S. I. Park[23], S. Park[260], S. Park[373], W. Park[23],

A. Parker[323], B. Parker[13], C. Parkes[343], V. Parma[33], Z. Parsa[13], R. Partridge[251], S. Pastor[108], E. Paterson[251], M. Patra[83], J. R. Patterson[43], M. Paulini[20], N. Paver[129], S. Pavy-Bernard[167], B. Pawlik[268], A. Pérez Vega-Leal[285], B. Pearson[355], J. S. Pedersen[211], A. Pedicini[397], S. Pedraza López[108], G. Pei[102], S. Pei[102], G. Pellegrini[32], A. Pellegrino[213], S. Penaranda[286], H. Peng[364], X. Peng[102], M. Perelstein[43], E. Perez[112], M. A. Perez-Garcia[284,114], M. Perez-Victoria[281], S. Peris[293], D. Perret-Gallix[164], H. Perrey[47], T. M. Perry[384], M. E. Peskin[251], P. Petagna[33], R. Y. Peters[64,47], T. C. Petersen[211], D. P. Peterson[43], T. Peterson[58], E. Petrakou[210], A. A. Petrov[403], A. Petrukhin[89,107], S. Pfeiffer[47], H. Pham[94], K. H. Phan[395,71], N. Phinney[251], F. Piccinini[124], A. Pich[108], R. Pichai[85], J. Piedra[283], J. Piekarski[401], A. Pierce[348], P. Pierini[119], N. Pietsch[333,47], A. Pineda[293], J. Pinfold[312,152], A. Piotrowski[263], Y. Pischalnikov[58], R. Pittau[281], M. Pivi[251], W. Placzek[132], T. Plehn[296], M. A. Pleier[13], M. Poelker[269], L. Poggioli[167], I. Pogorelsky[13], V. Poireau[164], M. E. Pol[30], I. Polak[105], F. Polci[169], M. Polikarpov[107], T. Poll[316,259], M. W. Poole[258,40], W. Porod[145], F. C. Porter[17], S. Porto[333], J. Portolés[108], R. Pöschl[167], S. Poss[33], C. T. Potter[356], P. Poulose[86], K. T. Pozniak[401], V. Prahl[47], R. Prepost[384], C. Prescott[251], D. Price[87], T. Price[315], P. S. Prieto[58], D. Protopopescu[332], D. Przyborowski[2], K. Przygoda[263], H. Przysiezniak[164], F. Ptochos[325], J. Puerta-Pelayo[31], C. Pulvermacher[90], M. Purohit[366], Q. Qin[102], F. Qiu[102], H. Qu[102], A. Quadt[64], G. Quast[90], D. Quirion[32], M. Quiros[88], J. Rademacker[316], R. Rahmat[350], S. Rai[67], M. Raidal[205], S. Rakshit[84], M. Ramilli[333], F. Rarbi[170], P. Ratoff[176], T. Raubenheimer[251], M. Rauch[91], L. Raux[93], G. Raven[400,213], P. Razis[325], V. Re[124], S. Redford[33], C. E. Reece[269], I. Reichel[177], A. Reichold[358,140], P. Reimer[105], M. Reinecke[47], A. Rekalo[100], J. Repond[7], J. Resta-Lopez[108], J. Reuter[47], J. T. Rhee[156], P. M. Ribeiro Cipriano[47], A. Ribon[33], G. Ricciardi[292,123], F. Richard[167], E. Richter-Was[132], G. Riddone[33], S. Riemann[48], T. Riemann[48], M. Rijssenbeek[257], K. Riles[348], C. Rimbault[167], R. Rimmer[269], S. D. Rindani[229], A. Ringwald[47], L. Rinolfi[33], I. Ripp-Baudot[94], I. Riu[294], T. G. Rizzo[251], P. Robbe[167], J. Roberts[140,33], A. Robson[332], G. Rodrigo[108], P. Rodriguez[251], P. Rodriguez Perez[112], K. Rolbiecki[111], P. Roloff[33], R. S. Romaniuk[401], E. Romero Adam[108], A. Ronzhin[58], L. Roos[169], E. Ros[108], A. Rosca[47], C. Rosemann[47], J. Rosiek[382], M. C. Ross[251], R. Rossmanith[90], S. Roth[235], J. Rouëné[167], A. Rowe[58], P. Rowson[251], A. Roy[115], L. Royer[168], P. Royole-Degieux[93], C. Royon[21], A. Rozanov[27], M. Ruan[172], D. L. Rubin[43], I. Rubinskiy[47], R. Rückl[145], R. Ruiz[361], R. Ruiz De Austri[108], P. Ruiz Valls[108], P. Ruiz-Femenía[108], A. Ruiz-Jimeno[110], R. Ruland[251], V. Rusinov[107], J. J. Russell[251], I. Rutkowski[401], V. Rybnikov[47], A. Ryd[43], A. Sabio Vera[111], B. Sabirov[142], J. J. Saborido Silva[112], H. F. W. Sadrozinski[322], T. Saeki[71], B. Safarzadeh[101], P. Saha[305], H. Sahoo[7], A. Sailer[33], N. Saito[71], T. Saito[272], T. Sakaguchi[13], H. Sakai[71], K. Sakai[71], K. Sakaue[402], K. Sakurai[47], R. Salerno[172], J. Salfeld-Nebgen[47], J. Salt[108], L. Sanchez[31], M. A. Sanchis Lozano[108], J. Sandweiss[405], A. Santa[377], A. Santagata[286], A. Santamaria[108], P. Santorelli[292], T. Sanuki[272], A. A. Sapronov[142], M. Sasaki[96], H. Sato[248], N. Sato[71], Y. Sato[272], M. Satoh[71], E. Sauvan[164], V. Saveliev[238,47], A. Savoy-Navarro[166,126], M. Sawabe[71], R. Sawada[116], H. Sawamura[402], L. Sawyer[181], O. Schäfer[300,47], R. Schäfer[47], J. Schaffran[47], T. Schalk[322,251], R. D. Schamberger[257], J. Scheirich[34], G. Schierholz[47], F. P. Schilling[90], F. Schirra[89], F. Schlander[47], H. Schlarb[47], D. Schlatter[33], P. Schleper[333], J. L. Schlereth[7], R. D. Schlueter[177], C. Schmidt[47], U. Schneekloth[47], S. Schnetzer[239], T. Schoerner-Sadenius[47], M. Schram[225], H. J. Schreiber[48], S. Schreiber[47], K. P. Schüler[47], D. Schulte[33], H. C. Schultz-Coulon[297], M. Schumacher[5], S. Schumann[64], B. A. Schumm[322], M. H. Schune[167], S. Schuwalow[333,48], C. Schwanda[224], C. Schwanenberger[343], F. Schwartzkopff[295], D. J. Scott[258,58], F. Sefkow[47,33], A. Segui[286], N. Seguin-Moreau[93], S. Seidel[352], Y. Seiya[222], J. Sekaric[338], K. Seki[197], S. Sekmen[33], S. Seletskiy[13], S. Sen[337], E. Senaha[158], K. Senyo[406], S. Senyukov[94], I. Serenkova[227],

D. A. Sergatskov[58], H. Sert[47,333], D. Sertore[119], A. Seryi[358,140], O. Seto[74], R. Settles[186], P. Sha[102], S. Shahid[301], A. Sharma[33], G. Shelkov[142], W. Shen[297], J. C. Sheppard[251], M. Sher[41], C. Shi[102], H. Shi[102], T. Shidara[71], W. Shields[237,139], M. Shimada[71], H. Shimizu[71], Y. Shimizu[272], M. Shimojima[195], S. Shimojima[276], T. Shindou[155], N. Shinoda[272], Y. Shinzaki[272], M. Shioden[78,71], I. Shipsey[233], S. Shirabe[161], M. J. Shirakata[71], T. Shirakata[71], G. Shirkov[142], T. Shishido[71], T. Shishido[71], J. G. Shiu[210], R. Shivpuri[326], R. Shrock[257], T. Shuji[71], N. Shumeiko[209], B. Shuve[228,188], P. Sicho[105], A. M. Siddiqui[133], P. Sievers[33], D. Sikora[401], D. A. Sil[86], F. Simon[186], N. B. Sinev[356], W. Singer[47], X. Singer[47], B. K. Singh[10], R. K. Singh[82], N. Sinha[103], R. Sinha[103], K. Sinram[47], T. Sinthuprasith[14], P. Skubic[355], R. Sliwa[167], I. Smiljanic[396], J. R. Smith[373,7], J. C. Smith[251,43], S. R. Smith[251], J. Smolík[105,45], J. Snuverink[237,139], B. Sobloher[47], J. Sola[311], C. Soldner[186,57], S. Soldner-Rembold[343], D. Son[23], H. S. Song[260], N. Sonmez[51], A. Sopczak[44], D. E. Soper[356], P. Spagnolo[126], S. Spannagel[47], M. Spannowsky[49], A. Sparkes[329], C. M. Spencer[251], H. Spiesberger[138], M. Spira[261], M. Stahlhofen[47], M. Stanescu-Bellu[48], M. Stanitzki[47], S. Stapnes[33], P. Starovoitov[209], F. Staufenbiel[48], L. Steder[47], M. Steder[47], A. Steen[89], G. Steinbrueck[333], M. Steinhauser[92], F. Stephan[48], W. Stephen[237], S. Stevenson[358], I. Stewart[185], D. Stöckinger[264], H. Stoeck[369], M. Strauss[355], S. Striganov[58], D. M. Strom[356], R. Stromhagen[47], J. Strube[33], A. Strumia[205], G. Stupakov[251], N. Styles[47], D. Su[251], F. Su[102], S. Su[313], J. Suarez Gonzalez[209], Y. Sudo[161], T. Suehara[116], F. Suekane[274], Y. Suetsugu[71], R. Sugahara[71], A. Sugamoto[220], H. Sugawara[71], Y. Sugimoto[71], A. Sugiyama[240], H. Sugiyama[377], M. K. Sullivan[251], Y. Sumino[272], T. Sumiyoshi[276,71], H. Sun[102], M. Sun[20], X. Sun[170], Y. Sun[102], Y. Susaki[197], T. Suwada[71], A. Suzuki[71], S. Suzuki[240], Y. Suzuki[71], Y. Suzuki[73], Z. Suzuki[272], K. Swientek[2], C. Swinson[13], Z. M. Szalata[251], B. Szcezepanski[47], M. Szelezniak[94], J. Szewinski[202], A. Sznajder[289], L. Szymanowski[202], H. Tabassam[329], K. Tackmann[47], M. Taira[71], H. Tajima[197,251], F. Takahashi[272], R. Takahashi[71], R. Takahashi[75], T. Takahashi[73], Y. Takahashi[197], K. Takata[71], F. Takayama[160], Y. Takayasu[75], H. Takeda[154], S. Takeda[71], T. Takeshita[248], A. Taketani[236], Y. Takeuchi[378], T. Takimi[262], Y. Takubo[71], Y. Tamashevich[47], M. Tamsett[181], M. Tanabashi[198,197], T. Tanabe[116], G. Tanaka[161], M. M. Tanaka[71], M. Tanaka[223], R. Tanaka[73], H. Taniuchi[377], S. Tapprogge[138], E. Tarkovsky[107], M. A. Tartaglia[58], X. R. Tata[334], T. Tauchi[71], M. Tawada[71], G. Taylor[347], A. M. Teixeira[168], V. I. Telnov[15,219], P. Tenenbaum[251], E. Teodorescu[206], S. Terada[71], Y. Teramoto[222], H. Terao[200], A. Terashima[71], S. Terui[71], N. Terunuma[71], M. Terwort[47], M. Tesar[186], F. Teubert[33], T. Teubner[340], R. Teuscher[376], T. Theveneaux-Pelzer[168], D. Thienpont[93,172], J. Thom-Levy[43], M. Thomson[323], J. Tian[71], X. Tian[366], M. Tigner[43], J. Timmermans[213], V. Tisserand[164], M. Titov[21], S. Tjampens[164], K. Tobe[197], K. Tobioka[150,319], K. Toda[278], M. Toda[71], N. Toge[71], J. Tojo[161], K. Tokushuku[71], T. Toma[49], R. Tomas[33], T. Tomita[161], A. Tomiya[223], M. Tomoto[197,198], K. Toms[352], M. Tonini[47], F. Toral[31], E. Torrence[356], E. Torrente-Lujan[33], N. Toumbas[325], C. Touramanis[339], F. Toyoda[161], K. Toyomura[71], G. Trahern[55], T. H. Tran[172], W. Treberspurg[224], J. Trenado[311], M. Trimpl[58], S. Trincaz-Duvoid[169], M. Tripathi[320], W. Trischuk[376], M. Trodden[360], G. Trubnikov[142], H. C. Tsai[39], J. F. Tsai[210], K. H. Tsao[336], R. Tschirhart[58], E. Tsedenbaljir[210], S. Y. Tseng[201], T. Tsuboyama[71], A. Tsuchiya[250], K. Tsuchiya[71], T. Tsukamoto[71], K. Tsumura[197], S. Tsuno[71], T. Tsurugai[189], T. Tsuyuki[96], B. Tuchming[21], P. V. Tyagi[68,254], I. Tyapkin[142], M. Tytgat[331], K. Uchida[295], F. Uchiyama[71], Y. Uchiyama[116], S. Uehara[71], H. Ueno[161], K. Ueno[71], K. Ueno[71], K. Ueshima[274], Y. Uesugi[73], N. Ujiie[71], F. Ukegawa[378], N. Ukita[378], M. Ullán[32], H. Umeeda[72], K. Umemori[71], Y. Unno[66], S. Uozumi[23], J. Urakawa[71], A. M. Uranga[111], J. Urresti[32], A. Ushakov[333], I. Ushiki[272], Y. Ushiroda[71], A. V[83], P. Vázquez Regueiro[112], L. Vacavant[27], G. Valencia[118], L. Valery[168], I. Valin[94], J. W. Valle[108], C. Vallee[27], N. Van Bakel[213], H. Van Der Graaf[213], N. Van Der Kolk[167],

E. Van Der Kraaij[33], B. Van Doren[338], B. Van Eijk[213], R. Van Kooten[87], W. T. Van Oers[279], D. Vanegas[108], P. Vanhoefer[186], P. Vankov[47], P. Varghese[58], A. Variola[167], R. Varma[85], G. Varner[334], G. Vasileiadis[162], A. Vauth[47], J. Velthuis[316], S. K. Vempati[83], V. Vento[108], M. Venturini[177], M. Verderi[172], P. Verdier[89], A. Verdugo[31], A. Vicente[171], J. Vidal-Perona[108], H. L. R. Videau[172], I. Vila[110], X. Vilasis-Cardona[299], E. Vilella[311], A. Villamor[32], E. G. Villani[259], J. A. Villar[286], M. A. Villarejo Bermúdez[108], D. Vincent[169], P. Vincent[169], J. M. Virey[28], A. Vivoli[58], V. Vogel[47], R. Volkenborn[47], O. Volynets[47], F. Von Der Pahlen[110], E. Von Toerne[295], B. Vormwald[47], A. Voronin[180], M. Vos[108], J. H. Vossebeld[339], G. Vouters[164], Y. Voutsinas[94,47], V. Vrba[105,45], M. Vysotsky[107], D. Wackeroth[256], A. Wagner[47], C. E. Wagner[7,52], R. Wagner[7], S. R. Wagner[324], W. Wagner[385], J. Wagner-Kuhr[90], A. P. Waite[251], M. Wakayama[197], Y. Wakimoto[276], R. Walczak[358,140], R. Waldi[300], D. G. E. Walker[251], N. J. Walker[47], M. Walla[47], C. J. Wallace[49], S. Wallon[171,393], D. Walsh[328], S. Walston[178], W. A. T. Wan Abdullah[342], D. Wang[102], G. Wang[102], J. Wang[251], L. Wang[251], L. Wang[52], M. H. Wang[251], M. Z. Wang[210], Q. Wang[102], Y. Wang[102], Z. Wang[24], R. Wanke[138], C. Wanotayaroj[356], B. Ward[8], D. Ward[323], B. Warmbein[47], M. Washio[402], K. Watanabe[71], M. Watanabe[212], N. Watanabe[71], T. Watanabe[155], Y. Watanabe[71], S. Watanuki[272], Y. Watase[71], N. K. Watson[315], G. Watts[383], M. M. Weber[90], H. C. Weddig[47], H. Weerts[7], A. W. Weidemann[251], G. Weiglein[47], A. Weiler[47], S. Weinzierl[138], H. Weise[47], A. Welker[138], N. Welle[47], J. D. Wells[33,348], M. Wendt[58,33], M. Wenskat[47], H. Wenzel[58], N. Wermes[295], U. Werthenbach[301], W. Wester[58], L. Weuste[186,57], A. White[373], G. White[251], K. H. Wichmann[47], M. Wielers[183], R. Wielgos[58], W. Wierba[202], T. Wilksen[47], S. Willocq[346], F. F. Wilson[259], G. W. Wilson[338], P. B. Wilson[251], M. Wing[309], M. Winter[94], K. Wittenburg[47], P. Wittich[43], M. Wobisch[181], A. Wolski[339,40], M. D. Woodley[251], M. B. Woods[251], M. Worek[385], S. Worm[33,259], G. Wormser[167], D. Wright[178], Z. Wu[251], C. E. Wulz[224], S. Xella[211], G. Xia[40,343], L. Xia[7], A. Xiao[7], L. Xiao[251], M. Xiao[102], Q. Xiao[102], J. Xie[7], C. Xu[102], F. Xu[210], G. Xu[102], K. Yagyu[201], U. A. Yajnik[85], V. Yakimenko[251], S. Yamada[71,116], S. Yamada[71], Y. Yamada[272], Y. Yamada[402], A. Yamaguchi[274], D. Yamaguchi[275], M. Yamaguchi[272], S. Yamaguchi[272], Y. Yamaguchi[375], Y. Yamaguchi[75], A. Yamamoto[71,375], H. Yamamoto[272], K. Yamamoto[222], K. Yamamoto[118], M. Yamamoto[71], N. Yamamoto[197], N. Yamamoto[71], Y. Yamamoto[71], Y. Yamamoto[375], T. Yamamura[375], T. Yamanaka[116], S. Yamashita[116], T. Yamashita[3], Y. Yamashita[214], K. Yamauchi[197], M. Yamauchi[71], T. Yamazaki[375], Y. Yamazaki[154], J. Yan[375,71], W. Yan[364], C. Yanagisawa[257,11], H. Yang[247], J. Yang[56], U. K. Yang[245,343], Z. Yang[24], W. Yao[177], S. Yashiro[71], F. Yasuda[375], O. Yasuda[276], I. Yavin[188,228], E. Yazgan[331], H. Yokoya[377], K. Yokoya[71], H. Yokoyama[375], S. Yokoyama[275], R. Yonamine[71], H. Yoneyama[240], M. Yoshida[71], T. Yoshida[62], K. Yoshihara[116,33], S. Yoshihara[116,33], M. Yoshioka[71,272], T. Yoshioka[161], H. Yoshitama[73], C. C. Young[251], H. B. Yu[348], J. Yu[373], C. Z. Yuan[102], F. Yuasa[71], J. Yue[102], A. Zabi[172], W. Zabolotny[401], J. Zacek[34], I. Zagorodnov[47], J. Zalesak[105,58], A. F. Zarnecki[381], L. Zawiejski[268], M. Zeinali[101], C. Zeitnitz[385], L. Zembala[401], K. Zenker[47], D. Zeppenfeld[91], D. Zerwas[167], P. Zerwas[47], M. Zeyrek[192], A. Zghiche[164], J. Zhai[102], C. Zhang[102], J. Zhang[102], J. Zhang[7], Y. Zhang[24,33], Z. Zhang[167], F. Zhao[102], F. Zhao[102], T. Zhao[102], Y. Zhao[251], H. Zheng[102], Z. Zhengguo[364], L. Zhong[24], F. Zhou[251], X. Zhou[364,102], Z. Zhou[102], R. Y. Zhu[17], X. Zhu[24], X. Zhu[102], M. Zimmer[47], F. Zomer[167], T. Zoufal[47], R. Zwicky[329] 1 Academia Sinica - 128 Sec. 2, Institute of Physics, Academia Rd., Nankang, Taipei 11529, Taiwan, R.O.C.

2 AGH University of Science and Technology, Akademia Gorniczo-Hutnicza im. Stanislawa Staszica w Krakowie, Al. Mickiewicza 30 PL-30-059 Cracow, Poland

3 Aichi Medical University, Nagakute, Aichi, 480-1195, Japan

4 Akita International University, Yuwa, Akita City, 010-1292, Japan

5 Albert-Ludwigs Universität Freiburg, Physikalisches Institut, Hermann-Herder Str. 3, D-79104 Freiburg, Germany

6 Ankara Üniversitesi Fen Fakültesi, Fizik Bölümü, Dögol Caddesi, 06100 Tandoĝan Ankara, Turkey

7 Argonne National Laboratory (ANL), 9700 S. Cass Avenue, Argonne, IL 60439, USA

8 Baylor University, Department of Physics, 101 Bagby Avenue, Waco, TX 76706, USA

9 Beijing University, Department of Physics, Beijing, China 100871

10 Benares Hindu University, Benares, Varanasi 221005, India

11 Borough of Manhattan Community College, The City University of New York, Department of Science, 199 Chambers Street, New York, NY 10007, USA

12 Brandenburg University of Technology, Postfach 101344, D-03013 Cottbus, Germany

13 Brookhaven National Laboratory (BNL), P.O.Box 5000, Upton, NY 11973-5000, USA

14 Brown University, Department of Physics, Box 1843, Providence, RI 02912, USA

15 Budker Institute for Nuclear Physics (BINP), 630090 Novosibirsk, Russia

16 Calcutta University, Department of Physics, 92 A.P.C. Road, Kolkata 700009, India

17 California Institute of Technology, Physics, Mathematics and Astronomy (PMA), 1200 East California Blvd, Pasadena, CA 91125, USA

18 California State University, Los Angeles, Dept. of Physics and Astronomy, 5151 State University Dr., Los Angeles, CA 90032, USA

19 Carleton University, Department of Physics, 1125 Colonel By Drive, Ottawa, Ontario, Canada K1S 5B6

20 Carnegie Mellon University, Department of Physics, Wean Hall 7235, Pittsburgh, PA 15213, USA

21 CEA Saclay, IRFU, F-91191 Gif-sur-Yvette, France

22 CEA Saclay, Service de Physique Théorique, CEA/DSM/SPhT, F-91191 Gif-sur-Yvette Cedex, France

23 Center for High Energy Physics (CHEP) / Kyungpook National University, 1370 Sankyuk-dong, Buk-gu, Daegu 702-701, Republic of Korea

24 Center for High Energy Physics (TUHEP), Tsinghua University, Beijing, China 100084

25 Center For Quantum Spacetime (CQUeST), Sogang University, 35 Baekbeom-ro, Mapo-gu, Seoul 121-742, Republic of Korea

26 Center for the Advancement of Natural Discoveries using Light Emission (CANDLE), Acharyan 31, 0040, Yerevan, Armenia

27 Centre de Physique des Particules de Marseille (CPPM), Aix-Marseille Université, CNRS/IN2P3, 163, Avenue de Luminy, Case 902, 13288 Marseille Cedex 09, France

28 Centre de Physique Theorique, CNRS - Luminy, Universiti d"Aix - Marseille II, Campus of Luminy, Case 907, 13288 Marseille Cedex 9, France

29 Centre Lasers Intenses et Applications (CELIA), Université Bordeaux 1 - CNRS - CEA, 351 Cours de la Libération, 33405 Talence Cedex, France

30 Centro Brasileiro de Pesquisas Físicas (CBPF), Rua Dr. Xavier Sigaud, n.150 22290-180, Urca - Rio de Janeiro, RJ, Brazil

31 Centro de Investigaciones Energéticas, Medioambientales y Tecnológicas, CIEMAT, Avenida Complutense 22, E-28040 Madrid, Spain

32 Centro Nacional de Microelectrónica (CNM), Instituto de Microelectrónica de Barcelona (IMB), Campus UAB, 08193 Cerdanyola del Vallès (Bellaterra), Barcelona, Spain

33 CERN, CH-1211 Genève 23, Switzerland

34 Charles University, Institute of Particle & Nuclear Physics, Faculty of Mathematics and Physics, V Holesovickach 2, CZ-18000 Prague 8, Czech Republic

35 Chiba University of Commerce, 1-3-1 Konodai, Ichikawa-shi, Chiba, 272-8512, Japan

36 Chonbuk National University, Division of Science Education, Jeonju 561-756, Republic of Korea

37 Chonbuk National University, Physics Department, Jeonju 561-756, Republic of Korea

38 Chubu University, 1200 Matsumoto-cho, Kasugai-shi, Aichi, 487-8501, Japan

39 Chung Yuan Christian University, Department of Physics, 200 Chung Pei Rd., Chung Li 32023 Taiwan, R.O.C

40 Cockcroft Institute, Daresbury, Warrington WA4 4AD, UK

41 College of William and Mary, Department of Physics, Williamsburg, VA, 23187, USA

42 Columbia University, Department of Physics, New York, NY 10027-6902, USA

43 Cornell University, Laboratory for Elementary-Particle Physics (LEPP), Ithaca, NY 14853, USA

44 Czech Technical University in Prague, Institute of Experimental and Applied Physics (IEAP), Horska 3a/22, 12800 Prague 2, Czech Republic

45 Czech Technical University, Faculty of Nuclear Science and Physical Engineering, Brehova 7, CZ-11519 Prague 1, Czech Republic

46 Departamento de Física Teórica, Facultad de Ciencias, Módulo 15 (antiguo C-XI) y Módulo 8, Universidad Autónoma de Madrid, Campus de Cantoblanco, 28049 Madrid, Spain 47 Deutsches Elektronen-Synchrotron DESY, A Research Centre of the Helmholtz Association, Notkestrasse 85, 22607 Hamburg, Germany (Hamburg site)

48 Deutsches Elektronen-Synchrotron DESY, A Research Centre of the Helmholtz Association, Platanenallee 6, 15738 Zeuthen, Germany (Zeuthen site)

49 Durham University, Department of Physics, Ogen Center for Fundamental Physics, South Rd., Durham DH1 3LE, UK

50 École Normale Supérieure de Lyon, 46 allée d'Italie, 69364 Lyon Cedex 07, France

51 Ege University, Department of Physics, Faculty of Science, 35100 Izmir, Turkey

52 Enrico Fermi Institute, University of Chicago, 5640 S. Ellis Avenue, RI-183, Chicago, IL 60637, USA

53 ETH Zürich, Institute for Particle Physics (IPP), Schafmattstrasse 20, CH-8093 Zürich, Switzerland

54 ETH Zürich, Institute for Theoretical Physics (ITP), Wolfgang-Pauli-Str. 27, Zürich, Switzerland

55 European Spallation Source ESS AB, Box 176, 221 00 Lund, Sweden

56 Ewha Womans University, 11-1 Daehyun-Dong, Seodaemun-Gu, Seoul, 120-750, Republic of Korea

57 Excellence Cluster Universe, Technische Universität München, Boltzmannstr. 2, 85748 Garching, Germany

58 Fermi National Accelerator Laboratory (FNAL), P.O.Box 500, Batavia, IL 60510-0500, USA

59 Florida State University, Department of Physics, 77 Chieftan Way, Tallahassee, FL 32306-4350, USA

60 Fujita Gakuen Health University, Department of Physics, Toyoake, Aichi 470-1192, Japan

61 Fukui University of Technology, 3-6-1 Gakuen, Fukui-shi, Fukui 910-8505, Japan

62 Fukui University, Department of Physics, 3-9-1 Bunkyo, Fukui-shi, Fukui 910-8507, Japan

63 Gangneung-Wonju National University, 210-702 Gangneung Daehangno, Gangneung City, Gangwon Province, Republic of Korea

64 Georg-August-Universität Göttingen, II. Physikalisches Institut, Friedrich-Hund-Platz 1, 37077 Göttingen, Germany

65 Global Design Effort

66 Hanyang University, Department of Physics, Seoul 133-791, Republic of Korea

67 Harish-Chandra Research Institute, Chhatnag Road, Jhusi, Allahabad 211019, India

68 Helmholtz-Zentrum Berlin für Materialien und Energie (HZB), Wilhelm-Conrad-Röntgen Campus, BESSY II, Albert-Einstein-Str. 15, 12489 Berlin, Germany

69 Helsinki Institute of Physics (HIP), P.O. Box 64, FIN-00014 University of Helsinki, Finland

70 Henan Normal University, College of Physics and Information Engineering, Xinxiang, China 453007

71 High Energy Accelerator Research Organization, KEK, 1-1 Oho, Tsukuba, Ibaraki 305-0801, Japan

72 Hiroshima University, Department of Physics, 1-3-1 Kagamiyama, Higashi-Hiroshima, Hiroshima 739-8526, Japan

73 Hiroshima University, Graduate School of Advanced Sciences of Matter, 1-3-1 Kagamiyama, Higashi-Hiroshima, Hiroshima 739-8530, Japan

74 Hokkai-Gakuen University, 4-1-40 Asahimachi, Toyohira-ku, Sapporo 062-8605, Japan

75 Hokkaido University, Department of Physics, Faculty of Science, Kita, Kita-ku, Sapporo-shi, Hokkaido 060-0810, Japan

76 Humboldt Universität zu Berlin, Fachbereich Physik, Institut für Elementarteilchenphysik, Newtonstr. 15, D-12489 Berlin, Germany

77 Hyogo University of Teacher Education, 942-1 Shimokume, Kato-city, Hyogo 673-1494, Japan

78 Ibaraki National College of Technology, 866 Nakane, Hitachinaka, Ibaraki 312-8508, Japan

79 Ibaraki University, College of Technology, Department of Physics, Nakanarusawa 4-12-1, Hitachi, Ibaraki 316-8511, Japan

80 Imperial College, Blackett Laboratory, Department of Physics, Prince Consort Road, London, SW7 2BW, UK

81 Indian Association for the Cultivation of Science, Department of Theoretical Physics and Centre for Theoretical Sciences, Kolkata 700032, India

82 Indian Institute of Science Education and Research (IISER) Kolkata, Department of Physical Sciences, Mohanpur Campus, PO Krishi Viswavidyalaya, Mohanpur 741252, Nadia, West Bengal, India

83 Indian Institute of Science, Centre for High Energy Physics, Bangalore 560012, Karnataka, India

84 Indian Institute of Technology Indore, IET Campus, M-Block, Institute of Engineering and Technology (IET), Devi Ahilya Vishwavidyalaya Campus, Khandwa Road, Indore - 452017, Madhya Pradesh, India

85 Indian Institute of Technology, Bombay, Powai, Mumbai 400076, India

86 Indian Institute of Technology, Guwahati, Guwahati, Assam 781039, India

87 Indiana University, Department of Physics, Swain Hall West 117, 727 E. 3rd St., Bloomington, IN 47405-7105, USA

88 Institucio Catalana de Recerca i Estudis, ICREA, Passeig Lluis Companys, 23, Barcelona 08010, Spain

89 Institut de Physique Nucléaire de Lyon (IPNL), Domaine scientifique de la Doua, Bâtiment Paul Dirac 4, rue Enrico Fermi, 69622 Villeurbanne, Cedex, France

90 Institut für Experimentelle Kernphysik, KIT,Universität Karlsruhe (TH), Wolfgang-Gaede-Str. 1, Postfach 6980, 76128 Karlsruhe, Germany

91 Institut für Theoretische Physik (ITP), Karlsruher Institut für Technologie (KIT), Fakultät für Physik, Postfach 6980, 76049 Karlsruhe, Germany

92 Institut für Theoretische Teilchenphysik, Campus Süd, Karlsruher Institut für Technologie (KIT), 76128 Karlsruhe, Germany

93 Institut National de Physique Nucleaire et de Physique des Particules, 3, Rue Michel- Ange, 75794 Paris Cedex 16, France

94 Institut Pluridisciplinaire Hubert Curien, 23 Rue du Loess - BP28, 67037 Strasbourg Cedex 2, France

95 Institute for Chemical Research, Kyoto University, Gokasho, Uji, Kyoto 611-0011, Japan

96 Institute for Cosmic Ray Research, University of Tokyo, 5-1-5 Kashiwa-no-Ha, Kashiwa, Chiba 277-8582, Japan

97 Institute for Mathematics, Astrophysics and Particle Physics (IMAPP), P.O. Box 9010, 6500 GL Nijmegen, Netherlands

98 Institute for Nuclear Research, Russian Academy of Sciences (INR RAS), 60-th October Anniversary Prospect 7a, 117312, Moscow, Russia

99 Institute for Particle and Nuclear Physics, Wigner Research Centre for Physics, Hungarian Academy of Sciences, P.O. Box 49, 1525 Budapest, Hungary

100 Institute for Scintillation Materials (ISMA), 60 Lenina Ave, 61001, Kharkiv, Ukraine

101 Institute for studies in fundamental sciences (IPM), Niavaran Square, P.O. Box 19395-5746, Tehran, Iran

102 Institute of High Energy Physics - IHEP, Chinese Academy of Sciences, P.O. Box 918, Beijing, China 100049

103 Institute of Mathematical Sciences, Taramani, C.I.T. Campus, Chennai 600113, India

104 Institute of Particle Physics, Canada

105 Institute of Physics, ASCR, Academy of Science of the Czech Republic, Division of Elementary Particle Physics, Na Slovance 2, CZ-18221 Prague 8, Czech Republic

106 Institute of Physics, Vietnam Academy of Science and Technology (VAST), 10 Dao-Tan, Ba-Dinh, Hanoi 10000, Vietnam

107 Institute of Theoretical and Experimetal Physics, B. Cheremushkinskawa, 25, RU-117259, Moscow, Russia

108 Instituto de Fisica Corpuscular (IFIC), Centro Mixto CSIC-UVEG, Edificio Investigacion Paterna, Apartado 22085, 46071 Valencia, Spain

109 Instituto de Física da Universidade Federal do Rio Grande do Sul (UFRGS), Av. Bento Gonçalves 9500, Caixa Postal 15051, CEP 91501-970, Porto Alegre, RS, Brazil

110 Instituto de Fisica de Cantabria, (IFCA, CSIC-UC), Facultad de Ciencias, Avda. Los Castros s/n, 39005 Santander, Spain

111 Instituto de Física Teórica UAM/CSIC, C/ Nicolás Cabrera 13-15, Universidad Autónoma de Madrid, Cantoblanco, 28049 Madrid, Spain

112 Instituto Galego de Fisica de Altas Enerxias (IGFAE,USC) Facultad de Fisica, Campus Sur E-15782 Santiago de Compostela, Spain

113 Instituto Tecnológico de Aragón (ITA), C/ María de Luna 7-8, 50018 Zaragoza, Spain

114 Instituto Universitario de Física Fundamental y Matemáticas de la Universidad de Salamanca (IUFFyM), Casas del Parque, 37008 Salamanca, Spain

115 Inter-University Accelerator Centre, Aruna Asaf Ali Marg, Post Box 10502, New Delhi 110067, India

116 International Center for Elementary Particle Physics, University of Tokyo, Hongo 7-3-1, Bunkyo District, Tokyo 113-0033, Japan

117 International Institute of Physics, Federal University of Rio Grande do Norte, Av. Odilon Gomes de Lima, 1722 - Capim Macio - 59078-400 - Natal-RN, Brazil

118 Iowa State University, Department of Physics, High Energy Physics Group, Ames, IA 50011, USA

119 Istituto Nazionale di Fisica Nucleare (INFN), Laboratorio LASA, Via Fratelli Cervi 201, 20090 Segrate, Italy

120 Istituto Nazionale di Fisica Nucleare (INFN), Sezione di Firenze, Via G. Sansone 1, I-50019 Sesto Fiorentino (Firenze), Italy

121 Istituto Nazionale di Fisica Nucleare (INFN), Sezione di Milano Bicocca, Piazza della Scienza 3, I-20126 Milano, Italy

122 Istituto Nazionale di Fisica Nucleare (INFN), Sezione di Milano, Via Celoria 16, I-20133 Milano, Italy

123 Istituto Nazionale di Fisica Nucleare (INFN), Sezione di Napoli, Complesso Universitá di Monte Sant'Angelo,via, I-80126 Naples, Italy

124 Istituto Nazionale di Fisica Nucleare (INFN), Sezione di Pavia, Via Bassi 6, I-27100 Pavia, Italy

125 Istituto Nazionale di Fisica Nucleare (INFN), Sezione di Perugia, Via A. Pascoli, 06123 Perugia, Italy

126 Istituto Nazionale di Fisica Nucleare (INFN), Sezione di Pisa, Edificio C - Polo Fibonacci Largo B. Pontecorvo, 3, I-56127 Pisa, Italy

127 Istituto Nazionale di Fisica Nucleare (INFN), Sezione di Roma, c/o Dipartimento di Fisica - Università degli Studi di Roma "La Sapienza", P.le Aldo Moro 2, I-00185 Roma, Italy

128 Istituto Nazionale di Fisica Nucleare (INFN), Sezione di Torino, c/o Universitá di Torino, facoltá di Fisica, via P Giuria 1, 10125 Torino, Italy

129 Istituto Nazionale di Fisica Nucleare (INFN), Sezione di Trieste, Padriciano 99, I-34012 Trieste (Padriciano), Italy

130 ITER Organization, Route de Vinon-sur-Verdon, 13115 St. Paul-lez-Durance, France

131 Iwate University, 4-3-5 Ueda, Morioka, Iwate, 020-8551, Japan

132 Jagiellonian University, Institute of Physics, Ul. Reymonta 4, PL-30-059 Cracow, Poland

133 Jamia Millia Islamia, Department of Physics, Jamia Nagar, New Delhi 110025, India

134 Japan Aerospace Exploration Agency, Sagamihara Campus, 3-1-1 Yoshinodai, Sagamihara, Kanagawa 220-8510 , Japan

135 Japan Atomic Energy Agency, 4-49 Muramatsu, Tokai-mura, Naka-gun, Ibaraki 319-1184, Japan

136 Japan Atomic Energy Agency, Tokai Research and Development Center, 2-4 Shirane Shirakata, Tokai-mura, Naka-gun, Ibaraki 319-1195, Japan

137 Japan Synchrotron Radiation Research Institute (JASRI), 1-1-1, Kouto, Sayo-cho, Sayo-gun, Hyogo 679-5198, Japan

138 Johannes Gutenberg Universität Mainz, Institut für Physik, 55099 Mainz, Germany

139 John Adams Institute for Accelerator Science at Royal Holloway University of London, Egham Hill, Egham, Surrey TW20 0EX, UK

140 John Adams Institute for Accelerator Science at University of Oxford, Denys Wilkinson Building, Keble Road, Oxford OX1 3RH, UK

141 Johns Hopkins University - Henry A. Rowland Department of Physics & Astronomy 3701 San Martin Drive, Baltimore, Maryland (MD) 21218, USA

142 Joint Institute for Nuclear Research (JINR), Joliot-Curie 6, 141980, Dubna, Moscow Region, Russia

143 Joint Institute for Power and Nuclear Research "Sosny" at National Academy of Sciences of Belarus, 99 Academician A.K.Krasin Str., Minsk BY-220109, Belarus

144 Jozef Stefan Institute, Jamova cesta 39, 1000 Ljubljana, Slovenia

145 Julius-Maximilians-Universität Würzburg, Fakultät für Physik und Astronomie, Am Hubland, 97074 Würzburg, Germany

146 Juntendo University, School of Medicine, Dept. of Physics, Hiraga-gakuendai 1-1, Inzai-shi, Chiba 270-1695, Japan

147 Justus-Liebig-Universität Gießen, II. Physikalisches Institut, Heinrich-Buff-Ring 16, 35392 Gießen, Germany

148 Kanazawa University, Institute for Theoretical Physics (KITP), School of Mathematics and Physics, College of Science and Engineering, Kakuma-machi, Kanazawa city, Ishikawa 920-1192, Japan

149 Kansas State University, Department of Physics, 116 Cardwell Hall, Manhattan, KS 66506, USA

150 Kavli Institute for the Physics and Mathematics of the Universe (Kavli IPMU), University of Tokyo, 5-1-5 Kashiwanoha, Kashiwa, 277-8583, Japan

151 King Saud University (KSU), Dept. of Physics, P.O. Box 2454, Riyadh 11451, Saudi Arabia

152 King's College London - Department of physics, Strand, London WC2R 2LS, London, UK

153 Kinki University, Department of Physics, 3-4-1 Kowakae, Higashi-Osaka, Osaka 577-8502, Japan

154 Kobe University, Department of Physics, 1-1 Rokkodai-cho, Nada-ku, Kobe, Hyogo 657-8501, Japan

155 Kogakuin University, Department of Physics, Shinjuku Campus, 1-24-2 Nishi-Shinjuku, Shinjuku-ku, Tokyo 163-8677, Japan

156 Konkuk University, 93-1 Mojin-dong, Kwanglin-gu, Seoul 143-701, Republic of Korea

157 Korea Advanced Institute of Science & Technology, Department of Physics, 373-1 Kusong-dong, Yusong-gu, Taejon 305-701, Republic of Korea

158 Korea Institute for Advanced Study (KIAS), School of Physics, 207-43 Cheongryangri-dong, Dongdaemun-gu, Seoul 130-012, Republic of Korea

159 Kyoto University, Department of Physics, Kitashirakawa-Oiwakecho, Sakyo-ku, Kyoto 606-8502, Japan

160 Kyoto University, Yukawa Institute for Theoretical Physics, Kitashirakawa-Oiwakecho, Sakyo-Ku, Kyoto 606-8502, Japan

161 Kyushu University, Department of Physics, 6-10-1 Hakozaki, Higashi-ku, Fukuoka 812-8581, Japan

162 L.P.T.A., UMR 5207 CNRS-UM2, Université Montpellier II, Case Courrier 070, Bât. 13, place Eugène Bataillon, 34095 Montpellier Cedex 5, France

163 Laboratoire Charles Coulomb UMR 5221 CNRS-UM2, Université Montpellier 2, Place Eugène Bataillon - CC069, 34095 Montpellier Cedex 5, France

164 Laboratoire d'Annecy-le-Vieux de Physique des Particules (LAPP) , Université de Savoie, CNRS/IN2P3, 9 Chemin de Bellevue, BP 110, F-74941 Annecy-Le-Vieux Cedex, France

165 Laboratoire d'Annecy-le-Vieux de Physique Theorique (LAPTH), Chemin de Bellevue, BP 110, F-74941 Annecy-le-Vieux Cedex, France

166 Laboratoire d'AstroParticules et Cosmologie (APC), Université Paris Diderot-Paris 7 - CNRS/IN2P3, Bâtiment Condorcet, Case 7020, 75205 Paris Cedex 13, France

167 Laboratoire de l'Accélérateur Linéaire (LAL), Université Paris-Sud 11, Bâtiment 200, 91898 Orsay, France

168 Laboratoire de Physique Corpusculaire de Clermont-Ferrand (LPC), Université Blaise Pascal, I.N.2.P.3./C.N.R.S., 24 avenue des Landais, 63177 Aubière Cedex, France

169 Laboratoire de Physique Nucléaire et des Hautes Energies (LPNHE), UPMC, UPD, IN2P3/CNRS, 4 Place Jussieu, 75005, Paris Cedex 05, France

170 Laboratoire de Physique Subatomique et de Cosmologie (LPSC), Université Joseph Fourier (Grenoble 1), CNRS/IN2P3, Institut Polytechnique de Grenoble, 53 rue des Martyrs, F-38026 Grenoble Cedex, France

171 Laboratoire de Physique Theorique, Université de Paris-Sud XI, Batiment 210, F-91405 Orsay Cedex, France

172 Laboratoire Leprince-Ringuet (LLR), École polytechnique – CNRS/IN2P3, Route de Saclay, F-91128 Palaiseau Cedex, France

173 Laboratoire Univers et Particules de Montpellier (LUPM) - UMR5299, Université de Montpellier II, Place Eugène Bataillon - Case courrier 72, 34095 Montpellier Cedex 05, France

174 Laboratori Nazionali di Frascati, via E. Fermi, 40, C.P. 13, I-00044 Frascati, Italy

175 Laboratório de Instrumentação e Física Experimental de Partículas (LIP LISBOA), Av. Elias Garcia 14 - 1°, 1000-149 Lisbon, Portugal

176 Lancaster University, Physics Department, Lancaster LA1 4YB, UK

177 Lawrence Berkeley National Laboratory (LBNL), 1 Cyclotron Rd, Berkeley, CA 94720, USA

178 Lawrence Livermore National Laboratory (LLNL), Livermore, CA 94551, USA

179 Lebedev Physical Institute, Leninsky Prospect 53, RU-117924 Moscow, Russia

180 Lomonosov Moscow State University, Skobeltsyn Institute of Nuclear Physics (MSU SINP), 1(2), Leninskie gory, GSP-1, Moscow 119991, Russia

181 Louisiana Tech University, Department of Physics, Ruston, LA 71272, USA

182 Ludwig-Maximilians-Universität München, Fakultät für Physik, Am Coulombwall 1, D - 85748 Garching, Germany

183 Lunds Universitet, Fysiska Institutionen, Avdelningen för Experimentell Högenergifysik, Box 118, 221 00 Lund, Sweden

184 L'Université Hassan II, Aïn Chock, "Réseau Universitaire de Physique des Hautes Energies" (RUPHE), Département de Physique, Faculté des Sciences Aïn Chock, B.P 5366 Maarif, Casablanca 20100, Morocco

185 Massachusetts Institute of Technology (MIT), Laboratory for Nuclear Science, 77 Massachusetts Avenue, Cambridge, MA 02139, USA

186 Max-Planck-Institut für Physik (Werner-Heisenberg-Institut), Föhringer Ring 6, 80805 München, Germany

187 McGill University, Department of Physics, Ernest Rutherford Physics Bldg., 3600 University Street, Montreal, Quebec, H3A 2T8 Canada

188 McMaster University, Department of Physics & Astronomy, 1280 Main Street West, Hamilton, ON, L8S 4M1, Canada

189 Meiji Gakuin University, Department of Physics, 2-37 Shirokanedai 1-chome, Minato-ku, Tokyo 244-8539, Japan

190 Michigan State University, Department of Chemical Engineering & Materials Science, 2527 Engineering Building East Lansing, MI 48824-1226, USA

191 Michigan State University, Department of Physics and Astronomy, East Lansing, MI 48824, USA

192 Middle East Technical University, Department of Physics, TR-06531 Ankara, Turkey

193 Miyagi Gakuin Women's University, Faculty of Liberal Arts, 9-1-1 Sakuragaoka, Aoba District, Sendai, Miyagi 981-8557, Japan

194 MSU-Iligan Institute of Technology, Department of Physics, Andres Bonifacio Avenue, 9200 Iligan City, Phillipines

195 Nagasaki Institute of Applied Science, 536 Abamachi, Nagasaki-Shi, Nagasaki 851-0193, Japan

196 Nagoya University, Department of Materials Science and Engineering, Furo-cho, Chikusa-ku, Nagoya, 464-8603, Japan

197 Nagoya University, Department of Physics, School of Science, Furo-cho, Chikusa-ku, Nagoya, Aichi 464-8602, Japan

198 Nagoya University, Kobayashi-Maskawa Institute for the Origin of Particles and the Universe (KMI), Furo-cho, Chikusa-ku, Nagoya Aichi 464-8602, Japan

199 Nanjing University, Department of Physics, Nanjing, China 210093

200 Nara Women's University, High Energy Physics Group, Kitauoya-Nishimachi, Nara 630-8506, Japan

201 National Central University, High Energy Group, Department of Physics, Chung-li, Taiwan 32001, R.O.C

202 National Centre of Nuclear Research (NCBJ), ul. Andrzeja Soltana 7, 05-400 Otwock-Swierk, Poland

203 National Cheng Kung University, Physics Department, 1 Ta-Hsueh Road, Tainan, Taiwan 70101, R.O.C

204 National Chiao-Tung University, Institute of Physics, 1001 Ta Hsueh Rd, Hsinchu, Taiwan 300, R.O.C.

205 National Institute of Chemical Physics and Biophysics (NICPB), Ravala pst 10, 10143 Tallinn, Estonia

206 National Institute of Physics and Nuclear Engineering "Horia Hulubei" (IFIN-HH), Str. Reactorului no.30, P.O. Box MG-6, R-76900 Bucharest - Magurele, Romania

207 National Research Centre "Kurchatov Institute", 1 Akademika Kurchatova pl., Moscow, 123182, Russia

208 National Science Center - Kharkov Institute of Physics and Technology (NSC KIPT), Akademicheskaya St. 1, Kharkov, 61108, Ukraine

209 National Scientific & Educational Centre of Particle & High Energy Physics (NCPHEP), Belarusian State University, M.Bogdanovich street 153, 220040 Minsk, Belarus

210 National Taiwan University, Physics Department, Taipei, Taiwan 106, R.O.C

211 Niels Bohr Institute (NBI), University of Copenhagen, Blegdamsvej 17, DK-2100 Copenhagen, Denmark

212 Niigata University, Department of Physics, Ikarashi, Niigata 950-218, Japan

213 Nikhef, National Institute for Subatomic Physics, P.O. Box 41882, 1009 DB Amsterdam, Netherlands

214 Nippon Dental University School of Life Dentistry at Niigata, 1-8 Hamaura-cho, Chuo-ku, Niigata 951-1500, Japan

215 North Carolina A&T State University, 1601 E. Market Street, Greensboro, NC 27411, USA

216 Northeastern University, Physics Department, 360 Huntington Ave, 111 Dana Research Center, Boston, MA 02115, USA

217 Northern Illinois University, Department of Physics, DeKalb, Illinois 60115-2825, USA

218 Northwestern University, Department of Physics and Astronomy, 2145 Sheridan Road., Evanston, IL 60208, USA

219 Novosibirsk State University (NGU), Department of Physics, Pirogov st. 2, 630090 Novosibirsk, Russia

220 Ochanomizu University, Department of Physics, Faculty of Science, 1-1 Otsuka 2, Bunkyo-ku, Tokyo 112-8610, Japan

221 Orissa University of Agriculture & Technology, Bhubaneswar 751003, Orissa, India

222 Osaka City University, Department of Physics, Faculty of Science, 3-3-138 Sugimoto, Sumiyoshi-ku, Osaka 558-8585, Japan

223 Osaka University, Department of Physics, 1-1 Machikaneyama, Toyonaka, Osaka 560-0043, Japan

224 Österreichische Akademie der Wissenschaften, Institut für Hochenergiephysik, Nikolsdorfergasse 18, A-1050 Vienna, Austria

225 Pacific Northwest National Laboratory, (PNNL), PO Box 999, Richland, WA 99352, USA

226 Panjab University, Chandigarh 160014, India

227 Pavel Sukhoi Gomel State Technical University, ICTP Affiliated Centre & Laboratory for Physical Studies, October Avenue, 48, 246746, Gomel, Belarus

228 Perimeter Institute for Theoretical Physics, 31 Caroline Street North, Waterloo, Ontario N2L 2Y5, Canada

229 Physical Research Laboratory, Navrangpura, Ahmedabad 380 009, Gujarat, India

230 Pohang Accelerator Laboratory (PAL), San-31 Hyoja-dong, Nam-gu, Pohang, Gyeongbuk 790-784, Republic of Korea

231 Pontificia Universidad Católica de Chile, Avda. Libertador Bernardo OHiggins 340, Santiago, Chile

232 Princeton University, Department of Physics, P.O. Box 708, Princeton, NJ 08542-0708, USA

233 Purdue University, Department of Physics, West Lafayette, IN 47907, USA

234 Queen Mary, University of London, Mile End Road, London, E1 4NS, United Kingdom

235 Rheinisch-Westfälische Technische Hochschule (RWTH), Physikalisches Institut, Physikzentrum, Otto-Blumenthal-Straße, 52056 Aachen

236 RIKEN, 2-1 Hirosawa, Wako, Saitama 351-0198, Japan

237 Royal Holloway, University of London (RHUL), Department of Physics, Egham, Surrey TW20 0EX, UK

238 Russian Academy of Science, Keldysh Institute of Applied Mathematics, Muiskaya pl. 4, 125047 Moscow, Russia

239 Rutgers, The State University of New Jersey, Department of Physics & Astronomy, 136 Frelinghuysen Rd, Piscataway, NJ 08854, USA

240 Saga University, Department of Physics, 1 Honjo-machi, Saga-shi, Saga 840-8502, Japan

241 Saha Institute of Nuclear Physics, 1/AF Bidhan Nagar, Kolkata 700064, India

242 Salalah College of Technology (SCOT), Engineering Department, Post Box No. 608, Postal Code 211, Salalah, Sultanate of Oman

243 Saudi Center for Theoretical Physics, King Fahd University of Petroleum and Minerals (KFUPM), Dhahran 31261, Saudi Arabia

244 Seikei University, Faculty of Science and Technology, 3-3-1 Kichijoji-Kitamachi, Musashino-shi, Tokyo 180-8633, Japan

245 Seoul National University, San 56-1, Shinrim-dong, Kwanak-gu, Seoul 151-742, Republic of Korea

246 Setsunan University, Institute for Fundamental Sciences, 17-8 Ikeda Nakamachi, Neyagawa, Osaka, 572-8508, Japan

247 Shanghai Jiao Tong University, Department of Physics, 800 Dongchuan Road, Shanghai, China 200240

248 Shinshu University, 3-1-1, Asahi, Matsumoto, Nagano 390-8621, Japan

249 Shiv Nadar University, Village Chithera, Tehsil Dadri, District Gautam Budh Nagar, 203207 Uttar Pradesh, India

250 Shizuoka University, Department of Physics, 836 Ohya, Suruga-ku, Shizuoka 422-8529, Japan

251 SLAC National Accelerator Laboratory, 2575 Sand Hill Road, Menlo Park, CA 94025, USA

252 Society for Applied Microwave Electronics Engineering and Research (SAMEER), I.I.T. Campus, Powai, Post Box 8448, Mumbai 400076, India

253 Sokendai, The Graduate University for Advanced Studies, Shonan Village, Hayama, Kanagawa 240-0193, Japan

254 Spallation Neutron Source (SNS), Oak Ridge National Laboratory (ORNL), P.O. Box 2008 MS-6477, Oak Ridge, TN 37831-6477, USA

255 State University of New York at Binghamton, Department of Physics, PO Box 6016, Binghamton, NY 13902, USA

256 State University of New York at Buffalo, Department of Physics & Astronomy, 239 Franczak Hall, Buffalo, NY 14260, USA

257 State University of New York at Stony Brook, Department of Physics and Astronomy, Stony Brook, NY 11794-3800, USA

258 STFC Daresbury Laboratory, Daresbury, Warrington, Cheshire WA4 4AD, UK

259 STFC Rutherford Appleton Laboratory, Chilton, Didcot, Oxon OX11 0QX, UK

260 Sungkyunkwan University (SKKU), Natural Science Campus 300, Physics Research Division, Chunchun-dong, Jangan-gu, Suwon, Kyunggi-do 440-746, Republic of Korea

261 Swiss Light Source (SLS), Paul Scherrer Institut (PSI), PSI West, CH-5232 Villigen PSI, Switzerland

262 Tata Institute of Fundamental Research, School of Natural Sciences, Homi Bhabha Rd., Mumbai 400005, India

263 Technical University of Lodz, Department of Microelectronics and Computer Science, al. Politechniki 11, 90-924 Lodz, Poland

264 Technische Universität Dresden, Institut für Kern- und Teilchenphysik, D-01069 Dresden, Germany

265 Tel-Aviv University, School of Physics and Astronomy, Ramat Aviv, Tel Aviv 69978, Israel

266 Texas A&M University, Physics Department, College Station, 77843-4242 TX, USA

267 Texas Tech University, Department of Physics, Campus Box 41051, Lubbock, TX 79409-1051, USA

268 The Henryk Niewodniczanski Institute of Nuclear Physics, Polish Academy of Sciences (IFJ PAN), ul. Radzikowskiego 152, PL-31342 Cracow, Poland

269 Thomas Jefferson National Accelerator Facility (TJNAF), 12000 Jefferson Avenue, Newport News, VA 23606, USA

270 Tohoku Gakuin University, Department of Business Administration, 1-3-1 Tsuchitoi, Aoba-ku Sendai, Miyagi 980-8511, Japan

271 Tohoku Gakuin University, Faculty of Technology, 1-13-1 Chuo, Tagajo, Miyagi 985-8537, Japan

272 Tohoku University, Department of Physics, Aoba District, Sendai, Miyagi 980-8578, Japan

273 Tohoku University, Research Center for Electron Photon Science, Taihaku District, Sendai, Miyagi 982-0826, Japan

274 Tohoku University, Research Center for Neutrino Science, Aoba District, Sendai, Miyagi 980-8578, Japan

275 Tokyo Institute of Technology, Department of Physics, 2-12-1 O-Okayama, Meguro, Tokyo 152-8551, Japan

276 Tokyo Metropolitan University, Faculty of Science and Engineering, Department of Physics, 1-1 Minami-Osawa, Hachioji-shi, Tokyo 192-0397, Japan

277 Tokyo University of Agriculture Technology, Department of Applied Physics, Naka-machi, Koganei, Tokyo 183-8488, Japan

278 Toyama Prefectural University, Department of Mathematical Physics, 5180 Kurokawa Imizu-shi, Toyama, 939-0398, Japan

279 TRIUMF, 4004 Wesbrook Mall, Vancouver, BC V6T 2A3, Canada

280 Universidad Autónoma de San Luis Potosí, Alvaro Obregon 64, Col. Centro, San Luis Potosí, S.L.P. 78000, México
281 Universidad de Granada, Departamento de Física Teórica y del Cosmos, Campus de Fuentenueva, E-18071 Granada, Spain
282 Universidad de los Andes, Faculty of Science, Department of Physics, Carrera 1 18A-10, Bloque Ip. Bogotá, Colombia
283 Universidad de Oviedo, Departamento de Física, Campus de Llamaquique. C/ Calvo Sotelo, s/n 33005 Oviedo, Spain
284 Universidad de Salamanca, Departamento de Física Fundamental, Plaza de la Merced, s/n., 37008 Salamanca, Spain
285 Universidad de Sevilla, Escuela Técnica Superior de Ingeniería, Departamento Ingeniería Electrónica, Camino de los Descubrimientos s/n, 41092 Sevilla, Spain
286 Universidad de Zaragoza - Departamento de Física Teórica, Pedro Cerbuna 12, E-50009 Zaragoza, Spain
287 Universidad Nacional Autónoma de México, Instituto de Física, Circuito de la Investigación Cientifica s/n, Ciudad Universitaria, CP 04510 México D.F., Mexico
288 Universidad Nacional de La Plata, Departamento de Física, Facultad de Ciencias Exactas, C.C. N 67, 1900 La Plata, Argentina
289 Universidade do Estado do Rio de Janeiro (UERJ), Rio de Janeiro, RJ - Brasil 20550-900, Brazil
290 Universidade Federal de Pelotas, Instituto de Física e Matemática, Campus Universitário, Caixa Postal 354, 96010-900 Pelotas, RS, Brazil
291 Universidade Federal do Rio de Janeiro (UFRJ), Instituto de Física, Av. Athos da Silveira Ramos 149, Centro de Tecnologia - Bloco A, Cidade Universitária, Ilha do Fundão, Rio de Janeiro, RJ, Brazil
292 Universitá degli Studi di Napoli “Federico II”, Dipartimento di Fisica, Via Cintia, 80126 Napoli, Italy
293 Universitat Autònoma de Barcelona, Departament de Física, Edifici C, 08193 Bellaterra, Barcelona, Spain
294 Universitat Autònoma de Barcelona, Institut de Fisica d’Altes Energies (IFAE), Campus UAB, Edifici Cn, E-08193 Bellaterra, Barcelona, Spain
295 Universität Bonn, Physikalisches Institut, Nußallee 12, 53115 Bonn, Germany
296 Universität Heidelberg, Institut für Theoretische Physik, Philosophenweg 16, 69120 Heidelberg, Germany
297 Universität Heidelberg, Kirchhoff-Institut für Physik, Im Neuenheimer Feld 227, 69120 Heidelberg, Germany
298 Universitat Politècnica de Catalunya, Institut de Tècniques Energètiques, Campus Diagonal Sud, Edifici PC (Pavelló C). Av. Diagonal, 647 08028 Barcelona, Spain
299 Universitat Ramon Llull, La Salle, C/ Quatre Camins 2, 08022 Barcelona, Spain
300 Universität Rostock, 18051 Rostock, Germany
301 Universität Siegen, Naturwissenschaftlich-Technische Fakultät, Department Physik, Emmy Noether Campus, Walter-Flex-Str.3, 57068 Siegen, Germany
302 Universität Wien - Theoretische Physik Boltzmanngasse 5, A-1090 Vienna, Austria
303 Université catholique de Louvain, Centre for Cosmology, Particle Physics and Phenomenology (CP3), Institute of Mathematics and Physics, 2 Chemin du Cyclotron, 1348 Louvain-la-Neuve, Belgium
304 Université de Genève, Section de Physique, 24, quai E. Ansermet, 1211 Genève 4, Switzerland
305 Université de Montréal, Département de Physique, Groupe de Physique des Particules, C.P. 6128, Succ. Centre-ville, Montréal, Qc H3C 3J7, Canada
306 Université de Strasbourg, UFR de Sciences Physiques, 3-5 Rue de l’Université, F-67084 Strasbourg Cedex, France
307 Université Libre de Bruxelles, Boulevard du Triomphe, 1050 Bruxelles, Belgium
308 Universittà di Catania, Dipartimento di Fisica e Astronomia, Via Santa Sofia 64, 95123 Catania, Italy
309 University College London (UCL), High Energy Physics Group, Physics and Astronomy Department, Gower Street, London WC1E 6BT, UK
310 University College, National University of Ireland (Dublin), Department of Experimental Physics, Science Buildings, Belfield, Dublin 4, Ireland
311 University de Barcelona, Facultat de Física, Av. Diagonal, 647, Barcelona 08028, Spain
312 University of Alberta - Faculty of Science, Department of Physics, 4-181 CCIS, Edmonton AB T6G 2E1, Canada
313 University of Arizona, Department of Physics, 1118 E. Fourth Street, PO Box 210081, Tucson, AZ 85721, USA
314 University of Bergen, Institute of Physics, Allegaten 55, N-5007 Bergen, Norway
315 University of Birmingham, School of Physics and Astronomy, Particle Physics Group, Edgbaston, Birmingham B15 2TT, UK
316 University of Bristol, H. H. Wills Physics Lab, Tyndall Ave., Bristol BS8 1TL, UK
317 University of British Columbia, Department of Physics and Astronomy, 6224 Agricultural Rd., Vancouver, BC V6T 1Z1, Canada
318 University of California (UCLA), Los Angleles, CA 90095, US
319 University of California Berkeley, Department of Physics, 366 Le Conte Hall, #7300, Berkeley, CA 94720, USA
320 University of California Davis, Department of Physics, One Shields Avenue, Davis, CA 95616-8677, USA
321 University of California Irvine, Department of Physics and Astronomy, High Energy Group, 4129 Frederick Reines Hall, Irvine, CA 92697-4575 USA
322 University of California Santa Cruz, Institute for Particle Physics, 1156 High Street, Santa Cruz, CA 95064, USA
323 University of Cambridge, Cavendish Laboratory, J J Thomson Avenue, Cambridge CB3 0HE, UK
324 University of Colorado at Boulder, Department of Physics, 390 UCB, University of Colorado, Boulder, CO 80309-0390, USA
325 University of Cyprus, Department of Physics, P.O.Box 20537, 1678 Nicosia, Cyprus

326 University of Delhi, Department of Physics and Astrophysics, Delhi 110007, India
327 University of Delhi, S.G.T.B. Khalsa College, Delhi 110007, India
328 University of Dundee, Department of Physics, Nethergate, Dundee, DD1 4HN, Scotland, UK
329 University of Edinburgh, School of Physics, James Clerk Maxwell Building, The King's Buildings, Mayfield Road, Edinburgh EH9 3JZ, UK
330 University of Florida, Department of Physics, Gainesville, FL 32611, USA
331 University of Ghent, Department of Subatomic and Radiation Physics, Proeftuinstraat 86, 9000 Gent, Belgium
332 University of Glasgow, SUPA, School of Physics & Astronomy, University Avenue, Glasgow G12 8QQ, Scotland, UK
333 University of Hamburg, Physics Department, Luruper Chaussee 149, 22761 Hamburg, Germany
334 University of Hawaii, Department of Physics and Astronomy, HEP, 2505 Correa Rd., WAT 232, Honolulu, HI 96822-2219, USA
335 University of Helsinki, Department of Physical Sciences, P.O. Box 64 (Vaino Auerin katu 11), FIN-00014, Helsinki, Finland
336 University of Illinois at Chicago, Department Of Physics, 845 W Taylor St., Chicago IL 60607, USA
337 University of Iowa, Department of Physics and Astronomy, 203 Van Allen Hall, Iowa City, IA 52242-1479, USA
338 University of Kansas, Department of Physics and Astronomy, Malott Hall, 1251 Wescoe Hall Drive, Room 1082, Lawrence, KS 66045-7582, USA
339 University of Liverpool, Department of Physics, Oliver Lodge Lab, Oxford St., Liverpool L69 7ZE, UK
340 University of Liverpool, Division of Theoretical Physics, Department of Mathematical Sciences, Chadwick Building, Liverpool L69 3BX, UK
341 University of Ljubljana, Faculty of Mathematics and Physics, Jadranska ulica 19, 1000 Ljubljana, Slovenia
342 University of Malaya, Faculty of Science, Department of Physics, 50603 Kuala Lumpur, Malaysia
343 University of Manchester, School of Physics and Astronomy, Schuster Lab, Manchester M13 9PL, UK
344 University of Maribor, Faculty of Chemistry and Chemical Engineering (FKKT), Smetanova ulica 17, 2000 Maribor, Slovenia
345 University of Maryland, Department of Physics and Astronomy, Physics Building (Bldg. 082), College Park, MD 20742, USA
346 University of Massachusetts - Amherst, Department of Physics, 1126 Lederle Graduate Research Tower (LGRT), Amherst, MA 01003-9337, USA
347 University of Melbourne, School of Physics, Victoria 3010, Australia
348 University of Michigan, Department of Physics, 500 E. University Ave., Ann Arbor, MI 48109-1120, USA
349 University of Minnesota, 148 Tate Laboratory Of Physics, 116 Church St. S.E., Minneapolis, MN 55455, USA
350 University of Mississippi, Department of Physics and Astronomy, 108 Lewis Hall, PO Box 1848, Oxford, Mississippi 38677-1848, USA
351 University of Missouri – St. Louis, Department of Physics and Astronomy, 503 Benton Hall One University Blvd., St. Louis Mo 63121, USA
352 University of New Mexico, New Mexico Center for Particle Physics, Department of Physics and Astronomy, 800 Yale Boulevard N.E., Albuquerque, NM 87131, USA
353 University of North Carolina at Chapel Hill, Department of Physics and Astronomy, Phillips Hall, CB #3255, 120 E. Cameron Ave., Chapel Hill, NC 27599-3255, USA
354 University of Notre Dame, Department of Physics, 225 Nieuwland Science Hall, Notre Dame, IN 46556, USA
355 University of Oklahoma, Department of Physics and Astronomy, Norman, OK 73071, USA
356 University of Oregon, Department of Physics, 1371 E. 13th Ave., Eugene, OR 97403, USA
357 University of Oslo, Department of Physics, P.O box 1048, Blindern, 0316 Oslo, Norway
358 University of Oxford, Particle Physics Department, Denys Wilkinson Bldg., Keble Road, Oxford OX1 3RH England, UK
359 University of Pavia, Department of Physics, via Bassi 6, I-27100 Pavia, Italy
360 University of Pennsylvania, Department of Physics and Astronomy, 209 South 33rd Street, Philadelphia, PA 19104-6396, USA
361 University of Pittsburgh, Department of Physics and Astronomy, 100 Allen Hall, 3941 O'Hara St, Pittsburgh PA 15260, USA
362 University of Regina, Department of Physics, Regina, Saskatchewan, S4S 0A2 Canada
363 University of Rochester, Department of Physics and Astronomy, Bausch & Lomb Hall, P.O. Box 270171, 600 Wilson Boulevard, Rochester, NY 14627-0171 USA
364 University of Science and Technology of China, Department of Modern Physics (DMP), Jin Zhai Road 96, Hefei, China 230026
365 University of Silesia, Institute of Physics, Ul. Uniwersytecka 4, PL-40007 Katowice, Poland
366 University of South Carolina, Department of Physics and Astronomy, 712 Main Street, Columbia, SC 29208, USA
367 University of Southampton, School of Physics and Astronomy, Highfield, Southampton S017 1BJ, England, UK
368 University of Southern California, Department of Physics & Astronomy, 3620 McClintock Ave., SGM 408, Los Angeles, CA 90089-0484, USA
369 University of Sydney, Falkiner High Energy Physics Group, School of Physics, A28, Sydney, NSW 2006, Australia
370 University of Tartu, Institute of Physics, Riia 142, 51014 Tartu, Estonia
371 University of Texas at Austin, Department of Physics, 1 University Station C1600, Austin, Texas 78712, USA
372 University of Texas at Dallas, Department of Physics, 800 West Campbell Road, Richardson, Texas 75080, USA

373 University of Texas, Center for Accelerator Science and Technology, Arlington, TX 76019, USA

374 University of Tokushima, Institute of Theoretical Physics, Tokushima-shi 770-8502, Japan

375 University of Tokyo, Department of Physics, 7-3-1 Hongo, Bunkyo District, Tokyo 113-0033, Japan

376 University of Toronto, Department of Physics, 60 St. George St., Toronto M5S 1A7, Ontario, Canada

377 University of Toyama, Department of Physics, 3190 Gofuku, Toyama 930-8555, Japan

378 University of Tsukuba, Faculty of Pure and Applied Sciences, 1-1-1 Ten'nodai, Tsukuba, Ibaraki 305-8571, Japan

379 University of Victoria, Department of Physics and Astronomy, P.O.Box 3055 Stn Csc, Victoria, BC V8W 3P6, Canada

380 University of Virginia, Department of Physics, 382 McCormick Rd., PO Box 400714, Charlottesville, VA

381 University of Warsaw, Institute of Experimental Physics, Ul. Hoza 69, PL-00 681 Warsaw, Poland

382 University of Warsaw, Institute of Theoretical Physics, Ul. Hoza 69, PL-00 681 Warsaw, Poland

383 University of Washington, Department of Physics, PO Box 351560, Seattle, WA 98195-1560, USA

384 University of Wisconsin, Physics Department, Madison, WI 53706-1390, USA

385 University of Wuppertal, Gaußstraße 20, D-42119 Wuppertal, Germany

386 Università degli Studi di Milano, Dipartimento di Fisica, Via Celoria 16, 20133 Milano, Italy

387 Università degli Studi di Roma "La Sapienza", Dipartimento di Fisica, Piazzale Aldo Moro 5, 00185 Roma, Italy

388 Università degli Studi di Trieste, Dipartimento di Fisica, via A. Valerio 2, I-34127 Trieste, Italy

389 Università dell'Insubria in Como, Dipartimento di Scienze CC.FF.MM., via Vallegio 11, I-22100 Como, Italy

390 Università di Milano-Bicocca, Dipartimento di Fisica"G. Occhialin", Piazza della Scienza 3, 20126 Milano, Italy

391 Università di Pisa, Departimento di Fisica "Enrico Fermi", Largo Bruno Pontecorvo 3, I-56127 Pisa, Italy

392 Universiy of Huddersfield, International Institute for Accelerator Applications, Queensgate Campus, Huddersfield HD1 3DH, UK

393 UPMC Univ. Paris 06, Faculté de Physique (UFR 925), 4 Place Jussieu, 75252 Paris Cedex 05, France

394 Vietnam National University, Laboratory of High Energy Physics and Cosmology, Faculty of Physics, College of Science, 334 Nguyen Trai, Hanoi, Vietnam

395 Vietnam National University, University of Natural Sciences, 227 Nguyen Van Cu street, District 5, Ho Chi Minh City, Vietnam

396 VINCA Institute of Nuclear Sciences, Laboratory of Physics, PO Box 522, YU-11001 Belgrade, Serbia

397 Virginia Commonwealth University, Department of Physics, P.O. Box 842000, 701 W. Grace St.,Richmond, VA. 23284-2000, USA

398 Virginia Polytechnic Institute and State University, Physics Department, Blacksburg, VA 2406, USA

399 Vrije Universiteit Brussel, Pleinlaan 2, 1050 Brussels, Belgium

400 Vrije Universiteit, Department of Physics, Faculty of Sciences, De Boelelaan 1081, 1081 HV Amsterdam, Netherlands

401 Warsaw University of Technology, The Faculty of Electronics and Information Technology, ul. Nowowiejska 15-19, 00-665 Warsaw, Poland

402 Waseda University, Advanced Research Institute for Science and Engineering, Shinjuku, Tokyo 169-8555, Japan

403 Wayne State University, Department of Physics, Detroit, MI 48202, USA

404 Weizmann Institute of Science, Department of Particle Physics, P.O. Box 26, Rehovot 76100, Israel

405 Yale University, Department of Physics, New Haven, CT 06520, USA

406 Yamagata University, 1-4-12 Kojirakawa-cho, Yamagata-shi, Yamagata, 990-8560, Japan

407 Yerevan Physics Institute, 2 Alikhanyan Brothers St., Yerevan 375036, Armenia

408 Yonsei University, Department of Physics, 134 Sinchon-dong, Sudaemoon-gu, Seoul 120-749, Republic of Korea

ISSUED BY international linear collider

WWW.LINEARCOLLIDER.ORG